%% file: main.tex
\documentclass[a4paper,12pt]{book}

\usepackage[bottom=2.8cm, top=2.5cm, inner=2.0cm, outer=2cm, headheight=14.5pt]{geometry}

\usepackage{graphicx}	  		     
\usepackage{amsfonts,amssymb,amstext,amsmath,amsthm,verbatim,cancel,epsfig,epigraph,dirtytalk,physics} 
\usepackage[british]{babel}
\usepackage[utf8]{inputenc}
\usepackage[T1]{fontenc} 
\usepackage{fancyhdr}
\usepackage[toc,page]{appendix}
\usepackage{tcolorbox}
\tcbuselibrary{listings,theorems,skins,xparse,breakable}

\newtcbox{\compbox}[1]{colback=white,colbacklower=white,colbacklower=white,colframe=yellow!10!green!5!brown,fonttitle=\bfseries,title=#1}
\newcounter{computation}
\newcommand{\comp}{%
  \refstepcounter{computation}%
  \textbf{Computation~\thecomputation :}\,
}

\newtcbox{\remarkbox}[1]{colback=green!5!blue!5!white,colbacklower=white,colframe=green!15!blue!80!yellow,fonttitle=\bfseries,title=#1}
\newcounter{remark}
\newcommand{\rem}{%
  \refstepcounter{remark}%
  \textbf{Remark~\theremark :}\,
}

\newtcbox{\resultbox}[1]{colback=red!10!white,colbacklower=white,colframe=yellow!5!green!5!red,fonttitle=\bfseries,title=#1}
\newcounter{resultmark}
\newcommand{\result}{%
  \refstepcounter{resultmark}%
  \textbf{Result~\theresultmark :}\,
}

\newtcbox{\definitionbox}[1]{colback=violet!4!white,colbacklower=white,colframe=violet!65!black,fonttitle=\bfseries,title=#1}
\newcounter{definitionmark}
\newcommand{\defin}{%
  \refstepcounter{definitionmark}%
  \textbf{Definition~\thedefinitionmark :}\,
}

\usepackage{mathtools,tensor}
\usepackage{mathrsfs} 
\usepackage{setspace}
\usepackage[english]{varioref}
\usepackage[hidelinks]{hyperref}
\DeclareGraphicsExtensions{.eps}
\usepackage{float}
\usepackage{slashed}
\usepackage{caption}
\usepackage{comment}
\usepackage{subcaption}
\usepackage{tikz,pgffor}
\usetikzlibrary{positioning,shadows,shapes,arrows,decorations.pathmorphing,backgrounds,automata}
\usetikzlibrary{decorations.pathreplacing,decorations.markings}
\usetikzlibrary{patterns}
\usetikzlibrary{calc}
\tikzset{>=latex}

\usepackage{wrapfig,framed,caption}

\usepackage[autostyle=false,style=english]{csquotes} 
\usepackage[backend=biber,style=numeric-comp,doi=true,url=true,sorting=none,sortcites=true]{biblatex}
\usepackage{ragged2e}

\usepackage{multirow}
\usepackage{stackengine}

\usepackage[normalem]{ulem}

\usepackage{titlesec}

\titleformat{\chapter}[display]
  {\normalfont\Large\bfseries}
  {\textsc{Chapter \thechapter}} 
  {1ex}                          
  {\Huge}                       

\usepackage{etoolbox}
\patchcmd{\tableofcontents}
  {\contentsname}
  {\textsc{\contentsname}}
  {}{}

\DefineBibliographyStrings{english}{
  bibliography = {\textsc{Bibliography}},
}

\graphicspath{{./Figures/}{./Figures/Pictures/}}

\renewcommand*\contentsname{Table of Contents}

\newcommand{\be}{\begin{equation}}
\newcommand{\ee}{\end{equation}}

\newcommand{\gnewton}{G_\mathrm{N}}

\newcommand{\rmin}{r_\mathrm{min}}
\newcommand{\rps}{r_\mathrm{ps}}
\newcommand{\robs}{r_\mathrm{obs}}
\newcommand{\ellp}{\ell_\mathrm{P}}

\newcommand{\bsh}{b_\mathrm{sh}}

\newcommand{\xh}{x_H}
\newcommand{\rd}{\kappa}

\newcommand{\si}[1]{s_{#1}}

\newcommand{\zhp}{r_{H,+}}
\newcommand{\zhm}{r_{H,-}}
\newcommand{\zhpm}{r_{H,\pm}}

\newcommand{\zbr}{r_{\text{BR},H}}

\newcommand{\dho}{d_{H}}
\newcommand{\dhp}{d_{H,+}}

\newcommand{\zi}[1]{\mathfrak{p}_{#1}}

\newcommand{\fn}[1]{f^{(#1)}_{H}}
\newcommand{\hn}[1]{h^{(#1)}_{H}}

\newcommand{\fid}[1]{\Phi^{(#1)}_{H}}

\newcommand{\ma}{M}

\newcommand{\ha}[1]{\widehat{a}_{#1}}
\newcommand{\scheme}{EMD}
\newcommand{\rootk}{\varpi}

\newcommand{\kc}[1]{K_{#1}}
\newcommand{\sig}[1]{\sigma_{#1}}

\newcommand{\gen}{\mathcal{X}}
\newcommand{\wgen}{\widetilde{\mathcal{X}}}
\newcommand{\cgen}[1]{x_{#1}}
\newcommand{\wcgen}[1]{\widetilde{x}_{#1}}

\newcommand{\tilxi}[1]{\widetilde{\xi}_{#1}}
\newcommand{\tiltheta}[1]{\widetilde{\theta}_{#1}}

\newcommand{\wtq}[1]{\widetilde{q}_{#1}}
\newcommand{\tp}[1]{\tau_{#1}}
\newcommand{\uva}[1]{u_{#1}}

\newcommand{\sit}[1]{\widetilde{s}_{#1}}

\newenvironment{linesmall}[1]{\arraycolsep=3pt\def\arraystretch{0.75}\fontsize{8}{9.5}\selectfont\array{#1}}{\endarray} 

\newcommand{\co}{r}
\newcommand{\zh}{r_H}
\newcommand{\hr}{r_H}

\newcommand{\newom}{\Omega}
\newcommand{\newgam}{\Gamma}

\colorlet{MDPteal}{teal}

\def\bcond{\left\vert_{\begin{array}{l}\scriptstyle r=\hr \end{array}}\right.\hskip-0.1cm}

\newcommand{\ols}[1]{\mskip.5\thinmuskip\overline{\mskip-.5\thinmuskip {#1} \mskip-.5\thinmuskip}\mskip.5\thinmuskip} 
\newcommand{\olsi}[1]{\,\overline{\!{#1}}} 
\newcommand\closure[1]{
  \tctestifnum{\count@stringtoks{#1}>1} 
  {\ols{#1}} 
  {\olsi{#1}} 
}

\usepackage{array}
\newcolumntype{P}[1]{>{\centering\arraybackslash}p{#1}} 

\theoremstyle{definition}

\newcommand{\tah}{\tau_{H}}
\newcommand{\zs}{r_{S}}

\newcommand{\zp}{r_{P}}
\newcommand{\zgam}{r_{\gamma}}

\newcommand{\ang}{\beta}
\newcommand{\imp}{b}
\newcommand{\risco}{r_{\rm ISCO}}

\newcommand{\sa}[1]{{\sigma^{(1)}_{#1}}}
\newcommand{\se}[1]{{\sigma^{(2)}_{#1}}}
\newcommand{\sd}[1]{{\sigma^{(2)}_{#1}}}

\title{Black Holes beyond General Relativity: the Effective Metric Description}

\begin{document}
\typeout{START baseline=\the\baselineskip, jot=\the\jot}
\frontmatter

\begin{titlepage}
\centering
\vspace*{2.2cm}

{\color{black!60}\rule{0.85\textwidth}{1.2pt}}\\[0.15em]
{\color{black!60}\rule{0.85\textwidth}{0.4pt}}

\vspace{2.2em}

{\LARGE\bfseries\scshape Black Holes beyond General Relativity}

\vspace{1.4em}

{\Large\itshape The Effective Metric Description}

\vspace{2.2em}

{\color{black!60}\rule{0.85\textwidth}{0.4pt}}\\[0.15em]
{\color{black!60}\rule{0.85\textwidth}{1.2pt}}

\vspace{3.5em}

{\scshape Manuel Del Piano$^{\,1}$ \;\textcolor{black!50}{$\bullet$}\; Stefan Hohenegger$^{\,2}$ \;\textcolor{black!50}{$\bullet$}\; Francesco Sannino$^{\,1,3,4,5}$}

\vspace{2.8em}

\begin{minipage}{0.86\textwidth}
\centering\small\itshape
$^{1}$\,Quantum Theory Center ($\hbar$QTC) \& D-IAS, University of Southern Denmark,\\
Campusvej 55, 5230 Odense M, Denmark\\[0.5em]
$^{2}$\,Universit\'e Lyon 1, CNRS, IP2I, UMR 5822, Villeurbanne, France\\[0.5em]
$^{3}$\,Scuola Superiore Meridionale, Largo S.~Marcellino 10, 80138 Napoli, Italy\\[0.5em]
$^{4}$\, Institute for Particle Physics Phenomenology (IPPP), Durham Univ., South Road, DH1 3LE, Durham, United Kingdom
\\[0.5em]
$^{5}$\,Dipartimento di Fisica ``E.~Pancini'', Universit\`a di Napoli Federico II,\\
via Cintia, 80126 Napoli, Italy
\end{minipage}

\vfill

{\large\scshape November 2026}

\vspace*{1.5cm}
\end{titlepage}
\chapter*{Abstract}
Black holes provide a privileged window on gravity in the strong-field regime, and deviations from the classical solutions of General Relativity (GR) may encode information about quantum gravity and other extensions of the theory. The abundance of proposed deformed black hole geometries, however, makes model-by-model comparison with observations impractical. This review presents a unified account of the Effective Metric Description (EMD), a model-independent framework in which deformations of static, spherically symmetric black hole spacetimes are parametrized by physical, coordinate-invariant quantities, such as the proper radial distance from the event horizon or the proper time of a free-falling observer. After reviewing the classical black hole solutions and the open issues that motivate going beyond GR, we develop the EMD as a series expansion near the horizon for a given physical quantity. We provide the general solution of the resulting self-consistency conditions compatible with the symmetries of the geometry, derive universal regularity conditions ensuring a finite Hawking temperature and curvature invariants, establish the equivalence of descriptions based on different physical quantities, and extend the framework to charged, extremal and lower-dimensional black holes. Padé resummation extends the description away from the horizon, enabling model-independent computations of the black hole shadow and, in the eikonal limit, of quasi-normal mode frequencies in the test-field approximation. The framework is tested on a broad set of geometries used to model physics beyond GR, and compared with other model-independent parametrizations through explicit transformation maps. As an effective description of deformations beyond GR, the EMD provides a coherent bridge between fundamental black hole physics, quantum-gravity-inspired deformations, and astrophysical observables.
\tableofcontents

\mainmatter

\include{subfiles/1-Intro}

\include{subfiles/2-BHRecap}

\include{subfiles/3-EMD}

\include{subfiles/4-Applications}

\include{subfiles/5-Shadow}

\include{subfiles/GeodesicMotion}

\include{subfiles/6-Params}

\include{subfiles/7-Conclusions}

\include{subfiles/Appendix}

\newpage
\leavevmode\thispagestyle{empty}
 
\printbibliography[heading=bibintoc]

\end{document}

%% file: subfiles/1-Intro.tex
\typeout{START baseline=\the\baselineskip, jot=\the\jot}
\chapter{Introduction}\label{chap:intro}

Black holes (BHs) are among the most intriguing and fundamental objects predicted by Einstein's theory of General Relativity (GR)~\cite{Schwarzschild:1916uq,Wald:1984rg,Chandrasekhar:1985kt}. Defined by regions of spacetime from which no signal can escape, black holes provide a unique stage to probe gravitational physics in extreme regimes. 
The most general classical solution in vacuum GR is the Kerr solution \cite{Kerr:1963ud}, which is characterised by its mass and angular momentum. This simplicity, often encapsulated in the \emph{no-hair conjecture}~\cite{Wheeler:1971ba,Israel:1967wq,Carter:1971zc,Robinson:1975bv,Bekenstein:1972ky,Bekenstein:1972ny,Heusler:1996ft,Chrusciel:2012jk,Herdeiro2015}, belies the profound richness of their physical and mathematical properties. For instance, Kerr black holes exhibit inner and outer horizons, whose coalescence defines an extremal configuration with vanishing surface gravity~\cite{Weyl,Nordstrom,Carroll:2009maa}. The rotation introduces frame-dragging effects and a more intricate causal structure.
Coupling GR to Maxwell fields, the static and spherically symmetric BH solution is the Reissner-Nordstr\"om metric~\cite{Reissner}, which is characterised by its mass and electromagnetic charge, see some of the excellent textbooks on GR \cite{misner1973gravitation}.

Astrophysical observations have now reached a level of accuracy that allows for the direct probing of black hole properties. The Event Horizon Telescope (EHT) has provided unprecedented imaging of the supermassive black holes M$87^*$ and Sgr A$^*$, revealing their shadows and constraining their mass and spin parameters~\cite{EventHorizonTelescope:2019dse, EventHorizonTelescope:2021dqv, EventHorizonTelescope:2022wkp, EventHorizonTelescope:2022xqj}. Simultaneously, gravitational wave observatories, including LIGO, Virgo, and KAGRA, have detected numerous BH binary mergers, providing detailed information about their mass, spin distributions, and merger dynamics~\cite{LIGOScientific:2016aoc,LIGOScientific:2020tif,LIGOScientific:2020stg}. These observations not only confirm the existence of black holes as predicted by GR but also open the door to probing potential deviations from classical solutions, thereby providing a unique window into physics beyond GR and hints of quantum gravity effects.  

Despite these advances, a fully consistent and predictive theory of quantum gravity remains elusive. Consequently, understanding the modifications of black hole spacetimes due to quantum or otherwise beyond-GR effects poses a significant challenge. Many approaches have been proposed, ranging from concrete quantum gravity models to phenomenological modifications inspired by symmetry or regularity principles. The resulting plethora of individual models~\cite{Hayward_2006,bardeen1968,Dymnikova:1992ux,Dymnikova:2004qg,Garcia95,Bjerrum-Bohr:2002fji,Donoghue:2001qc,Kirilin:2006en,Frolov:2016pav,Calmet:2017qqa,Calmet:2019eof,Simpson:2018tsi,Simpson:2019mud,Nicolini:2019irw,Calmet:2021lny,Battista:2023iyu,Carballo-Rubio:2025ntd,Carballo-Rubio:2025fnc,Emparan:2020znc,Emparan:2021hyr} often employ distinct metric deformations, making direct comparison with observational data and cross-model generalization difficult \cite{Vagnozzi:2022moj}.

To address this challenge, an \emph{Effective Metric Description} (EMD) framework has been developed \cite{Binetti:2022xdi,DelPiano:2023fiw,DelPiano:2024gvw,DamiaPaciarini:2025xjc}. The core idea is to construct a model-independent and systematic representation of black hole deformations in terms of physical quantities, \emph{i.e.} suitable observer-independent invariants of the spacetime geometry. In the case of non-extremal, static, and spherically symmetric black holes, a natural choice for such a quantity is the proper radial distance from the event horizon. For extremal black holes, where this invariant is not well-defined, one can for example use the proper time for a free-falling observer to reach the horizon.  

In the EMD framework, the metric is expanded in a series around the horizon in terms of these physical quantities. Regularity conditions, such as finiteness of the Hawking temperature and curvature scalars, impose constraints on the coefficients of these expansions. These constraints are coordinate-independent and provide universal conditions for physically admissible black hole spacetimes \cite{DelPiano:2024nrl,DAlise:2023hls} (while ruling out others). By formulating deformations in this manner, the EMD allows a coherent description of a wide class of black holes, whether classical, regular, or quantum-deformed, in a way that is both physically meaningful and well-defined.

Parameterizing physically relevant observables is at the heart of the EMD. The black hole shadow, corresponding to the apparent shape of the photon capture region as seen by a distant observer, probes the geometry near the photon sphere and provides relevant insights into the spacetime structure. In order to describe the shadow, the EMD needs to be adapted to describe the black hole geometry beyond an infinitesimally small region around the horizon. This was eventually achieved by the employment of Padé approximants \cite{bender78:AMM}, which enjoy large applications in very different fields \cite{Dondi:2019ivp,Dondi2020,Gruber:2013wua,Sultan:2025mvc,Mroue:2008fu}. The successful extension of the EMD to observables away from the event horizon \cite{DelPiano:2024nrl}, sparked the idea of applying such approach to other interesting physical quantities, such as the orbital frequency and Lyapunov exponent of photon orbits which determine the real and imaginary parts of quasi-normal mode (QNM) frequencies \cite{Berti:2004md,Berti:2022xfj,Berti:2009kk,Konoplya2017,Konoplya:2016jvv} of massless scalar test-fields in the eikonal limit \cite{Stefanov:2010xz,Cardoso:2008bp,Dolan:2010wr}. As a result, the shadow radius and eikonal QNMs are tightly linked to the underlying spacetime geometry and can be computed within the EMD framework to assess deviations from classical predictions.  

One of the core principles of the EMD is to parametrize metric deformations beyond GR in terms of physical parameters, such as (measurable) distances or (proper) times. This is crucial in order to give a physical interpretation to observables computed within such a space-time and make more direct contact to physics beyond GR. In turn, however, this also implies that the deformation coefficients of the EMD must be directly accessible by physical measurements. This was demonstrated in \cite{Hohenegger:2025dic}, where a Gedankenexperiment was proposed that allows to extract these coefficients directly from electromagnetic signals of probes falling into the black hole. These signals are accessible from observers outside of the black hole, thus allowing to reconstruct the geometry. It was also observed that the deformation functions can be expressed in terms of (the invariant eigenvalues of) an effective energy momentum tensor. The approach was illustrated by the Hayward space-time as an example of a spherically symmetric and static black hole \cite{Hayward_2006}.

Many other parametrizations exist in the literature. Johannsen–Psaltis \cite{Johannsen:2010xs,Johannsen:2011dh} introduced a quadrupole moment independent of mass and spin, facilitating tests of the no-hair theorem through ISCO and lensing observations. Cardoso, Pani and Rico \cite{Cardoso:2014rha} extended this approach to include additional parameters affecting both weak- and strong-field regimes. Rezzolla–Zhidenko \cite{Rezzolla:2014mua,Konoplya:2016jvv} introduced a continued-fraction expansion for static, spherically symmetric spacetimes, providing superior convergence properties compared to Taylor series. Our EMD framework complements these approaches by focusing on invariants, series expansions near the horizon, and the connection to directly observable quantities. Establishing a map between these frameworks allows a coherent comparison of different parametrizations, revealing their respective strengths and limitations \cite{DeFalco:2025xxx}.

Several works have applied the EMD framework to explore different aspects of black hole physics. Li \emph{et al.}~\cite{Li:2023djs} analyzed timelike and lightlike geodesics in a quantum black hole described within the EMD, studying the periapsis precession of bounded orbits and angular shadow size. They found that quantum corrections enhance both the precession and the shadow relative to the classical case, particularly for smaller black holes. Belfiglio \emph{et al.}~\cite{Belfiglio:2024qsa} investigated entanglement entropy for a massive non-minimally coupled scalar field in several quantum black hole models, showing a significant deviation from the classical area law near the origin, with convergence to classical results at larger distances. D’Alise \emph{et al.}~\cite{DAlise:2023hls} explored positivity conditions and asymptotic stress-energy tensors, demonstrating how physicality requirements constrain metric deformations, with results applicable to both classical and quantum black holes.

These works collectively demonstrate the versatility of the EMD, highlighting its ability to:  
\begin{itemize}
    \item Capture universal features of black hole physics across models,  
    \item Provide a systematic parametrization of quantum or regularized deformations,  
    \item Enable computation of physically observable quantities such as Hawking temperature, shadows, and geodesic precession in a model-independent fashion,  
    \item Connect theoretical parameters to observable quantities measurable by external observers.  
\end{itemize}  
 
Moreover, modifications of near-horizon metrics can significantly affect gravitational wave emission, particularly during the ringdown phase after perturbations. The EMD provides a practical approach to compute these corrections systematically, enabling predictions for potential deviations in future gravitational wave detections. By connecting these observables to series expansion coefficients of the EMD, one can, in principle, constrain or measure quantum corrections using astrophysical data.

This review aims to provide a unified view of model-independent black hole parametrizations, with a particular focus on the EMD framework. We explore its construction, theoretical consistency, and practical applications, while reviewing and extending its previous uses in the literature. Special attention is given to physical observables, including shadows and eikonal QNMs, and their sensitivity to quantum or regularized deformations. The EMD framework is not merely a theoretical tool, but a bridge connecting fundamental black hole physics, quantum gravity-inspired modifications, and observable astrophysical signatures. It provides a systematic, physically meaningful, and model-independent approach to study the rich phenomenology of black holes, paving the way for rigorous confrontation with current and future observations. To facilitate navigation, the notion of black hole adopted throughout is fixed in a boxed \emph{Definition}, the main results are collected in boxed \emph{Results}, while longer derivations and conceptual comments are presented in separate \emph{Computation} and \emph{Remark} boxes.  \\

The review is organized as follows. Chapter~\ref{chap2} recalls the classical black hole solutions of GR, including the $2+1$-dimensional BTZ geometry, together with the physical quantities used throughout the review: geodesic motion and the black hole shadow, quasi-normal modes (QNMs) and black hole thermodynamics. It then summarizes the open issues of classical black holes and the main frameworks beyond classical GR, from effective field theory and asymptotic safety to phenomenological regular models, and closes with the swampland program as a bottom-up approach to effective field theories of gravity. Chapter~\ref{chap3} introduces the EMD. Building on~\cite{Binetti:2022xdi,DelPiano:2023fiw}, we construct the EMD based on the proper distance and derive its self-consistency and regularity conditions at the outer and inner horizons. Following~\cite{DamiaPaciarini:2025qhq}, these conditions are then extended to the BH core. The chapter further discusses how EMDs formulated in terms of different physical quantities are related~\cite{DelPiano:2024gvw}, the generalization to charged and extremal black holes~\cite{DamiaPaciarini:2025xjc}, and the extension to $2+1$ dimensions~\cite{DamiaPaciarini:2024ssf}. In Chapter~\ref{chap examples}, the framework is applied to a number of black hole models, including asymptotically safe, regular and holographic geometries whose horizons and cores are analysed within the EMD, and it is used to discuss the weak gravity conjecture. Chapter~\ref{chap4}, based on~\cite{DelPiano:2024nrl}, extends the near-horizon parametrization with Padé approximants and derives a model-independent expression for the black hole shadow. Chapter~\ref{chapGeo} describes the Gedankenexperiment proposed in~\cite{Hohenegger:2025dic}, which links the EMD deformations to the telemetric signals of probes falling into the black hole, accessible to external observers. Chapter~\ref{chap5}, based on~\cite{DeFalco:2025xxx}, compares EMDs near the horizon and at large distances with the Johannsen--Psaltis and Rezzolla--Zhidenko parametrizations, and tests all frameworks by computing eikonal QNMs for different black hole models. Finally, Chapter~\ref{chap:conclusions} summarizes the main results, emphasizing the consistency of the EMD as a model-independent description of classical, regular and quantum-deformed black holes, and outlines directions for future research, including rotating black holes, the interior structure of black holes and further observables.

Throughout the review we adopt the natural unit convention $c = \gnewton = \hbar = k_{\rm B}= 1$.

%% file: subfiles/2-BHRecap.tex
\typeout{START baseline=\the\baselineskip, jot=\the\jot}
\chapter{Black holes in General Relativity}\label{chap2}

This chapter introduces the essential theoretical background of black hole physics forming the basis of the effective metric description. We review the main BH solutions of general relativity (GR), \emph{i.e.} Schwarzschild, Reissner–Nordström, and briefly Kerr, highlighting their geometric structure, causal properties, and the implications of the no-hair theorems. We then discuss key thermodynamic aspects such as surface gravity, Hawking radiation, and the area law, which link gravity, quantum theory, and thermodynamics. The chapter also examines light propagation and the formation of the BH shadow, followed by an introduction to BH perturbations and their quasi-normal modes (QNMs), which  appear in the ringdown phase of gravitational-wave emission. Altogether, these topics provide the conceptual and mathematical foundation for the effective metric framework developed in the following chapters.

General Relativity is a dynamical theory of the spacetime,  which in 4 dimensions is captured by the Einstein-Hilbert action. The total system action composed of matter and spacetime reads 
\begin{equation}\label{total action}
    S_{\rm tot} = S_{\rm EH} + S_m \qq{with} S_\mathrm{EH} =  \frac{1}{16 \pi}\int \dd^4 x \sqrt{-g}\, R \qq{and} S_m= \int \dd^4 x \sqrt{-g} \, \mathcal{L}_m \ ,
\end{equation}
where $g$ is the determinant of the metric tensor $g_{\mu \nu}$, $R$ is the Ricci scalar and $\mathcal{L}_m$ is the matter Lagrangian density. A possible additional term in $S_{\rm EH}$ is the cosmological constant $\Lambda$ which models the spacetime dynamics driven by the presence of dark energy in the system. The latter mostly changes physics at cosmological scales and we shall hence neglect it in the action for the time being. Variation of the total action~\eqref{total action} with respect to the metric tensor yields the second-order non-linear field equations
\begin{equation}\label{chap2 Einstein Eqs}
    R_{\mu \nu} - \frac{1}{2}R\, g_{\mu \nu} = 8 \pi \, T_{\mu \nu} \ ,
\end{equation}
 known as the Einstein equations. Here $R=g^{\mu\nu}R_{\mu\nu}$ is the Ricci scalar, which is defined as the contraction of the Ricci tensor $R_{\mu \nu}$ (with the inverse metric $g^{\mu\nu}$), while the latter in turn is obtained as the contraction of the Riemann curvature tensor, \emph{i.e.} $R_{\mu \nu}={R^\rho}_{\mu\rho\nu}$.   Concretely, upon choosing coordinates $x^{\mu}$ (with $\mu\in\{0,1,2,3\}$ and $\partial_\mu=\frac{\partial }{\partial x^\mu}$) we have
\begin{align}
{R^\rho}_{\sigma\mu\nu}=\partial_\mu\,{\Gamma^\rho}_{\nu\sigma}-\partial_\nu\,{\Gamma^\rho}_{\mu\sigma}+{\Gamma^\rho}_{\mu\lambda}{\Gamma^\lambda}_{\nu\sigma}-{\Gamma^\rho}_{\nu\lambda}{\Gamma^\lambda}_{\mu\sigma}\,,
\end{align}
with the Christoffel-symbols
\begin{align}
{\Gamma^\mu}_{\nu\rho}=\frac{1}{2}\,g^{\mu\sigma}\,\left(\partial_\rho g_{\sigma\nu}+\partial_\nu\,g_{\sigma\rho}-\partial_\sigma g_{\nu\rho}\right)\,,
\end{align}
Furthermore, $T_{\mu \nu}$ in (\ref{chap2 Einstein Eqs})
 is the energy-momentum tensor defined by the matter Lagrangian density as 
\begin{equation}
    T_{\mu \nu} \coloneqq -\frac{2}{\sqrt{-g}} \fdv{ \left( \sqrt{-g}\, \mathcal{L}_m\right)}{g^{\mu \nu}} \ .
\end{equation}
This tensor acts as the source term in the Einstein equations~\eqref{chap2 Einstein Eqs}, allowing us to distinguish between vacuum and non-vacuum solutions.

\section{The Schwarzschild solution}\label{Sect:Schwarzschild}

 Firstly, let us examine the case of vacuum solutions, namely  we study solutions of Eq.~\eqref{chap2 Einstein Eqs} with
\begin{equation}\label{chap2 vacuum cond}
    T_{\mu \nu}=0 \ , 
\end{equation}
in some region of the spacetime. Typically, the choices for $T_{\mu\nu}$ depend on the type of matter distribution and the most used is a spatially limited distribution and a vacuum exterior space.  For $T_{\mu\nu}=0$, contracting \eqref{chap2 Einstein Eqs} with $g^{\mu\nu}$ leads to $R=0$, such that the Einstein vacuum equation reduces to
\begin{equation}\label{Einstein vacuum equation}
    R_{\mu \nu} = 0 \ .
\end{equation}
We can further simplify the problem by imposing symmetries of the spacetime.
The simplest case is the static and spherically symmetric spacetime, which is described in spherical coordinates $(t,r,\theta,\varphi)$ by the following line element
\begin{equation}\label{spherical static metric}
    \dd s^2 = - h(r) \dd t^2 + f(r)^{-1} \dd r^2 + r^2 \dd \theta^2 + r^2 \sin^2 \theta \dd \varphi ^2 \ .
\end{equation}
The Ricci scalar in terms of $f$ and $h$ is
\begin{equation}\label{eq: ricci scalar h e f}
R = \frac{ f (h^{(1)})^2}{2 h^2}-\frac{\left(r f^{(1)}+4 f\right) h^{(1)}}{2 r h} - \frac{f \, h^{(2)}}{h}  -\frac{2\big(r f^{(1)}+f-1\big)}{r^2} \ .
\end{equation}

\begin{center}
\remarkbox{\rem\label{rem:shorthandderivatives}{Shorthand Notation for Derivatives}}{\parbox{16cm}{
From here on out the notation $F^{(n)}$ is used to indicate the $n$-th derivative of a function $F$ with respect to its argument. Concretely, let
\begin{align}
F:\,\mathbb{R}&\longrightarrow \mathbb{R}\nonumber\\
x&\longmapsto F(x)\,,\nonumber
\end{align}
be $n$-times differentiable, then we denote the value of its $n$th derivative at $r\in\mathbb{R}$ as
\begin{align}
F^{(n)}(r):=\left.\dv[n]{F}{x}\right|_{x=r}\,.\label{Deffid}
\end{align}
The first derivative is sometimes simply denoted with a prime, \emph{i.e.} $F^{(1)}(r)=F'(r)$. Furthermore, we shall use the subscript $_H$ to denote the evaluation of a quantity at the horizon $r =\zh$, \emph{e.g.} $F^{(n)}_{H}:=F^{(n)}(r=\zh)$.
}}
\end{center}

\noindent
Substituting the metric (\ref{spherical static metric}) into the field equations~\eqref{chap2 Einstein Eqs} with the condition~\eqref{chap2 vacuum cond}, and imposing asymptotic flatness as well as the weak field limit as boundary conditions, one obtains the celebrated Schwarzschild solution~\cite{Schwarzschild:1916uq}
\begin{equation}\label{chap2 Schwarzschild sol}
    h(r) \equiv f(r) = f_{\rm S}(r)= 1 - \frac{2  M}{r} \ .
\end{equation}
 It describes the exterior (\emph{i.e.} for $r>0$) gravitational field of an object of mass $M$ that is located at the origin $r=0$.
We remark that \eqref{chap2 Schwarzschild sol} is the unique spherically symmetric and static vacuum solution (\emph{i.e.} with $T_{\mu\nu}=0$) of the Einstein equation~(\ref{Einstein vacuum equation}), which is the contents of Birkhoff's theorem \cite{Carroll}. Thus,  even in the case of a time-evolving matter distribution, for instance in a radially pulsating or collapsing spherical star, the exterior region is described by the Schwarzschild metric.

Let us remark that the metric elements $ - g_{tt}(r) = g_{rr}^{-1} =  1-2 M/r $ have two singular points: 
\begin{itemize}
\item origin $r=0$, with
\begin{align}
&\lim_{r\to 0}g_{tt}\to\infty\,,&&\lim_{r\to 0}g_{rr}\to 0^-\,.
\end{align}
\item $r=2M$, with
\begin{align}
&\lim_{r\to 2M}g_{tt}\to 0\,,&&\begin{array}{l}\lim_{r\to (2M)^+}g_{rr}\to \infty\,,\\[4pt]\lim_{r\to (2M)^-}g_{rr}\to -\infty\end{array}
\end{align}
\end{itemize}
 However, both the physical and mathematical nature of these points is very different: while the origin represents a curvature singularity, the singularity at $r=2M$ is only a consequence of our choice of coordinates. In order to understand the difference, we need to study curvature invariants such as the Ricci scalar $R$ in Eq.~\eqref{eq: ricci scalar h e f}. However, since the latter is identically vanishing in the Schwarzschild geometry, a more insightful choice is the Kretschmann scalar, which is defined as  
\begin{equation}
    K \coloneqq \tensor{R}{_\rho _\sigma _\mu _\nu}\tensor{R}{^\rho ^\sigma ^\mu ^\nu}\ ,\label{KretschDef}
\end{equation}
where $\tensor{R}{_\rho _\sigma _\mu _\nu}=\tensor{g}{_\rho _\lambda} \tensor{R}{^\lambda _\sigma _\mu _\nu}$ is the fully covariant Riemann tensor. For the metric \eqref{spherical static metric} with \eqref{chap2 Schwarzschild sol} we find explicitly 
\begin{equation}
    K  = \frac{48 M^2}{r^6} \ ,
\end{equation}
which exhibits a pole only at $r=0$. Hence, the singularity at $r=2  M$ is called \emph{coordinate singularity}, because it can be removed with some appropriate coordinate transformations, for example the Eddington-Finkelstein in-falling or out-going null coordinate sets. Nonetheless, the surface at $r=2 M$ has remarkable features. Considering the hypersurface $\Sigma$
\begin{equation}
    \Sigma = r-2  M = 0 \ ,
\end{equation}
the dual normal vector is given by the gradient
\begin{equation}
    n_\alpha = \partial_\alpha \Sigma \ ,
\end{equation}
and whose norm is 
\begin{equation}
    n_\alpha n^\alpha = g^{\alpha \beta}n_{\alpha}n_{\beta} = g^{rr} (\partial_r\Sigma )^2= 1 - \frac{2  M}{r} \ .
\end{equation}
Hence, we can examine three cases:
\begin{itemize}
    \item Any sphere with radius $r > 2  M$ is a collection of timelike points where signals can be sent towards the origin $r \to 0$ and the spatial infinity $r \to \infty$, hence they are all causally connected.
    \item The spheres with radii $r< 2  M$ are spacelike, and any light signal is forced to converge towards the origin, therefore any observer located in this region is unable to communicate with observers at $r > 2  M$ .
    \item The 2-sphere $r= 2  M$ is a lightlike surface that separates the two causally disconnected regions described above.  It is called the \textbf{event horizon}. 
\end{itemize}
Due to these properties, the Schwarzschild solution including the horizon at $r=2M$, is said to describe a {\bf black hole} (BH), \emph{i.e.} an extremely compact object, which creates such a strong gravitational field that not even light can escape to asymptotic infinity.
This intuitive characterisation can be made precise in terms of the causal structure of the spacetime. Since the notion of BH will be used throughout this review, also for geometries beyond GR, we adopt the following definition, stated as in~\cite{Booth:2005qc} (see also~\cite{Hawking:1973uf,Wald:1984rg}).

\begin{center}
\definitionbox{\defin\label{def:BlackHole}{Black hole and event horizon}}{\parbox{16cm}{
Let $\mathcal{M}$ be a manifold equipped with a Lorentzian metric $g$, describing an asymptotically flat (more precisely, strongly asymptotically predictable~\cite{Hawking:1973uf,Wald:1984rg}) spacetime with future null infinity $\mathscr{I}^+$, and let $J^-(\mathscr{I}^+)$ denote the causal past of $\mathscr{I}^+$, \emph{i.e.} the set of all events from which a future-directed causal curve reaches $\mathscr{I}^+$. The spacetime contains a \textbf{black hole} if the complement of $J^-(\mathscr{I}^+)$ is non-empty. This complement,
\begin{align}
\mathcal{B}:=\mathcal{M}\setminus J^-(\mathscr{I}^+)\,,\label{DefBHregion}
\end{align}
is the \emph{black hole region}: the set of events from which no signal, not even light, can reach $\mathscr{I}^+$. Its boundary $\mathcal{H}^+:=\partial J^-(\mathscr{I}^+)$ is the \textbf{event horizon}, a null hypersurface.\\[6pt]
For the static and spherically symmetric metrics~\eqref{spherical static metric} considered in this review, in GR and beyond, $\mathcal{H}^+$ is the sphere $r=\zh$, where $\zh$ is the largest radius at which both metric functions vanish,
\begin{align}
&f(\zh)=h(\zh)=0\,,&&\text{with}&&f(r)>0\quad\text{and}\quad h(r)>0\quad\forall\, r>\zh\,.\label{DefBHhorizon}
\end{align}
This sphere is a Killing horizon of the static Killing vector $\chi^t=\partial_t$. If the zero in~\eqref{DefBHhorizon} is simple, the BH is non-extremal and the spheres just inside the horizon are trapped ($f<0$); if it is of higher order, the BH is extremal. Further zeros of $f$ and $h$ at $r<\zh$, as in the Reissner--Nordstr\"om solution or in regular BHs, are \emph{inner} horizons. A geometry in which $f$ and $h$ have no common zero is horizonless and, however compact, is \emph{not} a black hole.
}}
\end{center}

\noindent
The same conclusions can be drawn from the investigation of the timelike Killing vector 
\begin{equation}
(\chi^{t})^{\mu} = (1, 0, 0 ,0) \ ,
\end{equation}
which is such that the Lie derivative vanishes, \emph{e.g.} $\mathcal{L}_{\chi^{t}}g_{\mu \nu} = \partial_{t} g_{\mu \nu} = 0$, reflecting the time-translational invariance of the metric (\emph{i.e.} the fact that the spacetime is static). A timelike \emph{Killing horizon} is defined as the collection of points where the norm $(\chi^{t})_{\mu} (\chi^{t})^{\mu}=-f_{\rm S}(r)$ vanishes. Hence, for the Schwarzschild spacetime the event horizon coincides with the timelike Killing horizon. 

Two comments on Definition~\ref{def:BlackHole} are in order. First, the event horizon is a global and \emph{teleological} notion: whether an event belongs to $\mathcal{B}$ depends on the entire future evolution of the spacetime, so that $\mathcal{H}^+$ cannot be located by local measurements~\cite{Booth:2005qc}. In dynamical situations, such as gravitational collapse or BH evaporation, which are relevant for quantum-corrected BHs, it is therefore more appropriate to characterise BHs quasi-locally in terms of trapped surfaces, \emph{e.g.} through apparent, trapping~\cite{Hayward:1993wb}, isolated or dynamical horizons~\cite{Ashtekar:2004cn} (see~\cite{Booth:2005qc} for a review). For the static geometries studied in this review, the marginally trapped spheres sit at the zeros of $f$, so that these quasi-local horizons are located at the same radius $r=\zh$ as the event horizon; we shall therefore simply refer to $r=\zh$ as the (event) horizon. Second, for spacetimes which are not asymptotically flat, such as the asymptotically anti-de Sitter BTZ geometry of Section~\ref{ClassicalBTZ}, $\mathscr{I}^+$ is replaced by the conformal boundary of the spacetime.

\section{The Kerr solution}
For the sake of completeness of vacuum GR solutions, let us now examine the case in which we relax the spherical symmetry of the spacetime and introduce the metric of a rotating black hole with mass $M$ and angular momentum $J$, known as Kerr BH~\cite{Kerr:1963ud}. As an isolated rotating object, we can find a coordinate frame in which we can align one axis along the conserved angular momentum. The most commonly used coordinates are the Boyer-Lindquist coordinates $(t,r,\theta,\varphi)$ and the line element reads
\begin{multline}\label{kerr metric}
    \dd s^2 = - \left( 1- \frac{2  M \, r}{\Sigma}\right) \dd t^2 + \frac{\Sigma}{\Delta} \dd r^2 + \Sigma \, \dd \theta^2 + \\ +\left( r^2 + a^2 + \frac{2  M a^2 \, r}{\Sigma}\sin^2 \theta\right) \sin^2 \theta \, \dd \varphi^2 - \frac{4  M a\, r}{\Sigma} \sin^2 \theta \, \dd t \, \dd \varphi \ .
\end{multline}
Here $a \coloneqq J/M$ and we have introduced the auxiliary functions
\begin{equation}
    \Sigma \coloneqq r^2 + a^2 \cos^2 \theta \qq{and} \Delta \coloneqq r^2 - 2M \,r + a^2 \ .
\end{equation}
We note the appearance of the cross term $g_{t\phi} \propto a$ in the line element, which vanishes when the angular momentum is switched off and the entire metric~\eqref{kerr metric} reduces to the Schwarzschild one. Rotation leads to two distinct surfaces of interest. The roots of $\Delta$ determine the inner and outer horizons
\begin{equation}\label{kerr horizons}
    r_\pm = M \pm \sqrt{M^2 - a^2} \, .
\end{equation}
The stationary limit surface is defined by $g_{tt}=0$, \emph{i.e.}
\begin{equation}
    r_{\rm ergo}(\theta) = M + \sqrt{M^2 - a^2\cos^2\theta} \ ,
\end{equation}
which encloses the ergoregion, where no static observer can exist and the angular velocity of the BH horizon is
\begin{equation}
    \Omega_H \coloneqq \left.\frac{g_{t\varphi}}{g_{\varphi \varphi}}\right|_{r=r_+}= \frac{a}{2Mr_+} \, .
\end{equation}
The distinction between the ergoregion boundary and the horizon, $r_{\rm ergo}(\theta) > r_+$ for $0<\theta<\pi$, enables extraction of rotational energy, which is an important feature for the Penrose process~\cite{Penrose:1969pc,Bardeen:1972fi} and for the effect of superradiance of fields around the BH~\cite{Press:1972zz,Damour:1976jd,Brito:2015oca}.\footnote{Kerr supports the superradiant scattering of classical scalar fields if the field frequency $\omega$ is such that $\omega < m \Omega_H$
leading to amplification of waves with azimuthal number $m$. If the wave is trapped, a superradiant instability can arise, constraining ultralight fields.}


From Eq.~\eqref{kerr horizons}, the existence of two real and distinct roots requires $a < M$ ensuring that the metric describes a rotating black hole. The limiting case $a=M$ corresponds to an \emph{extremal} Kerr black hole, for which the inner and outer horizons coincide. If instead $a>M$, the function $\Delta$ has no real roots and the curvature singularity at $r=0$ becomes exposed to distant observers, resulting in a \emph{naked singularity}~\cite{Penrose:1973um}. Such configurations would violate the \emph{cosmic censorship conjecture}~\cite{Penrose:1964wq}, which posits the existence of an event horizon that shields singularities from the exterior universe. Consequently, the bound
\begin{equation}
a \leq M \ ,
\end{equation}
is enforced as a necessary condition for physically acceptable Kerr spacetimes.

\section{The Reissner-Nordström solution}\label{Sect:RNBH}

Before studying the presence of matter content in the system, we remark that the cosmological constant $\Lambda$ behaves like and effective geometrical source in vacuum Einstein equations, \emph{e.g.} $R_{\mu \nu} = \Lambda \, g_{\mu \nu}$, realizing the condition for Einstein spaces. The solution to such equations are known as de Sitter ($\Lambda > 0$) or Anti-de Sitter ($\Lambda < 0$) which are maximally symmetric spaces.\\
While vacuum black holes already capture essential aspects of strong-gravity physics, realistic astrophysical environments and fundamental interactions demand the inclusion of matter fields. 
In this Section, we investigate the class of solutions in GR that are supported by electromagnetic charges.

The classical Reissner-Nordström (RN) geometry~\cite{Reissner,Weyl,Nordstrom,Baker} represents a static and spherical solution to the Einstein field equations~\eqref{chap2 Einstein Eqs} in presence of an electromagnetic field. In the region outside of the event horizon, the source term in the field equation reads
\begin{align}
    T_{\mu \nu} = \frac{1}{4 \pi}\left(\tensor{F}{_\mu^\rho} F_{\nu \rho} - \frac{1}{4} g_{\mu \nu} F^{\rho\sigma} F_{\rho\sigma}\, \right)\ , \label{EFE}
\end{align}
which is the energy-momentum tensor related to a $U(1)$ gauge vector field $A^\mu$, with $F_{\mu \nu}=-F_{\nu \mu}= \nabla_\mu A_\nu - \nabla_\nu A_\mu$ being the field strength tensor. This gauge field is subject to the Maxwell equations in the curved background as well as the Bianchi identity:
\begin{align}
 g^{\rho \mu} \nabla_{\rho} F_{\mu \nu} = 0 \qq{and} \nabla_{[ \mu} F_{\nu \rho ]} = 0 \ . \label{MFE}
\end{align}
Here $\nabla_\mu$ is the covariant derivative obtained through the metric $g_{\mu \nu}$. Assuming spherical symmetry and time invariance, as in the metric~\eqref{spherical static metric}, we notably obtain the Reissner-Nordström solution, that takes the following form \cite{Carroll}:
\begin{align}\label{RNMetric}     
&h(r) = f(r) = 1 - \frac{2M}{r} + \frac{Q^2}{r^2} \ , 
\end{align}
in which $M$ is the mass of the black hole and $Q^2=q^2 + m^2$, where $q$ and $m$ are its electric and magnetic charges respectively. We also remark that the limit $Q \rightarrow 0$, leads to the Schwarzschild space-time. On this metric, the solution to the electromagnetic field equations~\eqref{MFE} are given by the point-like charge Coulomb electric field and the magnetic monopole field
\begin{equation}\label{classical RN em fields}
    E_r = F_{rt} = \frac{q}{r^2} \qq{and} B_r = \frac{F_{\theta \varphi}}{r^2 \sin^2 \theta} =\frac{m}{r^2} \ .
\end{equation}
Similarly as in the Kerr BH, we have two roots for the metric function 
\begin{equation}
    r_\pm = M \pm \sqrt{M^2 - Q^2} \ ,
\end{equation} 
representing the inner and outer horizons. They are real and distinct if $Q < M$, they coincide in the case of $Q=M$ giving an extremal RN BH and if $Q > M$ we will have a naked singularity. However, many studies have proven that the inner horizon (also known as a Cauchy horizon) is unstable under perturbations ~\cite{Matzner:1979zz,Poisson:1989zz,Chandrasekhar1982,Hollands:2019whz}. 

\subsection{Uniqueness and the No-Hair Theorems}\label{sec:uniqueness}

Within classical general relativity, the family of stationary, asymptotically flat black hole solutions in four dimensions is remarkably restricted to the ones discussed above. According to the \emph{black hole uniqueness theorems}~\cite{Israel:1967wq,Carter:1971zc,Israel:1967za,Robinson:1975bv}, any such solution of the Einstein–Maxwell equations is completely characterized by only three global conserved quantities: mass $M$, angular momentum $J$, and electric charge $Q$. This implies that, under reasonable physical assumptions (such as the presence of an event horizon and the absence of naked singularities), the only admissible stationary black holes are the Schwarzschild, Reissner–Nordström, and Kerr (or Kerr–Newman in the charged and rotating case~\cite{Newman:1965my}) solutions. This simplicity is encapsulated in the \emph{no-hair theorem}~\cite{Wheeler:1971ba,Bekenstein:1972ky,Bekenstein:1972ny,Heusler:1996ft,Chrusciel:2012jk,Herdeiro2015}, which states that all higher multipole moments of the gravitational and electromagnetic fields are determined by these three parameters alone. Consequently, any additional structure or deviation from these metrics must arise from new physics beyond classical general relativity, such as modifications of the gravitational action, coupling to non-linear matter fields, or quantum-gravity effects.

\section{GR in 2+1 dimensions}\label{ClassicalBTZ}
Moving down to the case of general relativity in 2+1 dimensions, we see that the Newton's gravitational constant acquires mass dimensions $[G_3]=M^{-1}$. This means that the theory is still non-renormalizable by power counting, as in the 3+1 dimensions, but the product $G_3M$ is now dimensionless, signaling that there is no physical characteristic scale. 
Examining thoroughly the geometrical features, we notice that the curvature tensors are algebraically constrained in vacuum. This can be shown by the following argument: let us count the degrees of freedom of the $d$-dimensional Riemann curvature and Ricci tensors
\begin{equation}
    N_{\rm Riem}^{(d)}=\frac{d^2(d^2-1)}{12} \qq{and} N_{\rm Ric}^{(d)}=\frac{d(d+1)}{2} \ .
\end{equation}
Hence, the number of degrees of freedom carried by the Weyl tensor can be obtained as
\begin{equation}
    N_{\rm Weyl}^{(d)} = N_{\rm Riem}^{(d)} - N_{\rm Ric}^{(d)} \quad \overset{d=3}{\longrightarrow} \quad N_{\rm Weyl}^{(3)}  = 0 \ .
\end{equation}
This leads us to conclude that the Weyl tensor is identically vanishing in 3-dimensional spacetimes and, hence, that the Riemann tensor can be decomposed solely in terms of $R_{\mu \nu}$ and $R$ as follows
\begin{equation}\label{3d riemann}
    R_{\mu \nu \rho \sigma} = g_{\mu \rho}R_{\nu \sigma}-g_{\mu \sigma}R_{\nu \rho}+g_{\nu \sigma}R_{\mu \rho} - g_{\nu \rho}R_{\mu \sigma} - \frac{R}{2}(g_{\mu \rho}g_{\nu \sigma} - g_{\mu \sigma}g_{\nu \rho}) \ .
\end{equation}
Now, using the Einstein equations with the presence of a cosmological constant
\begin{equation}
    R_{\mu \nu} - \frac{1}{2}R\, g_{\mu \nu} + \Lambda g_{\mu \nu} = 0 \quad \to \quad R = 6 \Lambda \ , \quad R_{\mu \nu} = 2 \Lambda g_{\mu \nu} \ ,
\end{equation}
we can insert it into the Riemann tensor decomposition~\eqref{3d riemann} and obtain
\begin{equation}
    R_{\mu \nu \rho \sigma} = \Lambda (g_{\mu \rho}g_{\nu \sigma} - g_{\mu \sigma}g_{\nu \rho}) \ ,
\end{equation}
concluding that any vacuum solution in 3-dimensional spacetimes must be maximally symmetric.
Moreover, the general number of graviton polarizations is $d(d-3)/2$, which clearly vanishes for 3-dimensional spacetimes, signaling that all the metric components are just pure gauge. Such argument suffices to state that pure gravity in 2+1 dimensions is finite and any higher order curvature operator is renormalizable.\\

Therefore, nontrivial spacetimes can arise through global structure, not local curvature, and this is the reason why the authors in~\cite{Banados:1992wn,Banados:1992gq} successfully obtained a metric that shares the ``general features'' of a BH spacetime while being locally isomorphic to an AdS space. The natural way to introduce the BTZ metric is via the quotient construction of the universal covering space $\widetilde{ \rm AdS}_3$ by some group of isometries~\cite{Carlip:1995qv}. Let us introduce the $\widetilde{ \rm AdS}_3$ space with radius $\ell$ as embedded in the ambient space $\mathbb{R}^{2,2}$
\begin{equation}
    -T_1^2-T_2^2+X_1^2+X_2^2=-\ell^2 \ ,
\end{equation}
with flat metric
\begin{equation}\label{R22flatmetric}
    \dd s^2 = - \dd T_1^2 - \dd T_2^2 + \dd X_1^2 + \dd X_2^2 \ .
\end{equation}
To find a representation of the isometries, we can collect the coordinates into a $2\times 2$ matrix
\begin{equation}
    \mathbf{X} = \frac{1}{\ell} \begin{pmatrix}
        T_1 + X_1 & T_2+X_2 \\ -T_2+X_2 & T_1 - X_1
    \end{pmatrix} \ ,
\end{equation}
such that $\det (\mathbf{X})= 1$, so $\mathbf{X}\in SL(2,\mathbb{R})$. We can now define the isometries as $(\rho_L,\rho_R) \in SL(2,\mathbb{R})\times SL(2,\mathbb{R})/\mathbb{Z}_2 \sim SO(2,2)$, such that $\mathbf{X} \mapsto \rho_L \mathbf{X} \rho_R$ and $(\rho_L,\rho_R) \sim (-\rho_L, - \rho_R)$. The universal covering space $\widetilde{ \rm AdS}_3$ can be now described by the adapted coordinates
\begin{subequations}\label{Ads3Emdedding}
\begin{align}
    X_1 &= \ell \frac{r}{r_+} \sinh\left(\frac{r_+}{\ell} \phi \right) \\ X_2 &= \ell \frac{r}{r_+} \cosh\left(\frac{r_+}{\ell^2}t  \right) \\ T_1 &= \ell \frac{r}{r_+} \cosh\left(\frac{r_+}{\ell} \phi \right) \\ T_2 &= \ell \frac{r}{r_+} \sinh\left(\frac{r_+}{\ell^2}t \right) \, .
\end{align}
\end{subequations}
where $\phi \in [- \infty, + \infty]$ and $t \in [-\infty, + \infty]$. Clearly, $\phi$ is not compact as a proper angular variable, and to enforce it we have to realize the periodic identification $\phi \sim \phi + 2 \pi$ through the quotient. To do this, we take the boosts in the $X_1$-$T_1$ plane given with parameter $\alpha$ by the following group elements
\begin{equation}
    \rho_L(\alpha) = \rho_R (\alpha)= \begin{pmatrix}
        e^{\alpha r_+/\ell} & 0 \\ 0 & e^{-\alpha r_+/\ell}
    \end{pmatrix} \ ,
\end{equation}
which induces the transformation $\phi \mapsto \phi + 2 \alpha$. Then, we take the quotient $\widetilde{\rm AdS}_3/\langle(\rho_L(\pi), \rho_R(\pi))\rangle$, in which $\langle(\rho_L(\pi), \rho_R(\pi))\rangle$ indicates the orbits generated by $(\rho_L(\pi), \rho_R(\pi))$. Now, with the correct identification of $\phi$, we can take the coordinate transformation in Eqs.~\eqref{Ads3Emdedding} and transform the metric~\eqref{R22flatmetric}, yielding
\begin{equation}
    \dd s^2 = - \frac{r^2-r_+^2}{\ell^2}\dd t^2 + \frac{\ell^2}{r^2-r_+^2} \dd r^2 + r^2 \dd \phi^2 \ .
\end{equation}
The parameter $r_+$ has now the clear interpretation as the event horizon radius and is related to the ``mass'' $M$ via the asymptotics $r_+ = 2\ell \sqrt{2G_3M}$. The case can be generalized to a rotating metric around the angle $\phi$ with the introduction of the inner horizon $r_-$ in the boosts $\rho_R(\alpha) \neq \rho_L(\alpha)$, which reflects the $SL(2,\mathbb{R}) \times SL(2,\mathbb{R})$ structure of the algebra of isometries of AdS$_3$. Having modified the boosts along the $X_2$-$T_2$ plane and keep the coordinate $\phi$ as an angular variable and stationarity of the metric~\cite{Carlip:1995qv} we obtain the rotating BTZ metric in terms of the mass $M$ and angular momentum $J$ \footnote{Such procedure realizes the isometries, i.e. the Killing vectors $\partial_t$ and $\partial_\varphi$, of the spacetime as derivation of identifications and quotients of the embedding of AdS$_3$ in $\mathbb{R}^{2,2}$. For more insights see~\cite{Carlip:1995qv}.}
\begin{equation}
    \dd s^2 = - \left(- 8 G_3 M - \frac{r^2}{\ell^2}+\frac{J^2}{4 r^2} \right) \dd t^2 + \left(- 8 G_3 M - \frac{r^2}{\ell^2}+\frac{J^2}{4 r^2}   \right)^{-1}\dd r^2 + r^2 \left(-\frac{J}{2 r^2}\dd t + \dd \varphi \right)^2 \ ,
\end{equation}
where the inner and outer event horizons are given by
\begin{equation}\label{btzhorizons}
    r_\pm^2 = \frac{8 G_3 M \ell^2}{2} \left(1 \pm \sqrt{1- \left(\frac{J}{8G_3M \ell}\right)^2} \right) \ ,
\end{equation}
which are related to the conserved quantities
\begin{equation}
    M= \frac{r_+^2 + r_-^2}{8G_3\ell^2} \qq{and} J=\frac{2 r_+r_-}{\ell} \ .
\end{equation}
The bound on $J$ in order the square root to be real in~\eqref{btzhorizons} reads
\begin{equation}
    J \leq 8 G_3 M \ell \ ,
\end{equation}
in which the equality enforces an extremal configuration $r_+=r_-$. Remarkably, this metric can be seen as the result of a gravitational matter collapse, which has no curvature singularity at the center, while sharing the causal structure and thermodynamical features of a typical 3+1-dimensional black hole.

\section{BH Physics}

\subsection{Geodesic motion around spherically symmetric black holes}\label{Sect:PhotonRadius}
We study the motion of point-like uncharged probes in the background \eqref{spherical static metric} by analyzing its geodesic structure. The dynamics is governed by the worldline action 
\begin{equation}\label{worldlineaction}
    S = \frac{1}{2}\int \dd \tau \, g_{\mu \nu}\dot{x}^\mu \dot{x}^\nu \ ,
\end{equation}
and the corresponding Lagrangian
\begin{equation}\label{pointlike particle lagrangian}
    \mathcal{L}=\frac{1}{2}g_{\mu \nu} \dv{x^\mu}{\tau} \dv{x^\nu}{\tau} \equiv \frac{u}{2}\ ,
\end{equation}
where $x^\mu$ is the coordinate of the particle, $\tau$ is an affine parameter and the constant $u=+1,0,-1$ identifies respectively space-like (tachyons), light-like (or null) and time-like (massive) particles and it originates from the normalization of the four-velocity $u^\mu = \dd x^\mu / \dd \tau$.
It is remarkable that the symmetries of the spacetime become also symmetries of the particle's motion. We can see it by writing out the Lagrangian~\eqref{pointlike particle lagrangian} and, for convenience and in virtue of the spherical symmetry, set on the equatorial plane $\theta=\pi/2$
\begin{equation}\label{spelled out lagrangian}
    \mathcal{L}= -\frac{h(r)}{2} \left(\dv{t}{\tau} \right)^2+\frac{1}{2 f(r)} \left(\dv{r}{\tau} \right)^2+\frac{r^2}{2} \left(\dv{\varphi}{\tau} \right)^2 \ .
\end{equation}
We see that the Lagrangian does not depend explicitly on the time coordinate $t$ and azimuthal angle $\varphi$, hence their conjugate momenta $p_t$ and $p_\varphi$ are constants of motion
\begin{subequations}
    \begin{align}
        p_t &\coloneqq \pdv{\mathcal{L}}{(\dd t/ \dd \tau)} = -h(r) \dv{t}{\tau} \equiv E \ \longrightarrow \dv{t}{\tau}=-\frac{E}{h(r)} \ , \label{geodesic energy}\\
        p_\varphi & \coloneqq \pdv{\mathcal{L}}{(\dd \varphi/ \dd \tau)}= r^2 \dv{\varphi}{\tau} \equiv L \ \longrightarrow \dv{\varphi}{\tau} = \frac{L}{r^2}\ , \label{geodesic angular momentum}
    \end{align}
\end{subequations}
where we have defined the total energy of the particle $E$ and the angular momentum $L$ as conserved quantities. Using the relations~\eqref{geodesic energy} and~\eqref{geodesic angular momentum} into the Lagrangian in Eq.~\eqref{pointlike particle lagrangian}, we obtain
\begin{equation}
    -\frac{E^2}{h(r)} + \frac{1}{f(r)} \left(\dv{r}{\tau} \right)^2 + \frac{L^2}{r^2} = u \ ,
\end{equation}
which can be re-arranged in the form of a one-dimensional point particle moving in an effective potential
\begin{equation} \label{geodesic eq u}
    \left(\dv{r}{\tau} \right)^2 = \frac{f(r)}{h(r)} \left[E^2 - h(r) \left(\frac{L^2}{r^2}-u \right)  \right] \ ,
\end{equation}
This expression is general and is valid for space-like, light-like and time-like particles. Let us now specialize to the physically relevant cases: the light-like case first and then examine the time-like one.
\subsubsection{Massless particles}
Dividing both members of Eq.~\eqref{geodesic eq u} by~\eqref{geodesic angular momentum} for $u=0$, we have the radial equation of massless particles in the geometry (\ref{spherical static metric}) 
\begin{align}\label{effradeq}
\left(\dv{r}{\varphi} \right)^2 = r^2 f(r)\left( \frac{r^2}{b^2 h(r)}-1 \right) \qq{with} b \coloneqq L/E \ ,
\end{align}
where $b$ is the impact parameter. Equation \eqref{effradeq} can be read as a one-dimensional energy conservation law \cite{Perlick:2021aok,Carroll:2004st}  in the sense $(\dd r/\dd \varphi)^2 + V_\text{eff}=0$, with the effective potential $ V_\text{eff}$ given by the right hand side in (\ref{effradeq}). The particle follows a circular orbit (\emph{i.e.} with $r=\rps=$const.) for $(\rps,b)$ that satisfy 
\begin{align}
 &V_\text{eff} = 0 \qq{and} \dv{V_\text{eff}}{r}=0 &&\text{such that} &&\dv{r}{\varphi}\bigg|_{r=\rps} = \dv[2]{r}{\varphi}\bigg|_{r=\rps} = 0\ ,\label{PhotonRadius}
\end{align}
It is convenient to re-formulate the problem in terms of the minimal radius distance $\rmin$ of a massless particle on a general open trajectory  \cite{Perlick:2021aok,Carroll:2004st,misner1973gravitation} (see Figure~\ref{fig:bhshadow}), such that (\ref{effradeq}) becomes~\cite{Bozza:2002zj}
\begin{align}
&\left(\dv{r}{\varphi}\right)^2 =  r^2 f(r)\left( \frac{U(\rmin)^2}{U(r)^2} - 1 \right) \ ,&&\text{with} &&U(r)^2=\frac{h(r)}{r^2}\,,\label{PhotonPotentialU}
\end{align}
where the new effective potential satisfies $U(\rmin)=b^{-1}$. Furthermore, the inclination angle $\alpha$ measured by a static observer located at a (large) radial distance $\robs$ is \cite{Synge1966,Perlick:2015vta,Perlick:2021aok,PhysRevD.65.103004}
\begin{align}
&\cot \alpha = \frac{1}{\sqrt{r^2f(r)}} \left. \dv{r}{\varphi} \right|_{r=\robs} &&\text{such that} &&  \sin^2 \alpha = \frac{U(\robs)^2}{U(\rmin)^2}\ ,
\end{align}
from which we can obtain the angular aperture of the black hole shadow as an observer-dependent quantity
\begin{equation}\label{bphdef}
    \sin^2 \alpha_\text{sh} = U(\robs)^2 \bsh^2 \qq{with} \bsh := U(\rps)^{-1} \ .
\end{equation}
Here $\bsh$ is usually called the \emph{radius of the black hole shadow} \cite{Perlick:2021aok} and, as before, $\rps$ is the radius of the photon sphere, which is computed from (\ref{PhotonRadius}).Using the potential in (\ref{PhotonPotentialU}), the latter condition becomes
\begin{equation}\label{rpheq}
    \left.\dv{U(r)^2}{r}\right|_{r=\rps}=0 \implies \rps h^\prime(\rps) = 2 h(\rps) \hspace{0.2cm}\ ,
\end{equation}
where the prime indicates differentiation w.r.t. $r$.
In principle, there may be more than one solution to this equation, allowing for the presence of several photon rings, some of which are stable (around which photons oscillate) and some unstable (towards which photons spiral) \cite{Gan:2021pwu}. However, following \cite{Cunha:2020azh} and \cite{Claudel:2000yi}, we expect at least one local maximum for the effective potential $U$ outside of the event horizon.\\

Using Eqs.~\eqref{geodesic energy} and~\eqref{geodesic angular momentum}, together with the circular conditions~\eqref{PhotonRadius}, the photons orbit at the photon sphere with an angular frequency given by \cite{Perlick:2021aok}
\begin{equation}\label{photon-sphere-frequency}
    \Omega_{\rm ps} \coloneqq \left.\dv{\varphi}{t}  \right|_{r = \rps} = \frac{1}{\bsh} \ .
\end{equation}

The photon sphere can be further characterized in terms of the Lyapunov exponent, which quantifies the instability of such orbit. Let us perturb the photon sphere radius as
\begin{equation}
    r(\tau) = \rps + \delta r(\tau) \ ,
\end{equation}
and insert it into Eq.~\eqref{geodesic eq u} for $u=0$, obtaining
\begin{equation}\label{Lyapunov-mid-step}
    \left(\dv{\delta r}{\tau} \right)^2 = -\frac{f(\rps)}{2h(\rps)} V_{\rm eff}^{\prime \prime}(\rps) \delta r^2 \ ,
\end{equation}
where we defined the effective potential 
\begin{equation}
    V_{\rm eff}(r) \coloneqq h(r) \frac{L^2}{r^2} \ ,
\end{equation}
and used the circular orbit conditions~\eqref{PhotonRadius} in the expansion of the r.h.s. of Eq.~\eqref{geodesic eq u}. We can now differentiate Eq.~\eqref{Lyapunov-mid-step} w.r.t. $\tau$ and divide by $2 \dd \delta r/ \dd \tau$,  obtaining the linearized equation of motion
\begin{equation}
    \dv[2]{\delta r}{\tau} = - \frac{f(\rps)}{2h(\rps)} V_{\rm eff}^{\prime \prime}(\rps) \delta r \ ,
\end{equation}
which is solved for $\delta r (\tau) \sim e^{\lambda_\tau \, \tau}$, where $\lambda_\tau$ is the Lyapunov exponent associated to the affine parameter $\tau$, given by
\begin{equation}
    \lambda_\tau^2 = - \frac{f(\rps)}{2h(\rps)} V_{\rm eff}^{\prime \prime}(\rps) \ .
\end{equation}
The Lyapunov exponent with respect to the coordinate time $t$ can be obtained using the last expression in Eq.~\eqref{geodesic energy}, namely
\begin{equation}
    \lambda_t = \dv{\tau}{t} \lambda_\tau = \frac{h(\rps)}{E} \lambda_\tau \ ,
\end{equation}
and using the circular orbit conditions~\eqref{PhotonRadius}, yields
\begin{equation}\label{Lyapunov-t}
    \lambda_t^2 = - \frac{\rps^2 f(\rps)}{2}\left.\dv[2]{U(r)^2}{r}\right|_{r=\rps} \ ,
\end{equation}
where $U^2(r)$ is given in Eq.~\eqref{PhotonPotentialU}.
We note that if the photon sphere is unstable (stable), namely $U_{\rm eff}^{\prime \prime}(\rps) < 0$ ($U_{\rm eff}^{\prime \prime}(\rps)>0$), then $\lambda_t >0$ ($\lambda_t<0$). The Lyapunov exponent of the photon sphere for the Schwarzschild BH, is $\lambda_t = \bsh^{-1} = 1/3\sqrt{3}M$.

\begin{figure}[t]
    \centering
\tikzset{every picture/.style={line width=0.75pt}} 
\resizebox{0.65\columnwidth}{!}{
\begin{tikzpicture}[x=0.75pt,y=0.75pt,yscale=-1,xscale=1]

\draw  [draw opacity=0][fill={rgb, 255:red, 0; green, 0; blue, 0 }  ,fill opacity=1 ] (86,140) .. controls (86,121.22) and (101.22,106) .. (120,106) .. controls (138.78,106) and (154,121.22) .. (154,140) .. controls (154,158.78) and (138.78,174) .. (120,174) .. controls (101.22,174) and (86,158.78) .. (86,140) -- cycle ;
\draw  [color={rgb, 255:red, 208; green, 2; blue, 27 }  ,draw opacity=1 ] (65,140) .. controls (65,109.62) and (89.62,85) .. (120,85) .. controls (150.38,85) and (175,109.62) .. (175,140) .. controls (175,170.38) and (150.38,195) .. (120,195) .. controls (89.62,195) and (65,170.38) .. (65,140) -- cycle ;
\draw [color={rgb, 255:red, 255; green, 255; blue, 255 }  ,draw opacity=1 ]   (120,140) -- (135.07,167.09) ;
\draw [shift={(136.53,169.71)}, rotate = 240.91] [fill={rgb, 255:red, 255; green, 255; blue, 255 }  ,fill opacity=1 ][line width=0.08]  [draw opacity=0] (5.36,-2.57) -- (0,0) -- (5.36,2.57) -- cycle    ;
\draw [color={rgb, 255:red, 208; green, 2; blue, 27 }  ,draw opacity=1 ]   (120,140) -- (79.77,172.95) ;
\draw [shift={(77.45,174.85)}, rotate = 320.68] [fill={rgb, 255:red, 208; green, 2; blue, 27 }  ,fill opacity=1 ][line width=0.08]  [draw opacity=0] (5.36,-2.57) -- (0,0) -- (5.36,2.57) -- cycle    ;
\draw [color={rgb, 255:red, 74; green, 144; blue, 226 }  ,draw opacity=1 ]   (120,140) -- (100.82,72.88) ;
\draw [shift={(100,70)}, rotate = 74.05] [fill={rgb, 255:red, 74; green, 144; blue, 226 }  ,fill opacity=1 ][line width=0.08]  [draw opacity=0] (5.36,-2.57) -- (0,0) -- (5.36,2.57) -- cycle    ;
\draw [color={rgb, 255:red, 74; green, 144; blue, 226 }  ,draw opacity=1 ] [dash pattern={on 4.5pt off 4.5pt}]  (60,10) -- (362.82,140) ;
\draw  [dash pattern={on 0.84pt off 2.51pt}]  (120,140) -- (362.82,140) ;
\draw [color={rgb, 255:red, 150; green, 144; blue, 226 }  ,draw opacity=1]   (100,70) .. controls (181.33,54.91) and (350.33,135.92) .. (362.82,140) ;
\draw    (363.82,144.83) .. controls (361.4,143.93) and (361.09,136.67) .. (363.82,135.67) ;
\draw    (370,140.69) .. controls (367.84,139.79) and (364.39,139.79) .. (362.52,132.39) ;
\draw    (370,140.69) .. controls (367.84,141.59) and (364.39,141.59) .. (362.52,148.99) ;
\draw [fill={rgb, 255:red, 0; green, 0; blue, 0 }  ,fill opacity=1 ]   (362.04,143.04) .. controls (363.32,142.46) and (363.49,137.79) .. (362.04,137.14) ;

\draw [color={rgb, 255:red, 74; green, 144; blue, 226 }  ,draw opacity=1 ] [dash pattern={on 4.5pt off 4.5pt}]  (280,140) -- (362.82,140) ;
\draw [color={rgb, 255:red, 150; green, 144; blue, 226 }  ,draw opacity=1 ]   (10.61,177) .. controls (16.82,148.14) and (38.65,85.85) .. (100,70) ;
\draw [color={rgb, 255:red, 74; green, 144; blue, 226 }  ,draw opacity=1 ] [dash pattern={on 4.5pt off 4.5pt}]  (280,140) -- (280,106.11) ;
\draw  [draw opacity=0][dash pattern={on 1.69pt off 2.76pt}][line width=1.5]  (280,140) .. controls (280,139.94) and (280,139.87) .. (280,139.81) .. controls (280,128.48) and (282.93,117.83) .. (288.08,108.59) -- (344.21,139.81) -- cycle ; \draw  [color={rgb, 255:red, 74; green, 144; blue, 226 }  ,draw opacity=1 ][dash pattern={on 1.69pt off 2.76pt}][line width=1.5]  (280,140) .. controls (280,139.94) and (280,139.87) .. (280,139.81) .. controls (280,128.48) and (282.93,117.83) .. (288.08,108.59) ;  
\draw [color={rgb, 255:red, 74; green, 144; blue, 226 }  ,draw opacity=1 ]   (120,140) -- (156.4,54.62) ;
\draw [shift={(157.57,51.86)}, rotate = 113.09] [fill={rgb, 255:red, 74; green, 144; blue, 226 }  ,fill opacity=1 ][line width=0.08]  [draw opacity=0] (7.14,-3.43) -- (0,0) -- (7.14,3.43) -- (4.74,0) -- cycle    ;

\draw (131,142.4) node [anchor=north west][inner sep=0.75pt]  [font=\scriptsize,color={rgb, 255:red, 255; green, 255; blue, 255 }  ,opacity=1 ]  {$\hr$};
\draw (85,168) node [anchor=north west][inner sep=0.75pt]  [font=\scriptsize,color={rgb, 255:red, 208; green, 2; blue, 27 }  ,opacity=1 ]  {$\rps$};
\draw (285.83,116.17) node [anchor=north west][inner sep=0.75pt]  [color={rgb, 255:red, 74; green, 144; blue, 226 }  ,opacity=1 ]  {$\alpha $};
\draw (298.38,144.76) node [anchor=north west][inner sep=0.75pt]  [font=\scriptsize,color={rgb, 255:red, 74; green, 144; blue, 226 }  ,opacity=1 ]  {$\sqrt{g_{rr}} \dd r$};
\draw (226.06,114.63) node [anchor=north west][inner sep=0.75pt]  [font=\scriptsize,color={rgb, 255:red, 74; green, 144; blue, 226 }  ,opacity=1 ]  {$\sqrt{g_{\varphi \varphi }} \dd \varphi $};
\draw (106.4,72) node [anchor=north west][inner sep=0.75pt]  [font=\scriptsize,color={rgb, 255:red, 74; green, 144; blue, 226 }  ,opacity=1 ]  {$\rmin$};
\draw (149.83,74.46) node [anchor=north west][inner sep=0.75pt]  [font=\scriptsize,color={rgb, 255:red, 74; green, 144; blue, 226 }  ,opacity=1 ]  {$b$};
\draw (358.97,149.89) node [anchor=north west][inner sep=0.75pt]  [font=\scriptsize,color={rgb, 255:red, 74; green, 144; blue, 226 }  ,opacity=1 ]  {$r_\text{O}$};

\end{tikzpicture}
}

\caption{Observer-dependent inclination angle $\alpha$ and the minimal radial distance $\rmin$ for a photon on an open trajectory around the black hole. The impact parameter $b$ is measured by an observer at the (large) distance $\robs$. The red circle $\rps$ represents the photon ring (\emph{i.e.} a circular orbit for massless particles around the black hole) and the associated critical impact parameter represent the radius of the black-hole cross-section for photon capture.}
    \label{fig:bhshadow}
\end{figure}
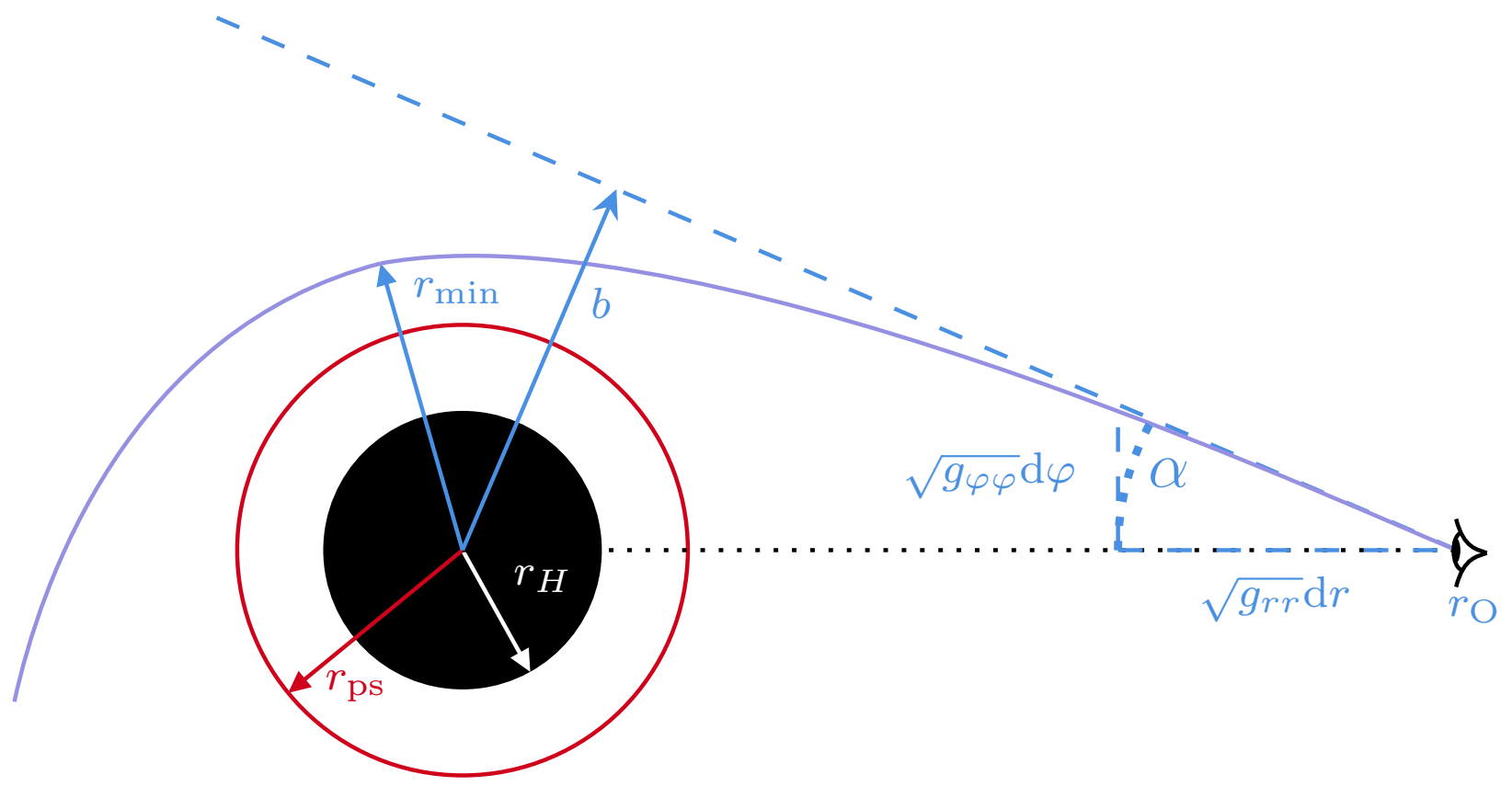
\begin{figure}
    \centering
    \includegraphics[width=0.7\linewidth]{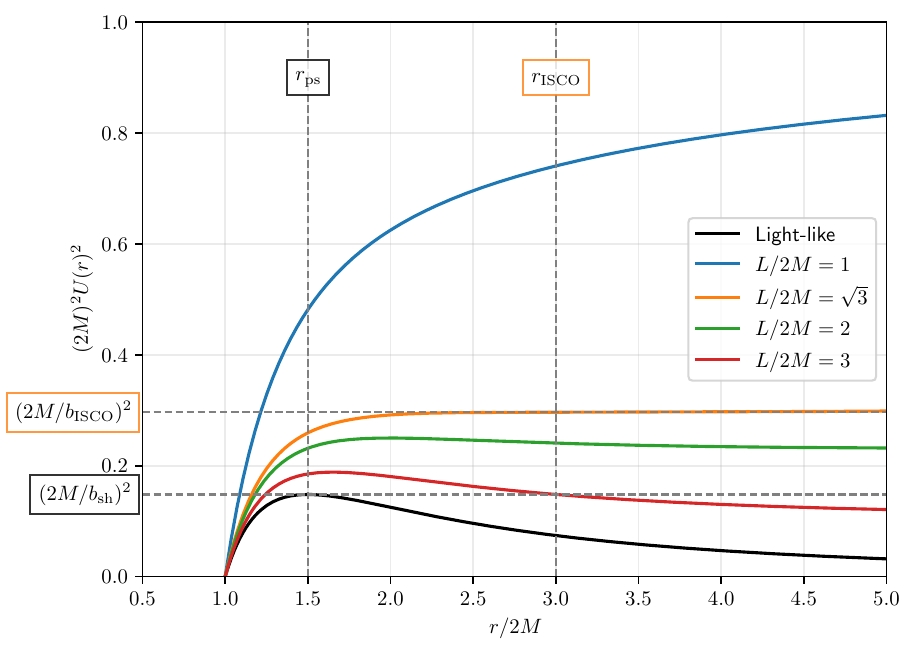}
    \caption{Effective potential of a Schwarzschild black hole for null geodesics given in Eq.~\eqref{PhotonPotentialU} (solid black line) and timelike geodesics, Eq.~\eqref{timelike-effective-potential} for different values of the angular momentum (colored solid lines). All quantities are normalized with respect to the Schwarzschild radius. The positions of the photon sphere $r_{\rm ps} = 3M$, the inverse shadow radius $b_{\rm sh}^{-1}=1/3\sqrt{3}M$, the innermost stable circular orbit $r_{\rm ISCO}=6M$, and the inverse of the corresponding impact parameter $b_{\rm ISCO}^{-1}=\sqrt{2}/3\sqrt{3}M$ are highlighted.}
    \label{fig:schwarzschild-effective.potentials}
\end{figure}

\subsubsection{Massive particles}
Let us now consider the case of timelike (massive) particles, so in Eq.~\eqref{geodesic eq u} we set $u=-1$, yielding the one-dimensional equation
\begin{equation}
    \left( \dv{r}{\tau}\right)^2 = \frac{f(r)}{h(r)}\left[E^2 -  U(r)^2 \right] \ ,
\end{equation}
where the timelike effective potential is 
\begin{equation}\label{timelike-effective-potential}
    U_{\rm eff}(r)^2 = h(r)\left( \frac{L^2}{r^2}+1\right) \ .
\end{equation}
Circular orbits are defined by the conditions:
\begin{equation}
    \left.\dv{r}{\tau}\right|_{r = r_c} = 0 \qq{and} \left.\dv{r}\left(\dv{r}{\tau}\right)^2\right|_{r = r_c} = 0 \ ,
\end{equation}
which are given in terms of the conserved quantities as
\begin{equation}\label{timelike-circular-orbits-E-L}
    E_c^2 = \frac{2 h(r_c)^2}{2h(r_c)-r_c h^\prime(r_c)} \qq{and} L_c^2  =  \frac{r_c^3 h^\prime(r_c)}{2h(r_c)-r_c h^\prime(r_c)} \ ,
\end{equation}
where the prime denotes differentiation w.r.t. the coordinate $r$. We note that the expressions in Eq~\eqref{timelike-circular-orbits-E-L} are independent on $f(r)$.
A timelike circular orbit can be either stable ($V_{\rm eff}^{\prime \prime}(r_c)>0$) or unstable ($V_{\rm eff}^{\prime \prime}(r_c)<0$), with the limiting case given by $V_{\rm eff}^{\prime \prime}(\risco) = 0$, which reads into the algebraic equation for the radius of the Innermost Stable Circular Orbit (ISCO)
\begin{equation}
    \risco h(\risco) h^{\prime \prime}(\risco) - 2 \risco h^\prime(\risco)^2 + 3 h(\risco)h^\prime(\risco)= 0 \ .
\end{equation}
The associated impact parameter (which is not critical in the same sense for the photon sphere) is
\begin{equation}
    b_{\rm ISCO}^2 = \frac{\risco^3 h^\prime(\risco)}{2 h(\risco)^2} \ .
\end{equation}
Using Eqs.~\eqref{geodesic energy} and~\eqref{geodesic angular momentum}, together with the conditions~\eqref{timelike-circular-orbits-E-L}, a massive particle orbits at the ISCO with an angular frequency given by
\begin{equation}
    \Omega_{\rm ISCO} \coloneqq \left.\dv{\varphi}{t}  \right|_{r = \risco} = \sqrt{\frac{h^\prime(\risco)}{2\,  \risco}} \ .
\end{equation}
The ISCO plays a central role in the physics of accreting black holes, as it determines the innermost stable radius at which matter can maintain a nearly circular orbit before plunging towards the horizon. In the standard thin-disk picture of accretion \cite{Shakura:1972te,Novikov:1973kta}, the ISCO is identified with the inner edge of the disk, and its location controls the efficiency of the conversion of gravitational binding energy into radiation. In particular, the radiative efficiency is given by
\begin{equation}
    \eta = 1-E_{\rm ISCO} \ ,
\end{equation}
where $E_{\rm ISCO}$ is the conserved energy per unit mass evaluated at the ISCO. Therefore, deviations from the Schwarzschild geometry can modify both the ISCO radius and the associated orbital frequency, leaving imprints on electromagnetic observables such as the thermal spectrum of accretion disks and the iron $K\alpha$ emission line \cite{Bambi:2015kza,Tripathi:2021rqs}. 

Furthermore, the ISCO frequency represents a characteristic dynamical scale of the innermost region of the accretion flow and can be related to quasi-periodic oscillations observed in X-ray binaries \cite{Stella:1997tc}. In the context of tests of strong-field gravity, the measurement of ISCO-related observables provides a complementary probe to black hole shadow measurements and gravitational-wave observations, allowing one to constrain possible deviations from general relativity in the near-horizon regime.

In Fig.~\ref{fig:schwarzschild-effective.potentials} we compare the effective potentials for lightlike particles (Eq.~\eqref{PhotonPotentialU}) and for timelike particles (Eq.~\eqref{timelike-effective-potential}) in a Schwarzschild background, at different values of the angular momentum. We highlighted the position of the photons sphere and ISCO and their associated (critical) impact parameters, (normalized w.r.t. the Schwarzschild radius).

\subsection{Quasi-Normal Modes}\label{sec:QNM}

Perturbations of BH spacetimes play a central role in probing their stability and dynamical properties. When a BH is perturbed, the spacetime geometry responds by emitting gravitational radiation that carries information about its intrinsic properties such as mass, charge, and angular momentum and eventually settles back to equilibrium. The intermediate stage of this relaxation process is characterized by a discrete set of damped oscillations known as \emph{quasi-normal modes} (QNMs)~\cite{Vishveshwara:1970zz,Press:1971wr,Chandrasekhar:1975zza,Kokkotas:1999bd,Nollert:1999ji,Berti:2009kk}. These modes are intrinsic to the black hole itself and depend solely on its parameters (mass, charge, angular momentum), being independent of the details of the perturbation.

The analysis of perturbations around a static, spherically symmetric metric can be traced back to the seminal work of Regge and Wheeler~\cite{Regge:1957td}, who studied the stability of the Schwarzschild geometry under odd-parity (axial) perturbations. Later, Zerilli~\cite{Zerilli:1970se} extended this analysis to even-parity (polar) perturbations, deriving the corresponding effective potential and wave equation.

Consider a generic field perturbation $\Psi(t,r,\theta,\varphi)$ of spin $s$ ($s=0,1,2$ respectively for scalar, electromagnetic and gravitational perturbations) on a static, spherically symmetric background of the form as in Eq.~\eqref{spherical static metric}.
After decomposing the perturbation into spherical harmonics and separating variables
\begin{equation}
    \Psi(t,r,\theta,\varphi) = \sum_{\ell\geq 0} e^{-i \omega t}\frac{\psi_\ell(r)}{r}{}_s Y_\ell^m(\theta ,\varphi) \ ,
\end{equation}
the linearized response of the BH to such perturbations is governed by a one-dimensional Schrödinger-like equation for the radial component $\psi_\ell(r)$~\cite{Regge:1957td,Zerilli:1970se,Kokkotas:1999bd,Berti:2009kk,Konoplya:2011qq}
\begin{equation}\label{eq:QNM-wave}
    \frac{\dd^2 \psi_\ell}{\dd r_*^2} + \left[\omega^2 - V_{\ell, s}(r)\right] \psi_\ell = 0 \ ,
\end{equation}
where $r_*$ is the \emph{tortoise coordinate} defined by $\dd r_*/ \dd r \coloneqq \sqrt{-g_{rr}/g_{tt}} = \sqrt{f(r)/h(r)}$, and $V_{\ell, s}(r)$ is an effective potential determined by the background metric, the multipole moment $\ell$ and the spin $s$ of the perturbation.

The quasi-normal modes correspond to the solutions of Eq.~\eqref{eq:QNM-wave} satisfying the boundary conditions of \emph{purely ingoing waves} at the event horizon and \emph{purely outgoing waves} at spatial infinity (or at the cosmological horizon in de Sitter backgrounds):
\begin{equation}
    \psi(r_*) \sim
    \begin{cases}
        e^{-i \omega r_*}  & r_* \to -\infty \ (r \to r_H) \ ,\\[4pt]
        e^{+i \omega r_*}  & r_* \to +\infty \ (r \to \infty) \ .
    \end{cases}
\end{equation}
These boundary conditions quantize the complex frequency spectrum $\omega = \omega_R + i \omega_I$, where $\omega_R$ governs the oscillation frequency and $\omega_I < 0$ describes the damping rate due to energy loss through gravitational radiation. 

The QNM spectrum encodes the linear response of the spacetime and provides a characteristic \emph{ringdown} signal in gravitational-wave observations~\cite{LIGOScientific:2016aoc,Abbott:2016nmj}. In particular, the detection of ringdown frequencies allows one to test the no-hair theorem by verifying that the observed modes are consistent with those predicted for a Kerr black hole~\cite{Maselli:2023khq}. Moreover, deviations from the general-relativistic spectrum may signal the presence of new physics, such as quantum-gravity corrections, extra dimensions, or modified theories of gravity~\cite{Cardoso:2016rao,Berti:2015itd,Barack:2018yly}.

Analytically, QNMs can be computed using various approximation methods such as the Pöschl-Teller potential~\cite{Blome:1981azp}, the WKB expansion~\cite{Schutz:1985km,Iyer:1986np,Konoplya:2003ii,Matyjasek:2017psv}, continued fractions~\cite{Leaver:1985ax,Nollert:1999ji}, or numerical integration of the wavelike equations~\cite{Gundlach:1993tp,Chandrasekhar:1975zza}.

\subsubsection{Eikonal limit}
\label{sec:eikonal_limit}

In the semiclassical eikonal limit (namely the limit of large angular momentum $\ell \gg 1$) or high overtones, the spectrum exhibits universal scaling properties closely linked to the structure of the effective potential barrier and, in the eikonal limit, to the properties of unstable null geodesics near the photon sphere~\cite{Cardoso:2008bp,Berti:2025hly}.
In this limit, the QNM frequency reads
\begin{equation}\label{eq:eikonal}
        \omega_{n \ell}=\Omega \,\ell-i\left(n+\frac{1}{2}\right)|\lambda_t|\ ,
\end{equation}
where $n \in \mathbb{N}$ is the overtone number, $\Omega$ is the orbital frequency of light rays at the photon-sphere radius $r_{\rm ps}$, and $\lambda_t$ is the coordinate-time Lyapunov exponent~\eqref{Lyapunov-t} \cite{Cardoso:2008bp}.

The eikonal limit establishes a correspondence between the QNMs and the properties of the photon sphere, under the following two criteria \cite{Konoplya2017}: (\emph{i}) the perturbation potential is positive definite (to avoid instabilities), single peaked, and decays to zero at the boundaries; and (\emph{ii}) the perturbation is a test scalar field or other field minimally coupled to gravity. This implies that the eikonal QNMs associated with gravitational perturbations may not, in general, be directly related to the properties of the unstable photon orbit. For this reason, we will restrict our analysis to perturbations of test scalar and electromagnetic fields. 

Given a static and spherically symmetric spacetime in Eq.~\eqref{spherical static metric}, the photon dynamics can be described using the effective potential $U$ defined in Eq.~\eqref{PhotonPotentialU}. The radius, orbital frequency~\eqref{photon-sphere-frequency} \cite{Bardeen:1972fi}, and Lyapunov exponent of the photon sphere are then given by \cite{Cardoso:2008bp}\footnote{Note the change of the metric signature with respect to eqs.~(35),~(37), and~(40) reported in Ref.~\cite{Cardoso:2008bp}.} 
\begin{subequations}\label{eq:EIKONAL-QUANTITIES}
\begin{align}
&\rps \, g_{tt}^\prime(\rps)=2 \, g_{tt}(\rps) \ ,\\
\Omega_{\rm ps}^2 & = U^2(r_{\rm ps})= -\frac{g_{tt}(r)}{r^2}\bigg|_{r=\rps} \ , \label{eq:orbital_frequency}\\
\lambda_t^2 & = -\frac{\rps^2}{2  \, g_{rr}(\rps)}\dv[2]{U^2(r)}{r}\bigg|_{r=\rps} \ ,\label{eq:Lyapunov_exponent}
\end{align}          
\end{subequations} 
respectively. 
\footnote{In Ref. \cite{Cardoso:2008bp}, the authors use a different potential $V_r(r)$ instead of $U^2(r)$.}

Altogether, quasi-normal modes represent the “spectroscopic fingerprint” of black holes, encoding information about their geometry and offering a unique observational window into the strong-field regime of gravity.

\section{Black hole thermodynamics } \label{hawking-classical-intro}


We consider a static, spherically symmetric metric of the form in Eq.~\eqref{spherical static metric} and assume an non-extremal event horizon at \(r=r_H\) where \(h(r_H)=f(r_H)=0\). The \emph{surface gravity} \(\kappa\) associated with the horizon generator
\((\chi^t)^\mu=\delta^{\mu0}\) can be defined geometrically by
\begin{equation}
    (\chi^t)^\nu \, \nabla_\nu (\chi^t)^\mu = \kappa\,(\chi^t)^\mu \qq{on} r=r_H \ .
\end{equation}
Equivalently, for the metric \eqref{spherical static metric} one obtains the convenient expression
\begin{equation}\label{eq:kappa-general}
    \kappa \;=\; \frac{1}{2}\,\sqrt{\hn{1}\,\fn{1}} \ .
\end{equation}
In most cases one has \(h(r)=f(r)\) which yields \(\kappa=\tfrac{1}{2}\fn{1}\).

The Hawking temperature follows from the requirement of regularity of the Euclidean section of the black-hole geometry. Introduce the Euclidean time with a Wick rotation \(\tau = i t\) and expand the metric~\eqref{spherical static metric} near the horizon. Let \(r=r_H + \delta r\) with \(\delta r\ll r_H\). To leading order
\begin{equation}
    h(r) \simeq \hn{1}\,\delta r \qq{and} f(r) \simeq \fn{1}\,\delta r \ .
\end{equation}
Define a radial coordinate \(\rho\) by
\begin{equation}
    \rho \coloneqq 2\sqrt{\frac{\delta r}{\fn{1}}} \quad\Longrightarrow\quad
    \delta r = \tfrac{1}{4} \fn{1}\,\rho^2 \ .
\end{equation}
Substituting into the Euclidean metric gives, to leading order,
\begin{equation}
    \dd s_E^2 \simeq \rho^2 \left(\frac{\hn{1}\fn{1}}{4}\right)\dd \tau^2 + \dd \rho^2 + r_H^2 \dd\Omega^2 \ .
\end{equation}
Using \eqref{eq:kappa-general} this becomes the flat polar form
\begin{equation}
    \dd s_E^2 \simeq \rho^2 \kappa^2 \dd \tau^2 + \dd \rho^2 + r_H^2 \dd \Omega^2 \ .
\end{equation}
Absence of a conical singularity at \(\rho=0\) (the horizon) requires the Euclidean time \(\tau\) to be periodic with period
\begin{equation}
    \beta \;=\; \frac{2\pi}{\kappa}\ .
\end{equation}
Interpreting \(\beta\) as the inverse temperature of the quantum field state regular on the horizon yields the Hawking temperature \cite{Bogoliubov1958,Wald:1984rg,Sarkar:2007uz,Hawking:1975vcx,birrell_davies_1982}
\begin{equation}\label{Hawkingtempdef}
    T_{\!H} \;=\; \frac{1}{\beta} \;=\; \frac{\kappa}{2\pi} \ .
\end{equation}

\begin{center}
\remarkbox{\rem\label{rem:classicaltemperatures}{Noteworthy Examples}}{\parbox{16.25cm}{
the Hawking temperatures for the Schwarzschild, Reissner-Nordström and Kerr metrics read respectively
\begin{equation}\label{classical temperatures}
    T_{\rm H}^{\rm Schw} = \frac{1}{8 \pi M} \ , \quad T_{\rm H}^{\rm RN} = \frac{\sqrt{M^2-Q^2}}{2 \pi(M + \sqrt{M^2-Q^2})^2} \qq{and} T_{\rm H}^{\rm Kerr} = \frac{\sqrt{M^2-a^2}}{2 \pi\left[(M + \sqrt{M^2-a^2})^2 + a^2\right]}
\end{equation}
Notice that in all cases the temperatures decrease with increasing mass, implying that large astrophysical BHs are extremely cold.
}}
\end{center}

\noindent
Bekenstein first proposed that BHs possess an entropy proportional to the area of their event horizon \cite{Bekenstein:1973ur,Hawking:1975vcx,Wald:1993nt}, leading to the \emph{Bekenstein–Hawking entropy}
\begin{equation}\label{BekHaw entropy}
    S_{\rm BH} = \frac{A_{H}}{4} \ ,
\end{equation}
where $A_{H}$ is the event horizon area measured in units of the Planck length $\ellp$. This entropy–area relation establishes BHs as the most entropic objects for a given mass.

These results complete the analogy between BH mechanics and standard thermodynamics:
\begin{itemize}
    \item The surface gravity plays the role of temperature.
    \item The horizon area plays the role of entropy.
    \item The ADM mass serves as internal energy.
\end{itemize}
and culminate in the \emph{first law of BH thermodynamics}
\begin{equation}
    \dd M = T_{\rm H}\, \dd S + \Omega_H \, \dd J + \Phi_H \, \dd Q\ ,
\end{equation}
holding for the most general stationary BH with angular momentum $J$ and electric charge $Q$, whose conjugate variables are respectively the angular velocity at the horizon $\Omega_H$ and the scalar potential at the horizon $\Phi_H$.
Notably, for static neutral and non-rotating BHs, one can read the entropy as
\begin{equation}\label{EntropyDef}
    S=\int \frac{\dd M}{T_{\rm H}} \ ,
\end{equation}
which, for the Schwarzschild case in Eq.~\eqref{classical temperatures}, yields the Bekenstein-Hawking entropy~\eqref{BekHaw entropy} $S=4\pi M^2$.

In the context of the AdS/CFT correspondence~\cite{Maldacena:1997re,Gubser:1998bc,Witten:1998qj,Hubeny:2007xt,Nishioka:2009un,Rangamani:2016dms,Emparan:2021hyr}, the Bekenstein–Hawking area law finds a natural realization in the Ryu–Takayanagi prescription~\cite{Ryu:2006bv,Ryu:2006ef}. There, the von Neumann entropy (also called entanglement entropy) of a quantum subsystem in the boundary conformal field theory (CFT) is given by the area of an extremal codimension-two surface in the bulk spacetime,
\begin{equation}
S_{\rm EE} = \frac{{\rm Area}(\gamma_A)}{4} \ ,
\end{equation}
where $\gamma_A$ is the minimal surface homologous to the boundary region $A$. This result establishes a direct link between quantum entanglement and spacetime geometry, providing a microscopic interpretation of the gravitational entropy and reproducing the area law as a manifestation of boundary entanglement.

Several approaches to quantum gravity, however, predict logarithmic corrections to the entropy–area law~\cite{Fursaev:1994te, Xiao:2021zly, Solodukhin:2011gn, Solodukhin:1994yz, Sen:2012dw, El-Menoufi:2015cqw, Cai:2009ua, Carlip:2000nv, Banerjee:2008cf, Banerjee:2008fz, Kaul:2000kf}. In a complementary direction, the seminal work by Jacobson~\cite{Jacobson:1995ab,Jacobson:1993vj,Eling:2006aw} interpreted the Einstein field equations as an equation of state: by requiring that the Clausius relation $\delta Q = T\,\delta S$ hold for all local Rindler horizons, one can recover the Einstein equations as a thermodynamic consistency condition. This thermodynamics-of-spacetime viewpoint suggests that gravity may emerge from microscopic degrees of freedom whose coarse-grained behaviour is encoded by horizon thermodynamics. The program of quantum phenomenological gravitational dynamics~\cite{Jacobson:2015hqa,Alonso-Serrano:2020dcz,Alonso-Serrano:2025spn} extends this insight by investigating how modifications of the entropy–area relation affect the gravitational field equations and the dynamics of spacetime.

Quantum emission implies BHs are not eternally stable: they lose mass via Hawking radiation and eventually evaporate. This process raises deep questions regarding information conservation and the microscopic origin of BH entropy. For extremal configurations ($Q=M$ for RN and $a=M$ for Kerr) the temperatures in Eq.~\eqref{classical temperatures} vanish implying that the BH does not emit thermal radiation, reaching a state that is known as \emph{remnant}, which is speculated to be the possible endpoint of the evolution of a BH that would preserve information~\cite{Page:1993up,Bianchi:2018mml,Chen:2014jwq}. Altogether, Hawking radiation and the entropy–area law establish BHs as legitimate thermodynamic systems, bridging general relativity, quantum theory, and statistical mechanics.

\section{Open issues}\label{sec:open-issues}

The classical picture reviewed so far, once confronted with quantum theory, leaves a number of fundamental questions unanswered. In this Section we briefly collect the open issues that are most relevant for the remainder of this review: they motivate the study of deformed black hole geometries and, at the same time, delimit the regime of validity of any effective metric approach.

\subsection{BH regularity at the origin}

The singularity theorems~\cite{Penrose:1964wq} establish that, under rather general assumptions on the causal structure and on the energy content of spacetime, gravitational collapse in classical GR inevitably produces geodesically incomplete spacetimes. In the Schwarzschild, Reissner--Nordstr\"om, and Kerr solutions this incompleteness manifests itself as a curvature singularity hidden behind the event horizon, in accordance with the cosmic censorship conjecture~\cite{Penrose:1969pc}. It is widely expected that quantum gravity effects regularize this singular behavior. In the absence of a complete theory, a large class of \emph{regular} black hole models has been proposed, in which the central singularity is replaced by a de Sitter-like core or by a smooth transition to another region of spacetime; prominent examples are the Bardeen~\cite{bardeen1968}, Dymnikova~\cite{Dymnikova:1992ux}, Hayward~\cite{Hayward_2006}, Frolov~\cite{Frolov:2016pav}, and Simpson--Visser~\cite{Simpson:2018tsi,Simpson:2019mud} geometries, several of which are analyzed within the EMD framework in Chapter~\ref{chap examples}. Regularity, however, comes at a price: these geometries generically feature an inner horizon, whose exponential blueshift instability, known as \emph{mass inflation}~\cite{Poisson:1990eh}, raises serious concerns about their viability as endpoints of gravitational collapse~\cite{Carballo-Rubio:2018pmi,Carballo-Rubio:2021bpr,Carballo-Rubio:2025ntd,Carballo-Rubio:2025fnc}. A systematic classification of regular cores, based on their asymptotic behavior at the origin and on geodesic completeness, has recently been carried out in~\cite{DamiaPaciarini:2025qhq}. Whether a fully consistent, dynamically stable regular black hole exists remains an open question, and any model-independent parametrization of the near-horizon and interior geometry, such as the EMD, provides a natural language in which to formulate it.

\subsection{Positivity conditions}\label{subsec:positivity}

Classical GR theorems, from the singularity theorems to the laws of black hole mechanics, rely on \emph{energy conditions}, namely positivity requirements on contractions of the stress-energy tensor such as the null, weak, strong, and dominant energy conditions (see~\cite{Curiel:2014zba,Kontou:2020bta} for comprehensive reviews). Quantum fields violate these pointwise conditions: the Hawking flux itself carries negative energy across the horizon, and regular black hole cores necessarily violate the strong energy condition. This tension raises the question of which positivity requirements a physically sensible deformed black hole geometry should still satisfy, and in which averaged or asymptotic form. Within the effective metric approach, this question can be made quantitative: interpreting the deformed geometry as a solution of Einstein's equations with an effective stress-energy tensor, positivity requirements translate into concrete constraints on the deformation functions and their expansion coefficients. This program was carried out in~\cite{DAlise:2023hls}, where positivity conditions and the asymptotic structure of the effective stress-energy tensor were shown to significantly restrict the space of admissible metric deformations, for both classical and quantum-corrected black holes. The systematic incorporation of such constraints into the EMD parameter space, alongside the regularity conditions derived in Chapter~\ref{chap3}, is an ongoing effort.

\subsection{Hawking radiation and the information paradox}

Hawking's discovery that black holes emit thermal radiation~\cite{Hawking:1975vcx} promoted the analogy between black hole mechanics and thermodynamics to a physical equivalence, but it simultaneously created the deepest open problem of the field. If the radiation is exactly thermal throughout the evaporation, an initially pure state evolves into a mixed one, in conflict with unitarity of quantum mechanics~\cite{Hawking:1976ra}. Unitarity instead requires the entanglement entropy of the radiation to follow the \emph{Page curve}, rising during the first half of the evaporation and decreasing thereafter~\cite{Page:1993wv}. Recent progress based on entanglement wedge reconstruction and the island prescription has shown that semiclassical gravity, refined by replica wormhole contributions to the gravitational path integral, is able to reproduce the Page curve~\cite{Penington:2019npb,Almheiri:2019psf,Almheiri:2020cfm}. These developments strongly suggest that the resolution of the paradox involves subtle, non-local corrections to the near-horizon physics, without necessarily requiring large deviations from the classical geometry far from the horizon. Precisely because such corrections are expected to be small and concentrated near the horizon, model-independent parametrizations of the near-horizon geometry, of the type developed in this review, offer a natural framework to quantify their imprint on observables.

\subsection{Black holes as quantum systems}

The modern viewpoint, often referred to as the \emph{central dogma}~\cite{Almheiri:2020cfm}, regards a black hole, as seen from the outside, as an ordinary quantum system with a finite number $e^{S_{\rm BH}}$ of degrees of freedom evolving unitarily, where $S_{\rm BH}$ is the Bekenstein--Hawking entropy~\cite{Bekenstein:1973ur,Hawking:1975vcx}. Identifying the microscopic degrees of freedom that account for this entropy remains an outstanding challenge for any theory of quantum gravity. A concrete corpuscular realization is provided by the quantum $N$-portrait~\cite{Dvali:2011aa}, in which the black hole is described as a self-sustained condensate of $N \sim M^2/M_{\rm P}^2$ soft gravitons at the critical point of a quantum phase transition. In this picture, Hawking radiation arises from the depletion of the condensate, and semiclassical results are recovered in the $N \to \infty$ limit, while at finite $N$ the geometry receives $1/N$ corrections. Such corrections, although minuscule for astrophysical black holes, accumulate over the evaporation time and can qualitatively alter the late-time behavior. From the perspective of this review, the lesson is that the metric description itself is an effective, coarse-grained notion: the EMD parametrizes precisely the deviations from the classical geometry that such underlying quantum descriptions may generate, without committing to a specific microscopic model.

\subsection{Memory burden}

A quantitative handle on the breakdown of the semiclassical description is provided by the \emph{memory burden} effect~\cite{Dvali:2018xpy,Dvali:2020wft}. Systems of enhanced memory storage capacity, of which black holes are the prime gravitational example, are universally stabilized by the backreaction of the information they carry: the energy cost of rewriting the stored quantum information suppresses the decay of the system. Applied to black holes, this implies that the self-similar Hawking evaporation cannot persist beyond, at the latest, the half-decay time, after which the black hole either becomes extremely long-lived or undergoes a qualitatively new decay process~\cite{Dvali:2020wft}. This conclusion has striking phenomenological consequences: the standard evaporation constraints on light primordial black holes are lifted, reopening the mass window below $\sim 10^{10}\,$g in which primordial black holes can constitute the entirety of dark matter~\cite{Alexandre:2024nuo,Thoss:2024hsr}. The memory burden effect underlines that quantum deviations from the classical metric are not necessarily confined to Planckian curvatures, but can become order-one for sufficiently old or light black holes. A model-independent description of the associated geometry deformations, and of their observational signatures, is a natural target for the EMD framework developed in the following chapters.

\section{Beyond Classical GR}\label{sec:beyond-gr}

The open issues discussed above have motivated a wide spectrum of extensions of classical GR, each predicting characteristic deformations of the black hole solutions reviewed in this chapter. Without any pretense of completeness (see~\cite{Berti:2015itd,Clifton:2011jh} for extensive reviews), in this Section we outline the main frameworks whose black hole geometries will serve as testing grounds for the EMD in the following chapters. The purpose of this overview is twofold: to illustrate the physical mechanisms through which the classical geometries get deformed, and to emphasize that the resulting corrections, while very different in origin, can all be cast in the form of modified metric functions, which is precisely the starting point of the effective metric approach.

\subsection{String theory and higher-curvature corrections}

String theory provides an ultraviolet-complete framework for quantum gravity in which the graviton arises as a massless excitation of the closed string. At energies well below the string scale, the dynamics is captured by a low-energy effective action that supplements the Einstein--Hilbert term with an infinite tower of higher-curvature operators, suppressed by powers of the Regge slope $\alpha'$, together with additional light fields, most notably the dilaton and antisymmetric tensor fields. Already the first correction, proportional to the Gauss--Bonnet invariant in the heterotic string~\cite{Zwiebach:1985uq}, modifies the Schwarzschild geometry and its thermodynamics: the $\alpha'$-corrected black holes of Callan, Myers and Perry~\cite{Callan:1988hs} exhibit shifted Hawking temperature and entropy, with corrections growing as the horizon approaches the string scale. The dilaton has an even more dramatic effect for charged solutions. In Einstein--Maxwell-dilaton theory, obtained as a consistent truncation of the heterotic string, the charged black holes of Gibbons--Maeda and Garfinkle--Horowitz--Strominger~\cite{Gibbons:1987ps,Garfinkle:1991bv} differ qualitatively from Reissner--Nordstr\"om: the dilaton coupling removes the inner horizon, the extremal limit becomes singular and, for the string value of the coupling, the Hawking temperature remains finite at extremality. These features make dilatonic black holes an instructive stress test for any parametrization designed around the classical two-horizon structure. A two-function axion-dilaton geometry~\cite{Garcia95} is analyzed in detail within the EMD framework in Section~\ref{Sect:DilatonBH}, where it provides an example in which the two metric deformation functions are genuinely independent, $h \neq f$.

\subsection{Effective field theory of gravity}

A conservative and model-independent path beyond classical GR is to treat it as an effective field theory (EFT), valid below the Planck scale, in which quantum effects are organized in a derivative expansion~\cite{Donoghue:1994dn}. While the ultraviolet completion is unknown, the infrared dynamics of gravitons is dictated by the symmetries of GR, and the leading long-distance quantum corrections are therefore calculable and unambiguous: graviton loops generate non-analytic contributions to the gravitational potential and to the metric surrounding a static source, translating into corrections to the Schwarzschild and Reissner--Nordstr\"om geometries of the schematic form $G^2 M \hbar/r^3$~\cite{Donoghue:2001qc,Bjerrum-Bohr:2002fji,Kirilin:2006en}. A complementary strand encodes quantum effects in local and non-local higher-curvature operators added to the Einstein--Hilbert action, whose coefficients can be matched to specific ultraviolet scenarios; the corresponding corrected black hole metrics, temperatures, and shadows have been computed in~\cite{Calmet:2017qqa,Calmet:2019eof,Calmet:2021lny,Battista:2023iyu}. The EFT viewpoint is conceptually important for this review: it demonstrates that quantum-gravitational corrections to black hole geometries are not confined to Planckian curvature regions but leave (albeit tiny) imprints at macroscopic distances, and it produces concrete deformation functions that can be directly accommodated, and compared with other approaches, within the EMD parametrization.

\subsection{Asymptotically safe gravity}

The asymptotic safety scenario posits that gravity is nonperturbatively renormalizable, its ultraviolet behavior being controlled by a non-trivial fixed point of the renormalization group at which Newton's coupling becomes scale-invariant. Substantial evidence for such a fixed point has accumulated from functional renormalization group studies~\cite{Reuter:1996cp}. In this framework, black hole physics is commonly explored through renormalization-group improvement: the classical geometry is dressed by replacing Newton's constant with its running counterpart $G(k)$, evaluated at a physical cutoff scale $k$ identified with an inverse distance from the center. The prototype of this construction is the Bonanno--Reuter black hole~\cite{Bonanno:2000ep}, in which the running of $G$ regularizes the short-distance behavior of the Schwarzschild geometry, producing an inner horizon, a critical minimal mass below which no horizon exists, and a finite temperature that vanishes in the extremal limit. Refinements of the scale-setting procedure and iterated improvements have been explored extensively~\cite{Platania:2019kyx,Platania:2023srt}. The Bonanno--Reuter spacetime plays a special role in this review: as shown in Section~\ref{Sect:ExampleBonannoReuter}, treated self-consistently it violates the EMD regularity conditions derived in Chapter~\ref{chap3}, leading to an ill-defined Hawking temperature and a divergent Ricci scalar at the horizon. It thereby illustrates how the EMD consistency conditions can discriminate among quantum-gravity-inspired geometries. Which identifications of the running scale can avoid this obstruction is discussed in Remark~\ref{rem:BRalternatives}.

\subsection{Loop quantum gravity}

Loop quantum gravity (LQG) is a background-independent, nonperturbative quantization of GR in which geometric operators such as areas and volumes possess discrete spectra. The existence of a minimal area gap has profound consequences for black hole interiors: the classical singularity is generically replaced by a quantum bounce. Effective descriptions of the Schwarzschild interior, treated as a homogeneous Kantowski--Sachs cosmology with holonomy corrections, as well as full loop quantizations of the spherically symmetric sector~\cite{Gambini:2013ooa}, show the resolution of the $r=0$ singularity into a transition surface connecting a trapped to an anti-trapped region. In the quantum-corrected Kruskal spacetime of Ashtekar, Olmedo and Singh~\cite{Ashtekar:2018lag}, this transition surface joins the black hole interior to a white-hole-type region, while curvature invariants remain bounded by universal, mass-independent values, realizing the limiting-curvature expectation in a concrete quantum-gravity setting. Outside the horizon, the effective LQG geometries reduce to deformations of the Schwarzschild metric functions with corrections controlled by the area gap, which fall squarely within the class of spherically symmetric, static effective metrics parametrized by the EMD. Their near-horizon expansion coefficients are therefore directly constrained by the regularity and consistency conditions of Chapter~\ref{chap3}.

\subsection{\texorpdfstring{$f(R)$}{f(R)} and scalar-tensor theories}

A large body of work modifies GR already at the classical level by enlarging its gravitational degrees of freedom. The simplest deformation promotes the Einstein--Hilbert Lagrangian to a function of the Ricci scalar, $f(R)$; the resulting theories propagate an additional scalar mode and are dynamically equivalent to Brans--Dicke theories with a potential (see~\cite{Sotiriou:2008rp} for a comprehensive review). Historically, the quadratic model of Starobinsky~\cite{Starobinsky:1980te} demonstrated that curvature-squared terms can drive inflation, establishing higher-curvature gravity as a phenomenologically serious option. Static, spherically symmetric black holes in $f(R)$ gravity include deformed Schwarzschild and charged solutions~\cite{Mazharimousavi:2012ca_FofR}, with modifications of both the horizon structure and the thermodynamics. More generally, scalar-tensor theories with second-order field equations are classified by the Horndeski Lagrangian~\cite{Horndeski:1974wa}, which encompasses galileons and covers most single-scalar dark-energy models. Whether such theories admit black holes different from Kerr is governed by no-hair theorems, such as the one for galileons~\cite{Hui:2012qt}, and by their known evasions through specific couplings, leading to genuinely scalarized black hole solutions. For the purposes of this review, the essential point is that entire families of classical modified-gravity black holes are described by metric functions $h \neq f$ with regular horizons, exactly the setting for which the two-function EMD of Chapter~\ref{chap3} was developed.

\subsection{Non-commutative geometry}

Non-commutative geometry implements a minimal length directly in the structure of spacetime by promoting coordinates to non-commuting operators, $[x^\mu, x^\nu] = i \theta^{\mu\nu}$, with $\sqrt{\theta}$ setting the scale below which the notion of a point loses meaning. In gravitational applications, the leading effect can be captured by replacing point-like matter distributions with smeared, Gaussian-like sources of width $\sqrt{\theta}$. The resulting non-commutative-geometry-inspired Schwarzschild black hole~\cite{Nicolini:2005vd} is regular at the origin, where the geometry approaches a de Sitter core, and exhibits a characteristic thermodynamic behavior: the Hawking temperature reaches a maximum and then decreases to zero as the horizon shrinks toward a finite-size extremal remnant, thereby avoiding the divergent final flash of the semiclassical picture. Generalizations to charged, rotating, and higher-dimensional configurations have been developed, together with refinements of the underlying smearing prescription (see~\cite{Nicolini:2019irw} and references therein). From the EMD perspective, these geometries constitute a one-parameter family of regular deformations of the classical solutions, with $\sqrt{\theta}$ controlling the near-horizon expansion coefficients, and they share with the phenomenological regular models discussed below the generic appearance of an inner horizon, with the attendant stability concerns raised in Section~\ref{sec:open-issues}.

\subsection{Braneworld and holographic constructions}

In braneworld scenarios of the Randall--Sundrum type~\cite{Randall:1999vf}, our four-dimensional universe is a brane embedded in a higher-dimensional anti-de Sitter bulk, and gravity on the brane is recovered at long distances while receiving calculable short-distance modifications. Holography gives these constructions a powerful reinterpretation: through the AdS/CFT correspondence, classical bulk dynamics encodes the quantum backreaction of a strongly coupled conformal field theory on the brane geometry. Black holes localized on the brane are therefore \emph{quantum} black holes, incorporating the exact backreaction of the CFT to all orders in $\hbar$ at large central charge. While the four-dimensional problem remains formidable, an exact solution exists one dimension lower: the holographic quantum BTZ (quBTZ) black hole~\cite{Emparan:2020znc,Emparan:2021hyr}, constructed from a black string in AdS$_4$ braneworlds, describes a three-dimensional black hole dressed by the full quantum backreaction of a holographic CFT, with computable corrections to the horizon structure, temperature, and entropy, including the generation of a genuine horizon for the otherwise horizonless conical geometries. The quBTZ spacetime is discussed in Chapter~\ref{chap examples} as a rare example of an exactly known quantum-corrected geometry, and the lower-dimensional EMD developed in Chapter~\ref{chap3} provides the natural language for its near-horizon parametrization.

\subsection{Phenomenological regular models}

Finally, a large and influential class of geometries is not derived from an underlying action but engineered to implement physical principles, chiefly regularity at the origin, while reproducing the classical solutions at large distances. The Bardeen spacetime~\cite{bardeen1968} historically provided the first example of a singularity-free black hole, later reinterpreted as an exact solution of GR coupled to nonlinear electrodynamics~\cite{AyonBeato:1998ub,AyonBeato:2000zs}. The Dymnikova geometry~\cite{Dymnikova:1992ux} replaces the singularity with a gravitating vacuum core interpolating between a de Sitter interior and the Schwarzschild exterior, while the Hayward metric~\cite{Hayward_2006} realizes the same infrared/ultraviolet interpolation with a particularly simple rational profile, designed to model the formation and evaporation process. The Simpson--Visser family~\cite{Simpson:2018tsi,Simpson:2019mud} interpolates, through a single parameter, between the Schwarzschild black hole, a regular black hole, and a traversable wormhole, and the Frolov spacetime~\cite{Frolov:2016pav} embeds the Hayward-type core into a charged geometry with an extremal configuration. All these models share the generic features anticipated in Section~\ref{sec:open-issues}: a de Sitter-like core, an inner horizon with its associated instability concerns, and violations of the strong energy condition. Precisely because they are simple, explicit, and widely used as benchmarks, they constitute the primary testing ground for the EMD framework: their near-horizon expansions, consistency conditions, shadows, and quasi-normal modes are worked out in Chapters~\ref{chap examples},~\ref{chap4} and~\ref{chap5}.

\subsection{The swampland program}\label{sec:swampland}

The frameworks discussed so far proceed from the top down: a specific theory, or at least a specific mechanism, is assumed, and the corresponding black hole geometries are derived from it. The \emph{swampland program}~\cite{Vafa:2005ui,Ooguri:2006in,Palti:2019pca,vanBeest:2021lhn,Agmon:2022thq} follows the opposite route. Its goal is to identify general principles that any consistent theory of quantum gravity must satisfy, and to use them to constrain low-energy EFTs without committing to a specific microscopic completion. The starting observation is that an EFT which is perfectly consistent in the absence of gravity may fail to admit an ultraviolet completion once it is coupled to dynamical gravity. EFTs that admit such a completion form the \emph{landscape}, while those that do not are said to belong to the \emph{swampland}. Although the program originated from patterns observed in string compactifications, its most robust criteria can be motivated by bottom-up arguments that do not rely on string theory~\cite{Basile:2025zjc}.

Black holes play a central role in these arguments. The conjectured absence of global symmetries in quantum gravity~\cite{Banks:2010zn,Harlow:2018tng} follows from the observation that Hawking radiation is blind to global charges: a black hole that has absorbed an arbitrarily large global charge could evaporate back to its initial mass without releasing it, leading either to remnants or to a conflict with the finite Bekenstein--Hawking entropy. The weak gravity conjecture~\cite{Arkani-Hamed:2006emk,Harlow:2022ich} requires instead that extremal charged black holes be able to decay, and hence the existence of states whose charge-to-mass ratio exceeds the extremal one. In its black hole formulation, it is tied to the sign of the higher-derivative corrections to the extremality bound of charged black holes~\cite{Kats:2006xp,Cheung:2018cwt}, \emph{i.e.} to small deformations of the extremal Reissner--Nordstr\"om geometry of Section~\ref{Sect:RNBH}. Other conjectures, such as the distance conjecture~\cite{Ooguri:2006in}, constrain instead the geometry of the space of EFTs itself.

\begin{figure}[t]
\centering
\begin{tikzpicture}[x=1cm,y=1cm,
  uv/.style={draw,rounded corners=2pt,inner sep=3pt,font=\footnotesize,fill=white},
  lbl/.style={font=\scriptsize,align=center},
  flow/.style={-latex,thick,decorate,decoration={snake,amplitude=.6mm,segment length=4mm,post length=2mm}}]
\draw[thick,fill=cyan!8] plot[smooth cycle,tension=.75] coordinates
  {(-4.3,0.1) (-3.7,1.45) (-1.6,1.95) (1.0,1.85) (3.4,1.55) (4.35,0.3) (3.6,-1.55) (0.6,-2.0) (-2.4,-1.85) (-3.9,-1.2)};
\node[lbl,text=cyan!40!black] at (-0.1,-1.25) {\bf\large\emph{swampland}};
\draw[thick,violet,dashed,fill=violet!12,fill opacity=.8] (-1.25,0) ellipse (1.75 and 0.75);
\draw[thick,blue!70!black,fill=blue!10,fill opacity=.8] (1.45,0) ellipse (1.75 and 0.68);
\node[lbl,text=violet!80!black] at (-1.4,0) {compatible with\\observations};
\node[lbl,text=blue!70!black] at (1.5 ,0) {QG landscape};
\node[lbl] at (0.12,0) {?};
\node[font=\footnotesize\bfseries,anchor=north] at (0,-2.05) {space of gravitational EFTs};
\node[uv,draw=blue!70!black,text=blue!70!black]     (AS) at (1.65,3.9) {Quantum Gravity};
\draw[flow,blue!70!black] (AS.south) -- (1.55,0.65);
\node[lbl,text=gray!60!black] (UVl) at (-4.6,3.9) {UV};
\node[lbl,text=gray!60!black] (IRl) at (-4.6,0.0) {IR};
\draw[-latex,gray!60!black,thick] (UVl.south) -- node[lbl,rotate=90,above=1pt] {RG flow} (IRl.north);
\begin{scope}[xshift=8.9cm]
\draw[thick,rounded corners=10pt,fill=orange!6] (-2.4,-1.95) rectangle (2.4,1.95);
\node[lbl,anchor=north] at (0,1.88) {EMD parameters $\big(\zh,\{\xi_n,\theta_n\}\big)$};
\draw[thick,orange!70!black,fill=orange!15] (0,-0.2) ellipse (2.05 and 1.35);
\node[lbl,text=orange!50!black] at (0,0.25) {regular horizon, $T_{\text{H}}>0$,\\ finite curvature};
\draw[thick,dashed,red!70!black,fill=red!10] (0.15,-0.8) ellipse (1.25 and 0.52);
\node[lbl,text=red!60!black] at (0.15,-0.8) {shadow, QNM\\bounds};
\node[font=\footnotesize\bfseries,anchor=north,align=center] at (0,-2.05) {static, spherically symmetric\\black hole geometries};
\end{scope}
\draw[-latex,thick] (4.25,0.8) to[bend left=25] node[lbl,above] {black hole\\solutions} (6.5,1.3);
\draw[-latex,thick,dashed] (6.5,-1.3) to[bend left=25] node[lbl,below] {model-independent\\constraints} (4.0,-1.25);
\end{tikzpicture}
\caption{Schematic view of the swampland program and of its relation to the EMD, inspired by Fig.~1 of~\cite{Basile:2025zjc}. \emph{Left:} each theory of Quantum Gravity (such as string theory and asymptotic safety) gives rise, through the RG flow to the infrared, to a landscape of consistent gravitational EFTs, while the remaining EFTs belong to the swampland. Whether the landscapes of different completions overlap, is an open question. Only a subset of EFTs is compatible with observations. \emph{Right:} the static, spherically symmetric black holes of a given EFT correspond to points in the EMD parameter space, which is restricted by the consistency conditions of Chapter~\ref{chap3} and by observations of the shadow and of the quasi-normal modes (Chapters~\ref{chap4} and~\ref{chap5}). These constraints hold irrespective of the underlying theory.}
\label{fig:swampland}
\end{figure}
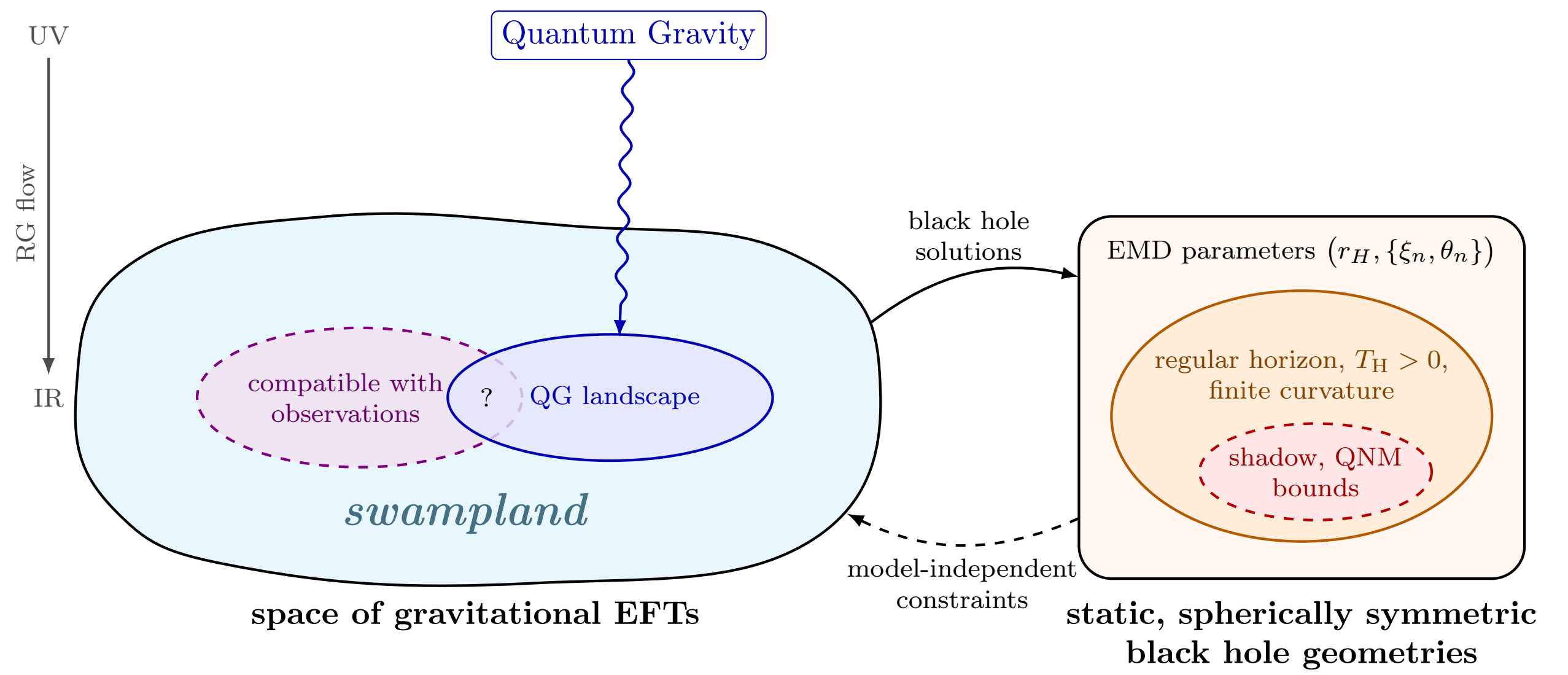

A central open question is whether the landscape is unique, namely whether all consistent EFTs descend from string theory, or whether inequivalent theories of quantum gravity give rise to distinct and possibly overlapping landscapes. This question has been addressed in particular for asymptotically safe gravity, where the intersection between its landscape and swampland bounds such as the weak gravity conjecture has been investigated~\cite{deAlwis:2019aud,Basile:2021xxx,Knorr:2024xxx,Eichhorn:2024xxx}. A recent conceptual assessment~\cite{Basile:2025zjc} identified spacetime topology change and black hole thermodynamics as the key points of tension. It concludes that either gravity admits no fundamental description as a quantum field theory, or the corresponding swampland conditions do not apply universally; in a consistent fundamental realization of asymptotic safety, black hole entropy would require an interpretation different from the standard statistical one. These logical possibilities are illustrated schematically in the left part of Figure~\ref{fig:swampland}.

\begin{center}
\remarkbox{\rem\label{rem:SwamplandEMD}{Swampland and EMD: bottom-up at two different levels}}{\parbox{16cm}{
The swampland program and the EMD share the same bottom-up philosophy, but they act at different levels. The swampland program constrains effective \emph{actions}, by requiring consistency with general principles of quantum gravity. The EMD constrains effective \emph{geometries}: it parametrizes static and spherically symmetric black holes in terms of coordinate-invariant quantities, and restricts the parameters through physical consistency conditions (Chapter~\ref{chap3}) and through observations (Chapters~\ref{chap4} and~\ref{chap5}). The static, spherically symmetric black holes of any EFT, provided they satisfy the assumptions of Chapter~\ref{chap3}, correspond to points in the EMD parameter space, so that constraints derived within the EMD apply to all of them at once. Since many swampland arguments rely on black holes, and in particular on near-extremal charged ones, the EMD provides a natural, model-independent language in which such statements can be formulated. A first application to the weak gravity conjecture is presented in Section~\ref{sec:WGC-EMD}.
}}
\end{center}

\vspace{0.5em}
The diversity of these frameworks, each with its own fields, scales, and expansion parameters, is precisely what makes model-by-model comparisons with observations impractical, and motivates the model-independent effective metric approach developed in the next chapter.

%% file: subfiles/3-EMD.tex
\typeout{START baseline=\the\baselineskip, jot=\the\jot}
\chapter{The Effective Metric Description}\label{chap3}

This chapter is based on and extends the results presented in~\cite{DelPiano:2023fiw,DelPiano:2024gvw,DamiaPaciarini:2025xjc,Binetti:2022xdi}, where the Effective Metric Deformation (EMD) framework was first introduced and applied to static, spherically symmetric spacetimes. Building on those foundations, we further develop the formalism by constructing the proper-distance EMD and exploring its mapping to curvature-based schemes such as the Ricci and Kretschmann EMDs. We then extend the analysis from the horizons to the entire spacetime and derive the conditions under which the geometry is regular everywhere, including at its core, following~\cite{DamiaPaciarini:2025qhq}. The final sections are devoted to the EMD for charged BHs, including the analysis of the extremal limit and the introduction of a proper-time EMD suited for dynamical observers.

\section{Definitions and Review}
This Section serves the purpose of illustrating what is the Effective Metric Description (EMD) and how it is used to describe (quantum) deformed Schwarzschild black holes.

The starting point is the metric of a spherically symmetric and static black hole of mass $\ma$ written in terms of the Schwarzschild coordinates $(t,\co,\theta,\varphi)$
\begin{align}
&\dd s^2=-h(\co) \dd t^2+\frac{\dd \co^2}{f(\co)}+\co^2 \dd \theta^2+\co^2 \sin^2\theta \, \dd\varphi^2\ , \label{SpStMetric}
\end{align}
where
\begin{equation}\label{h-f-emd-def}
     h(\co)=1-\frac{\Psi}{\co} \qq{with} f(\co)=1-\frac{\Phi}{\co} \ .
\end{equation}
where $\Psi$ and $\Phi$ parametrize deformations of the Schwarzschild encoding corrections due to physical effects beyond GR, either classical or quantum in nature.

In order to describe a black hole, it is assumed that the metric functions $h(r)$ and $f(r)$ possess a simple zero at some finite radius $\hr > 0$, corresponding to the position of the (outer) event horizon. In the following, Eq.~\eqref{SpStMetric} will be considered only for $r \geq \hr$, where both $h(r)$ and $f(r)$ are taken to be positive definite. 

To ensure the regularity of the geometry and its derivatives in the vicinity of the horizon, the metric functions can be expanded in Taylor series around $r = \hr$ as
\begin{equation}\label{f-h-taylor-r}
    f(r) = \sum_{n \geq 1} \frac{\fn{n}}{n!}\,(r - \hr)^n,
    \qquad 
    h(r) = \sum_{n \geq 1} \frac{\hn{n}}{n!}\,(r - \hr)^n \, .
\end{equation}
These expansions make explicit that the first derivatives $\fn{1}$ and $\hn{1}$ are finite and non-vanishing, thereby guaranteeing a regular and non-degenerate horizon. 

Then we require that the geometry is asymptotically flat. Concretely, one assumes that the geometry approaches the Schwarzschild metric (with mass parameter $M$) for very large distances from the origin
 \begin{equation}\label{asymptotic flatness}
 \lim_{r \to \infty}\Psi(r) = 2M =\lim_{r \to \infty} \Phi(r)\ . 
\end{equation}
Furthermore, in order for coordinate transformations of the undeformed space-time to be also realized in the deformed case, we demand that $\Psi$ and $\Phi$ are invariant quantities. This can be achieved by writing them as functions of a physical quantity, for which in \cite{Bonanno:2000ep,Binetti:2022xdi} the proper distance from the origin was proposed.\footnote{One can choose a more general geodesic distance to prove a Taylor theorem for general tensor calculus  shown in \cite{ruse1931,synge1931,dewitt1960radiation}. However, as mentioned in these works, explicit computations become rapidly involved.}\\
An Effective Metric Description is largely determined by identifying the latter with a physical quantity $\mathcal{X}$, which is invariant under reparameterization and which is a monotonic function of the Schwarzschild coordinate $\co$. Hence, we can denote the deformation functions as $\Psi^{(\mathcal{X})}$ and $\Phi^{(\mathcal{X})}$ when they are parametrized in terms of a physical quantity $\mathcal{X}$. \footnote{The (slightly abusive) notation $\Psi^{(\mathcal{X})}(r)$ and $\Phi^{(\mathcal{X})}(r)$ will be used to signify $\Psi^{(\mathcal{X})}(\mathcal{X}(r))$ and $\Phi^{(\mathcal{X})}(\mathcal{X}(r))$. Furthermore, since the goal is to compare \scheme s defined by different physical quantities, various different deformation functions will be introduced and the superscript $(\mathcal{X})$ simply serves as a naming convention.}

To formalize the discussion, an \scheme\, for describing a (quantum) deformed spherically symmetric and static black hole is characterized by the following set of quantities:
\begin{enumerate}
\item[\emph{(i)}] a defining {\bf physical quantity} $\mathcal{X}$. The symbol $\mathcal{X}$ is mostly used to loosely label the \scheme, \emph{e.g.} the EMD outlined below is simply called the \emph{distance}-\scheme. Furthermore, additional {\bf (normalisation) condition(s)} may be required that supplement (and sharpen) the definition of the physical quantity: these can act as boundary conditions for the self-consistency condition (point \emph{(iii)} below), or are related to the classical limit (in which the metric deformation is switched off). In the example of the distance \scheme, this condition is the convention to measure the distance from the horizon of the black hole \emph{i.e.} $\rho(\zh)=0$ as well as the choice of the sign $v=+1$ in (\ref{Coeffsa}) (which allows for the recovery of the Schwarzschild geometry in the limit $\xi_{2n}\to 0$ $\forall n\geq 1$).   

\item[\emph{(ii)}] a {\bf set of {{free parameters}}} which are used to deform the functions $h$ and $f$ in the metric (\ref{SpStMetric}). More concretely, one typically refers to the expansion coefficients of $\Psi^{(\mathcal{X})}$ and $\Phi^{(\mathcal{X})}$ close to the horizon (as \emph{e.g.} in \eqref{SeriesDistanceRho}).

\item[\emph{(iii)}] a {\bf self-consistency equation} which encodes how $\mathcal{X}$ is computed from the metric. This equation usually takes the form of a (non-linear) differential equation (see (\ref{DiffEqDistance}) in the case of the distance \scheme).
\end{enumerate}

Once the \scheme\, has been defined, physical quantities related to the black hole geometry, can be computed in terms of the free parameters. For example, as we will show in Sec.~\ref{EMD propdist ricci temp}, the Hawking temperature (cf. Sec.~\ref{hawking-classical-intro}) can be parametrized as\footnote{This expression takes into account that here we only consider a single metric deformation function $h=f$.}
\begin{align}
T_H=\frac{1}{4\pi}\,\frac{\dd f}{\dd \co}\bigg|_{\co=\zh}=\frac{1+\varpi}{8\pi\zh}\,,\label{HawkingTempFirst}
\end{align}
that is indeed only a function of $\zh$ and $\varpi$ (or equivalently $\xi_{0,2}$), which will be defined in the following Section.

\begin{figure}
\centering
\tikzset{every picture/.style={line width=0.75pt}} 

\begin{tikzpicture}[x=0.75pt,y=0.75pt,yscale=-1,xscale=1]

\draw [color={rgb, 255:red, 74; green, 144; blue, 226 }  ,draw opacity=1 ][line width=1.5]    (50,220) .. controls (50,145.67) and (113.67,108) .. (180,80) ;
\draw [color={rgb, 255:red, 208; green, 2; blue, 27 }  ,draw opacity=1 ][line width=1.5]    (50,220) .. controls (51,171.67) and (139,134) .. (220,110) ;
\draw [line width=1.5]    (50,220) -- (50,74) ;
\draw [shift={(50,70)}, rotate = 90] [fill={rgb, 255:red, 0; green, 0; blue, 0 }  ][line width=0.08]  [draw opacity=0] (11.07,-5.32) -- (0,0) -- (11.07,5.32) -- (7.35,0) -- cycle    ;
\draw [line width=1.5]    (50,220) -- (226,220) ;
\draw [shift={(230,220)}, rotate = 180] [fill={rgb, 255:red, 0; green, 0; blue, 0 }  ][line width=0.08]  [draw opacity=0] (11.07,-5.32) -- (0,0) -- (11.07,5.32) -- (7.35,0) -- cycle    ;
\draw  [dash pattern={on 0.84pt off 2.51pt}]  (50,120) -- (220,120) ;
\draw  [dash pattern={on 0.84pt off 2.51pt}]  (190,220) -- (190,80) ;
\draw  [dash pattern={on 0.84pt off 2.51pt}]  (105,220) -- (105,80) ;

\draw (30,109.07) node [anchor=north west][inner sep=0.75pt]    {$\rho $};
\draw (98,221.73) node [anchor=north west][inner sep=0.75pt]    {$r_{1}$};
\draw (182.33,222.73) node [anchor=north west][inner sep=0.75pt]    {$r_2$};
\end{tikzpicture}

\caption{Since the proper distance from the horizon $\rho$ can be the only physically and unambiguous quantity that an observer measures, one should relate such observable to the shape of the spacetime. Given two models of deforming the spacetime (\textcolor[RGB]{74,144,226}{$\Phi_1$} and \textcolor[RGB]{208,2,27}{$\Phi_2$}), we have that in the Schwarzschild coordinates, in order to obtain the same physical distance $\rho$, one has to evaluate each model in each own corresponding value of $r$ (namely $r_1$ and $r_2$). The advantage of parameterizing the deformations in the proper distance (such as in this case), is that the ambiguity about the position with respect to the black hole, at which one performs experiments and measurements, is lifted and the coefficients that appear in the expansions can be readily read off from the observational point of view.}
\end{figure}
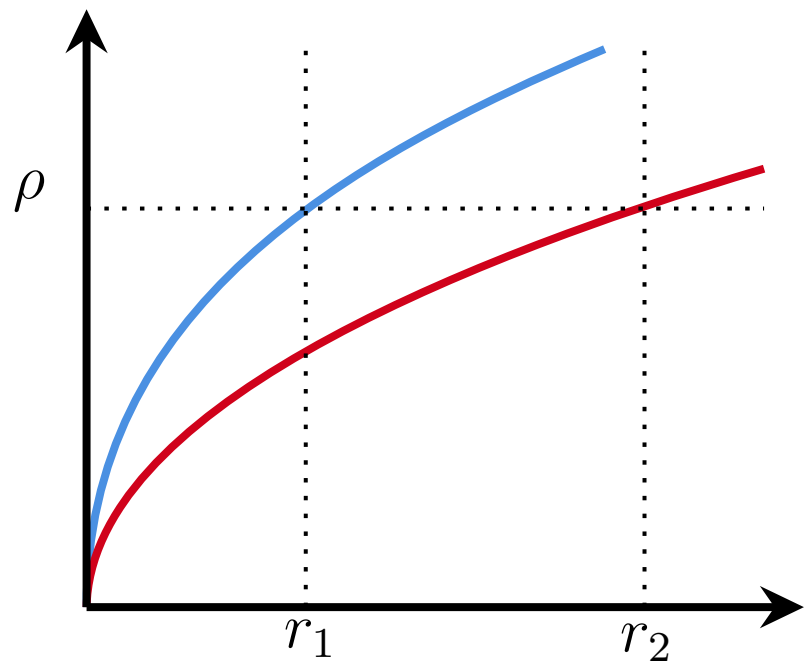

The motivation for introducing the Effective Metric Deformation lies in the need for a coordinate-independent and physically transparent way to parametrize deviations from classical spacetimes. In conventional approaches, metric deformations are expressed as functions of coordinate variables such as the Schwarzschild radius $r$, which, although geometrically convenient, lack diffeomorphism-invariant meaning. The functional form $\Phi(r)$ is not preserved under coordinate transformations, while only $\Phi(\mathcal{X})$, with $\mathcal{X}$ an invariant scalar constructed from the spacetime, is truly physical. This issue becomes critical when defining the notion of \emph{smallness} of a deformation, as what appears small in one coordinate system may not be so in another. The EMD framework addresses this by parametrizing metric deformations directly in terms of observable and invariant quantities, ensuring that both the expansion coefficients and the definition of proximity to the classical theory are physically meaningful and coordinate-independent. 

In spherical symmetry, the areal radius $r$ already provides an invariant definition, related to the quadratic Casimir of the rotation group and to the Misner-Sharp mass, making it a useful testing ground for the formalism. However, from an operational point of view, observables such as the proper distance from the horizon offer a more direct and measurable way to quantify deviations, free from coordinate ambiguities. The EMD thus provides a general and flexible framework in which quantum or phenomenological corrections to black hole geometries can be encoded in terms of physically interpretable, observer-independent quantities, offering a robust bridge between theoretical parametrizations and measurable effects, which is a property sought after by other QG approaches~\cite{Held:2021vwd}.

\section{The proper distance \scheme}\label{Sect:ReviewDistance}
To explain the framework in more detail, as in the references \cite{Binetti:2022xdi,DelPiano:2023fiw}, the physical quantity is chosen to be the proper distance $\rho$ measured from the horizon of the black hole, \emph{i.e.} $\Phi^{(\rho)}(r)$. In writing $f(\co)$ in \eqref{SpStMetric}, we implicitly understand $\rho$ to be a (monotonic) function of the Schwarzschild coordinate $\co$, which is fixed through the first order differential equation
\begin{align}\label{DiffEqDistance}
    \dv{\rho}{r}=\frac{1}{\sqrt{f}}=\left(1-\dfrac{\Phi^{(\rho)}(\rho)}{r}\right)^{-1/2} \qq{with} \rho(\hr)=0 \ .
\end{align}
This differential equation is non-linear, since the right hand side through $\Phi^{(\rho)}$ is also a function of the distance $\rho$.
The surface gravity and the Ricci scalar are fundamental quantities that must be well-defined at and near the horizon (cf. Sec.~\ref{hawking-classical-intro}), while here we give the definition of the Ricci scalar as a geometric quantity, which directly follows from Eq.~\eqref{SpStMetric}: it is a scalar quantity which appears in the equation of motion of the gravitational field (\emph{i.e.} the Einstein equations). Therefore, its regularity (notably at the horizon) in the deformed metric ensures that in the deformed case no additional singularities arise beyond the classical ones. 
We mainly focus on the finiteness of the Ricci scalar defined in eq~\eqref{eq: ricci scalar h e f} at the horizon but, in some examples, we shall also examine the behavior of the Kretschmann scalar in eq.~\eqref{KretschDef}.\\

In~\cite{DelPiano:2023fiw} two equivalent approaches were proposed to find regularity conditions of the curvature scalars at the horizon and solve Eq.~\eqref{DiffEqDistance} in a self-consistent way. The first method uses an exponential definition of the deformation function and used the Taylor expansions in Eqs.\eqref{f-h-taylor-r} to find regularity of the first and second derivatives at the horizon. In this review, the second approach, which is based on self-consistent methods, will be discussed, in light of the further applications of such parametrization.
To solve Eq.~\eqref{DiffEqDistance} in a self-consistent fashion as a series expansion close to the horizon, the following series expansions are defined
\begin{align}
\Psi^{(\rho)}(\rho)=\hr+2M\sum_{n=0}^\infty \theta_n\left(\frac{\rho}{2M}\right)^n \qq{and} \Phi^{(\rho)}(\rho)=\hr + 2M \sum_{n=0}^\infty \xi_n\left(\frac{\rho}{2M}\right)^n\ ,\label{SeriesDistanceRho}
\end{align}
such that the deformation of the classical Schwarzschild geometry is fully encoded in $\hr$ and the real dimensionless parameters $\{\theta_n, \xi_n\}_{n\in\mathbb{N}_0}$. The series in \eqref{SeriesDistanceRho} is assumed to have a non-vanishing radius of convergence { {(see \cite{ruse1931,synge1931,dewitt1960radiation} where a Taylor theorem for more general tensor calculus, based on (various) geodesic distances, was developed)}}.

\subsection{Solving the distance function}\label{Sect:DistanceFunctionExpansion}
\subsubsection{Series expansion of the radial coordinate}\label{Sect:SeriesExpansion}
Using the expansion \eqref{SeriesDistanceRho} allows to express the distance $\rho$ as a function of the Schwarzschild coordinate $\co$ and vice-versa
\begin{align}
&\co(\rho)=\zh+2M \sum_{n=1}^\infty a_n\, \left(\frac{\rho}{2M}\right)^n \qq{and} \rho(\co)=2M \sum_{n=1}^\infty b_n\,\left(\frac{r - \hr}{2M}\right)^{n/2} \ .\label{SeriesDistance}
\end{align}
We can solve for the coefficients $a_n$ in terms of the $\xi_n$ by inserting the series on the left of Eq.~\eqref{SeriesDistance} into the differential equation~\eqref{DiffEqDistance}, that we write in the form
\begin{equation}\label{DistanceDiffEq}
    r(\rho) \left[1 - \left( \dv{r}{\rho}\right)^2 \right] = \Phi^{(\rho)}(\rho) \ .
\end{equation}
We can expand the square of the series using the Cauchy product of two series as shown in Computation~\ref{appendix power of a series}, then use again the Cauchy product to expand the left-hand side in series of $\rho$ while the right-hand side is expanded as in Eq.~\eqref{SeriesDistanceRho}.

\begin{center}
\compbox{\comp\label{appendix power of a series}{Cauchy Product of a Power Series}}{\parbox{16cm}{Let $S:\,\mathbb{R}\to\mathbb{R}$ be a function which affords the following series expansion 
\begin{align}
&S(x)=\sum_{k=0}^\infty c_k\,x^k \ ,  && \forall |x|\leq x_c \ ,\label{CauchyInitialSum}
\end{align}
where $x_c>0$ is a (finite) radius convergence and $c_k\in\mathbb{R}$ are a set of (known) expansion coefficients. For $n\in\mathbb{N}$, we are interested in the series expansion of the power of $S(x)$
\begin{align}
(S(x))^n=\sum_{k=0}^\infty \mathfrak{c}_k\,x^k\,,\label{PowerPowerExpansion}
\end{align}
\emph{i.e.} we shall express the coefficients $\mathfrak{c}_k$ as functions of the $\{c_k\}_{k\in\mathbb{N}^*}$ in (\ref{CauchyInitialSum}). In order to calculate $\mathfrak{c}_p$ (for given $p\in\mathbb{N}$), we start from
\begin{align}
(S(x))^n=\left(\sum_{k=0}^p c_k\,x^k\right)^n+\mathcal{O}(x^{p+1})=\sum_{\ell_0=0}^n\, \left(\begin{linesmall}{@{}c@{}} \\[-6pt]n \\ \ell_0 \end{linesmall}\right)\,c_0^{n-\ell_0}\,x^{\ell_0}\,\left(\sum_{k=1}^p c_k x^{k-1}\right)^{\ell_0}+\mathcal{O}(x^{p+1})\,.\nonumber
\end{align}
Using iteratively the binomial formula, we may write 
\begin{align}
(S(x))^n=\sum_{\ell_0=0}^n\sum_{\ell_1=0}^{\ell_0}&\ldots\sum_{\ell_{p-1}=0}^{\ell_{p-2}}\, \left(\begin{linesmall}{@{}c@{}} \\[-6pt]n \\ \ell_0 \end{linesmall}\right)\left(\begin{linesmall}{@{}c@{}} \\[-6pt]\ell_0 \\ \ell_1 \end{linesmall}\right)\ldots \left(\begin{linesmall}{@{}c@{}} \\[-6pt]\ell_{p-2} \\ \ell_{p-1} \end{linesmall}\right)\nonumber\\
&\times x^{\ell_0+\ell_1+\ldots+\ell_{p-1}}\,c_0^{n-\ell_0}\,c_1^{\ell_0-\ell_1}\ldots c_{p-1}^{\ell_{p-2}-\ell_{p-1}}\,c_p^{\ell_{p-1}}+\mathcal{O}(x^{p+1})\,.
\end{align}
We can therefore read off the coefficients $\mathfrak{c}_p$ in (\ref{PowerPowerExpansion})
\begin{align}
\mathfrak{c}_0&=c_0^n\,,\nonumber\\
\mathfrak{c}_p&=\sum_{\genfrac{}{}{0pt}{}{0\leq \ell_{p-1}\leq \ell_{p-2}\leq \ldots\leq \ell_0\leq n}{\ell_0+\ell_1+\ldots+\ell_{p-1}=p}}\,\frac{n!\,c_0^{n-\ell_0}c_1^{\ell_0-\ell_1}\ldots c_{p-1}^{\ell_{p-2}-\ell_{p-1}}c_p^{\ell_{p-1}}}{(n-\ell_0)!(\ell_0-\ell_1)!\ldots (\ell_{p-2}-\ell_{p-1})!\ell_{p-1}!}\ ,&&\forall p\geq 1\,.
\end{align}
We remark that for $c_0=0$, the summation over $\ell_0$ only has the term $\ell_0=n$, such that in this case
\begin{align}
\mathfrak{c}_0&=0\,,\hspace{1cm}\mathfrak{c}_1=c_1\,\delta_{n,1}\,,\nonumber\\
\mathfrak{c}_p&=\sum_{\genfrac{}{}{0pt}{}{0\leq \ell_{p-1}\leq \ell_{p-2}\leq \ldots\leq \ell_1\leq n}{\ell_1+\ldots+\ell_{p-1}=p-n}}\,\frac{n!\,c_1^{n-\ell_1}\ldots c_{p-1}^{\ell_{p-2}-\ell_{p-1}}c_p^{\ell_{p-1}}}{(n-\ell_1)!\ldots (\ell_{p-2}-\ell_{p-1})!\ell_{p-1}!}\,,&&\forall p\geq 2\,.\label{DefRedFormPower}
\end{align}
}}
\end{center}

\begin{center}
\compbox{\comp\label{appendix power of a series_Part}{Power of a power series}}{\parbox{16cm}{Let $\rho\in[0,\dho)$ (with $\dho>0$) and $n\in\mathbb{N}$, then we have the following series expansion
\begin{align}
&\left(\sum_{k=0}^\infty\left(-\tfrac{\rho}{\dho}\right)^k\right)^n=\sum_{p=0}^\infty \mathfrak{c}_p\,\rho^p\,,&&\text{with} &&\mathfrak{c}_p=\left(-\tfrac{1}{\dho}\right)^p\,\left(\begin{linesmall}{@{}c@{}}n+p-1\\p\end{linesmall}\right)=\left(-\tfrac{1}{\dho}\right)^p\frac{(n+p-1)!}{p!(n-1)!}\,.\label{PowerExpansionPower}
\end{align} 
In order to demonstrate this relation, we first remark that for $p=0$, we indeed have $\mathfrak{c}_0=1$. To show (\ref{PowerExpansionPower}) for $p \in\mathbb{N}$, we furthermore consider
\begin{align}
\left(\sum_{k=0}^{p-\ell}\left(-\tfrac{\rho}{\dho}\right)^k\right)^\ell\bigg|_{\rho^{p-\ell}}=\left(-\tfrac{1}{\dho}\right)^{p-\ell}\,\frac{(p-1)!}{(p-\ell)!(\ell-1)!}\,,&& \begin{array}{l}\forall p\in\mathbb{N}\,,\\\forall \ell\in\{0,\ldots,p\}\,.\end{array}\label{LemmaExpansion}
\end{align}
This relation can explicitly be verified for all finitely many values of $\ell\in\{0,\ldots,p\}$ (with $p\geq 1$):
\begin{itemize}
\item for $\ell=0$ both sides are  vanishing, since we use the convention $1/((-1)!)\to 0$.
\item for $\ell=1$ we find directly $\sum_{k=0}^{p}\left(-\tfrac{\rho}{\dho}\right)^k\big|_{\rho^{p-1}}=\left(-\tfrac{1}{\dho}\right)^{p-1}$, which indeed agrees with (\ref{LemmaExpansion}).
\item for general $2\leq \ell\leq p$ we have
\begin{align}
&\left(\sum_{k=0}^{p-\ell}\left(-\tfrac{\rho}{\dho}\right)^k\right)^\ell\big|_{\rho^{p-\ell}}=\sum_{u_{2,\ldots,\ell}=0}^{p-\ell} \left(\sum_{k_1=0}^{p-\ell-u_2-\ldots-u_\ell}\left(-\tfrac{\rho}{\dho}\right)^{k_1}\big|_{\rho^{p-\ell-u_2-\ldots-u_\ell}}\right)\nonumber\\
&\hspace{0.2cm}\times \left(\sum_{k_2=0}^{u_2}\left(-\tfrac{\rho}{\dho}\right)^{k_2}\bigg|_{\rho^{u_2}}\right)\times\ldots\times  \left(\sum_{k_\ell=0}^{u_\ell}\left(-\tfrac{\rho}{\dho}\right)^{k_\ell}\big|_{\rho^{u_\ell}}\right)=\left(-\tfrac{1}{\dho}\right)^{p-\ell}\,\sum_{\genfrac{}{}{0pt}{}{u_{2,\ldots,\ell}=0}{u_2+\ldots+u_\ell\leq p-\ell}}^{p-\ell}1\,.\nonumber
\end{align}
For the last summation, we write
\begin{align}
\sum_{\genfrac{}{}{0pt}{}{u_{2,\ldots,\ell}=0}{u_2+\ldots+u_\ell\leq p-\ell}}^{p-\ell}1=\sum_{k_1=0}^{p-\ell}\sum_{\genfrac{}{}{0pt}{}{u_{3,\ldots,\ell}=0}{u_3+\ldots+u_\ell\leq k_1}}1=\sum_{k_1=0}^{p-\ell}\sum_{k_2=0}^{k_1}\ldots\sum_{k_{\ell-1}=0}^{k_{\ell-2}}1=\sum_{k_1=0}^{p-\ell}\sum_{k_2=0}^{k_1}\ldots\sum_{k_{\ell-2}=0}^{k_{\ell-3}}\left(\genfrac{}{}{0pt}{}{k_{\ell-2}+1}{ k_{\ell-2}}\right)\,,\nonumber
\end{align}
such that, with the help of the identity $\sum_{u=0}^k\left(\begin{linesmall}{@{}c@{}}{u+s}\\u\end{linesmall}\right)=\left(\begin{linesmall}{@{}c@{}}{k+s+1}\\{k}\end{linesmall}\right)$ ($\forall s\in\mathbb{N}$), we obtain
\begin{align}
\sum_{\genfrac{}{}{0pt}{}{u_{2,\ldots,\ell}=0}{u_2+\ldots+u_\ell\leq p-\ell}}^{p-\ell}1=\left(\genfrac{}{}{0pt}{}{p-1}{p-\ell}\right)=\frac{(p-1)!}{(p-\ell)!(\ell-1)!}\,.
\end{align}
This result therefore indeed demonstrates (\ref{LemmaExpansion}).
\end{itemize}
}}
\end{center}

Thus, we have
\begin{multline}\label{propdistseriesexp}
    \hr \left(1-a_1 \right)^2 + \sum_{p=1}^\infty \left[2M(1-a_1^2)a_p - \hr \sum_{m=1}^{p+1}m(p-m+2)a_m a_{p-m+2} \, + \right.\\
    \left. -\, 2M \sum_{n=1}^{p-1}\sum_{m=1}^{n+1}m(n-m+2) a_{p-n}a_m a_{n-m+2}\right]\left(\frac{\rho}{2M}\right)^p =\hr + 2M \sum_{p=1}^\infty \xi_p \left(\frac{\rho}{2M}\right)^p   \ .
\end{multline}
Identifying the coefficients order by order for $p \in \{0,1,2 \}$ leads to
\begin{subequations}\label{InitialConditionsGenericExpansion}
    \begin{align}
    \hr &= \hr (1- a_1^2) \ ,\\
    \xi_1 &= (1-a_1^2)a_1 - \frac{2 \hr}{M} a_1 a_2 \ ,\\
    \xi_2 &= (1-a_1^2)a_2 - \frac{\hr}{M}(3 a_1 a_3 + 2 a_2^2) - 4 a_1^2 a_2 \ .
    \end{align}
\end{subequations}
The first of these equations requires $a_1=0$, which implies that $\xi_1 = 0$. The last equation becomes
\begin{equation}
    \xi_2 = \left( 1  - \frac{2 \hr}{M} a_2 \right) a_2 \ ,
\end{equation}
which is solved for
\begin{equation}\label{a_2-exactsol}
    a_2 = \frac{M(1 + v\,  \varpi)}{4 \hr} \qq{with} \varpi  \coloneqq \sqrt{1 - \frac{8\hr}{M}\xi_2 }  \qq{and} v \coloneqq \pm 1 \ .
\end{equation}
The choice of the sign parameter $v=+1$ picks the solution that is smoothly connected to the Schwarzschild case when turning off the deformation parameter, \emph{i.e.} when $\xi_2 \to 0$. Moreover, the requirement for $\varpi$ to be real imposes an upper bound on $\xi_2 \leq M/8 \hr$.
Equating the remaining powers of $(\rho/ 2M)^p$, for $p \geq 3$, Eq.~\eqref{propdistseriesexp} becomes a recursive equation, which allows to express $a_p$ in terms of $\xi_p$ and the $a_{q}$ with $q < p$:
\begin{align}
&a_p=\frac{\xi_p+\zh/2M \, \sum_{n=3}^{p-1}(p-n+2)\,n\,a_n\,a_{p-n+2}+\sum_{n=2}^{p-2}\sum_{m=2}^n(n-m+2)\,m\,a_{p-n}\,a_m\,a_{n-m+2}}{1-2p \, \zh \, a_2/M}\ .\label{Coeffsa}
%
\end{align}
The coefficients $b_n$ in Eq.~\eqref{SeriesDistance} follow from inversion of the series (\ref{SeriesDistance}) \footnote{Notice that for $\xi_2\leq\frac{M}{8\zh}$, the coefficient $a_2\neq 0$.}
\begin{align}
&b_1=\frac{1}{a_2^{1/2}}\,,&&b_2=-\frac{a_3}{2a_2^2}\,,&&b_n=\frac{(-1)^{n-1}}{2^{n-1}\,n!\, a_2^{n/2}}\,\text{det}(\mathcal{M}_n)\hspace{0.5cm}\forall n\geq 3\ .\label{Coeffsb}
\end{align}
where $\mathcal{M}_n$ is a $(n-1)\times (n-1)$ matrix explicitly written as
\begin{equation}
\scalebox{0.65}{$\mathcal{M}_n =
\begin{pmatrix}
n\frac{a_3}{a_2} & 1 & 0 & 0 & 0 & 0 & 0 & 0 & \cdots & 0\\
4n\frac{a_4}{a_2} & (n+2)\frac{a_3}{a_2} & 2 & 0 & 0 & 0 & 0 & 0 & \cdots & 0\\
12n\frac{a_5}{a_2} & (4n+4)\frac{a_4}{a_2} & (n+4)\frac{a_3}{a_2} & 3 & 0 & 0 & 0 & 0 & \cdots & 0\\
32n\frac{a_6}{a_2} & (12n+8)\frac{a_5}{a_2} & (4n+8)\frac{a_4}{a_2} & (n+6)\frac{a_3}{a_2} & 4 & 0 & 0 & 0 & \cdots & 0\\
80n\frac{a_7}{a_2} & (32n+16)\frac{a_6}{a_2} & (12n+16)\frac{a_5}{a_2} & (4n+12)\frac{a_4}{a_2} & (n+8)\frac{a_3}{a_2} & 5 & 0 & 0 & \cdots & 0\\
\vdots & \vdots & \vdots & \vdots & \vdots & \ddots & \ddots & \cdots & \cdots & \vdots\\
2^{k-1}kn\frac{a_{k+2}}{a_2} & 2^{k-2}((k-1)n+2)\frac{a_{k+1}}{a_2} & 2^{k-3}((k-2)n+4)\frac{a_k}{a_2} & \cdots & \cdots & \cdots & (n+2^{k-1})\frac{a_3}{a_2} & k & \cdots & 0\\
\vdots & \vdots & \vdots & \vdots & \vdots & \vdots & \vdots & \ddots & \ddots & \vdots\\
2^{n-3}n(n-2)\frac{a_{n}}{a_2} & \vdots & \vdots & \vdots & \vdots & \vdots & \vdots & \ddots & \ddots & n-2\\
2^{n-2}n(n-1)\frac{a_{n+1}}{a_2} & 2^{n-3}\big(n(n-2)+2\big)\frac{a_n}{a_2} & \cdots & \cdots & \cdots & \cdots & \cdots & \cdots & \cdots & (n+2^{n-2})\frac{a_3}{a_2}
\end{pmatrix} \ .$}
\end{equation}
Concretely, we find for the first few coefficients with $n\geq 3$
\begin{align}\label{first-few-bs-in-temrs-of-as}
    b_3 &= \frac{5a_3^2-4a_2a_4}{8a_2^{7/2}} \ , \qquad b_4 = \frac{3a_2a_3a_4 - 2 a_3^3-a_2^2a_5}{2 a_2^5} \ , \notag \\
    b_5 &= \frac{231 a_3^4 - 504a_2 a_3^2 a_4 + 224 a_2^2 a_3 a_5 + 112 a_2^2 a_4^2 - 64 a_2^3 a_6}{129 a_2^{13/2}} \ .
\end{align}
Remarkably, expanding the above expressions, one finds that
\begin{align}
&b_{2n-1}=\frac{2^{2n-1}\zh^{n+1}\sqrt{\frac{4M(1+\varpi)}{\zh}}\,\xi_{2n}}{M^{n+1}(1+\varpi)^{n+1}((n-1)+n\varpi)}+\widetilde{b}_{2n-1}(\xi_2,\ldots,\xi_{2n-2})&&\forall n\geq 2\ , \label{ConditionExpansionBs}
\end{align}
where notably $\widetilde{b}_{2n-1}$ is independent of $\xi_{2k}$ for $k>n$. 
For completeness, in Appendix~\ref{App:SerNonZeroXi} we analyze the impact of $\xi_1 \neq 0$.


\subsection{Metric functions and curvature scalars}\label{SectMetricFunctCurv}
In the previous Subsection, the differential equation~\eqref{DistanceDiffEq} was solved assuming an integer series expansion of the radial coordinate $r$ as a function of the distance (from the event horizon) $\rho$. In this Subsection, this result is used to compute the Ricci scalar. 

Using~\eqref{SeriesDistanceRho} and~\eqref{SeriesDistance}, as well as the following expansion (see App.~\ref{app:inverse r to rho})
\begin{align} \label{inverse-z-rho-exp}
\frac{1}{r(\rho)} = \sum_{n=0}^\infty \mathfrak{p}_n \left(\frac{\rho}{2M}\right)^n \ ,&&\text{with}&& \begin{array}{l} \mathfrak{p}_0 = \frac{1}{r_H}, \quad \mathfrak{p}_1 = -\frac{2M a_1}{r_H^2}, \quad \mathfrak{p}_2 = \frac{ 2M\left(2Ma_1^2-a_2 r_H\right)}{r_H^3} \ , \\[20pt]
    \mathfrak{p}_p = - \frac{2M \, a_p}{r_H^2} - \frac{2M}{r_H} \sum_{n=1}^{p-1} \mathfrak{p}_n a_{p-n} \ , \quad \forall p \geq 3 \ ,\end{array}
\end{align}
(with $a_0\equiv \hr$) we can express the metric functions as power series in $\rho$
\begin{align}
&f=1-\sum_{k=0}^\infty\left(\frac{\rho}{2M}\right)^k\sum_{n=0}^k\,\zi{n}\,\xi_{k-n}\,,&&\text{and} &&h=1-\sum_{k=0}^\infty\left(\frac{\rho}{2M}\right)^k\sum_{n=0}^k\,\zi{n}\,\theta_{k-n}\,,&&\forall \rho\geq 0\,,\label{fhSerRho}
\end{align}
which reads explicitly to leading orders 
\begin{subequations}
\begin{align}
f&=\frac{a_2-\xi_2}{2M \zh}\,\rho^2+\frac{a_3-\xi_3}{4 M^2 \zh}\,\rho^3+\mathcal{O}(\rho^4)\ ,\label{frho}\\
h&=-\frac{\theta_1}{\zh}\,\rho+\frac{a_2-\theta_2}{2M \zh}\,\rho^2+\frac{(a_3-\theta_3)\zh+a_2\theta_1}{4M^2\zh^2}\,\rho^3+\mathcal{O}(\rho^4)\ .\label{hrho}
\end{align}
\end{subequations}
We can equivalently write these as expansions in $(r-\zh)$
\begin{subequations}\label{SeriesFormaInt}
\begin{align}
f&=\frac{a_2-\xi_2}{a_2\,\zh}\,(r-\zh)+\frac{a_3\xi_2-a_2\xi_3}{(2M)^{1/2}(a_2)^{5/2}\zh}\,(r-\zh)^{3/2}+\mathcal{O}((r-\zh)^2)\,,\label{fzExpansion}\\
h&=-\frac{\theta_1}{\zh}\sqrt{\frac{2M}{a_2}}\,(r-\zh)^{1/2}+\frac{2a_2^2+a_3\theta_1-2a_2\theta_2}{2a_2^2\,\zh}\,(r-\zh)+\mathcal{O}((r-\zh)^{3/2})\,,\label{hzExpansion}
\end{align}
\end{subequations}
which shows certain similarities with the Taylor series expansions in Eq.~\eqref{f-h-taylor-r}, but also some differences: on the one hand, comparing the expansion for the function $f$ in Eq.~\eqref{fzExpansion} with the first in Eq.~\eqref{f-h-taylor-r}, we can read off the first derivative at the horizon in terms of the parameters
\begin{align}
\frac{a_2-\xi_2}{a_2\,\zh}=4\,a_2=\frac{M(1+\rootk)}{\zh}=2M \, f^{(1)}_H \ .
\end{align}
On the other hand, \eqref{fzExpansion} contains a term of order $(r-\zh)^{3/2}$ (and \eqref{frho} a term of order $\rho^3$), which is absent in \eqref{f-h-taylor-r}. Indeed \eqref{fzExpansion} is not an integer series expansion and therefore more general than \eqref{f-h-taylor-r}. Similarly, (since the condition $\theta_1= 0$ is not imposed), the function $h$ even starts from a term $(r-\zh)^{1/2}$ in \eqref{hzExpansion} (order $\rho$ in \eqref{hrho}), which is absent in the second equation in Eq.~\eqref{f-h-taylor-r}. The series from Eqs.~\eqref{SeriesDistanceRho} and~\eqref{fhSerRho} are therefore more general than the Taylor expansion in $r-\hr$ in Eq.~\eqref{f-h-taylor-r}. However, it is notable that for generic values of $\xi_3$ and $\theta_1$ one has
\begin{align}
&\lim_{r\to \zh}f^{(2)}(r)\to\infty\,,&&\text{and} &&\lim_{r\to \zh}h^{(1)}(r)\to\infty\,.\label{SingDers}
\end{align}
Indeed, the finiteness of the derivatives of the metric functions at the horizon was not the initial assumption and in the next Subsubsection the absence of a physical curvature singularity instead as a more general condition shall be imposed. Before doing so, however, it is stressed that the conditions for the absence of the singularities~\eqref{SingDers} from~\eqref{fzExpansion} and~\eqref{hzExpansion} can be deduced in a straightforward manner:
\begin{align}\label{xi_3-theta_3_cond}
&\theta_1=0\,,&&\text{and} &&a_3\xi_2=a_2\xi_3\,,&&\text{and} &&a_3\theta_2=a_2\theta_3\,.
\end{align}
Using Eq.~\eqref{Coeffsa}, the second of these relations implies 
\begin{align}
\left( \frac{M(1+\varpi)}{4 \hr} + \frac{2 \xi_2}{1+3\varpi}\right)\,\xi_3=0\,,
\end{align}
which has as only solution $\xi_3=0$, which further implies $a_3=0$ and so, from the third relation in \eqref{xi_3-theta_3_cond}, yields $\theta_3=0$. Notice that the conditions 
\begin{align}
\xi_1=\xi_3=\theta_1=\theta_3=0\ ,\label{ConditionsfhRegular}
\end{align}
which guarantee existence of the first and second derivative of $f$ and $h$ for $r=\zh$ (which are necessary for the expansions in Eq.~\eqref{f-h-taylor-r}). 
The upper bound $\xi_2\leq M/8 \zh$ was already obtained previously to guarantee reality of $\rootk$ in Eq.~\eqref{a_2-exactsol}.

\subsubsection{Ricci scalar and Hawking temperature}\label{EMD propdist ricci temp}
Next, the Ricci scalar is considered, however, for simplicity, we shall work out its series expansions only to leading order. Moreover, we shall start out by only assuming $\xi_1=0$ (which is required for the consistent expansion \eqref{SeriesFormaInt}), but the remaining conditions in \eqref{ConditionsfhRegular} shall not be assumed. Let us notably first consider the case of $\theta_1\neq 0$. Inserting \eqref{fzExpansion} and \eqref{hzExpansion} into the Ricci scalar in Eq.~\eqref{eq: ricci scalar h e f}, we obtain the following series expansion (for $r \geq \zh$)
\begin{align}
R=\frac{a_2-\xi_2}{8a_2\zh}\,(r-\zh)^{-1}+\frac{(2a_2^2+a_3\theta_1-2a_2\theta_2)(a_2-\xi_2)}{8\sqrt{2M}a_2^{5/2}\theta_1 \zh}\,(r-\zh)^{-1/2}+\mathcal{O}((r-\zh)^0)\ .\label{RicciSeriesExpansionThetaneq}
\end{align}
Using \eqref{Coeffsa} the coefficient of the leading term becomes
\begin{align}
\frac{a_2-\xi_2}{8a_2\zh}=\frac{\sqrt{1- 8\hr \xi_2 /M}}{8\zh}\ ,
\end{align}
which is non-vanishing for all values of $\xi_2\leq M/8\zh$ and therefore signals a curvature singularity at the event horizon. To avoid the latter, we now impose $\theta_1=0$, which also leads to a well defined derivative $\hn{1}$:
\begin{align}
\hn{1}=\frac{a_2-\theta_2}{a_2\zh}\ ,\label{h1hSerExp}
\end{align}
Moreover, the condition $\theta_1=0$ also changes the series expansion in (\ref{RicciSeriesExpansionThetaneq}), which now becomes
\begin{align}
&R=\frac{\sqrt{2}\sqrt{\zh(1+\rootk)}\left((1+3\rootk)\theta_3+2\xi_3\right)}{M^4(1+3\rootk)(1+\varpi- 4 \zh\theta_2/M)}\,(r-\zh)^{-1/2}+\mathcal{O}((r-\zh)^{0}) \ .
\end{align}
We have furthermore verified that the singularity of $R$ at $r=\zh$ cannot be removed if $1+\varpi-4\zh\theta_2/M=0$ and we, therefore, require $\theta_2 \neq a_2$.\footnote{This condition is compatible with $\theta_2<\frac{M(1+\varpi)}{4\zh}$ which guarantees that $\hn{1}>0$ in \eqref{h1hSerExp}. The latter is necessary such that $h(r)>0$ for $r > \zh$ (with a simple zero at $r=\zh$).} In this case, the necessary condition for regularity of $R$ at the horizon is $(1+3\rootk)\theta_3+2\xi_3=0$. We have verified that under the same condition also the Kretschmann scalar is finite at the horizon. The consistency conditions for the approach outlined in Section~\ref{Sect:SeriesExpansion}, the conditions for the absence of a singularity of the Ricci scalar, and the bounds for positive metric functions $f$ and $h$ for $r > \zh$ are collected in Result~\ref{res:NeutralRegularity}.

\begin{center}
\resultbox{\result\label{res:NeutralRegularity}{Regularity of the proper-distance EMD}}{\parbox{16cm}{
The self-consistent near-horizon solution of \eqref{DistanceDiffEq} exists, the metric functions $f$ and $h$ are positive for $r>\zh$ with a simple zero at $r=\zh$, and the Ricci and Kretschmann scalars are finite at the horizon, provided the EMD coefficients in \eqref{SeriesDistanceRho} satisfy
\begin{align}
&\xi_1=0\,,&&\theta_1=0\,,&&\xi_3=-\frac{1}{2}\,(1+3\rootk)\,\theta_3\,,&&\xi_2\leq \frac{M}{8\zh}\,,&&\theta_2< \frac{M(1+\rootk)}{4\zh}\ ,
\label{ConRicciFinite}
\end{align}
with $\varpi$ defined in \eqref{a_2-exactsol}. The Hawking temperature then depends only on $\zh$, $\xi_2$ and $\theta_2$, see \eqref{NeutralTH}.
}}
\end{center}

\noindent

We further remark that the series coefficients \eqref{Coeffsa} and \eqref{Coeffsb} along with the expansions \eqref{fhSerRho} allow to compute the (finite) value of $R$ at the horizon. While the general form is rather complicated, here we only give the expression in the particular case $f=h$ (\emph{i.e.} $\xi_n=\theta_n$ $\forall n\geq 1$) with $\xi_1=\xi_3=0$ 
\begin{align}
R\big|_{r=\zh}&=\frac{(1-\rootk^2)(2M a_2^2 - \hr \, a_4)}{8 \zh^3 a_2^3}+\frac{\xi_4}{M \zh a_2^2}= \notag \\ &=\frac{(1-\varpi)(33 \varpi^4+14 \varpi^3+2 \varpi-1)}{16 \hr^2 \varpi^3 (1+2 \varpi)} + \frac{24 \hr \, \xi_4}{M^3 (1+\varpi)(1+2 \varpi)}\ .
\end{align}
which we shall use in the examples in Chapter~\ref{chap examples}.

Before closing this Section, we also provide the expression for the Hawking temperature
\begin{align}
T_{\text{H}}=\frac{\sqrt{\fn{1}\hn{1}}}{4\pi}=\frac{\sqrt{1+\varpi-4\zh \theta_2/M}}{4\sqrt{2}\,\pi\,\zh}\ .\label{NeutralTH}
\end{align}

The upper bound on $\theta_2$ in \eqref{ConRicciFinite} guarantees that $T_{\text{H}}>0$. Indeed, $T_{\text{H}}=0$ would require $\fn{1}=0$ and/or $\hn{1}=0$, which translate into $a_2=\xi_2$ and $a_2=\theta_2$ respectively. The former has no real solution, while the latter leads to a singularity of the Ricci scalar at the horizon. We remark, however, that black hole solutions with $T_{\text{H}}=0$ are possible upon choosing the solution with $v=-1$ giving $a_2=\frac{M(1-\varpi)}{4\zh}$ in Eq.~\eqref{a_2-exactsol}.

\subsection{Regular higher order derivatives}\label{App:HigherOrderDerF}
In the previous subsections we have derived the conditions \eqref{ConRicciFinite} by assuming that the first and second derivatives of the metric functions $f$ and $h$ are finite at the horizon $\zh$. In this subsection, we explore further conditions that stem from assuming that also higher derivatives (\emph{i.e.} beyond the second) are finite. For simplicity, we shall focus on the function $f$, while the same considerations also apply to $h$. Concretely, let $N\in\mathbb{N}$ and let us assume that all derivatives $\fn{k}$ for $k\in\{1,\ldots,N\}$ at $r=\zh$ are finite. This allows us to go beyond the second order in Eq.~\eqref{f-h-taylor-r} and write 
\begin{align}
&f(z)=\sum_{k=1}^N \frac{\fn{k}}{k!} (r-\zh)^k+\mathcal{O}((r-\zh)^{n+1})\,, &&\text{with} &&\fn{k}:=\left.\dv[k]{f}{r}\right|_{z=\zh}\ .\label{Taylorf}
\end{align}  
\subsubsection{Distance function}
We can truncate the series for the proper distance $\rho$ in series of $r-\hr$ in Eq.~\eqref{SeriesDistance} at order $N$
\begin{align}
\rho=\sum_{k=1}^{2N-1}b_k\,(r-\zh)^{k/2}+\mathcal{O}\left((r-\zh)^{N}\right)\ ,\label{TaylorPolyRho}
\end{align}
Note that the coefficients $b_k$ of this series are re-scaled by absorbing some powers of $2M$ and they can be found as the solutions of the differential equation~\eqref{DiffEqDistance} up to order $N$, which we re-write in the form
\begin{align}
\left(\dv{r}{\rho}\right)^2=\frac{1}{f(r)}\ .\label{QuadDistExpansion}
\end{align}
The right-hand side of this equation has a simple pole at $r=\zh$ and we can write the Laurent series expansion
\begin{align}
&\frac{1}{f(r)}=\sum_{m=-1}^{N-2}\ell_m\,(r-\zh)^{m}+\mathcal{O}((r-\zh)^{N-1}) \qq{with} \ell_m\in\mathbb{R}\ .\label{InverseFexpansion}
\end{align} 
Multiplying both sides of this equation by $(r-\zh)$ and taking the limit $r\to \zh$, we find for the leading coefficient $\ell_{-1}=1/\fn{1}$. To extract the remaining coefficients, we consider the relation
\begin{align}
f(r)\,\dv{r}\frac{1}{f(r)}=-\frac{1}{f(r)}\,\dv{f}{r} \ ,
\end{align}
and expand both sides in powers of $(r-\zh)$. Comparing order by order (see Appendix~\ref{App:SerExpansions}) we then find
\begin{align}
\ell_0&=-\frac{\ell_{-1}\fn{2}}{2\fn{1}}=-\frac{\fn{2}}{2(\fn{1})^2}\,,\nonumber\\
\ell_p&=\frac{-1}{(p+1)\fn{1}}\left[\frac{(p+1)\ell_{-1}\,\fn{p+2}}{(p+2)!}+\sum_{k=1}^{p-1}\frac{k\ell_k \fn{p-k+1}}{(p-k+1)!}+\sum_{k=2}^{p+1}\frac{\ell_{p-k+1}\fn{k}}{(k-1)!}\right]\,,\hspace{0.2cm}1\leq p\leq N-2\,,
\end{align}
which fixes the coefficients $\ell_p$ iteratively (in terms of the $\fn{n}$). Inserting (\ref{InverseFexpansion}) into (\ref{QuadDistExpansion}), we find the following recursive structure for the coefficients $b_p$ 
\begin{align}
&b_1=2\sqrt{\ell_{-1}}=\frac{2}{\sqrt{\fn{1}}} \ , \hspace{2cm}b_2=0\ ,\nonumber\\
&b_p=\left\{\begin{array}{lcl}\frac{2}{(2s+3)b_1}\left[\ell_p-\sum_{n=2}^{2s+2}\frac{n(2s-n+4)}{4}\,b_n\,b_{2s-n+4}\right] &\text{if}& p=2s+3\in\mathbb{N}_{\text{odd}}\ , \\ -\frac{1}{(s+1)b_1}\sum_{n=2}^{2s+1}\frac{n(2s-n+3)}{4}b_nb_{2s-n+3} &\text{if}& p=2s+2\in\mathbb{N}_{\text{odd}}\ ,\end{array}\right.\hspace{0.1cm}3\leq p\leq 2N-1\,,\label{RecBF}
\end{align} 
which allows to fix them iteratively in terms of the $\fn{n}$. Since $b_2=0$, the recursive structure in (\ref{RecBF}) implies that $b_{2s}=0$ for $s\in\{1,\ldots, N-2\}$. For the odd coefficients, we find for the first few instances (for sufficiently large $N$)
\begin{align}
&b_1=\frac{2}{(\fn{1})^{1/2}}\ ,&&b_3=-\frac{\fn{2}}{6(\fn{1})^{3/2}}\ ,&&b_5=\frac{9(\fn{2})^2-8\fn{1}\fn{3}}{240(\fn{1})^{5/2}}\ .
\end{align}
These results are indeed compatible with the expressions in Eqs.~\eqref{first-few-bs-in-temrs-of-as}.

\subsubsection{Conditions for the regularity of higher derivatives \texorpdfstring{$f^{(n)}$}{fn}}\label{App:AllHigherOrders}
The results of the previous subsection can be used to derive necessary conditions for the function $\Phi^{(\rho)}$ such that the first $N$ derivatives of the function $f$ (with respect to $r$) are finite, which is required for~\eqref{Taylorf}. Assuming an expansion of the latter of the form \eqref{SeriesDistanceRho} (rescaling the coefficients $\xi_n \mapsto (2M)^{n-1} \xi_n$ and $\xi_0\equiv \hr$), we have with \eqref{TaylorPolyRho}
\begin{align}
f(r)=1-\frac{1}{r}\sum_{n=0}^{2N}\xi_n\,\left(\sum_{k=1}^{2N-1}b_k\,(r-\zh)^{k/2}\right)^n+\mathcal{O}\left((r-\zh)^N\right)\ ,\label{fFormExpansion}
\end{align}
which is only a function of $r$. Due to the fact that this expression contains half-integer powers of $(r-\zh)$, derivatives of $f$ can contain negative powers unless certain conditions for the coefficients $\xi_n$ are satisfied. To understand these conditions, we first re-write the summation in eq.~\eqref{fFormExpansion}, taking into account that $b_{2s}=0$ for $s\in\{1,\ldots,N-2\}$
\begin{align}
f(r)=1&-\frac{1}{r}\sum_{m=0}^N\xi_{2m}(r-\zh)^m\left(\sum_{k=0}^{N-1}b_{2k+1}\,(r-\zh)^{k}\right)^{2m}\nonumber\\
&\qquad -\frac{1}{r}\sum_{m=1}^{N}\xi_{2m-1}\left(\sum_{k=0}^{N-1}b_{2k-1}\,(r-\zh)^{k+1/2}\right)^{2m-1}+\mathcal{O}\left((r-\zh)^N\right)\ .\label{fFunSplit}
\end{align}
Re-writing the above result by expressing the Cauchy product with the expansion of $1/r$ in power series in $r-\hr$
\begin{align}
\frac{1}{r}=\frac{1}{\zh}\sum_{n=0}^N\frac{(-1)^n}{\zh^n}\,(r-\zh)^n+\mathcal{O}\left((r-\zh)^{N+1}\right)\,,
\end{align}
it is clear that the terms in the first line of \eqref{fFunSplit} contain no half-integer powers of $r-\zh$ and thus cannot contribute to singularities of derivatives of $f$ at $r=\zh$. The terms in the second line of \eqref{fFunSplit}, however, contain terms that lead to negative powers of $r-\zh$ for derivatives of $f$. The conditions to eliminate these singular terms for all $\fn{k}$ for $k\in\{1,\ldots,N\}$ are therefore
\begin{align}
\xi_{2n-1}=0\hspace{1cm} \forall n\in \{1,\ldots,N\}\ .\label{XiHigherRel}
\end{align}
To see this, let $s\in\{1,\ldots,N\}$ and assume that $\xi_{2s-1}\neq 0$, while $\xi_{2m-1}=0$ $\forall m\in\{1,\ldots,s-1\}$. In this case, $\fn{s}$ has a singularity of the form $(r-\zh)^{-1/2}$
\begin{align}
\dv[s]{r}f(r)&=-\frac{\xi_{2s-1}}{\zh}\,b_1^{2s-1}\,\dv[s]{r}(r-\zh)^{s-1/2}+\mathcal{O}((r-\zh)^0)\nonumber\\
&=-\frac{(2s-1)!!}{2^s\zh}\,\xi_{2s-1}\,b_1^{2s-1}\,(r-\zh)^{-1/2}+\mathcal{O}((r-\zh)^0)\,.
\end{align}
Absence of this singular contribution therefore requires $\xi_{2s-1}=0$.
Generalizing the procedure for $N\to \infty$, similar conclusions can be drawn for the function $\Psi^{(\rho)}$ in $h$, which read
\begin{equation}
    \theta_{2n+1}=0 \qq{with} \forall n\in \mathbb{N}_0 \ .\label{ThetaHigherRel}
\end{equation}
Using these conditions into the recursive relation in Eq.~\eqref{Coeffsa} and repeat the procedure for Eq.~\eqref{Coeffsb}, we have that 
\begin{equation}
    a_{2n+1}=0=b_{2n} \qq{with} n \in \mathbb{N}_0 \ . \label{ABcondition}
\end{equation}

\begin{center}
\resultbox{\result\label{res:SmoothHorizon}{Smoothness of the metric at the horizon}}{\parbox{16cm}{
All derivatives $\fn{n}$ and $\hn{n}$ of the metric functions at the horizon are finite if and only if all odd EMD coefficients vanish,
\begin{align}
&\xi_{2n-1}=0\,,&&\theta_{2n-1}=0\,,&&\forall n\in\mathbb{N}\,.\nonumber
\end{align}
Necessity follows from \eqref{XiHigherRel} and \eqref{ThetaHigherRel}. Conversely, if all odd coefficients vanish, then \eqref{ABcondition} holds, $\rho^2$ is a power series in integer powers of $r-\zh$, and $f$ and $h$, which depend on $\rho$ only through $\rho^2$, admit Taylor expansions of the form \eqref{f-h-taylor-r}. In this case the deformation functions are even functions of $\rho$, and the EMD is parametrised by $\zh$ and $\{\xi_{2n},\theta_{2n}\}_{n\in\mathbb{N}}$.
}}
\end{center}

\subsection{Conditions for regularity at inner horizons}\label{App:InnerHorizon} 
The sufficient conditions summarized in Eq.~\eqref{ConRicciFinite} at the end of Section~\ref{EMD propdist ricci temp} for regularity of the Ricci tensor at $\zh$, have been derived assuming that the latter is the position of the outer horizon: notably, we have assumed in various instances that $f(r)>0$ for $r >\zh$. Generalized Schwarzschild BHs, however, may have further horizons, (for example, the BHs with regularized cores \cite{Hayward_2006,bardeen1968,Dymnikova:1992ux,Frolov:2016pav,Bonanno:2000ep}) which are characterized by a vanishing of the function $f(r)$ in \eqref{SpStMetric}, such that the derivative of the distance~\eqref{DiffEqDistance} diverges. The latter can (in the same way as discussed in Section~\ref{EMD propdist ricci temp}) lead to curvature singularities that are physically not acceptable. Assuming that the metric functions $h$ and $f$ are of the form
\begin{equation}
   h = 1- \frac{\Psi^{(d)}}{r} \qq{and} f=1-\frac{\Phi^{(d)}}{r} \ ,
\end{equation}
where the full proper distance $d$ is defined as
\begin{equation}\label{complete prop dist def}
    \dv{d}{r} \coloneqq \frac{1}{\sqrt{|f(r)|}} \ ,
\end{equation}
and the presence of an inner horizon, therefore, puts further conditions on the functions $\Phi$ and $\Psi$. We can obtain these conditions by straightforwardly generalizing the discussion of Section~\ref{EMD propdist ricci temp}. Here we shall briefly sketch them, assuming a black hole with $f=h$ and two (simple) horizons at $\zhpm$ (with $\zhp>\zhm$).\footnote{Here $\zhp$ is understood to be the position of the outer horizon, which we denote by $\zh$ throughout the remainder of this review.}
At $r= \zhm$, one can introduce the series expansion analogous to Eq.~\eqref{SeriesDistanceRho} as
\begin{equation}
    \Phi^{(\zeta)}= \zhm + 2M \sum_{n=0} \overline{\xi}_n \left(\frac{\zeta}{2M} \right)^n \ ,
\end{equation}
where $\zeta$ is the the proper distance from the inner event horizon $\zhm$. The solution of the series expansion of $\zeta (r)$ (or equivalently of $r(\zeta)$) in power series of $r - \zhm$ from Eq.~\eqref{complete prop dist def}, leads to the condition $\overline{\xi}_1 = 0$ which makes the first derivative of $f$ regular at $\zhm$. Investigating the regularity of the second derivative through the Ricci scalar, one finds that $\overline{\xi}_3 = 0$, in complete analogy with the conditions summarized in Eqs.~\eqref{ConRicciFinite} for the external horizon.


\subsection{EMD classification of regular cores}\label{Sect:FullRegularity}
The consistency conditions derived so far concern the immediate vicinity of the horizons.\footnote{At a horizon located at $r_P$, the derivatives of the deformation functions have to be taken separately on its two sides, since there $r-r_P\propto\pm(d-d_P)^2$. For $f\neq h$, the third-order condition then takes the general form $\Phi^{(3)}_P+\big(3\,r_P\,f'(r_P)-1\big)\Psi^{(3)}_P=0$, with $\Phi^{(n)}_P$ and $\Psi^{(n)}_P$ the (one-sided) derivatives with respect to $d$ at the horizon. It reproduces the last condition in~\eqref{ConRicciFinite} at the outer horizon and gives $\Phi^{(3)}_P=0$ for $f=h$, cf. Section~\ref{App:InnerHorizon}.} A geometry that is well behaved there may nevertheless be singular elsewhere, most notably at the centre $r=0$, where the Schwarzschild solution has its curvature singularity. Since the removal of this singularity is one of the main motivations for studying deformed black holes (cf. Section~\ref{sec:open-issues}), we now discuss, following~\cite{DamiaPaciarini:2025qhq}, the conditions under which the EMD describes a geometry that is regular at its core and geodesically complete, together with the resulting classification of regular cores. Applications to concrete geometries are discussed in Section~\ref{Sect:CoreExamples}.

To describe the interior of the black hole, including its centre, we use the deformation functions of the full proper distance $d$ introduced in~\eqref{complete prop dist def}, with $d(0)=0$, and write $\Phi^{(d)}\equiv \Phi$ and $\Psi^{(d)}\equiv\Psi$. Their $n$-th derivatives with respect to $d$ at the origin are denoted by $\Phi^{(n)}_0$ and $\Psi^{(n)}_0$. For the metric~\eqref{SpStMetric}, the Riemann tensor in an orthonormal frame adapted to $(t,r,\theta,\varphi)$ has four independent components $\mathcal{K}_{ab}\coloneqq R^{\hat{a}\hat{b}}{}_{\hat{a}\hat{b}}$ (no summation), and the Kretschmann scalar~\eqref{KretschDef} is a sum of their squares, $K=4\mathcal{K}_{tr}^2+8\mathcal{K}_{t\theta}^2+8\mathcal{K}_{r\theta}^2+4\mathcal{K}_{\theta\varphi}^2$. All curvature invariants are therefore finite at a point if and only if these four components are. At the origin, $\mathcal{K}_{\theta\varphi}=(1-f)/r^2=\Phi/r^3$ is finite only if $\Phi$ vanishes at least as $r^3$, while $\mathcal{K}_{t\theta}=-f\,h'/(2rh)$ requires $h$ to be finite, non-vanishing and free of a linear term. These requirements lead to
\begin{align}
&\Phi^{(0)}_0=\Phi^{(1)}_0=\Phi^{(2)}_0=0\,,&&\Psi^{(0)}_0=\Psi^{(2)}_0=0\,,&&\Psi^{(1)}_0\neq1\,,\label{reg:OriginCond}
\end{align}
which are also sufficient for all curvature invariants to be finite at the centre. The Schwarzschild solution, with $\Phi=\Psi=2M$, clearly violates the first of these conditions.

\subsubsection{Geodesic completeness}\label{Sect:FullGeodesic}
Finite curvature invariants and geodesic completeness are logically independent notions of regularity~\cite{Hawking:1973uf,Geroch:1968ut}: the Taub--NUT spacetime~\cite{Taub:1950ez,Newman:1963yy} has bounded curvature but is geodesically incomplete, whereas geodesically complete spacetimes with divergent curvature can be found among the Kundt spacetimes~\cite{Kundt:1961xxx,Podolsky:2004xxx}. A spacetime is geodesically complete if every geodesic can be extended to arbitrary values of its affine parameter, \emph{i.e.} if the trajectories of free test particles never end abruptly. For the metric~\eqref{SpStMetric}, the radial equation of geodesic motion is Eq.~\eqref{geodesic eq u} of Section~\ref{Sect:PhotonRadius}, which reveals two obstructions to the extension of geodesics. First, if $f$ vanishes at some $r_P>0$ while $h$ does not, the right-hand side of~\eqref{geodesic eq u} is proportional to $r-r_P$ and changes sign at $r_P$. This point is reached at finite affine parameter, $\tau\propto|r-r_P|^{1/2}$, and beyond it the metric changes signature, such that the geodesics cannot be continued. Second, radial null geodesics ($L=u=0$) obey $(\dv*{r}{\tau})^2=E^2f/h$, which approaches a constant at a regular origin, so that the centre is reached at finite affine parameter. Their continuation through $r=0$ requires $h(0)>0$, since otherwise both congruences of radial null geodesics focus at the origin~\cite{DamiaPaciarini:2025qhq}.

\begin{center}
\resultbox{\result\label{res:FullGeodesic}{Necessary conditions for geodesic completeness}}{\parbox{16cm}{
Besides the finiteness of all curvature invariants, geodesic completeness requires
\begin{align}
&\Phi(d_P)=r_P\quad\Longleftrightarrow\quad\Psi(d_P)=r_P\quad\forall\,d_P>0\,,&&\text{and}&&\Psi^{(1)}_0<1\,,\label{reg:GeodesicCond}
\end{align}
\emph{i.e.} the zeros of $f$ and $h$ must coincide, and $h(0)=1-\Psi^{(1)}_0$ must be positive. For a regular origin and simple zeros, the second condition follows from the first (see Remark~\ref{rem:EvenHorizons}). The conditions are necessary but not sufficient: the continuation of geodesics through inner horizons, as in the Reissner-Nordstr\"om spacetime, requires analytic extensions of the geometry that lie beyond the present analysis.
}}
\end{center}

\begin{center}
\remarkbox{\rem\label{rem:EvenHorizons}{Regular black holes have an even number of horizons}}{\parbox{16cm}{
At a regular origin, \eqref{reg:OriginCond} implies $f(0)=1$, while $f\to1$ at spatial infinity. If all zeros of $f$ are simple, $f$ changes sign at each of them, so their number must be even, and by~\eqref{reg:GeodesicCond} the same holds for $h$, which is therefore positive at the origin. Black holes with a finite curvature everywhere thus possess an even number of horizons, typically an inner and an outer one. Conversely, geometries with an odd number of horizons, such as the Schwarzschild black hole, necessarily have a singular core. This is consistent with the minimal deformation model of Section~\ref{Sect:MinimalModelInside}, where a regular origin was found to require an additional inner horizon.
}}
\end{center}

\subsubsection{Classification of regular cores}\label{Sect:CoreClassification}
When the conditions~\eqref{reg:OriginCond} hold, the proper distance close to the origin is $d=r+\Phi^{(3)}_0\,r^3/36+\mathcal{O}(r^4)$, and the metric functions behave as
\begin{align}
&f(r)=1-\frac{\Phi^{(3)}_0}{6}\,r^2+\mathcal{O}(r^3)\,,&&h(r)=1-\Psi^{(1)}_0-\frac{1}{6}\left(\Psi^{(3)}_0+\frac{\Phi^{(3)}_0\Psi^{(1)}_0}{6}\right)r^2+\mathcal{O}(r^3)\,.\label{reg:CoreExpansion}
\end{align}
The four components of the Riemann tensor at the centre are then
\begin{align}
&\mathcal{K}_{r\theta}=\mathcal{K}_{\theta\varphi}=\frac{\Phi^{(3)}_0}{6}\,,&&\mathcal{K}_{tr}=\mathcal{K}_{t\theta}=\frac{1}{6\,(1-\Psi^{(1)}_0)}\left(\Psi^{(3)}_0+\frac{\Phi^{(3)}_0\Psi^{(1)}_0}{6}\right)\qquad\text{at}\quad r=0\,.\label{reg:CoreCurvature}
\end{align}
The core is locally maximally symmetric if all four components coincide, in which case the Riemann tensor at the origin is that of a space of constant curvature $\Lambda/3$. This leads to the following classification.

\begin{center}
\resultbox{\result\label{res:Cores}{Classification of regular cores}}{\parbox{16cm}{
A regular core, satisfying~\eqref{reg:OriginCond}, is locally maximally symmetric if and only if
\begin{align}
\Psi^{(3)}_0=\left(1-\frac{7}{6}\,\Psi^{(1)}_0\right)\Phi^{(3)}_0\,,\label{reg:MaxSymCore}
\end{align}
with effective cosmological constant $\Lambda=\Phi^{(3)}_0/2$. The core is then
\vspace{5pt}
\begin{itemize}\centering
\item \emph{de Sitter} for $\Phi^{(3)}_0>0$,
\item \emph{anti-de Sitter} for $\Phi^{(3)}_0<0$,
\item \emph{Minkowski} for $\Phi^{(3)}_0=\Psi^{(3)}_0=0$,
\end{itemize}
\vspace{5pt}
while it is anisotropic (\emph{mixed}) if~\eqref{reg:MaxSymCore} is violated. For $h(0)=1$, \emph{i.e.} $\Psi^{(1)}_0=0$, the condition reduces to $\Psi^{(3)}_0=\Phi^{(3)}_0$.
}}
\end{center}

\begin{center}
\remarkbox{\rem\label{rem:CoreConventions}{Relation to the conventions of~\cite{DamiaPaciarini:2025qhq}}}{\parbox{16cm}{
In~\cite{DamiaPaciarini:2025qhq} the deformation functions are dimensionless, $h=1-2M\hat{\Psi}(d)/r$ and $f=1-2M\hat{\Phi}(d)/r$, so that $\Psi=2M\hat{\Psi}$ and $\Phi=2M\hat{\Phi}$. For instance, $\Lambda=M\hat{\Phi}^{(3)}_0$ and $h(0)>0$ becomes $\hat{\Psi}^{(1)}_0<1/(2M)$. Moreover, cores with $\Psi^{(1)}_0\neq0$ are classified there as mixed. Such cores have $h(0)\neq1$, \emph{i.e.} a constant redshift between the centre and spatial infinity (where $t$ is normalized). Since a constant can be absorbed into a local rescaling of $t$, Result~\ref{res:Cores} classifies the core by its local curvature alone, and the two classifications coincide for $\Psi^{(1)}_0=0$.
}}
\end{center}

\section{EMD redefinitions}\label{Sect:ShiftOfFunction}
Before discussing \scheme s  based on different physical quantities (that is curvature scalars rather than the physical distance), we shall first consider \scheme\,redefinitions, in which the physical quantity $\gen$ is replaced by simple functions of it: indeed, consider the starting point 
\begin{align}
\Phi^{(\gen)}(\gen)=\sum_{n=0}^\infty \cgen{n}\,\gen^n\,&&\text{with}&&\cgen{n}\in\mathbb{R}\hspace{0.2cm}\forall n\geq0\,,\label{SeriesGenPhys}
\end{align}
which we assume to have finite radius of convergence $\gen_c$, and replace $\gen$ by a continuous, monotonic function of $\gen$, {{which we denote}} $\sigma:\,[0,\gen_c]\longrightarrow \mathbb{R}$. Then we define $\wgen=\sigma(\gen)$, along with the deformation function
\begin{align}
\Phi^{(\wgen)}(\wgen)=\sum_{n=0}^\infty \wcgen{n}\,\wgen^n\,&&\text{with}&&\wcgen{n}\in\mathbb{R}\hspace{0.2cm}\forall n\geq0\,.\label{SeriesGenPhysMod}
\end{align}
A very simple example of this idea was already used in \cite{DAlise:2023hls}, where a physical distance shifted by a constant was considered, \emph{i.e.} $\widetilde{\gen}=\sigma(\gen)=\gen+c$ for $c\in\mathbb{R}$. In this case, the two sets of coefficients $\{\cgen{n}\}_{n\in\mathbb{N}_0}$ and $\{\wcgen{n}\}_{n\in\mathbb{N}_0}$ are related
\begin{align}
&\cgen{n}=\sum_{\ell=n}^\infty\frac{\wcgen{\ell}\,c^{\ell-n}}{(\ell-n)!}\,,&&\wcgen{n}=\sum_{\ell=n}^\infty\frac{\cgen{\ell}\,(-c)^{\ell-n}}{(\ell-n)!}\,,&&\forall n\in\mathbb{N}\,.\label{ConstShiftRel}
\end{align}
Notice that in this case, any single $\cgen{n}$ depends on infinitely many of the $\{\wcgen{n'}\}_{n'\geq n}$ and vice-versa. 

In order to describe the relation between \eqref{SeriesGenPhys} and \eqref{SeriesGenPhysMod} for a more general function $\sigma$, we assume that the latter can be expanded 
\begin{align}
&\sigma(\gen)=\sum_{k=1}^\infty \sig{k}\,\gen^k\,,&&\text{with} &&\sig{k}\in\mathbb{R}\,,\label{SeriesExpandL}
\end{align}
which has a radius of convergence $\gen'_c\geq \gen_c$.\footnote{Since \eqref{ConstShiftRel} already describes the relation for a shift by a constant, we have assumed $\sig{0}=0$. Furthermore, in the following we shall implicitly assume $\sig{1}\neq 0$.}
By demanding $\Phi^{(\gen)}(\gen)=\Phi^{(\wgen)}(\wgen)$
\begin{align}
\Phi^{(\wgen)}(\wgen)=\sum_{n=0}^\infty \wcgen{n}\,\left(\sum_{k=1}^\infty \sig{k}\,\gen^k\right)^n
=\Phi^{(\gen)}(\gen)\ ,
\end{align}
and using \eqref{DefRedFormPower} in Appendix~\ref{App:SerExpansions}, we obtain the following relation
\begin{align}
&\cgen{0}=\wcgen{0}\,,\hspace{1cm}\cgen{1}=\sig{1}\,\wcgen{1}\,,\nonumber\\
&\cgen{m}=\sig{m}\,\wcgen{1}+\sum_{p=2}^m \wcgen{p}   \sum_{\genfrac{}{}{0pt}{}{0\leq \ell_{p-1}\leq \ell_{p-2}\leq \ldots\leq \ell_1\leq n}{\ell_1+\ldots+\ell_{p-1}=p-n}}\,\frac{n!\,\sig{1}^{n-\ell_1}\ldots \sig{p-1}^{\ell_{p-2}-\ell_{p-1}}\sig{p}^{\ell_{p-1}}}{(n-\ell_1)!\ldots (\ell_{p-2}-\ell_{p-1})!\ell_{p-1}!}\ .\label{FormOmegRel}
\end{align}
This identification is a linear relation, which expresses $\cgen{m}$ as a linear combination of \emph{finitely} many $\wcgen{n}$ with $n\leq m$. Formally, this relation can be written as a linear transformation with an infinite dimensional lower triangular matrix
\begin{align}
\left(\begin{array}{c}\cgen{0} \\ \cgen{1} \\ \cgen{2} \\ \cgen{3} \\ \cgen{4} \\ \cgen{5}\\ \vdots\end{array}\right)=\left(\begin{array}{ccccccc}1 & 0 & 0 & 0 & 0& 0 & \cdots \\ 0 & \sig{1} & 0 & 0 & 0 & 0 & \cdots \\ 0 & \sig{2} & \sig{1}^2 & 0 & 0 & 0 & \cdots \\ 0 & \sig{3} & 2\sig{1}\sig{2} & \sig{1}^3 & 0 & 0 & \cdots \\ 0 & \sig{4} & \sig{2}^2+2\sig{1}\sig{3} & 3\sig{1}^2 \sig{2} & \sig{1}^4 &  0 & \cdots \\ 0 & \sig{5} & 2(\sig{1}\sig{4}+\sig{2}\sig{3}) & 3(\sig{1}\sig{2}^2+\sig{1}^2\sig{3}) & 4\sig{1}^3 \sig{2} & \sig{1}^5 & \cdots\\ \vdots & \vdots & \vdots & \vdots &\vdots & \vdots & \ddots\end{array}\right)\cdot\left(\begin{array}{c}\wcgen{0} \\ \wcgen{1} \\ \wcgen{2} \\ \wcgen{3} \\ \wcgen{4} \\ \wcgen{5}\\ \vdots\end{array}\right)\,.\label{GenLinTransform}
\end{align}
Combining (\ref{ConstShiftRel}) with (\ref{GenLinTransform}) allows to describe and connect to more general \scheme\,redefinitions, such as the Ricci- and Kretschmann-EMDs in App.~\ref{Sect:RicciScalar\scheme} and~\ref{Sect:KretschmannScalar}.

\section{EMD for charged black holes}\label{Sect:ChargedEMD}
Up to this point, the EMD has been formulated for deformations of the Schwarzschild geometry. In this Section we show that the same logic can be applied when the classical limit is the Reissner-Nordstr\"om (RN) solution reviewed in Sec.~\ref{Sect:RNBH}, following the analysis of~\cite{DamiaPaciarini:2025xjc}. Two aspects distinguish the charged case from the neutral one. First, the classical metric functions contain two structurally different terms, the mass term $\propto M/r$ and the charge term $\propto Q^2/r^2$, each of which may be modified in its own way by physics beyond GR. Second, already the classical RN spacetime generically possesses an inner and an outer horizon, which merge into a single degenerate horizon in the extremal limit $|Q|=M$. The first aspect can be handled by a simple reorganisation of the deformation functions. The second one, instead, forces us to reconsider which physical quantity is best suited to define the EMD.

The logic of this Section is as follows. We first introduce the deformed RN metric, parametrised in terms of the proper distance to the outer horizon, and solve the corresponding self-consistency condition as a near-horizon series (Sec.~\ref{Sect:EMDDistance}). We then derive the constraints on the deformation parameters that follow from a regular geometry and a well-defined Hawking temperature (Sec.~\ref{Sect:Charged_Curvature}), and we discuss the regime of small charge (Sec.~\ref{Sect:SmallCharge}) as well as the electromagnetic field supported by the deformed metric (Sec.~\ref{Sect:MaxwellDeformed}). Finally, we show that extremal configurations call for an EMD based on the proper time of a free-falling observer, which we construct and analyse in Sec.~\ref{Sect:Extremal}.

\paragraph{Deformed Reissner-Nordstr\"om metric.} We retain the static and spherically symmetric form~\eqref{RNMetric} of the line element and deform the mass and the charge contributions to the two metric functions independently
\begin{equation}
    \label{DeformedFunctions}
    h(r) = 1-\frac{2 M }{r} \Psi ^{(\rho)}\left(\rho\right)  + \frac{Q^2}{r^2}  X^ {\left(\rho\right)} (\rho)\ , \qq{and}
    f(r) = 1-\frac{2 M}{r}  \Phi^ {\left(\rho\right)}(\rho)  + \frac{Q^2}{r^2}  \Upsilon ^{\left(\rho\right)}(\rho) \ ,
\end{equation}
As in Sec.~\ref{Sect:ReviewDistance}, the four functions $\Psi^{(\rho)}$, $X^{(\rho)}$, $\Phi^{(\rho)}$ and $\Upsilon^{(\rho)}$ depend on the Schwarzschild coordinate only through the proper radial distance $\rho$ from the outermost horizon, which is located at $r=\zh$,\footnote{Both \eqref{DeformedFunctions} and \eqref{DefinitionRadialProperDistance} could be continued to $\rho<0$, as long as the interior of the black hole is still described by a metric of the same form. We shall not need this continuation, since we only describe the exterior region.}
\begin{align}
&\rho(r) \coloneqq \int_{\zh}^{r} \frac{ \dd r' }{ \sqrt{ |f(r')|} } \qq{for} r\geq \zh\ .\label{DefinitionRadialProperDistance}
\end{align}
Together with $\zh$, these four functions define the charged $\rho$-EMD. Because $\rho$ is a coordinate-independent quantity that can be evaluated in any given model, charged black holes stemming from different theories beyond GR can all be mapped into the same set of EMD parameters and compared directly, without reference to the details of the underlying theory. In the following we only describe the region $r\geq\zh$ (\emph{i.e.} $\rho\geq 0$), which is accessible to an external observer.

A priori, the four deformation functions are unrelated. Two requirements, however, restrict them. The spacetime should be asymptotically flat and approach the classical RN geometry at large distances, which requires
\begin{equation}
\lim_{\rho\to\infty}\Phi(\rho) = \lim_{\rho\to\infty}X(\rho) = \lim_{\rho\to\infty}\Psi (\rho) = \lim_{\rho\to\infty}\Upsilon (\rho) = 1\ , \qquad \forall Q\in\mathbb{R}  \ .
\end{equation}
Moreover, both metric functions have to vanish at $\rho=0$, so that the latter indeed represents a horizon. Denoting with a subscript $_H$ the value of the deformation functions at $\rho=0$, this amounts to
\begin{align}
&\Phi_H-\frac{Q^2}{2M\zh}\,\Upsilon_H = \Psi_H - \frac{Q^2}{2 M \zh}\,X_H=\frac{\zh}{2M}\ , \qquad \forall  Q \in\mathbb{R}   \ .\label{ChargedHorizonCond}
\end{align}
Under this condition, the time-like Killing vector $(\chi^t)^\mu = \delta^{0\mu}$ becomes null at $\rho=0$,
\begin{equation}
    \left.(\chi^t)^{\mu} (\chi^t)_{\mu}\right|_{r=\hr} = \left. g_{00}\right|_{r=r_H} = - h(\zh) = 0 \ ,
\end{equation}
such that $\rho=0$ is a Killing horizon.

\subsection{Self-consistent near-horizon solution}\label{Sect:EMDDistance}
We now make the charged EMD explicit in a neighbourhood of the outer horizon. The strategy is the same as in the neutral case (Sec.~\ref{Sect:DistanceFunctionExpansion}): the defining relation \eqref{DefinitionRadialProperDistance} is solved order by order in a series expansion around $\rho=0$.

A first simplification concerns the number of deformation functions. The metric functions in \eqref{DeformedFunctions} depend on the four deformation functions only through two combinations. Since close to the horizon $r$ is an invertible function of $\rho$, these combinations can themselves be regarded as functions of $\rho$, and we can equivalently write
\begin{align}
    f(r) = 1 - \left( \frac{2 M }{r} - \frac{Q^2}{r^2} \right) \Gamma ^{(\rho)}\left( \rho \right), \quad
    h(r) = 1- \left(\frac{2 M }{r}  - \frac{Q^2}{r^2} \right)  \Omega ^{(\rho)}\left(\rho \right) \ ,\label{MetricDefGen}
\end{align}
with
\begin{align}
    \Gamma ^{(\rho)}\left( \rho \right) = \frac{ 2 M \, r(\rho)\, \Phi^{(\rho)} \left( \rho \right)  - Q^2  \Upsilon ^{(\rho)} \left( \rho \right)}{2 M \,  r (\rho) - Q^2},  \quad
    \Omega ^{(\rho)}\left( \rho \right) = \frac{ 2 M \, r(\rho) \, \Psi ^{(\rho)} \left( \rho \right)  - Q^2  X^{(\rho)} \left( \rho \right)}{2 M \, r(\rho) - Q^2} \ .
\end{align}

\begin{center}
\remarkbox{\rem\label{rem:ReductionDeformations}{Mass versus charge deformations}}{\parbox{16cm}{
The functions $\Gamma^{(\rho)}$ and $\Omega^{(\rho)}$ are weighted combinations of the mass and charge deformations, with weights $2Mr$ and $-Q^2$, normalised such that $\Gamma^{(\rho)}=1$ for $\Phi^{(\rho)}=\Upsilon^{(\rho)}=1$ (and similarly for $\Omega^{(\rho)}$). Only these combinations enter the metric, so that any deformation of the charge term can be traded for a suitable deformation of the mass term and vice-versa. Furthermore, as we shall see in Sec.~\ref{Sect:MaxwellDeformed}, also the electromagnetic field strength compatible with the deformed metric is completely fixed by $f$ and $h$. Within the setup considered here, the split of a given deformation into a \emph{mass} and a \emph{charge} part is therefore not observable, and $\Gamma^{(\rho)}$ and $\Omega^{(\rho)}$ can be regarded as the genuine deformation functions of the charged EMD. This also makes the computations considerably more compact.
}}
\end{center}

\noindent
For sufficiently small $\rho$, we assume that $\Gamma^{(\rho)}$ and $\Omega^{(\rho)}$ admit convergent power series expansions
\begin{align}
\label{SeriesExpDefFunc}
    &\Gamma \left(\rho \right) = \sum_{n = 0} \xi_n \rho^n \,,&&\text{and}&& \Omega \left(\rho \right) = \sum_{n = 0} \theta_n \rho^n\,.
\end{align}
The requirement $f(\zh)=0=h(\zh)$ then fixes their leading coefficients
\begin{equation}\label{GammaHorizonCondition}
\xi_0=\Gamma_H=\frac{\zh^2}{2M\,  \zh - Q^2}=\Omega_H=\theta_0 \ ,
\end{equation}
which is the counterpart of \eqref{ChargedHorizonCond} for the combined deformation functions. Locally, the charged EMD is thus characterised by $\zh$ (along with $M$ and $Q$) and by the coefficients $\{\xi_n,\theta_n\}_{n\in\mathbb{N}}$, which we regard as the physical input parameters that specify the near-horizon geometry.

In order to solve \eqref{DefinitionRadialProperDistance} self-consistently, we restrict ourselves to non-extremal black holes. In other words, we assume that $r=\zh$ is a simple zero of both metric functions
\begin{align}
\label{DeformedNearH}
 &f(r) = f^{(1)}_H (r-r_H) + \order{(r-r_H)^{3/2}} \,,&&\text{and} &&h(r)=h^{(1)}_H (r-r_H) + \order{(r-r_H)^{3/2}} \,,
\end{align}
with non-vanishing coefficients $f_H^{(1)}\neq 0\neq h_H^{(1)}$.\footnote{The extremal case $f_H^{(1)}=h_H^{(1)}=0$ requires a different treatment, which is presented in Section~\ref{Sect:Extremal}.} This assumption guarantees that the integrand in \eqref{DefinitionRadialProperDistance} has an integrable singularity at $r=\zh$, such that the proper distance is finite and behaves as
\begin{align}
\rho(r) = \frac{2 \sqrt{r-r_H}}{\sqrt{f^{(1)}_H}} + \order{(r-r_H)} \ .\label{dExpansion}
\end{align}
The inverse relation, $r(\rho) = r_H + \frac{f^{(1)}_H}{4} \rho^2 +  \mathcal{O}(\rho^3)$, suggests to parametrise the radial coordinate, in analogy with \eqref{SeriesDistance}, as
\begin{equation}
\label{zExpansionInRho}
    r = r_H + \sum_{n=1}^{\infty} a_n \rho^n \ .
\end{equation}
where the coefficients $\{a_n\}_{n\in\mathbb{N}}$ have to be determined in terms of the EMD parameters. Differentiating \eqref{DefinitionRadialProperDistance} with respect to $\rho$ and using \eqref{MetricDefGen}, the self-consistency condition takes the form of the non-linear differential equation
\begin{equation}
\label{NonExtremeDiffEq}
     r \left( 1- \left( \dv{r}{\rho} \right)^2 \right)  =  \left(2 M - \frac{Q^2 }{r}\right) \Gamma (\rho ) \ .
\end{equation}
The steps of its order-by-order solution are summarised in Computation~\ref{comp:ChargedSeries}.

\begin{center}
\compbox{\comp\label{comp:ChargedSeries}{Near-horizon series solution of the charged EMD}}{\parbox{16cm}{
Inserting \eqref{SeriesExpDefFunc} and \eqref{zExpansionInRho} into \eqref{NonExtremeDiffEq} and collecting powers of $\rho$ gives
\begin{align}
\label{Self-ConsistentCoefficients}
    &\sum_{p=0}^\infty \left( 2 M\, \xi_p - Q^2 \sum_{n=0}^{p} \xi_n \mathfrak{p}_{p-n} \right) \rho^p = r_H (1- a_1^2) \nonumber \\
    & \qquad+ \sum_{p=1}^\infty \left[ (1-a_1^2) a_p - r_H \sum_{m=1}^{p+1} (p-m+2) m a_m a_{p-m+2} \right. \nonumber \\
    &\qquad \qquad \left. - \sum_{n=1}^{p-1} \sum_{m=1}^{n+1} (n-m+2) m a_{p-n} a_m a_{n-m+2}\right] \rho^p \ ,
\end{align}
where $\mathfrak{p}_n$ are the coefficients of the expansion of $1/r$ in powers of $\rho$, defined in analogy with~\eqref{inverse-z-rho-exp}. The three lowest orders of \eqref{Self-ConsistentCoefficients} read
\begin{subequations}\label{FirstThreeCoeff}
\begin{align}
    r_H^2(1-a_1^2) &= r_H 2 M\, \xi_0 - Q^2 \xi_0 \ ,  \\
    -2 a_1 r_H \left(2 a_2 r_H+a_1^2-1\right) &= 2M \, r_H \,\xi_1 + 2M\, a_1\, \xi_0 - Q^2 \xi_1 \ ,  \\
    a_1^2 \left(1-10 a_2 r_H\right)-6 a_3 a_1 &r_H^2+2 a_2 r_H \left(1-2 a_2 r_H\right)-a_1^4-a_2 a_1^3 \nonumber \\& \qquad= 2M \, r_H \, \xi_2 + a_1  2 M\, \xi_1 + 2 M\, a_2 \xi_0 - Q^2 \xi_2\ .
\end{align}
\end{subequations}
Because of \eqref{GammaHorizonCondition}, the first relation is solved only by $a_1=0$. With this, the remaining two relations simplify to
\begin{align}
\label{ConstraintOnXi1}
   & \left( 2 M\, \zh  - Q^2 \right) \xi_1 = 0 \ ,  &&\text{and} &&
    2 a_2 r_H \left(1-2 a_2 r_H\right) = \left(2M\, r_H  - Q^2\right) \xi_2 + 2 M\, a_2\, \xi_0 \ .
\end{align}
}}
\end{center}

\begin{center}
\remarkbox{\rem\label{rem:NoClassicalLimit}{Black holes with $Q^2=2M\zh$}}{\parbox{16cm}{
Throughout this Section we exclude deformed black holes with $Q^2= 2M\zh$. For such configurations the leading coefficients $\xi_0=\theta_0$ in \eqref{GammaHorizonCondition} are not defined, and they cannot be reached as small deformations of a classical solution. Indeed, for the RN black hole $\zh=M(1+\sqrt{1-\mathfrak{q}^2})$ with $\mathfrak{q}=|Q|/M\in[0,1]$, such that
\begin{align}
2M\zh-Q^2=M^2\left(2+2\sqrt{1-\mathfrak{q}^2}-\mathfrak{q}^2\right)>0\,,\nonumber
\end{align}
and the condition $Q^2=2M\zh$ is never realised in GR. The same argument applies to the EMD based on the proper time introduced in Sec.~\ref{Sect:EMDProperTime}.
}}
\end{center}

\noindent
Since $Q^2\neq 2M\zh$ (see Remark~\ref{rem:NoClassicalLimit}), the first relation in \eqref{ConstraintOnXi1} implies $\xi_1=0$. As in the neutral case, this is a genuine constraint on the EMD coefficients. The second relation in~\eqref{ConstraintOnXi1} is a quadratic equation for $a_2$, whose solutions are
\begin{align}
\label{a2Coefficient}
    a_2 =  \frac{ r_H - M \,\xi _0  +\sqrt{ \left(r_H-M\,\xi _0 \right)^2 + 4 r_H^2 \, \xi_2  \left(Q^2 - 2 M\,  r_H \right)} }{4 r_H^2} \ ,
\end{align}
Here we have chosen the positive sign in front of the square root, since only this branch reproduces the classical RN result in the limit $\Gamma^{(\rho)}\to 1$ (\emph{i.e.} $\xi_0\to 1$ and $\xi_{n\geq 1}\to 0$). Equation~\eqref{a2Coefficient} is the charged generalisation of~\eqref{a_2-exactsol}. All remaining coefficients of \eqref{zExpansionInRho} are then uniquely fixed by the recursion
\begin{multline}
    a_p = \frac{1}{1-4 p \, r_H a_2} \left[ 2 M \xi_p -  Q^2 \sum_{n=0}^{p} \xi_n \mathfrak{p}_{p-n} + r_H \sum_{n=3}^{p-1}(p-n+2) n a_n a_{p-n+2} +\right. \\
    \left. + \sum_{n=2}^{p-2} \sum^{n}_{m=2}(n-m+2)m a_{p-n} a_m a_{n-m+2} \right] \qq{for} p \geq 3 \ .\label{Arelations}
\end{multline}
which expresses $a_p$ in terms of $\xi_p$ and of coefficients $a_{q}$ with $q<p$.

\begin{center}
\remarkbox{\rem\label{rem:ChargedNeutralLimit}{Uncharged limit and normalisation of the coefficients}}{\parbox{16cm}{
In the charged EMD the deformation functions multiply $(2M/r-Q^2/r^2)$ and are therefore dimensionless, while their expansion coefficients carry dimensions. For $Q\to 0$ one has $\Gamma^{(\rho)}=\Phi^{(\rho)}/(2M)$ and $\Omega^{(\rho)}=\Psi^{(\rho)}/(2M)$, with $\Phi^{(\rho)}$ and $\Psi^{(\rho)}$ normalised as in \eqref{SeriesDistanceRho}. The two sets of conventions are thus related by
\begin{align}
&\xi_n\big|_{\text{charged}}\longleftrightarrow\frac{\xi_n}{(2M)^n}\,,&&\theta_n\big|_{\text{charged}}\longleftrightarrow\frac{\theta_n}{(2M)^n}\,,&&a_n\big|_{\text{charged}}\longleftrightarrow\frac{a_n}{(2M)^{n-1}}\,.\nonumber
\end{align}
With this identification, \eqref{a2Coefficient} reduces to \eqref{a_2-exactsol} with $v=+1$, the recursion \eqref{Arelations} reduces to \eqref{Coeffsa}, and all regularity conditions derived below reduce to their counterparts in Sec.~\ref{EMD propdist ricci temp}.
}}
\end{center}

\subsection{Regularity conditions and Hawking temperature}\label{Sect:Charged_Curvature}
The condition $\xi_1=0$ guarantees the existence of the self-consistent series solution, but it does not yet ensure that the resulting geometry is physically acceptable. As in the neutral case, we further require that
\begin{enumerate}
\item[\emph{(i)}] the Hawking temperature~\eqref{Hawkingtempdef} is well defined, real and positive,
\item[\emph{(ii)}] the Ricci scalar~\eqref{eq: ricci scalar h e f} remains finite at the horizon.
\end{enumerate}
Both requirements lead to conditions on the leading EMD coefficients.

To analyse them, we first write the metric functions as power series in $\rho$, assuming convergence in a neighbourhood of the horizon. Using the expansion of $1/r(\rho)$ introduced above, we find
\begin{subequations}\label{ExpansionsInRhof&h}
\begin{align}
    f(\rho) &= 1 - \left(\frac{2 M}{r(\rho)} - \frac{Q^2}{r(\rho)^2} \right) \Gamma^{(\rho)} (\rho) = 1 + \sum_{n=0}^{\infty} \left(\sum_{k=0}^{n}    \xi_k \left( \sum_{l=0}^{n-k} Q^2 \mathfrak{p}_l \mathfrak{p}_{n-k-l} - 2 M \, \mathfrak{p}_{n-k}\right)  \right) \rho^n \ , \\
    h(\rho) &= 1 - \left(\frac{2 M}{r(\rho)} - \frac{Q^2}{r(\rho)^2} \right) \Omega ^{(\rho)}(\rho) = 1 + \sum_{n=0}^{\infty} \left(\sum_{k=0}^{n}   \theta_k \left( \sum_{l=0}^{n-k} Q^2 \mathfrak{p}_l \mathfrak{p}_{n-k-l} - 2 M\,  \mathfrak{p}_{n-k}\right)  \right) \rho^n  \ .
\end{align}
\end{subequations}
Re-expanding with the help of~\eqref{zExpansionInRho} in powers of $r-\zh$, the leading terms are
\begin{subequations}\label{DeformationGeneralExpansionsZ}
\begin{align}
    f(r) &= \frac{ r_H \xi_2 \left( Q^2  -2 M \,  r_H \right) + a_2 \left(  r_H^2 - Q^2 \xi _0 \right) }{ a_2 r_H^3} \left(r - r_H\right)+\order{(r - r_H)^{3/2}}\,, \\
    h(r) &= \frac{\theta _1  \left(Q^2-2 M \,  r_H\right)}{\sqrt{a_2} r_H^2} \sqrt{r-r_H} + \left( \frac{\theta _2 r_H \left(Q^2-2 M\, r_H\right)- 2 a_2 \theta _0 \left(Q^2- M\,  r_H\right)}{a_2 r_H^3}+ \right.\nonumber \\
    & \quad \left. - \frac{a_3 \theta _1 \left(Q^2- 2 M \,  r_H\right)}{2 a_2^2 r_H^2} \right) (r-r_H)+\order{(r-r_H)^{3/2}}\,.
\end{align}
\end{subequations}
where the coefficients $a_p=a_p\left(\{\xi_{n}\}_{2\leq n \leq p}\right)$ are given by~\eqref{Arelations}. For compactness we do not substitute them explicitly.

The structure of~\eqref{DeformationGeneralExpansionsZ} immediately reveals a difference between the two metric functions. While $f$ starts linearly in $r-\zh$, the expansion of $h$ contains a term proportional to $\sqrt{r-\zh}$, whose derivative diverges at the horizon. The Hawking temperature~\eqref{Hawkingtempdef} depends on $h^{(1)}_H$, so condition \emph{(i)} requires this term to be absent. For $Q^2\neq 2M\zh$ this imposes $\theta_1=0$. Once this condition holds, $T_{\text{H}}$ is well defined and its square is given by
\begin{align}\label{Temp-EMD-RN}
    T_{\text{H}}^2  = \frac{\left(r_H^2-Q^2 \xi_0 \right) \left(r_H - M\,  \xi_0 +\sqrt{4 r_H^2\, \xi_2 \left(Q^2-2 M \,  r_H\right)+\left(r_H - M \,  \xi_0 \right)^2}\right)+4 r_H^3\, \theta_2 \left(Q^2 - 2 M \, r_H\right)}{16 \pi^2 \, r_H^5  } \ .
\end{align}
After using~\eqref{GammaHorizonCondition} to eliminate $\xi_0$, the temperature depends on the deformation only through the three parameters $\zh$, $\xi_2$ and $\theta_2$. Requiring $T_{\text{H}}\in\mathbb{R}_+$ then leads to upper bounds on $\xi_2$ and $\theta_2$, which are listed in Result~\ref{res:ChargedRegularity}.

Condition \emph{(ii)} is more restrictive. Even after imposing all the constraints discussed so far, the expansion of the Ricci scalar around the horizon still contains a term $R\propto\rho^{-1}$. The explicit form of its coefficient is not needed: it suffices to observe that it vanishes if the coefficients of order three satisfy the relation given in~\eqref{ChargedRicciCond}. This relation is the charged analogue of the last condition in \eqref{ConRicciFinite}, to which it reduces smoothly for $Q\to 0$ (with the identification of Remark~\ref{rem:ChargedNeutralLimit}). All conditions are collected below.

\begin{center}
\resultbox{\result\label{res:ChargedRegularity}{Regularity of the charged proper-distance EMD}}{\parbox{16cm}{
Consider a non-extremal, static and spherically symmetric charged black hole with $Q^2\neq 2M\zh$, described by the $\rho$-EMD \eqref{MetricDefGen} with deformation functions \eqref{SeriesExpDefFunc}. The near-horizon series solution exists, the Hawking temperature \eqref{Temp-EMD-RN} is real and positive, and the Ricci scalar is finite at the horizon, provided
\begin{align}
&\xi_0=\theta_0=\frac{\zh^2}{2M\zh-Q^2}\,,\hspace{2cm}\xi_1=\theta_1=0\,,\nonumber\\[4pt]
&\xi_2 \leq \frac{\left(r_H- M\, \xi_0\right)^2}{4 r_H^2 \left(2 M\, r_H - Q^2\right)} \ ,\nonumber\\[4pt]
&\theta_2 < \frac{\left(r_H - M\,  \xi_0 + \sqrt{4 r_H^2 \, \xi_2 \left(Q^2-2 M\,   r_H\right)+\left(r_H - M \, \xi_0 \right)^2}\right) \left(r_H^2 - Q^2 \xi_0 \right)}{4 r_H^3 \left(2 M \, r_H - Q^2\right)} \ ,\label{ConditionsOnSecondDerivatives}\\[4pt]
&\xi_{3} = \frac{1}{2} \theta _3 \left(\frac{3 \varpi  \left(Q^2-2 M \, r_H\right)}{ 2(M\,  r_H - Q^2)}-1\right)\ ,\hspace{0.5cm}\text{with} \hspace{0.5cm}
    \varpi = 2\sqrt{\frac{ 4 \xi _2 \left(Q^2-2 M \, r_H\right)^3+\left(Q^2- M\,   r_H\right)^2}{\left(Q^2 - 2 M\,  r_H\right)^2}} \ .\label{ChargedRicciCond}
\end{align}
The bound on $\xi_2$ ensures a real square root in \eqref{a2Coefficient} and \eqref{Temp-EMD-RN} (for $2M\zh>Q^2$, see Remark~\ref{rem:NoClassicalLimit}), while the bound on $\theta_2$ ensures $T_{\text{H}}^2>0$. For $Q\to 0$ these conditions reduce to \eqref{ConRicciFinite}.
}}
\end{center}

\subsection{Hawking temperature at small charge}\label{Sect:SmallCharge}
The general expression~\eqref{Temp-EMD-RN} becomes more transparent in the regime of small charge, which is also the regime in which semi-classical results are available for comparison. We denote with a subscript $_*$ quantities evaluated at $Q=0$ and assume that, to leading order, the charge shifts the horizon in the same way as in the classical RN geometry, namely
\begin{align}
\label{SmallQH}
    r_H = r_{\ast} - \frac{Q^2}{2 M} + \mathcal{O} (Q^3) \ .
\end{align}
Inserting \eqref{SmallQH} into \eqref{Temp-EMD-RN} and expanding in powers of $Q$ leads to the result summarised below.

\begin{center}
\resultbox{\result\label{res:SmallCharge}{Hawking temperature of a weakly charged deformed black hole}}{\parbox{16cm}{
Under the assumption \eqref{SmallQH}, the Hawking temperature of the charged $\rho$-EMD receives no correction of order $Q^2$:
\begin{align}
\label{SmallQT}
    T_{\text{H}} = T_{\ast} -  \frac{ \left(8 M\, r_{\ast} \left(\xi_{2\ast} \left(32M\,  \theta_{2\ast} \, r_{\ast} - 10 \varpi_* - 6\right) - \theta_{2\ast}\right)+ 3 \varpi_* \right)}{512 \pi^2  M ^2 \, r_{\ast}^4  \varpi_* T_{\ast}  }Q^4 + \mathcal{O}(Q^5) \ .
\end{align}
Here $T_*$ is the temperature of the uncharged deformed black hole, which follows from the $Q\to 0$ limit of \eqref{Temp-EMD-RN},
\begin{align}
    T_{\ast} = \frac{1}{4\pi\, r_{\ast}}\sqrt{\frac{1+\varpi_*}{2}- 8 M\,  \theta_{2\ast} r_{\ast}} \ , \qquad \varpi_*= \sqrt{1- 32 M\, \xi_{2*} \, r_* } \ .\label{SmallQTstar}
\end{align}
With the identification of Remark~\ref{rem:ChargedNeutralLimit}, $T_*$ coincides with the temperature of Sec.~\ref{EMD propdist ricci temp}. The leading effect of the charge thus appears at order $Q^4$, but its coefficient depends non-trivially on the deformation parameters.
}}
\end{center}

\noindent
The absence of a correction of order $Q^2$ is in agreement with semi-classical analyses~\cite{Balbinot:2007kr}, while the dependence of the $Q^4$ coefficient on $\xi_{2*}$ and $\theta_{2*}$ is a genuine effect of the deformation. The result, however, depends on the assumption \eqref{SmallQH} about the position of the horizon, as the following comparison shows.

\begin{center}
\remarkbox{\rem\label{rem:EFTcharged}{Comparison with effective field theory}}{\parbox{16cm}{
Corrections to the Hawking temperature of charged black holes have also been computed with effective field theory techniques~\cite{Campos_Delgado_2022}. There, higher-curvature terms in the action shift the position of the horizon in a way that is not of the form \eqref{SmallQH}. As a consequence, the temperature obtained in~\cite{Campos_Delgado_2022} is already modified at order $\mathcal{O}(Q^2)$, in contrast to \eqref{SmallQT}. Within the EMD, such a scenario corresponds to a different dependence of $\zh$ on $Q$, and the small-charge behaviour of $T_{\text{H}}$ is therefore a probe of how the underlying theory modifies the horizon.
}}
\end{center}

\subsection{Electromagnetic field on the deformed background}\label{Sect:MaxwellDeformed}
A charged black hole is characterised not only by its geometry, but also by the electromagnetic field it sources. Having constructed a regular deformed metric, we now determine the field strength that is compatible with it. We assume that the new physics responsible for the deformation preserves spherical symmetry and staticity, so that the field strength can only differ from its classical counterpart by functions of the radial coordinate
\begin{equation}
\label{FST}
    F_{\mu \nu} = - F_{\nu \mu} = F^\text{cl}_{\mu \nu} + F^\text{def}_{\mu \nu} \ .
\end{equation}
Here $F^{\rm cl}_{\mu\nu}$ describes the classical radial electric and magnetic fields given in Eq.~\eqref{classical RN em fields}, with $Q^2 = q^2 + m^2$ as in Sec.~\ref{Sect:RNBH}. The deformation is parametrised by two unknown functions $w(r)$ and $k(r)$,
\begin{align}
  &E^\text{def}_r = F^\text{def}_{rt}  \coloneqq w(r) \ , &&\text{and} && B^\text{def}_r = \frac{F^\text{def}_{\theta \phi}}{r^2 \sin \theta} \coloneqq k(r) \ ,
\end{align}
which are fixed by the Maxwell equations and the Bianchi identity on the deformed background, as shown in Computation~\ref{comp:MaxwellDeformed}.

\begin{center}
\compbox{\comp\label{comp:MaxwellDeformed}{Maxwell equations on the deformed background}}{\parbox{16cm}{
The source-free Maxwell equations~\eqref{MFE} read
\begin{align}
&    \partial_{\mu} \left( \sqrt{-g} F^{\mu \nu}\right) = 0  \ ,&&\text{with} &&g \coloneqq \det g_{\mu \nu} = -\frac{h(r)}{f(r)} r^4 \sin^2 \theta \ .\nonumber
\end{align}
For the ansatz \eqref{FST}, the only non-trivial components are
\begin{align}
\label{SimplifiedME}
    \left(\partial_r \sqrt{-g}\right) F^{rt} +  \sqrt{-g} \left( \partial_r F^{rt}\right) &=0\ , \\
    \label{magneticME}\left(\partial_\theta \sqrt{-g}\right) F^{\theta \phi} +  \sqrt{-g} \left( \partial_\theta F^{\theta \phi}\right) &= 0  \ .
\end{align}
Eq.~\eqref{SimplifiedME} is solved by $w(r) = -\frac{q}{r^2} + \frac{c_1}{r^2} \sqrt{\frac{h(r)}{f(r)}}$, with an integration constant $c_1$. Demanding $w=0$ when the deformation is switched off ($f=h$) fixes $c_1 = q$. Eq.~\eqref{magneticME} is identically satisfied, for any $k(r)$. The Bianchi identity
\begin{equation}
    \partial_\mu F_{\nu \rho} + \partial_\nu F_{\rho \mu} + \partial_\rho F_{\mu \nu} = 0 \nonumber
\end{equation}
has only one non-trivial component, $\partial_r F_{\theta \phi} = 0$, which implies $\partial_r [r^2 \,k(r)] = 0$ and hence $k(r)= c_2/r^2$. Since $k$ does not depend on the deformation at all, recovering the classical magnetic field requires $c_2 = 0$.
}}
\end{center}

\begin{center}
\resultbox{\result\label{res:MaxwellDeformed}{Electromagnetic field of the deformed charged black hole}}{\parbox{16cm}{
The non-vanishing components of the field strength compatible with the deformed metric \eqref{MetricDefGen} are
\begin{align}
    &-F_{tr} = F_{rt} = \frac{q}{r^2} \sqrt{\frac{h(r)}{f(r)}} =  \frac{q}{r^2} \frac{ \sqrt{Q^2 \Omega^{(\rho)} \left(\rho \right)- 2 M \, r \,\Omega^{(\rho)} \left(\rho \right)+r^2}}{\sqrt{Q^2 \Gamma ^{(\rho)}\left(\rho\right) - 2 M\,  r \,  \Gamma ^{(\rho)}\left(\rho\right) + r^2}} \ , &&-F_{\phi \theta}=F_{\theta \phi} = m \sin \theta \ .\label{DeformedFieldStrength}
\end{align}
The radial electric field is rescaled by the factor $\sqrt{h/f}$, while the magnetic field coincides with the classical one.
}}
\end{center}

\noindent
In particular, if the two metric functions are deformed in the same way ($\Gamma^{(\rho)}=\Omega^{(\rho)}$, \emph{i.e.} $f=h$), the classical Coulomb field is recovered even though the geometry is deformed. A modification of the electric field is therefore a direct measure of the difference between the two deformation functions. The magnetic field, instead, is fixed by the Bianchi identity alone and is insensitive to the deformation of the (static) geometry.

\subsection{Extremal black holes and the proper-time EMD}\label{Sect:Extremal}
We finally turn to extremal charged black holes. In the present context, extremality means that $r=\zh$ is a zero of higher order of the metric functions $f$ and $h$ (regarded as functions of $r-\zh$), \emph{i.e.} that the coefficients $f^{(1)}_H$ and $h^{(1)}_H$ in~\eqref{DeformedNearH} vanish. Physically, this is equivalent to a vanishing surface gravity~\eqref{eq:kappa-general}, and therefore to a vanishing Hawking temperature.

For the construction of the previous Subsections, extremality is fatal. If $f$ has a double zero at $r=\zh$, the integrand of \eqref{DefinitionRadialProperDistance} behaves as $(r-\zh)^{-1}$ close to the horizon. The singularity is no longer integrable and the proper distance from any point outside the black hole to the horizon diverges logarithmically. As a consequence, the expansion~\eqref{dExpansion}, and everything built upon it, breaks down: the spatial distance is not a suitable physical quantity to define an EMD in the immediate vicinity of an extremal horizon. We therefore need a different quantity, which remains finite at the horizon also in the extremal case. As we show below, the proper time measured by an observer that falls freely towards the black hole has this property. The corresponding EMD was developed in detail for uncharged black holes in~\cite{Hohenegger:2025dic}, and it is also the natural language for the reconstruction of the metric from observations discussed in Chapter~\ref{chapGeo}.

\subsubsection{Proper time of a radially infalling observer}\label{Sect:EMDProperTime}
\begin{figure}[ht]
\begin{center}
\includegraphics[width=7.5cm]{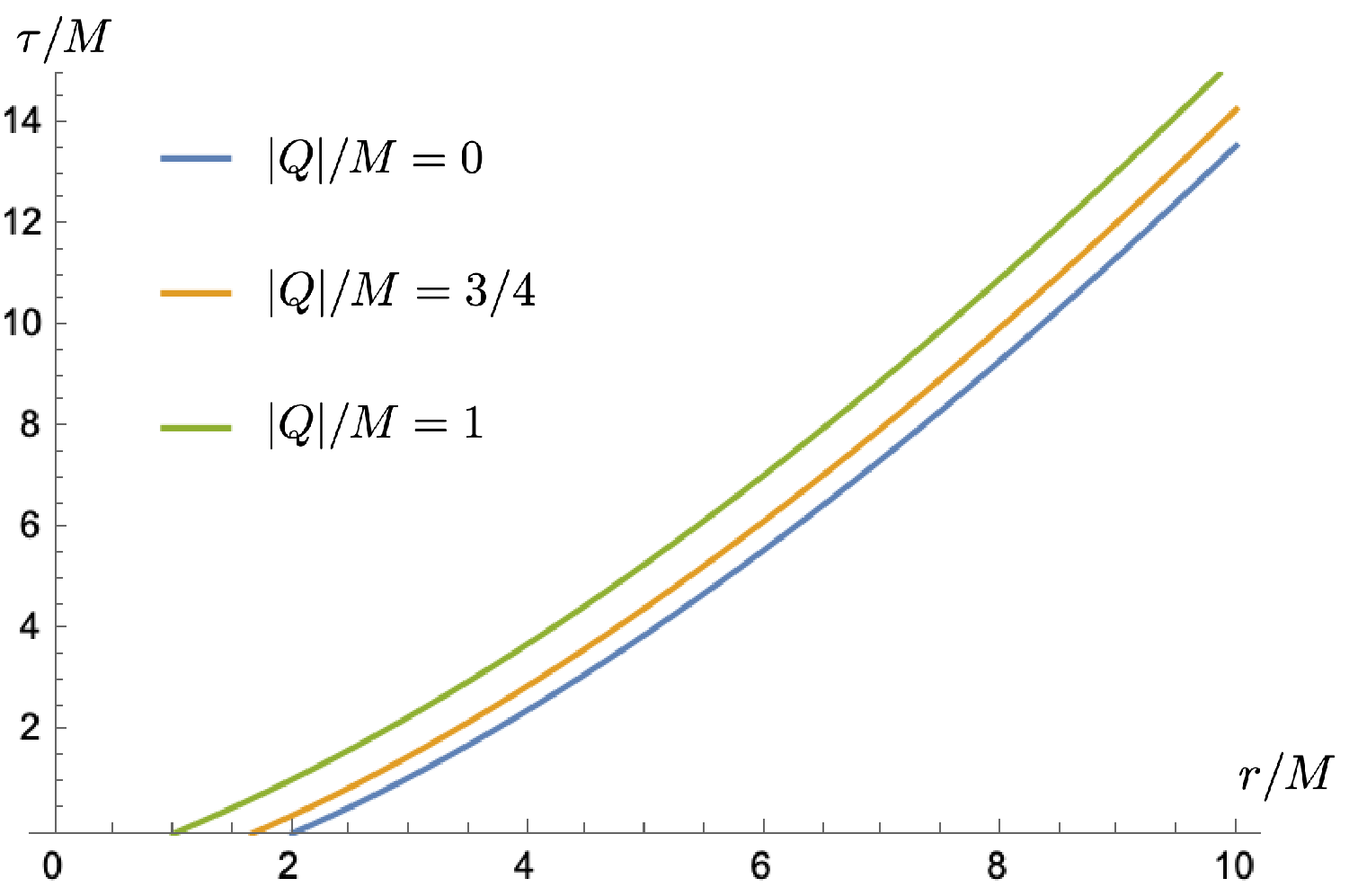}
\end{center}
\caption{Proper time (\ref{AnalyticRN}) measured by a time-like observer falling along a radial geodesic from a space-time point labelled by $r$ to the horizon of a Reissner-Nordstr\"om black hole, for different values of $\mathfrak{q}=|Q|/M$. The extremal case $\mathfrak{q}=1$ is still described by a finite function.}
\label{Fig:RNtime}
\end{figure}

We consider a generic static and spherically symmetric metric
\begin{align}
    \dd s^2 = - h(r) \dd t^2 + \frac{\dd r^2}{f(r)} + r^2 \dd \theta^2 + r^2 \sin^2 \theta \dd \phi^2 \ ,\label{MetricForm}
\end{align}
For purely radial motion in the equatorial plane $\theta = \pi/2$, the line element along a geodesic can be written as
\begin{align}\label{defddtau}
    -\dd \tau^2 = \frac{ u \, h(r)}{f(r)} \frac{\dd r^2 }{(1 + u \, h(r))} \ ,
\end{align}
where $u = -1$ corresponds to the time-like geodesic of an observer with unit energy per unit mass (\emph{i.e.} falling from rest at spatial infinity), while $u=+1$ describes a space-like geodesic. Integrating \eqref{defddtau} for $u=-1$, the proper time that elapses for the observer while falling from a point labelled by $r$ to the event horizon at $\zh$ is
\begin{align}
\label{FreeFallingTime}
     \tau(r) = \int_{r_H}^{r} \sqrt{\frac{h(r')}{f(r')}} \frac{\dd r' }{\sqrt{1-h(r')}} \ .
\end{align}
where the overall sign has been chosen such as to describe an infalling motion. For the classical RN geometry, $f(r) = h(r) = 1 - \frac{2 M}{r} + \frac{ Q^2}{r^2}$, the integral can be performed in closed form
\begin{align}
    &\tau(r)=\frac{M}{3} \left(\mathfrak{q}^2 \left(\sqrt{2 \mathfrak{r}-\mathfrak{q}^2}-\sqrt{2 \mathfrak{r}_H-\mathfrak{q}^2}\right)+\mathfrak{r} \sqrt{2\mathfrak{r}-\mathfrak{q}^2}-\mathfrak{r}_H \sqrt{2 \mathfrak{r}_H-\mathfrak{q}^2}\right)\,,\label{AnalyticRN}
\end{align}
where we have introduced the dimensionless quantities
\begin{align}
    &\mathfrak{q}=\frac{|Q|}{M}\,,&&\mathfrak{r}=\frac{r}{M}\,,&&\mathfrak{r}_H=\frac{\zh}{M}=1+\sqrt{1-\mathfrak{q}^2}\ .
\end{align}
The Schwarzschild black hole corresponds to $\mathfrak{q}=0$, while the extremal configuration is reached for $\mathfrak{q}=1$.\footnote{For $\mathfrak{q}>1$ the geometry does not describe a black hole, but a naked singularity, which is generally considered unphysical (cf. Sec.~\ref{Sect:RNBH}).} As shown in Figure~\ref{Fig:RNtime}, $\tau(r)$ is a finite, monotonic function for all $\mathfrak{q}\in[0,1]$, which vanishes at the horizon $\zh= M(1+\sqrt{1-\mathfrak{q}^2})$, including the extremal case.

\begin{center}
\remarkbox{\rem\label{rem:TauVsRho}{Why the proper time remains finite at an extremal horizon}}{\parbox{16cm}{
The different behaviour of $\rho$ and $\tau$ can be understood directly from their integrands. If $f$ and $h$ have zeros of the same order $k\geq 1$ at $r=\zh$, then close to the horizon
\begin{align}
&\frac{1}{\sqrt{f}}\sim (r-\zh)^{-k/2}\,,&&\text{while}&&\sqrt{\frac{h}{f}}\,\frac{1}{\sqrt{1-h}}\longrightarrow \sqrt{\frac{h^{(k)}_H}{f^{(k)}_H}}\,.\nonumber
\end{align}
The integrand of $\rho$ in \eqref{DefinitionRadialProperDistance} is integrable only for $k=1$, whereas the integrand of $\tau$ in \eqref{FreeFallingTime} is finite at the horizon for any $k$. Hence $\tau$ is well defined, and linear in $r-\zh$ close to the horizon, for both non-extremal and extremal black holes. Both quantities are monotonic functions of $r$ in the vicinity of the horizon, so that in the non-extremal case the EMDs based on $\rho$ and on $\tau$ are equivalent and can be expressed in terms of each other. This equivalence breaks down in the extremal case, where only the $\tau$-EMD exists.
}}
\end{center}

\noindent
Motivated by these observations, we parametrise the metric functions in~\eqref{MetricForm} in terms of $\tau$ as
\begin{align}
    f(r)= 1- \left( \frac{2 M}{r} -\frac{Q^2}{r^2}\right) {\Gamma}^{(\tau)} (\tau) \qq{and} 
    h(r)=1- \left( \frac{2 M}{r} -\frac{Q^2}{r^2}\right) {\Omega}^{(\tau)} (\tau)\ .\label{fhFunctions}
\end{align}
The functions ${\Gamma}^{(\tau)}$ and ${\Omega}^{(\tau)}$, along with $\zh$, define the EMD based on the proper time. To distinguish it from the distance-based EMD of Sec.~\ref{Sect:EMDDistance}, we denote the coefficients of their expansions in powers of the proper time with a tilde
\begin{align}
&{\Gamma}^{(\tau)} (\tau) = \sum_{n=0}^\infty \tilxi{n}\tau^n \ , && {\Omega}^{(\tau)} (\tau) = \sum_{n=0}^\infty \tiltheta{n}\tau^n \ ,&&\text{with} &&\tilxi{n}\,, \tiltheta{n}\in\mathbb{R}\,.\label{zExpansiontau}
\end{align}
In order to solve~\eqref{FreeFallingTime} iteratively, we furthermore introduce the expansions
\begin{align}
&r = r_H + \sum_{n=1}^{\infty} \widetilde{a}_n \tau^n \ ,&&\tau(r)=\sum_{n=1}^\infty \widetilde{b}_n\,(r-\zh)^n\,,&&\text{with} &&\widetilde{a}_n\,,\widetilde{b}_n\in\mathbb{R}\,,\label{TauExpansion}
\end{align}
The coefficients $\{\widetilde{a}_n\}_{n\in\mathbb{N}}$ are implicitly determined by \eqref{FreeFallingTime}, and the $\{\widetilde{b}_n\}_{n\in\mathbb{N}}$ follow from them by series inversion~\cite{Whittaker,DelPiano:2023fiw,DelPiano:2024gvw}. The first few read (see also~\cite{Hohenegger:2025dic})
\begin{align}
&\widetilde{b}_1=\frac{1}{\widetilde{a}_1}\,,&&\widetilde{b}_2=-\frac{\widetilde{a}_2}{\widetilde{a}_1^3}\,,&&\widetilde{b}_3=\frac{2\widetilde{a}_2^2-\widetilde{a}_1\widetilde{a}_3}{\widetilde{a}_1^5}\,,&&\widetilde{b}_4=\frac{5\widetilde{a}_1\widetilde{a}_2\widetilde{a}_3-5\widetilde{a}_2^3-\widetilde{a}_1^2\widetilde{a}_4}{\widetilde{a}_1^7}\,.\label{SeriesInversion}
\end{align}
For $r\to\zh$, which is equivalent to $\tau\to 0$, both metric functions must vanish. Inserting \eqref{zExpansiontau} and \eqref{TauExpansion} into \eqref{fhFunctions} gives
\begin{align}
&f(r)=\left(1-\frac{\tilxi{0}(2M\, \zh - Q^2)}{\zh^2}\right)+\mathcal{O}(r-\zh)\,,&&h(r)=\left(1-\frac{\tiltheta{0}(2M \, \zh-Q^2)}{\zh^2}\right)+\mathcal{O}(r-\zh)\,,\label{fhzExpansion}
\end{align}
so that the leading coefficients are fixed to
\begin{align}
\tilxi{0}=\frac{\zh^2}{2M \, \zh-Q^2}=\tiltheta{0}\,.\label{Condxi0}
\end{align}
where, as before, we exclude the case $2 M \, \zh = Q^2$ (see Remark~\ref{rem:NoClassicalLimit}). The remaining parameters, namely $\{\tilxi{n}\}_{n\in\mathbb{N}}$ and $\{\tiltheta{n}\}_{n\in\mathbb{N}}$ together with $\zh$, $Q$ and $M$, are regarded as the physical input parameters that define the black hole geometry outside the horizon. The coefficients $\{\widetilde{a}_n\}_{n\in\mathbb{N}}$, and through \eqref{SeriesInversion} also $\{\widetilde{b}_n\}_{n\in\mathbb{N}}$, are determined by expanding \eqref{FreeFallingTime} order by order in $(r-\zh)$. Taking into account \eqref{Condxi0}, the leading order $\mathcal{O}(r-\zh)$ yields
\begin{align}
\widetilde{a}_1=\sqrt{\frac{\tilxi{1}(Q^2-2M\, \zh)^2+2\widetilde{a}_1\,\zh(Q^2-M\, \zh)}{\tiltheta{1}(Q^2-2M\, \zh)^2+2\widetilde{a}_1\,\zh(Q^2-M\, \zh)}}\ ,\label{DiffEq1}
\end{align}
Here we have assumed that the right-hand side is well defined. As will become clear in the next Subsection, this requires the terms of order $\mathcal{O}(r-\zh)$ in \eqref{fhzExpansion} to be non-vanishing, \emph{i.e.} $r=\zh$ to be a \emph{simple} horizon. In this case, $\widetilde{a}_1$ is fixed in terms of $(\tilxi{1}\,,\tiltheta{1}\,,Q\,,\zh\,,M)$.\footnote{Eq.~(\ref{DiffEq1}) has several solutions, but generically only one of them is real.} At higher orders $\mathcal{O}((r-\zh)^p)$ with $p>1$, the expansion of \eqref{FreeFallingTime} only involves $\widetilde{a}_{p'}$ with $p'\leq p$, and it is linear in $\widetilde{a}_p$. These relations therefore fix the expansion \eqref{TauExpansion} uniquely, and with it the black hole geometry (see~\cite{Hohenegger:2025dic} for a more detailed discussion in the uncharged case).

\begin{center}
\resultbox{\result\label{res:ProperTimeEMD}{Proper-time EMD of a non-extremal charged black hole}}{\parbox{16cm}{
The metric \eqref{fhFunctions}, with deformation functions expanded as in \eqref{zExpansiontau}, describes a non-extremal charged black hole with a simple horizon at $r=\zh$ for any choice of the parameters $\left(\{\tilxi{n}\}_{n\geq 1}\,,\{\tiltheta{n}\}_{n\geq 1}\,,Q\,,\zh\,,M\right)$ with $2M\zh\neq Q^2$, subject only to $\tilxi{0}=\tiltheta{0}$ given in \eqref{Condxi0} and to the inequalities
\begin{align}
&(Q^2-2M\, \zh)\,\tilxi{1}>-2\widetilde{a}_1\,\zh\,\frac{Q^2-M\, \zh}{Q^2-2M\, \zh}\,,&&\text{and} &&(Q^2-2M\, \zh)\,\tiltheta{1}>-2\widetilde{a}_1\,\zh\,\frac{Q^2-M\, \zh}{Q^2-2M\,\zh}\ ,\label{ProperTimeInequalities}
\end{align}
with $\widetilde{a}_1$ determined by \eqref{DiffEq1}. These inequalities guarantee $\widetilde{a}_1\in\mathbb{R}$ as well as $f(r)>0$ and $h(r)>0$ for $r>\zh$. Unlike the $\rho$-EMD, the finiteness of curvature invariants such as the Ricci and Kretschmann scalars poses no further constraint, since the second derivatives of $f$ and $h$ are automatically finite at the horizon.
}}
\end{center}

\subsubsection{Parameter space of extremal configurations}
\paragraph{General charged extremal black holes}\label{Sect:GeneralCharge}
The proper-time EMD parametrises charged black holes in terms of $\left(\{\tilxi{n}\}_{n\in\mathbb{N}}\,,\{\tiltheta{n}\}_{n\in\mathbb{N}}\,,Q\,,\zh\,,M\right)$. We now identify the regions of this parameter space that correspond to \emph{extremal} black holes. For the latter, $r=\zh$ is a double zero of $f$ and $h$ in \eqref{fhFunctions}. In addition to \eqref{Condxi0}, we therefore have to impose that the terms of order $\mathcal{O}(r-\zh)$ in \eqref{fhzExpansion} vanish:
\begin{align}
&\frac{\tilxi{1}(Q^2-2M\, \zh)^2+2\widetilde{a}_1\,\zh(Q^2-M\,\zh)}{\widetilde{a}_1\,\zh^2 (Q^2-2M\, \zh)}=0\,,&&\text{and} &&\frac{\tiltheta{1}(Q^2-2M\, \zh)^2+2\widetilde{a}_1\,\zh(Q^2-M\, \zh)}{\widetilde{a}_1\,\zh^2 (Q^2-2M\, \zh)}=0\,.\label{ExtremConstraint}
\end{align}
With these conditions, the right-hand side of \eqref{DiffEq1} is no longer defined, and the leading order $\mathcal{O}(r-\zh)$ of \eqref{FreeFallingTime} takes instead the form
\begin{align}
\widetilde{a}_1=\sqrt{\frac{\widetilde{a}_1^2\, Q^4+(Q^2-2M\, \zh)\left(\tilxi{2}(Q^2-2M\, \zh)^2+2\widetilde{a}_2\zh(Q^2-M\, \zh)\right)}{\widetilde{a}_1^2\, Q^4+(Q^2-2M\,\zh)\left(\tiltheta{2}(Q^2-2M\,\zh)^2+2\widetilde{a}_2\zh(Q^2-M\,\zh)\right)}}\ .\label{ExtremalDiffEq}
\end{align}
This relation differs qualitatively from \eqref{DiffEq1}, since it also involves $\widetilde{a}_2$ and the second-order deformation parameters. As a consequence, the structure of the solutions of \eqref{ExtremConstraint} and \eqref{ExtremalDiffEq} depends sensitively on the remaining parameters. Still assuming $Q^2\neq 2M\, \zh $, three cases have to be distinguished (see Appendix~\ref{Appx:A} for more details):
\begin{enumerate}
\item[\emph{(i.a)}] $Q^2\neq M\,\zh$ and $\tilxi{2}=\tiltheta{2}$. The only solution of \eqref{ExtremalDiffEq} is $\widetilde{a}_1=1$, and \eqref{ExtremConstraint} then fixes the first-order deformation parameters. The terms $\mathcal{O}((r-\zh)^p)$ with $p>1$ in the expansion of \eqref{FreeFallingTime} are linear in $\widetilde{a}_p$ (and only contain $\widetilde{a}_{p'}$ with $p'\leq p$), so that the expansion \eqref{TauExpansion} is uniquely determined.
\item[\emph{(i.b)}] $Q^2\neq M\,\zh$ and $\tilxi{2}\neq \tiltheta{2}$. In this case $\widetilde{a}_2$ does not drop out of \eqref{ExtremalDiffEq}. Moreover, the terms $\mathcal{O}((r-\zh)^p)$ with $p>1$ in the expansion of \eqref{FreeFallingTime} also acquire a linear dependence on $\widetilde{a}_{p+1}$, while remaining independent of all $\widetilde{a}_{p'}$ with $p'> p+1$.\footnote{This linear dependence on $\widetilde{a}_{p+1}$ has been checked explicitly for all orders $p\leq 5$, for $\tilxi{2}\neq \tiltheta{2}$, $Q^2\neq M\, \zh$ and $Q^2\neq 2M\, \zh$.} Condition \eqref{ExtremConstraint} is then used to fix $\tiltheta{1}$ and $\widetilde{a}_1$ in terms of $\tilxi{1}$, which remains a free parameter. Eq.~\eqref{ExtremalDiffEq} determines $\widetilde{a}_2$, and the higher orders fix each $\widetilde{a}_{p+1}$ uniquely, without further constraints on the deformation parameters.
\item[\emph{(ii)}] $Q^2= M\, \zh$. Condition \eqref{ExtremConstraint} is solved by $\tilxi{1}=0=\tiltheta{1}$, while \eqref{ExtremalDiffEq} fixes $\widetilde{a}_1$, irrespective of whether $\tilxi{2}=\tiltheta{2}$ or not. Of the two resulting solutions, the one with the positive sign in \eqref{Sola1GnDef} reduces to the classical value $\widetilde{a}_1=1$ for $\tilxi{2},\tiltheta{2}\to 0$, while the other one tends to $\widetilde{a}_1=0$. Also in this case the higher orders in the expansion of \eqref{FreeFallingTime} are linear in $\widetilde{a}_{p}$ and impose no further constraints on the $\tilxi{n}$ and $\tiltheta{n}$.
\end{enumerate}
The resulting parametrisations are summarised in the following Result.

\begin{center}
\resultbox{\result\label{res:ExtremalEMD}{EMD of extremal charged black holes}}{\parbox{16cm}{
Extremal charged black holes (with $Q^2\neq 2M\zh$) are described by the proper-time EMD \eqref{fhFunctions} in three distinct regions of parameter space:
\begin{itemize}
\item[\emph{(i.a)}] $Q^2\neq M\zh$, $\tilxi{2}=\tiltheta{2}$: \quad $\widetilde{a}_1=1$ and $\displaystyle\tilxi{1}=\tiltheta{1}=-\frac{2\zh(Q^2-M\,\zh)}{(Q^2-2M\,\zh)^2}$. \\ Free parameters: $\left(\{\tilxi{n}\}_{n\geq 2}\,,\{\tiltheta{n}\}_{n\geq 2}\,,Q\,,\zh\,,M\right)$.
\item[\emph{(i.b)}] $Q^2\neq M\zh$, $\tilxi{2}\neq\tiltheta{2}$: \quad $\tiltheta{1}=\tilxi{1}$ and $\displaystyle\widetilde{a}_1=\tilxi{1}\left(2M-\frac{Q^4}{2\zh(Q^2-M\,\zh)}\right)$. \\ Free parameters: $\left(\{\tilxi{n}\}_{n\geq 1}\,,\{\tiltheta{n}\}_{n\geq 2}\,,Q\,,\zh\,,M\right)$.
\item[\emph{(ii)}] $Q^2=M\zh$: \quad $\tilxi{1}=\tiltheta{1}=0$ and
\begin{align}
\widetilde{a}_1=\frac{1}{\sqrt{2}}\,\sqrt{1+M\,\zh\,\tiltheta{2}\pm\sqrt{(1+ M\, \zh\, \tiltheta{2})^2-4M\, \zh\tilxi{2}}}\ ,\label{Sola1GnDef}
\end{align}
which is real (and non-negative) provided $\tilxi{2}\leq \frac{(1+M\, \zh\tiltheta{2})^2}{4M\,\zh}$, and $\tiltheta{2}\geq -\frac{1}{M\,\zh}$ if $\tilxi{2}>0$. \\ Free parameters: $\left(\{\tilxi{n}\}_{n\geq 2}\,,\{\tiltheta{n}\}_{n\geq 2}\,,\zh\,, M\right)$.
\end{itemize}
In all cases, the remaining free parameters are only restricted by the positivity of the metric functions,
\begin{align}
&f(r)>0\,,&&\text{and} &&h(r)>0\,,&&\forall r > \zh\,,\label{PositiveSignature}
\end{align}
while curvature invariants such as the Ricci scalar impose no further constraints, since the second derivatives of $f$ and $h$ remain finite at the horizon.
}}
\end{center}

\paragraph{Small deformations of the extremal Reissner-Nordstr\"om geometry}\label{DeformedComputations}
In full generality, the inequalities \eqref{PositiveSignature} are cumbersome (see Appendix~\ref{Appx:A}). They become tractable for black holes that are small deformations of the extremal RN geometry. To parametrise this situation, we introduce
\begin{align}
&\epsilon:=\frac{M-\zh}{\zh}\,,&&\tilxi{n}:=\frac{x_n}{M^n}\,,&&\tiltheta{n}:=\frac{t_n}{M^n}\ ,\label{DefSmall}
\end{align}
where $\epsilon>-1$ plays the role of a small deformation parameter, defined in analogy with the Rezzolla-Zhidenko parametrisation~\cite{Rezzolla:2014mua} that will be discussed in Chapter~\ref{chap5}. In particular, $\zh=\frac{M}{1+\epsilon}$. The two regions \emph{(i)} and \emph{(ii)} of Result~\ref{res:ExtremalEMD} behave rather differently in this respect: in region \emph{(i)} a small deformation requires $\epsilon$ itself to be small, and we work to leading order in $\epsilon$, whereas region \emph{(ii)} can be analysed compactly to all orders for $\epsilon>-1$.
\begin{enumerate}
\item[\emph{(i)}] For a black hole with $Q^2\neq M\, \zh$ to be close to the extremal RN geometry, all deformation parameters must vanish together with $\epsilon$. To leading order we therefore assume
\begin{align}
&x_n=\epsilon\,x_n^{(1)}+\mathcal{O}(\epsilon^2)\,,&&\text{and} &&t_n=\epsilon\,t_n^{(1)}+\mathcal{O}(\epsilon^2)\,,\hspace{0.2cm}\forall n\geq 1\,,&&|Q|= M(1+ q\,\epsilon)+\mathcal{O}(\epsilon^2)\,,\label{DeformEps}
\end{align}
with
\begin{align}
    x_n^{(p)}\coloneqq \left.\dv[p]{x_n}{\epsilon}\right|_{\epsilon=0} \ , \quad t_n^{(p)}\coloneqq \left.\dv[p]{t_n}{\epsilon}\right|_{\epsilon=0} \qq{and} q \coloneqq \left.\dv{(|Q|/M)}{\epsilon}\right|_{\epsilon=0}\ .
\end{align}
Correspondingly, the coefficients in (\ref{TauExpansion}) take the form
\begin{align}
&\widetilde{a}_n=\widetilde{a}_n^{(0)}+\epsilon\,\widetilde{a}_n^{(1)}+\mathcal{O}(\epsilon^2)\,,&&\text{with} &&\left\{\begin{array}{l}\widetilde{a}_1^{(0)} = 1\,, \\ \widetilde{a}_2^{(0)}=0\,, \\ \widetilde{a}_3^{(0)}=-\frac{1}{6M^2}\,.
\end{array}\right.
\end{align}
To leading order in $\epsilon$, eqs.~(\ref{Condxi0}) and (\ref{ExtremConstraint}) impose
\begin{align}
&\tilxi{0}=\tiltheta{0}=1+2q\,\epsilon+\mathcal{O}(\epsilon^2)\,,&&\text{and} &&t_1^{(1)}=-2(1+2q)=x_1^{(1)}\,,
\end{align}
respectively. For arbitrary values of $x_2^{(1)}$ and $t_2^{(1)}$, eq.~(\ref{ExtremalDiffEq}) and the higher orders in the expansion of \eqref{FreeFallingTime} then fix the leading deformations of the coefficients $\widetilde{a}_n$:
\begin{align}
&\widetilde{a}_1^{(1)}=\frac{t_2^{(1)}-x_2^{(1)}}{2}\,,\hspace{2cm}\widetilde{a}_2^{(1)}=\frac{2(t_2^{(1)}-x_2^{(1)})+t_3^{(1)}-x_3^{(1)}}{4M}\,,\nonumber\\
&\widetilde{a}_3^{(1)}=-\frac{36+48q+t_2^{(1)}-7x_2^{(1)}-12(t_3^{(1)}-x_3^{(1)})-6(t_4^{(1)}-x_4^{(1)})}{36M^2}\,.
\end{align}
At this order, the positivity conditions (\ref{PositiveSignature}) reduce to
\begin{align}
&1+(6+8q-x_2^{(1)})\epsilon>0\,,&&\text{and} &&1+(6+8q-t_2^{(1)})\epsilon>0\,.\label{CondLeading}
\end{align}

\item[\emph{(ii)}] For the RN geometry, extremality corresponds to $Q^2=M^2$, which implies $\zh=M$ ($\epsilon = 0$) and hence $Q^2= M\, \zh $. Region \emph{(ii)} can thus be viewed as a deformation of the classical extremal configuration for any value of $\epsilon>-1$. Since classically $\widetilde{a}_1=1$, we select the positive sign in~(\ref{Sola1GnDef}), which in terms of the variables \eqref{DefSmall} reads
\begin{align}
\widetilde{a}_1=\frac{\sqrt{1+\epsilon+t_2+\sqrt{(1+\epsilon+t_2)^2-4(1+\epsilon)x_2}}}{\sqrt{2(1+\epsilon)}}\ .
\end{align}
The conditions (\ref{PositiveSignature}) can then be formulated as
\begin{align}
&\left\{\begin{array}{lcl}x_2<0 & \text{if} & t_2\leq -(1+\epsilon)\,, \\ x_2\leq \frac{(1+\epsilon+t_2)^2}{4(1+\epsilon)} & \text{if} & -(1+\epsilon)<t_2<1+\epsilon\,, \\ x_2<t_2 & \text{if} & t_2\geq 1+\epsilon\,.\end{array}\right.\label{ExtremalRegionii}
\end{align}
The corresponding allowed region is shown in green in Figure~\ref{Fig:TypeiiExtremal}: all points in its interior satisfy~\eqref{PositiveSignature}, while on its boundary only the solid black line is allowed.
\begin{figure}[ht]
\begin{center}
\includegraphics[width=8cm]{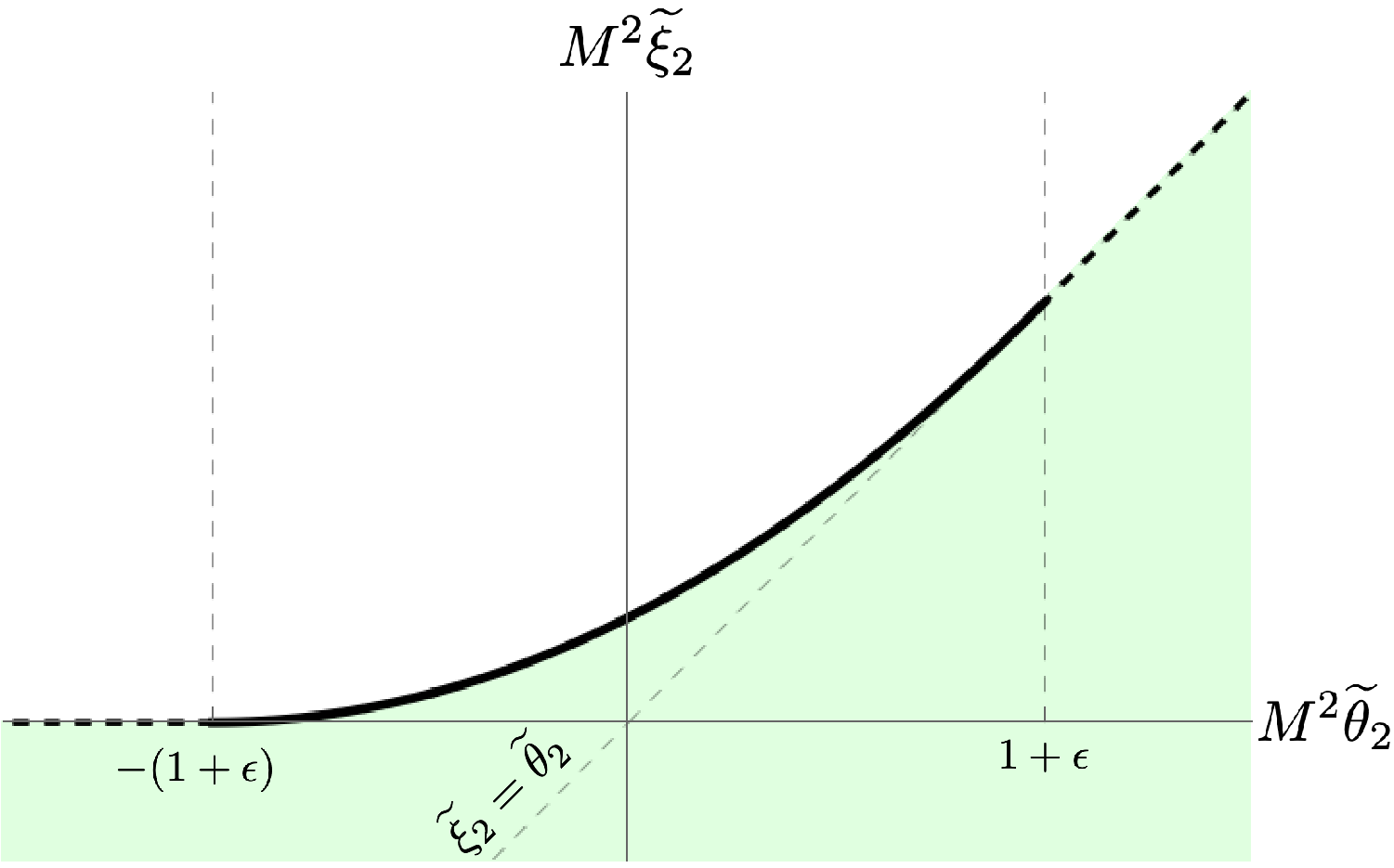}
\end{center}
\caption{Allowed region of the parameters $\tilxi{2}$ and $\tiltheta{2}$ for extremal black holes with $Q^2=M\, \zh$. Points in the interior of the green region lead to extremal geometries that satisfy all relevant conditions. On its boundary, only the solid black line is allowed, while the dashed black line is excluded.}
\label{Fig:TypeiiExtremal}
\end{figure}
\end{enumerate}

\section{Extension to \texorpdfstring{$2+1$}{2+1}-dimensions}\label{Sect:3Dim}

In Sec.~\ref{ClassicalBTZ} we introduced the classical BTZ geometry~\cite{Banados:1992wn,Banados:1992gq}. The EMD construction is extended to $2+1$-dimensional, stationary, asymptotically AdS black holes, taking as classical reference the rotating BTZ solution~\cite{DamiaPaciarini:2024ssf},
\begin{equation}
\dd s^{2} = -g(r)\,\dd t^{2} + \frac{\dd r^{2}}{f(r)} + r^{2}\big(h(r)\,\dd t + \dd\phi\big)^{2} \ ,
\end{equation}
with
\begin{equation}
g(r) = f(r) = -M - r^{2}\Lambda + \frac{J^{2}}{4r^{2}}, \qquad h(r) = -\frac{J}{2r^{2}}, \qquad \sqrt{-\Lambda}\,|J| \le M \ .
\end{equation}
Unlike the static, spherically symmetric $3+1$D EMD, the $2+1$D construction incorporates rotation \emph{ab initio}, through the additional metric function $h(r)$.
Deformations are introduced by dressing the mass and angular-momentum terms with three functions $\Phi,\Psi,\Omega$ of the physical (coordinate-invariant) proper distance $d(r)$:
\begin{subequations}
\begin{align}
f(r) &= -M - \left(r^{2}\Lambda - \frac{J^{2}}{4r^{2}}\right)\Phi(d(r)) \ , \\
g(r) &= -M - \left(r^{2}\Lambda - \frac{J^{2}}{4r^{2}}\right)\Psi(d(r)) \ , \\
h(r) &= -\frac{J}{2r^{2}}\Omega(d(r)) \ , \\
d(r) &= \int_{0}^{r} \frac{dz}{\sqrt{|f(z)|}}\ .
\end{align}
\end{subequations}
Asymptotic AdS behavior requires $\Phi,\Psi,\Omega \to 1$ as $d\to\infty$. As in the $3+1$D case, the physical content of the deformation is encoded in the local Taylor/Laurent coefficients of $\Phi,\Psi,\Omega$ around three regions: the outer horizon, spatial infinity, and the origin.
 
\subsection{Near-Horizon Expansion}
 
Expanding in the proper distance from the horizon, $\rho = d(r)-d_H$,
\begin{equation}
\Phi(\rho) = \Phi_H + \sum_{n=1}^{\infty}\frac{\Phi_H^{(n)}}{n!}\rho^{n} \ , \qquad (\text{and similarly for } \Psi,\Omega) \ ,
\end{equation}
a smooth horizon requires
\begin{equation}
\Phi_H = \Psi_H = \frac{4Mr_H^{2}}{J^{2}-4r_H^{4}\Lambda} \ ,
\end{equation}
while finiteness of the Ricci and Kretschmann scalars at $r_H$ forces the first- and third-order coefficients to vanish,
\begin{equation}
\Phi_H^{(1)} = \Psi_H^{(1)} = \Omega_H^{(1)} = 0 \ , \qquad
\Phi_H^{(3)} = \Psi_H^{(3)} = \Omega_H^{(3)} = 0 \ ,
\end{equation}
and imposes inequality constraints on the second derivatives,
\begin{equation}
\frac{\Phi_H^{(2)}}{\Phi_H} \ge -\frac{M}{2r_H^{2}}\vartheta_H^{2} \ , \qquad \frac{\Psi_H^{(2)}}{\Psi_H} > -\frac{M}{r_H^{2}}\vartheta_H(\varkappa+\vartheta_H) \ , \qquad \vartheta_H > -\varkappa \ ,
\end{equation}
where $\vartheta(r) \equiv (4r^{4}\Lambda+J^{2})/(4r^{4}\Lambda-J^{2})$~.
 
\subsection{Large-Distance and Origin Expansions}
 
Far from the horizon, $\Phi,\Psi,\Omega$ are expanded in inverse powers of $d$,
\begin{equation}
\Phi(d) = 1+\sum_{n=1}^{\infty}\frac{\omega_n}{d^{n}}, \qquad
\Psi(d) = 1+\sum_{n=1}^{\infty}\frac{\gamma_n}{d^{n}}, \qquad
\Omega(d) = 1+\sum_{n=1}^{\infty}\frac{\sigma_n}{d^{n}},
\end{equation}
and convergence up to the horizon translates the near-horizon regularity conditions into sum rules on the rescaled coefficients $\bar\omega_n,\bar\gamma_n,\bar\sigma_n$, e.g.
\begin{equation}
\sum_{n=1}^{\infty} n\bar\omega_n = \sum_{n=1}^{\infty} n\bar\gamma_n = \sum_{n=1}^{\infty} n\bar\sigma_n = 0, \qquad
\sum_{n=1}^{\infty} n^{2}(n+3)\bar\omega_n = \cdots = 0 \ .
\end{equation}
At the origin, assuming analyticity of $\Phi,\Psi,\Omega$ \ , matching Laurent expansions in $r$ and $d$ restricts the metric functions to
\begin{equation}
f(r) = f_0^{(-2)}\frac{1}{r^{2}} + f_0 + f_0^{(2)}r^{2} + \mathcal{O}(r^{3}) \ ,
\end{equation}
i.e.\ odd powers and a simple $1/r$ pole are excluded. Regularity of the curvature invariants at $r=0$ then reduces to the single algebraic condition
\begin{equation}\label{EMD-BTZ-origin}
\Omega_0^{2} = \Psi_0 \ ,
\end{equation}
a mixed condition intrinsically tied to the presence of rotation.
 
\subsection{Thermodynamics}
 
The Hawking temperature, $T = \kappa/2\pi = \tfrac{1}{4\pi}\sqrt{f_H^{(1)}g_H^{(1)}}$, is expressed entirely in terms of the physical expansion coefficients,
\begin{equation}\label{EMD-BTZ-temperature}
T = \frac{1}{2\pi}\sqrt{\frac{M^{2}}{2r_H^{2}}\vartheta_H(\vartheta_H+\varkappa) + \frac{M}{2}\frac{\Psi_H^{(2)}}{\Psi_H}} \ ,
\end{equation}
with an equivalent form in terms of the large-distance coefficients $\bar\gamma_n$; the classical BTZ temperature is recovered as a limiting case. The first law, $T\,\dd S = \dd M - \omega_H\,\dd J$ with $\omega_H = J\Omega_H/2r_H^2$, is stated, but obtaining $S(M,J)$ requires integrating along a path in $(M,J)$-space whose independence from the choice of path is not established in general.

The rotating sector is built into the formalism from the outset (through $\Omega$ and the mixed origin condition~\eqref{EMD-BTZ-origin}), which is the main structural difference from the static $3+1$D construction and directly addresses the goal of learning how to extend the higher-dimensional EMD beyond staticity.

%% file: subfiles/4-Applications.tex
\typeout{START baseline=\the\baselineskip, jot=\the\jot}
\chapter{From the EMD to black hole models}\label{chap examples}
In this Chapter, we illustrate applications of the approach presented in the previous Sections, in particular, the conditions~\eqref{ConRicciFinite} which are sufficient for a finite Ricci scalar at the horizon, we shall consider five concrete examples: the first one is the specifically constructed metric to satisfy~\eqref{ConRicciFinite} in a minimal fashion. The second one is Bonanno-Reuter~\cite{Bonanno:2000ep} black hole arising from the Asymptotically safe quantum gravity program. The third and fourth examples are the the Hayward black hole \cite{Hayward_2006} and the Dymnikova space-time \cite{Dymnikova:1992ux}, and although we are aware that these examples do not exhibit any divergent physical quantities, it is still interesting to demonstrate how our approach can be applied to deformation functions that explicitly depend on coordinates other than the proper distance. The last application is the electrically charged Frolov spacetime~\cite{Frolov:2016pav} in the case of extremal BH. Finally, we discuss a connection of our framework to the  Weak Gravity Conjecture within the Swampland program \cite{Palti:2019pca,Vafa:2005ui,Arkani-Hamed:2006emk} and discuss the efficiency of the approach to describe the proper distance within some models.

\section{Minimal metric deformation}
\begin{center}
    \begin{figure}[!t]
        \centering
        \includegraphics[width=0.75\textwidth]{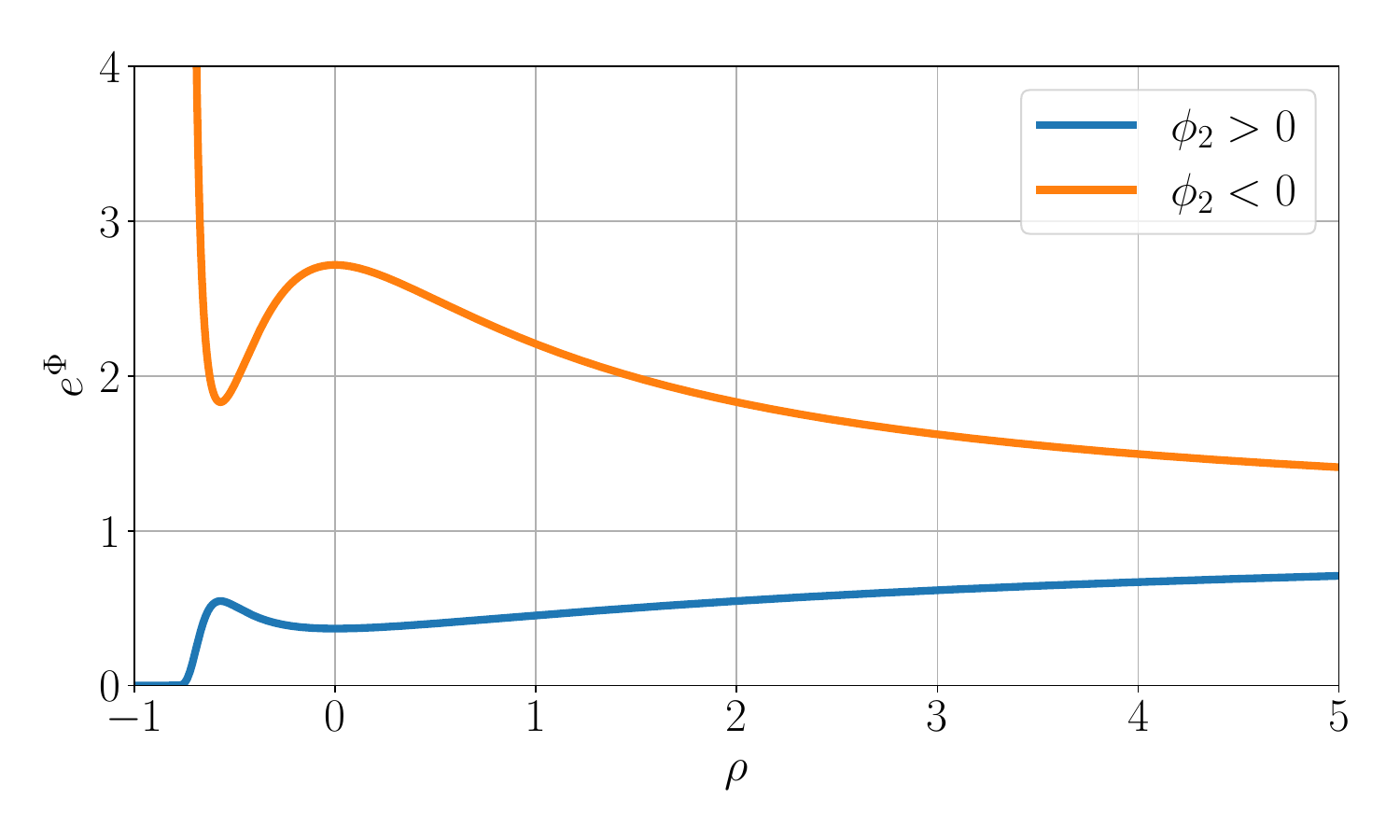}
        \caption{Function $e^{\Phi}$ in eq.~\eqref{FormExpPhiExample} with $\phi_0=-3\phi_2/2\dho^2$ and $\dho=1$.}
        \label{Fig:MinimalExample}
    \end{figure}
\end{center}

As a first example, we introduce a novel model which is an \emph{ad hoc} construction based on the constraints~\eqref{ConRicciFinite}. The physical motivations behind such toy model are irrelevant as we aim at showing the utility of the framework a model-building tool. We consider the following form of the metric function
\begin{equation}
    h_{\rm MM}(r) = f_{\rm MM}(r)= 1 - \frac{2M}{r}e^{\Phi\left( \frac{1}{d_H+\rho(r)}\right)} \ ,
\end{equation}
We can model the deformation $\Phi$ in such a way to abide by a minimal set of  conditions. Namely, the requirements that the Ricci scalar is finite at the horizon, summarized in eq.~\eqref{ConRicciFinite}, and that all the derivatives of $\Phi$ on the horizon of order higher than the third vanish translate into the following conditions:
\begin{align}
&\Phi_H=\phi_0\ , \quad \fid{1}=0 \ ,\quad \fid{2}=\phi_2\ , \quad \fid{3}=-6\dho \phi_2  \qq{and} \fid{n}=0 \quad \forall n \geq 4\ .\label{Phifuns}
\end{align}
The solution to~\eqref{Phifuns} can be written compactly in the form
\begin{align}
\Phi\left(\frac{1}{\dho+\rho}\right)=\phi_0+\frac{\rho^2(\dho+3\rho)\phi_2}{2\dho^2(\dho+\rho)^3}\,.\label{FormExpPhiExample}
\end{align}
Here we consider $\dho$ as a parameter of the model, which encodes information about the interior of the black hole. Moreover, $\phi_0,\phi_2\in\mathbb{R}$ are arbitrary parameters, which, however, are not independent: indeed, in order for the metric to asymptotically approach the Schwarzschild one (with mass parameter $M$), we require the asymptotic limit 
\begin{align}
\lim_{\rho\to\infty}e^{\Phi\left(\frac{1}{\dho+\rho}\right)}=e^{\phi_0+\frac{3\phi_2}{2\dho^2}}\overset{!}{=}  1 \quad \implies  \quad \phi_0 = -\frac{3\phi_2}{2\dho^2}\ .\label{ExampleLimits}
\end{align}
Choosing $\fid{n}\neq 0$ for some $n\geq 4$ in (\ref{Phifuns}) would yield higher modifications of $e^{\Phi}$ of order $\mathcal{O}(\rho^4)$, which are negligible close to the horizon. The position of the latter is located at 
\begin{align}
\zh=2M\, e^{\phi_0}=2M\,\text{exp}\left(-\frac{3\phi_2}{2\dho^2}\right)\,.
\end{align}
A schematic plot of $e^\Phi$ as a function of $\rho$ is shown in Figure~\ref{Fig:MinimalExample}. Here we have plotted $e^\Phi$ in the entire space-time (\emph{i.e.} also inside of the event horizon for $\rho<0$): following our general philosophy, we shall discuss the metric function (and all associated quantities) only outside of the horizon (\emph{i.e.} for $\rho\geq 0$) and only make a few brief remarks on the physics inside of the black hole in Section~\ref{Sect:MinimalModelInside}.
The leading coefficients $\xi_n$ (stemming from the series expansion of $2M\,e^\Phi$ in Eq.~(\ref{SeriesDistanceRho})) are exhibited in (\ref{CoefsMinModel}). In particular, due to the choice (\ref{Phifuns}), we find $\xi_1=0=\xi_3$, which therefore satisfies (\ref{ConRicciFinite}). 


\subsubsection{Curvature and temperature}
As a first step to calculating physical quantities for the spacetime metric characterized by (\ref{Phifuns}), we compute the first derivative of the function $f$ at the horizon $\zh=2 M e^{\phi_0}$
\begin{align}
&\fn{1}=\hn{1}=\frac{e^{-\phi_0}}{4M}\left(1+\sqrt{1-\frac{32 e^{2\phi_0}M^2\phi_2}{\dho^4}}\right) \qq{with} \phi_0=-\frac{3\phi_2}{2\dho^2}\ ,\label{ModelDerivativefh}
\end{align}
which imposes the condition $\phi_2\leq \frac{\dho^4}{32\,e^{2\phi_0}\,M^2}$, which with (\ref{ExampleLimits}) therefore becomes the non-linear relation for $\phi_2$
\begin{align}
\phi_2\,e^{3\phi_2/\dho^2}\leq \frac{\dho^4}{32M^2}\,.
\end{align}
The Ricci scalar at the horizon is finite in this model and takes the value
\begin{align}
&R\big|_{\zh}=\frac{3+(2-5\rootk)\rootk}{2\zh^2(1+2\rootk)}+\frac{24 \zh^2 \,\phi_2(\phi_2-12\dho^2)}{\dho^8(1+\rootk)(1+2\rootk)}\,,&&\text{with} &&\rootk=\sqrt{1-\frac{8\zh^2 \phi_2 }{\dho^4}}\ ,
\end{align}
which vanishes when $\phi_2 \to 0$ ($\varpi \to 1$), recovering the classical case.
Finiteness of the derivatives (\ref{ModelDerivativefh}) (at the event horizon) is a necessary requirement for well-behaved thermodynamical properties of the black hole. Indeed, the Hawking temperature is given by
\begin{equation}\label{ExampleTemp}
    T_\mathrm{H}=\frac{\fn{1}}{4\pi}=\frac{1}{8 \pi \zh}\left(1+\sqrt{1-\frac{8 \zh^2\phi_2}{\dho^4}}\right)\ .
\end{equation}
Determining the entropy using the law of thermodynamics \eqref{EntropyDef}, requires specifying the $M$-dependence of $\dho$ and $\phi_2$ and thus requires further refinement of the model. To give a concrete example, we shall consider $M\gg 1$ along with  
\begin{align}
&\phi_2=\mathcal{O}(M^0) \qq{and} \dho=\pi M+ \mathcal{O}(M^0)\,,\label{InteriorMassDependence}
\end{align}
that is, we are assuming only sub-leading corrections to the classical Schwarzschild metric. We, therefore, find the horizon position
\begin{align}
\zh=2 M-\frac{3\phi_2}{\pi^2 M}+\mathcal{O}(M^{-2})\,,
\end{align}
and thus for the Hawking temperature 
\begin{align}
T_\mathrm{H}=\frac{1}{8\pi M}\left[1+\frac{(3\pi^2 -16)\phi_2}{2\pi^4 M^2}+\mathcal{O}(M^{-3})\right]\,,
\end{align}
and the entropy
\begin{align}
    S = 4 \pi  M ^2 \left[1-\frac{ \left(3 \pi ^2-16\right) \phi _2 }{3 \pi^2 M^2 }\log (M^2 ) + \mathcal{O} \left( M^{-4}\right)\right] + \mathrm{const}\ .\label{FormEntropyExample}
\end{align}
This approximation exhibits a logarithmic correction, compatible with previous results in the literature \cite{Fursaev:1994te, Xiao:2021zly, Solodukhin:2011gn, Solodukhin:1994yz, Sen:2012dw, El-Menoufi:2015cqw, Cai:2009ua, Carlip:2000nv, Banerjee:2008cf, Banerjee:2008fz, Kaul:2000kf}. We remind the reader, however, that (\ref{FormEntropyExample}) is based on the assumptions (\ref{InteriorMassDependence}), which are related to the interior of the black hole solution. 

\subsubsection{Extending the metric inside the horizon}\label{Sect:MinimalModelInside}
Following our general philosophy, so far we have considered the metric function only outside of the black hole horizon and have used the horizon distance $\dho$ (and $\zh$) as the only additional input that is required for the interior of the black hole. Indeed, due to its definition
\begin{equation}\label{propd}
    d(r)=\int_0^r\frac{\dd r'}{\sqrt{|f(r')|}} \ ,
\end{equation}
the calculation of $\dho$ requires knowing the metric in the interior of the black hole, which we have not specified up to this point. In the current example, since the function (\ref{FormExpPhiExample}) can in fact be extended to the entire space-time (\emph{i.e.} also for $\rho<0$), as is showcased in Figure~\ref{Fig:MinimalExample}, one can contemplate the possibility to use (\ref{FormExpPhiExample}) as a model for the entire space-time. Although this discussion is generally outside of the scope of this review, here we shall nevertheless make a few remarks regarding this possibility. For concreteness, we shall focus on $\phi_2>0$ in the entire Subsubsection.

As a first question, we may ask the behavior of the metric at the origin. In order to get a better understanding of the space-time described by (\ref{FormExpPhiExample}) for $r \ll 1$, we may use a right rectangular approximation for the integral of the proper distance (\ref{propd})
\begin{align}
&d(r)\simeq \frac{r}{\sqrt{|f(r)|}} \qquad \forall r\ll M\ ,\label{ApproximationOrigin}
\end{align}
which is an algebraic equation. Numerical analysis suggests that there are no real positive values for $(r,d(r))$ that satisfy this relation for small $r$ when $f(r)<0$. We have, however, obtained the following solution of (\ref{ApproximationOrigin}) for $r(d)$ in the case $f(r)>0$
\begin{align}
\frac{r}{d}&=\frac{\sqrt{d}\left(3^{1/3}d^{2/3}+\left(-9e^{\Phi}M+\sqrt{-3d^2+81 e^{2\Phi}M^2}\right)\right)^{2/3}}{3^{2/3}\left(-9e^{\Phi}M+\sqrt{-3d^2+81e^{2\Phi}M^2}\right)^{1/3}}\,\nonumber\\
&=\cos\left[\frac{1}{3}\,\arcsin\left(\frac{3\sqrt{3} e^{\Phi}M}{d}\right)\right]-\frac{1}{\sqrt{3}}\sin\left[\frac{1}{3}\,\arcsin\left(\frac{3\sqrt{3} e^{\Phi}M}{d}\right)\right]\ .\label{zSerDef1}
\end{align}
where $e^\Phi$ is understood to be a function of $d=d_H+\rho$.
The function $f$, therefore, approaches $+1$ for $d\to 0$, which hints towards the absence of a singularity at the origin. To verify this, we realize that the function $r$ in (\ref{zSerDef1}) cannot be expanded in a Taylor series for $d$. However, noting that (for $\Phi$ given in (\ref{FormExpPhiExample})) for $\phi_2>0$ we have the limit $\displaystyle \lim_{\rho\to -\dho}\,e^{\Phi}$ in an exponential fashion, we can (formally) write $r/d$ as a series expansion in powers of $e^\Phi$
\begin{align}
&\frac{r}{d}=1-\sum_{n=1}^\infty \frac{2^{n-1}\,\Gamma\left(\frac{3n-1}{2}\right)}{n!\,\Gamma\left(\frac{n+1}{2}\right)}\,\left(\frac{M\, e^\Phi}{d}\right)^n \qq{for} \frac{M\,e^\Phi}{d}\in\left[0,\tfrac{1}{3\sqrt{3}}\right)\,.\label{Taylorzd}
\end{align}
We then find the Ricci scalar close to the origin 
\begin{align}
R=\frac{2M \,e^\Phi\, \phi_2}{\dho^2 d^9}\left[\dho^2 d^3(7d-6\dho)+(3\dho-4d)^2(d-\dho)^2\phi_2\right]+\mathcal{O}(e^{2\Phi})\,,\label{RicciOrigin}
\end{align}
which is indeed finite in the limit $d\to 0$, due to the exponential suppression of $e^\Phi$.\footnote{We have also verified that the Kretschmann scalar is finite at the origin in this model.}

While the above result of the absence of a curvature singularity at the origin is very encouraging conceptually, it also highlights another problem: indeed, if $f(r)>0$ close to the origin and $f(r)>0$ for $r>\zh$, with a simple zero at $\zh$, $f$ necessarily has (at least) one further zero in the interval $r\in(0,\zh)$, more concretely for $\rho\leq -\dho/3$. In other words, the space-time described by (\ref{FormExpPhiExample}) has at least one more inner horizon. At the latter, we have to verify again if all necessary conditions are met for the absence of a curvature singularity. However, since (\ref{Phifuns}) is only a minimal solution of the conditions (\ref{ConRicciFinite}) (which are tailored to remove unphysical singularities at the outer event horizon), we cannot guarantee the absence of a curvature singularity at this inner horizon. This problem can be circumvented by allowing some of the $\fid{n}$ (for $n\geq 4$) to be non-zero, as we have shown in Section~\ref{App:InnerHorizon} how to derive local conditions for the absence of curvature singularities at an inner horizon. 


\section{Bonanno-Reuter asymptotically safe black hole}\label{Sect:ExampleBonannoReuter}

The perturbative quantisation of gravity based on Steven Weinberg’s seminal proposal leads to insurmountable divergences due to the negative mass‑dimension of Newton’s constant. In his 1979 work~\cite{Weinberg:1979}, Weinberg argued that a quantum theory of gravity might still be consistent and predictive if the renormalisation group (RG) flow of all dimensionless couplings approaches a non‑Gaussian fixed point in the ultraviolet (UV). In other words, one requires a finite number of relevant directions so the theory remains “safe” from unphysical divergences as the momentum scale \(k \to \infty\).

More formally, let \(\{g_i(k)\}\) denote the set of dimensionless couplings in the theory; asymptotic safety demands
\[
\beta_{g_i}(g_j) \;=\; k\,\frac{\dd g_i}{\dd k} \;\to\; 0 \quad\text{for} \; k \to \infty,
\]
with \(g_i(k) \to g_i^*\) finite and the number of UV‑relevant directions finite.

In the gravitational context this idea materialises most concretely within the Functional Renormalisation Group (FRG) formalism for the effective average action \(\Gamma_k\). The authors of~\cite{Reuter:1996cp} obtained the Newton's constant beta function using some approximate truncations of the gravitational action, thus supporting the idea of a non‑perturbatively renormalisable quantum field theory of the metric.

Key reviews of the scenario (see~\cite{Niedermaier:2006wt,Percacci:2007sz,Eichhorn:2018yfc,Eichhorn:2022bgu,Reichert:20200Q}) underline its conceptual import and technical challenges.  
Several phenomenological consequences follow: in the UV the theory becomes scale‑invariant and predictive from a finite number of input parameters; in the infrared it must reproduce classical General Relativity; and deviations might appear at Planckian or near‑Planckian scales, potentially influencing black‑hole physics, cosmology and the structure of singularities.

While not yet providing a complete theory of quantum gravity, the asymptotic‑safety programme remains one of the viable frameworks consistent with the basic field‑theoretic paradigm, offering a minimal and field‑theory–based route to a UV completion of gravity.

Let us consider as an example the black hole metric introduced by Bonanno-Reuter~\cite{Bonanno:2000ep} as a renormalization group improved generalization of the Schwarzschild space-time. Indeed, in~\cite{Bonanno:2000ep}, it has been proposed to replace the (dimensionful) Newton constant $G_{\text{Newton}}$ by a running Newton constant
\begin{align}
G_{\text{Newton}}\longrightarrow G(k)=\frac{G_0}{1+\omega\,G_0\,k^2} \ ,
\end{align}
where $\omega\in\mathbb{R}$ is a constant and $k$ a (position dependent) scale (with reference scale $k=0$, at which we measure the Newton's constant $G_0\equiv G(k=0)$, whose length dimension is $[G_0] = [L]^2$). The choice of the latter is ambiguous, but it has been proposed in \cite{Bonanno:2000ep} to use an inverse physical distance from the center of the black hole\footnote{Other options discussed in \cite{Bonanno:2000ep} include distances of the form $\int_{\mathcal{C}}\sqrt{|\dd s^2|}$, for different choices of contours $\mathcal{C}$ (for example the world-line of a free-falling observer).} and here we present the re-scaled version with respect to $G_0$
\begin{align}
k(r)=(G_0)^{-1/2}\, \xi\, /\, \mathcal{D}(r)
\end{align}
with $\xi$ a suitable dimensionless constant and $\mathcal{D}(r)$ the dimensionless proper distance function. Such a distance has physical meaning, independent of a specific choice of coordinates. The metric proposed in \cite{Bonanno:2000ep} can be written in the form 
\begin{align}
    h(r)=f(r)=f_{\text{BR}}(r):=1-\frac{2G_0M}{r}\frac{1}{1+\widetilde{\omega}/\mathcal{D}(r)^2} \, ,\label{BRmetricfunction}
\end{align}
where $\widetilde{\omega}=\omega \, \xi^2$ is a dimensionless constant. In \cite{Bonanno:2000ep} the concrete value $\widetilde{\omega}=\frac{118}{15\pi}$ was given, however, subsequent works in the literature \cite{Donoghue:1993eb,Donoghue:1994dn,Hamber:1995cq,Bjerrum-Bohr:2002gqz,Donoghue:2022chi} potentially point towards different values (and a different sign). In the following, we shall consider $\widetilde{\omega}$ a generic parameter and discover marked differences between positive and negative values. Furthermore, we shall consider (\ref{BRmetricfunction}) to be valid only outside of the event horizon, which is located at $d_{\text{BR},H}$ (which we take as an input of the model.\footnote{Here we are allowing for the possibility that the metric inside of the black hole is different from (\ref{BRmetricfunction}) in which case $d_{\text{BR},H}$ would need to be computed as a separate input to the model. As we shall see, our conclusions will be entirely independent of this choice and thus the concrete value of $d_{\text{BR},H}$.}) In order for $f_{\rm BR}$ to remain well defined at the horizon, we shall assume $\widetilde{\omega}+d^2_{\text{BR},H}=\frac{2G_0M\, d_{\text{BR},H}^2}{\zh}\neq 0$.

For concrete computations, a choice for the physical distance needs to be made. Here we shall discuss three different possibilities that lead to a geometry with an (outer) event horizon, for which we can verify whether the conditions (\ref{ConRicciFinite}) are satisfied and whether therefore the Ricci scalar is finite, namely: \emph{(i)} the proper distance computed from the metric~(\ref{BRmetricfunction}); \emph{(ii)} the proper distance computed from the Schwarzschild metric; \emph{(iii)} an interpolating function. We shall discuss all three possibilities in the following:
\begin{itemize}
\item[\emph{(i)}] choosing $\mathcal{D}(r)$ as the \emph{proper distance} $d_{\rm BR}(r)$:
\begin{align}
&d_{\rm BR}(r)=(G_0)^{-1/2}\int_0^r \frac{\dd r'}{\sqrt{|f_{\text{BR}}(r')|}}\ ,&&\forall r \geq 0 \ .\label{BRselfDistance}
\end{align}
This is a self-consistent choice in the sense that the proper distance is compatible with the metric~\eqref{BRmetricfunction}. As discussed in~\cite{Binetti:2022xdi}, this guarantees that the modified metric exhibits the same diffeomorphism invariance as the (classical) Schwarzschild black hole. However, explicitly computing the distance becomes more involved (since~\eqref{BRselfDistance} is an implicit definition). For negative values of $\widetilde{\omega}$, a series expansion close to the horizon is developed in Appendix~\ref{Sect:BRDistanceComputations}. However, for our purposes, this is not in fact required, since we can simply verify the regularity conditions developed in the previous Sections, \emph{i.e.} eq.~\eqref{ConRicciFinite}. For the concrete function~(\ref{BRmetricfunction}) with $d(r)=d_{\text{BR}}(r)=d_{\text{BR},H}+ G_0^{-1/2}\rho(r)$, we obtain
\begin{align}
&\xi_1=\theta_1=\frac{4G_0 M \, \widetilde{\omega}\, d_{\text{BR},H}}{(\widetilde{\omega}+d_{\text{BR},H}^2)^2}\neq 0 \ ,\label{BRbreakingRelation}
\end{align}
such that Eq.~(\ref{ConRicciFinite}) is not satisfied. In fact, (\ref{BRbreakingRelation}) implies that already the first derivative of $f_{\text{BR}}$
\begin{align}
    f_{\text{BR}}^{(1)}(r)=(1-f_{\text{BR}}(r))\left(\frac{1}{r}-\frac{2 \widetilde{\omega}}{d_{\text{BR}}(r)(d_{\text{BR}}(r)^2+\widetilde{\omega})}\frac{1}{\sqrt{f_{\text{BR}}(r)}} \right)\,,&&\forall r \geq r_{\text{BR},H} \ , \nonumber
\end{align}
diverges at the horizon
\begin{align}
\lim_{\epsilon\to 0^+}  f_{\text{BR}}^{(1)}(r_{\text{BR},H}+\epsilon) \to \infty \ .
\end{align}
This is due to the fact that $f_{\text{BR}}(r_{\text{BR},H})=0$. Following the discussion of Section~\ref{hawking-classical-intro}, this poses problems with the interpretation of the black hole's thermodynamical properties, notably the Hawking temperature's definition. Furthermore, it also leads to a curvature singularity at the horizon, since for example, the Ricci scalar becomes 
\begin{align}
     &R= \frac{\widetilde{\omega}  (f_{\text{BR}}-1) \left( d_{\text{BR}}(d_{\text{BR}}^2+ \widetilde{\omega}) (1-5 f_{\text{BR}})f_{\text{BR}}^{1/2}  + r \, (6 d_{\text{BR}}^2 f_{\text{BR}}-2 \widetilde{\omega} )\right)}{ r\, d_{\text{BR}}^2  \left(d_{\text{BR}}^2+ \widetilde{\omega}\right)^2 f_{\text{BR}}^2} \ ,&&\forall r \geq \zh \ .\nonumber
\end{align}
This expression diverges at the event horizon due to the factor $f_{\text{BR}}^{2}$ in the denominator (while the numerator at the horizon assumes the finite value $2\widetilde{\omega}^2 r_H$).

Finally, we remark since $\xi_1\neq 0$, the results of the series expansion approach developed in Section~\ref{Sect:ReviewDistance} are not directly applicable. In appendix~\ref{App:SerNonZeroXi}, we show how it can be generalised in this case, and the consequences for the Bonanno-Reuter space-time for $\widetilde{\omega}<0$, are discussed in appendix~\ref{Sect:BRDistanceComputations} (confirming further our above conclusions in this case).

\item[\emph{(ii)}] choosing $\mathcal{D}(r)$ as the \emph{proper distance} of the (classical) \emph{Schwarzschild geometry} $d_{\text{S}}(r)$, \emph{i.e.}
\begin{align}
&d_{\text{S}}(r)=(G_0)^{-1/2}\int_0^r\frac{\dd r'}{\sqrt{\left|1-\frac{2G_0M}{r'}\right|}}=\\
&=(G_0)^{-1/2} \times\left\{\begin{array}{lcl}\pi G_0 M-\sqrt{r(2 G_0 M-r)}-2 G_0 M \, \arctan\sqrt{\frac{2 G_0 M}{r}-1} & \text{if} & 0 < r \leq 2 G_0 M\ , \\[2pt] \pi G_0 M+\sqrt{r(r-2 G_0 M)} + 2 G_0 M \, \text{arctanh}\sqrt{1-\frac{2 G_0 M}{r}} & \text{if} & r \geq 2 G_0M \ . \end{array}\right.\label{SchwarzschildDistance}
\end{align}
This option was initially advocated in \cite{Bonanno:2000ep} and has the advantage that it can be computed as a closed expression in terms of $r$. This geometry possesses a horizon, whose position is corrected by $\widetilde{\omega}$, \emph{e.g.} for large mass $M$ of the black hole
\begin{align}
r_{\text{BR},H}=2 G_0 M -\frac{2}{\pi^2}\,\frac{\widetilde{\omega}}{M}+\mathcal{O}\left(|\widetilde{\omega}|^{3/2}/ M^2\right)\ .
\end{align}
For $f_{\text{BR}}$ to be well defined at the horizon, we assume that $\widetilde{\omega}+d_{\text{S}}^2(r=r_{\text{BR},H})\neq 0$. The derivative of $f$ in this case becomes
\begin{align}
f^{(1)}_{\text{BR}}(r)=\frac{2G_0M}{r^2\left(1+\frac{\widetilde{\omega}}{d_{\text{S}}(r)^2}\right)}-\frac{4 G_0 M \, \widetilde{\omega} }{r\,d_{\text{S}}(r)^3 \left|1-\frac{2 G_0 M}{r}\right| \left(1+\frac{\widetilde{\omega}}{d_{\text{S}}(r)^2}\right)^2}\,,
\end{align}
which is finite at the horizon $r_{\text{BR},H}$, but diverges for $r=2 G_0 M$ (which for $\widetilde{\omega}>0$ lies outside the horizon of the black hole). Because of this, while the Ricci scalar is finite at $r=r_{\text{BR},H}$, the geometry has a curvature singularity at $r=2 G_0 M$.

\item[\emph{(iii)}] Although a closed function of $r$, the Schwarzschild proper distance (\ref{SchwarzschildDistance}) is still difficult to work with for concrete computations. Therefore in \cite{Bonanno:2000ep} the following approximation for (\ref{SchwarzschildDistance}) was proposed
\begin{equation}
d_{\text{S}}(r)\simeq \mathfrak{d}_{\text{S}}(r)=\left(\frac{r^3}{G_0(r+\gamma\,G_0M)}\right)^{1/2}\ ,\qq{such that} \begin{array}{l} \displaystyle \lim_{r \to 0} \mathfrak{d}_{\text{S}}(r)\sim (\gamma G_0 M)^{-1/2}r^{3/2}\,,\\ \displaystyle\lim_{r \to \infty} \mathfrak{d}_{\text{S}}(r)\sim (G_0)^{-1/2}\,r\ . \end{array}\label{BRdistanceApprox}
\end{equation}  
Here $\gamma\in\mathbb{R}$ is a constant, which in order to mimic the same behaviour as (\ref{SchwarzschildDistance}) at the origin ($r \to 0$) needs to be chosen as $\gamma=9/2$. The function $\mathfrak{d}_{\text{S}}$ has no inflection points, and thus does not feature the same behaviour as a proper distance at a horizon for any value of $r$. 

Identifying $\mathcal{D}$ in (\ref{BRmetricfunction}) with $\mathfrak{d}_{\text{S}}$ leads to a zero of $f_{\text{BR}}$ at 
\begin{align}
r_{\text{BR},H} = 2 G_0 M - \frac{2+\gamma}{4}\,\frac{\widetilde{\omega}}{  M}+\mathcal{O}(\widetilde{\omega}^2/ M^2) \ .
\end{align} 
At this horizon, both the first and second derivatives of $f_{\text{BR}}$ are finite implying that the Ricci scalar takes a finite value.

\end{itemize}
To summarise, treating (\ref{BRmetricfunction}) in a \emph{self-consistent fashion} by identifying $\mathcal{D}$ by the proper distance $d_{\rm BR}$ calculated from $f_{\text{BR}}$ itself leads to a divergent first derivative of the metric function at the horizon, which in turn leads to significant problems for physical quantities. Notably, it poses problems for defining a finite Hawking temperature and leads to a curvature singularity at the (outer) event horizon. This is in line with the results of Section~\ref{Sect:ReviewDistance}, due to the fact that the function $\Phi^{(\rho)}(\rho)=\frac{2G_0 M}{1+\widetilde{\omega}/(d_{\text{BR},H}+ G_0^{-1/2}\rho)^2}$ does not satisfy the conditions (\ref{ConRicciFinite}), independent of the geometry of the black hole inside the event horizon (\emph{i.e.} independent of the value of $d_{\text{BR},H}$). The choices \emph{(ii)} and \emph{(iii)} constitute a departure from the original idea presented in \cite{Bonanno:2000ep} (proposed as an approximation in this work) by replacing $\mathcal{D}$ by a function of the radial coordinate $r$, which does \emph{not} represent a physical distance that has been consistently calculated from the metric characterised by (\ref{BRmetricfunction}). 

While \emph{(ii)}, depending on the sign of $\widetilde{\omega}$, may have a curvature singularity outside of the event horizon, choice \emph{(iii)} is at least well-behaved from this perspective. However, from the point of view of the original motivation to deform the Schwarzschild metric by a function of a (consistently calculated) distance function, the choice \emph{(iii)} corresponds to a different deformation function than (\ref{BRmetricfunction}). Using the approach outlined in Section~\ref{Sect:DistanceFunctionExpansion}, we can determine this modified deformation function by reverse engineering the coefficients $\xi_n$: indeed, by integrating eq.~\eqref{DiffEqDistance} we find the coefficients $b_n$ in eq.~\eqref{SeriesDistanceRho}
\begin{align}
&b_1=\frac{\sqrt{2}\left|\zbr^3+\widetilde{\omega}G_0(\zbr+\gamma\, G_0 M)\right|}{\sqrt{\zbr G_0 M (\zbr^3-\widetilde{\omega}G_0(\zbr+2\gamma G_0 M))}}\,,\hspace{1cm}b_2=0\,,\nonumber\\
&b_3=\frac{b_1^3 G_0 M \left(\zbr^6-3\zbr^4 G_0\widetilde{\omega}-7\gamma \zbr^3 \, G_0^2 M \, \widetilde{\omega}+ (\gamma G_0^2 M \, \widetilde{\omega})^2\right)}{24\sqrt{2}\left(\zbr^3+\widetilde{\omega}G_0(\gamma G_0 M+\zbr)\right)^3}\,,\hspace{1cm}b_4=0\,.
\end{align}
Therefore with
\begin{align}
&a_1=0\,,&&a_2=\frac{1}{b_1^2}\,,&&a_3=0\,,&&a_4=-\frac{2b_3}{b_1^5}\,,
\end{align}
we obtain for the leading coefficients of $\xi_n$ (which are equal to $\theta_n$) 
\begin{align}
&\xi_1=0\,,&&\xi_2=\frac{2\zbr^2 \widetilde{\omega}\,  G_0 M (2\zbr+3\gamma \, G_0 M)}{b_1^2(\zbr^3+ \widetilde{\omega} \, G_0 (\zbr+\gamma \, G_0 M))^2}\,,&&\xi_3=0\,.
\end{align}
These indeed satisfy the conditions (\ref{ConRicciFinite}). Therefore, the ``approximation'' to use (\ref{BRdistanceApprox}) for $\mathcal{D}$ in (\ref{BRmetricfunction}) instead of the self-consistently calculated proper distance, corresponds to changing the metric function (\ref{BRmetricfunction}), in a way characterised by the above expansion coefficients.

We stress that the analysis above is a no-go result for the specific identification of the RG scale proposed in~\cite{Bonanno:2000ep}, rather than for the RG improvement of the Schwarzschild geometry as such: identifying $k$ self-consistently with the inverse proper distance from the centre is incompatible with a regular horizon, while the options \emph{(ii)} and \emph{(iii)} restore regularity only by giving up a consistently computed physical distance. This problem can be addressed in two possible ways, as summarized in Remark~\ref{rem:BRalternatives}: \emph{(a)}  modify the functional form of the deformations on the scale or \emph{(b)} choose another scale identification. Nevertheless, the EMD makes the origin of the obstruction transparent and indicates which identifications can avoid it, as summarised in Remark~\ref{rem:BRalternatives}.

\begin{center}
\remarkbox{\rem\label{rem:BRalternatives}{Towards a consistent RG improvement}}{\parbox{16cm}{
\begin{itemize}
\item[\emph{(a)}] \emph{The obstruction is not specific to $k\propto 1/\mathcal{D}$.} Consider any identification $k=k(d)$, with $d=d_{\text{BR},H}+G_0^{-1/2}\rho$ the self-consistently computed proper distance from the centre, such that $\Phi^{(\rho)}(\rho)=2M\,G\big(k(d)\big)$. Then
\begin{align}
\xi_1=\left.\dv{\Phi^{(\rho)}}{\rho}\right|_{\rho=0}\propto\left.\dv{G}{k}\right|_{k=k_H}\,\left.\dv{k}{d}\right|_{d=d_{\text{BR},H}}\,,\nonumber
\end{align}
which is non-vanishing whenever $G$ runs at the scale $k_H$ and $k$ is a monotonic function of $d$, in conflict with~\eqref{ConRicciFinite}. More generally, since $\rho\propto\sqrt{r-\zh}$ close to a non-extremal horizon, odd powers of $\rho$ in the deformation generate half-integer powers of $r-\zh$ in $f$ and $h$. A consistent identification must therefore depend on $\rho$ only through even powers (to the order relevant for Result~\ref{res:NeutralRegularity}), \emph{i.e.} it must be a smooth function of $r$ across the horizon.
\item[\emph{(b)}] \emph{Candidates.} An identification of the scale is compatible with a regular horizon if it is a smooth function of $r$ across the horizon and, as required by self-consistency, it is computed from the RG-improved geometry itself. Two classes of physical quantities satisfy both requirements: the proper time $\tau$ of a radially free-falling observer, which is linear in $r-\zh$ close to the horizon (Remark~\ref{rem:TauVsRho}), such that the corresponding proper-time EMD is automatically regular at the horizon (Result~\ref{res:ProperTimeEMD}); and local curvature invariants, \emph{e.g.} $k\propto K^{1/4}$, which can be treated self-consistently near the horizon with the Kretschmann scalar EMD of Appendix~\ref{Sect:KretschmannScalar}.
\item[\emph{(c)}] \emph{Open questions.} Whether these identifications lead to geometries that are also regular at the inner horizon and in the core (Section~\ref{Sect:CoreExamples}), how the sign and value of $\widetilde{\omega}$ are fixed, and to what extent RG-improved metrics capture solutions of a quantum effective action~\cite{Knorr:2022kqp,Platania:2023srt} are left for future work.
\end{itemize}
}}
\end{center}

\noindent
Let us comment on these candidates in relation to the literature. The free-fall proper time corresponds to one of the contours $\mathcal{C}$ already mentioned in~\cite{Bonanno:2000ep}: the identification $k\propto 1/(\tau_H+\tau)$, with $\tau_H$ the proper time needed to reach the centre from the horizon (playing the role of $d_{\text{BR},H}$) and $\tau$ computed from the deformed metric functions as in Section~\ref{Sect:EMDProperTime}, is self-consistent and compatible with a regular horizon for any choice of parameters subject to~\eqref{ProperTimeInequalities}. Curvature-based identifications, $k\propto K^{1/4}$, have been adopted in~\cite{Held:2019xde,Eichhorn:2021iwq,Ishibashi:2021kmf_ImprovedRN}. In these models, however, $K$ is evaluated on the classical (Schwarzschild, Kerr or Reissner--Nordstr\"om) geometry rather than on the improved one. The resulting metric functions are smooth in $r$ and hence regular at the horizon (\emph{cf.} Section~\ref{Sect:CoreExamples}), but, as for the options \emph{(ii)} and \emph{(iii)} above, the physical quantity is replaced by a fixed function of $r$: the scale is not set by the curvature of the geometry it generates, so that this description is not self-consistent. A self-consistent identification requires $K$ to be computed from the improved metric, which turns it into a differential equation for $f$; close to the horizon, this is precisely the problem solved order by order by the Kretschmann scalar EMD of Appendix~\ref{Sect:KretschmannScalar}. A self-consistent construction of a different kind was proposed in~\cite{Platania:2019kyx}: there the scale is identified with the effective energy density $\varepsilon$, $k^2\propto G_0\,\varepsilon$, of the geometry obtained at the previous step of an iterative RG improvement, and the fixed point of the iteration is a regular geometry of Dymnikova type (see Section~\ref{Sect:Dymnikova}).

\section{Hayward black hole}\label{sec:hayward}

The Hayward black hole~\cite{Hayward_2006} is a well-known example of a static and spherically symmetric regular black hole solution, designed to eliminate the central singularity while preserving the essential causal structure of the Schwarzschild geometry. The metric was introduced as a simple, phenomenological modification of the Schwarzschild solution that replaces the singularity with a regular de Sitter core, characterized by a finite curvature and energy density at the origin, without committing to any specific modification of General Relativity. The Hayward spacetime smoothly interpolates between a de Sitter region near $r=0$ and an asymptotically flat Schwarzschild-like behavior at large radii, with the transition governed by a length scale parameter associated with quantum-gravitational effects. Owing to its minimal and physically transparent form, the Hayward model has become a standard prototype in the study of regular black holes, black hole evaporation, and possible quantum modifications of classical horizons.
The metric function for the Hayward space-time can be written as
\begin{align}
&f(r)=h(r)=f_{\text{Hay}}(r):=1-\frac{2M r^2}{r^3+2M \gamma} \ ,
\end{align}
where $\gamma$ is a free parameter that determines the scale at which the departure from the classical Schwarzschild solution becomes significant \cite{Frolov:2016pav}. Here we assume $f_{\text{Hay}}$ to hold in the entire space-time.
From the metric element, we can directly deduce the form of the deformation function, which explicitly depends on the coordinate $r$ rather than the proper distance $d$
\begin{equation}
    \Phi\left(d(r)\right) := \frac{2M r^3}{r^3+2M\gamma}\ .
\end{equation}
The Hayward space-time exhibits two event horizons, indicated by the existence of two zeroes of the function $f_\mathrm{Hay}$. The position of the outer horizon is:
\begin{align}
\zhp=\frac{2M}{3}\,\left(1+2\cos\left[\frac{1}{3}\,\arccos\left(1-\frac{27\gamma}{8M^2}\right)\right]\right)\,.
\end{align}
Furthermore, the proper distance can be computed using the integral expression \eqref{propd}:
\begin{equation}\label{HaywardPropDist}
    d_\mathrm{Hay}(r)= \int_0^r \sqrt{\abs{\frac{\tilde{r}^3+2M \gamma}{ \tilde{r}^3 - 2 M (\tilde{r}^2 + \gamma)} }}\dd \tilde{r} \ .
\end{equation}

\begin{figure}[!ht]
    \centering
    \includegraphics[width=0.75\textwidth]{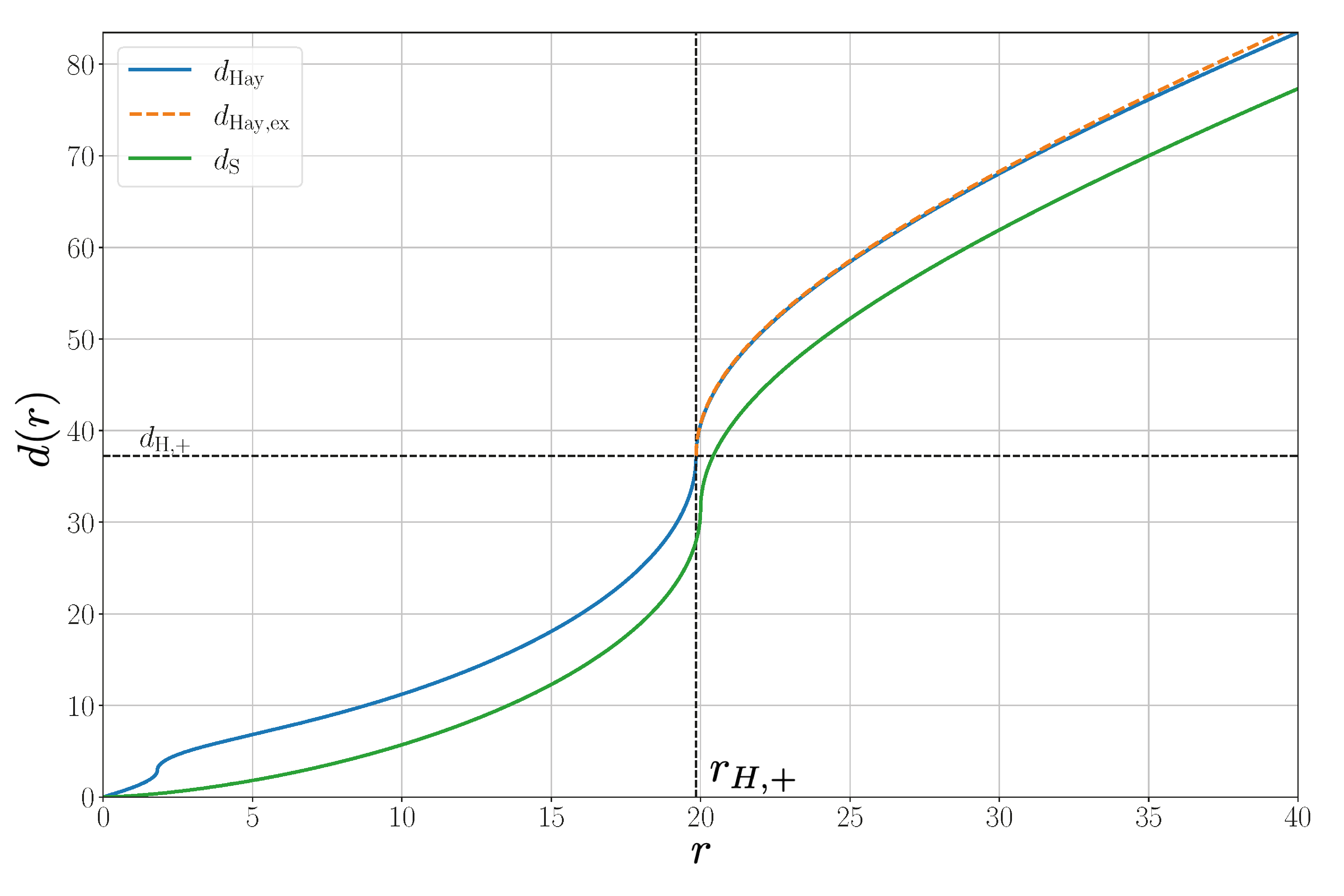}
    \caption{Comparison between the proper distance in the Hayward space-time (blue, calculated using equation \eqref{HaywardPropDist}) and the proper distance in the Schwarzschild space-time $d_\mathrm{S}$ (green) for specific parameter values: $M=10$ and $\gamma=3$. The dashed orange line corresponds to the Taylor series expansion of the proper distance from the event horizon, obtained from equation \eqref{DistHayExpansion}.}
    \label{HaywardDistance}
\end{figure}

Numerical evaluation of this integral suggests the following form of the distance of the horizon for large values of $M$
\begin{align}
&\dhp=\pi M+c_1\, \gamma^{c_2}\,\ln M+\mathcal{O}(M^0)\,,&&\text{with} &&\begin{array}{l}c_1=0.3334\pm 0.0002\,, \\ c_2=0.4999\pm 0.0002\,.\end{array}
\end{align}
More importantly, the distance (for a generic point outside of the horizon) can be expanded in powers of $(r-\zhp)$, which takes the explicit form
\begin{align}
d_{\text{Hay,ex}}&=\dhp+\frac{\sqrt{2}(\zhp^3+2\gamma M) \sqrt{r-\zhp}}{\sqrt{\zhp M(\zhp^3-4\gamma M)}}+\frac{M(\zhp^6-14\zhp^3\gamma M + 4\gamma^2 M^2)}{3\sqrt{2}(\zhp M (\zhp^3-4\gamma M))^{3/2}}\,(r-\zhp)^{3/2}\nonumber\\
&\hspace{1.35cm}+\mathcal{O}((r-\zhp)^{5/2})\,.\label{DistHayExpansion}
\end{align}
A comparison between the numerical solution for the distance function \eqref{HaywardPropDist} and \eqref{DistHayExpansion} is shown in Figure~\ref{HaywardDistance}.
Using expansion \eqref{SeriesDistance} we can read off the coefficients $b_n$ from eq.~\eqref{DistHayExpansion} and express them in terms of the $a_n$'s using the recursive relation listed in~\eqref{Coeffsb}
\begin{equation}
   a_2^{-1/2}=\frac{\sqrt{2}(\zhp^3+2\gamma M)}{\sqrt{\zhp M(\zhp^3-4\gamma M)}}\, , \,  - \frac{a_3}{2 a_2^2} = 0\, , \, \frac{5 a_3^2 - 4 a_2 a_4}{8 a_2^{7/2}} = \frac{M(\zhp^6-14\zhp^3\gamma M+4\gamma^2 M^2)}{3\sqrt{2}(\zhp M(\zhp^3-4\gamma M))^{3/2}}\, .
\end{equation}
Solving this system yields the solutions for $a_2$, $a_3$ and $a_4$ 
\begin{align}
    a_2 &= \frac{\zhp M (\zhp^3 - 4 \gamma  M)}{2(\zhp^3 + 2 \gamma M)^2} \, , \qquad a_3 = 0 \, , \\ a_4 &= \frac{-M ^2 \zhp^{10}+18 \gamma  M ^3 \zhp^7-60 \gamma ^2 M ^4 \zhp^4+16 \gamma ^3 M ^5 \zhp}{384 \gamma ^5 M ^5+12 \zhp^{15}+120 \gamma  M  \zhp^{12}+480 \gamma ^2 M ^2 \zhp^9+960 \gamma ^3 M ^3 \zhp^6+960 \gamma ^4 M  ^4 \zhp^3}\ .
\end{align}
Plugging these coefficients into the series \eqref{SeriesDistance} allows us to write $r$ as a power series in $\rho$ and then expand around $\rho=0$
\begin{multline}
\frac{2 M r(\rho)^3}{r(\rho)^3+2 M\gamma} = \zhp + \frac{6 \gamma  M ^3 \zhp^3 \left(\zhp^3-4 \gamma  M \right)}{\left(2 \gamma  M +\zhp^3\right)^4} \rho^2 +\\+ \frac{\gamma  M ^4 \left(-7 \zhp^{12}+72 \gamma  M  \zhp^9-204 \gamma ^2 M ^2 \zhp^6+112 \gamma ^3 M ^3 \zhp^3\right)}{\left(2 \gamma  M +\zhp^3\right)^7}\rho^4 + \mathcal{O}(\rho^5)\ ,
\end{multline}
from which we can read the coefficients $\xi_n$. In particular, we obtain 
\begin{equation}
  \xi_1=0\, , \quad \xi_2 = \frac{6 \gamma  M ^3 \zhp^3 \left(\zhp^3-4 \gamma  M \right)}{\left(2 \gamma  M +\zhp^3\right)^4} \qq{and} \xi_3 = 0\ .
\end{equation}
Alternatively, we can use the last relation in equation \eqref{InitialConditionsGenericExpansion} and the recursive relation in equation \eqref{Coeffsa} for $p=3$ to determine the values of the coefficients $\xi_2$ and $\xi_3$. By doing so, we can establish that the Hayward space-time satisfies the condition \eqref{ConditionsfhRegular}. Consequently, it is not surprising that the Ricci scalar and the Hawking temperature in this space-time are well-defined and free from singularities at the event horizon $\zhp$.

\section{Dymnikova black hole}\label{Sect:Dymnikova}

The Dymnikova black hole~\cite{Dymnikova:1992ux,Dymnikova:2004qg} represents one of the earliest and most studied examples of a regular black hole, that is, a spherically symmetric and asymptotically flat solution of Einstein’s equations free from curvature singularities at the center. The solution arises from coupling general relativity to a specific form of anisotropic vacuum-like matter with a de Sitter core, which replaces the classical singularity with a smooth region of finite energy density. The metric interpolates between a de Sitter geometry near the origin and a Schwarzschild-like behavior at large radii, yielding a horizon structure similar to that of the Reissner–Nordström spacetime. The Dymnikova model thus provides a self-consistent and physically motivated realization of a nonsingular black hole, serving as a valuable prototype for studying quantum-gravity inspired regular geometries and their thermodynamic or quasi-normal mode properties.

The metric function in the Dymnikova space-time is given by
\begin{align}
f(r)=h(r)=f_{\text{Dymn}}(r):=1-\frac{2 M }{r}\,\left(1-e^{-r^3/\ell_*^3}\right) \qq{with} \ell_\ast^3 = 2 M \ell_0^2 \ ,
\end{align}
which (as for the Hayward black hole) we take to hold for the entire space-time. Also, similar to the Hayward black hole, the parameter $\ell_0$ ensures the regularity of the solution near the origin. Additionally, the density profile of the effective energy-momentum tensor is chosen such that the metric possesses a de Sitter core.
Furthermore, the Dymnikova space-time possesses two horizons, which for large masses are located at
\begin{equation}
    \zhp = 2 M \left(1 - \mathcal{O}\left( e^{-4 M^2/\ell_0^2} \right) \right) \qq{and} \zhm = 2 M \left(1 - \mathcal{O} \left(\frac{\ell_0}{8 M} \right) \right)\ .
\end{equation}

\begin{figure}[!ht]
    \centering
    \includegraphics[width=0.75\textwidth]{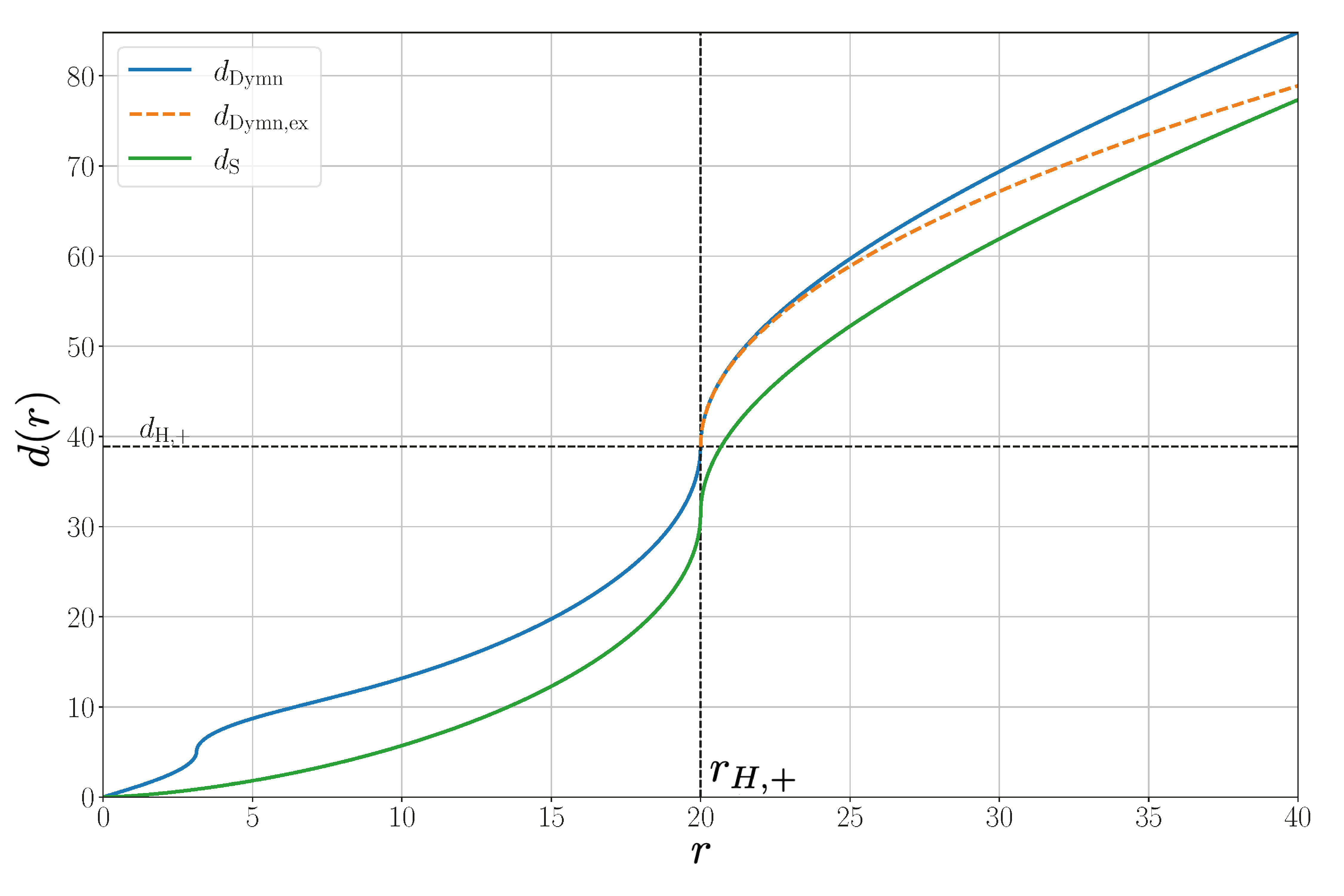}
    \caption{Comparison of the proper distance in the Dymnikova space-time, computed using the definition \eqref{propd} (blue), with the proper distance in the Schwarzschild space-time $d_\mathrm{S}$ (green) for $M=10$ and $\ell_0=3$. The dashed orange line represents the Taylor series expansion of the proper distance from the event horizon, as given by equation \eqref{DistDymnExpansion}.}
    \label{figDymnikova}
\end{figure}

We find from \eqref{SeriesDistance} that the power series expansion of the proper distance has the following form
\begin{multline}
    d_\mathrm{Dymn,ex} = \dhp + \frac{2 \zhp \sqrt{r-\zhp}}{\sqrt{2 M - e^{-\zhp^3/2 M  \ell_0^2} \left(2 M + 3 \zhp^3/\ell_0^2 \right)}}  + \\ + \frac{e^{-\zhp^3/ 2 M  \ell_0^2} \left(8 M ^2 \ell_0^4 \left(e^{ \zhp^3/ 2 M  \ell_0^2}-1 \right)-9 \zhp^6\right)}{24 M  \ell_0^4 \left(2 M +e^{- \zhp^3/ 2 M  \ell_0^2} \left(-2 M - {3 \zhp^3}/{\ell_0^2}\right)\right)^{3/2}}(r-\zhp)^{3/2} + \mathcal{O}((r-\zhp)^2)\ .\label{DistDymnExpansion}
\end{multline}
In Figure \ref{figDymnikova} we provide a numerical approximation of the proper distance using the definition \eqref{propd} and compare this expansion for $r > \zhp$. Furthermore, we can recover the coefficients $a_2$, $a_3$ and $a_4$
\begin{align}
    &a_2 = \frac{e^{-\zhp^3/2 M  \ell_0^2} \left(2 M  \ell_0^2 \left(e^{\zhp^3/2 M  \ell_0^2}-1\right)-3 \zhp^3\right)}{4 \ell_0^2 \zhp^2}\, , \qquad a_3=0 \, ,\\
    & a_4 = \frac{e^{-\zhp^3/2 M  \ell_0^2} \left(2 M  \ell_0^2 \left(e^{\zhp^3/2 M  \ell_0^2}-1\right)-3 \zhp^3\right) \left(9 \zhp^6-8 M ^2 \ell_0^4 \left(e^{\zhp^3/2 M  \ell_0^2}-1\right)\right)}{384 M  \ell_0^6 \zhp^5}\ .
\end{align}
As for the Hayward space-time, we can expand the function $2M(1- e^{-r(\rho)^3/\ell_\ast^3})$ around $\rho=0$
\begin{equation}
    2M(1- e^{-r (\rho)^3/\ell_\ast^3}) = \zhp + \frac{3 e^{-\zhp^3/ M  \ell_0^2 } \left(2 M  \ell_0^2 \left(e^{\zhp^3/ 2 M  \ell_0^2 }-1\right)-3 \zhp^3\right)}{4 \ell_0^4} \rho^2 + \mathcal{O}(\rho^4)
\end{equation}
from which we read off the coefficients $\xi_1$, $\xi_2$ and $\xi_3$
\begin{equation}
    \xi_1 = 0\ , \quad \xi_2 = \frac{3 e^{-\zhp^3/ M  \ell_0^2 } \left(2 M  \ell_0^2 \left(e^{\zhp^3/ 2 M  \ell_0^2 }-1\right)-3 \zhp^3\right)}{4 \ell_0^4} \qq{and} \xi_3 = 0\ .
\end{equation}
We observe that the conditions stated in equation \eqref{ConditionsfhRegular} are satisfied in the Dymnikova space-time as well, leading to the same conclusions as those drawn for the Hayward black hole. These conditions ensure that the Ricci scalar and the Hawking temperature remain well-defined and free from singularities at the event horizons, which is indeed the case for this space-time. 

\section{Regularity of the core}\label{Sect:CoreExamples}
We now apply the conditions of Section~\ref{Sect:FullRegularity} to the geometries considered in~\cite{DamiaPaciarini:2025qhq}: the Schwarzschild black hole and four regular models, namely the Hayward (Section~\ref{sec:hayward}), Dymnikova (Section~\ref{Sect:Dymnikova}), Bardeen and Simpson-Visser~II black holes. All of them satisfy $f=h$, which simplifies the analysis considerably, as we first discuss in general.

\paragraph{Geometries with \texorpdfstring{$f=h$}{f=h}.} For $f=h$ the two deformation functions coincide, $\Phi=\Psi$, and the first condition in~\eqref{reg:GeodesicCond} is automatically satisfied. Moreover, if $f$ is a smooth function of $r$ with simple zeros, then on each side of a zero $r-r_P$ is a power series in $(d-d_P)^2$. All odd one-sided derivatives of $\Phi$ then vanish, so that the regularity conditions at the horizons hold as well. This is the reason why the coefficients $\xi_1$ and $\xi_3$ were found to vanish for the Hayward and Dymnikova geometries in the previous Sections. The only non-trivial information is therefore contained in the behaviour at the origin. Writing $f=1-\Phi(d)/r$ and using $\Phi=r\,(1-f)$, all regular models below have $\Phi=\mathcal{O}(r^3)$, hence $f=1+\mathcal{O}(r^2)$ and $d=r+\mathcal{O}(r^3)$. This implies $\Phi^{(1)}_0=\Phi^{(2)}_0=0$, while $\Phi^{(3)}_0$ coincides with the third derivative of $r\,(1-f)$ with respect to $r$ at $r=0$. Since $\Psi=\Phi$, also $\Psi^{(1)}_0=0$ and $\Psi^{(3)}_0=\Phi^{(3)}_0$, such that, by Result~\ref{res:Cores}, the core is locally maximally symmetric with $\Lambda=\Phi^{(3)}_0/2$. The results are summarized in Table~\ref{tab:CoreExamples}.

\begin{table}[ht]
\centering
\renewcommand{\arraystretch}{2.1}
\begin{tabular}{|c|c|c|c|c|}\hline
model & $f(r)=h(r)$ & $\Phi^{(0)}_0$ & core & $\Lambda=\Phi^{(3)}_0/2$ \\\hline\hline
Schwarzschild & $\displaystyle 1-\frac{2M}{r}$ & $2M$ & singular & -- \\[5pt]\hline
Hayward~\cite{Hayward_2006} & $\displaystyle 1-\frac{2Mr^2}{r^3+2M\gamma}$ & $0$ & de Sitter & $\displaystyle\frac{3}{\gamma}$ \\[5pt]\hline
Dymnikova~\cite{Dymnikova:1992ux} & $\displaystyle 1-\frac{2M}{r}\left(1-e^{-r^3/(2M\ell_0^2)}\right)$ & $0$ & de Sitter & $\displaystyle\frac{3}{\ell_0^2}$ \\[5pt]\hline
Bardeen~\cite{bardeen1968} & $\displaystyle 1-\frac{2Mr^2}{(r^2+l_{\rm B}^2)^{3/2}}$ & $0$ & de Sitter & $\displaystyle\frac{6M}{l_{\rm B}^3}$ \\[5pt]\hline
Simpson-Visser~II~\cite{Simpson:2019mud} & $\displaystyle 1-\frac{2M}{r}\,e^{-a/r}$ & $0$ & Minkowski & $0$ \\[5pt]\hline
\end{tabular}
\caption{Core structure of the static, spherically symmetric geometries discussed in Section~\ref{Sect:CoreExamples}, obtained from~\eqref{reg:OriginCond} and Result~\ref{res:Cores}. All models satisfy $f=h$, so that $\Psi^{(n)}_0=\Phi^{(n)}_0$.}
\label{tab:CoreExamples}
\end{table}

\paragraph{Schwarzschild black hole.} For the classical solution $\Phi=\Psi=2M$ is constant, so that $\Phi^{(0)}_0=2M\neq0$ and condition~\eqref{reg:OriginCond} is violated. Indeed, inside the horizon $|f|\simeq2M/r$, the distance behaves as $d\propto r^{3/2}$ and the Kretschmann scalar~\eqref{KretschDef} shows the familiar divergence $K=48M^2/r^6$. The single horizon is consistent with Remark~\ref{rem:EvenHorizons}.

\paragraph{Hayward and Dymnikova black holes.} Both geometries have two simple horizons for small enough values of the regularization parameter ($\gamma<16M^2/27$ for the Hayward black hole), at which the regularity conditions are satisfied, as discussed above. Their cores are of de Sitter type, with $\Lambda=3/\gamma$ and $\Lambda=3/\ell_0^2$, respectively, in agreement with their construction as interpolations between a de Sitter interior and a Schwarzschild exterior.

\paragraph{Bardeen black hole.} The Bardeen geometry~\cite{bardeen1968}, already used in Chapter~\ref{chap5} with the metric function~\eqref{gtt Bardeen}, can be interpreted as the field of a magnetic monopole in non-linear electrodynamics~\cite{AyonBeato:2000zs}, with $l_{\rm B}$ playing the role of the magnetic charge. For $l_{\rm B}^2<16M^2/27$ it has two simple horizons, which merge into an extremal one for $l_{\rm B}^2=16M^2/27$, where $d$ ceases to be finite. Expanding $\Phi=2Mr^3/(r^2+l_{\rm B}^2)^{3/2}$ around the origin gives $\Phi^{(3)}_0=12M/l_{\rm B}^3$, \emph{i.e.} a de Sitter core with $\Lambda=6M/l_{\rm B}^3$, in agreement with~\cite{DamiaPaciarini:2025qhq}.

\paragraph{Simpson-Visser II black hole.} This geometry~\cite{Simpson:2019mud}, already encountered in Chapters~\ref{chap4} and~\ref{chap5} (with $a=2M\eta$ in the notation of Table~\ref{subs models table}), has two simple horizons for $a<2M/e$. Its metric function is not analytic at $r=0$, since $e^{-a/r}$ has an essential singularity there. Nevertheless, $f$ is infinitely differentiable for $r\geq0$, and all derivatives of $\Phi=2M\,e^{-a/r}$ vanish as $r\to0^+$. Hence all coefficients $\Phi^{(n)}_0$ vanish and the core is of Minkowski type, with $f=1+\mathcal{O}(r^n)$ for any $n$. Since the conditions of Section~\ref{Sect:FullRegularity} only involve (one-sided) derivatives of the deformation functions, this example illustrates that the EMD also applies to geometries which are smooth, but not analytic.

\vspace{0.5em}
In all four regular models the regularization parameter controls both the core and the horizon structure. As it increases, the two horizons approach each other and eventually merge, in agreement with Remark~\ref{rem:EvenHorizons}. Beyond this point the geometry describes a horizonless compact object. At the extremal configuration the proper distance to the horizon diverges, and the analysis has to be carried out with the proper-time EMD of Section~\ref{Sect:EMDProperTime}.

\section{Extremal Frolov space-time}

The Frolov spacetime~\cite{Frolov:2016pav} represents a class of regular black hole solutions designed to capture the essential features of quantum-gravity–induced modifications to the classical black hole interior. In this model, the singularity is replaced by a smooth transition region connecting the trapped and anti-trapped domains through a de Sitter-like core, ensuring that all curvature invariants remain finite. The geometry can be interpreted as an effective semiclassical description of a spherical black hole undergoing evaporation, and it incorporates a bounce that prevents the formation of a curvature singularity. The Frolov metric thus provides a physically motivated and analytically tractable framework to explore non-singular black hole evolution, causal structure, and information recovery scenarios in quantum gravity.

In this Subsection we shall discuss the Frolov spacetime as the extremal configuration of the deformed Reissner-Nordstr\"om geometry, or equivalently as an extremal charged Hayward BH (cf. Sec.~\ref{sec:hayward}). 
As a concrete example of the EMD based on the proper time (notably in the extremal case), we shall consider here the following metric (using the same notation as in \eqref{MetricForm}), first introduced by Frolov in~\cite{Frolov:2016pav}
\begin{align}
\label{FrolovMetric}
    h(r)=f(r)= 1 - \left(\frac{2M}{r}-\frac{Q^2}{r^2} \right)\widetilde{\Gamma} \ , \qq{where} \widetilde{\Gamma}\coloneqq\frac{r^4}{r^4 + (2 M r + Q^2) \eta^2} \ .
\end{align}
If $Q=0$, we recover the Hayward metric \cite{Hayward_2006}. Inspecting the two limits
\begin{align}
    f(r) &\overset{r\to \infty}{=} 1 - \frac{2 M}{r} + \frac{Q^2}{r^2} + \eta^2 \mathcal{O}(r^{-4}) &\text{and} & & f(r) &\overset{r \to 0}{=} 1 + \frac{r^2}{\eta^2} + \mathcal{O}(r^6) \ ,
\end{align}
the metric (\ref{FrolovMetric}) features asymptotic flatness, while reproducing the Reissner-Nordström metric for $\eta\to 0$. This latter parameter can be interpreted as the radius of an Anti de-Sitter core at the origin. Finally, the position of the event horizon satisfies $f(\zh)=0$, which has 4 solutions: we shall understand $\zh$ to be the largest (real) value among these.

\begin{figure}[ht]
\begin{center}
\includegraphics[width=0.45\textwidth]{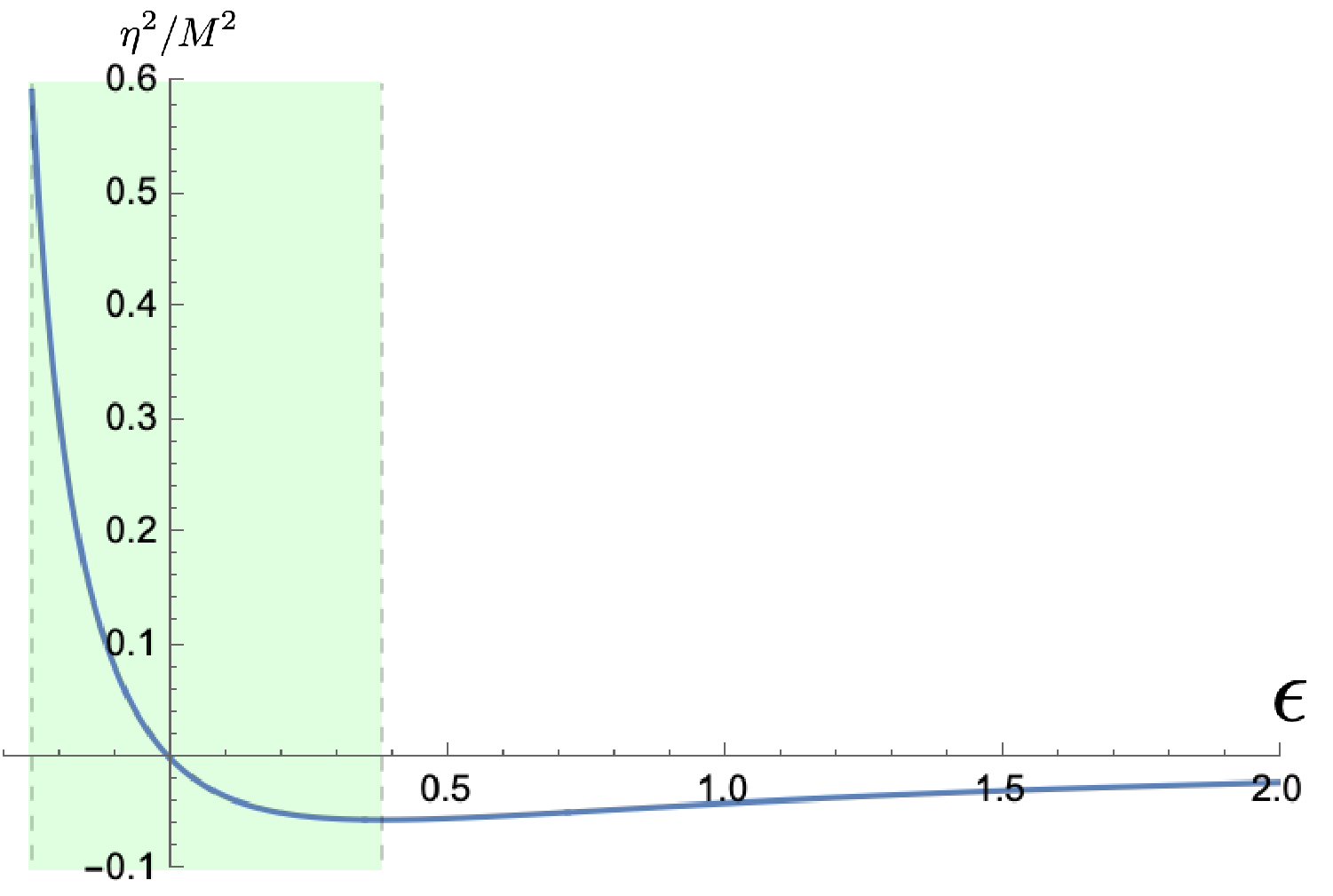} \includegraphics[width=0.45\textwidth]{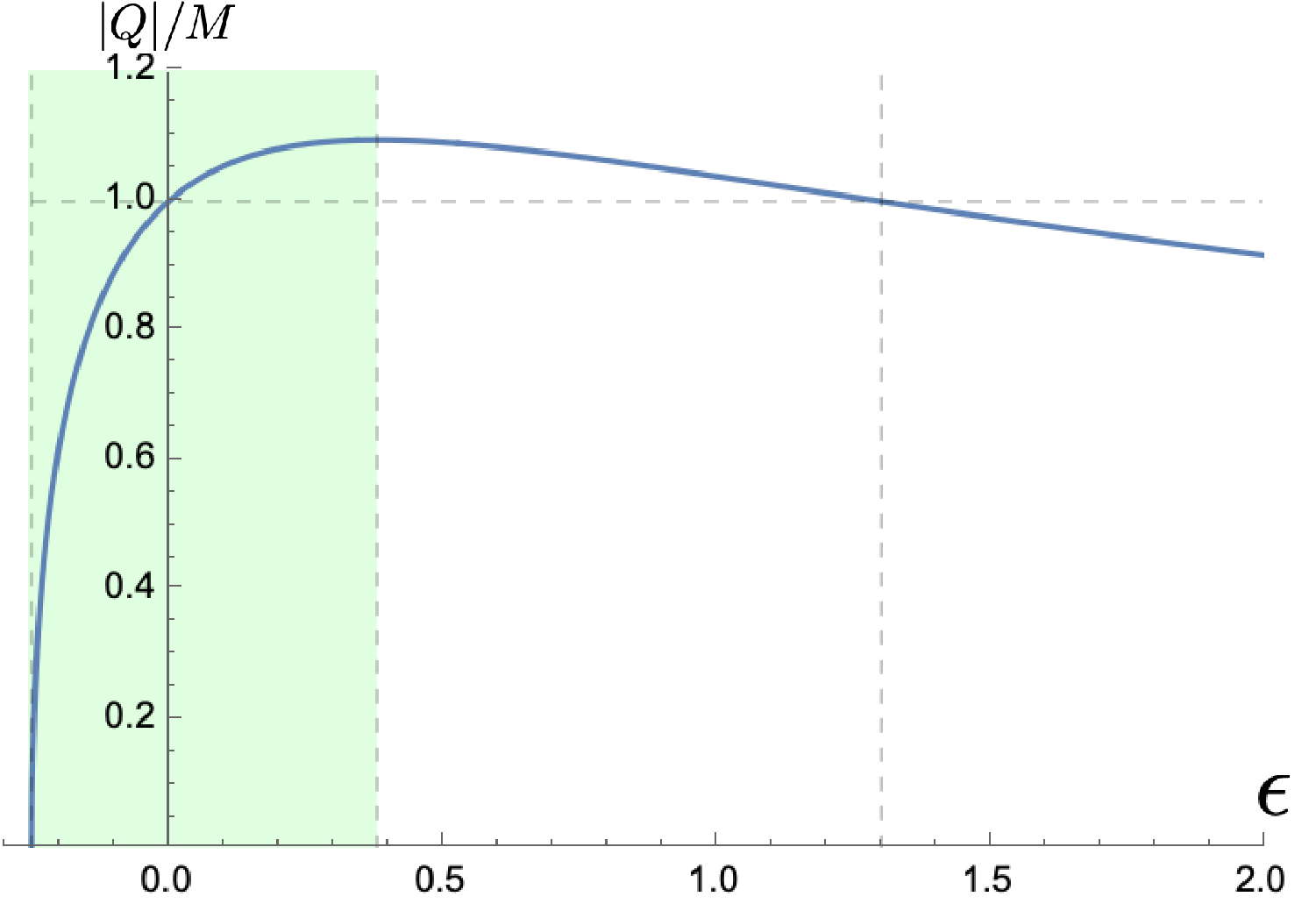}
\end{center}
\caption{Parametrisation of the radius of the Anti de-Sitter core (left) and mass-to-charge ratio (right) of the extremal Frolov-space-time as a function of the deformation parameter $\epsilon$ (see eq.~(\ref{FormFrolovQeta})). The green shaded region leads to a geometry with the correct signature outside of the event horizon. Notice, however, that $\epsilon>0$ leads to $|Q|>M$ and $\eta^2<0$, such that the latter is no longer the radius of a Anti de-Sitter core.}
\label{Fig:ExtremalFrolov}
\end{figure}

In the following, we shall be interested in the extremal Frolov space-time. Indeed, extremality ({\emph{i.e.}} $\frac{\dd f}{\dd r}(\zh)=0$) imposes a condition among the three parameters $(Q,M,\eta)$. In order to characterise this condition, we shall consider $\eta$ to be parametrically small, by introducing the parameter $\epsilon$ as in (\ref{DefSmall}) to capture small deformations of the Reissner-Nordström geometry
\begin{align}
&\epsilon:=\frac{M-\zh}{\zh}&&\text{such that}&&\zh=\frac{M}{1+\epsilon}\,,
\end{align}
where $\zh=M$ corresponds to the classical event horizon of the (extremal) Reissner-Nordstr\"om geometry. The conditions for a (deformed) extremal black hole are then solved by
\begin{align}
&\frac{|Q|}{M}=\frac{\sqrt{\epsilon+\sqrt{1+5\epsilon+5\epsilon^2}}}{1+\epsilon}\,,&&\frac{\eta^2}{M^2}=\frac{1+2\epsilon-\sqrt{1+5\epsilon+5\epsilon^2}}{(1+\epsilon)^3}\,.\label{FormFrolovQeta}
\end{align}
Reality of $Q$ furthermore requires $\epsilon\geq -\frac{1}{4}$, while $f>0$ for $r>\zh$ requires $\epsilon\leq\frac{5+\sqrt{105}}{40}$. These results are schematically shown in Figure~\ref{Fig:ExtremalFrolov}, where the green shaded region $\epsilon\in[-1/4,\frac{5+\sqrt{105}}{40}]$ denotes the parameter space that gives rise to an extremal black hole with the correct space-time signature outside of the event horizon. 
We remark that $\eta^2>0$ for $\epsilon\in[-1/4,0]$, which correspond to a horizon radius that is larger than its classical counterpart. Allowing $\eta^2<0$, we find $|Q|\geq M$ for $\epsilon\in\left[0,\frac{\sqrt{13}-1}{2}\right]$, thus circumventing the classical bound for the charge of the black hole, known as {\emph{super-extremal}}. We remark, however, that $\eta^2<0$, no longer represents the radius of an Anti de-Sitter core at the origin, which, however, is a priori not visible only from outside of the event horizon. Notice, also that $\epsilon=\frac{5+\sqrt{105}}{40}$ corresponds to the largest possible value of $|Q|/M$ and the smallest possible value of $\eta^2/M^2$ (\emph{i.e.} the extrema of the expressions in (\ref{FormFrolovQeta})). Furthermore we remark that $Q^2=2 M\,\zh$ for $\epsilon=3/4>\frac{5+\sqrt{105}}{40}$, which is outside of the green shaded region in Figure~\ref{Fig:ExtremalFrolov}, thus justifying the assumption made in Section~\ref{Sect:GeneralCharge}.

\begin{figure}[ht]
\begin{center}
\includegraphics[width=0.45\textwidth]{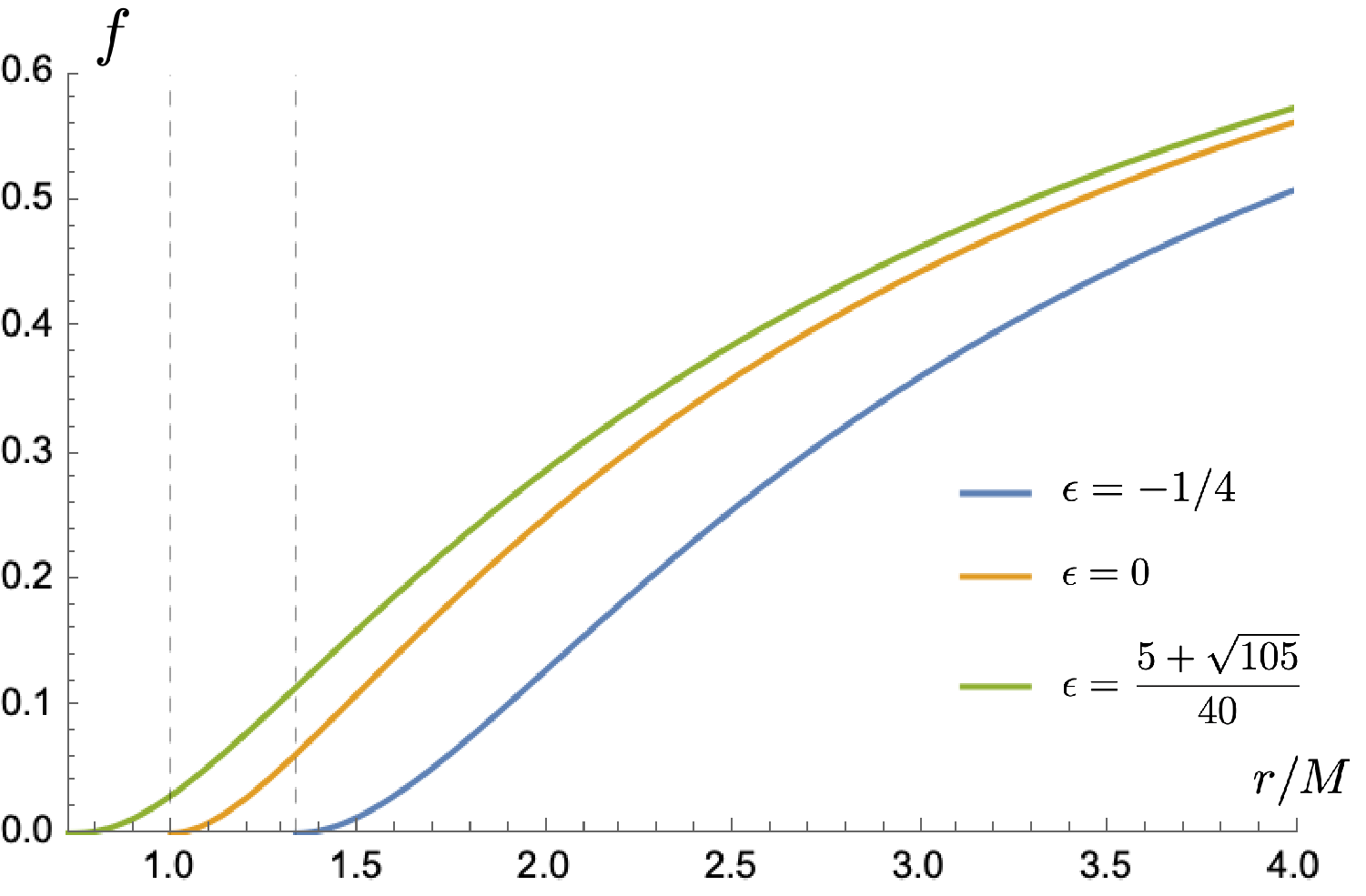} \hspace{0.0cm}\includegraphics[width=0.5\textwidth]{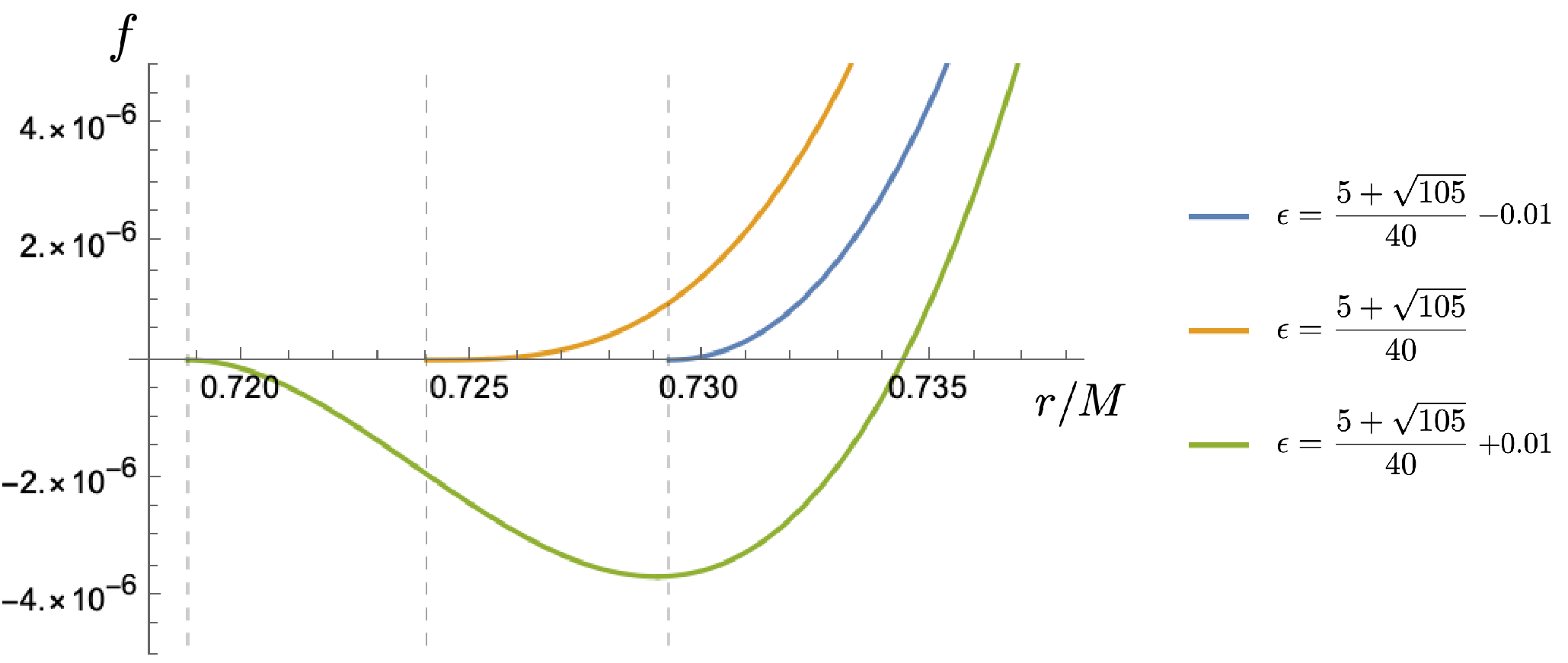}
\end{center}
\caption{Left: plot of the function $f$ outside of the horizon for different values of $\epsilon\in\left[-\tfrac{1}{4},\tfrac{5+\sqrt{105}}{40}\right]$ (\emph{i.e.} the green shaded region in Figure~\ref{Fig:ExtremalFrolov}). Notice that indeed each plot starts with a horizontal tangent, indicating the second order zero at the horizon. Right: for $\epsilon>\frac{5+\sqrt{105}}{40}$, $\zh=\frac{M}{1+\epsilon}$ no longer represents the outermost horizon of the geometry.}
\label{Fig:PlotsF}
\end{figure}

Figure~\ref{Fig:PlotsF} depicts the function $f$ for different values of $\epsilon$ outside of the 
event horizon. The right panel of this Figure indicates that indeed for $\epsilon>\frac{5+\sqrt{105}}{40}$, the function $f$ is negative outside of the double horizon located at $\frac{M}{1+\epsilon}$: moreover, it develops a further single horizon at a position $r> M$.

\begin{figure}[ht]
\begin{center}
\includegraphics[width=0.75 \textwidth]{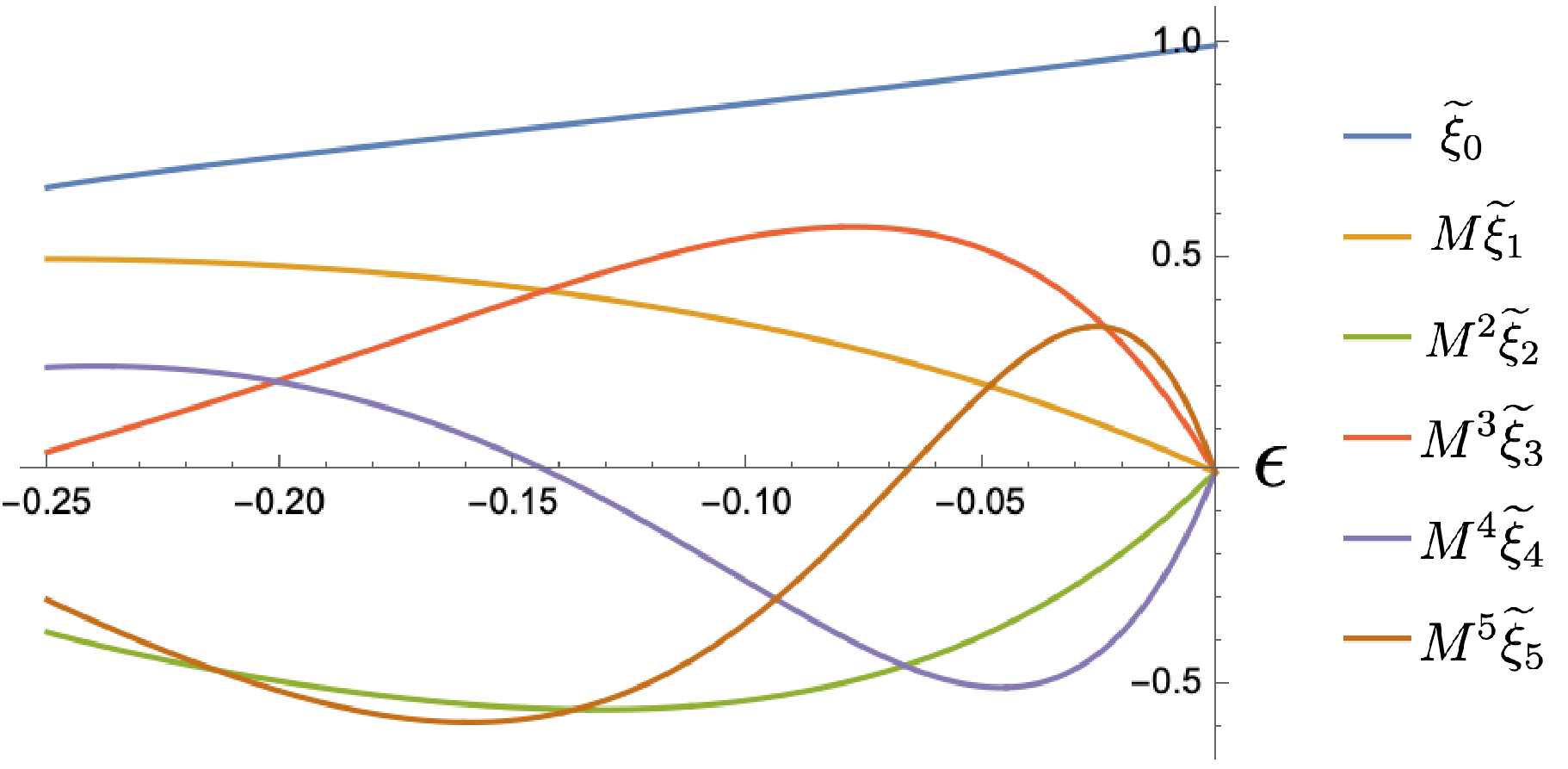}
\end{center}
\caption{Leading Coefficients $\tilxi{n}(\epsilon)$ for $\epsilon\in\left[-\tfrac{1}{4},0\right]$.}
\label{Fig:PlotsXiCoefs}
\end{figure}

Finally, we also provide the expressions for the first few coefficients $(\tilxi{n}=\tiltheta{n})_{n\in\mathbb{N}}$ as functions of $\epsilon$
\begin{align}
&\tilxi{0}=\frac{1}{2+\epsilon -\sqrt{1+5 \epsilon  (1+\epsilon )}}=1+\frac{3\epsilon}{2}+\mathcal{O}(\epsilon^2)\,,\nonumber\\
&\tilxi{1}=\frac{2(1+\epsilon)(1-\sqrt{1+5\epsilon(1+\epsilon)})}{(2+\epsilon-\sqrt{1+5\epsilon(1+\epsilon)})^2\,M}=-\frac{5\epsilon}{M}+\mathcal{O}(\epsilon^2)\,,\nonumber\\
&\tilxi{2}=\,\frac{2(1+\epsilon)^2(1+10\epsilon+8\epsilon^2-(1+2\epsilon)\sqrt{1+5\epsilon(1+\epsilon))})}{(2+\epsilon-\sqrt{1+5\epsilon(1+\epsilon)})^3\,M^2}=\frac{11\epsilon}{M^2}+\mathcal{O}(\epsilon^2)\,.
\end{align}
which are plotted (for $\epsilon\in[-1/4,0]$) in Figure~\ref{Fig:PlotsXiCoefs}.

In order to compare to the leading order computations in Section~\ref{DeformedComputations} (\emph{i.e.} for the case \emph{(i)} of the parameter space), we need to expand (\ref{FormFrolovQeta}) to obtain $\frac{|Q|}{M}=1+\frac{3\epsilon}{4}+\mathcal{O}(\epsilon^2)$, such that $q=\frac{3}{4}$ in (\ref{DeformEps}). For small enough values of $\epsilon$ (such that the higher orders in $\epsilon$ can indeed be neglected), this is compatible with (\ref{CondLeading}).

\section{Weak gravity conjecture}\label{sec:WGC-EMD}
The \emph{Swampland program}~\cite{Vafa:2005ui,Ooguri:2006in,Palti:2019pca,Agmon:2022thq} aims to delineate the subset of effective field theories (EFTs) that can arise from a consistent theory of quantum gravity, such as string theory. While effective theories are ubiquitous in particle physics and cosmology, only a small fraction of them can be consistently embedded in a UV-complete gravitational framework, the so-called \emph{string landscape}, whereas the remaining inconsistent ones belong to the \emph{Swampland}. The program formulates a series of \emph{Swampland conjectures}, such as the Distance conjecture and the de Sitter conjecture, providing quantitative criteria that low-energy EFTs must satisfy to admit a quantum-gravitational completion. These conjectures have profound implications for cosmology, black hole physics, and the nature of spacetime itself, offering a bridge between phenomenological models and the fundamental constraints imposed by quantum gravity.

The weak gravity conjecture \cite{Palti:2019pca, Harlow:2022ich,Antipin:2021rsh} states that the emission spectrum requires a particle with charge-to-mass ratio larger than the black hole's charge-to-mass ratio. In particular, in Planck units, the following relation holds:
\begin{align}
\label{CtMRatio}
    \frac{M}{q} \geq \left. 
    \left(\frac{\mathfrak{m}}{\mathfrak{q}} \right) \right|_\text{min} \ , 
\end{align}
where $\mathfrak{m}$ and $\mathfrak{q}$ are the mass and charge of the particle with the largest charge-to-mass ratio of the emission spectrum. Subsequent works in the literature have focused on refinements (or corrections) to (\ref{CtMRatio}), by studying different effects in (effective) quantum theories. As a recent example, we mention \cite{Antonelli:2025zwc}, which has analysed the impact of selected higher order operators in the effective action. To gain insights on $\tilxi{0}$ in our framework, we consider the decay of an electrically charged black hole. From the definition of horizon, the ratio between the black hole mass $M$ and its charge $q$ is
\begin{align}
    \frac{M}{q} = \frac{r_H^2 + q^2 \tilxi{0}}{2  \, q \,r_H \tilxi{0}} \ .
\end{align}
Close to the classical extremality condition, $\zh$ and the deformation function $\tilxi{0}$ take the following form:
\begin{align}
\label{NearExtremality}
    r_H = q (1+\gamma) \ , \quad \tilxi{0} =1 + \lambda \ ,
\end{align}
where $\lambda$ and $\gamma$ play the role of the quantum corrections, and $\gamma \sim \order{\lambda}$. 
Using equation \eqref{NearExtremality}, we can obtain the following upper limit:
\begin{align}
    \lambda \leq 2 - 2 \left. \left(\frac{\mathfrak{m}}{\mathfrak{q}}\right) \right| _{\text{min}} \ ,
\end{align}
which relates the inside of the black hole with its emission spectrum, and may allow to grasp some information of the interior structure of the black hole.

\section{Holographic quBTZ black hole}
The framework is tested against the non-rotating quantum BTZ black hole obtained from braneworld holography \cite{Emparan:2020znc,Emparan:2021hyr} ,
\begin{equation}
f(r) = g(r) = \frac{r^{2}}{\ell_3^{2}} - F_1 - \frac{\ell F_2}{r} \, .
\end{equation}
The extra $1/r$ term is incompatible with the analyticity condition at the origin derived in Eq.~\ref{EMD-BTZ-origin}, consistent with the known essential singularity of quBTZ at $r=0$; the origin expansion therefore does not apply, while the horizon remains regular. Matching gives
\begin{equation}
\Phi(d(r)) = \Psi(d(r)) = 1 - 8\,\frac{\nu\ell_3^{3}z^{4}(1+z^{2})(1+\nu z)^{2}}{r^{3}(1+3z^{2}+2\nu z^{3})^{3}} \, ,
\end{equation}
and substitution into the EMD temperature formula reproduces the known quBTZ result,
\begin{equation}
T = \frac{z}{2\pi\ell_3}\,\frac{2+3\nu z+\nu z^{3}}{1+3z^{2}+2\nu z^{3}} \, ,
\end{equation}
providing a non-trivial cross-check of the formalism.

\section{Computing the proper distance}
To showcase the approach developed in Section~\ref{Sect:DistanceFunctionExpansion} to compute the proper distance as a function of the deformation parameters, we consider three simple examples, corresponding to different choices of the coefficients $\xi_n$.

\subsubsection{Schwarzschild distance}
\begin{figure}[htbp]
\begin{center}
\includegraphics[width=8cm]{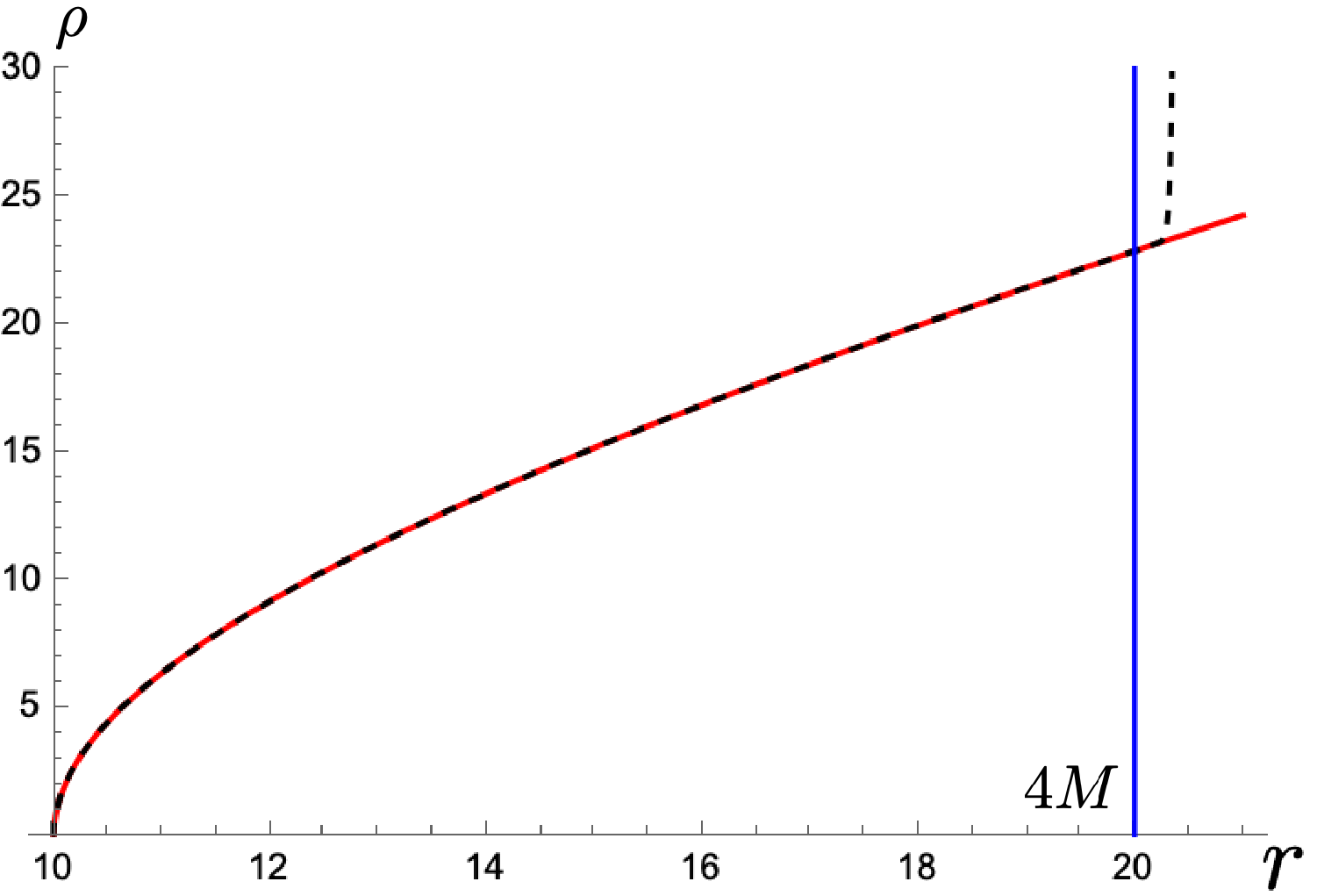}
\end{center}
\caption{Comparison of the series expansion (\ref{SeriesDistance}) up to order $\mathcal{O}((r-\zh)^{500-1/2})$ (dashed black line) to the analytic result (\ref{AnalyticRhoSchwarzschild}) (red curve) for $M=5$. The blue line $4M=20$ denotes the boundary of the interval of convergence for the series expansion.}
\label{Fig:SchwarzschildExDist}
\end{figure}

The simplest choice is to set $r_H =2M$ and so $\xi_0=1$ and $\xi_n=0$ $\forall n>0$, which corresponds to the Schwarzschild black hole (\emph{i.e.} $\Phi=2M$). In this case, we also have $\dho=\pi M$. Using (\ref{a_2-exactsol}) and (\ref{Coeffsa}), the first few coefficients $a_{2n}$ (and their reversions $b_n$) can be tabulated as follows 
\begin{center}
\begin{tabular}{|c||c|c|c|c|c|c|}\hline
  & $n=1$ & $n=2$  & $n=3$ & $n=4$ & $n=5$ & $n=6$ \\ \hline\hline
 &&&&&&\\[-8pt]
 $a_{2n}$ & $\tfrac{1}{8}$ & $-\tfrac{1}{48}$ & $\tfrac{11}{2880}$ & $-\tfrac{73}{80640}$ & $\tfrac{887}{3628800 M^9}$ & $-\tfrac{136883}{1916006400}$ \\[6pt]\hline
 &&&&&&\\[-8pt]
 $b_{2n-1}$ & $2$ & $\tfrac{1}{3}$ & $-\tfrac{1}{20}$ & $\tfrac{1}{56}$& $-\tfrac{5}{576}$ & $\tfrac{7}{1408}$ \\[6pt]\hline
\end{tabular}
\end{center}
which allows to compute a series expansion of the distance function, as in eq.~(\ref{SeriesDistance}). These coefficients have previously been obtained in \cite{Meissel,Foong_2008} (see also \cite{Peetre1997ErnstMA}) in a different context, namely solutions for free-falling bodies in Newtonian gravity. We have verified up to $(r-2M)^{500-1/2}$ that these coefficients follow the pattern
\begin{align}
b_{2n-1}=\frac{(-1)^n(2n-5)!!}{2^{n-2}(2n-1)(n-1)!}\,,
\end{align} 
such that the series expansion of $\rho$ in terms of $(r-\zh)^{1/2}$ has an interval of convergence of $r\in [0,4M]$. A graphical example (for $M=5$) is shown in Figure~\ref{Fig:SchwarzschildExDist}, with an expansion up to order $(r-\zh)^{500-1/2}$. We have furthermore verified that the coefficients $b_{n}$ agree with a series expansion of $\rho(r)$: indeed, in the case of the Schwarzschild geometry, the  distance $\rho$ can in fact be computed in closed form as a function of $r$ (see eq.~(\ref{SchwarzschildDistance}))
\begin{align}
&\rho(r)=\sqrt{r(r-2M)}+2M\, \text{arctanh}\sqrt{1-\frac{2M}{r}} \qq{for} r>2M\,,\label{AnalyticRhoSchwarzschild}
\end{align}
allowing us to verify whether (\ref{SeriesDistance}) is indeed a good representation of the distance function (see Figure~\ref{Fig:SchwarzschildExDist}).

\subsubsection{Bonanno-Reuter black hole}\label{Sect:BRDistanceComputations}
\begin{figure}[htbp]
\begin{center}
\includegraphics[width=8cm]{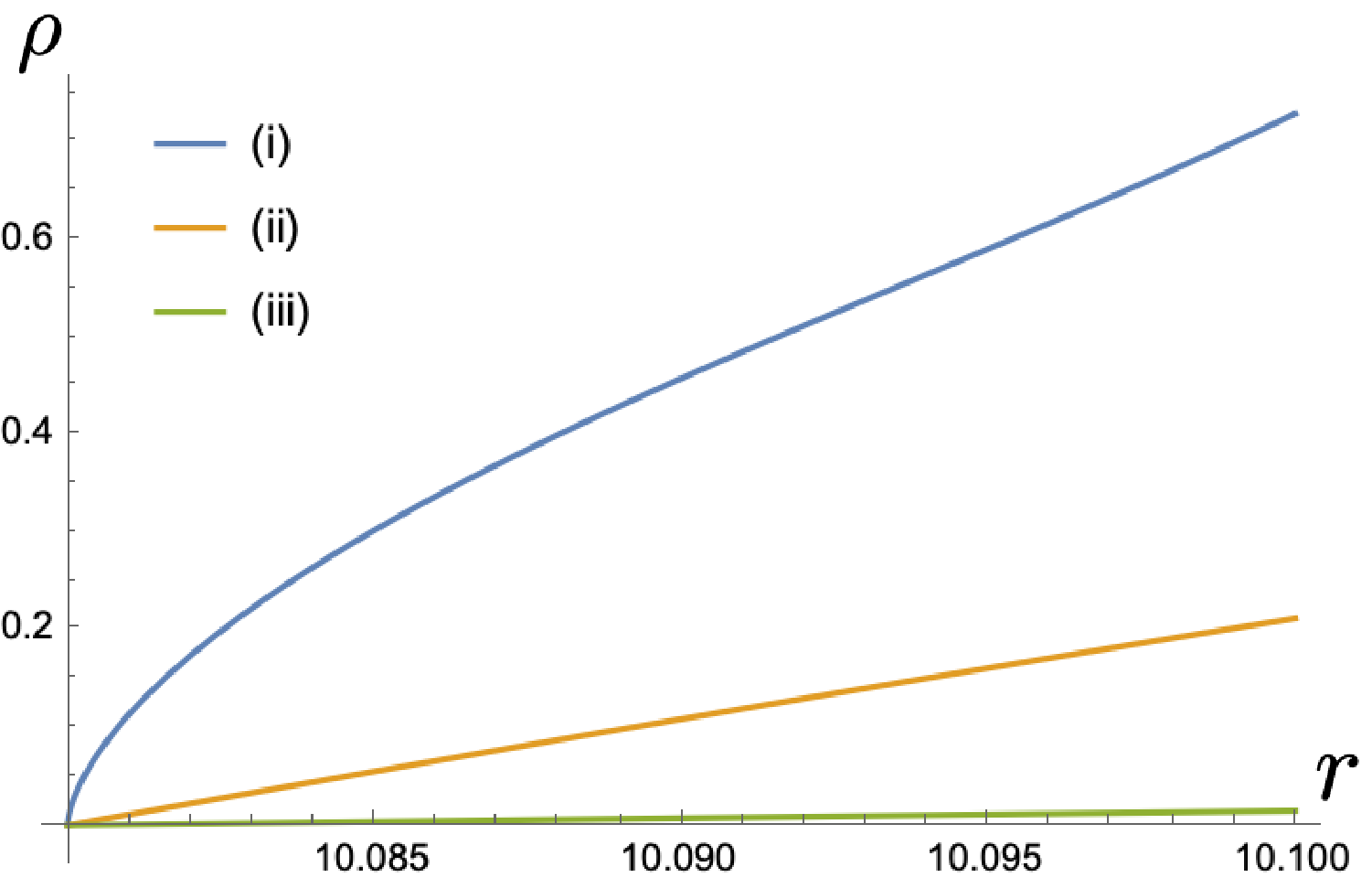}
\end{center}
\caption{Comparison of three different distance functions for the Bonanno-Reuter space-time: \emph{(i)} $\rho_{\text{BR}}$ (represented by the expansion (\ref{BRSeriesExpansion})), \emph{(ii)} the Schwarzschild proper distance $\rho_{\text{S}}=d_{\text{S}}-\dho$ and \emph{(iii)} the approximating function $\mathfrak{d}_{\text{S}}-\dho$ in eq.~(\ref{BRdistanceApprox}). Here we have chosen $\widetilde{\omega}=-1$, $M=5$ and $\zh=10.08$ and $\dho=12.96$.}
\label{Fig:BRdistance}
\end{figure}

For the choice (\ref{BRmetricfunction}) of the metric deformation (along with $f=h$), the coefficients $\xi_n$ in the expansion (\ref{SeriesDistanceRho}) are given by
\begin{align}
\xi_0&=\frac{\dho^2}{\widetilde{\omega}+\dho^2}\,,\nonumber\\
\xi_n&=\frac{(-1)^n\,(2M)^n}{(\widetilde{\omega}+\dho^2)^{n+1}}\,\sum_{k=1}^{\lfloor\frac{n}{2}+1\rfloor}(-1)^k\,\left(\begin{array}{c}n+1 \\ 2k-1\end{array}\right)\,\dho^{n+2-2k}\,\widetilde{\omega}^k\,,\ \forall n\geq 1\,.
\end{align}
Here the interval of convergence (for $\widetilde{\omega}<0$) is given by $\rho\in\left[0,\frac{|\dho^2+\widetilde{\omega}|}{\dho+\sqrt{-\widetilde{\omega}}}\right)$. Moreover, since $\xi_1=\frac{4M \widetilde{\omega}\,\dho}{(\widetilde{\omega}+\dho^2)^2}\neq 0$, the results of Section~\ref{Sect:DistanceFunctionExpansion} cannot be directly applied. However, as explained in Appendix~\ref{App:SerNonZeroXi}, for $\widetilde{\omega}<0$ (such that $\xi_1<0$), this approach can be adapted, leading to a series expansion (\ref{SeriesOdd}) with coefficients $\ha{n}$ in (\ref{LowSolutionsHa}) and (\ref{SolutionsHa}). This expansion provides a solution of the differential equation (\ref{DiffEqDistance}) (albeit with a divergent first derivative $\fn{1}$ of $f$ at the horizon). Inversion of the series (\ref{SeriesOdd}) leads to an expansion of the proper distance to the horizon of the form
\begin{align}
\rho_{\text{BR}}(r)=\sum_{n=0}^\infty\,\widehat{b}_n\,(r-\zh)^{\frac{2+n}{3}}\,,\label{BRSeriesExpansion}
\end{align}
where for concreteness, we provide explicitly the first few coefficients
\begin{align}
&\widehat{b}_0=\frac{1}{\widehat{a}_1^{2/3}}=\frac{3^{2/3} \zh^{1/3}(\dho+\widetilde{\omega})^{2/3}}{2^{4/3}(-\dho M \widetilde{\omega})^{1/3}}\,,&&\widehat{b}_1=\frac{2\widehat{a}_2}{3\widehat{a}_1^2}=\frac{(\dho+\widetilde{\omega})^2}{16\dho M \widetilde{\omega}}\,,&&\widehat{b}_2=\frac{7\widehat{a}_2^2-6\widehat{a}_1\widehat{a}_3}{9 \widehat{a}_1^{10/3}}\,.
\end{align}
A numerical plot of the three different distance functions used in Section~\ref{Sect:ExampleBonannoReuter} for the Bonanno-Reuter black hole is shown in Figure~\ref{Fig:BRdistance}.

\subsubsection{Minimal example}
As a last example, we consider the metric function characterised by the minimal choice (\ref{Phifuns}), leading to the function $\Phi$ in eq.~(\ref{FormExpPhiExample}). 
The first few $\xi_n$ read explicitly
\begin{align}
&\xi_0=\,e^{-\frac{3\phi_2}{2\dho^2}}\,,&&\xi_1=0\,,&&\xi_2=\frac{(2M)^2}{\dho^4}\,\phi_2\,e^{-\frac{3\phi_2}{2\dho^2}}\,,&&\xi_3=0\,, \label{CoefsMinModel}
\end{align}
which satisfy the conditions (\ref{ConditionsfhRegular}), as expected. The leading coefficients $a_n$ in the expansion of $r(\rho)$ in (\ref{SeriesDistance}) therefore become
\begin{align}
&a_2=\frac{M(1+\rootk)}{4\zh}\,,&&a_3=-\frac{2M \zh \phi_2}{\dho^4(1+3\rootk)}\,,&&a_4=-\frac{8M^3}{1+2\rootk}\left(\frac{(1+\rootk)^3}{256\zh^3}+\frac{9\zh^3 \phi_2^2}{\dho^8(1+3\rootk)^2}\right)\,,
\end{align}
where we have again used the shorthand notation $\rootk=\sqrt{1-8\zh\xi_2/M}$. The series inversion yields the following coefficients for the distance function (\ref{SeriesDistance})
\begin{align}
&b_1=\frac{2\sqrt{\zh}}{\sqrt{M(1+\rootk)}}\,,\qquad b_2=\frac{16\zh^3\phi_2}{M\dho^4(1+\rootk)^2(1+3\rootk)}\,,\notag \\
&b_3=\sqrt{\frac{M}{\hr}}\frac{\dho^8(1+\rootk)^4(1+3\rootk)^3+256\zh^6(19+29\rootk)\phi_2}{2\dho^8(1+\rootk)^{7/2}(1+2\rootk)(1+3\rootk)^2}\,.
\end{align}

%% file: subfiles/5-Shadow.tex
\typeout{START baseline=\the\baselineskip, jot=\the\jot}
\chapter{Stepping away from the abyss}\label{chap4}
The EMD of Chapter~\ref{chap3} is, by construction, a near-horizon description. The deformation functions are specified through their Taylor coefficients at $\rho=0$, and the regularity conditions derived there are statements about the geometry in an arbitrarily small neighbourhood of the horizon. This is precisely what is needed for horizon quantities such as the Hawking temperature, but it is not sufficient for many observables of astrophysical interest. The photon sphere, the black hole shadow and the quasi-normal modes in the eikonal regime are all determined by the geometry at a finite distance from the horizon, where the near-horizon series may converge slowly or not at all.

In this Chapter, which reviews and extends the results of~\cite{DelPiano:2024nrl}, we show how this limitation can be overcome. The key tool is the Padé approximant (see Appendix~\ref{App:DefPadeApproximant}): a rational function that reproduces a prescribed number of Taylor coefficients, but typically remains a good approximation well beyond the radius of convergence of the underlying series. We first explain how Padé approximants of the metric and of the proper distance can be organised such that the physical meaning of the EMD coefficients is preserved (Sec.~\ref{Sect:PadeMetric}). The same idea, applied to the effective potential of massless particles, then leads to a compact and model-independent expression for the leading deformation of the black hole shadow (Sec.~\ref{Sect:PadePhoton}). The method is tested on several concrete geometries in Sec.~\ref{subs example models}, and the convergence assumption on which it relies is examined in Sec.~\ref{Sect:Convergence}. The extension to the eikonal quasi-normal modes is discussed in Chapter~\ref{chap5}.

\section{Extending the EMD beyond the local limit}\label{Sect:ReviewMetric}
We work with the proper-distance EMD of Sec.~\ref{Sect:ReviewDistance}. The line element is \eqref{SpStMetric}, with metric functions \eqref{h-f-emd-def} whose deformation functions $\Psi$ and $\Phi$ depend on the proper distance $\rho\geq 0$ from the (simple) horizon at $r=\zh$. Since we are interested in geometries that are smooth at the horizon, we impose the conditions of Result~\ref{res:SmoothHorizon}. All odd coefficients then vanish, and the deformation functions take the form
\begin{equation}
\Psi(\rho)=\zh + 2\ma\sum_{n=1}^\infty \theta_{2n}\,\left(\frac{\rho}{2M}\right)^{2n}  \qq{and} \Phi(\rho)=\zh + 2\ma\sum_{n=1}^\infty \xi_{2n}\,\left(\frac{\rho}{2M}\right)^{2n} \ ,
\label{SeriesExpansionMetric}
\end{equation}
where $\ma$ is the ADM mass (in units of the Planck mass) seen by an observer at spatial infinity. The coefficients $\{\theta_{2n}\}_{n\in\mathbb{N}}$ and $\{\xi_{2n}\}_{n\in\mathbb{N}}$, together with $\zh$, are the effective parameters of the black hole. Where convenient, we also use the shorthand $\xi_0=\frac{\zh}{2\ma}=\theta_0$, and it is understood that $\xi_{2n+1}=\theta_{2n+1}=0$ for all $n\geq 0$. The self-consistency condition \eqref{DiffEqDistance} is solved by the series \eqref{SeriesDistance}, with the coefficients derived in Sec.~\ref{Sect:DistanceFunctionExpansion}.

Two further representations of the metric functions, which are equivalent wherever they converge, will be useful. The first one is the Taylor expansion \eqref{f-h-taylor-r} in powers of $r-\zh$, whose coefficients $\{\fn{n}\}_{n\in\mathbb{N}}$ and $\{\hn{n}\}_{n\in\mathbb{N}}$ are functions of $\{\xi_{2n}\}_{n\in\mathbb{N}}$ and $\{\theta_{2n}\}_{n\in\mathbb{N}}$. For the lowest orders one finds
\begin{subequations}\label{emd nh fh1 hh1 hh2}
\begin{align}
\fn{1}&=\frac{1+\varpi}{2\hr} \ , \label{f1h emd nh}\\
\hn{1}&=\frac{1}{\hr}-\frac{4\theta_2/M}{1+\varpi}\ ,\label{h1h emd nh}\\
\hn{2}&=-\frac{2}{\hr^2}+\frac{(\varpi +1) \left[4\,\theta_2  (\varpi +1) (3 \varpi +1)\hr/M-
   16\, \theta_4 (2 \varpi +1)\hr^3/M^3\right]-64  \,\theta_2 \,\xi_4\, \zh ^4/M^4 }{\zh ^2 (\varpi +1)^3 (2 \varpi +1)}\ ,
\end{align}
\end{subequations}
where, as in Eq.~\eqref{a_2-exactsol},
\begin{equation}
    \varpi \coloneqq \sqrt{1-\frac{8\hr \,\xi_2}{M}} \ .
\end{equation}
The second representation is the expansion \eqref{fhSerRho} in powers of $\rho$, which is obtained with the help of \eqref{inverse-z-rho-exp}.

\begin{center}
\remarkbox{\rem\label{rem:TriangularStructure}{Triangular structure of the horizon derivatives}}{\parbox{16cm}{
The map between the EMD coefficients and the horizon derivatives of the metric is triangular. The $n$-th derivative of $f$ at the horizon depends linearly on $\xi_{2n}$ and non-linearly on the lower-order parameters,
\begin{equation}
   \left(\fn{n}- (\fn{n})_{\rm class} \right) \, \propto \,  \xi_{2n}+\mathrm{n.l.}(\hr,\xi_2,\ldots,\xi_{2n-2}) \ ,\nonumber
\end{equation}
where $(\fn{n})_{\rm class}$ is the classical value (for the Schwarzschild black hole, $\hr=2M$ and $(\fn{n})_{\rm class} = (-1)^{n+1}\,n!\,(2M)^{-n}$) and $\mathrm{n.l.}$ denotes a non-linear function of $\hr$ and $\{\xi_{2p}\}_{1 \leq p < n}$. The same holds for $\hn{n}$ with respect to $\theta_{2n}$ and $(\hr,\{\theta_{2p}\}_{1 \leq p < n})$. Consequently, knowledge of the first $n$ horizon derivatives is equivalent to knowledge of the first $n$ even EMD coefficients, and a truncation at a given order is meaningful in both descriptions.
}}
\end{center}

\noindent
Both expansions (\ref{f-h-taylor-r}) and (\ref{fhSerRho}) generically have a finite radius of convergence around the horizon. They are indispensable for solving (\ref{DiffEqDistance}) self-consistently, and thus for determining the geometry, but they cannot in general be used to evaluate observables at a finite distance from the black hole. Even close to the horizon, their convergence may be too slow for practical purposes.\footnote{Quantities that only depend on the immediate vicinity of the horizon, such as the Hawking temperature, can instead be computed very efficiently from (\ref{SeriesExpansionMetric}), (\ref{f-h-taylor-r}) and (\ref{fhSerRho}), as shown in Chapter~\ref{chap3}.} In the following we therefore construct approximations of the metric that do not suffer from these limitations. The observable we have in mind is the radius of the photon sphere reviewed in Sec.~\ref{Sect:PhotonRadius}, which determines the black hole shadow and, as shown in Chapter~\ref{chap5}, the eikonal quasi-normal modes.

\begin{center}
\remarkbox{\rem\label{rem:PadeNotation}{Notation: Padé orders and the mass}}{\parbox{16cm}{
Following Appendix~\ref{App:DefPadeApproximant}, $U_{N,M}$ denotes the Padé approximant of a function $U$ with a numerator of degree $N$ and a denominator of degree $M$. Whenever $M$ appears as a subscript of an approximant, in a pair $(N,M)$ labelling an order of approximation, or in combinations such as $\lfloor\frac{M+N}{2}\rfloor$, it refers to this degree. In all other expressions, $M\equiv\ma$ is the ADM mass of the black hole.
}}
\end{center}

\noindent
Before constructing the approximations, we specify how the size of the deformation is organised. Since $\Psi$ and $\Phi$ encode corrections to GR, they are expected to be small in an appropriate sense. Indeed, many models contain a dedicated parameter that controls the size of the deformation (called $\eta$ in Sec.~\ref{subs example models} below) and in terms of which approximations can be organised. To make contact with such models, we write the horizon radius as
\begin{align}
&\zh=:2\ma\left(1+\mathfrak{c}\right)\ ,\label{RescaledMetricFunctions}
\end{align}
and treat $\mathfrak{c}$ as well as the coefficients $\{\theta_{2n}\,,\xi_{2n}\}$ as small deformations. For definiteness, and in order to illustrate the conceptual points below, we assume the scaling
\begin{align}
&\mathfrak{c}\ll 1&&\text{and}&&\frac{\theta_{2n}}{\mathfrak{c}}\gg \mathfrak{c}\,,&&\text{and} &&\frac{\xi_{2n}}{\mathfrak{c}}\gg \mathfrak{c}\,,&&\forall n\geq 1\,,\label{IllumScaling}
\end{align}
\emph{i.e.} all deformation coefficients are at least of the same order as $\mathfrak{c}$. Assuming observables to be analytic functions of $\mathfrak{c}$, we present them as expansions in powers of $\mathfrak{c}$ and mostly focus on the leading correction $\mathcal{O}(\mathfrak{c})$, which we call the \emph{leading quantum correction}.

As the examples of Sec.~\ref{subs example models} will show, the photon sphere at $\rps>\zh$ typically lies in a region where the series (\ref{SeriesExpansionMetric}) and (\ref{SeriesDistance}) either do not converge or converge very slowly. They can therefore not be used directly to solve the photon sphere condition (\ref{rpheq}), nor to compute the impact parameter and the shadow $\bsh$. This is where Padé approximants (see Appendix~\ref{App:DefPadeApproximant}, see also~\cite{Maharana:2025lbm}) enter. An important ingredient in what follows is that each approximation comes with an \emph{order}, defined as the number of EMD coefficients $\{\xi_{2n},\theta_{2n}\}_{n\in\mathbb{N}}$ that it treats as genuine and independent input. This order is directly linked to the order of the Padé approximant.

\subsection{Padé Approximant of the Metric and Horizon Distance}\label{Sect:PadeMetric}
The most direct way to extend the metric beyond the convergence domain of (\ref{SeriesExpansionMetric}) is to replace the deformation functions by functions that are defined for all $\rho>0$, but still reproduce $\Psi$ and $\Phi$ accurately close to the horizon. Concretely, we approximate the deformation functions by their Padé approximants
\begin{align}
&\Psi(\rho)\sim\zh+\Psi_{N,M}(\rho)\,,&&\text{and} && \Phi(\rho)\sim\zh+ \Phi_{N,M}(\rho)\,.\label{PadeApproximation}
\end{align}
where $\Psi_{N,M}$ and $\Phi_{N,M}$ are the Padé approximants of $\Psi-\zh$ and $\Phi-\zh$, regarded as functions of $\rho/2\ma$. By construction, their expansions around $\rho=0$ reproduce the exact coefficients up to order $N+M$. All higher coefficients, instead, are replaced by functions of the lower ones,
\begin{align}
&\theta_{2n}\longrightarrow \widetilde{\theta}_{2n}(\theta_2,\ldots,\theta_{2L})&&\text{and}&&\xi_{2n}\longrightarrow \widetilde{\xi}_{2n}(\xi_2,\ldots,\xi_{2L})\,,&&\forall n>L:=\left\lfloor\frac{M+N}{2}\right\rfloor\,.\label{ReplacementCoefficients}
\end{align}
In this sense, a Padé approximation of order $(N,M)$ treats $\theta_2,\ldots,\theta_{2L}$ and $\xi_2,\ldots,\xi_{2L}$ as the physically significant input, while the replacement (\ref{ReplacementCoefficients}) is interpreted as the approximation.

The approximation (\ref{PadeApproximation}) also affects the proper distance. According to (\ref{SeriesDistance}) and~\eqref{Coeffsa}, the coefficient $a_{2n}(\xi_2,\ldots,\xi_{2n})$ does not depend on $\xi_{2m}$ with $m>n$. The same is true for the coefficients $b_{2n-1}(a_2,\ldots,a_{2n})$ in (\ref{SeriesDistance}) with respect to $a_{2m}$. Replacing the deformation function therefore leaves the first $L$ coefficients of the distance unchanged and modifies the higher ones, which leads to a new solution $\widetilde{\rho}$ of (\ref{DiffEqDistance}) of the form
\begin{align}
&\widetilde{\rho}(\co)=2M\sum_{k=1}^{L}b_{2k-1}\,\left(\frac{\co-\zh}{2M}\right)^{\frac{2k-1}{2}}+2M\sum_{k=L+1}^{\infty}\widetilde{b}_{2k-1}\,\left(\frac{\co-\zh}{2M}\right)^{\frac{2k-1}{2}}\ ,\label{ExactApproximation}
\end{align}
The coefficients $\widetilde{b}_{2k-1}(\xi_2,\ldots, \xi_{2L})$ are implicit functions of the input parameters $\{\xi_2,\ldots,\xi_{2L}\}$ and follow by series reversion from
\begin{align}
\co=\zh+2M\sum_{k=1}^{L} a_{2k}\,\left( \frac{\rho}{2M}\right)^{2k}+2M\sum_{k=L+1}^{\infty} \widetilde{a}_{2k}\,\left( \frac{\rho}{2M}\right)^{2k}\ ,
\end{align}
where the modified coefficients obey the recursion
\begin{align}
\widetilde{a}_p&=\frac{\widetilde{\xi}_p+\zh/2M\sum_{n=3}^{p-1}(p-n+2)\,n\,\widetilde{a}_n\,\widetilde{a}_{p-n+2}+\sum_{n=2}^{p-2}\sum_{m=2}^n(n-m+2)\,m\,\widetilde{a}_{p-n}\,\widetilde{a}_m\,\widetilde{a}_{n-m+2}}{1-2\,p\,a_2\,\zh\,/M} \ ,\label{ModCoeffsA}
\end{align}
for $p > 2L$, with $\widetilde{\xi}_{p} = \xi_p$ and $\widetilde{a}_{k}=a_{k}$ for $k\leq 2L$ (the latter given in (\ref{Coeffsa})). Equivalently, $\widetilde{\rho}$ is the exact solution of the differential equation with the approximated metric function,
\begin{align}
\dv{\widetilde{\rho}}{\co}=\frac{1}{\sqrt{\widetilde{f}(\co)}} \qq{with} \widetilde{f}(\co)=1-\frac{\zh+\Phi_{N,M}}{\co}\ .\label{ReducedDiff}
\end{align}
For $k>L$ one generally has $a_{2k}\neq \widetilde{a}_{2k}$, since $\widetilde{\xi}_{2k'}\neq \xi_{2k'}$ for $L<k'\leq k$. If the differences $|a_{2k}-\widetilde{a}_{2k}|$ are small, which depends on the $\{\xi_{2n}\}_{n\in\mathbb{N}}$ and will be examined for several models in Section~\ref{subs example models}, then (\ref{ExactApproximation}) is a good approximation of the exact proper distance (\ref{SeriesDistance}). However, (\ref{ExactApproximation}) is still a series in powers of $\sqrt{\co-\zh}$, and it need not converge better than the original series (\ref{SeriesDistance}). We therefore introduce a second approximation, namely the Padé approximant of the distance function itself
\begin{align}
\widehat{\rho}:=\rho_{N,M}(\sqrt{\co-\zh})=2M\sum_{k=1}^{L}b_{2k-1}\,\left(\frac{\co-\zh}{2M}\right)^{\frac{2k-1}{2}}+2M\sum_{k=L+1}^{\infty}\widehat{b}_{2k-1}\,\left(\frac{\co-\zh}{2M}\right)^{\frac{2k-1}{2}}\ ,\label{PadeDistance}
\end{align}
It is determined by the coefficients $\{b_1,\ldots,b_{2L-1}\}$, hence by $\{a_2,\ldots,a_{2L}\}$ and ultimately by $\{\xi_2,\ldots,\xi_{2L}\}$. Like $\widetilde{\rho}$, the approximant $\widehat{\rho}$ carries no information about $\xi_{2k}$ with $k>L$, and in this sense it is an approximation of the same order as (\ref{ExactApproximation}). In general, however, $\widetilde{b}_{2k-1}\neq \widehat{b}_{2k-1}$ for $k>L$, and $\widehat{\rho}$ solves (\ref{DiffEqDistance}) and (\ref{ReducedDiff}) only up to order $(\co-\zh)^{\frac{2L-1}{2}}$. The simplest non-trivial case is worked out in Computation~\ref{comp:Pade22}.

\begin{center}
\compbox{\comp\label{comp:Pade22}{Padé approximants of order $(2,2)$}}{\parbox{16cm}{
For $N=M=2$ the approximants of the deformation functions are
\begin{align}
&\Psi_{2,2}(\rho)=2\ma\,\frac{\theta_2(\rho/2M)^2}{1-\frac{\theta_4 (\rho/2M)^2}{\theta_2}} \ , &&\Phi_{2,2}(\rho)=2\ma\,\frac{\xi_2(\rho/2M)^2}{1-\frac{\xi_4 (\rho/2M)^2}{\xi_2}}\ ,\nonumber
\end{align}
which follow from the determinant formula of Appendix~\ref{App:DefPadeApproximant}. Their expansion shows that the replacement (\ref{ReplacementCoefficients}) takes the form
\begin{align}
&\widetilde{\theta}_{2n}=\frac{\theta_4^{n-1}}{\theta_2^{n-2}}\,,
&&\text{and} &&\widetilde{\xi}_{2n}=\frac{\xi_4^{n-1}}{\xi_2^{n-2}}\,,&& \forall n>2\,. \label{exampleP22}
\end{align}

For $\theta_4/\theta_2>0$ ($\xi_4/\xi_2>0$) the function $\Psi_{2,2}$ ($\Phi_{2,2}$) has a pole at $(\rho/2M)^2=\theta_2/\theta_4$ ($(\rho/2M)^2=\xi_2/\xi_4$), and it is well defined elsewhere for $\rho\geq 0$. Accordingly, the Taylor series of $\Psi_{2,2}$ and $\Phi_{2,2}$ in the variable $\rho/2M$ have radii of convergence $\sqrt{|\theta_2/\theta_4|}$ and $\sqrt{|\xi_2/\xi_4|}$, respectively. For the coefficients of the radial coordinate one finds from (\ref{ModCoeffsA})
\begin{align}
&a_2=\frac{M(1+ \varpi)}{4 \zh}\,,&&a_4=-\frac{\xi_4+\frac{\left(1+\varpi\right)^3M^3}{16 \zh^3}}{1+2\varpi}\,,&&\text{and}&&\widetilde{a}_{2n}=\frac{a_4^{n-1}}{a_2^{n-2}}\hspace{0.5cm}\forall n>2\,,\nonumber
\end{align}
and the Padé approximant (\ref{PadeDistance}) of the distance reads
\begin{align}
&\widehat{\rho}\,(\co)=\rho_{2,2}(\co)=2M\frac{b_1 (\frac{\co-\zh}{2M})^{1/2}}{1-\frac{b_3}{b_1}\,\frac{\co-\zh}{2M}}\,,&&\text{with} &&b_1=\frac{1}{\sqrt{a_2}}\,,\quad b_3=-\frac{a_4}{2a_2^{5/2}}\ .\nonumber
\end{align}
}}
\end{center}

\noindent
To quantify how much $\widehat{\rho}$ deviates from $\rho$, we expand the coefficients $\widehat{b}_{2n-1}$ and $b_{2n-1}$ in powers of $\mathfrak{c}$, using the scaling (\ref{IllumScaling}). For the first coefficient beyond the input order one finds
\begin{subequations}
\begin{align}
\widehat{b}_5&=\frac{1}{18}+\frac{1}{108}\left(-9+28 \,\frac{\xi_2}{\mathfrak{c}}+192\,\frac{\xi_4}{\mathfrak{c}}\right)\,\mathfrak{c}+\mathcal{O}(\mathfrak{c}^2)\ , \\
b_5&=-\frac{1}{20}+\frac{1}{5}\left(\frac{3}{8}+\frac{7}{90}\,\frac{\xi_2}{\mathfrak{c}}+\frac{56}{3}\,\frac{\xi_4}{\mathfrak{c}}+64\,\frac{\xi_6}{\mathfrak{c}}\right)\,\mathfrak{c}+\mathcal{O}(\mathfrak{c}^2)\,.
\end{align}
\end{subequations}
The two expressions differ already at order $\mathfrak{c}^0$, as well as in their dependence on the deformation parameters. These differences have two conceptually distinct origins, which are worth distinguishing carefully.

\begin{figure}[t]
    \centering
    \includegraphics[width=0.65\textwidth]{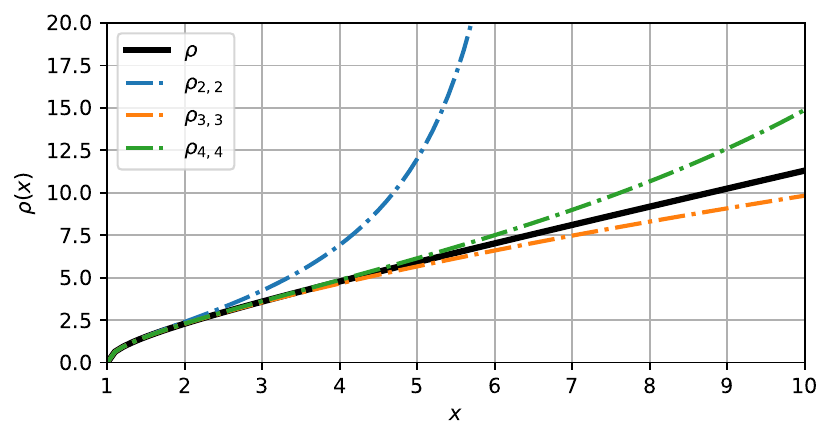}
    \caption{Proper distance $\rho$ to the horizon of the Schwarzschild black hole ($\Psi=\Phi=2\ma$, $x\coloneqq r/2M$), compared with several of its Padé approximants. The approximants differ from $\rho$ even in the absence of deformations, which illustrates the deviations of type \emph{(i)} of Remark~\ref{rem:TwoErrors}.}
    \label{fig:disSchwarz}
\end{figure}

\begin{center}
\remarkbox{\rem\label{rem:TwoErrors}{Two sources of deviation in Padé-based approximations}}{\parbox{16cm}{
\begin{enumerate}
\item[\emph{(i)}] \emph{Approximation of the classical geometry.} The functions $\rho$ and $\widehat{\rho}$ are defined differently. This affects the full form of $\widehat{b}_5$, but in particular also its coefficient of order $\mathcal{O}(\mathfrak{c}^{0})$, which is independent of $\{\theta_{2n}\,,\xi_{2n}\}$: the two functions differ even for the undeformed Schwarzschild black hole, as illustrated in Figure~\ref{fig:disSchwarz}.
\item[\emph{(ii)}] \emph{Truncation of the physical input.} Being obtained from a Padé approximation, $\widehat{b}_5$ depends only on $\mathfrak{c}$, $\xi_2$ and $\xi_4$, whereas $b_5$ also depends on $\xi_6$. The coefficient $\widetilde{\xi}_6$ entering $\widehat{b}_5$ is a function of $\xi_2$ and $\xi_4$, which contributes to the numerical differences at order $\mathcal{O}(\mathfrak{c})$. Such modifications only affect terms of order~$\mathcal{O}(\mathfrak{c})$.
\end{enumerate}
Deviations of type \emph{(ii)} are unavoidable, but they are pushed to higher and higher orders in the distance expansion as $N$ and $M$ are increased. Deviations of type \emph{(i)}, instead, can be eliminated altogether for observables that, for the undeformed black hole, are rational functions of a suitable variable: for such observables the Padé approximant of the Schwarzschild result becomes exact beyond a certain order.
}}
\end{center}

\subsection{Padé Approximation for the Photon Radius}\label{Sect:PadePhoton}
We now apply these ideas to the photon sphere. With the series (\ref{SeriesExpansionMetric}) and (\ref{SeriesDistance}), the effective potential (\ref{PhotonPotentialU}) of massless particles can be expanded in powers of the proper distance,
\begin{align}
U^2&=\sum_{p=1}^\infty \uva{2p}\,\left(\frac{\rho}{2M}\right)^{2p}
=\sum_{p=1}^\infty\ \left[\mathfrak{q}_{2p}-2\ma \sum_{m=0}^p \sum_{\ell=0}^{p-m}\theta_{2(p-m-\ell)}\, \mathfrak{p}_{2\ell}\, \mathfrak{q}_{2m}\right]\left(\frac{\rho}{2M}\right)^{2p}\,,\label{PotentialDistanceExpansion}
\end{align}
where $\mathfrak{p}_{2n}$ are the coefficients in \eqref{inverse-z-rho-exp}, while the $\mathfrak{q}_{2p}$ are the coefficients of the analogous expansion of $1/r^2$,
\begin{align}
\mathfrak{q}_{2p}&=\sum_{\ell = 0}^p \mathfrak{p}_{2(p-\ell)} \, \mathfrak{p}_{2\ell} \ .
\end{align}
The first few coefficients of $U^2$ are
\begin{subequations}\label{uva coeffs}
\begin{align}
\uva{2}&=\frac{M^2(1-4\zh \theta_2/M+\varpi)}{2\zh^4}\ ,\\
\uva{4}&=-\frac{M^4(1+\varpi)^2(7+13\varpi)}{8\zh^6(1+2\varpi)}+\frac{3\ma^3\theta_2(1+\varpi)}{\zh^5}-2\ma\frac{\xi_4+(1+2\varpi)\theta_4}{\zh^3(1+2\varpi)}\ ,\\
\uva{6}&=-\frac{16 \xi_4^2}{(3 \varpi +2) (1+2 \varpi)^2\hr^2}-\frac{1}{\zh^4} \left(\frac{12\ma^2 \theta_2 \xi_4}{2 \varpi
   +1}+2\ma\zh \left(\theta_6+\frac{\xi_6}{3 \varpi +2}\right)\right)+ \nonumber\\
   &-\frac{3 \ma^5\theta_2
   (\varpi +1)^2 (9 \varpi +5)}{4 (2 \varpi +1) \zh^7} +\frac{\ma^3 (\varpi +1) \left[6 \theta_4 (3
   \varpi +2) (2 \varpi +1)^2+ \xi_4 (13 \varpi  (6 \varpi +7)+25)\right]}{2 (2 \varpi +1)^2 (3 \varpi +2) \zh^5}+\nonumber \\
   &+\frac{M^6(\varpi +1)^3 \left[\varpi  (\varpi  (368 \varpi +655)+386)+75\right]}{32 (2 \varpi +1)^2 (3
   \varpi +2) \zh^8}\ .
\end{align}
\end{subequations}
For every finite $p>0$, the coefficient $\uva{2p}=\uva{2p}(\zh,\xi_{2,4,\ldots,2p},\theta_{2,4,\ldots,2p})$ depends on finitely many EMD coefficients only.

In analogy with the proper distance, the series (\ref{PotentialDistanceExpansion}) can be used to construct Padé approximants of $U^2$, whose extrema can be determined without worrying about convergence. At order $(N,M)=(2,4)$, for example,
\begin{align}
U^2_{2,4}=\frac{\uva{2}^3(\rho/2M)^2}{\uva{2}^2+\uva{4}^2(\rho/2M)^4-\uva{2}(\rho/2M)^2(\uva{4}+\uva{6}(\rho/2M)^2)}\,,
\end{align}
which is maximal at
\begin{align}
\rho_\text{ps}=2M\frac{\sqrt{\uva{2}}}{(\uva{4}^2-\uva{2}\uva{6})^{1/4}}\,.
\end{align}
Assuming again the scaling (\ref{IllumScaling}), this becomes
\begin{align}
\frac{\rho_{\text{ps}}}{2\ma}=\left({\frac{15}{43}}\right)^{1/4}\left[2+43\mathfrak{c}\left(86-\frac{88}{3}\,\frac{\theta_2}{\mathfrak{c}}-440\,\frac{\theta_4}{\mathfrak{c}}-480\,\frac{\theta_6}{\mathfrak{c}}+\frac{864}{5}\,\frac{\xi_2}{\mathfrak{c}}-8\,\frac{\xi_4}{\mathfrak{c}}-96\,\frac{\xi_6}{\mathfrak{c}}\right)+\mathcal{O}(\mathfrak{c}^2)\right]\,,\label{PhotonRadiusPade}
\end{align}
This result is not exact even in the absence of deformations. Classically, the photon sphere of the Schwarzschild black hole is located at
\begin{align}
&\rps^{\text{class}}=3\,\ma\,,&&\text{or equivalently} &&\rho_{\text{ps}}^{\text{class}}=(\sqrt{3}+2\text{arccoth}\sqrt{3})\,\ma\sim 3.04901\,\ma\,,\label{ClassSchwarzschildRad}
\end{align}
whereas (\ref{PhotonRadiusPade}) gives $4 \left({\frac{15}{43}}\right)^{1/4}\ma\sim 3.07408\,\ma$ for $\mathfrak{c}=0$. The mismatch is a deviation of type \emph{(i)} in the classification of Remark~\ref{rem:TwoErrors}: it is caused by the Padé approximation itself. It can be reduced by increasing $(N,M)$, but at the price of a growing computational complexity, since more and more coefficients $\{\theta_{2n}\}$ and $\{\xi_{2n}\}$ contribute to the terms of order $\mathcal{O}(\mathfrak{c})$.

For the photon radius, fortunately, a better strategy exists, which removes deviations of type \emph{(i)} entirely and only retains those of type \emph{(ii)}, which are due to the deformation of the geometry. It relies on choosing the expansion variable appropriately.

\begin{center}
\remarkbox{\rem\label{rem:ExactPade}{Choice of the expansion variable}}{\parbox{16cm}{
For the undeformed geometry ($\Psi=\Phi=2\ma$) the potential of massless particles is
\begin{align}
U^2_{\text{class}}=\frac{1-\frac{2\ma}{\co}}{\co^2}=\frac{\co-2\ma}{\co^3}\,,\nonumber
\end{align}
which is a rational function of $\co-\zh=\co-2\ma$, with a numerator of degree one and a denominator of degree three. Any Padé approximant of $U^2_{\text{class}}$ in the variable $\co-\zh$ with $N\geq 1$ and $M\geq 3$ is therefore exact. For deformed black holes, Padé approximants of $U^2$ in the variable $\co-\zh$ of at least this order are thus free of deviations of type \emph{(i)}, and all remaining corrections are of type \emph{(ii)}, depending on finitely many of the physical coefficients $\{\theta_{2n}\}_{n\in\mathbb{N}^*}$ and $\{\xi_{2n}\}_{n\in\mathbb{N}^*}$. By contrast, $U^2_{\text{class}}$ is not a rational function of $\rho$, which is why the approximant (\ref{PhotonRadiusPade}) is not exact.
}}
\end{center}

\noindent
Instead of (\ref{PotentialDistanceExpansion}), we therefore consider the expansion of the potential in powers of $\co-\zh$,
\begin{align}
&U^2=\sum_{n=1}^\infty v_n\,(\co-\zh)^n\,,&&\text{with} &&\begin{array}{l} v_1=\frac{\hn{1}}{\zh^2}\,, \\[8pt] v_2=\frac{\hn{2}\zh-4\hn{1}}{2\zh^3}\,, \\[8pt] v_3=\frac{18\hn{1}-6\zh \hn{2}+\zh^2 \hn{3}}{6\zh^4}\,,\end{array}\label{PadePotential}
\end{align}
where $\hn{n}$ are the Taylor coefficients (\ref{f-h-taylor-r}) of the metric function, which implicitly depend on the physical coefficients $\{\theta_{2k}\}_{k\leq n}$ and $\{\xi_{2k}\}_{k\leq n}$. From (\ref{PadePotential}) we compute the Padé approximant $(U^2)_{N,M}$ in the variable $\co-\zh$ and determine the position of its maximum. With the scaling (\ref{IllumScaling}), the result takes the form
\begin{align}\label{rps expans}
\frac{\rps}{\ma}=\sigma_0+\sigma_1\,\mathfrak{c}+\mathcal{O}(\mathfrak{c}^2)\,.
\end{align}
The coefficients $\sigma_{0,1}$ depend on the order $(N,M)$ of the approximation and are listed in Table~\ref{Tab:PositionPade}. For finite $(N,M)$, $\sigma_1$ depends only on $\mathfrak{c}$ and on $\theta_{2L}$ up to $2L=2(N+M)$, so that the counting of physically relevant coefficients discussed above is respected. For $N\geq 2$ and $M=3$, the classical value $\sigma_0=3$ is reproduced exactly, as anticipated in Remark~\ref{rem:ExactPade}.

\begin{table}
\begin{center}
\resizebox{\textwidth}{!}{%
\begin{tabular}{|c|c||c|c|c|}\hline
&&&\\[-10pt]
$N$ & $M$ & $\sigma_0$ & $\sigma_1$ \\[2pt]\hline
&&&\\[-10pt]
2 & 1 & $(1+\sqrt{3})$ & $(1+\sqrt{3})+\frac{4(8-15 \sqrt{3})}{135}\,\frac{\theta_2}{\mathfrak{c}}+\left(\frac{8}{9}-\frac{16}{\sqrt{3}}\right)\,\frac{\theta_4}{\mathfrak{c}}-\frac{32}{3}\,\frac{\theta_6}{\mathfrak{c}}$\\[4pt]\hline
&&&\\[-8pt]
2 & 2 & $\frac{23+9\sqrt{3}}{13}$ & $\frac{23+9\sqrt{3}}{13}-\frac{44(28+33\sqrt{3})}{4095}\frac{\theta_2}{\mathfrak{c}}-\frac{8(3359+1272\sqrt{3})}{2535}\frac{\theta_4}{\mathfrak{c}}-\frac{32(281-24\sqrt{3})}{169}\frac{\theta_6}{\mathfrak{c}}-\frac{768(14-3\sqrt{3})}{169}\frac{\theta_8}{\mathfrak{c}}$\\[6pt]\hline
&&&\\[-8pt]
2 & 3 & $3$ & $3-\frac{13471}{14175}\frac{\theta_2}{\mathfrak{c}}-\frac{56128 }{2835}\frac{\theta_4}{\mathfrak{c}}-\frac{3104 }{45}\frac{\theta_6}{\mathfrak{c}}-\frac{1088}{9}\frac{\theta_8}{\mathfrak{c}}-\frac{256 }{3}\frac{\theta_{10}}{\mathfrak{c}}$\\[4pt]\hline
&&&\\[-8pt]
3 & 3 & $3$ & $3-\frac{443939}{467775}\frac{\theta_2}{\mathfrak{c}}-\frac{2625104}{127575}\frac{\theta_4}{\mathfrak{c}}-\frac{732448}{8505}\frac{\theta_6}{\mathfrak{c}}-\frac{92608}{405}\frac{\theta_8}{\mathfrak{c}}-\frac{9472}{27}\frac{\theta_{10}}{\mathfrak{c}}-\frac{2048 }{9}\frac{\theta_{12}}{\mathfrak{c}}$\\[4pt]\hline
\end{tabular}
} 
\end{center}
\caption{Position of the maximum of the Padé approximant $(U^2)_{N,M}$ of the potential, for several orders $(N,M)$ and with the scaling~(\ref{IllumScaling}).}
\label{Tab:PositionPade}
\end{table}

The position $\rps$ determines the radius of the black hole shadow via (\ref{bphdef}). With the scaling (\ref{IllumScaling}) we write
\begin{align}
\frac{\bsh}{\ma}= \tau_0+\tau_1\,\mathfrak{c}+\mathcal{O}(\mathfrak{c}^2)\,.\label{ShadowExpans}
\end{align}
The coefficients $\tau_{0,1}$ for different orders of the Padé approximation are given in Table~\ref{Tab:ValuePade}. The leading corrections $\tau_1$ depend on the same number of physically relevant deformation coefficients as $\sigma_1$.

\begin{table}
\begin{center}
\resizebox{\textwidth}{!}{
\begin{tabular}{|c|c||c|c|c|}\hline
&&&\\[-10pt]
$N$ & $M$ & $\tau_0$ & $\tau_1$ \\[2pt]\hline
&&&\\[-8pt]
2 & 1 & $2(1+\sqrt{3})$ & $2(1+\sqrt{3})+\frac{8(82+75\sqrt{3})}{135}\,\frac{\theta_2}{\mathfrak{c}}+\frac{16(8+3\sqrt{3})}{9}\,\frac{\theta_4}{\mathfrak{c}}+\frac{64}{3}\,\frac{\theta_6}{\mathfrak{c}}$\\[4pt]\hline
&&&\\[-8pt]
2 & 2 & $2\sqrt{\frac{10}{3}+2\sqrt{3}}$ & \parbox{15cm}{$2\sqrt{\frac{10}{3}+2\sqrt{3}}+\frac{4}{945}\sqrt{\frac{12051454}{3}+2415638\sqrt{3}}\frac{\theta_2}{\mathfrak{c}}+\frac{88}{45}\sqrt{\frac{265}{3}+59\sqrt{3}}\frac{\theta_4}{\mathfrak{c}}+\frac{32}{3}\sqrt{38\sqrt{3}-\frac{146}{3}}\frac{\theta_6}{\mathfrak{c}}+ 128\sqrt{\sqrt{3}-\frac{5}{3}}\frac{\theta_8}{\mathfrak{c}}$}\\[6pt]\hline
&&&\\[-8pt]
2 & 3 & $3\sqrt{3}$ & $3\sqrt{3}+\frac{1}{\sqrt{3}}\left(\frac{98837}{4725 }\frac{\theta_2}{\mathfrak{c}}+\frac{136096}{2835 }\frac{\theta_4}{\mathfrak{c}}+\frac{4384}{45 }\frac{\theta_6}{\mathfrak{c}}+\frac{1216}{9} \frac{\theta_8}{\mathfrak{c}}+\frac{256 }{3}\frac{\theta_{10}}{\mathfrak{c}}\right)$\\[4pt]\hline
&&&\\[-8pt]
3 & 3 & $3\sqrt{3}$ & $3\sqrt{3}+\frac{1}{\sqrt{3}}\left(\frac{652294}{31185}\frac{\theta_2}{\mathfrak{c}}+\frac{2066276}{42525}\frac{\theta_4}{\mathfrak{c}}+\frac{62528}{567}\frac{\theta_6}{\mathfrak{c}}+\frac{29152}{135}\frac{\theta_8}{\mathfrak{c}}+\frac{2560}{9}\frac{\theta_{10}}{\mathfrak{c}}+\frac{512}{3}\frac{\theta_{12}}{\mathfrak{c}}\right)$\\[4pt]\hline
\end{tabular}
}
\end{center}
\caption{Radius of the black hole shadow obtained by replacing the potential with its Padé approximant, $U^2\sim (U^2)_{N,M}$, for several orders $(N,M)$ and with the scaling (\ref{IllumScaling}).}
\label{Tab:ValuePade}
\end{table}

For approximants of the form $(U^2)_{N,3}$, the result can be cast in an even more compact form. To this end, we parametrise $\tau_1$ as
\begin{align}
\tau_1(N)=3\sqrt{3}+\sum_{k=1}^{3+N}\beta_{2k}(N)\,\frac{\theta_{2k}}{\mathfrak{c}}\,,\label{TauParametrisation}
\end{align}
where the $\beta_{2k}$ are numerical coefficients (see Table~\ref{Tab:ValuePade} for $N=2,3$). For sufficiently large $N$, each $\beta_{2k}$ changes very little when $N$ is increased further.
\begin{figure}
    \centering
\includegraphics[width=0.65\textwidth]{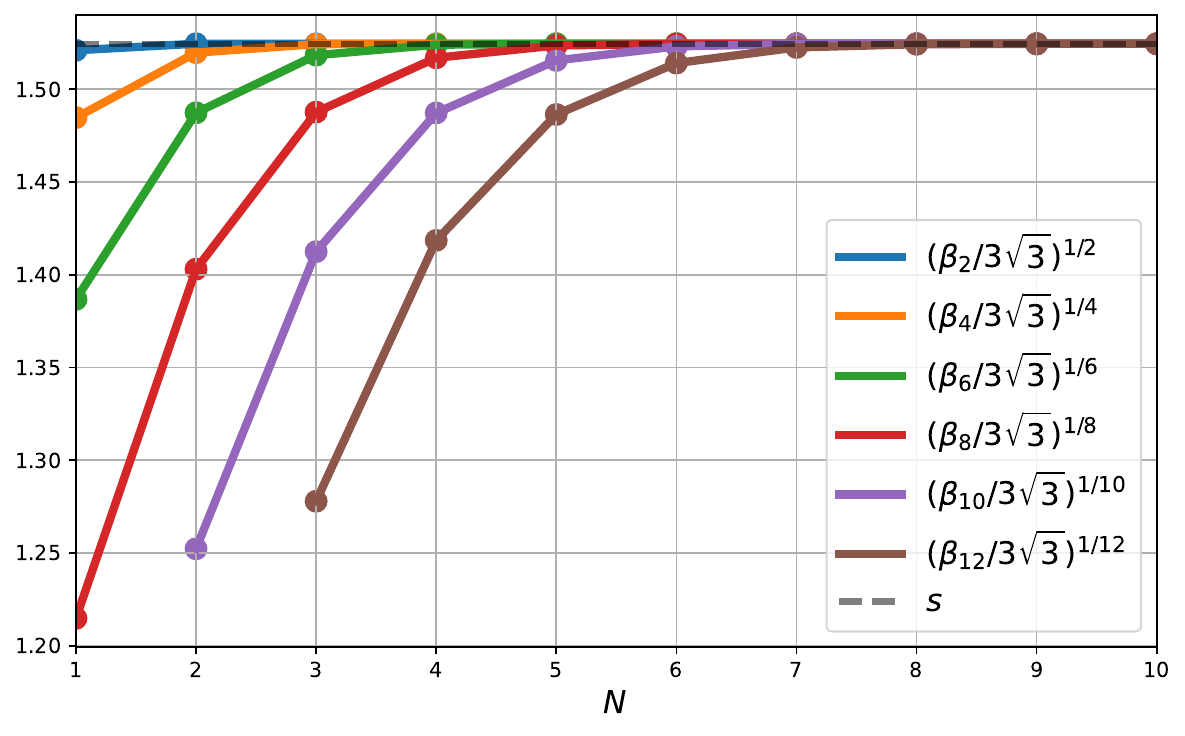}
    \caption{Coefficients $\beta_{2n}$ of the leading deformation $\tau_1$ of the shadow, see (\ref{TauParametrisation}), for Padé approximants of order $(N,3)$. For large $N$, suitable powers of the $\beta_{2n}$ approach a common value, cf.~(\ref{Svalue}), which allows the resummation \eqref{ResumTau1} of the asymptotic value of $\tau_1$.}
    \label{fig:ConvergenceCoefs}
\end{figure}
As shown numerically in Figure~\ref{fig:ConvergenceCoefs}, the coefficients approach the asymptotic values
\begin{align}
&\lim_{N\to \infty}\beta_{2k}(N)\to 3\sqrt{3}\,\mathfrak{s}^{2k}\hspace{0.5cm}\forall k\geq 1\,,&&\text{with} &&\mathfrak{s}\sim 1.5245\sim \frac{\rho^{\text{class}}_{\text{ps}}}{2\ma}\ .\label{Svalue}
\end{align}
Remarkably, $2\ma\mathfrak{s}$ is numerically very close to the proper distance of the photon sphere of the classical Schwarzschild black hole, see (\ref{ClassSchwarzschildRad}). Assuming that (\ref{Svalue}) holds for all $k\geq 1$, the asymptotic behaviour of $\tau_1(N)$ can be resummed,
\begin{align}
\lim_{N\to \infty}(\mathfrak{c}\,\tau_1(N))=3\sqrt{3}\,\mathfrak{c}+3\sqrt{3}\,\sum_{k=1}^\infty\,\theta_{2k}\,\mathfrak{s}^{2k}=3\sqrt{3}\,\mathfrak{c}+3\sqrt{3}\,\sum_{k=1}^\infty\theta_{2k}\,\left(\frac{\rho^{\text{class}}_{\text{ps}}}{2\ma}\right)^{2k}\,,\label{ResumTau1}
\end{align}
where we have used (\ref{IllumScaling}). With the series expansion (\ref{SeriesExpansionMetric}), this can be formally written as
\begin{align}
\lim_{N\to \infty}(\mathfrak{c}\,\tau_1(N))=3\sqrt{3}\,\mathfrak{c}+\frac{3\sqrt{3}}{2\ma}\,\left(\Psi(2\ma\,\mathfrak{s})-\zh\right)\,,
\end{align}
provided $2\ma\,\mathfrak{s}$ lies inside the radius of convergence of (\ref{SeriesExpansionMetric}).\footnote{Equivalently, the assumption is that $\displaystyle \lim_{N\to \infty} \Psi_{N,3}(2\ma\mathfrak{s})=\Psi(2\ma\mathfrak{s})-\zh$. Section~\ref{Sect:Convergence} tests this assumption in several models of quantum black holes: it holds at least as long as the deformation remains small.} Combining this with (\ref{RescaledMetricFunctions}) and (\ref{ShadowExpans}) leads to one of the central results of this Chapter.

\begin{center}
\resultbox{\result\label{res:ShadowEMD}{Leading deformation of the black hole shadow}}{\parbox{16cm}{
To leading order in the deformation, the radius of the shadow of a static, spherically symmetric black hole described by the proper-distance EMD is determined by the deformation function $\Psi$ evaluated at a single, universal value of the proper distance from the horizon:
\begin{align}
\lim_{N\to \infty}\frac{\bsh(N)}{\ma}=\frac{\bsh}{\ma}\sim\frac{3\sqrt{3}}{2\ma}\,\Psi(2\ma\,\mathfrak{s})\,,\hspace{1cm}\text{with}\hspace{0.5cm}\mathfrak{s}\sim 1.5245\,.\label{CompactCorrectionShadow}
\end{align}
The result assumes that $\rho=2\ma\mathfrak{s}$ lies within the radius of convergence of the series (\ref{SeriesExpansionMetric}) for $\Psi$. It is independent of $\Phi$ at this order, and it reproduces the Schwarzschild value $\bsh=3\sqrt{3}\,\ma$ for $\Psi=2\ma$.
}}
\end{center}

\noindent
Conversely, (\ref{CompactCorrectionShadow}) can be used to bound the deformation function with an observation of the shadow. Assuming the absence of rotation, which is compatible with the data, the observations of M87$^\ast$ by the Event Horizon Telescope give at 1-$\sigma$ confidence level~\cite{EventHorizonTelescope:2021dqv}
\begin{equation}
    \tilde{b}_\text{sh}^{\text{EHT M}87^\ast \, (\pm)} = 3 \sqrt{3}M_{\text{M}87^\ast}(1 \pm 0.17) \ ,
\end{equation}
where $M_{\text{M}87^\ast}$ is the mass of M87$^\ast$. If the latter is modelled as a spherically symmetric black hole, (\ref{CompactCorrectionShadow}) translates into a bound on its deviation from the Schwarzschild geometry.

\begin{center}
\resultbox{\result\label{res:EHTbound}{Model-independent bound from the shadow of M87$^\ast$}}{\parbox{16cm}{
Modelling M87$^\ast$ as a static, spherically symmetric black hole, the EHT measurement of its shadow implies for the deformation function of the proper-distance EMD
\begin{equation}
 \abs{\frac{\Psi(2\ma_{\text{M}87^\ast}\,\mathfrak{s})}{2M_{\text{M}87^\ast}} - 1} \leq 0.17 \ .\label{ExperimentalBound}
\end{equation}
This is a rough bound on the value of $\Psi$ at the single point $\rho=2\ma_{\text{M}87^\ast}\mathfrak{s}$, which holds independently of the model that generates the deformation.
}}
\end{center}

\section{Simple Examples of Corrected Schwarzschild Geometries}\label{subs example models}
We now test the formalism on several models of static and spherically symmetric black holes proposed in the literature~\cite{Hayward_2006,Simpson:2018tsi,Simpson:2019mud,Dymnikova:1992ux,Garcia95}. They have been selected because they are compatible with the scaling (\ref{IllumScaling}), such that the results of the previous Section can be applied at leading order in the deformation. At the same time, they cover a broad range of functional dependences on the Schwarzschild coordinate $r$, and each of them contains an explicit parameter that controls the size of the deviation from GR. In line with the philosophy of the EMD, the theory of quantum gravity in whose context a given model was proposed plays no role in the analysis.

\subsection{Metrics and Deformation Parameters}
We first consider four geometries~\cite{Hayward_2006,Simpson:2018tsi,Simpson:2019mud,Dymnikova:1992ux} that deform the Schwarzschild solution~\cite{Wald:1984rg}. We call them \emph{simple}, in the sense that $f(\co)=h(\co)$ for all $\co \geq \zh$ in (\ref{SpStMetric}).

To treat the four models in parallel, as far as possible, we introduce a common notation. Besides the mass $\ma$, each model contains one additional parameter, which we call $\eta$, that describes the deformation of the Schwarzschild spacetime. Although $\eta$ has a different meaning in each model, it can be traded for a small parameter that is defined in the same way for all of them. To this end, we first absorb the mass scale into the radial coordinate,
\begin{equation}\label{x definition}
x:=\frac{\co}{2\ma} \qq{and} \xh:=\frac{\zh}{2\ma}\ .
\end{equation}
and then define the fractional deviation of the event horizon radius from the Schwarzschild radius as in~\cite{Rezzolla:2014mua},\footnote{For $\epsilon\ll 1$ this choice guarantees that the metric is expanded around the outer horizon, which is expected to lie close to the classical Schwarzschild radius $\xh^{\text{class}}=1$, and not around the inner horizon that arises in these models from the regularisation of the central singularity.}
\begin{equation}
    \epsilon:= \frac{2M-\hr}{\hr} = \frac{1-\xh}{\xh} \ \longleftrightarrow \ \xh = \frac{1}{1+\epsilon}\ .\label{EpsEq}
\end{equation}
Table~\ref{subs models table} lists the metric functions of the four models, the relation between $\eta$ and $\epsilon$, which can be rational, exponential or non-analytic at $\eta = 0$, and, anticipating the results below, the leading corrections to the position of the photon sphere and to the shadow.

\begin{table}[!ht]
\centering
\resizebox{\textwidth}{!}{
\begin{tabular}{c|c|c|c|c} 
\hline
Model          & $\displaystyle f=h$                             & $\displaystyle \eta(\epsilon)$                                                                  & $ \rps/M-3$  & $\bsh /M - 3\sqrt{3}$\\[3pt]
\hline
Hayward \cite{Hayward_2006}    & $\displaystyle 1-\frac{x^2}{x^3+\eta}$        & $\displaystyle\frac{\epsilon}{(1+\epsilon)^3}$                                                   & $-\dfrac{16}{9}\epsilon + \order{\epsilon^2}$        & $-\dfrac{8}{3\sqrt{3}}\epsilon + \order{\epsilon^2} $         \\[8pt]
Simpson-Visser (I) \cite{Simpson:2018tsi} & $\displaystyle1-\frac{1}{\sqrt{x^2+\eta}}$    & $\displaystyle\frac{\epsilon\,(2+\epsilon)}{(1+\epsilon)^2}$                                             & $-\dfrac{20}{9}\epsilon + \order{\epsilon^2}$       & $-\dfrac{4}{\sqrt{3}}\epsilon + \order{\epsilon^2}$             \\[11pt]
Simpson-Visser (II) \cite{Simpson:2019mud} & $\displaystyle1-\frac{1}{x}e^{-\eta / x}$    & $\displaystyle\frac{\log(1+\epsilon)}{1+\epsilon}$                                             & $-\dfrac{8}{3}\epsilon + \order{\epsilon^2}$            & $-2\sqrt{3} \, \epsilon + \order{\epsilon^2}$        \\[8pt]
Dymnikova \cite{Dymnikova:1992ux}     & $\displaystyle1-\frac{1-e^{-x^3 / \eta}}{x}$ & $\displaystyle-\left[(1+\epsilon)^3 \log \left(\frac{\epsilon}{1+ \epsilon} \right)\right]^{-1}$ & $ \left(\dfrac{81}{8}\log \epsilon - 3 \right) \epsilon^{27/8} + \order{\epsilon^{33/8}}$ & $- 3  \sqrt{3} \, \epsilon^{27/8} + \order{\epsilon^{33/8}}$\\[8pt]
\hline
\end{tabular}
}
\caption{Overview of the four models considered in this Section. From left to right: name of the model (with reference), metric function as in (\ref{SpStMetric}) depending on the deformation parameter $\eta$, relation between $\eta$ and the fractional shift $\epsilon$ of the horizon radius (eq.~(\ref{EpsEq})), and leading corrections in $\epsilon$ to the photon radius and to the shadow.} 
\label{subs models table}
\end{table}

In all these models, the deformation coefficients $\{\xi_{2n}\}_{n\in\mathbb{N}}$ in (\ref{SeriesExpansionMetric}), which coincide with the $\{\theta_{2n}\}_{n\in\mathbb{N}}$, do not depend explicitly on the mass. Indeed, (\ref{DiffEqDistance}) implies $\rho(r)=2\,\ma\,\underline{\rho}(x,\epsilon)$, where $\underline{\rho}(x,\epsilon)$ is independent of $\ma$, and with (\ref{h-f-emd-def}) and (\ref{SeriesExpansionMetric}) we have
\begin{align}
f=h=1-\frac{\xh+\sum_{n=1}^\infty \xi_{2n}\,\underline{\rho}(x,\epsilon)^{2n}}{x}\ .
\end{align}
Moreover, (\ref{EpsEq}) gives $\mathfrak{c}=-\frac{\epsilon}{1+\epsilon}$, and we shall see that in all cases $\xi_{2n}=\mathcal{O}(\epsilon)$, such that $\frac{\xi_{2n}}{\mathfrak{c}}=\mathcal{O}(1)$, in agreement with (\ref{IllumScaling}).

\begin{figure}[t]
\begin{center}
\includegraphics[width=\textwidth]{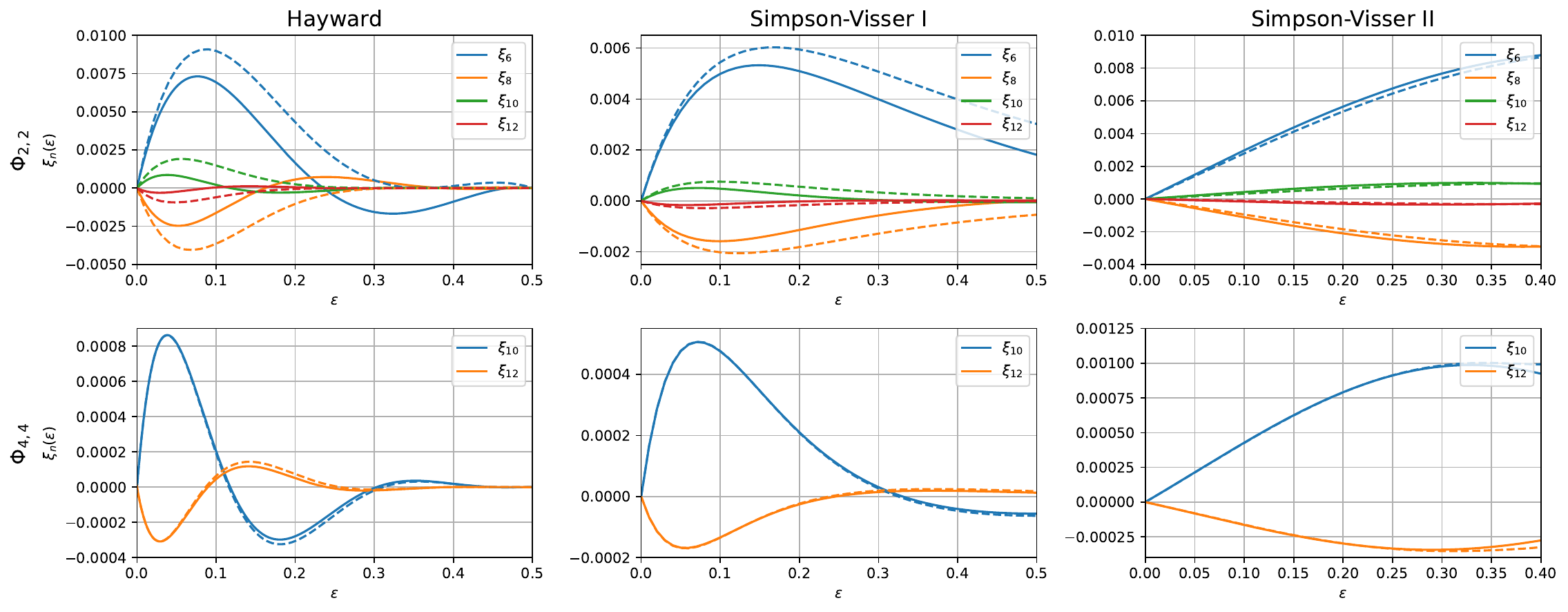}
\end{center}
\caption{Exact deformation coefficients $\xi_{2n}$ (solid lines) and the mass-rescaled coefficients (\ref{ReplacementCoefficients}) implied by a Padé approximant of $\Phi$ (dashed lines), as functions of $\epsilon$, for the Hayward (left), Simpson-Visser I (middle) and Simpson-Visser II (right) geometries. Upper row: $\xi_{6,8,10,12}$ compared with the predictions of $\Phi_{2,2}$. Lower row: $\xi_{10,12}$ compared with the predictions of $\Phi_{4,4}$.}
\label{Fig:XiReplacementHayward}
\end{figure}

\begin{figure}[!ht]
\centering
\includegraphics[width=\textwidth]{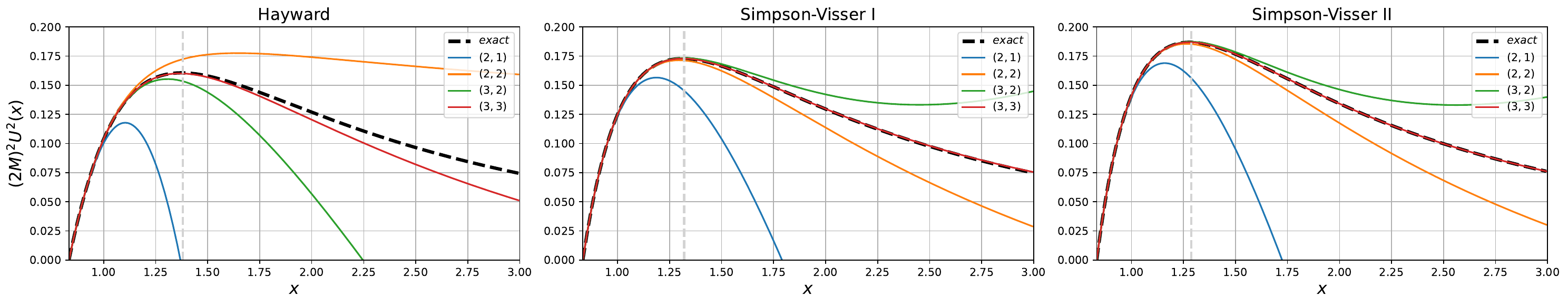}
\caption{Exact potential $U^2$ and Padé approximants \eqref{PadePotential} of several orders for the Hayward (left), Simpson-Visser I (middle) and Simpson-Visser II (right) geometries, at $\epsilon=0.2$. The dashed gray vertical line marks the maximum of the exact potential.} 
\label{Fig:PotentialPlots}
\end{figure}

\begin{figure}[t]
\centering
\includegraphics[width=\textwidth]{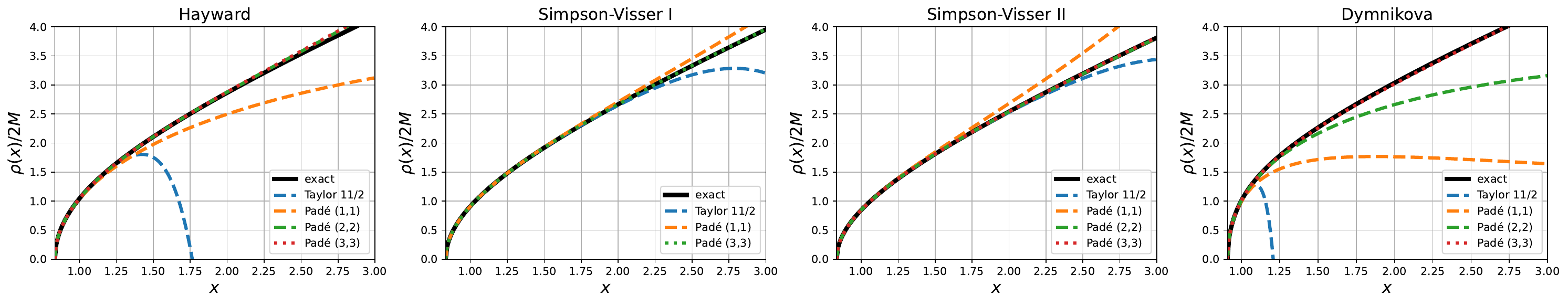}
\caption{Proper distance as a function of $x$ for the Hayward, Simpson-Visser I and Simpson-Visser II geometries ($\epsilon=0.2$) and for the Dymnikova geometry ($\epsilon=0.1$). Solid black: numerical solution of (\ref{DiffEqDistance}). Dashed blue: Taylor series (\ref{SeriesDistance}) including terms up to $n=6$. Coloured dashed and dotted: Padé approximants (\ref{PadeDistance}).} 
\label{Fig:DistanceHayward}
\end{figure}

For the Hayward and the two Simpson-Visser geometries, the quality of the approximations is illustrated by three sets of plots. Figure~\ref{Fig:XiReplacementHayward} compares the exact coefficients $\xi_{2n}$ with the coefficients $\widetilde{\xi}_{2n}$ implied by a Padé approximant $\Phi_{N,M}$ ($=\Psi_{N,M}$) as in (\ref{ReplacementCoefficients}), for $(N,M)=(2,2)$ and $(N,M)=(4,4)$. Figure~\ref{Fig:PotentialPlots} compares the exact potential $U^2$ with Padé approximants of various orders, and Figure~\ref{Fig:DistanceHayward} compares the numerically integrated proper distance with its Taylor expansion (\ref{SeriesDistance}) and with Padé approximants (\ref{PadeDistance}). For each model, we also list the leading corrections to the photon radius and to the shadow obtained from Tables~\ref{Tab:PositionPade} and~\ref{Tab:ValuePade}.

\paragraph{Hayward black hole.} The first model is the regular black hole introduced in~\cite{Hayward_2006}. Rescaling the deformation functions as in (\ref{RescaledMetricFunctions}), the leading coefficients (with $\theta_{2n}=\xi_{2n}$ $\forall n\geq 1$) are
\begin{align}
&\xi_2=\frac{3 \epsilon (1-2 \epsilon )  }{4 (1+\epsilon )}\, \qquad \xi_{4}=-\frac{\epsilon  (1-2 \epsilon ) (\epsilon  (7 \epsilon -22)+7)}{16 (1+\epsilon )}\,,\notag \\
&\xi_6=\frac{\epsilon  (1-2 \epsilon ) \left[\epsilon  (\epsilon  (2 \epsilon  (61 \epsilon -899)+3225)-1600)+221\right]}{960 (1+\epsilon )}\ .
\end{align}
The left column of Figure~\ref{Fig:XiReplacementHayward} shows that the approximation $(N,M)=(2,2)$ captures at least the rough dependence of the higher coefficients on $\epsilon$, while at order $(N,M)=(4,4)$ the $\widetilde{\xi}_{2n}$ remain good approximations even for finite values of $\epsilon$. The left panels of Figures~\ref{Fig:PotentialPlots} and~\ref{Fig:DistanceHayward} (for $\epsilon=0.2$) highlight the finite radius of convergence of the Taylor series (\ref{SeriesDistance}), and show that Padé approximants of fairly low order already provide viable approximations. The leading corrections to the photon radius and to the shadow are
\begin{center}
\begin{tabular}{|c|c||c|c|c|}\hline
&&&\\[-10pt]
$N$ & $M$ & $\sigma_1\,\mathfrak{c}$ & $\tau_1\,\mathfrak{c}$ \\[2pt]\hline
&&&\\[-8pt]
2 & 1 & $-\frac{11 - 3 \sqrt{3}}{3}\,\epsilon+\mathcal{O}(\epsilon^2)$ &  $\frac{1-3\sqrt{3}}{3}\,\epsilon+\mathcal{O}(\epsilon^2)$\\[4pt]\hline
&&&\\[-10pt]
2 & 2 & $\frac{3}{169}(-131+16\sqrt{3})\,\epsilon+\mathcal{O}(\epsilon^2)$ & $-\frac{1}{6}\sqrt{-\frac{965}{3}+233\sqrt{3}}\,\epsilon +\mathcal{O}(\epsilon^2)$\\[4pt]\hline
&&&\\[-8pt]
2 & 3 & $-\frac{16}{9}\,\epsilon+\mathcal{O}(\epsilon^2)$ & $-\frac{8}{3\sqrt{3}}\,\epsilon+\mathcal{O}(\epsilon^2)$ \\[4pt]\hline
&&&\\[-8pt]
3 & 3 & $-\frac{16}{9}\,\epsilon+\mathcal{O}(\epsilon^2)$ & $-\frac{8}{3\sqrt{3}}\,\epsilon+\mathcal{O}(\epsilon^2)$\\[4pt]\hline
\end{tabular}
\end{center}
\paragraph{Simpson-Visser I black hole.} The second example is the geometry introduced in~\cite{Simpson:2018tsi}. Its leading deformation coefficients are
\begin{equation}
    \xi_2 = \frac{\epsilon (2 + \epsilon)}{4 (1+\epsilon)^3} \  , \ \xi_4 = \frac{\epsilon(2 + \epsilon)(\epsilon (2 + \epsilon)-11)}{96(1 + \epsilon)^5} \ , \
    \xi_6 = \frac{\epsilon  (\epsilon +2) \left[\epsilon  (\epsilon +2) (\epsilon  (\epsilon +2)-121)+292\right]}{5760 (\epsilon +1)^7} \ .
\end{equation}
The middle columns and panels of Figures~\ref{Fig:XiReplacementHayward}, \ref{Fig:PotentialPlots} and~\ref{Fig:DistanceHayward} lead to conclusions similar to those for the Hayward geometry.\footnote{The Padé approximant of the distance of order $(N,M)=(2,2)$ develops a vertical tangent at some point $\tilde{x} \geq \xh$ outside of the horizon, both for this model and for the Simpson-Visser II geometry below. This is an artefact of the low order of the approximation and disappears for a suitable choice of the order.} The leading corrections to the photon radius and to the shadow are
\begin{center}
\begin{tabular}{|c|c||c|c|c|}\hline
&&&\\[-10pt]
$N$ & $M$ & $\sigma_1\,\mathfrak{c}$ & $\tau_1\,\mathfrak{c}$ \\[2pt]\hline
&&&\\[-8pt]
2 & 1 & $-\frac{13}{6}\,\epsilon+\mathcal{O}(\epsilon^2)$ &  $-\left(\frac{2}{3}+\sqrt{3}\right)\,\epsilon+\mathcal{O}(\epsilon^2)$\\[4pt]\hline
&&&\\[-10pt]
2 & 2 & $-\frac{15}{338}(47+2\sqrt{3})\,\epsilon+\mathcal{O}(\epsilon^2)$ & $-\frac{1}{6}\,\sqrt{\frac{265}{3}+59\sqrt{3}}\,\epsilon +\mathcal{O}(\epsilon^2)$\\[4pt]\hline
&&&\\[-8pt]
2 & 3 & $-\frac{20}{9}\,\epsilon+\mathcal{O}(\epsilon^2)$ & $-\frac{4}{\sqrt{3}}\,\epsilon+\mathcal{O}(\epsilon^2)$ \\[4pt]\hline
&&&\\[-8pt]
3 & 3 & $-\frac{20}{9}\,\epsilon+\mathcal{O}(\epsilon^2)$ & $-\frac{4}{\sqrt{3}}\,\epsilon+\mathcal{O}(\epsilon^2)$\\[4pt]\hline
\end{tabular}
\end{center}

\begin{center}
\remarkbox{\rem\label{rem:OtherRationalModels}{Other models with rational deformations}}{\parbox{16cm}{
The black hole geometries proposed by Bardeen~\cite{bardeen1968} and Frolov~\cite{Frolov:2016pav}, as well as the one proposed in~\cite{Cadoni:2022vsn}, depend on $x$ in a way similar to the Hayward and Simpson-Visser I geometries. We therefore expect results comparable to the ones reported above for these models. The Bardeen black hole is analysed explicitly, in a different context, in Chapter~\ref{chap5}.
}}
\end{center}

\paragraph{Simpson-Visser II black hole.} The third model, introduced in~\cite{Simpson:2019mud} (see also~\cite{Cunha:2020azh,Culetu:2014lca}), has a deformation function that depends exponentially on the inverse radial coordinate. It is constructed such that the core of the black hole is Minkowski, rather than de Sitter. Its leading deformation coefficients are
\begin{align}
&\xi_2 = -\frac{\overline{\epsilon}}{4}  (\log (\overline{\epsilon})-1) \log (\overline{\epsilon}) \  , \notag \\
&\xi_4 = \frac{(\overline{\epsilon})^3}{96}  (\log (\overline{\epsilon})-1) \log (\overline{\epsilon}) (\log (\overline{\epsilon}) (4 \log (\overline{\epsilon})-13)+8) \ , \nonumber \\
&\xi_6 = -\frac{(\overline{\epsilon})^5}{5760} (\log (\overline{\epsilon})-1) \log (\overline{\epsilon}) \left\{\log (\overline{\epsilon}) \left[\log (\overline{\epsilon}) (\log (\overline{\epsilon}) (34 \log (\overline{\epsilon})-263)+626)-568\right]+172\right\}\ ,
\end{align}
with $\overline{\epsilon} \coloneqq \epsilon+1$. The corresponding comparisons are shown in the right columns and panels of Figures~\ref{Fig:XiReplacementHayward} and~\ref{Fig:PotentialPlots}, and in the third panel of Figure~\ref{Fig:DistanceHayward}. The leading corrections to the photon radius and to the shadow are
\begin{center}
\begin{tabular}{|c|c||c|c|c|}\hline
&&&\\[-10pt]
$N$ & $M$ & $\sigma_1\,\mathfrak{c}$ & $\tau_1\,\mathfrak{c}$ \\[2pt]\hline
&&&\\[-8pt]
2 & 1 & $-\frac{2}{3}\,(2+\sqrt{3})\,\epsilon+\mathcal{O}(\epsilon^2)$ &  $-\frac{4}{3}\,(1+\sqrt{3})\,\epsilon+\mathcal{O}(\epsilon^2)$\\[4pt]\hline
&&&\\[-10pt]
2 & 2 & $-\frac{6}{13}\,(4+\sqrt{3})\,\epsilon+\mathcal{O}(\epsilon^2)$ & $-\frac{4}{3}\,\sqrt{\frac{10}{3}+2\sqrt{3}}\,\epsilon +\mathcal{O}(\epsilon^2)$\\[4pt]\hline
&&&\\[-8pt]
2 & 3 & $-\frac{8}{3}\,\epsilon+\mathcal{O}(\epsilon^2)$ & $-2\,\sqrt{3}\,\epsilon+\mathcal{O}(\epsilon^2)$ \\[4pt]\hline
&&&\\[-8pt]
3 & 3 & $-\frac{8}{3}\,\epsilon+\mathcal{O}(\epsilon^2)$ & $-2\,\sqrt{3}\,\epsilon+\mathcal{O}(\epsilon^2)$\\[4pt]\hline
\end{tabular}
\end{center}

\begin{figure}[t]
    \centering
    \includegraphics[width=\textwidth]{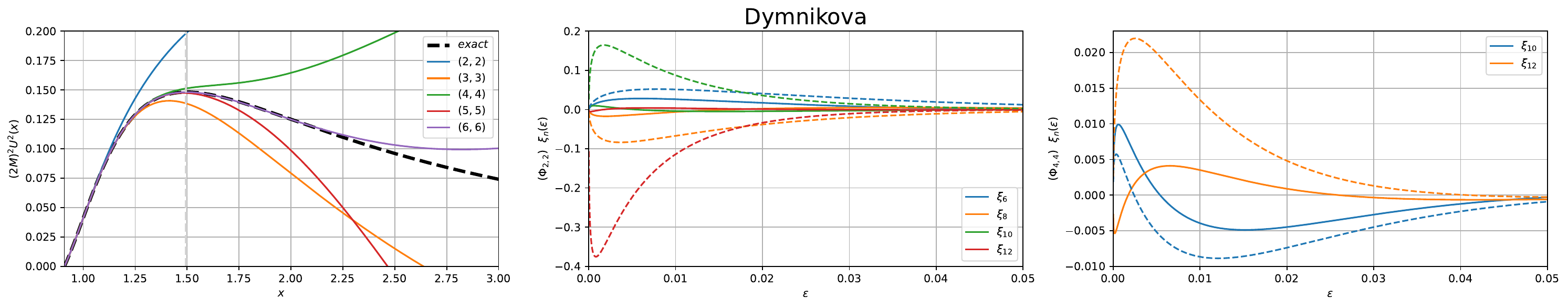}
    \caption{Dymnikova geometry. Left: exact potential $U^2$ and Padé approximants \eqref{PadePotential} at $\epsilon=0.1$, with the maximum of the exact potential marked by the dashed gray vertical line. Middle and right: exact coefficients $\xi_{2n}$ (solid lines) and the mass-rescaled coefficients (\ref{ReplacementCoefficients}) (dashed lines) implied by $\Phi_{2,2}$ (middle) and $\Phi_{4,4}$ (right), as functions of $\epsilon$.}
    \label{fig:dymnpot}
\end{figure}

\paragraph{Dymnikova black hole.} The Dymnikova spacetime~\cite{Dymnikova:1992ux} requires special care, since its deformation function is not analytic at $\eta=0$. To deal with this, we use a double expansion in powers of $u := e^{-1/\eta}$ and $\log u$. For example, the outer event horizon is located at $\xh^{(\text{Dymn})} = 1 - u + 3u^2 \log u + \order{u^3}$. As a consequence, the deformations of the photon radius and of the shadow are not analytic in $\mathfrak{c}$, and the results of Tables~\ref{Tab:PositionPade} and~\ref{Tab:ValuePade} cannot be used directly. Padé approximants are nevertheless an efficient tool to describe the geometry and the observables computed from it. As shown in the right panel of Figure~\ref{Fig:DistanceHayward}, already at order $(3,3)$ the Padé approximant is an excellent approximation of the numerically computed distance function (for $\epsilon=0.1$). As functions of $\epsilon$, the first two deformation coefficients (\ref{RescaledMetricFunctions}) are
\begin{align}
&\xi_2= -\frac{3 \epsilon}{4}  (\epsilon+1) \log \left(\frac{\epsilon }{\epsilon+1}\right) \left[3 \epsilon  \log \left(\frac{\epsilon }{\epsilon+1}\right)+1\right] \ , \notag \\
&\xi_4 = -\frac{\epsilon}{32}  (\epsilon+1)^3 \log \left(\frac{\epsilon }{\epsilon+1}\right) \left[3 \epsilon  \log \left(\frac{\epsilon }{\epsilon+1}\right)+1\right] \left[9 \log \left(\frac{\epsilon }{\epsilon+1}\right) \left(2 \epsilon +4 \epsilon  \log \left(\frac{\epsilon }{\epsilon+1}\right)+1\right)+4\right] \ .
\end{align}
The middle and right panels of Figure~\ref{fig:dymnpot} compare the higher coefficients with their Padé-based approximations of order $(N,M)=(2,2)$ and $(N,M)=(4,4)$, respectively. Expanded around $\epsilon=0$, the coefficients behave as $\xi_{2n} \propto \epsilon \log^{n}(\epsilon)$, which causes numerical instabilities. Nevertheless, as the plots in Figure~\ref{fig:dymnpot} show, they tend to zero for $\epsilon\to 0$, albeit steeply.

\begin{figure}[t]
    \centering
    \includegraphics[width=0.65\textwidth]{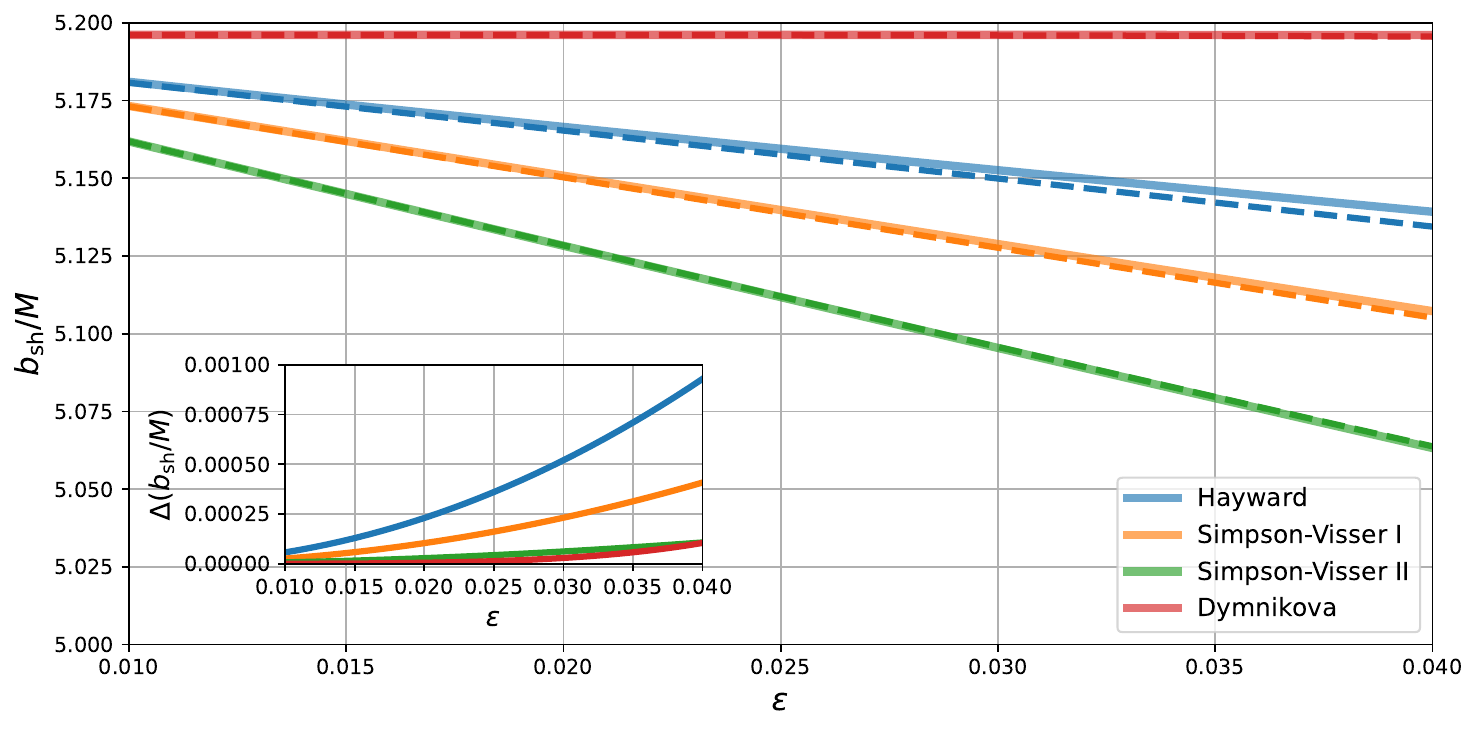}
    \caption{Shadow radius from the effective expression (\ref{CompactCorrectionShadow}) (dashed lines) and from the full metric (solid lines) for the models of Table~\ref{subs models table}. The inset shows the relative difference defined in the text.}
    \label{fig: shadow comparison}
\end{figure}

\paragraph{Shadow.} Finally, all examples can be used to test the effective expression (\ref{CompactCorrectionShadow}) for the leading deformation of the shadow. Figure~\ref{fig: shadow comparison} compares, as functions of (small) $\epsilon$, the approximation $\bsh^{\text{app}}/\ma$ obtained from (\ref{CompactCorrectionShadow}) (dashed lines) with the numerical value $\bsh^{\text{exact}}/\ma$ computed from the full metric (solid lines). The inset shows the relative difference $\Delta (\bsh/\ma)=\left|\frac{\bsh^{\text{exact}}-\bsh^{\text{app}}}{\bsh^{\text{exact}}}\right|$, which is below $0.1\%$ for small $\epsilon$ in all cases. This confirms that (\ref{CompactCorrectionShadow}) correctly captures the leading contribution to the shadow. Notably, this also holds for the Dymnikova black hole, for which approximations at finite order of the Padé approximant are not expected to be analytic in $\mathfrak{c}$ (or $\epsilon$). The intermediate results of Table~\ref{Tab:ValuePade} can therefore not be used in this case, but the final expression (\ref{CompactCorrectionShadow}) apparently re-packages the leading contributions correctly.

\subsection{Dilaton black hole}\label{Sect:DilatonBH}
All models of the previous Subsection have a single deformation function, \emph{i.e.} they satisfy $f=h$ in (\ref{SpStMetric}). We now consider an example without this restriction. The \emph{dilaton-axion} system studied in~\cite{Garcia95} admits a black hole solution which, in the absence of the axion field and of rotation, has the line element
\begin{equation}
    \dd s^2 = - \frac{\rd - 2 \mu}{\rd + 2 b} \dd t^2 + \frac{\rd + 2 b}{\rd - 2 \mu} \dd \rd^2 + (\rd^2 + 2 b \rd) \dd \Omega_2 \qq{with} r^2 = \rd^2 + 2 b \rd \qq{and} M=\mu + b \ .\label{AxioDil}
\end{equation}
In the coordinates of (\ref{SpStMetric}), with $r=2\ma x$, the metric functions are
\begin{align}
&h=1+\frac{\eta-\sqrt{\eta^2+4x^2(1+\eta)^2}}{2x^2(1+\eta)}\,,&&f=\frac{\eta^2+4x^2(1+\eta)^2}{4x^2(1+\eta)^2}\,\left[1+\frac{\eta-\sqrt{\eta^2+4x^2(1+\eta)^2}}{2x^2(1+\eta)}\right]\ , \notag\\
&\text{with}\qquad  \eta=\frac{b}{\mu}=\epsilon(2+\epsilon)\ ,\label{MetDilatonBH}
\end{align}
where $\epsilon$ is defined as in (\ref{EpsEq}). The coefficients (\ref{RescaledMetricFunctions}) are
\begin{align}
&\theta_2=\frac{\epsilon  (\epsilon +2) (\epsilon  (\epsilon +2)+2) \left(\sqrt{(\epsilon  (\epsilon +2)+2)^2}-\epsilon  (\epsilon +2)\right)}{16 (\epsilon +1)   \sqrt{(\epsilon  (\epsilon +2)+2)^2}}=\frac{\epsilon}{4}-\frac{\epsilon^2}{8}+\mathcal{O}(\epsilon^3)\ ,\notag\\ &
\theta_4=-\frac{\epsilon}{12}+\frac{\epsilon^2}{96}+\mathcal{O}(\epsilon^3)\ ,\nonumber\\
&\xi_2=\frac{\epsilon  (\epsilon +2) (\epsilon  (\epsilon +2)+2) \left(3 \epsilon  (\epsilon +2) \left(\epsilon  (\epsilon +2)-\sqrt{(\epsilon  (\epsilon
   +2)+2)^2}+2\right)+4\right)}{64 (\epsilon +1)^3} =\frac{\epsilon}{4}-\frac{3\epsilon^2}{8}+\mathcal{O}(\epsilon^3) \ ,\nonumber\\
&\xi_4=-\frac{\epsilon}{12}+\frac{5\epsilon^2}{32}+\mathcal{O}(\epsilon^3)\ .
\end{align}
\begin{figure}[t]
    \centering
    \includegraphics[width=0.9\textwidth]{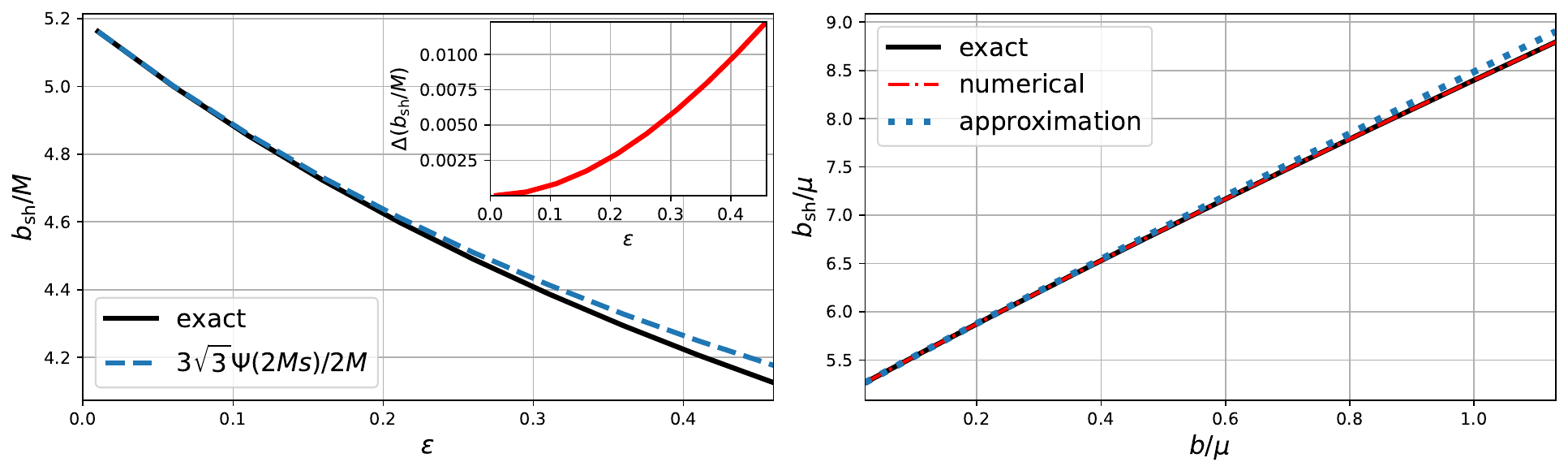}
    \caption{Dilaton black hole (\ref{MetDilatonBH}). Left: shadow radius from (\ref{CompactCorrectionShadow}) (dashed blue) and from the full metric (solid black) as functions of $\epsilon$, with their relative difference in the inset. Right: analytic result (\ref{DilatonAnalyticShadow}) (solid black), numerical computation (red) and approximation (\ref{CompactCorrectionShadow}) (dashed blue) over a wide range of the parameter $b/\mu$ of (\ref{MetDilatonBH}).}
    \label{fig: dilaton plots}
\end{figure}
All of them are of order $\mathcal{O}(\epsilon)$, in agreement with the scaling (\ref{IllumScaling}). Tables~\ref{Tab:PositionPade} and~\ref{Tab:ValuePade} can therefore be used directly, and give for the photon radius and the shadow
\begin{center}
\begin{tabular}{|c|c||c|c|c|}\hline
&&&\\[-10pt]
$N$ & $M$ & $\sigma_1\,\mathfrak{c}$ & $\tau_1\,\mathfrak{c}$ \\[2pt]\hline
&&&\\[-8pt]
2 & 1 & $-\frac{2}{3}\,(2+\sqrt{3})\,\epsilon+\mathcal{O}(\epsilon^2)$ &  $-\frac{4}{3}\,(1+\sqrt{3})\,\epsilon+\mathcal{O}(\epsilon^2)$\\[4pt]\hline
&&&\\[-10pt]
2 & 2 & $-\frac{6}{13}\,(4+\sqrt{3})\,\epsilon+\mathcal{O}(\epsilon^2)$ & $-\frac{4}{3}\,\sqrt{\frac{10}{3}+2\sqrt{3}}\,\epsilon +\mathcal{O}(\epsilon^2)$\\[4pt]\hline
&&&\\[-8pt]
2 & 3 & $-\frac{8}{3}\,\epsilon+\mathcal{O}(\epsilon^2)$ & $-2\,\sqrt{3}\,\epsilon+\mathcal{O}(\epsilon^2)$ \\[4pt]\hline
&&&\\[-8pt]
3 & 3 & $-\frac{8}{3}\,\epsilon+\mathcal{O}(\epsilon^2)$ & $-2\,\sqrt{3}\,\epsilon+\mathcal{O}(\epsilon^2)$\\[4pt]\hline
\end{tabular}
\end{center}
These values coincide with those found for the Simpson-Visser II black hole. The reason is that the two geometries agree to the relevant order,
\begin{align}
&\theta_{2n}^{\text{axion-dilaton}}-\theta_{2n}^{\text{Simpson-Visser II}}=\mathcal{O}(\epsilon^3)\,,&&\text{and} &&\xi_{2n}^{\text{axion-dilaton}}-\theta_{2n}^{\text{Simpson-Visser II}}=\mathcal{O}(\epsilon^2)\,,\label{DilatonSimilarity}
\end{align}
and therefore give identical results at leading order in $\epsilon$. An analytic expression for the shadow of the geometry (\ref{AxioDil}) was derived in~\cite{Wei:2013kza},
\begin{align}
\frac{\bsh}{\ma}=\frac{1}{\sqrt{2}(1+\eta)}\,\sqrt{27 + 36 \eta + 8 \eta^2 + (9 + 8 \eta)^{3/2}}=3\sqrt{3}-2\sqrt{3}\,\epsilon+\mathcal{O}(\epsilon^2)\,,\label{DilatonAnalyticShadow}
\end{align}
in agreement with the approximations of Table~\ref{Tab:ValuePade} for sufficiently large $(N,M)$. A numerical computation of the shadow, based on a different effective approximation, can be found in~\cite{Rezzolla:2014mua}. As shown in Figure~\ref{fig: dilaton plots}, the effective expression (\ref{CompactCorrectionShadow}) approximates the shadow very well over a large range of $\epsilon$.

\begin{figure}[ht]
    \centering
    \includegraphics[width=0.4\textwidth]{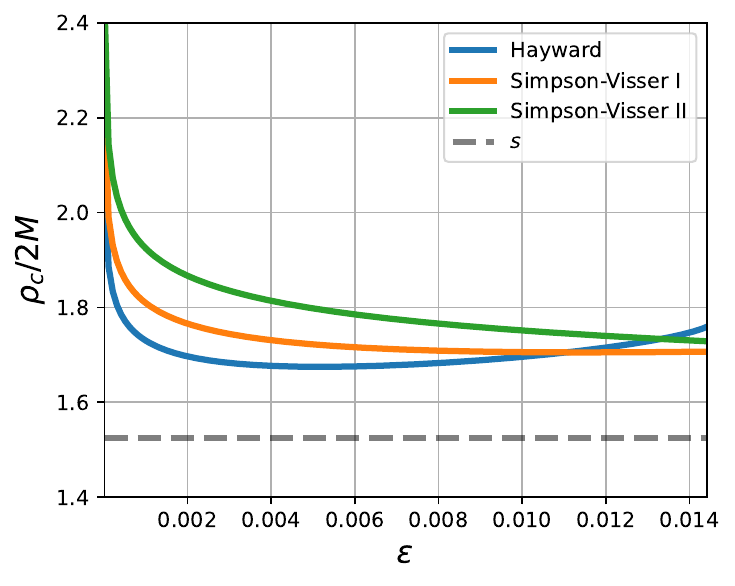}
    \caption{Estimated radius of convergence of the series (\ref{SeriesExpansionMetric}) for the Hayward, Simpson-Visser I and Simpson-Visser II black holes, as a function of $\epsilon$. The dashed black line indicates the value of $\mathfrak{s}$ in (\ref{Svalue}).}
    \label{fig:ShadowLimit}
    \includegraphics[width=\textwidth]{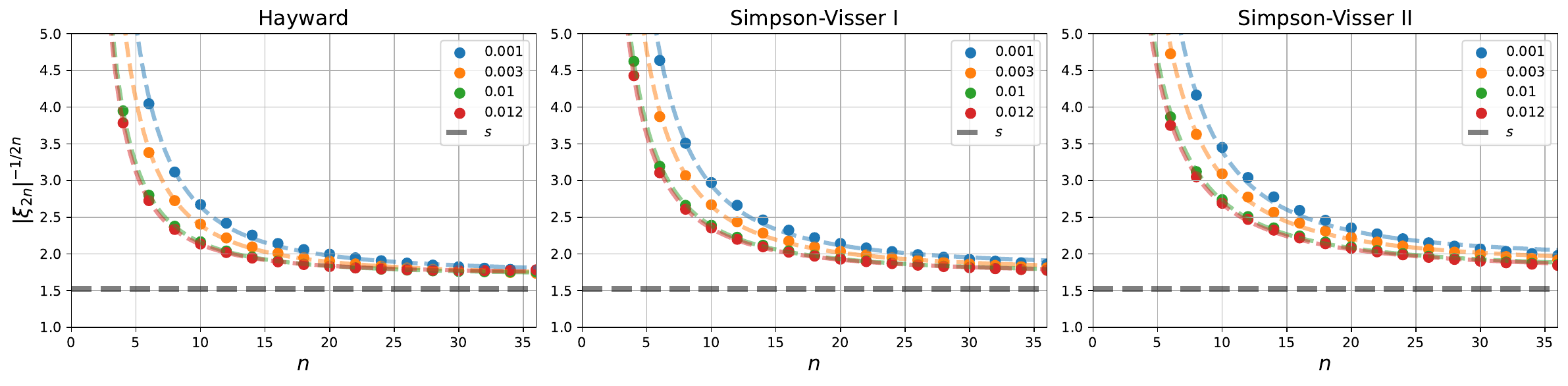}
    \caption{Behaviour of $|\xi_{2n}|^{-\frac{1}{2n}}$ for large $n$ and several values of $\epsilon$, for the Hayward (left), Simpson-Visser I (middle) and Simpson-Visser II (right) black holes. Dashed coloured lines: fits of the form $a+\frac{b}{x^c}$. Dashed black line: value of $\mathfrak{s}$ in (\ref{Svalue}).}
    \label{fig:ConvergenceExamples}
\end{figure}

\section{Convergence}\label{Sect:Convergence}
The derivation of the effective expression (\ref{CompactCorrectionShadow}) in Section~\ref{Sect:PadePhoton} implicitly assumed that the Taylor series (\ref{SeriesExpansionMetric}) of the deformation function $\Psi$ converges at $\rho=2\ma\mathfrak{s}$. To assess whether this is justified, we estimate numerically the radii of convergence for the Hayward and the two Simpson-Visser geometries. Figure~\ref{fig:ConvergenceExamples} shows $|\xi_{2n}|^{-\frac{1}{2n}}$ up to $n=36$ for several values of $\epsilon$. The dashed lines are numerical fits of the form $a+\frac{b}{x^c}$ (with $a,b,c\in\mathbb{R}$ and $c>0$), such that $a$ provides an estimate of the radius of convergence $\rho_c$ of $\Psi$. The result of a more systematic analysis is shown in Figure~\ref{fig:ShadowLimit}, where $\rho_c$ is plotted as a function of $\epsilon$ and compared with $\mathfrak{s}$ (dashed line). At least for small $\epsilon$, $\mathfrak{s}$ lies within the radius of convergence, which justifies the resummation leading to~(\ref{CompactCorrectionShadow}). Because of the similarity (\ref{DilatonSimilarity}) between the axion-dilaton and the Simpson-Visser II geometries at leading order in $\epsilon$, we expect the same conclusion to hold for the example of Section~\ref{Sect:DilatonBH}.

%% file: subfiles/GeodesicMotion.tex
\typeout{START baseline=\the\baselineskip, jot=\the\jot}
\chapter{Geodesic Motion and EMD Coefficients}\label{chapGeo}
In Chapter~\ref{chap4}, we have discussed how to extend the EMD framework to compute observables at a certain distance to the (simple) event horizon of a static and spherically symmetric black hole. As a concrete example, we study the black hole shadow in Section~\ref{Sect:PadePhoton}, which is related to the photon radius that encodes information about the motion of massless particles around the black hole. We demonstrate that $b_{\text{sh}}$, which can be determined from observations at a large distance to the black hole, allows to reconstruct (approximately) the metric deformation function $\Psi$ in eq.~(\ref{h-f-emd-def}) at a particular distance to the horizon. While very encouraging, this result raises the question whether further aspects of the deformed metric can be reconstructed from experimental measurements from outside of the black hole. In this Section we concretely analyse the question whether the coefficients $\theta_n$ and $\xi_n$ in (\ref{SeriesDistanceRho}) (or their equivalents in some other scheme) can be determined from physical measurements outside the horizon of the black hole. In order to answer this question, we review a Gedankenexperiment proposed in \cite{Hohenegger:2025dic}, that allows to extract the physical deformation coefficients of an EMD, describing a static, spherically symmetric black hole geometry with a simple horizon. 
\section{Metric Deformations}
\subsection{Gedankenexperimente}
\begin{figure}[htbp]
\begin{center}
\begin{tikzpicture}
\draw[dashed] (0,0) -- (8,0);
\node at (0,0) {$\bullet$};
\node at (0,0.3) {\footnotesize center of the BH};
\node at (8,0) {$\bullet$};
\node at (8,0.3) {\footnotesize spacecraft};
\draw [domain=-20:20] plot ({2*cos(\x)}, {2*sin(\x)});
\node at (2.6,0.75) {\footnotesize horizon};
\node at (4,0.3) {\footnotesize probe};
\draw[thick,red] (4,0) -- (4.1,0);
\draw[->,thick] (4,0) -- (3.5,0); 
\node at (4,0) {$\bullet$};
\node at (0,-0.3) {\footnotesize $0$};
\node at (2.25,-0.2) {\footnotesize $\zh$};
\node at (4,-0.3) {\footnotesize $\zp$};
\node at (8,-0.3) {\footnotesize $\zs$};
\end{tikzpicture}
\end{center}\caption{Schematic depiction of the Gedankenexperiment discussed in this Section: a spacecraft (at $\zs$) releases a probe (at position $\zp$) that is freely falling towards the center of the black hole. (adapted from~\cite{Hohenegger:2025dic})
}
\label{Fig:ProbeFalling}
\end{figure}
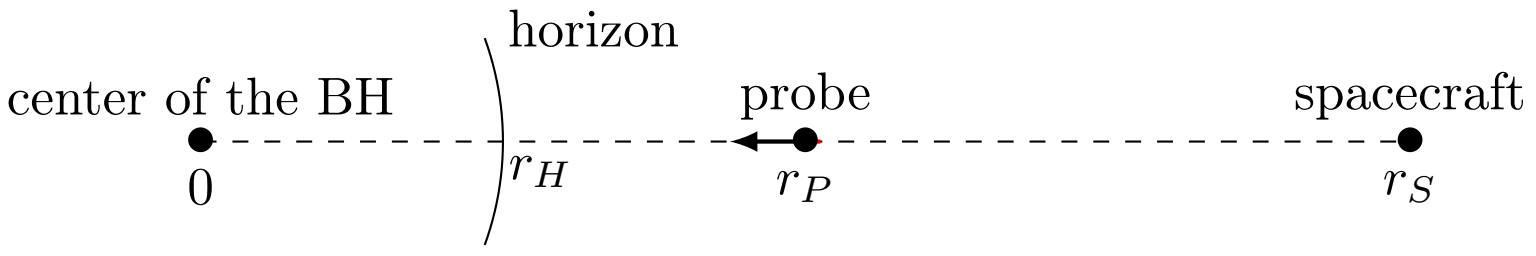

In order to extract the expansion coefficients of the metric functions $\Phi$ and $\Psi$ we need to devise measurements that are susceptible to the local metric in the vicinity of the black hole. As we have reviewed in Section~\ref{Sect:PhotonRadius}, the geodesic motion of particles (both massive and massless) around the black hole is indeed highly sensitive to the metric in general and the functions $f$ and $h$ in particular (in the spherically symmetric and static case). Our strategy to extract the expansion coefficients from physical measurements, is to continuously observe suitable particle movements in the gravitational field of the black hole: for example, in \cite{Herrera-Aguilar:2015kea,Herrera-Aguilar:2012jhe,Becerril:2016qxf} it was proposed to determine parameters of Kerr black holes (notably the mass) through red-shift data from photons escaping the black hole (see \cite{ShankarKuniyal:2017sxc,Uniyal:2017yll,Sheoran:2017dwb,Herrera-Aguilar:2018uxd,Kraniotis:2019ked,Becerril:2020fek,Banerjee:2022him} for similar discussions in the context of other geometries). In \cite{Hohenegger:2025dic}, this idea was further formalised by devising a concrete Gedankenexperiment which allows to reproduce series expansion of the EMD metric functions from two concrete measurements. The schematic setup is shown in Figure~\ref{Fig:ProbeFalling}: a spacecraft, positioned at the fixed radial distance $\zs$ to the black hole, releases a probe, which follows a radial, free-falling geodesic towards the center of the black hole (with radial position $\zp$). During its descend to the horizon, the probe transmits various telemetric data back to the spacecraft, which allow to reconstruct the metric along the path of the probe.

In the following, we shall review the details of the computation in \cite{Hohenegger:2025dic}: due to the nature of the setup in Figure~\ref{Fig:ProbeFalling}, we prefer to parametrise the reconstructed metric functions $\Psi$ and $\Phi$ (see (\ref{h-f-emd-def})) in terms of the proper time $\tau$ of an observer that is freely falling towards the center of the black hole. This was first defined in eq.~(\ref{FreeFallingTime}) in the context of charged spherically symmetric black holes.\footnote{In \cite{Hohenegger:2025dic}, it is explained how to express our results in terms of an expansion of $\Psi$ and $\Phi$ in powers of the radial distance to the horizon, as in (\ref{SeriesDistanceRho}).} Furthermore, to be precise, following the framework outlined in Section~\ref{chap3}, we shall assume that the functions $f$ and $h$ in (\ref{h-f-emd-def}) have simple zeroes at the position of the (outermost) event horizon $\zh$ of the black hole and that $f(r)>0$ and $h(r)>0$ for $r>\zh$.
\subsection{Free-Falling Probe and Observer}\label{Sect:FreeFallingObserver}
We start by describing and parametrising the setup in Figure~\ref{Fig:ProbeFalling}: the spacecraft remains at a fixed position $\zs>\zh$ outside of the horizon of the black hole. Since we assume it to be stationary throughout the experiment, we can write its 4-velocity in the following form\footnote{We use the same notation as in Section~\ref{Sect:Schwarzschild} and \cite{Hohenegger:2025dic}, \emph{i.e.} $x^\mu=(x^t,x^r,x^\theta,x^\varphi)$.}
\begin{align}
&u_S^\mu=\left(\frac{1}{\sqrt{h(\zs)}},0,0,0\right)\,.\label{FourVelSpacecraft}
\end{align}
While the value of $\zs$ (or the corresponding radial distance to the horizon) does not play a role in the following, the factor $\sqrt{h(\zs)}$ can be determined through the redshift factor of a photon (of known frequency) sent from the spacecraft to a point at spatial infinity (\emph{e.g.} to an observer on earth). We shall therefore consider it to be known in the computations below.

The spacecraft launches a (massive) probe towards the center of the black hole, which is capable of emitting photons back to the spacecraft that can be used as communication. To be precise, the probe is simply released (\emph{i.e.} without initial velocity) and follows a radial time-like geodesic towards the center of the black hole. At the moment of its release, the probe starts an internal clock to measure its proper time $\tau$, which, as mentioned before, shall be used to parametrise its trajectory: we denote $\zp(\tau)$ (in terms of Schwarzschild coordinates) the position of the probe and its 4-velocity is given by (see Section~\ref{Sect:PhotonRadius})
\begin{align}
u_P^\mu=\left(\frac{1}{h(\zp)}\,,-\sqrt{\frac{f(\zp)}{h(\zp)}(1-h(\zp))}\,,0\,,0\right)\,.\label{ProbeVelGeo}
\end{align}
Here the sign in the component $u_P^r$ is compatible with a massive particle falling \emph{into} the black hole. Furthermore, the time coordinate is chosen such that $u^t_P\,h(\zp)=1$. The probe shall continuously communicate through light signals its proper time $\tau$ (\emph{i.e.} the reading of its internal clock) and other telemetric data to the spacecraft. These will reach the spacecraft, as long as the probe is outside of the horizon. 

The position of the probe can be linked to the metric deformation function $\Psi$: indeed, in the current case, the differential equation (\ref{defddtau}) for the proper time takes the form
\begin{align}
(\zp-\Phi)\,\Psi=\left(\frac{\dd\zp}{\dd\tau}\right)^2\,\zp\,(\zp-\Psi)\,,\label{DiffEqExpand}
\end{align}
which is a relation between the functions $\Phi$ and $\Psi$ and the position of the probe $\zp$ (which is a priori unknown). As in previous examples, our strategy in solving (\ref{DiffEqExpand}) is to consider series expansions in proximity to the horizon. To this end, we first introduce slightly different coordinates: let $\tah$  denote the proper time at which the probe crosses the horizon and its light signals can no longer reach the spacecraft. The latter can therefore reconstruct $\tah$ as the (asymptotically)\footnote{The light signals emitted from the event horizon take an infinitely long time to reach the spacecraft. We shall therefore assume in the following that the latter can observe the signals stemming from the probe for a sufficiently long time to reliably extrapolate the telemetry. We shall see this in a concrete example later on.} last proper time that has been submitted by the probe. We then introduce the following series expansions in powers of the \emph{remaining} (proper) time $x$ for the probe to cross the horizon
\begin{align}
&\zp=\zh+\sum_{n=1}^\infty \alpha_n\, x^n\,,&&\Phi=\zh+\sum_{n=1}^\infty \varphi_n\, x^n\,,&&\Psi=\zh+\sum_{n=1}^\infty \psi_n\, x^n\,,&&\text{with} &&x:=\tah-\tau\,,\label{SeriesExpansionsDeformations}
\end{align}
with $\{\alpha_n\}$, $\{\varphi_n\}$ and $\{\psi_n\}$ as well as $\zh$ unknown. We implicitly assume these series to have finite radius of convergence, such that they represent well-defined functions sufficiently close to the horizon. Substituting (\ref{SeriesExpansionsDeformations}) into (\ref{DiffEqExpand}) we find 
{\allowdisplaybreaks
\begin{align}
\sum_{p=1}^\infty \zh &(\alpha_p-\varphi_p)x^p+\sum_{p=2}^\infty x^p\sum_{n=1}^{p-1}(\alpha_n-\varphi_n)\psi_{p-n}\nonumber\\
&=\sum_{p=1}^\infty x^p\sum_{r=0}^{p-1}\sum_{n=1}^{r+1}n(r-n+2)\zh\alpha_n\alpha_{r-n+2}(\alpha_{p-r}-\psi_{p-r})\nonumber\\
&\hspace{0.5cm}+\sum_{p=2}^\infty x^p\sum_{r=0}^{p-2}\sum_{n=1}^{r+1}\sum_{\ell=1}^{p-r-1} n(r-n+2)\alpha_n\alpha_{r-n+2}\alpha_{p-r-\ell}(\alpha_\ell-\psi_\ell)\,.\label{MasterSeriesProperTime}
\end{align}
}
At each order in $x$, this equation represents relations for the unknown coefficients. For example, matching terms to order $\mathcal{O}(x^1)$, we require
\begin{align}
(\alpha_1-\varphi_1)=\alpha_1^2(\alpha_{1}-\psi_{1})\,,\label{RelAlpha1}
\end{align}
which allows to express $\alpha_1$ in terms of $\varphi_1$ and $\psi_1$. For general values of $\varphi_1$ and $\psi_1$,\footnote{We briefly remark that identifying $\varphi_1=\psi_1$ (which is a necessary condition for $f=h$) imposes $\alpha_1=\pm 1$ and $\psi_1\neq \alpha_1$ and $\varphi_1\neq \alpha_1$ in general.} (\ref{RelAlpha1}) is a cubic equation for $\alpha_1$, but not all solutions represent meaningful black hole geometries: as explained in more detail in Remark~\ref{rem:finitecurvature}, the finiteness of the Ricci- and Kretschmann scalar at the horizon (imposed by the finiteness of the first and second derivative of $\tau$) require the stronger condition (\ref{SectStrongerEq}) (compatible with (\ref{RelAlpha1})).

\begin{center}
\remarkbox{\rem\label{rem:finitecurvature}{Finiteness of Curvature Invariants}}{\parbox{16cm}{
In various places (see \emph{e.g.} Section~\ref{SectMetricFunctCurv} or Section~\ref{Sect:Charged_Curvature}) of this review, we have used the finiteness of scalar curvature invariants, such as the Ricci- or Kretschmann scalar (see eq.~(\ref{eq: ricci scalar h e f}) and (\ref{KretschDef}) respectively) outside of the horizon of the black hole, as conditions for the consistency of the EMD description. Notably for EMDs based on the proper radial distance $\rho$ to the horizon, this yields non-trivial conditions, since the derivative of $\rho$ with respect to the radial coordinate $r$ (see (\ref{DiffEqDistance})) is singular at the horizon. In contrast, assuming that both metric functions $f$ and $h$, have simple zeroes at the horizon $r=\zh$, $\frac{\dd\tau}{\dd r}\big|_{r=\zh}$ is in general finite, which, however, is not necessarily true for the the second derivative.
Indeed, using (\ref{SeriesExpansionsDeformations}) (and implicitly $\varphi_1\neq\alpha_1\neq\psi_1$) we find to leading order in~$x$
\begin{align}
\frac{\dd^2\tau}{\dd z^2}=\frac{\left(1+\alpha_1\sqrt{\frac{\alpha_1-\psi_1}{\alpha_1-\varphi_1}}\right)(\psi_1-\varphi_1)}{2(\alpha_1-\psi_1)^2\sqrt{\frac{\alpha_1-\psi_1}{\alpha_1-\varphi_1}} \,x}+\mathcal{O}(x^0)\,.
\end{align}
The singular leading term is absent for
\begin{align}
&\alpha_1\sqrt{\tfrac{\alpha_1-\psi_1}{\alpha_1-\varphi_1}}=-1&&\text{or} &&\varphi_1=\psi_1\,.\label{SectStrongerEq}
\end{align}
The condition (\ref{RelAlpha1}), stemming to leading order from (\ref{DiffEqExpand}), is a consequence of the first relation, while it implies $\alpha_1=\pm1$ for the second relation. Since finiteness of the first and second derivative of $\tau$ implies finiteness of the Ricci- and Kretschmann scalar at the horizon, we do not find further conditions on the coeffficients $\varphi_1$ or $\psi_1$. Instead, finiteness of the curvature scalars fixes the coefficient $\alpha_1$ in a fashion which is compatible with (\ref{RelAlpha1}).
}}
\end{center}

\noindent
Higher orders $\mathcal{O}(x^p)$ in (\ref{MasterSeriesProperTime}) (\emph{i.e.} $p\geq 2$) leads to further relations beyond (\ref{RelAlpha1}), namely
\begin{align}
\zh(\alpha_p-\varphi_p)+&\sum_{n=1}^{p-1}(\alpha_n-\varphi_n)\psi_{p-n}=\sum_{r=0}^{p-1}\sum_{n=1}^{r+1}n(r-n+2)\zh\alpha_n\alpha_{r-n+2}(\alpha_{p-r}-\psi_{p-r})\nonumber\\
&+\sum_{r=0}^{p-2}\sum_{n=1}^{r+1}\sum_{\ell=1}^{p-r-1} n(r-n+2)\alpha_n\alpha_{r-n+2}\alpha_{p-r-\ell}(\alpha_\ell-\psi_\ell)\,.
\end{align}
We can re-arrange this expression by writing all terms containing $\alpha_p$ on the left and all terms with $\alpha_{p'}$ (for $p'<p$) on the right
\begin{align}
\alpha_p&\,\zh\left(1-2p \alpha_{1}(\alpha_{1}-\psi_1)-\alpha_1^2\right)\nonumber\\
&=\zh (\varphi_p-\alpha_1^2\psi_p)-\sum_{n=1}^{p-1}(\alpha_n-\varphi_n)\psi_{p-n}+\sum_{n=2}^{p-1}n(p-n+1)\zh\alpha_n\alpha_{p-n+1}(\alpha_{1}-\psi_{1})\nonumber\\
&\hspace{0.5cm}+\sum_{r=1}^{p-2}\sum_{n=1}^{r+1}n(r-n+2)\zh\alpha_n\alpha_{r-n+2}(\alpha_{p-r}-\psi_{p-r})\nonumber\\
&\hspace{0.5cm}+\sum_{r=0}^{p-2}\sum_{n=1}^{r+1}\sum_{\ell=1}^{p-r-1} n(r-n+2)\alpha_n\alpha_{r-n+2}\alpha_{p-r-\ell}(\alpha_\ell-\psi_\ell)\hspace{2cm}\forall p\geq 2\,.\label{HigherAlphaOrders}
\end{align}
Since this relation determines $\alpha_p$ in terms of finitely many coefficients $\alpha_{p'}$ with lower index ($p'<p$), it allows to iteratively compute all $\alpha_p$ for $p\geq 2$.
\subsection{Transmission of Telemetry}
Relations (\ref{RelAlpha1}) and (\ref{HigherAlphaOrders}) (and (\ref{SectStrongerEq})) express the coefficients $\{\alpha_n\}$ determining the position of the probe $\zp$ in terms of the expansion coefficients of the metric functions $\Phi$ and $\Psi$. In order to obtain information about the latter (\emph{i.e.} the coefficients $\{\varphi_n\}$ and $\{\psi_n\}$), the probe needs to determine information about the geometry back to the spacecraft. Indeed, we consider the case that the probe can emit photons and transmit information about their initial conditions to the latter. Concretely, we consider two particular signals (called telemetry signals):

\begin{figure}[htbp]
\begin{center}
\begin{tikzpicture}
\draw[dashed] (0,0) -- (8,0);
\node at (0,0) {$\bullet$};
\node at (0,0.3) {\footnotesize center of the BH};
\node at (8,0) {$\bullet$};
\node at (8,0.3) {\footnotesize spacecraft};
\draw [domain=-20:20] plot ({2*cos(\x)}, {2*sin(\x)});
\node at (2.6,0.75) {\footnotesize horizon};
\node at (4,0.3) {\footnotesize probe};
\draw[thick,red] (4,0) -- (4.1,0);
\draw[decorate, decoration={coil, aspect=0},thick,red] (4.1,0) -- (5.9,0);
\draw[->,thick,red] (5.9,0) -- (6,0);
\node[red,rotate=45] at (5.95,0.85) {\footnotesize emitted};
\node[red,rotate=45] at (6.2,0.6) {\footnotesize photon};
\draw[->,thick] (4,0) -- (3.5,0); 
\node at (4,0) {$\bullet$};
\node at (0,-0.3) {\footnotesize $0$};
\node at (2.25,-0.2) {\footnotesize $\zh$};
\node at (4,-0.3) {\footnotesize $\zp$};
\node[red] at (6,-0.3) {\footnotesize $\zgam$};
\node at (8,-0.3) {\footnotesize $\zs$};
\end{tikzpicture}
\end{center}
\caption{Schematic depiction of the telemetry signal $S_1$: the free-falling probe sends a photon along a radial trajectory to the spacecraft. (adapted from~\cite{Hohenegger:2025dic})}
\label{Fig:RadialPhoton}
\end{figure}
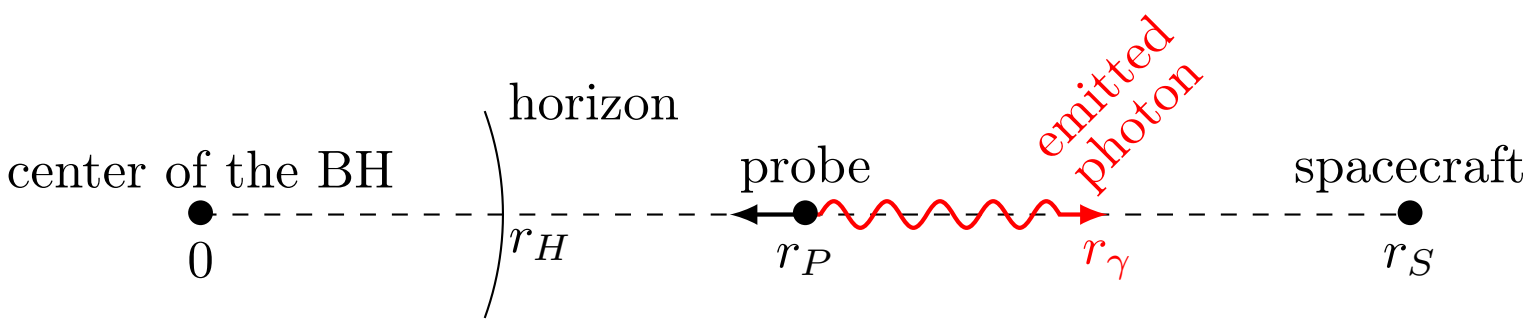

\begin{figure}[htbp]
\begin{center}
\begin{tikzpicture}
\draw[dashed] (0,0) -- (8,0);
\node at (0,0) {$\bullet$};
\node at (0,0.3) {\footnotesize center of the BH};
\draw [domain=-20:20] plot ({2*cos(\x)}, {2*sin(\x)});
\node at (2.6,0.75) {\footnotesize horizon};
\node at (4,0.3) {\footnotesize probe};
\draw[thick,red] (4,0) -- (4.1,0);
\draw[thick,red,dashed] (5.1,0.25) to [out=10,in=235] (7.72741, 2.07055);
\draw[decorate, decoration={coil, aspect=0},thick,red] (4.1,0) -- (5,0.225);
\draw[thick,red,->] (5,0.225) -- (5.1,0.25);
\node[red] at (5.2,0.9) {\footnotesize emitted};
\node[red] at (5.15,0.55) {\footnotesize photon}; 
\node at (4,0) {$\bullet$};
\node at (0,-0.3) {\footnotesize $0$};
\node at (2.25,-0.2) {\footnotesize $\zh$};
\node at (4,-0.3) {\footnotesize $\zp$};
\node[red] at (6.85,0.6) {\footnotesize $\zgam$};
\node at (7.7,-0.3) {\footnotesize $\zs$};
\draw [domain=-5.5:20] plot ({8*cos(\x)}, {8*sin(\x)});
\draw [domain=-5.5:12,dashed] plot ({6.25*cos(\x)}, {6.25*sin(\x)});
\node at ({8*cos(15)}, {8*sin(15)}) {$\bullet$};
\node at (9.225,2.1) {\footnotesize photon detected};
\node at (9,1.7) {\footnotesize by spacecraft};
\node at (6.225,-0.85) {\footnotesize photon sphere};
\draw[ultra thick,->] ({8*cos(15)}, {8*sin(15)}) -- ({8*cos(19)}, {8*sin(19)});
\draw[->,thick] (4,0) -- (3.5,0); 
\end{tikzpicture}
\end{center}
\caption{Schematic depiction of the telemetry signal $S_2$: the probe, still following a radial free-falling geodesic, emits a photon at an angle, which is detected by the spacecraft on a circular orbit (with fixed radial coordinate $\zs$). (adapted from~\cite{Hohenegger:2025dic})}
\label{Fig:AnglePhoton}
\end{figure}
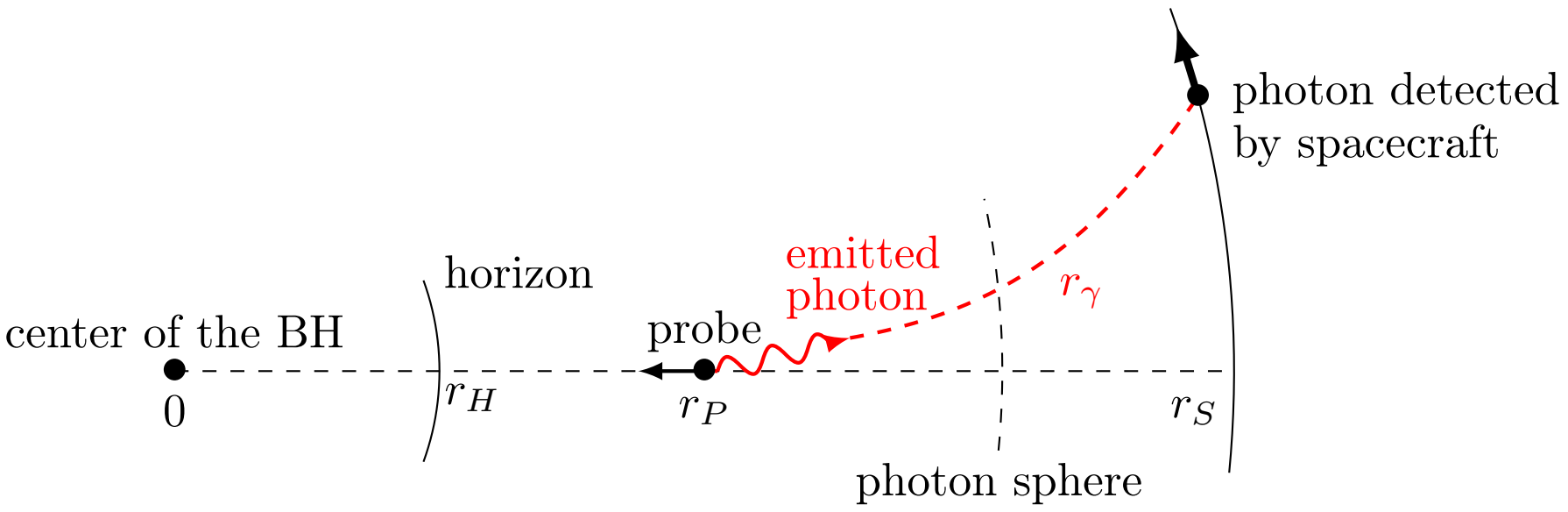

\begin{itemize}
\item {\bf telemetry signal $S_1$:} As schematically shown in Figure~\ref{Fig:RadialPhoton}, we shall first assume that the probe (following a free-falling radial geodesic at $\zp$) emits a photon along a radial geodesic (parametrised by $\zgam$) towards the spacecraft at $\zs$. Along with this photon, the probe transmits the proper time $\tau$ (which can be converted into $x$ by the spacecraft), at which the photon was emitted along with is emission frequency $\omega_e$. The signal $S_1$ (see (\ref{Telemetry1}) below), then consists of the quotient of the detected frquency by the emitted one (corrected by the redshift factor of the spacecraft). 
\item {\bf telemetry signal $S_2$:}  Since, as we shall see in detail below, the signal $S_1$ only encodes information about the function $\Psi(\zp)$ (i.e. the coefficients $\{\psi_n\}$), a second signal is required to determine the function $\Phi(\zp)$. To this end, as schematically shown in Figure~\ref{Fig:AnglePhoton} we consider the probe (following a free-falling radial geodesic at $\zp$) emitting photons in the equatorial plane $\theta=\pi/2$ of the black hole at non-trivial angles relative to the radial direction. Such photons can be recovered by the spacecraft (which is on a circular orbit with radial coordinate $\zs$) only up to a limiting emission angle. Indeed, if the latter is too large, the photons cannot leave the gravitational field of the black hole and are forced on a circular orbit around the black hole (or fall back into it). The telemetry signal $S_2$ (see (\ref{Telemetry3})) consists of the critical angle under which photons can still reach the spacecraft.

\end{itemize}

\subsubsection{Telemetry Signal \texorpdfstring{$S_1$}{S1}: Radial Emission and Redshift Factor}
We shall denote the radial coordinate of the emitted photon $\zgam$ and, since it is emitted with no angular momentum ($L_\gamma=0$ in (\ref{ConservedQuantities})) the associated 4-velocity is (see Computation~\ref{comp:Photon4Velocity})
\begin{align}
u_\gamma^\mu(\zgam)=\left(\frac{E_\gamma}{h(\zgam)}\,,\sqrt{\frac{f(\zgam)}{h(\zgam)}}\,E_\gamma\,,0\,,0\right)\,.\label{ugRadial}
\end{align}

\begin{center}
\compbox{\comp\label{comp:Photon4Velocity}{4-Velocity of a Massless Particle}}{\parbox{16cm}{We consider a massless (test-)particle with world-line $x_\gamma^\mu(\lambda)$, with $\lambda$ some suitable affine parameter. Following Section~\ref{Sect:PhotonRadius}, the geodesic equation reads
\begin{align}
&\frac{\dd u_\gamma^\mu}{\dd\lambda}=-{\Gamma^\mu}_{\nu\rho}\,u_\gamma^\nu\,u_\gamma^\rho\,,&&\text{with} &&u_\gamma^\mu=\frac{\dd x_\gamma^\mu}{\dd\lambda}\,,&&\text{and} &&g_{\mu\nu}\,u_\gamma^\mu\,u_\gamma^\nu=0\,,\label{GeodesicEquation}
\end{align}
where $u_\gamma^\mu$ is the 4-velocity of the particle. As explained in eq.~(\ref{geodesic energy}) and (\ref{geodesic angular momentum}), the components for $\mu=t$ and as well as the angular equations can be solved by setting
\begin{align}
&E_\gamma=u^t_\gamma\,h(r_\gamma)\,,&&\text{and} &&L_\gamma=r_\gamma^2\,u_\gamma^\varphi\,,\label{ConservedQuantities}
\end{align}
where $E_\gamma$ is the energy and $L_\gamma$ the angular momentum of the particle respectively. In order to solve the remaining radial geodesic equation, we consider a trajectory in the equatorial plane $\theta=\pi/2$, such that
\begin{align}
&0=-h\,(u^t_\gamma)^2+\frac{(u^r_\gamma)^2}{f}+z^2\,(u_\gamma^\varphi)^2\,,
\end{align}
such that we find the following expression for the 4-velocity
\begin{align}
u_\gamma^\mu=\left(\frac{E_\gamma}{h}\,,E_\gamma\,\sqrt{\frac{f}{h}\,\left(1-\frac{L^2 h}{r^2 E_\gamma^2}\right)}\,,0\,,\frac{L_\gamma}{r^2}\right)\,.\label{DefFourVelocityPhoton}
\end{align}
 
}}
\end{center}

\noindent
The frequency of a photon with 4-velocity $u_\gamma^\mu$ emitted (or detected) by a source with 4-velocity $v^\mu$ is (see \emph{e.g.} \cite{Herrera-Aguilar:2012jhe})
\begin{align}
\omega=-g_{\mu\nu}\,v^\mu\,u_\gamma^\nu\,.\label{DefFrequencyDetect}
\end{align}
such that we can compute the frequencies upon emission (by the probe) and detection (by the spacecraft) of the photon: 
\begin{enumerate}
\item[\emph{(i)}] {\bf emission} at $\zgam=\zp$: an observer co-moving with the probe measures the following emission frequency (\ref{DefFrequencyDetect}):
\begin{align}
\omega_e=h(\zp)\,u^t_P\,u^t_\gamma-\frac{1}{f(\zp)}\,u_P^z\,u_\gamma^z=\frac{E_\gamma}{h(\zp)}\,\left(1+\sqrt{1-h(\zp)}\right)\,,\label{EmissionFrquency}
\end{align}
which (as a function of $\tau$) is continuously transmitted to the spacecraft
\item[\emph{(ii)}] {\bf detection} at $\zgam=\zs$: the space-craft measures the detection frequency:
\begin{align}
\omega_d=h(\zs)\,u_S^t\,u_\gamma^t=\frac{E_\gamma}{\sqrt{h(\zs)}}\,.
\end{align} 
\end{enumerate}
An observer at the spacecraft can therefore compute the following quotient
\begin{align}
S_1=\sqrt{h(\zs)}\,\,\frac{\omega_d}{\omega_e}=\frac{h(\zp)}{1+\sqrt{1-h(\zp)}}\,,\label{Telemetry1}
\end{align}
which is the first telemetry signal. Since, as mentioned before, $\sqrt{h(\zs)}$ can be determined by a separate measurement, $S_1$ can be fully determined and expressed in terms of the metric function $h(\zp)$.

\begin{center}
\remarkbox{\rem\label{rem:receptiontime}{Time at which $S_1$ is received}}{\parbox{16cm}{
The proper time at which the spacecraft receives the telemetry $S_1$ is
\begin{align}
\tau^S_d=-\sqrt{h(\zs)}\int_{\zs}^{\zp}\frac{\dd z}{\sqrt{f(z)\,h(z)}}\left[1+\frac{1}{\sqrt{1-h(z)}}\right]\,.\label{ReceiveTime}
\end{align}
As $f$ and $h$ have a simple zero for $r=\zh$, (\ref{ReceiveTime}) diverges for $\zp=\zh$. This means that the telemetry emitted by the probe from the horizon reaches the spacecraft only after an infinitely long time after the launch of the probe.
}}
\end{center}

\subsubsection{Telemetry Signal \texorpdfstring{$S_2$}{S2}: Equatorial Emission and Critical Angle}\label{Sect:AngleEmission}
Compared to (\ref{ugRadial}), we now consider a photon being emitted with $L_\gamma\neq 0$, such that  $u^\mu_\gamma=\left(u_\gamma^t,u_\gamma^z,0,u_\gamma^\varphi\right)$. To define the angle under which this photon is emitted by the probe, we define the tangent frame to the world-line of the probe by introducing the vierbein
\begin{align}
&e_t^0=\sqrt{h(\zp)}\,,&&e_r^1=\frac{1}{\sqrt{f(\zp)}}\,,&&e_\theta^2=\zp\,,&&e_\varphi^3=\zp\,,
\end{align}
and all remaining components equal to zero. In this frame, the 4-velocity of the photon becomes
\begin{align}
&u_\gamma^a=\left(\frac{E_\gamma}{\sqrt{h(\zp)}}\,,E_\gamma\,\sqrt{\frac{1}{h(\zp)}\left(1-\frac{L^2 h(\zp)}{\zp^2 E_\gamma^2}\right)},0,\frac{L}{\zp}\right)\,,&&\text{with} &&\eta_{ab}\,u_\gamma^a\,u_\gamma^b=0\,.
\end{align}
This allows to define the angle of the emitted photon relative to the radial direction\footnote{As was remarked in \cite{Hohenegger:2025dic}, an equivalent definition of the angle (\ref{CosAngle}) is provided in \cite{Synge1966}.}
\begin{align}
&\parbox{3.5cm}{\begin{tikzpicture}
\node at (-0.25,1.8) {{\footnotesize tangent frame:}};
\draw[dashed] (0,0) -- (2,0);
\node at (0,0) {$\bullet$}; 
\node at (-0.25,-0.25) {$\zp$};
\draw[thick,dashed] (0,0) -- (1.5,1.5);
\draw[decorate, decoration={coil, aspect=0},thick,red,->] (0,0) -- (1.5,1.5);
\draw[thick] (1.5,0) -- (1.5,1.5);
\draw[thick] (1.5,0) -- (0,0);
\draw [domain=0:45] plot ({0.9*cos(\x)}, {0.9*sin(\x)});
\node at (0.6,0.25) {$\ang$};
\node at (0.5,1.1) {$u_\gamma^0$};
\node at (1.85,0.75) {$u_\gamma^3$};
\node at (0.9,-0.3) {$u_\gamma^1$};
\end{tikzpicture}
}&&\text{with} &&\sin\ang=\frac{u_\gamma^3}{u_\gamma^0}=\frac{L}{\zp}\,\frac{\sqrt{h(\zp)}}{E_\gamma}\,\,.\label{CosAngle}
\end{align}
Knowledeg of the angle $\beta$ and the frequency of the photon allows to determine $E_\gamma$ and $L_\gamma$. In the following, however, we shall only require the combination
\begin{align}
\imp:=\frac{L_\gamma}{E_\gamma}=\frac{\zp\,\sin\ang}{\sqrt{h(\zp)}}\,,
\end{align}
which depends on $\zp$. We next consider detection of the photons emitted with different $\beta$ by the spacecraft. The angle (in the reference frame of the black hole), at which the photon trajectory intersects the circular orbit of the spacecraft (at $r=\zs$) can be obtained from
\begin{align}
&\frac{\dd\varphi}{\dd z}=\frac{u_\gamma^\varphi}{u_\gamma^r}\,,&&\text{which integrates to} &&\varphi=\int_{\zp}^{\zs}\frac{\imp}{z^2}\,\frac{\dd z}{\sqrt{\frac{f}{h}\,\left(1-\frac{\imp^2 h}{z^2}\right)}}\,.\label{FormAngleSpacecraft}
\end{align}
In principle, the angle $\varphi$ for a photon emitted under a generic angle $\beta$ can be measured by the spacecraft. However, to express the latter in terms of the black hole geometry (\emph{i.e.} evaluating (\ref{FormAngleSpacecraft})) would require knowledge of the functions $f$ and $h$ for the entire geodesic of the photon between the probe and the spacecraft. In order to avoid this complication, we consider a different observable. We first remark that $\varphi$ in (\ref{FormAngleSpacecraft}) is well-defined only if 
\begin{align}
&\sin^2\ang< \frac{r^2}{h(r)}\,\frac{h(\zp)}{\zp^2}\,,&&\forall r\in[\zp,\zs]\,,\label{RealCond}
\end{align}
Here $U(r)^2=\frac{h(r)}{r^2}$ is the same potential defined in (\ref{PhotonPotentialU}), whose extrema define the photon orbit $r_{\text{ps}}$ and which is related to the black hole shadow $b_{\text{sh}}$ as in (\ref{bphdef}). We assume that $r_{\text{ps}}$ denotes the outermost photon orbit and that $\zp<r_{\text{ph}}<\zs$. The condition (\ref{RealCond}) then requires emission of the photon below a critical angle $\ang_c(\zp)<\tfrac{\pi}{2}$ for the photon to reach the spacecraft and not to continue orbiting the black hole at $r_{\text{ph}}$ (or fall back into the black hole). Assuming that the spacecraft can determine this critical angle $\ang_c$ as a function of $x$ (and that the black hole shadow $b_{\text{sh}}$ can be measured by a distant observer, see \cite{Synge1966,Perlick:2021aok,Cunha:2020azh,DelPiano:2024nrl}), we define the second telemetry signal as
\begin{align}
S_2:=\frac{h(z_{\text{ph}})}{z_{\text{ph}}^2}\,\sin^2\ang_c=\frac{\sin^2\ang_c}{b_{\text{sh}}}
=\frac{h(\zp)}{\zp^2}\,.\label{Telemetry3}
\end{align}

\subsection{Series Expansions and Self-Consistent Solution}
It was demonstrated in \cite{Hohenegger:2025dic} that the telemetry data $S_1$ in (\ref{Telemetry1}) and $S_2$ in (\ref{Telemetry3}) (as functions of $x$), allow to reconstruct the metric of the black hole order by order. To this end, we also expand the telemetry data 
\begin{align}
&S_1=\sum_{n=1}^\infty \sa{n}\,x^n\,,&&\text{and} &&S_2=\sum_{n=1}^\infty \sd{n}\,x^n\,,
\end{align}
with \emph{known} coefficients $\{\sa{n}\}$ and $\{\sd{n}\}$ and use (\ref{SeriesExpansionsDeformations}) with \emph{unknown} coefficients $\{\alpha_n\}$, $\{\varphi_n\}$ and $\{\psi_n\}$ (related by (\ref{RelAlpha1}) and (\ref{HigherAlphaOrders})).

\begin{center}
\compbox{\comp\label{comp:telemetry2}{Conditions from the Telemetry Signals}}{\parbox{16cm}{
{\bf telemetry signal $S_1$:} Assuming implicitly $\Psi>0$ and using (\ref{h-f-emd-def}), we can write $\Psi=\zp\,(S_1-1)^2$. Substituting the series (\ref{SeriesExpansionsDeformations}), we find
\begin{align}
&\zp\,(S_1-1)^2=\left(\zh+\sum_{n=1}^\infty \alpha_n x^n\right)\left(-1+\sum_{n=1}^\infty \sa{n}\,x^n\right)^2\nonumber\\
&=\zh+(\alpha_1-2\zh\sa{1})\,x+\left(\alpha_2-2\zh\sa{2}+(\zh\sa{1}-2\alpha_1)\sa{1}\right)\,x^2\nonumber\\
&\hspace{0.5cm}+\sum_{p=3}^\infty x^p\left[ \alpha_p-2\zh\sa{p}+\sum_{\ell=1}^{p-1}(\zh\,\sa{\ell}\sa{p-\ell}-2\alpha_\ell\,\sa{p-\ell})+\sum_{n=1}^{p-2}\alpha_n\sum_{\ell=1}^{p-n-1}\sa{\ell}\sa{p-n-\ell}\right]\,,\nonumber
\end{align}
and therefore at each order $\mathcal{O}(x^p)$
\begin{align}
&p=1:&&\psi_1=\alpha_1-2\zh\sa{1}\,,\label{TelemetryEq1}\\
&p=2:&&\psi_2=\alpha_2-2\zh\sa{2}+(\zh\sa{1}-2\alpha_1)\sa{1}\,,\nonumber\\
&p>2:&&\psi_p=\alpha_p-2\zh\sa{p}+\sum_{\ell=1}^{p-1}(\zh\,\sa{\ell}\sa{p-\ell}-2\alpha_\ell\,\sa{p-\ell})+\sum_{n=1}^{p-2}\alpha_n\sum_{\ell=1}^{p-n-1}\sa{\ell}\sa{p-n-\ell}\,.\nonumber
\end{align}
{\bf telemetry signal $S_2$:}  using relation~(\ref{SeriesExpansionsDeformations}), we find $\zp^3\,S_2-\zp+\Psi=0$ which leads to
{\allowdisplaybreaks
\begin{align}
0=&\,x\left(\zh^3\,\sd{1}+\psi_1-\alpha_1\right)+x^2\left(\zh^3\sd{2}+\psi_2-\alpha_2+3\zh^2\alpha_1\sd{1}\right)\nonumber\\
&+x^3\left(\zh^3\sd{3}+\psi_3-\alpha_3+3\zh^2\left(\alpha_1\sd{2}+\alpha_2\sd{1}\right)+3\zh \alpha_1^2\sd{1}\right)\nonumber\\
&+\sum_{p=4}^\infty x^p\bigg[\zh^3\sd{p}+\psi_p-\alpha_p+3\zh^2\sum_{\ell=1}^{p-1}\alpha_\ell\sd{p-\ell}+3\zh \sum_{k=1}^{p-2}\sum_{n=1}^{p-k-1}\alpha_n\alpha_{p-k-n}\,\sd{k}\nonumber\\
&\hspace{1.8cm}+\sum_{k=1}^{p-3}\sum_{n=1}^{p-k-2}\sum_{\ell=1}^{p-k-n-1}\alpha_\ell\alpha_{p-k-n-\ell}\alpha_n\,\sd{k}\bigg]\,.\label{StreamlineTel2}
\end{align}}
and therefore at each order $\mathcal{O}(x^p)$
\begin{align}
&p=1: &&0=\zh^3\,\sd{1}+\psi_1-\alpha_1\,,\nonumber\\
&p=2: &&0=\zh^3\sd{2}+\psi_2-\alpha_2+3\zh^2\alpha_1\sd{1}\,,\nonumber\\
&p=3: &&0=\zh^3\sd{3}+\psi_3-\alpha_3+3\zh^2\left(\alpha_1\sd{2}+\alpha_2\sd{1}\right)+3\zh \alpha_1^2\sd{1}\,,\nonumber\\
&p\geq 4: &&0=\zh^3\sd{p}+\psi_p-\alpha_p+3\zh^2\sum_{\ell=1}^{p-1}\alpha_\ell\sd{p-\ell}+3\zh \sum_{k=1}^{p-2}\sum_{n=1}^{p-k-1}\alpha_n\alpha_{p-k-n}\,\sd{k}\nonumber\\
& &&\hspace{0.75cm}+\sum_{k=1}^{p-3}\sum_{n=1}^{p-k-2}\sum_{\ell=1}^{p-k-n-1}\alpha_\ell\alpha_{p-k-n-\ell}\alpha_n\,\sd{k}\,.\label{TelemetryEq2}
\end{align}
}}
\end{center}
The infinitely many equations (\ref{RelAlpha1}), (\ref{HigherAlphaOrders}), (\ref{TelemetryEq1}) and (\ref{TelemetryEq2}) fix the coefficients $\{\psi_n\}$, $\{\alpha_n\}$ and $\zh$ in terms of the telemetry coefficients $\{\se{n}\}$ and $\{\sd{n}\}$: indeed, for $\zh$ (and $\psi_1$, $\varphi_1$ ) we have from the leading equation of each set (\emph{i.e.} (\ref{RelAlpha1}) and $p=1$ in (\ref{TelemetryEq1}) and (\ref{TelemetryEq2}))
\begin{align}
&\zh=\sqrt{\frac{2\sa{1}}{\sd{1}}}\,,&&\varphi_1=\alpha_1-2\zh\,\alpha_1^2\,\sa{1}\,,&&\psi_1=\alpha_1-2\zh\,\sa{1}\,.\label{Solzh}
\end{align}
Substituting these expressions into the second order (\emph{i.e.} $p=2$ in (\ref{HigherAlphaOrders}), (\ref{TelemetryEq1}) and (\ref{TelemetryEq2})) fixes
\begin{align}
&\varphi_1=-\frac{\left((\sa{1})^2 \sd{1}+2\sa{1} \sd{2}-2 \sa{2} \sd{1}\right)
   \left(\sd{1} \left((\sa{1})^3-2\sa{1} \sa{2}+\sd{1}\right)+2(\sa{1})^2
   \sd{2}\right)}{2 \sqrt{2} \sqrt{\sa{1}} (\sd{1})^{7/2}}\,,\nonumber\\
&\psi_1=-\frac{9\left(\sa{1}\right)^2 \sd{1}+2\sa{1} \sd{2}-2 \sa{2} \sd{1}}{2 \sqrt{2}
   \sqrt{\sa{1}} \left(\sd{1}\right)^{3/2}}\,.\label{LeadingPsiVarphi}
\end{align}
Similarly, order 2 determines $\varphi_2$ and $\psi_2$ as functions of $\alpha_2$ (which still needs to satisfy the order 3 (\emph{i.e.} $p=3$) equations (\ref{HigherAlphaOrders}), (\ref{TelemetryEq1}) and (\ref{TelemetryEq2})). Continuing iteratively, the set of equations (\ref{HigherAlphaOrders}), (\ref{TelemetryEq1}) and (\ref{TelemetryEq2}) for $p>2$ are linear in $\psi_p$, $\varphi_p$ and $\alpha_{p-1}$. With (\ref{Solzh}) and (\ref{LeadingPsiVarphi}), they fix $\{\varphi_n\}$ and $\{\psi_n\}$ (and $\zh$) in terms of the telemetry data $\{\sa{n}\}$ and $\{\sd{n}\}$ for all $n\in\mathbb{N}$. This therefore completely fixes the geometry.

\subsection{Examples}
We discuss two examples of spherically symmetric black holes (the Schwarzschild and the Hayward geometry) to illustrate our findings. 
\subsubsection{Schwarzschild Black Hole}
\begin{figure}[htbp]
\begin{center}
\parbox{7.5cm}{\includegraphics[width=7.5cm]{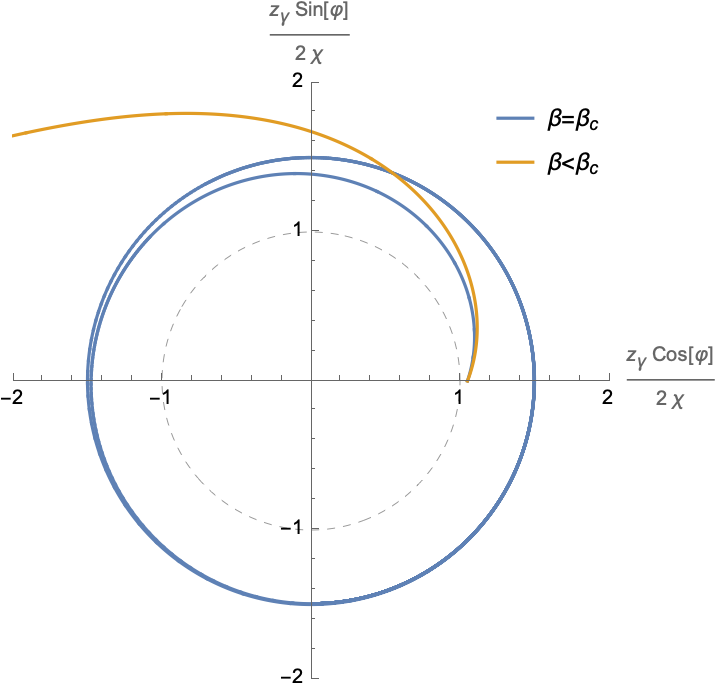}}\hspace{1cm}\parbox{7.5cm}{\includegraphics[width=7.5cm]{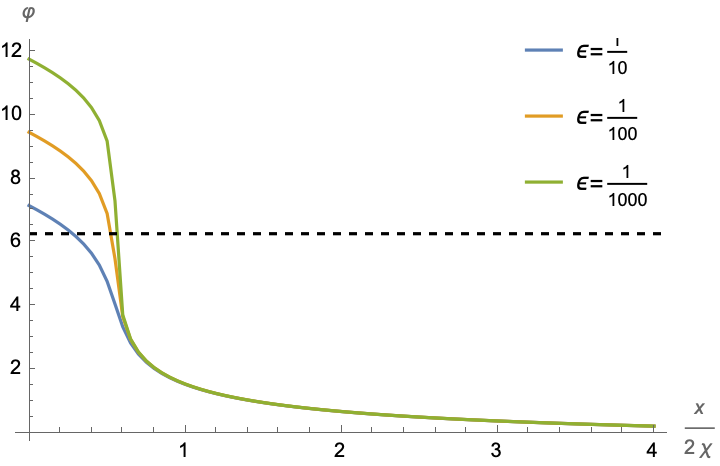}}
\end{center}
\caption{Left panel: the blue trajectory is emitted under a critical angle $\beta=\beta_c$ and ends on an orbit around the black hole, while the orange trajectory allows the photon to escape all the way to the spacecraft (both photons are emitted at $\zp=2.1\,M$) and the dashed line represents the horizon. Right panel: Angle $\varphi$ under which photons emitted with angle $\beta=\beta_c-\epsilon$ are captured by the spacecraft as a function of the time when it has been emitted by the probe (from~\cite{Hohenegger:2025dic}).
}
\label{Fig:SchwarzschildPhotonTrajectories}
\end{figure}

As the simplest example, we consider the Schwarzschild geometry \cite{Schwarzschild:1916uq}, $f(r)=h(r)=f_S(r)=1-\frac{2M}{r}$. This means that the deformation functions $\Phi$ and $\Psi$ in (\ref{h-f-emd-def}) are constant, which allows to perform some of the computations in closed form. For example, the proper time of the probe is given by
\begin{align}
&\tau=\frac{\sqrt{2}\left(\zs^{3/2}-\zp^{3/2}\right)}{3\sqrt{M}}\,,&&\text{with}&&\tah=\frac{\sqrt{2}\zs^{3/2}}{3\sqrt{M}}-\frac{4M}{3}\,.\label{Schwarzschildtau}
\end{align}
With $x=\tah-\tau$, the radial position of the probe 
\begin{align}
\zp=M\,\left(2\sqrt{2}+\frac{3x}{\sqrt{2}M}\right)^{2/3}\,.\label{Schwarzschildzp}
\end{align}
We can also explicitly study the geodesics of various emitted photons: Figure~\ref{Fig:SchwarzschildPhotonTrajectories} shows photons with different angles $\beta$: the left panel shows the difference between a photon escaping to the spacecraft and one that is caught in a circular orbit around the black hole. The right panel plots the angle under which photons are detected by the spacecraft, as a function of the positon where they have been emitted.

\noindent
With (\ref{Schwarzschildtau}) and (\ref{Schwarzschildzp}), the telemetry signals $S_{1,2}$ in (\ref{Telemetry1}) and (\ref{Telemetry3}) are
\begin{align}
&S_1(x)=1-\frac{2^{2/3}}{\left(4+\frac{3x}{M}\right)^{1/3}}\,,&&\text{and} &&S_2=\frac{\left(\frac{6x}{M} +8\right)^{2/3}-4}{\left(\frac{3x}{M} +4\right)^2 M ^2}\,.\label{SchwarzschildTelemetryExp}
\end{align}

\begin{figure}[htbp]
\begin{center}
\includegraphics[width=7.5cm]{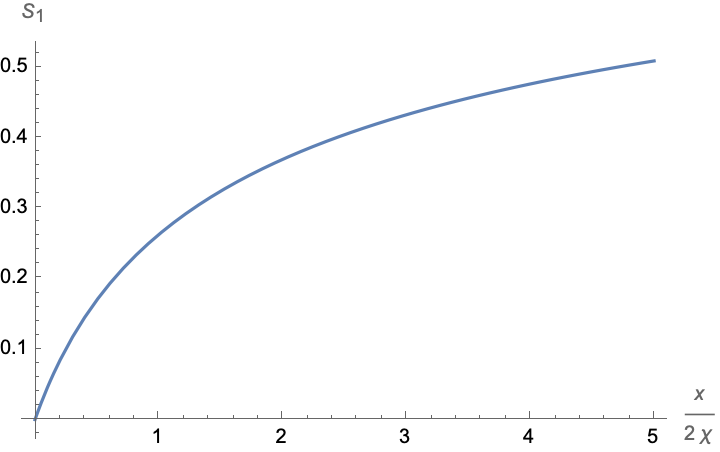}\hspace{1cm}\includegraphics[width=7.5cm]{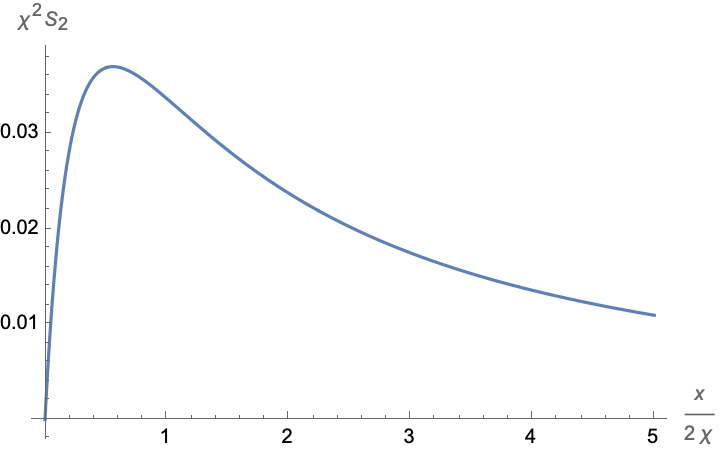}
\end{center}
\caption{Telemetry signals $S_1$ (left panel) and $S_2$ (right panel) as functions of $x$. (from~\cite{Hohenegger:2025dic})
}
\label{Fig:SchwarzschildTelemetry}
\end{figure}

\noindent
They are shown in Figure~\ref{Fig:SchwarzschildTelemetry}. Using (\ref{SchwarzschildTelemetryExp}), we find for the first few coefficients $\{\sa{n}\}$ and $\{\sd{n}\}$
\begin{align}
&\sa{1}=\frac{1}{4M}\,,&&\sa{2}=-\frac{1}{8M^2}\,,&&\sa{3}=\frac{7}{96 M^3}\,,&&\sa{4}=-\frac{35}{768M^4}\,,\nonumber\\
&\sd{1}=\frac{1}{8M^3}\,,&&\sd{2}=-\frac{13}{64M^4}\,,&&\sd{3}=\frac{23}{96M^5}\,,&&\sd{4}=-\frac{95}{384M^6}\,.\label{TelemetrySchwarzschild}
\end{align}
which yields
\begin{align}
&\zh=2M\,,&&\alpha_1=1\,,&&\alpha_2=-\frac{1}{8M}\,,&&\alpha_3=\frac{1}{24M^2}\,,&&\alpha_4=-\frac{7}{384M^3}\,,&&\varphi_{1,2,3,4}=0=\psi_{1,2,3,4}\,,\nonumber
\end{align}
which indeed correspond to the undeformed Schwarzschild geometry. 
\subsubsection{Hayward Black Hole}
As a non-trivial example, we consider the Hayward black hole \cite{Hayward_2006} and for which we shall use a similar parametrisation as in~\cite{Rezzolla:2014mua,DelPiano:2024nrl}
\begin{align}
&f(r)=h(r)=f_{\text{Hay}}(r):=1-\frac{2M r^2}{r^3+\frac{(2M)^3\gamma}{(1+\gamma)^3}}\,,&&\text{such that} &&\begin{array}{l}\Phi=\Psi=\Phi_{\text{Hay}}:=\frac{2M r^3}{r^3+\frac{(2M)^3\gamma}{(1+\gamma)^3}}\,,\\
\zh=\frac{2M}{1+\gamma}\,,\end{array}\label{HorizonHayward}
\end{align}
where $\gamma$ is a (small) real, non-negative parameter. The position of the probe $\zp$ in (\ref{SeriesExpansionsDeformations}), has the following coefficients
\begin{align}
&\alpha_1=1\,,&&\alpha_2=-\frac{1-2\gamma}{8M}\,,&&\alpha_3=\frac{1-7\gamma+\gamma^2}{24M^2}\,,&&\alpha_4=-\frac{7-105 \gamma +75 \gamma ^2-2 \gamma ^3}{384M^3}\,.
\end{align}
The expansion coefficients of the telemetry signal $S_1$ are given by
\begin{align}
&\sa{1}=\frac{1-2\gamma}{4M}\,,\hspace{1cm}\sa{2}=-\frac{1-7\gamma+\gamma^2}{8M^2}\,,\hspace{1cm}\sa{3}=\frac{7-105 \gamma +75 \gamma ^2-2 \gamma ^3}{96 M^3}\,,\nonumber\\
&\sa{4}=-\frac{35-910 \gamma +1560 \gamma ^2-328 \gamma ^3+2 \gamma ^4}{768 M^4}\,,\label{TelemetryHayward1}
\end{align}
and for $S_2$ by
\begin{align}
&\sd{1}=\frac{(1-2\gamma)(1+\gamma)^2}{8M^3}\,,\hspace{0.2cm}\sd{2}=\frac{(\gamma +1)^2 \left(8 \gamma ^2+40 \gamma -13\right)}{64 M ^4}\,,\hspace{0.2cm}\sd{3}=\frac{(\gamma +1)^2 \left(23-111 \gamma-3 \gamma ^2-4 \gamma ^3 \right)}{96 M ^5}\,,\nonumber\\
&\sd{4}=-\frac{(\gamma +1)^2 \left(380-2755 \gamma+1164 \gamma ^2-172 \gamma ^3-16 \gamma ^4\right)}{1536 M ^6}\,.\label{TelemetryHayward2}
\end{align}
%

\noindent
Substituting these into (\ref{Solzh}), we indeed recover $\zh$ in (\ref{HorizonHayward}). We finally obtain from (\ref{LeadingPsiVarphi}) (and the further higher solutions of (\ref{HigherAlphaOrders}), (\ref{TelemetryEq1}) and (\ref{TelemetryEq2}))
\begin{align}
&\varphi_1=\psi_1=\frac{3\gamma}{1+\gamma}\,,\hspace{1cm}\varphi_2=\psi_2=-\frac{9\gamma(3-2\gamma)}{8M (1+\gamma)}\,,\hspace{1cm}\varphi_3=\psi_3=\frac{9\gamma(3-6\gamma+\gamma^2)}{8M^2 (1+\gamma)}\,,\nonumber\\
&\varphi_4=\psi_4=-\frac{27 \gamma  \left(15-60 \gamma+33 \gamma ^2-2 \gamma ^3\right)}{128 M^3 (\gamma +1)}\,,\label{HaywardMetricCoefficients}
\end{align}
which indeed correctly reproduces the near-horizon geometry of the Hayward black hole. 


%% file: subfiles/6-Params.tex
\typeout{START baseline=\the\baselineskip, jot=\the\jot}
\chapter{Comparison of Parametrized Black Hole Metrics}\label{chap5}

Astrophysical black holes are the natural laboratories to probe gravity in the strong-field regime. Within GR, stationary black holes are completely characterised by their mass, charge and angular momentum, as expressed by the \emph{no-hair conjecture} (cf. Sec.~\ref{sec:uniqueness}). Current observations, however, cannot yet exclude sizeable departures from the Kerr or Schwarzschild geometries, so that this simple picture still awaits a precise test. This has motivated the construction of several \emph{agnostic} frameworks, which parametrise departures from the classical solutions without committing to a specific theory, while preserving covariance and regularity.

The EMD is one such framework, but it is not the only one. In this Chapter, based on~\cite{DeFalco:2025xxx}, we compare it with two widely used alternatives, namely the Johannsen-Psaltis (JP) and the Rezzolla-Zhidenko (RZ) parametrisations. The comparison has two parts. In the first one (Secs.~\ref{sec:comparison} and~\ref{sec:parametrization-comparison}) we derive explicit maps between the coefficients of the different parametrisations, including both the near-horizon EMD of the previous Chapters and a new large-distance version of the EMD. These maps make the relations and the complementarities between the frameworks explicit. In the second part we assess how efficiently each parametrisation reproduces physical observables of deformed black holes, focusing on the \emph{eikonal quasi-normal modes} (cf. Sec.~\ref{sec:eikonal_limit}) of three representative models.

\section{Agnostic parametrizations}
\label{sec:comparison}
We begin by summarising the JP~\cite{Johannsen:2010xs} (Sec.~\ref{sec:JP-parametrization}) and RZ~\cite{Rezzolla:2014mua} (Sec.~\ref{sec:RZ-parametrization}) parametrisations, and we briefly recall the near-horizon EMD in the language needed for the comparison (Sec.~\ref{sec:EMD-parametrization}), using the results of Sections~\ref{Sect:ReviewDistance} and~\ref{Sect:ReviewMetric}. We then introduce a formulation of the EMD in terms of asymptotic series at large distance from the black hole (Sec.~\ref{sec:EMD-large}), an idea first used in~\cite{Binetti:2022xdi} and further explored in~\cite{DelPiano:2023fiw}. The transformation maps connecting all these parametrisations are derived in Sec.~\ref{sec:parametrization-comparison}.

Throughout, we consider a generic static and spherically symmetric spacetime. In Schwarzschild coordinates $(t,r,\theta,\varphi)$ its line element reads
\begin{equation}\label{eq:generic-metric}
\dd s^2=g_{tt}(r)\dd t^2+g_{rr}(r)\dd r^2+r^2\left(\dd\theta^2+\sin^2\theta\dd\varphi^2\right) \ ,
\end{equation}
where $g_{tt}$ and $g_{rr}$ only depend on the radial coordinate. In each parametrisation, the weak-field limit is confronted with the experimental bounds on deviations from GR in the solar system, encoded in the parametrised post-Newtonian (PPN) parameters $\beta$ and $\gamma$~\cite{Will2014kxa,Williams:2004qba}.

\subsection{Johannsen-Psaltis parametrization}
\label{sec:JP-parametrization}
The JP parametrisation was designed to test the no-hair conjecture with electromagnetic observations of black holes. It describes a spacetime whose quadrupole moment is not fixed by the mass and the spin~\cite{Johannsen:2010xs}, such that any deviation from GR shows up as an anomalous contribution to the quadrupole moment~\cite{Glampedakis:2005cf,Johannsen:2010xs}. In the notation of~\eqref{eq:generic-metric}, the original proposal reads
\begin{equation}\label{eq:JP-metric}
g_{tt}^{\rm JP}=-[1+g(r)]\left(1-\frac{2\tilde{M}}{r}\right) \, , \quad g_{rr}^{\rm JP}=\frac{1+g(r)}{1-\frac{2\tilde{M}}{r}} \ ,
\end{equation}
with the deformation function
\begin{equation} \label{eq:g-JP}
    g(r) \coloneqq \sum_{n=1}^\infty\epsilon_n \left(\frac{\tilde{M}}{r}\right)^n = \sum_{n=1}^\infty\frac{\epsilon_n}{2^n} \left(\frac{\hr}{r}\right)^n \ ,
\end{equation}
Here the $\epsilon_n$ are real constants to be determined from observations. The parameter $\tilde{M}$ fixes the position of the horizon, $\hr=2\tilde{M}$, and is related to the gravitational mass $M$ by
\begin{equation}
    M=\tilde{M}\left(1-\frac{\epsilon_1}{2}\right) \ .
\end{equation}
Some shortcomings of this form were addressed in~\cite{Cardoso:2014rha}, where the two metric components were given independent deformation functions,
\begin{equation}\label{eq:revised-JP}
    g_{tt}^{\rm JP}= - \left[1+ g^t(r) \right]\left(1-\frac{2\tilde{M}}{r}\right) \, , \quad g_{rr}^{\rm JP}=\frac{1+g^r(r)}{1-\frac{2\tilde{M}}{r}} \ ,
\end{equation}
with
\begin{equation} \label{eq:new g-JP}
    g^i(r) = \sum_{n=1}^\infty\epsilon_n^i \left(\frac{\tilde{M}}{r}\right)^n  \qq{with} i=t,r \ ,
\end{equation}
The number of free parameters $\{ \epsilon_i^t,\epsilon^r_i \}_{i\geq 1}$ is thus doubled with respect to the original version. Matching the weak-field limit with the PPN expansion yields
\begin{subequations} \label{eq:relation-JPmass}
    \begin{align}
        M &= \tilde{M} \left( 1 - \frac{\epsilon_1^t}{2} \right)\ , \label{eq:mass}\\
        \epsilon_1^r &=  \gamma \,(2-\epsilon_1^t)-2 \ , \\
        2\epsilon_2^t &= (\beta - \gamma)(\epsilon_1^t-2)^2 + 4 \epsilon_1^t \ .
    \end{align}
\end{subequations}

\begin{center}
\remarkbox{\rem\label{rem:JPPPN}{Parameters left free by the PPN constraints}}{\parbox{16cm}{
The relations \eqref{eq:relation-JPmass} fix the mass and two combinations of the lowest JP coefficients in terms of the PPN parameters. The parameters $\epsilon_1^t$ and $\epsilon_2^r$, as well as all $\{ \epsilon_i^t,\epsilon^r_i \}_{i\geq 3}$, remain unconstrained, even if the PPN parameters take their GR values $\beta=\gamma=1$. Solar-system tests therefore leave most of the JP parameter space open, and the information on the strong-field regime has to come from observations close to the black hole.
}}
\end{center}

\subsection{Rezzolla-Zhidenko parametrization}
\label{sec:RZ-parametrization}
The RZ parametrisation aims at a model-independent description of black hole spacetimes in generic metric theories of gravity~\cite{Rezzolla:2014mua}. Its line element is written as (cf. eq.~\eqref{eq:generic-metric})
\begin{align}\label{eq:RZ-metric}
g_{tt}^{\rm RZ}&=-N^2(r) \qq{and} g_{rr}^{\rm RZ}=\frac{B^2(r)}{N^2(r)} \ .
\end{align}
with the event horizon located at $r = \hr > 0$, where $N(\hr)=0$. A key ingredient is the compactification of the radial coordinate through the dimensionless variable $x := 1- \hr / r$, which maps the horizon to $x=0$ and spatial infinity to $x=1$.

Writing $N^2= x\, A(x)$, with $A(x)>0$ for $0\leq x\leq1$, the functions $A$ and $B$ are expressed through the parameters $\epsilon$, $a_0$ and $b_0$ as
\begin{subequations}\label{asympfix}
\begin{align}
\label{asympfix_1}
A(x)&=1-\epsilon (1-x)+(a_0-\epsilon)(1-x)^2+{\tilde A}(x)(1-x)^3 \ ,
\\
\label{asympfix_2}
B(x)&=1+b_0(1-x)+{\tilde B}(x)(1-x)^2 \ ,
\end{align}
\end{subequations}
The parameter $\epsilon \coloneqq \frac{2M - \hr}{\hr}$ measures the deviation of $\hr$ from the Schwarzschild radius. The functions ${\tilde A}$ and ${\tilde B}$ encode the near-horizon geometry (for $x \simeq 0$) and are finite both at the horizon and at spatial infinity ($x \simeq 1$). To obtain a rapid convergence, they are represented by Padé approximants in the form of continued fractions,
\begin{subequations}
\label{contfrac}
\begin{align}
\label{contfrac_1}
{\tilde A}(x)=\frac{a_1}{\displaystyle 1+\frac{\displaystyle
    a_2x}{\displaystyle 1+\frac{\displaystyle a_3x}{\displaystyle
      1+\ldots}}}\ ,\\
\label{contfrac_2}
{\tilde B}(x)=\frac{b_1}{\displaystyle 1+\frac{\displaystyle
    b_2x}{\displaystyle 1+\frac{\displaystyle b_3x}{\displaystyle
      1+\ldots}}}\ ,
\end{align}
\end{subequations}
where $a_1, a_2, a_3\ldots$ and $b_1, b_2, b_3\ldots$ are dimensionless constants to be fixed by observations. Comparing the large-distance behaviour of~\eqref{eq:RZ-metric} with the PPN expansion constrains the first coefficients,
\begin{subequations}
\begin{align}\label{PPN bounds}
         a_0 &= \displaystyle \frac{(\beta - \gamma)(1+\epsilon)^2}{2} \lesssim10^{-5} \ ,\\
         b_0 &= \displaystyle \frac{(\gamma -1)(1+\epsilon)}{2} \lesssim 10^{-5} \ .
    \end{align}
\end{subequations}

\subsection{Effective Metric Description parametrization}
\label{sec:EMD-parametrization}
In the EMD, the deformations of the Schwarzschild metric are functions of spacetime invariants, which observers can measure independently of the choice of coordinates, and which respect the symmetries of the classical solution. As discussed extensively in Sec.~\ref{Sect:ReviewDistance} and summarised in Sec.~\ref{Sect:ReviewMetric}, a natural choice is the proper radial distance to the horizon. The metric components are
\begin{align}\label{eq:EMD-metric}
g_{tt}^{\rm EMD}&=-h(r) \qq{and} g_{rr}^{\rm EMD}=\frac{1}{f(r)} \ , 
\end{align}
with $h$ and $f$ as in eq.~\eqref{h-f-emd-def},
\begin{equation}\label{h-f-emd-nh}
    h(r) = 1 - \frac{\Psi^{(\rho)}(\rho)}{r} \qq{and} f(r) = 1 - \frac{\Phi^{(\rho)}(\rho)}{r} \ ,
\end{equation}
where $\rho$ is the proper distance from the event horizon, defined through eq.~\eqref{DiffEqDistance}. The functions in eq.~\eqref{h-f-emd-nh} are positive for $r>r_{H}$, and, being functions of an invariant, $\Psi$ and $\Phi$ do not change under coordinate transformations. They admit the Taylor expansions~\eqref{SeriesDistanceRho} with real coefficients $\xi_n$ and $\theta_n$. As in Chapter~\ref{chap4}, we assume that these series have a non-vanishing radius of convergence and that all derivatives of $h$ and $f$ with respect to $r$ are finite at $\hr$. By Result~\ref{res:SmoothHorizon}, this implies $\xi_{2n-1}=\theta_{2n-1}=0$ for all $n\in\mathbb{N}$, while Result~\ref{res:NeutralRegularity} requires $\xi_2\leq M/(8 \hr)$. The proper distance $\rho(r)$ and its inverse $r(\rho)$ follow from eq.~\eqref{SeriesDistance}, and the derivatives of the metric functions at the horizon are given in terms of $\{\xi_{2n},\theta_{2n}\}_{n\geq1}$ in eqs.~\eqref{emd nh fh1 hh1 hh2}.

\subsubsection{EMD at large distance from the BH}
\label{sec:EMD-large}
\begin{figure}[!ht]
    \begin{center}
        
\tikzset{every picture/.style={line width=0.75pt}} 

\begin{tikzpicture}[x=0.75pt,y=0.75pt,yscale=-1,xscale=1]

\draw  [color={rgb, 255:red, 0; green, 0; blue, 255 }  ,draw opacity=1 ][fill={rgb, 255:red, 105; green, 105; blue, 252 }  ,fill opacity=1 ] (93.99,64.29) -- (353.82,64.29) -- (353.82,96.33) -- (93.99,96.33) -- cycle ;
\draw  [color={rgb, 255:red, 6; green, 166; blue, 6 }  ,draw opacity=1 ][fill={rgb, 255:red, 121; green, 213; blue, 121 }  ,fill opacity=1 ] (96.37,69.92) -- (180.6,69.92) -- (180.6,91.57) -- (96.37,91.57) -- cycle ;
\draw [line width=1.5]  [dash pattern={on 5.63pt off 4.5pt}]  (83.16,37.44) .. controls (92.04,54.54) and (93.77,60.61) .. (93.99,80.74) ;
\draw [line width=1.5]  [dash pattern={on 5.63pt off 4.5pt}]  (83.16,124.05) .. controls (92.04,106.94) and (93.77,100.88) .. (93.99,80.74) ;

\draw  [color={rgb, 255:red, 208; green, 2; blue, 27 }  ,draw opacity=1 ][fill={rgb, 255:red, 229; green, 101; blue, 112 }  ,fill opacity=1 ] (244.58,69.92) -- (350.03,69.92) -- (350.03,91.57) -- (244.58,91.57) -- cycle ;
\draw [line width=1.5]    (7.37,80.74) -- (340,80.74) ;
\draw [shift={(343,80.74)}, rotate = 180] [color={rgb, 255:red, 0; green, 0; blue, 0 }  ][line width=1.5]    (9.95,-4.46) .. controls (6.32,-2.09) and (3.01,-0.61) .. (0,0) .. controls (3.01,0.61) and (6.32,2.09) .. (9.95,4.46)   ;
\draw [shift={(7.37,80.74)}, rotate = 0] [color={rgb, 255:red, 0; green, 0; blue, 0 }  ][fill={rgb, 255:red, 0; green, 0; blue, 0 }  ][line width=1.5]      (0, 0) circle [x radius= 3.05, y radius= 3.05]   ;

\draw (365,72) node [anchor=north west][inner sep=0.75pt]  [font=\normalsize]  {$r,\ d,\ \rho $};
\draw (235.31,104) node [anchor=north west][inner sep=0.75pt]  [font=\normalsize,color={rgb, 255:red, 208; green, 2; blue, 27 }  ,opacity=1 ]  {$\displaystyle 2M+2M \sum _{n=1}^{\infty} \sum_{m=0}^{n-1}\frac{\Omega_{n,m}M^n}{d^{n}}\log^m\left(\frac{d}{2M} \right)$};
\draw (90,104) node [anchor=north west][inner sep=0.75pt]  [font=\normalsize,color={rgb, 255:red, 6; green, 166; blue, 6 }  ,opacity=1 ]  {$\displaystyle  2M \sum _{n=0}^{\infty } \xi_{n} (\rho/2M)^{n}$};
\draw (180.4,35) node [anchor=north west][inner sep=0.75pt]  [font=\normalsize,color={rgb, 255:red, 0; green, 0; blue, 255 }  ,opacity=1 ]  {$\displaystyle \Phi^{(d)}( d)$};
\draw (65,-10) node [anchor=north west][inner sep=0.75pt]  [font=\normalsize]  {$ \begin{array}{c}
\zh ,\dho \\
\rho = 0
\end{array}$};

\end{tikzpicture}

    \end{center}
    \caption{The three regions of the exterior spacetime relevant for the EMD (only $f$ is shown, $h$ being analogous). The dashed curve marks the event horizon. Blue: the whole exterior, where the metric functions take their full form \eqref{h-f-emd-def}. Green: the vicinity of the horizon, where the deformation is a convergent series in the proper distance $\rho$ from the horizon, eq.~\eqref{SeriesDistanceRho}. Red: the asymptotic region, where the deformation is expanded in inverse powers (and logarithms) of the proper distance, eqs~\eqref{phipsilogs}.}
    \label{ranges}
\end{figure}
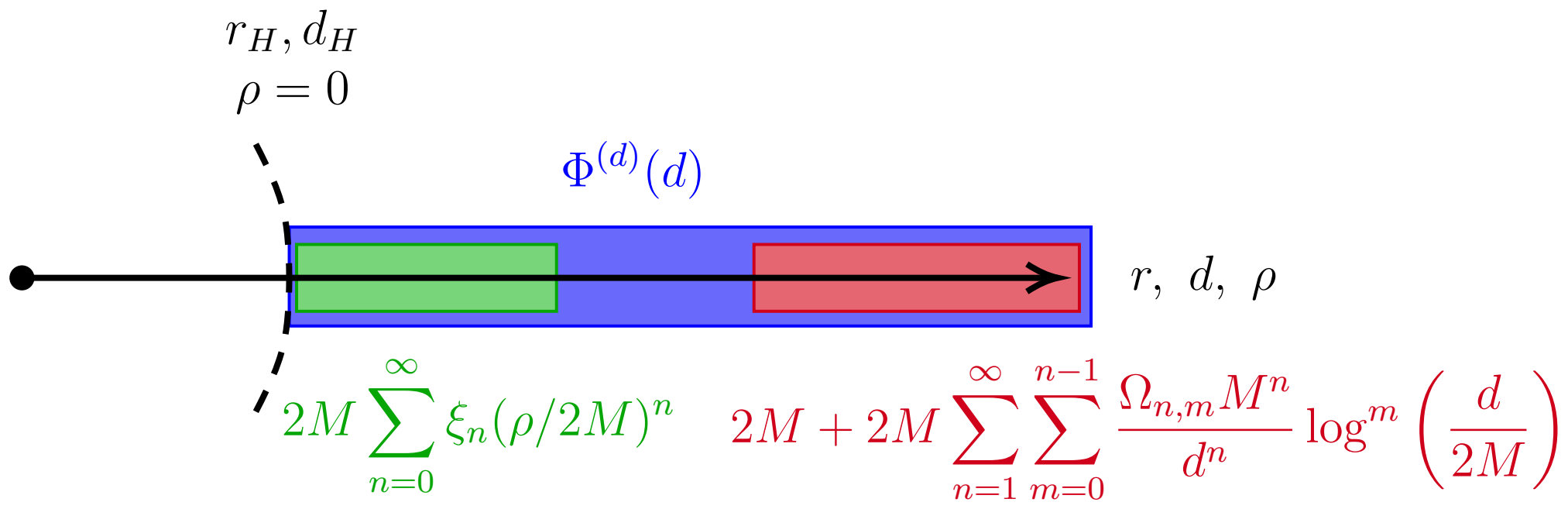

A genuinely new element of this Chapter is the formulation of the EMD far away from the horizon, where curvature is weak, but the gravitational field of the black hole is still felt (red region in Figure~\ref{ranges}). The strategy follows the one of Sec.~\ref{Sect:ReviewDistance}: the deformation functions are expanded asymptotically and the metric functions are determined self-consistently. The appropriate physical quantity is now the full radial proper distance $d(r)$, measured from $r=0$ and determined by
\begin{equation}\label{eqprop}
    \dv{d}{r}=\left|1-\dfrac{\Phi(d)}{r}\right|^{-1/2} \qq{with} d(0)=0 \ .
\end{equation}

A naive ansatz would expand $\Phi(d)$ and $\Psi(d)$ in powers of $1/d$. Solving eq.~\eqref{eqprop} with such an ansatz, however, generates logarithms of $r$ in the large-distance behaviour of $h(r)$ and $f(r)$. Such terms are absent in the post-Newtonian expansion, as well as in the other parametrisations discussed above. To avoid this mismatch, we allow for logarithms already in the deformation functions,
\begin{subequations}\label{phipsilogs}
    \begin{align}
        &\Phi(d) = 2M \left[ 1 + \sum_{n=1}^\infty \sum_{m=0}^{n-1} \frac{\newom_{n,m} M^n}{d^n} \log^{m}\left(\frac{d}{2M} \right)\right] \, ,\label{philogs} \\
        &\Psi(d) = 2M \left[ 1 + \sum_{n=1}^\infty \sum_{m=0}^{n-1} \frac{\newgam_{n,m} M^n}{d^n} \log^{m}\left(\frac{d}{2M} \right)\right]\, ,\label{psilogs}
    \end{align}
\end{subequations}
where $\newom_{n,m}$ and $\newgam_{n,m}$ are real constants, and we use the shorthand $\omega_n \equiv \newom_{n,0}$ and $\gamma_n \equiv \newgam_{n,0}$. The coefficients with $m\geq 1$ will be fixed by requiring that the logarithms cancel in the metric.

We work in the post-Newtonian (PN) regime $M/r\ll1$, which by the virial theorem corresponds to small velocities, $M/r\sim v^2/c^2$~\cite{Maggiore:2007ulw}. In this regime $1/(r^{i}d^j)\sim 1/r^{i+j}$, and the function $d(r)$ contains logarithmic corrections of the form $(M/r)^{j}\log^k(r/M)$, with $\{k,j\} \in \mathbb{N}_0$ and $0 < k \leq j-1$. Expanding eq.~\eqref{eqprop} with eq.~\eqref{phipsilogs} in this regime gives
\begin{align} \label{largedistexplogs}
    \frac{d(r)}{M}&=\frac{r}{M}+\log\left(\frac{r}{2M} \right) + k -\left(\frac{3}{2}+\omega_1\right)\frac{M}{r} 
    - \left[ \frac{5}{4}+\frac{\omega_1(5-2k)}{4} + \frac{\omega_2+ \newom_{2,1}}{2} \right] \frac{M^2}{r^2} \notag \\
    &\quad+\frac{M^2(\omega_1 - \newom_{2,1}) }{2r^2}\log\left(\frac{r}{2M} \right) - \left[\frac{35}{24} + \frac{(74 - 21 k + 9 k^2)\omega_1}{27} \right. \notag \\
    &\quad \quad \left.+ \frac{45 \omega_1^2 + 6 (7-6k) \omega_2 + 2(8+3k)\newom_{2,1}}{54 } + \frac{9\omega_3+3 \newom_{3,1} + 2\newom_{3,2}}{27} \right] \frac{M^3}{r^3} \notag \\
    &\quad \quad \quad +\Bigg{[}(7-6k) \omega_1 + 6 \omega_2-2(4-3k)\newom_{2,1}- 3\newom_{3,1}-2\newom_{3,2}\Bigg{]} \frac{M^3}{9r^3} \log\left( \frac{r}{2M}\right) \notag \\
    &\quad \quad \quad \quad - [ \omega_1 + 2 \newom_{2,1}-\newom_{3,2}]\frac{M^3}{3r^3} \log ^2\left( \frac{r}{2M}\right)+\order{\frac{M^4}{r^4}} \ , 
    \end{align}
where $k$ is a real integration constant, which we keep as a free parameter (see Remark~\ref{rem:GaugeK}). Inserting eq.~\eqref{largedistexplogs} into the metric functions with the deformations \eqref{phipsilogs}, the logarithms cancel up to a given order if
\begin{align}
    \newom_{2,1} &= \omega_1 \ , \quad \newom_{3,1} = -  \omega_1 + 2 \omega_2 \ ,   \quad \newom_{3,2} =  \omega_1 \ , \quad \newom_{4,3}=\omega_1 \ ,\quad  \newom_{4,2}=-\frac{5}{2}  \omega_1 + \frac{6}{2} \omega_2 \ ,\notag \\ 
     &\newom_{4,1} =\omega_1 -2  \omega_2 + 3  \omega_3 \ , \qquad   \newom_{5,4} = \omega_1 \ , \qquad \newom_{5,3} = -\frac{13}{3}  \omega_1 + \frac{12}{3} \omega_2 \ ,  \notag \\
     &\newom_{5,2} = \frac{9}{2} \omega_1 - \frac{14}{2} \omega_2 + \frac{12}{2} \omega_3 \ ,\qquad \newom_{5,1} = - \omega_1 + 2  \omega_2 - 3  \omega_3 + 4  \omega_4 \ , \notag
\end{align}
for the first few coefficients. A closed formula for general $n$ and $m$ is not available, but these conditions can be computed to any desired order, and analogous relations hold for the $\{ \newgam_{n,m} \}_{n\geq 1}^{0 \leq m \leq n-1}$. With these conditions, all logarithms disappear from the metric functions, which take the form
\begin{subequations}\label{eq:EMD-functions}
    \begin{align}
        f(r) &= 1 - \frac{2M}{r} - \frac{2M^2 \omega_1}{r^2} + (k \omega_1 - \omega_2)\frac{2M^3}{r^3} \notag \\
        & \qquad \qquad \qquad- \left[ \left(k^2  +  k   + \dfrac{3}{2}\right) \omega_1 + \omega_1^2 - 2 k \omega_2 +  \omega_3 \right]\frac{2M^4}{r^4}+ \order{\frac{M^5}{r^5}} \ , \\
        h(r) &= 1 - \frac{2M}{r} - \frac{2M^2 \gamma_1}{r^2} + (k \gamma_1 - \gamma_2)\frac{2M^3}{r^3}  \notag \\
        & \qquad \qquad \qquad - \left[ \left(k^2  +  k  + \dfrac{3}{2} \right) \gamma_1 + \gamma_1 \omega_1 - 2 k \gamma_2 +  \gamma_3 \right]\frac{2M^4}{r^4}+ \order{\frac{M^5}{r^5}} \ .
    \end{align}
\end{subequations}

\begin{center}
\remarkbox{\rem\label{rem:GaugeK}{The integration constant $k$}}{\parbox{16cm}{
The constant $k$ in \eqref{largedistexplogs} acts as a gauge choice for the distance $d$, since it fixes the point at which $d=0$. Shifting the proper distance by a constant induces a transformation of the coefficients $\{\Omega_{n,m} \}_{n \geq 1}^{0 \leq m \leq n-1}$, \emph{i.e.} an EMD redefinition of the type \eqref{ConstShiftRel} and~\eqref{GenLinTransform} discussed in Sec.~\ref{Sect:ShiftOfFunction}. Consistently, observable quantities do not depend on $k$: the combinations $k \omega_1-\omega_2$ and $k\gamma_1-\gamma_2$ appearing in \eqref{eq:EMD-functions}, as well as the analogous combinations at higher orders, are independent of the choice of this constant.
}}
\end{center}

\begin{center}
\remarkbox{\rem\label{rem:TruncationLargeDistance}{Truncation of the large-distance EMD}}{\parbox{16cm}{
The asymptotic series \eqref{eq:EMD-functions} have to be interpreted with care. If the parametrisation is truncated to a finite number of deformation parameters, say $(\omega_1,\ldots,\omega_n)$ and $(\gamma_1,\ldots,\gamma_n)$, only the terms up to order $1/r^{n+1}$ are reliable. Higher-order coefficients implicitly depend on parameters that have been set to zero by the truncation, so an apparent term $1/r^{m}$ with $m>n+1$ is an artefact of the cutoff rather than the true asymptotic behaviour of the full parametrisation. In the EMD the truncation therefore has to be imposed by hand, dropping every term beyond the order that the retained parameters actually control. This issue does not arise in the JP parametrisation, where truncating the expansion automatically removes all higher-order terms.
}}
\end{center}

\noindent
Up to order $\order{M^4/r^4}$, $f$ and $h$ in \eqref{eq:EMD-functions} have the same form, up to the exchange $\{\gamma_1, \gamma_2\}\leftrightarrow\{\omega_1, \omega_2\}$. Beyond this order the symmetry is lost, because of mixed terms proportional to $\gamma_1 \omega_1$. Finally, matching \eqref{eq:EMD-functions} with the PPN expansion leads to the bound
\begin{align}\label{bounds}
    |\gamma_1|=|\beta-\gamma|\lesssim 10^{-5}  \quad \text{with} \quad 
    \gamma= 1 \ .
\end{align}
At leading PPN order, only $\gamma_1$ can be mapped onto $\beta$ and $\gamma$, while $\omega_1$ remains unconstrained. Conversely, the expansion \eqref{eq:EMD-functions} only captures the subset of PPN corrections with ${\gamma=1}$.

\section{Transformation maps relating the different parametrizations}
\label{sec:parametrization-comparison}
We now derive the relations among the parametrisations introduced above. A useful observation is that, although the event horizon is defined in a different way in each framework and always depends on free parameters, all of them have to agree on its location $\hr$. This common reference point simplifies the computations that follow.

\subsubsection{JP to EMD (near horizon) parametrizations}
\label{sec:JP-EMD}
We start with the temporal components $g_{tt}$ of the EMD \eqref{eq:EMD-metric} and JP \eqref{eq:revised-JP} parametrisations. Using the Taylor expansion of $h$ around the horizon, we require
\begin{align}\label{eq:h-g}
g_{tt}^{\rm EMD}&=-\sum_{n=1}^\infty \frac{h^{(n)}_H}{n!}(r-\hr)^n=-\frac{1+g^t(r)}{r}(r-\hr)=g_{tt}^{\rm JP} \ .
\end{align}
After division by $(r-\hr)$ this becomes
\begin{align}\label{EMD_JP}
        &\sum_{n=1}^\infty \frac{h^{(n)}_H}{n!}(r-\hr)^{n-1}=\frac{1+g^t(r)}{r}=\frac{1}{r}\left[1+\sum_{j=1}^\infty\frac{\epsilon_j^t}{2^j} \left(\frac{\hr}{r}\right)^j\right]\equiv  \sum_{j=0}^\infty\frac{\epsilon_j^t }{2^j} \frac{\hr^j}{r^{j+1}}\ ,
    \end{align}
where we have included the additional coefficient $\epsilon_0^t=1$ in the sum. At $r=\hr$, eq.~\eqref{EMD_JP} gives for $n=1$
\begin{equation} \label{eq:h1}
        h^{(1)}_H=\frac{1+g^t(\hr)}{\hr} = \frac{1}{\hr}\sum_{j=0}^\infty\frac{\epsilon_j^t}{2^j}  \ .
    \end{equation}
while the higher coefficients $h^{(n)}_H$ follow by differentiating eq.~\eqref{EMD_JP} $n$ times and evaluating the result at $r=\hr$:
\begin{align}
        h^{(n)}_H&=n \dv[n]{r} \left(\sum_{j=1}^{\infty}\frac{\epsilon_j^t}{2^j} \frac{\hr^j}{r^{j+1}}\right)\bigg|_{r=\hr}=\frac{(-1)^n \, n}{\hr^{n+1}} \sum_{j=1}^{\infty}\frac{\epsilon_j^t }{2^j} \frac{(j+n)!}{j!}\ . \label{eq:hj}
    \end{align}
The radial components are treated in the same way,
\begin{align}
        g_{rr}^{\rm EMD}&=\frac{1}{f(r)}=\frac{1+g^r(r)}{1-\hr/r}=g_{rr}^{\rm JP}\ ,
     \end{align}
with
\begin{align}
         f(r)&= \sum_{n=1}^\infty\frac{f^{(n)}_H}{n!}(r-\hr)^n \ .
    \end{align}
from which one obtains
\begin{subequations} \label{eq:f_H}
    \begin{align}
        f^{(1)}_H&=\frac{1}{\hr[1+g^r(\hr)]} \ , \label{eq:f1} \\
        f^{(n)}_H&=n \dv[n]{r} \left[\frac{1}{r(1+g^r(r))}\right]\bigg|_{r=\hr} \ ,\label{eq:gr}
    \end{align}
\end{subequations}
Unlike for $h$, a closed form for the general coefficient $f_H^{(n)}$ is not easily obtained. Comparing eqs.~\eqref{eq:f1} and \eqref{f1h emd nh}, as well as eqs.~\eqref{eq:h1} and \eqref{h1h emd nh}, and using $\hr=2M/(1-\epsilon_1^t/2)$, leads to the following relations.

\begin{center}
\resultbox{\result\label{res:JPtoEMDnh}{Leading coefficients: JP $\to$ near-horizon EMD}}{\parbox{16cm}{
The leading near-horizon EMD coefficients are expressed in terms of the JP deformation functions evaluated at the horizon as
\begin{equation}
\xi_2 = \frac{g^r(\hr)(2-\epsilon_1^t)}{8[1+g^r(\hr)]^2} \ ,\quad
\theta_2 = \frac{g^t(\hr)(\epsilon_1^t-2)}{8[1+g^r(\hr)]} \ ,
\end{equation}
valid for $g^r(\hr) \neq -1$. For $g^r(\hr) = 1$ the bound $\xi_2 \leq  (2-\epsilon_1^t)/32$ is saturated, which coincides with the bound $\xi_2\leq M/(8\hr)$ of Result~\ref{res:NeutralRegularity}. Since $g^t(\hr)$ and $g^r(\hr)$ are infinite sums over the JP coefficients, each EMD coefficient depends on infinitely many JP parameters, and vice versa.
}}
\end{center}

\noindent
The expressions for the higher coefficients $\{\theta_{2n}, \xi_{2n} \}_{n\geq 2}$ are lengthy and we do not report them. Their computation is nevertheless straightforward, because each new horizon derivative is linear in the corresponding new parameter (see Remark~\ref{rem:TriangularStructure}). The inverse map, from EMD to JP coefficients, is obtained by inverting the linear system~\eqref{eq:hj}.

\subsubsection{JP to EMD (large distance) parametrizations}
At large distance, the JP metric behaves as
\begin{subequations}\label{JPexpinf}
\begin{align}
    g_{tt}^{\rm JP}&=-1+\frac{\tilde{M}(2-\epsilon_1^t)}{r}+\frac{\tilde{M}^2(2\epsilon_1^t-\epsilon_2^t)}{r^2}+\order{\tilde{M}^3/r^3} \ ,\\
    g_{rr}^{\rm JP}&=1+\frac{\tilde{M}(2+\epsilon_1^r)}{r}+\frac{\tilde{M}^2(\epsilon_2^r+2\epsilon_1^r+4)}{r^2}+\order{\tilde{M}^3/r^3} \ .
\end{align}
\end{subequations}
A comparison with the large-distance EMD \eqref{eq:EMD-functions}, taking into account the relation \eqref{eq:mass}, gives for the first coefficients
\begin{subequations}\label{JP to EMD fr}
    \begin{align}
        \epsilon_1^r & = - \epsilon_1^t \ , \\ 
        \omega_1 &= \frac{2[\epsilon_2^r+ \epsilon_1^t(2-\epsilon_1^t)]}{(\epsilon_1^t - 2)^2},\ \ \gamma_1 = \frac{2(2\epsilon_1^t-\epsilon_2^t)}{(\epsilon_1^t - 2)^2}, \\
        \omega_2 & =-\frac{2}{(\epsilon_1^t - 2)^3}\left\{2\epsilon_3^r+[k(\epsilon_1^t-2)-2\epsilon_1^t](\epsilon_1^t-2)\epsilon_1^t +\epsilon_2^r[4(\epsilon_1^t-1)+k(2-\epsilon_1^t)]\right\} \ , \\
        \gamma_2 &= \frac{2k(\epsilon_1^t-2)(2 \epsilon_1^t-\epsilon_2^t)-8\epsilon_2^t+4\epsilon_3^t}{(\epsilon_1^t - 2)^3} \ .    \end{align}
\end{subequations}
In contrast to the near-horizon case, this map is simple, since the asymptotic expansion of the JP metric is organised in the same way as the large-distance EMD: at each order, the coefficients of the two parametrisations are in one-to-one correspondence.

\subsubsection{JP to RZ parametrizations}
\label{sec:JP-RZ}
The map between the RZ parametrisation \eqref{eq:RZ-metric} and the original JP metric \eqref{eq:JP-metric} was already given in Ref.~\cite{Rezzolla:2014mua}. Here we adapt it to the revised form \eqref{eq:revised-JP}. A comparison with eqs.~\eqref{asympfix_1} and~\eqref{asympfix_2} gives
\begin{gather}
    -\epsilon \frac{\hr}{r}+(a_0-\epsilon)\left(\frac{\hr}{r}\right)^2 + \left(\frac{\hr}{r}\right)^3\tilde{A}\left(1-\frac{\hr}{r}\right) = g^t(r) \ , \\
    b_0 \frac{\hr}{r} + \left(\frac{\hr}{r}\right)^2\tilde{B}\left(1-\frac{\hr}{r}\right) =  g^{B}(r)\ .
\end{gather}
where we have defined the auxiliary function
\begin{equation}
g^{B}(r)\coloneqq\sqrt{[1+g^t(r)][1+g^r(r)]}-1 \ , 
\end{equation}
The lowest coefficients follow as
\begin{equation}
\label{JPrel}
\epsilon =-\frac{\epsilon_1^t}{2}\ , \ a_0=\frac{1}{2}\left( \frac{\epsilon_2^t}{2} - \epsilon_1^t\right)\ , \ b_0=\frac{\epsilon_1^t+\epsilon_1^r}{4}\ .
\end{equation}
Near the horizon, the matching expresses the RZ coefficients $\{a_n, b_n\}_{n \geq 1}$ algebraically through the JP coefficients $\{\epsilon_n^t, \epsilon_n^r \}_{n \geq 1}$. The lowest ones read
\begin{subequations}\label{as_bs_JP}
\begin{align}
a_1&=g^t(\hr)- a_0 + 2 \epsilon =\sum_{n=3}^{\infty}\frac{\epsilon_n^t}{2^n}\ ,\\ 
b_1&= g^B(\hr)-b_0 \ ,\\
a_2&= - \frac{(r\, g^t)^{\prime}+ g^t+\epsilon}{a_1}\Big|_{r=\hr}-1 =\displaystyle\sum_{n=4}^{\infty}\frac{\epsilon^t_n(n-3)}{2^n} \Big/\displaystyle\sum_{n=3}^{\infty}\frac{\epsilon^t_n}{2^n}\ , \\
b_2&=- \dfrac{(r \, g^B)^{\prime}}{b_1}\bcond -1 \ ,\label{a2_b2_JP}  \\
a_3&=\frac{(r^2 g^t)^{\prime \prime}-2a_1[1+a_2(a_2+2)]}{2 a_1 a_2}\Big|_{r=\hr}\ ,\\
b_3&=\dfrac{(r^2 g^B)''}{2b_1 b_2}\bcond - (b_2 + 1)  \ ,\label{a3_b3_JP}
\end{align}
\end{subequations}
where a prime denotes the derivative with respect to $r$. The extension of eq.~\eqref{as_bs_JP} to higher orders is straightforward.

\begin{center}
\remarkbox{\rem\label{rem:JPRZstructure}{Structure of the JP--RZ map}}{\parbox{16cm}{
The coefficients $a_1,a_2,a_3,\ldots$ do not depend on $\epsilon_1^t$ and $\epsilon_2^t$, whereas the $b_1,b_2,b_3,\ldots$ have a more involved dependence, which we do not display. In the simplest case, where $\epsilon_3^t$ is the only non-vanishing temporal JP coefficient beyond the second one ($\epsilon^t_n=0$ for $n>3$), one finds $a_2=0$, and the continued fraction reproduces the function $N(r)$ exactly. The inverse map follows from inverting the system~\eqref{as_bs_JP}.
}}
\end{center}

\subsubsection{RZ to EMD (near horizon) parametrizations}
\label{sec:RZ-EMD}
The last map relates the RZ and the near-horizon EMD parametrisations, for which we use the results of Sec.~\ref{sec:JP-RZ}. From eqs.~\eqref{eq:h-g} and \eqref{eq:gr} we obtain
\begin{subequations}
\begin{align}
g^t(r)&=r\left[\sum_{n=1}^\infty\frac{h_H^{(n)}}{n!}(r-\hr)^{n-1}\right]-1\ ,\\
g^r(r)&=\dfrac{1}{\sum_{n=1}^\infty\frac{f^{(n)}_H}{n!}(r-\hr)^{n-1}}-1\ .
\end{align}
\end{subequations}
The RZ coefficients can then be written, through eq.~\eqref{as_bs_JP}, in terms of the derivatives of $g^t(r)$ and $g^r(r)$, which are in turn determined by the EMD coefficients.

The leading EMD coefficients $(\theta_2,\xi_2)$ can be related directly to the RZ coefficients $(a_0,a_1,b_0,b_1)$ by using $h^{(1)}_H$ and $f^{(1)}_H$ (cf. eqs.~\eqref{eq:h1} and \eqref{eq:f1}) together with
\begin{equation}
    h(r) = N(r)^2 \qq{and} f(r)=\frac{N(r)^2}{B(r)^2} \ .
\end{equation}
With $y=r/(2M)$, the expansion of the RZ metric functions to first order in $y-y_H$ reads
\begin{subequations}\label{N2 and N2B2 series}
\begin{align}
    N(y)^2 &= \frac{(3+a_0+a_1)y_H - 2}{y_H^2} (y-y_H) + \mathcal{O}(y-y_H)^2 \ ,  \\
    \frac{N(y)^2}{B(y)^2} &= \frac{(3+a_0+a_1)y_H - 2}{(1+b_0+b_1)^2 \, y_H^2} (y-y_H) + \mathcal{O}(y-y_H)^2  \ .
\end{align}
\end{subequations}
Comparing these expansions with eqs.~\eqref{f1h emd nh} and~\eqref{h1h emd nh} and solving for $\xi_2$ and $\theta_2$ gives the result below.

\begin{center}
\resultbox{\result\label{res:RZtoEMDnh}{Leading coefficients: RZ $\to$ near-horizon EMD}}{\parbox{16cm}{
The leading near-horizon EMD coefficients depend on the RZ parameters only through $\epsilon$ and the combinations $\mathcal{A} \coloneqq a_0+a_1$ and $\mathcal{B} \coloneqq b_0+b_1$:
\begin{subequations}\label{RZtoxitheta}
    \begin{align}
        \xi_2 &= \frac{(1+\epsilon)(1+\mathcal{A}-2\epsilon)(\mathcal{B}(\mathcal{B}+2)-\mathcal{A}+2\epsilon)}{4(1+\mathcal{B})^4}\ , \label{RZtoxitheta-1}\\
        \theta_2 &= \frac{(1+\epsilon)(2\epsilon-\mathcal{A})}{8} \left( 1 + \frac{| 1+ 2\mathcal{A} - \mathcal{B}(\mathcal{B}+2)  - 4 \epsilon| }{(1+\mathcal{B})^2} \right) \label{RZtoxitheta-2} \ ,    
    \end{align}
\end{subequations}
In particular, the RZ coefficients $a_n$ and $b_n$ with $n > 1$ do not contribute to $\xi_2$ and $\theta_2$. Conversely, requiring that the Schwarzschild geometry is recovered for $\epsilon\to 0$,
\begin{subequations}
        \begin{align}
            \mathcal{A}&=\dfrac{1}{2\sqrt{2}\,\xi_2}\Big{[}8 \theta_2 \xi_2 + (1+\epsilon)(\xi_2- \theta_2)+\sqrt{1+\epsilon}\, (\xi_2-\theta_2) \sqrt{1+\epsilon-16 \xi_2}\Big{]}^{1/2}-1 \ ,\\
            \mathcal{B}&=2 \epsilon -\frac{\theta_2}{2 \xi_2}\left(1- \sqrt{1-\frac{16\xi_2}{1+\epsilon}}\right) \ .
        \end{align}            
        \end{subequations}
}}
\end{center}

\noindent
Additional constraints arise in the limit $b_0 = b_1 = 0$, \emph{i.e.} $\mathcal{B}=0$. In this case $B(y) = 1 + \order{(y-y_H)^2}$ at the horizon, such that the first derivatives in eqs.~\eqref{N2 and N2B2 series} must coincide with~\eqref{f1h emd nh} and~\eqref{h1h emd nh}. This implies $\xi_2 = \theta_2$, \emph{i.e.} equal first derivatives of $f$ and $h$ at the horizon. Moreover, for $\mathcal{B}=0$ the absolute value in eq.~\eqref{RZtoxitheta-2} requires
\begin{equation}
    1+2\mathcal{A} - 4 \epsilon > 0  \ \Rightarrow \ 2 \epsilon - a_0 - a_1 < \frac{1}{2}\ .
\end{equation}
The same procedure can be applied to the higher-order coefficients. Beyond first order, however, the expressions quickly become cumbersome, and we do not display them. Altogether, the coefficients of the two parametrisations are related directly and unambiguously.

\subsubsection{RZ to EMD (large distance) parametrization}
For the comparison at large distance, we expand the RZ metric components as
\begin{subequations}
\begin{align}
g_{tt}^{\rm RZ}&=-1+\frac{2 M}{r}-\frac{4 a_0 M^2}{ (\epsilon +1)^2 r^2}-\frac{8M^3 \left[\tilde{A}(1)-a_0 + \epsilon \right]}{ (\epsilon +1)^3r^3}+\mathcal{O}\left(\frac{M^4}{r^4}\right) \ ,\\
g_{rr}^{\rm RZ}&=1+\frac{2 M(2 b_0 + \epsilon +1)}{ (\epsilon +1)r}+\mathcal{O}\left(\frac{M^2}{r^2}\right) \ .
\end{align}    
\end{subequations}
Matching with eq.~\eqref{eq:EMD-functions} requires $b_0=0$, and gives for the first parameters
\begin{subequations}\label{RZ to EMD large distance}
    \begin{align}
        \omega_1 & =\frac{ 4 \tilde{B}(1)- 2 a_0}{(\epsilon +1)^2}, \quad \gamma_1 = -\frac{2 a_0}{(\epsilon +1)^2} \ , \quad \gamma_2 = -\frac{2 \left[a_0 (k(1+\epsilon) -2)+2 (\tilde{A}(1)+\epsilon )\right]}{(\epsilon +1)^3} \ , \\
        \omega_2 & =-\frac{2}{(\epsilon +1)^3} \Big{\{} a_0 [k (1+\epsilon)-2] +2 \Big{[}\epsilon + \tilde{A}(1)+ (2-k)(1+ \epsilon)\tilde{B}(1) + 2 \tilde{B}^\prime(1)\Big{]} \Big{\}} \ ,
    \end{align}
\end{subequations}
These relations involve the continued fractions \eqref{contfrac} evaluated at spatial infinity, $\tilde{A}(1)$ and $\tilde{B}(1)$, as well as their derivatives (a prime denotes here the derivative with respect to the compactified coordinate $x$ of Sec.~\ref{sec:RZ-parametrization}). As in the JP case, the coefficients of the RZ and of the large-distance EMD parametrisations are in one-to-one correspondence.

\begin{center}
\resultbox{\result\label{res:MapsStructure}{Structure of the transformation maps}}{\parbox{16cm}{
The maps derived in this Section fall into two classes:
\begin{itemize}
\item \emph{Near-horizon EMD $\leftrightarrow$ JP, RZ}: the leading EMD coefficients are determined by horizon values of the other parametrisations (Results~\ref{res:JPtoEMDnh} and~\ref{res:RZtoEMDnh}). Each near-horizon EMD coefficient depends on infinitely many JP coefficients and vice versa, while the leading EMD coefficients depend on the RZ parameters only through the combinations $a_0+a_1$ and $b_0+b_1$.
\item \emph{Large-distance EMD $\leftrightarrow$ JP, RZ}: the coefficients are related order by order in $M/r$, and the maps \eqref{JP to EMD fr} and \eqref{RZ to EMD large distance} are in one-to-one correspondence.
\end{itemize}
Within their common domain of validity, the maps can be inverted order by order. A deformation specified in one parametrisation can therefore be translated into any of the others, the main difference being the region of spacetime (near the horizon or asymptotically far) around which each parametrisation is organised.
}}
\end{center}

\subsection{Eikonal limit in different parametrizations}
As an application of these maps, we now study the eikonal limit of the QNMs (cf. Sec.~\ref{sec:eikonal_limit}) in black hole perturbation theory, within each of the parametrisations introduced in Sec.~\ref{sec:comparison}. For a generic parametrisation, its infinitely many coefficients are treated as independent. Once a specific model is chosen, however, they become functions of a finite number of free parameters.

Below we consider small deviations from the Schwarzschild solution in models that depend on the mass $M$ and on one additional length scale $l$, which quantifies the departure from GR. This parameter displaces the horizon from its Schwarzschild value $2M$. To describe this displacement uniformly, we use the perturbative parameter $\epsilon$ of the RZ parametrisation (see below eq.~\eqref{asympfix}), and we Taylor expand all quantities entering the eikonal limit in powers of $\epsilon$. Other choices, such as $\epsilon_1^t$ in the JP parametrisation, would be equally possible. For each parametrisation, the quantities in eq.~\eqref{eq:EIKONAL-QUANTITIES} are expanded up to linear order in $\epsilon$, and higher orders are used to estimate the truncation error.

\renewcommand{\arraystretch}{3.5}
\setlength{\tabcolsep}{10pt}
\begin{table*}[ht]
    \centering

\resizebox{\textwidth}{!}{
    \begin{tabular}{|c|c|c|c|c|}
        \hline
        \raisebox{0.5cm}{$N$} &  \raisebox{0.5cm}{$M$} &  \raisebox{0.51cm}{$\eta_0$} & \multicolumn{1}{c|}{ \raisebox{0.51cm}{$\eta_1$}} \\[-0.7cm]
        \hline
        \raisebox{0.2cm}{2} &\raisebox{0.2cm}{1} & \raisebox{0.22cm}{$ \dfrac{1}{4} \left(\sqrt{3}-1\right)$} 
          & \raisebox{0.21cm}{$-\dfrac{1}{4} \left(\sqrt{3}-1\right) 
          +\dfrac{1}{135} \left(61-68 \sqrt{3}\right) \dfrac{\theta_2}{\mathfrak{c}} 
          +\dfrac{2}{9} \left(2 \sqrt{3}-7\right) \dfrac{\theta_4}{\mathfrak{c}} 
          +\dfrac{8}{3} \left(\sqrt{3}-2\right) \dfrac{\theta_6}{\mathfrak{c}}$} \\[-0.1cm]
        \hline
      \multirow{2}{0.01\textwidth}{\raisebox{0.06cm}{2}} & \multirow{2}{0.01\textwidth}{ \raisebox{0.45cm}{2}} & \multirow{2}{0.2\textwidth}{\centering{\raisebox{0.27cm}{$\dfrac{1}{4} \sqrt{9 \sqrt{3}-15}$}} }& \multirow{2}{\textwidth}{\centering{\raisebox{0.1cm}{$-\dfrac{1}{4} \sqrt{9 \sqrt{3}-15}-\dfrac{1}{630} \sqrt{1274659 \sqrt{3}-\dfrac{6386663}{3}}\dfrac{\theta_2}{\mathfrak{c}}-\dfrac{11}{15} \sqrt{\dfrac{1}{6} \left(627 \sqrt{3}-1075\right)}\dfrac{\theta_4}{\mathfrak{c}}$}}\\

      \hspace{0.5cm}\raisebox{0.7cm}{$- \dfrac{2444\sqrt{2}}{\sqrt{\left(14 \sqrt{3}-9\right) \left(85675 \sqrt{3}+148401\right)}}\dfrac{\theta_6}{\mathfrak{c}}-\dfrac{16}{\sqrt{51 \sqrt{3}+\dfrac{265}{3}}}\dfrac{\theta_8}{\mathfrak{c}}$}} \\[0.1cm]
   & & &\\[-0.2cm]
        \hline
        \multirow{1}{0.01\textwidth} {1} & \multirow{1}{0.01\textwidth} {3} & \multirow{1}{0.2\textwidth} {\hspace{0.06cm}\centering{$\dfrac{1}{3 \sqrt{3}}$} }
          & \multirow{1}{0.69\textwidth} {\centering{ $-\dfrac{1}{3 \sqrt{3}} 
            -\dfrac{19676}{25515 \sqrt{3}}\dfrac{\theta_2}{\mathfrak{c}} 
            -\dfrac{656}{405 \sqrt{3}}\dfrac{\theta_4}{\mathfrak{c}} 
            -\dfrac{64}{27 \sqrt{3}}\dfrac{\theta_6}{\mathfrak{c}} 
        -\dfrac{128}{81 \sqrt{3}}\dfrac{\theta_8}{\mathfrak{c}}$}} \\
        \hline
       \multirow{1}{0.01\textwidth} {2} & \multirow{1}{0.01\textwidth} {3} & \multirow{1}{0.2 \textwidth}{ \centering{\hspace{0cm}$\dfrac{1}{3 \sqrt{3}}$}} 
          & \multirow{1}{\textwidth}{\centering{\raisebox{0.2cm}{$-\dfrac{1}{3 \sqrt{3}} 
            -\dfrac{98837}{127575 \sqrt{3}}\dfrac{\theta_2}{\mathfrak{c}} 
            -\dfrac{136096}{76545 \sqrt{3}}\dfrac{\theta_4}{\mathfrak{c}} 
            -\dfrac{4384}{1215 \sqrt{3}}\dfrac{\theta_6}{\mathfrak{c}} 
            -\dfrac{1216}{243 \sqrt{3}}\dfrac{\theta_8}{\mathfrak{c}} 
            -\dfrac{256}{81 \sqrt{3}}\dfrac{\theta_{10}}{\mathfrak{c}}$}}} \\
        \hline
         \multirow{2}{0.01\textwidth}{ \raisebox{0.6cm}{3}} & \multirow{2}{0.01\textwidth}{ \raisebox{0.6cm}{3}} & \multirow{2}{0.2\textwidth}{\hspace{-0.05cm}\centering{\raisebox{0.47cm}{$\dfrac{1}{3 \sqrt{3}}$}}} 
          & \raisebox{0.23cm}{$-\dfrac{1}{3 \sqrt{3}}-\dfrac{652294}{841995 \sqrt{3}}\dfrac{\theta_2}{\mathfrak{c}}-\dfrac{2066276}{1148175 \sqrt{3}}\dfrac{\theta_4}{\mathfrak{c}}-\dfrac{62528}{15309 \sqrt{3}}\dfrac{\theta_6}{\mathfrak{c}}-\dfrac{29152}{3645 \sqrt{3}}\dfrac{\theta_8}{\mathfrak{c}}$}\\
            & & &\raisebox{0.5cm}{$-\dfrac{2560}{{243 \sqrt{3}}}\dfrac{\theta_{10}}{\mathfrak{c}}-\dfrac{512}{81 \sqrt{3}}\dfrac{\theta_{12}}{\mathfrak{c}}$}\\[-0.35cm]
        \hline
        \hline        
        \raisebox{0.5cm}{$N$} & \raisebox{0.5cm}{$M$} & \raisebox{0.5cm}{$\Upsilon_0$} & \multicolumn{1}{c|}{\raisebox{0.5cm}{$\Upsilon_1$}} \\[-0.7cm]
        \hline
         \multirow{2}{0.01\textwidth}{ \raisebox{0.5cm}{2}} & \multirow{2}{0.01\textwidth}{ \raisebox{0.5cm}{1}} &\multirow{2}{0.2\textwidth}{\hspace{0.24cm}\raisebox{0.49cm}{ $\dfrac{\left(\sqrt{3}-1\right) \left(\sqrt{3}+1\right)^2}{16 \sqrt[4]{3}}$} 
          }& 
          
          \raisebox{0.1cm}{ $-\dfrac{\left(\sqrt{3}-1\right) \left(\sqrt{3}+1\right)^2}{16 \sqrt[4]{3}}+\dfrac{\left(13 \sqrt{3}-116\right) \left(\sqrt{3}+1\right)}{180 \sqrt[4]{3}}\dfrac{\theta_2}{\mathfrak{c}}+\dfrac{\left(\sqrt{3}-2\right) \left(\sqrt{3}+1\right)}{2 \sqrt[4]{3}}\dfrac{\theta_4}{\mathfrak{c}}$} \\
          & & & \raisebox{0.3cm}{ $-\dfrac{2 \left(\sqrt{3}-2\right) \left(\sqrt{3}+1\right)}{\sqrt[4]{3}}\dfrac{\theta_6}{\mathfrak{c}}$}\\[-0.2cm]
        \hline
         \multirow{2}{0.01\textwidth}{ \raisebox{0.45cm}{2}} & \multirow{2}{0.01\textwidth}{ \raisebox{0.45cm}{2}} &\multirow{2}{0.25\textwidth}{ \hspace{0.1cm}\raisebox{0.38cm}{$\dfrac{1}{24} \sqrt{\dfrac{1}{2} \left(963-529 \sqrt{3}\right)}$}}& \raisebox{0.15cm}{$
            -\dfrac{1}{24} \sqrt{\dfrac{1}{2} \left(963-529 \sqrt{3}\right)}
            -\dfrac{\sqrt{187519473-102837005 \sqrt{3}}}{3780}\dfrac{\theta_2}{\mathfrak{c}}
            +\dfrac{1}{60} \sqrt{162387-92399 \sqrt{3}}\dfrac{\theta_4}{\mathfrak{c}}$}\\
            & & &
             \raisebox{0.45cm}{+$\dfrac{78 \left(9 \sqrt{3}+23\right)}{\sqrt{61617 \sqrt{3}+106731}}\dfrac{\theta_6}{\mathfrak{c}}+4796 \sqrt{\dfrac{2}{241053 \sqrt{3}+417609}}\dfrac{\theta_8}{\mathfrak{c}}$ }\\[-1.5cm]
   & & &\\
        \hline
        \raisebox{0.16cm}{1} & \raisebox{0.16cm}{3} & \raisebox{0.2cm}{$\dfrac{1}{3 \sqrt{3}}$} 
          & \raisebox{0.19cm}{$-\dfrac{1}{3 \sqrt{3}} 
            -\dfrac{35396}{25515 \sqrt{3}}\dfrac{\theta_2}{\mathfrak{c}} 
            +\dfrac{112}{1215 \sqrt{3}}\dfrac{\theta_4}{\mathfrak{c}} 
            +\dfrac{448}{81 \sqrt{3}}\dfrac{\theta_6}{\mathfrak{c}} 
            +\dfrac{512}{81 \sqrt{3}}\dfrac{\theta_8}{\mathfrak{c}}$} \\
        \hline
       \multirow{1}{0.01\textwidth}{\raisebox{0.0cm}{2}}  & \multirow{1}{0.01\textwidth}{\raisebox{0.0cm}{3}} & \multirow{1}{0.2\textwidth}{\centering{\hspace{0.1cm}$\dfrac{1}{3 \sqrt{3}}$} 
         } &\multirow{1}{\textwidth}{ \centering{$-\dfrac{1}{3 \sqrt{3}} 
            -\dfrac{171496}{127575 \sqrt{3}}\dfrac{\theta_2}{\mathfrak{c}} 
            +\dfrac{10160}{5103 \sqrt{3}}\dfrac{\theta_4}{\mathfrak{c}} 
            +\dfrac{2752}{135 \sqrt{3}}\dfrac{\theta_6}{\mathfrak{c}} 
            +\dfrac{1280}{27 \sqrt{3}}\dfrac{\theta_8}{\mathfrak{c}} 
            +\dfrac{1024}{27 \sqrt{3}}\dfrac{\theta_{10}}{\mathfrak{c}}$}} \\
        \hline
         \multirow{2}{0.01\textwidth}{ \raisebox{0.35cm}{3}} & \multirow{2}{0.01\textwidth}{ \raisebox{0.35cm}{1}} &\multirow{2}{0.2\textwidth}{\centering{ \raisebox{0.1cm}{$\dfrac{1}{3 \sqrt{3}}$} }}
          & \raisebox{0.12cm}{$-\dfrac{1}{3 \sqrt{3}} 
            -\dfrac{5662841}{4209975 \sqrt{3}}\dfrac{\theta_2}{\mathfrak{c}} 
            +\dfrac{2857228}{1148175 \sqrt{3}}\dfrac{\theta_4}{\mathfrak{c}} 
            +\dfrac{2398688}{76545 \sqrt{3}}\dfrac{\theta_6}{\mathfrak{c}} 
            +\dfrac{423776}{3645 \sqrt{3}}\dfrac{\theta_8}{\mathfrak{c}}$} \\
            & & & \raisebox{0.32cm}{$+\dfrac{50432}{243 \sqrt{3}}\dfrac{\theta_{10}}{\mathfrak{c}} 
            +\dfrac{11776}{81 \sqrt{3}}\dfrac{\theta_{12}}{\mathfrak{c}}$} \\[4pt]\hline
    \end{tabular}
    }
    
    \caption{\justifying Coefficients $(\eta_0,\eta_1)$ of the orbital frequency and $(\Upsilon_0,\Upsilon_1)$ of the Lyapunov exponent of the photon sphere, computed from eqs. \eqref{eq:orbital_frequency} and \eqref{eq:Lyapunov_exponent} with Padé approximants $U^2_{N,M}$ of the potential.} 
    \label{Tab:potential_Padé}
\end{table*}

\subsubsection{Rezzolla-Zhidenko parametrization}
\label{EL:RZ}
In the RZ parametrisation, the expansion parameter $\epsilon$ appears naturally. We assume that the coefficients $a_n$ and $b_n$ admit power series in $\epsilon$,
\begin{equation}\label{RZ epsilon scaling}
    a_n(\epsilon) = \sum_{p=0} A_{n,p} \, \epsilon^p \qq{and} b_n(\epsilon) = \sum_{p=0} B_{n,p}\, \epsilon^p \ ,
\end{equation}
for $n\geq 0$, with $A_{0,0} = A_{1,0} = B_{0,0} = B_{1,0} = 0$. The latter condition ensures that the leading contributions of $a_0(\epsilon)$, $a_1(\epsilon)$, $b_0(\epsilon)$ and $b_1(\epsilon)$ to the metric functions vanish in the Schwarzschild limit $\epsilon \to 0$, such that the deformations in eqs.~\eqref{asympfix_1} and \eqref{asympfix_2} are switched off continuously.

The values of $A_{n,p}$ and $B_{n,p}$ depend on the model. In general, they enter the expansion of the quantities in eq.~\eqref{eq:EIKONAL-QUANTITIES} through non-linear combinations at each order in $\epsilon$. The expressions simplify considerably for $A_{2,0} = B_{2,0} = 0$, in which case
\begin{subequations}\label{eikonal q rz}
    \begin{align}
        \frac{\rps^{\rm RZ}}{M} &= 3 - \frac{4}{9}\left( 5+A_{0,1}+A_{1,1} \right)\, \epsilon + \order{\epsilon^2} \ , \\
        M\Omega^{\rm RZ} &=  \frac{1}{3 \sqrt{3}} + \frac{2}{81\sqrt{3}} \left( 6 +3 A_{0,1}+2A_{1,1} \right) \, \epsilon + \order{\epsilon^2}\  , \\
        M\lambda^{\rm RZ} &= \frac{1}{3 \sqrt{3}} + \frac{2}{81\sqrt{3}} \left( -4 + 7 A_{0,1}+ 4 A_{1,1} +  9 B_{0,1} - 6B_{1,1}\right) \, \epsilon + \order{\epsilon^2} \ . 
    \end{align}
\end{subequations}
In what follows we restrict to black hole metrics that satisfy this condition.

\subsubsection{Johannsen-Psaltis parametrization}
\label{EL:JP}
Since $g^t(r)$ and $g^r(r)$ have to disappear when the Schwarzschild geometry is recovered, all JP parameters $\{ \epsilon_n^t , \epsilon_n^r \}_{n\geq 1}$ scale homogeneously with $\epsilon$. In particular, eq.~\eqref{JPrel} gives exactly $\epsilon_1^t = -2\, \epsilon$. For $\epsilon \to 0$ and with the scaling \eqref{RZ epsilon scaling}, one generally has $\epsilon_1^r , \epsilon_2^t \sim \order{\epsilon^k}$ with $k \geq 1$. We therefore assume the power series
\begin{subequations}\label{JP epsilon scaling}
    \begin{align}
        \epsilon_n^t(\epsilon) = \sum_{p=1}(\epsilon^t_n)_p \, \epsilon^p \qq{for} n \geq 2 \ ,\\
        \epsilon_n^r(\epsilon) = \sum_{p=1}(\epsilon^r_n)_p \, \epsilon^p \qq{for} n \geq 1 \ ,
    \end{align}
\end{subequations}
which can equivalently be viewed as power series in $\epsilon_1^t$. The values of $(\epsilon^t_n)_p$ and $(\epsilon^r_n)_p$ are model dependent. Inserting eqs.~\eqref{JP epsilon scaling} into eq.~\eqref{eq:EIKONAL-QUANTITIES} written in the JP parametrisation, and truncating at $\epsilon_4^t$ and $\epsilon_4^r$ for brevity, we find
\begin{subequations}\label{eikonal q jp}
    \begin{align}
        \frac{\rps^{\rm JP}}{M} &= 3 - \left( \frac{8}{3} + \frac{(\epsilon_2^t)_1}{9}+\frac{(\epsilon_3^t)_1}{18} + \frac{2(\epsilon_4^t)_1}{81} \right) \epsilon + \order{\epsilon^2} \ , \\
        M\Omega^{\rm JP} &= \frac{1}{3\sqrt{3}} + \bigg( \frac{2}{9} + \frac{(\epsilon_2^t)_1}{54} +\frac{(\epsilon_3^t)_1}{162} + \frac{2(\epsilon_4^t)_1}{486} \bigg)\frac{ \epsilon}{\sqrt{3}} + \order{\epsilon^2}\  , \\
        M\lambda^{\rm JP} &= \frac{1}{3\sqrt{3}} + \bigg( \frac{5}{27} + \frac{2(\epsilon_2^t)_1}{81} + \frac{(\epsilon_3^t)_1}{162} + \frac{(\epsilon_4^t)_1}{1458} - \frac{(\epsilon_1^r)_1}{18} - \frac{(\epsilon_2^r)_1}{54} - \frac{(\epsilon_3^r)_1}{162}- \frac{(\epsilon_4^r)_1}{486} \bigg)\frac{ \epsilon}{\sqrt{3}}+ \order{\epsilon^2} \ . 
    \end{align}
\end{subequations}
As expected from eqs.~\eqref{as_bs_JP} and~\eqref{eikonal q rz}, all JP parameters enter linearly at first order in $\epsilon$.

\subsubsection{EMD (near horizon) parametrization}
\label{EL:EMD-near}
In Chapter~\ref{chap4} we have seen that the near-horizon EMD can be extended to the photon sphere with Padé approximants~\cite{bender78:AMM}. There, the position of the event horizon is written as
\begin{equation}\label{r_horizonEMD}
    \hr = 2M (1 + \mathfrak{c}) \ ,
\end{equation}
so that the first quantity in eq.~\eqref{eq:EIKONAL-QUANTITIES} takes the form \eqref{rps expans},
\begin{align}
        \frac{\rps^{\rm EMD(nh)}}{M} &= \sigma_0 + \sigma_1 \, \mathfrak{c} + \order{\mathfrak{c}^2}\ , \label{eikonal q rps emd nh}
    \end{align}
The parameter $\mathfrak{c}$, which shifts $\hr$ away from $2M$, is related to $\epsilon$ of the RZ parametrisation through
\begin{equation}\label{c_EMD}
    \mathfrak{c}=-\frac{\epsilon}{\epsilon+1} \ .
\end{equation}
The coefficients $(\sigma_0,\sigma_1)$ are linear combinations of the parameters $\{\theta_{2n},\xi_{2n}\}_{n \geq 1}$, whose explicit form depends on the order of the Padé approximant; the first orders are listed in Table~\ref{Tab:PositionPade}. A Padé approximant of order $(N,M)$ involves $N+M$ parameters in total. As we shall see for specific models, the effective parameters $\{\theta_{2n}, \xi_{2n}\}_{n \geq 1}$ still have to be expanded in powers of $\epsilon$.

In the near-horizon EMD, the potential $U^2(r)$ is expanded around $\hr$ as in eq.~\eqref{PadePotential},
\begin{equation}\label{u_series}
        U^2(r)=\sum_{n=1}^{\infty}v_n (r-\hr)^n,
    \end{equation}
with coefficients $v_n$ depending on $\{\theta_{2k}\}$ and $\{\xi_{2k}\}$. As shown in Chapter~\ref{chap4}, Padé approximants of sufficiently high order $(N,M)$ remain accurate up to the location of the photon sphere and beyond. This allows us to compute the shadow radius $\bsh$, the orbital frequency $\Omega$ and the Lyapunov exponent $\lambda$. For small deviations $\mathfrak{c} \ll 1$ of the horizon from its Schwarzschild position (see eq.~\eqref{r_horizonEMD}), we write $\Omega$ and $\lambda$ of eqs.~\eqref{eq:orbital_frequency} and~\eqref{eq:Lyapunov_exponent} in units of the mass $M$ as
\begin{subequations}\label{eq:omegal}
    \begin{align}
        M \Omega&=\eta_0+\eta_1\;\mathfrak{c}+\mathcal{O}(\mathfrak{c}^2) \ , \label{eikonal q omega emd nh}\\
        M \lambda&=\Upsilon_0+\Upsilon_1\;\mathfrak{c}+\mathcal{O}(\mathfrak{c}^2) \ . \label{eikonal q lyapunov emd nh}
    \end{align}        
    \end{subequations}
The coefficients $(\eta_0,\eta_1)$ and $(\Upsilon_0,\Upsilon_1)$ depend on the order $(N,M)$ of the Padé approximation and are given in Table~\ref{Tab:potential_Padé}. The Schwarzschild limit is reproduced from order $(1,3)$ on, in agreement with Remark~\ref{rem:ExactPade}. Through eq.~\eqref{eq:omegal}, the eikonal QNMs can be reconstructed from eq.~\eqref{eq:eikonal} at any order of the Padé approximation.
\begin{figure*}[ht]
        \centering
        \hbox{\includegraphics[width=\linewidth]{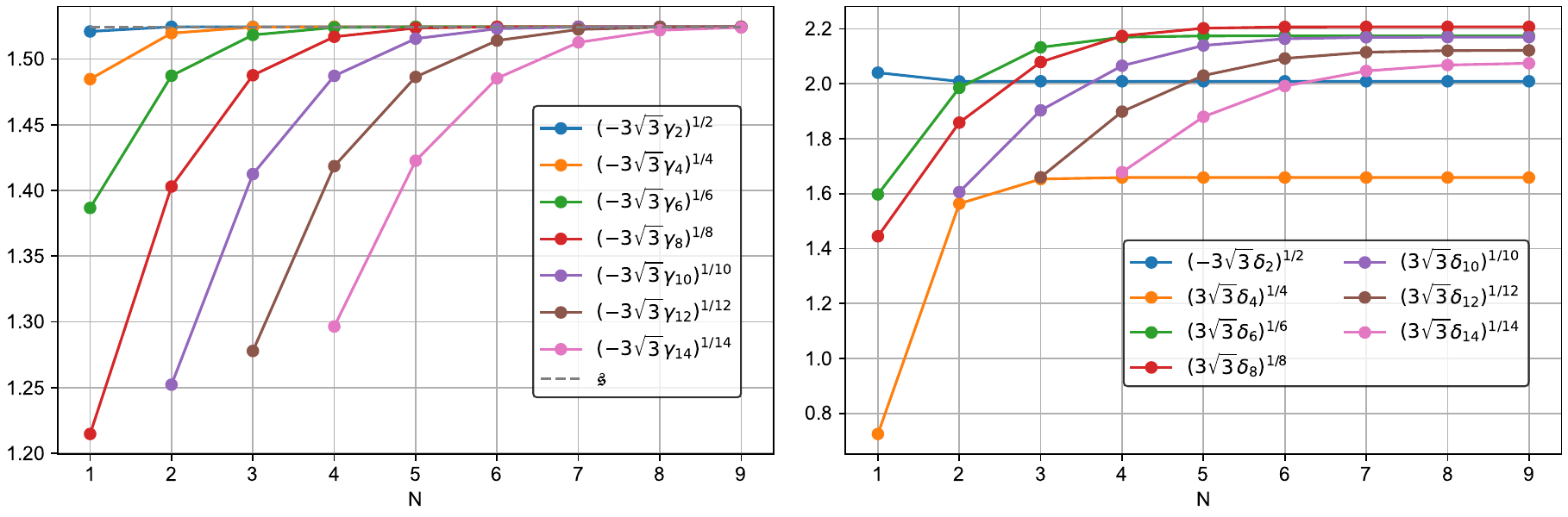}} 
        \caption{\justifying Coefficients $\gamma_{2k}$ (left) and $\delta_{2k}$ (right) as functions of the order $N$ of Padé approximants of type $(N,3)$.}
        \label{fig:QNM_frequency}
\end{figure*}

Since $\bsh$ and $\Omega$ are related, the large-$N$ analysis of Sec.~\ref{Sect:PadePhoton} for approximants of order $(N,3)$ can be extended to the orbital frequency. We recall the expansion \eqref{ShadowExpans} of the shadow radius in Sec.~\ref{Sect:PadePhoton},
\begin{equation}
          \frac{b_{\rm{sh}}}{M}=\tau_0+\tau_1\;\mathfrak{c}+\mathcal{O}(\mathfrak{c}^2)\ ,
    \end{equation}
together with the dependence \eqref{TauParametrisation} of $\tau_1$ on the deformation coefficients and on the order $N$,
\begin{align}
        \tau_1(N)&=3\sqrt{3}+\sum_{k=1}^{N+3} \beta_{2k}(N)\frac{\theta_{2k}}{\mathfrak{c}} \;\;\;\; \qq{with}   \;\;\;\;      \beta_{2k}(N)=\frac{\partial}{\partial \,\theta_{2k}}\frac{1}{U_{N,3}}\bigg|_{(\rps)_{N,3}}\ , 
    \end{align}
where $(\rps)_{N,3}$ and $U_{N,3}$ denote the photon sphere radius and the potential computed with the Padé approximant of order $(N,3)$. Similarly, we write the coefficient $\eta_1$ of the orbital frequency as
\begin{align}
        \eta_1(N)&=-\frac{1}{3\sqrt{3}}+\sum_{k=1}^{N+3} \gamma_{2k}(N)\frac{\theta_{2k}}{\mathfrak{c}}
        \;\;\;\; \qq{with}   \;\;\;\;         \gamma_{2k}(N)=\frac{\partial \,U_{N,3}}{\partial \,\theta_{2k}}\bigg|_{(\rps)_{N,3}}\ .
    \end{align}
At lowest order $U_{N,3}=1/(3\sqrt{3})$, such that the two sets of coefficients are related by
\begin{align}
        \beta_{2k}(N)&=\pdv{\theta_{2k}}\frac{1}{U_{N,3}}\bigg|_{(r_{\rm ps})_{N,3}}=-\frac{1}{U^2_{N,3}}\gamma_{2k}(N)=-(3\sqrt{3})^2\gamma_{2k}(N)\ ,
    \end{align}
The asymptotic behaviour of the $\gamma_{2k}(N)$ therefore follows from the $N\to \infty$ limit \eqref{Svalue} of the $\beta_{2k}(N)$,
\begin{equation}
        \lim_{N\to\infty} \frac{\beta_{2k}(N)}{3\sqrt{3}}\to\mathfrak{s}^{2k}\ , \qquad \mathfrak{s}\sim 1.5245.
    \end{equation}
This is compatible with Figure~\ref{fig:QNM_frequency}, where the first coefficients $\gamma_{2k}$ are shown up to $N=9$. The asymptotic value of $\eta_1(N)$ then becomes
\begin{align}
        \lim_{N\to\infty} \mathfrak{c}\,\eta_1(N)&=-\frac{\mathfrak{c}}{3\sqrt{3}}-\sum_{k=1}^\infty\frac{\mathfrak{s}^{2k}}{3\sqrt{3}}\theta_{2k}
        =-\frac{\mathfrak{c}}{3\sqrt{3}}-\frac{1}{3\sqrt{3}}\frac{1}{2M}[\Psi(2M\mathfrak{s})-\hr],
    \end{align}
which leads to the following compact expression.

\begin{center}
\resultbox{\result\label{res:OmegaEMD}{Leading deformation of the photon-sphere frequency}}{\parbox{16cm}{
In the limit $N\to\infty$ of Padé approximants of order $(N,3)$, the orbital frequency of the photon sphere of a black hole described by the near-horizon EMD is, to leading order in the deformation,
\begin{align}
       \lim_{N\to\infty} \Omega&=\lim_{N\to\infty}(\eta_0+\eta_1 \mathfrak{c})=
        \frac{2}{3\sqrt{3}}-\frac{1}{3\sqrt{3}}\frac{1}{2M}\Psi(2M\mathfrak{s}).
    \end{align}
in units $M=1$. As for the shadow (Result~\ref{res:ShadowEMD}), the leading correction depends only on the deformation function $\Psi$ at the single point $\rho=2M\mathfrak{s}$. It is consistent with $\Omega\simeq 1/\bsh$ and \eqref{CompactCorrectionShadow}.
}}
\end{center}

\noindent
The Lyapunov exponent is more intricate, since $\lambda$ depends not only on the photon sphere radius but also on second derivatives of the potential. By analogy with $\Omega$, we adopt the ansatz
\begin{equation}
        \Upsilon_1(N)=\frac{1}{3\sqrt{3}}+\sum_{k=1}^{N+3}\delta_{2k}(N)\frac{\theta_{2k}}{\mathfrak{c}},\label{LimFrequency}
    \end{equation}
where the numerical coefficients $\delta_{2k}$ depend on the order of the Padé approximation. Figure~\ref{fig:QNM_frequency} shows the $\delta_{2k}$ up to $k=7$ and $N=9$, which appear to approach constant values. A similar numerical analysis as in the case of the black hole shadow (Result~\ref{res:ShadowEMD}) or the photon-sphere frequency (Result~\ref{res:OmegaEMD}) suggests that asymptotically the coefficients $\delta_{2k}(N)$ approach\footnote{Indeed, for $N=11$, the difference between (\ref{AsymForm}) and the numerical value for $\delta_{2k}(11)$ up to $k=6$ is fully determined by the numerics.} the expression
\begin{align}
\lim_{N\to\infty}\,\delta_{2k}(N)=-\frac{1}{3\sqrt{3}}\left(\mathfrak{s}^{2k}-\frac{9}{8}\left(-\frac{2}{\sqrt{3}}\,2k\,\mathfrak{s}^{2k-1}+2k\,(2k-1)\,\mathfrak{s}^{2k-2}\right)\right)\,.\label{AsymForm}
\end{align}
Assuming this asymptotic form to be exact for $N\to\infty$ we may resum the Lyapunov exponent
\begin{align}
M\,\lambda&=\frac{1}{3\sqrt{3}}\bigg(1-\mathfrak{c}-\sum_{k=1}^\infty\,(2M\mathfrak{s})^{2k}\,\theta_{2k}-\frac{9}{4\sqrt{3}}\,2M\,\sum_{k=1}^\infty 2k\,(2M\mathfrak{s})^{2k-1}\,\theta_{2k}\nonumber\\
&\hspace{2cm}+\frac{9}{8}(2M)^2\sum_{k=1}^\infty2k(2k-1)\,(2M\mathfrak{s})^{2k-2}\,\theta_{2k}\bigg)\,,\label{FormLyap}
\end{align}
which leads to the following compact expression: 
\begin{center}
\resultbox{\result\label{res:lambdaEMD}{Leading deformation of the Lypaunov exponent}}{\parbox{16cm}{
In the limit $N\to\infty$ of Padé approximants of order $(N,3)$, the Lyapunov exponent of the photon sphere of a black hole described by the near-horizon EMD is, to leading order in the deformation,
\begin{align}
       \lim_{N\to\infty} M\,\lambda&=\frac{1}{3\sqrt{3}}\bigg(2-\frac{\Psi(2M\,\mathfrak{s})}{2M}-\frac{9}{4\sqrt{3}}\,\frac{\dd\Psi}{\dd\rho}(2M\,\mathfrak{s})+\frac{9}{8}(2M)\,\frac{\dd^2\Psi}{\dd\rho^2}(2M\,\mathfrak{s})\bigg)\,.
    \end{align}
In contrast to the shadow (Result~\ref{res:ShadowEMD}) or the photon radius frequency (Result~\ref{res:OmegaEMD}), the leading correction depends not only on the deformation function $\Psi$ at the single point $\rho=2M\mathfrak{s}$, but also its first and second derivative (with respect to the radial distance) evaluated at the same position. Thus, to leading order $\lambda$ is still determined by the metric deformation function at a radial distance $\rho=2M\mathfrak{s}$, however in a more intricate fashion. 
}}
\end{center}
Notice, upon defining the dimensionless quantities
\begin{align}
&\Psi_0(\rho_0)=\frac{1}{2M}\,\Psi(\rho/(2M))\,,&&\text{and} &&\rho_0=\frac{\rho}{2M}\,,
\end{align}
we can write $\lambda$ even more compactly
\begin{align}
M\,\lambda&=\frac{1}{3\sqrt{3}}\bigg(2-\Psi_0(\mathfrak{s})-\frac{9}{4\sqrt{3}}\,\Psi_0'(\mathfrak{s})+\frac{9}{8}\,\Psi_0''(\mathfrak{s})\bigg)\,,\label{LyapCompact}
\end{align}
where the primes denote derivatives with respect to $\rho_0$.

Finally, we comment on the expansion of the near-horizon EMD to second order in $\epsilon$. Since the $\epsilon$ dependence enters through $\mathfrak{c}$ in eq.~\eqref{r_horizonEMD}, one has to check whether this relation is modified at second order. For all models considered below, eq.~\eqref{r_horizonEMD} receives no corrections at $\order{\mathfrak{c}^2}$, so that $\hr$ acquires contributions of $\order{\epsilon^2}$ only through $\mathfrak{c}=-\epsilon/(1+\epsilon)$. The second-order expansion, on the other hand, contains many more terms than the first-order one: linear and quadratic terms in $\{\theta_n\}$ as well as mixed products of the linear coefficients. For a Padé approximant of order $(N,M)$, the number of coefficients at second order grows quadratically with $K=N+M$, as $(K^2+5K+6)/2$, and at order $\epsilon^N$ it grows as $\sim K^N$.

We also remark that the compact form of the linearised expressions (Result~\ref{res:ShadowEMD}) for the black hole shadow, (Result~\ref{res:OmegaEMD}) for the photon-sphere frequency and (Result~\ref{res:lambdaEMD}) for the Lyapunov exponent, have been the inspiration for a linear response formalism for general black hole observables~\cite{SHlinear}.

\subsubsection{EMD (at large distance) parametrization}
\label{EL:EMD-large}
The large-distance EMD must reduce to the Schwarzschild metric when all deformation parameters $\{\gamma_i, \omega_i \}_{i\geq 1}$ vanish. The position of the event horizon does not appear explicitly as an input in this parametrisation, so there is no direct link to the parameter $\epsilon$. Nevertheless, the relations \eqref{RZ to EMD large distance} together with the ansatz~\eqref{RZ epsilon scaling} (or, equivalently, eq.~\eqref{JP to EMD fr} together with the ansatz~\eqref{JP epsilon scaling}) show that the $\{\gamma_i, \omega_i \}_{i\geq 1}$ do admit power series expansions in $\epsilon$. We therefore write
\begin{equation}
    \gamma_n (\epsilon) = \sum_{p=1} \gamma_{n,p} \, \epsilon^p \qq{and} \omega_n (\epsilon) = \sum_{p=1} \omega_{n,p} \, \epsilon^p \ ,
\end{equation}
where the coefficients $\{\gamma_{n,p},\omega_{n,p}\}_{n\geq 1,\ p \geq 1}$ depend on the model. With the metric functions \eqref{eq:EMD-functions}, the quantities in eq.~\eqref{eq:EIKONAL-QUANTITIES} become
\begin{subequations}\label{eikonal q emd ld}
    \begin{align}
        \frac{\rps^{\rm EMD(ld)}}{M} &= 3 + \frac{1}{9}\Big[(15-3k+2k^2) \gamma_{1,1} + (5-4k)\gamma_{2,1}+2\gamma_{3,1} \Big] \, \epsilon + \order{\epsilon^2}\ , \\
        M \Omega^{\rm EMD(ld)} &= \frac{1}{3 \sqrt{3}} + \frac{1}{162\sqrt{3}} \Big[ (-2 k^2 + 4 k - 21) \gamma_{1,1} + (4k-6) \gamma_{2,1} - 2 \gamma_{3,1}\Big] \, \epsilon + \order{\epsilon^2} \  , \\
        M \lambda^{\rm EMD(ld)} &= \frac{1}{3 \sqrt{3}} + \frac{1}{162\sqrt{3}} \Big[ (6 k^2 - 4 k + 21) \gamma_{1,1} + 2 (5k-6) \gamma_{2,1} + 6 \gamma_{3,1} \notag \\
        & \qquad - (2k^2 - 4k + 21) \omega_{1,1} + (4k -6) \omega_{2,1} - 2 \omega_{3,1}\Big] \, \epsilon + \order{\epsilon^2} \  , 
    \end{align}
\end{subequations}
where the dependence on the gauge parameter $k$ has been kept to retain full generality.

\subsection{Application to BH models}
\begin{table*}[ht]
    \centering
    \renewcommand{\arraystretch}{2.5} 
    \resizebox{\textwidth}{!}{
    \begin{tabular}{|P{3cm}|P{3.5cm}|P{3.3cm}|P{3.5cm}|P{4.5cm}|}
        \hline
        \raisebox{0.07cm}{BH models} & \raisebox{0.07cm}{Parametrizations} & \raisebox{0.12cm}{$\dfrac{\rps}{M} - 3$} & \raisebox{0.12cm}{$M\Omega - \dfrac{1}{3\sqrt{3}}$} & \raisebox{0.12cm}{$M\lambda - \dfrac{1}{3\sqrt{3}}$} \\ 
        \hline
        \multirow{5}{*}{\raisebox{-0.22cm}{Hayward}}
        & From the model & $-\dfrac{16}{9}\epsilon + \dfrac{80}{27}\epsilon^2 $ 
                & \raisebox{0.015cm}{$\dfrac{8}{81 \sqrt{3}}\epsilon - \dfrac{40}{243 \sqrt{3}}\epsilon^2 $} 
                & \raisebox{0.015cm}{$-\dfrac{16}{81 \sqrt{3}}\epsilon + \dfrac{16}{243 \sqrt{3}} \epsilon^2$} \\ 
                & Rezzolla-Zhidenko & $-\dfrac{16}{9}\epsilon + \dfrac{28892}{9801} \epsilon^2$ & \raisebox{0.015cm}{$\dfrac{8}{81\sqrt{3}}\epsilon - \dfrac{4012}{24057\sqrt{3}} \epsilon^2$}& \raisebox{0.015cm}{$-\dfrac{16}{81\sqrt{3}} \epsilon + \dfrac{173960}{2910897 \sqrt{3}} \epsilon^2 $}\\ 
                & Johannsen-Psaltis &$-\dfrac{16}{9} \epsilon+ \dfrac{32}{9} \epsilon^2$ & \raisebox{0.015cm}{$\dfrac{8}{81 \sqrt{3}} \epsilon -\dfrac{440 }{2187 \sqrt{3}} \epsilon ^2$}& \raisebox{0.015cm}{$ -\dfrac{16 }{81 \sqrt{3}} \epsilon + \dfrac{32 }{243 \sqrt{3}} \epsilon ^2 $} \\
                & EMD near horizon & $-\dfrac{16}{9} \epsilon + \dfrac{799}{243} \epsilon^2$ & \raisebox{0.015cm}{$\dfrac{8}{81 \sqrt{3}} \epsilon- \dfrac{110}{729\sqrt{3}} \epsilon^2$} & \raisebox{0.015cm}{$-\dfrac{16}{81\sqrt{3}} \epsilon -\dfrac{107}{2187\sqrt{3}} \epsilon^2 $}\\ 
                & EMD large distance & $-\dfrac{16}{9} \epsilon + \dfrac{176 }{81} \epsilon ^2$ & \raisebox{0.015cm}{$\dfrac{8}{81 \sqrt{3}} \epsilon -\dfrac{296 }{2187 \sqrt{3}} \epsilon ^2$}& \raisebox{0.015cm}{$-\dfrac{16}{81\sqrt{3}} \epsilon -\dfrac{368 }{2187 \sqrt{3}} \epsilon ^2$} \\[0.2cm]
        \hline
        \multirow{5}{*}{\raisebox{-0.2cm}{Bardeen}} 
        & From the model & $-\dfrac{20}{9}\epsilon + \dfrac{650}{243}\epsilon^2$ 
                & $\dfrac{4}{27 \sqrt{3}}\epsilon - \dfrac{98}{729 \sqrt{3}}\epsilon^2 $ 
                & $ -\dfrac{8}{81 \sqrt{3}}\epsilon - \dfrac{68}{729 \sqrt{3}} \epsilon^2 $ \\ 
                & Rezzolla-Zhidenko & $-\dfrac{20}{9}\epsilon + \dfrac{650}{243}\epsilon^2$ & $\dfrac{4}{27 \sqrt{3}}\epsilon - \dfrac{98}{729 \sqrt{3}}\epsilon^2 $ & $ -\dfrac{8}{81 \sqrt{3}}\epsilon - \dfrac{68}{729 \sqrt{3}} \epsilon^2 $\\ 
                & Johannsen-Psaltis & $-\dfrac{20}{9}\epsilon + \dfrac{650}{243}\epsilon^2$ & $\dfrac{4}{27 \sqrt{3}}\epsilon - \dfrac{98}{729 \sqrt{3}}\epsilon^2 $ & $ -\dfrac{8}{81 \sqrt{3}}\epsilon - \dfrac{68}{729 \sqrt{3}} \epsilon^2 $ \\
                & EMD near horizon & $-\dfrac{20}{9}\epsilon+\dfrac{652 }{243}\epsilon ^2$ & $\dfrac{4}{27 \sqrt{3}} \epsilon- \dfrac{391}{2916\sqrt{3}} \epsilon^2$ & $-\dfrac{8}{81\sqrt{3}} \epsilon -\dfrac{421}{4374\sqrt{3}} \epsilon^2 $ \\ 
                & EMD large distance & $-\dfrac{20}{9} \epsilon + \dfrac{370 }{243} \epsilon ^2$ & $\dfrac{4}{27 \sqrt{3}} \epsilon -\dfrac{98 }{729 \sqrt{3}} \epsilon ^2$& $-\dfrac{8}{81\sqrt{3}} \epsilon - \dfrac{68}{729 \sqrt{3}} \epsilon ^2$ \\[0.2cm]
        \hline
        \multirow{1}{*}{\raisebox{-0.05cm}{Simpson-Visser II}} 
        &   \multirow{1}{*}{\raisebox{-0.05cm}{All parametrizations}} & \raisebox{0.05cm}{$-\dfrac{8}{3}\epsilon + \dfrac{74}{27}\epsilon^2 $} 
                & \raisebox{0.07cm}{$\dfrac{2}{9 \sqrt{3}}\epsilon - \dfrac{7}{81 \sqrt{3}}\epsilon^2 $} 
                & \raisebox{0.07cm}{$\dfrac{2}{27 \sqrt{3}}\epsilon - \dfrac{11}{81\sqrt{3}} \epsilon^2$} \\[0.1cm]
        \hline
    \end{tabular}
    }
      \caption{\justifying Photon sphere radius $\rps$, orbital frequency $\Omega$ and Lyapunov exponent $\lambda$ in units of the mass $M$, minus their Schwarzschild values, to second order in $\epsilon$. For the Hayward, Bardeen and Simpson-Visser II models, the exact expansion is compared with the four parametrisations, each truncated to four parameters. For Simpson-Visser II all parametrisations agree with the exact result, which is therefore the only one shown.}
      \label{tab:BH-models}
\end{table*}
\renewcommand{\arraystretch}{1}
\begin{table*}[ht]
    \centering
    \renewcommand{\arraystretch}{2} 

   \resizebox{0.85\textwidth}{!}{
\begin{tabular}{|P{5cm}|P{1cm}|P{1cm} |P{1cm} |P{1cm} |P{1.4cm} |P{1.4cm}| }
\hline
\raisebox{0.05cm}{BH models}  & \multicolumn{2}{c|}{\raisebox{0.07cm}{Hayward}} & \multicolumn{2}{c|}{\raisebox{0.07cm}{Bardeen}} & \multicolumn{2}{c|}{\raisebox{0.07cm}{Simpson-Visser II}} \\ 
\hline
\raisebox{0.07cm}{order} &\raisebox{0.07cm}{$\epsilon$} &\raisebox{0.07cm}{$\epsilon^2$} &  \raisebox{0.07cm}{$\epsilon$}         &      \raisebox{0.07cm}{$\epsilon^2$}         &       \raisebox{0.07cm}{$\epsilon$}    &        \raisebox{0.07cm}{$\epsilon^2$}      \\
\hline
Rezzolla-Zhidenko  &   2        &      7     &     2      &      4     &    1       &   1        \\
Johannsen-Psaltis  &     3      &   6        &      2  &      4     &     1      &       2    \\
EMD near horizon &     4      &      $>7$     &         4  &       $>7$     &   4        & 4        \\
EMD large distance &    3      &     6    &     2      &     4      &     1      &       2  \\[0.12cm]
\hline
\end{tabular}
}
    \caption{\justifying Minimal number of parameters with which each parametrisation reproduces the exact coefficients of order $\epsilon$ and $\epsilon^2$ of the three observables.}
     \label{tab:NumbParam}
\end{table*}
\begin{figure*}[ht]
    \centering
    \includegraphics[width=\textwidth]{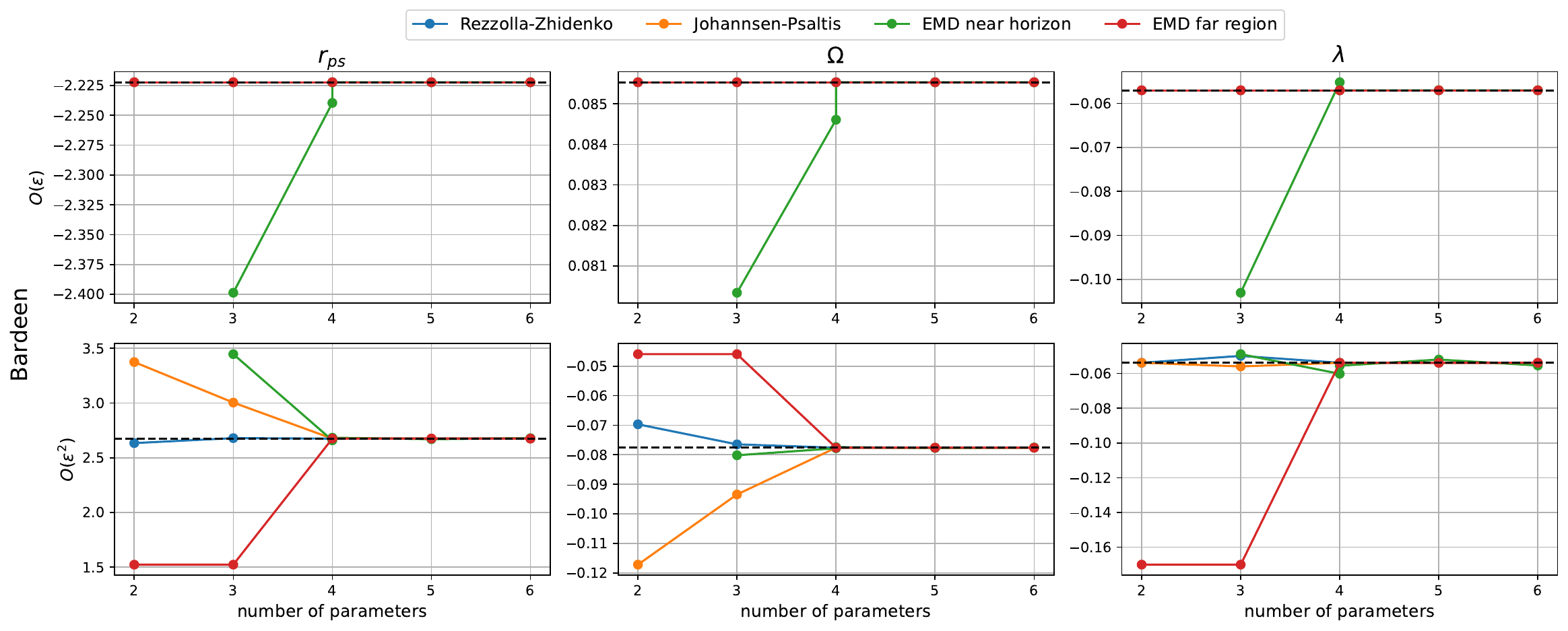}
    \caption{\justifying Bardeen black hole: convergence of the linear (upper row) and quadratic (lower row) coefficients as the number of parameters of each parametrisation is increased. The dashed black line is the exact value of Table~\ref{tab:BH-models}. For the near-horizon EMD with four parameters, both the $(2,2)$ and the more accurate $(1,3)$ Padé approximants are shown.}
    \label{fig:bardeen_conv}
\end{figure*}
\renewcommand{\arraystretch}{1}

To illustrate the formalism, we consider three regular black hole geometries: the Hayward~\cite{Hayward_2006}, the Bardeen~\cite{bardeen1968} and the Simpson-Visser II~\cite{Simpson:2019mud} black holes. They can be described as deformations of the Schwarzschild geometry and satisfy $g_{tt} \, g_{rr}=-1$. Their temporal metric components are
\begin{subequations}
    \begin{align}
        -g_{tt}^{\rm H}(r) &= 1 - \frac{2M r^2}{r^3 + 2 M \,l_{\rm H}^2} \ , \label{gtt Hayward} \\
        -g_{tt}^{\rm B}(r) &= 1 - \frac{2M r^2}{(r^2 + l_{\rm B}^2)^{3/2}} \ , \label{gtt Bardeen}\\
        -g_{tt}^{\rm SV2}(r) &= 1 - \frac{2M}{r}\exp{-\frac{l_{\rm SV2}}{r}} \ . \label{gtt SV2}
    \end{align}
\end{subequations}
The regularisation length scales $(l_{\rm H},l_{\rm B},l_{\rm SV2})$ have a different physical interpretation in each model and can vary over a finite range. In terms of $\epsilon$ they read
\begin{subequations}
    \begin{align}
        l_{\rm H} &= 2M \sqrt{\frac{\epsilon}{(1+\epsilon)^3}} \qq{with} l_{\rm H} \in [0,4M/3\sqrt{3}] \implies \epsilon \in [0,1/2] \ , \\
        l_{\rm B} &= 2M \frac{\sqrt{(1+\epsilon)^{2/3}-1}}{1+\epsilon} \qq{with} l_{\rm B} \in [0,4M/3\sqrt{3}] \implies \epsilon \in [0, (3/2)^{3/2}-1]  \ , \\
        l_{\rm SV2} &= 2M \frac{\log(1+\epsilon)}{1+\epsilon} \qq{with} l_{\rm SV2} \in [0,2M/e] \implies \epsilon \in [0, e-1]\ .
    \end{align}
\end{subequations}

For each model, we expand the quantities in Eqs.~\eqref{eq:EIKONAL-QUANTITIES} analytically around $\epsilon = 0$ up to quadratic order, and we compare the result with the Rezzolla-Zhidenko, Johannsen-Psaltis, near-horizon EMD and large-distance EMD parametrisations. In each case we fix the number of free parameters to four in Eqs.~\eqref{eikonal q rz}, \eqref{eikonal q jp}, \eqref{eikonal q rps emd nh}-\eqref{eikonal q omega emd nh}-\eqref{eikonal q lyapunov emd nh} and \eqref{eikonal q emd ld}. Table~\ref{tab:BH-models} summarises the comparison. At linear order, all parametrisations reproduce the exact result for all observables and all models. At quadratic order the picture is more varied. For the Simpson-Visser~II black hole all parametrisations trivially give identical expansions. For the Hayward black hole with four parameters, none of the parametrisations performs significantly better than the others, while for the Bardeen black hole the JP and RZ parametrisations already reproduce the exact expressions.

Table~\ref{tab:NumbParam} presents a complementary analysis. Instead of fixing the number of parameters beforehand, we fix the desired accuracy and determine how many parameters each parametrisation needs to reproduce the exact expressions. For the near-horizon EMD, this corresponds to a Padé approximation of order $(N,3)$, with the number of free parameters controlled by $N$. Also in this setting, the performance of each parametrisation depends on the model, and no parametrisation is favoured across all cases.

Finally, Figure~\ref{fig:bardeen_conv} shows, for the Bardeen black hole as a representative example, how the linear and quadratic coefficients converge as the number of parameters is increased. A similar behaviour is found for the other models. With a sufficiently large number of coefficients, all parametrisations perform comparably.

\begin{center}
\resultbox{\result\label{res:ParamComparison}{Performance of agnostic parametrisations for eikonal QNMs}}{\parbox{16cm}{
For the Hayward, Bardeen and Simpson-Visser II black holes, the JP, RZ, near-horizon EMD and large-distance EMD parametrisations with four parameters all reproduce the photon sphere radius, the orbital frequency and the Lyapunov exponent exactly at linear order in $\epsilon$ (Table~\ref{tab:BH-models}). The number of parameters needed to reproduce the higher orders depends on the parametrization  and the model (Table~\ref{tab:NumbParam}). The choice of parametrisation should therefore be guided by the observable, the required accuracy and how many parameters one is prepared to add, case by case. A distinctive feature of the near-horizon EMD is that its coefficients are defined through coordinate-invariant quantities and are directly tied to the regularity conditions at the horizon.
}}
\end{center}

%% file: subfiles/7-Conclusions.tex
\typeout{START baseline=\the\baselineskip, jot=\the\jot}
\chapter{Concluding remarks and outlooks}\label{chap:conclusions}

In this review we have given a unified account of the \emph{Effective Metric Description} (EMD), a model independent framework to describe black holes beyond GR, collecting and extending the results of~\cite{Binetti:2022xdi,DelPiano:2023fiw,DelPiano:2024gvw,DelPiano:2024nrl,DamiaPaciarini:2024ssf,DamiaPaciarini:2025xjc,DamiaPaciarini:2025qhq,Hohenegger:2025dic,DeFalco:2025xxx}. Focusing mainly on static and spherically symmetric black holes in 4 dimensions, deformations of metric functions arising as solutions of GR, are written as functions of a physical, coordinate-invariant quantity, such as the proper distance from the horizon, the proper time of a free-falling observer or a curvature invariant, which is in turn computed self-consistently from the deformed metric. A black hole is then characterised by its horizon radius and by the expansion coefficients of the deformation functions, which are physical parameters.\\

In Chapter~\ref{chap3} we solved the self-consistency problem through series expansions close to the horizon and derived the conditions for a regular horizon: working in an EMD based on the radial distance $\rho$ to the horizon, finiteness of the Hawking temperature and of the curvature invariants imposes conditions on the leading powers of $\rho$ in a series expansion of the metric functions (Result~\ref{res:NeutralRegularity}), while an analytic metric at the horizon requires all odd coefficients to vanish (Result~\ref{res:SmoothHorizon}). The same reasoning applied to inner horizons and to the origin yields necessary conditions for geodesic completeness and a classification of regular cores (Result~\ref{res:Cores}). We have shown that EMDs based on different physical quantities are equivalent to one another and related by explicit maps. Furthermore, we have extended the framework to charged black holes, to extremal ones, which are described by the proper time rather than the proper distance (Result~\ref{res:ExtremalEMD}), and to rotating black holes in $2+1$ dimensions.

In Chapter~\ref{chap examples} these results were tested and illustrated using a large number of black hole geometries that have previously been proposed in the literature: on a broad set of geometries: they can be used to construct new consistent deformations, they reproduce the properties of regular black holes (Hayward \cite{Hayward_2006}, Dymnikova \cite{Dymnikova:1992ux}, Bardeen \cite{bardeen1968}, Simpson--Visser \cite{Simpson:2018tsi,Simpson:2019mud}, Frolov \cite{Frolov:2016pav})  and of the holographic quBTZ black hole, they relate near-extremal deformations to the weak gravity conjecture (Section~\ref{sec:WGC-EMD}), and they show that the Bonanno--Reuter black hole~\cite{Bonanno:2000ep}, with the running scale identified self-consistently with the proper distance from the centre, has a singular horizon (Section~\ref{Sect:ExampleBonannoReuter}).

Chapter~\ref{chap4} extended the description away from the horizon by means of Pad\'e approximants: to leading order, the shadow depends only on the deformation function at a single, universal proper distance from the horizon (Result~\ref{res:ShadowEMD}), which the EHT image of M87$^\ast$ constrains at the level of $17\%$ (Result~\ref{res:EHTbound}). Chapter~\ref{chapGeo} showed, through a Gedankenexperiment, that the EMD coefficients can in principle be reconstructed from the signals of a probe falling into the black hole. Finally, Chapter~\ref{chap5} related the EMD to the Johannsen--Psaltis \cite{Johannsen:2010gy,Johannsen:2011dh} and Rezzolla--Zhidenko~\cite{Rezzolla:2014mua} parametrizations and showed that they are equivalent, for photon sphere radius, orbital frequency and eikonal QNMs (Result~\ref{res:ParamComparison}).\\

The results above illustrate three strengths of the approach. First, the EMD is agnostic: every static and spherically symmetric black hole (in the sense of Definition~\ref{def:BlackHole}) with a regular horizon corresponds to a point in the same parameter space, irrespective of the theory that produces it. A single method therefore applies to many models of black holes, and every constraint derived within the EMD, theoretical or observational, applies to all of them at once. The approach can be extended to more general geometries (\emph{e.g.} in Section~\ref{Sect:3Dim}, 3-dimensional black holes with angular momentum have been discussed). Second, the parameters are physical: they are defined through coordinate-invariant quantities and are, at least in principle, measurable, as illustrated by the shadow bound of Chapter~\ref{chap4} and by the Gedankenexperiment of Chapter~\ref{chapGeo}. Third, the EMD tests proposed geometries: self-consistency and regularity impose non-trivial constraints, strong enough to rule out certain proposals, as for the self-consistent Bonanno--Reuter black hole, and they can be supplemented by positivity conditions on the effective stress-energy tensor~\cite{DAlise:2023hls} (see Section~\ref{subsec:positivity}).

At the same time, the EMD is neither a fundamental nor a dynamical theory of gravity. It is an effective description of black hole \emph{solutions} of such a theory, and not of the theory itself: it delimits which geometries are (self-)consistent, but, in its current form, does not impose further consistency conditions that would single out \emph{physical} black hole models. Furthermore, the EMD is not derived from an action, so that its connection to an (effective) theory of the physics around the black hole remains indirect (see also Remark~\ref{rem:SwamplandEMD}). In particular, the deformation parameters are expected to be furnished by a theory of (quantum) gravity beyond GR, possibly through a dual description, and this step has not yet been carried out. The linear response formalism recently proposed in \cite{SHlinear} is in fact motivated by rendering this connection more direct. Another example is the quBTZ, which can be constructed from braneworld holography~\cite{Emparan:2020znc,Emparan:2021hyr}, as we have shown the EMD derived from this construction correctly captures the Hawking temperature, supporting the general approach of the EMD.\\

These considerations point to the following directions for future work.
\begin{itemize}
\item \emph{Computing the deformation parameters.} Extremal black holes are a natural starting point: their near-horizon region is an AdS$_2$ throat, so that AdS/CFT techniques~\cite{Maldacena:1997re} could be used to compute the EMD coefficients in the vicinity of the horizon. A similar strategy applies to three-dimensional black holes in asymptotically AdS spacetimes, where holography applies to the geometry as a whole, as for the quBTZ black hole. A complementary route is to match the EMD coefficients to the corrections computed in the effective field theory of gravity~\cite{Donoghue:2001qc,Calmet:2017qqa} (\emph{cf.} Remark~\ref{rem:EFTcharged}).

\item \emph{Physical constraints for parameters.} Study further conditions on the black hole geometries and turn them in physical constraints on the deformations functions. Possible conditions are the Quantum Inequalities~\cite{Emparan:2020znc,Frassino:2024bjg}, positivity conditions~\cite{DAlise:2023hls}, generalized~\cite{Gaiotto:2014kfa,Cordova:2022ruw,Benedetti:2021lxj,Hinterbichler:2022agn} and asymptotic~\cite{Bondi:1962px,Sachs:1962wk,Brown:1986nw,Strominger:2017zoo,Hawking:2016msc,Donnay:2015abr} symmetries.

\item \emph{Larger classes of geometries.} A phenomenologically interesting extension is to rotating black holes in four dimensions~\cite{Konoplya:2016jvv,AbhishekChowdhuri:2023gvu,BenAchour:2025uzp,Kumar:2024utz}, followed by dynamical and evaporating black holes, for which quasi-local horizons replace the event horizon (see the discussion of Definition~\ref{def:BlackHole}), by the black hole interior and its inner-horizon instabilities~\cite{Poisson:1990eh,Carballo-Rubio:2018pmi}, and by geometries that are not asymptotically flat.

\item \emph{Connections to experiments.} Formulate QNMs beyond the eikonal limit and gravitational-wave waveforms that are relevant for the upcoming experiments (such as LISA~\cite{LISA:2017pwj,LISA:2024hlh,LISA:2022kgy} and ET~\cite{Punturo:2010zz,Maggiore:2019uih,ET:2025xjr}) and imaging-related quantities for EHT~\cite{EventHorizonTelescope:2019dse,EventHorizonTelescope:2022wkp,EventHorizonTelescope:2022xqj,Johnson:2023ynn}. The combination of these observables would systematically sharpen the observational bounds on the EMD coefficients.

\item \emph{Cosmology.} Apply similar parametrization to cosmological geometries~\cite{Hu:2007pj,Gubitosi:2012hu,Bellini:2014fua} and constrain the cosmological observables~\cite{Clifton:2011jh,Ishak:2018his}.

\end{itemize}
Altogether, the EMD provides a common and physically meaningful language in which black hole geometries beyond GR can be constructed, tested against consistency requirements and confronted with observations, independently of the theory that produces them.\\
\vskip 1em

\section*{Acknowledgments}

We thank Mattia Damia Paciarini and Mikołaj Myszkowski for comments on the manuscript and longstanding collaboration on the various related projects. We further thank Emanuele Binetti, Salvatore Capozziello, Alessandra D’Alise, Vittorio De Falco, Ciro De Simone, Roberto Emparan, Giuseppe Fabiano, Antonia M. Frassino, Domenico Frattulillo, Aaron Held, Davide Iacobacci, Franco Pezzella, Luciano Rezzolla, Hugo Roussille and Vania Vellucci for collaborations and discussions.
This work was partially supported by Carlsberg Foundation, grant CF22-0922. 

\subsection*{AI declaration}
The authors used AI assistance for language editing, reviewed the resulting text, and take full responsibility for the final manuscript.

%% file: subfiles/Appendix.tex
\typeout{START baseline=\the\baselineskip, jot=\the\jot}
\appendix

\renewcommand{\thechapter}{} 

\chapter{Appendix}

\renewcommand{\thesection}{\Alph{section}}

\setcounter{equation}{0}
\renewcommand{\theequation}{\Alph{section}.\arabic{equation}}
\counterwithin*{equation}{section}

\setcounter{figure}{0}
\renewcommand{\thefigure}{\Alph{chapter}.\arabic{figure}}

\setcounter{table}{0}
\renewcommand{\thetable}{\Alph{chapter}.\arabic{table}}

\section{Ricci Scalar \scheme}\label{Sect:RicciScalar\scheme}
Let us note that the most general static space-time metric compatible with spherical symmetry is not considered here yet, but to streamline the discussion, a family of metrics, parametrized by a single function $\Phi^{(\mathcal{X})}$ of a single argument, are examined.
In this Section we shall discuss a new \scheme\,that is not based on the physical distance to the black hole (horizon), but instead on the Ricci scalar. Indeed, outside of the horizon of the black hole, the latter is expected to be a monotonic function of the Schwarzschild coordinate $\co$. Here we shall first discuss how to define such deformations in a self-consistent fashion and then explain how such descriptions can be linked to the distance \scheme\,that we have reviewed in the previous Section.

\subsection{Definition of the \scheme}
To define the Ricci scalar \scheme, we shall use for the deformation of the metric \eqref{SpStMetric}
a function $\Phi^{(R)}$ of the Ricci scalar:
\begin{align}
&f(\co)=1-\frac{\Phi^{(R)}(R(\co))}{\co}\ .
\end{align}
As before, we consider the outer horizon of the black hole to be located at $\co=\hr$, such that $f(\hr)=0$. The Ricci scalar $R$ is fixed in terms of the metric function in the following manner\footnote{In this work, primes denote derivatives with respect to the Schwarzschild coordinate $\co$. Furthermore we have used the shorthand notation $\frac{\dd \Phi^{(R)}}{\dd \co}=\frac{\dd \Phi^{(R)}}{\dd R}\,\frac{\dd R}{\dd \co}$, and similarly for the second derivative.}
\begin{align}
R(\co)=\frac{1}{\co^2}\,\left(2-2f -4\co f'-\co^2 f''\right)=\frac{2}{\co^2}\dv{\Phi^{(R)}}{\co}+ \frac{1}{\co} \dv[2]{\Phi^{(R)}}{\co}\ ,\label{FormRicciEq}
\end{align} 
which is a second order, non-linear differential equation for $R$ as a function of the Schwarzschild coordinate $\co$. Solving \eqref{FormRicciEq} in general in the regime $\co\geq \zh$ is a difficult task and we limit ourselves to studying a series expansion in the vicinity of the horizon. Concretely, we define the following series expansions
\begin{align}
&\Phi^{(R)}(R)=2M\sum_{n=0}^\infty \frac{q_n}{n!}\,R^n\,,&&\text{and} &&  R(\co)=\sum_{n=0}^\infty \frac{R_n}{n!}\,(\co-\hr)^n\ ,\label{SeriesExpansionszh}
\end{align}
which we assume to have a non-zero radius of convergence for $ r \geq \zh$. Notice that the sets of expansion coefficients $\{q_n\}_{n\in\mathbb{N}_0}$ and $\{R_{n}\}_{n\in\mathbb{N}_0}$ play very different roles: the $\{q_n\}$ implicitly define the function $\Phi^{(R)}$ and are therefore external input parameters that \emph{define} the metric deformation. They are therefore part of the {{free parameters}} that define the Ricci scalar \scheme. The coefficients $\{R_n\}$ are a priori not free, but instead fixed by \eqref{FormRicciEq} in terms of the $\{q_n\}$. However, since~\eqref{FormRicciEq} is a second order differential equation, its general solution depends on two further (unfixed) parameters. Thus, in order to fully describe the Ricci scalar \scheme, we first need to understand how to parametrise this freedom.

First of all, the condition of the horizon being located at $\zh$ (\emph{i.e.} $f(\hr)=0$) imposes
\begin{align}
\sum_{n=0}^\infty\frac{q_n}{n!}\,(R_0)^n= \frac{\hr}{2M}\ .
\end{align}
Furthermore, the expansion of $R$ in~\eqref{SeriesExpansionszh} allows to expand $\Phi^{(R)}$ in terms of the Schwarzschild coordinate 
\begin{align}\label{v exp r}
\Phi^{(R)}(r)=2M\sum_{n=0}^\infty \frac{q_n}{n!}\,\left(\sum_{k=0}^\infty \frac{R_k}{k!}\,(\co-\hr)^k\right)^n=:2M\sum_{p=0}^\infty \tp{p}\,(\co-\hr)^p\ .
\end{align}
This series can be developed in the following form (see Appendix~\ref{App:SerExpansions} for a discussion on the technical details of expanding powers of series)
\begin{align}
\Phi^{(R)}&=2M\sum_{n=0}^\infty \frac{q_n}{n!}\sum_{\ell_1=0}^n\sum_{\ell_2=0}^{\ell_1}\ldots\sum_{\ell_p=0}^{\ell_{p-1}}\begin{pmatrix} n \\ \ell_{1}\end{pmatrix} \begin{pmatrix} \ell_{1} \\ \ell_{2} \end{pmatrix} \ldots \begin{pmatrix} \ell_{p-1} \\ \ell_{p}\end{pmatrix}\,(R_0)^{n-\ell_1}\,\left(\tfrac{R_1}{1!}\right)^{\ell_1-\ell_2}\ldots \left(\tfrac{R_{p-1}}{(p-1)!}\right)^{\ell_{p-1}-\ell_p} \times \nonumber\\
&\hspace{1cm} \times \left(\tfrac{R_p}{p!}\right)^{\ell_p}\,(\co-\hr)^{\ell_1+\ell_2+\ldots \ell_{p-1}+\ell_p}+\mathcal{O}((\co-\hr)^{p+1})\ ,
\end{align}
such that we can read off the coefficients
\begin{align}\label{cp coeff}
\tp{0}&=\sum_{n=0}^\infty \frac{q_n}{n!}\,(R_0)^n\ \equiv \frac{\hr}{2M} \ ,\nonumber\\
\tp{p}&=\sum_{n=0}^\infty q_n\,\sum_{\genfrac{}{}{0pt}{}{0\leq \ell_p\leq \ell_{p-1}\leq\ldots\leq \ell_1\leq n}{\ell_1+\ell_2+\ldots+\ell_p=p}}\frac{(R_0)^{n-\ell_1}\,\left(\tfrac{R_1}{1!}\right)^{\ell_1-\ell_2}\ldots \left(\tfrac{R_{p-1}}{(p-1)!}\right)^{\ell_{p-1}-\ell_p}\left(\tfrac{R_p}{p!}\right)^{\ell_p}}{(n-\ell_1)!\,(\ell_1-\ell_2)!\ldots (\ell_{p-1}-\ell_p)!\ell_p!}\,,\hspace{0.25cm} \forall p\geq 1\ .
\end{align}
Furthermore, using the expansions
\begin{align}
&\frac{1}{\co}=\sum_{n=0}^\infty\frac{(-1)^n}{\hr^{n+1}}\,(\co-\hr)^n \,,&&\text{and} && \frac{1}{\co^2}=\sum_{n=0}^\infty\frac{(-1)^n(n+1)}{\hr^{n+2}}\,(\co-\hr)^n \ ,
\end{align}
we can further expand the two terms on the right hand-side of eq.~\eqref{FormRicciEq} by using the Cauchy product of two series
\begin{align}
\frac{1}{\co^2}\,\dv{\Phi^{(R)}}{\co}&=2M\sum_{p=0}^\infty (\co-\hr)^{p}\sum_{n=0}^p\frac{(-1)^n(n+1)(p+1-n)}{\hr^{n+2}}\,\tp{p+1-n}\ ,\\
\frac{1}{\co}\,\dv[2]{\Phi^{(R)}}{\co}&=2M\sum_{p=0}^\infty(\co-\hr)^p\sum_{n=0}^p\frac{(-1)^n(p+2-n)(p+1-n)}{\hr^{n+1}}\,\tp{p+2-n}\ .
\end{align}
Combining these results, (\ref{FormRicciEq}) can be expanded in powers of $(\co-\hr)$
\begin{align}
\sum_{p=0}^\infty\frac{R_p}{p!}\,(\co-\hr)^p&=2M\sum_{p=0}^\infty(\co-\hr)^p\sum_{n=0}^p\frac{(-1)^n}{\hr^{n+2}}\big[2\hr(n+1)(p+1-n)\,\tp{p+1-n}+\nonumber\\
&\hspace{5cm}+(p+2-n)(p+1-n)\,\tp{p+2-n}\big]\,,\label{LinRelRp}
\end{align}
such that, by comparing order by order, we find
\begin{align}
\frac{R_p}{p!}=2M\sum_{n=0}^p\frac{(-1)^n}{\hr^{n+2}}\left[2(n+1)(p+1-n)\,\tp{p+1-n}+\hr(p+2-n)(p+1-n)\,\tp{p+2-n}\right]\ .\label{RecursiveRp}
\end{align}
To understand this relation, we recall that the $\tp{p}$ in (\ref{cp coeff}) are functions of the coefficients $\{q_n\}$ and $\{R_n\}$. Therefore, a priori (\ref{RecursiveRp}) allows to express the latter in terms of the former. Concretely, however, (\ref{RecursiveRp}) can be seen as a recursive equation that allows to express iteratively the $\{R_n\}_{n\geq 2}$ in terms of $\{R_0,R_1,q_{n\geq 0}\}$. To see this, we re-write the second equation of \eqref{cp coeff} in the form
\begin{align}
\tp{p+2}=\frac{R_{p+2}}{(p+2)!}\sum_{n=1}^\infty\frac{q_n\,R_0^{n-1}}{(n-1)!}+\tau^{\text{sub-}R}_{p+2}(R_0,\ldots,R_{p+1},\{q_n\})\ ,\label{Leading1}
\end{align}
where $\tau^{\text{sub-}R}_{p+2}$ is notably independent of $R_{k\geq p+2}$
\begin{align}\label{tilde cp+2}
\tau^{\text{sub-}R}_{p+2}:=\sum_{n=0}^\infty q_n\,\sum_{\genfrac{}{}{0pt}{}{0\leq \ell_{p+1}\leq\ldots\leq \ell_1\leq n}{\ell_1+\ell_2+\ldots\ell_{p+1}=p+2}}\frac{R_0^{n-\ell_1}\,\left(\tfrac{R_1}{1!}\right)^{\ell_1-\ell_2}\ldots \left(\tfrac{R_{p+1}}{(p+1)!}\right)^{\ell_{p+1}}}{(n-\ell_1)!\,(\ell_1-\ell_2)!\ldots (\ell_{p}-\ell_{p+1})!\ell_{p+1}!}\, .
\end{align} 
Therefore (\ref{RecursiveRp}) is a linear equation in $R_{p+2}$ that allows to express it in terms of  $R_{p'}$, with $p^{\prime}<p+2$ and $\{q_n\}$
\begin{align}\label{Rp+2 iterative}
R_{p+2}&=\frac{\hr p!}{2M\sum_{k=1}^\infty\frac{q_k R_0^{k-1}}{(k-1)!}}\bigg[\frac{R_p}{p!}-\frac{ 2M(p+2)(p+1)}{\hr}\,\tau^{\text{sub }R}_{p+2}+\nonumber\\
&\qquad -4M\sum_{n=0}^p\frac{(-1)^n(n+1)(p+1-n)}{\hr^{n+2}}\tp{p+1-n}-2M\sum_{n=1}^p\frac{(-1)^n(p+2-n)(p+1-n)}{\hr^{n+1}}\,\tp{p+2-n}\bigg]\ .
\end{align}
Thus, we can choose as the {{free parameters}} to describe this \scheme\, the parameters $R_0$, $R_1$ and $\{q_n\}_{n\in\mathbb{N}_0}$. The discussion so far is summarised in Figure~\ref{Fig:Overview}, which formalises the definition of the Ricci scalar \scheme\, to describe (quantum) deformations of a spherically symmetric and static black hole. Notice that this \scheme\, a priori gives a different description of the black hole deformation as the distance \scheme\, discussed in the previous Section. Moreover, given the different sets of input parameters (\emph{i.e.} $\{\xi_n\}_{n\in\mathbb{N}_0}$ for the distance \scheme\, and $(R_0,R_1,\{q_n\})$ for the Ricci scalar \scheme) it is not clear whether these two descriptions are equivalent. In the following Subsection we shall therefore discuss how to connect the two \scheme s, that is, we shall try to formulate relations among the two sets of parameters that lead to identical deformations and therefore the same geometry.\footnote{Notice that we shall impose that the two deformations are indeed \emph{identical}. We shall not discuss the much more general possibility of the two space-times described by these deformations being equivalent and related to one another through coordinate transformations.}

\subsection{Relation to the Distance \scheme}
Our strategy in relating the distance \scheme\,to the Ricci scalar \scheme\,is to identify the deformation functions, \emph{i.e.} $\Phi^{(R)}=\Phi^{(\rho)}$. This can for example be achieved by expanding both functions in the Schwarzschild coordinate and identify the coefficients $\tp{p}$ (as defined in \eqref{v exp r}) order by order. However, a major technical complication on the Ricci scalar side is the fact that $\tp{p}$ for fixed $p$ generally depends on all $\{q_n\}_{n\in\mathbb{N}_0}$. We therefore first slightly re-define the Ricci scalar \scheme\,before relating it to the distance \scheme.

\subsubsection{Redefinition of the Ricci Scalar \scheme}
We may slightly re-define the Ricci scalar \scheme\,by choosing as physical quantity $\widetilde{R}(r):=R(r)-R_0$. Since $R_0$ is a constant (namely the value of the Ricci scalar at the horizon), $\widetilde{R}$ is a physical quantity and (outside of the horizon) a monotonic function of $r$. We thus consider the expansion of the modified deformation function
\begin{align}
&\Phi^{(\widetilde{R})}=2M\sum_{n=0}^\infty\frac{\wtq{n}}{n!}\,\widetilde{R}^n\,,&&\text{and} &&\widetilde{R}(\co)=\sum_{n=1}^\infty\frac{R_n}{n!}\,(\co-\zh)^n\ .\label{SeriesExpansionRtzh}
\end{align}
Following the same logic as explained in Section~\ref{Sect:ShiftOfFunction} (see \eqref{ConstShiftRel}), the coefficients $q_p$ in~\eqref{SeriesExpansionszh} and the $\wtq{p}$ in~\eqref{SeriesExpansionRtzh} can be related through
\begin{align}
&q_p=\sum_{n=p}^\infty\frac{\wtq{n}\,(-R_0)^{n-p}}{(n-p)!}\,,&&\text{and}&&\wtq{p}=\sum_{n=p}^\infty\frac{q_n\,R_0^{n-p}}{(n-p)!}\ .\label{SchemeReductionR}
\end{align}
Furthermore, the identification (\ref{SchemeReductionR}) allows to write
\begin{align}
\Phi^{(\widetilde{R})}(\widetilde{R})=2M \sum_{p=0}^\infty \tp{p}\,(\co-\zh)^p\,, \label{ExpansionDeformationFunction}
\end{align}
with the same coefficients $\tp{p}$ as in (\ref{v exp r}). Concretely, we find the following form for $\tp{p}$ as a function of $\{\wtq{n}\}_{n\in\mathbb{N}_0}$ and $\{R_n\}_{n\in\mathbb{N}}$
\begin{align}
\tp{0}&=\wtq{0}\,,\nonumber\\
\tp{p}&=\sum_{\genfrac{}{}{0pt}{}{0\leq \ell_{p-1}\leq\ldots\leq \ell_1\leq n\leq p}{\ell_1+\ell_2+\ldots+\ell_{p-1}+n=p}} \frac{\wtq{n}\,\left(\tfrac{R_1}{1!}\right)^{n-\ell_1}\left(\tfrac{R_2}{2!}\right)^{\ell_1-\ell_2}\ldots \left(\tfrac{R_{p-1}}{(p-1)!}\right)^{\ell_{p-2}-\ell_{p-1}}\left(\tfrac{R_{p}}{p!}\right)^{\ell_{p-1}}}{(n-\ell_1)!\,(\ell_1-\ell_2)!\ldots (\ell_{p-2}-\ell_{p-1})!\ell_{p-1}!}\hspace{0.5cm} \forall p\geq 1\,.\label{tpCoefsForm}
\end{align}
Inserting this relation in the self-consistency relation (\ref{RecursiveRp}) leads to a recursive relation that fixes $R_n$ for $n\geq 2$ in terms of $R_0$, $R_1$ and $\{\wtq{n}\}_{n\in \mathbb{N}_0}$:
\begin{align}\label{Rp+2 iterative 2}
R_{p+2}&=\frac{\hr p!}{2M \wtq{1}}\bigg[\frac{R_p}{p!}-\frac{2\ma(p+2)(p+1)}{\hr}\,\sum_{\genfrac{}{}{0pt}{}{0\leq \ell_{p}\leq\ldots\leq \ell_1\leq n\leq p+2}{\ell_1+\ell_2+\ldots\ell_{p}+n=p+2}}\frac{\wtq{n}\left(\tfrac{R_1}{1!}\right)^{n-\ell_1}\ldots \left(\tfrac{R_p}{p!}\right)^{\ell_{p-1}-\ell_p}\left(\tfrac{R_{p+1}}{(p+1)!}\right)^{\ell_p}}{(n-\ell_1)!\ldots (\ell_{p-1}-\ell_p)!\ell_p!}\nonumber\\
&-4M\sum_{n=0}^p\frac{(-1)^n(n+1)(p+1-n)}{\hr^{n+2}}\tp{p+1-n}-2M\sum_{n=1}^p\frac{(-1)^n(p+2-n)(p+1-n)}{\hr^{n+1}}\,\tp{p+2-n}\bigg]\ .
\end{align}
Together with (\ref{tpCoefsForm}) this solution allows to express the coefficients $\tp{n}$ in the expansion of the deformation function (\ref{ExpansionDeformationFunction}) in terms of $R_0$, $R_1$ and $\{\wtq{n}\}_{n\in \mathbb{N}_0}$. For the $\tau_{0,\ldots,4}$ we obtain
\begin{align}
&\tp{0}=\wtq{0}\, , \quad \tp{1}=R_1\,\wtq{1}\, , \quad \tp{2}=\frac{R_0 \zh}{4\ma}-\frac{ R_1}{\zh}\,\wtq{1}\, , \quad \tp{3}=\frac{ R_1}{\zh^2}\,\wtq{1}-\frac{R_0-\zh R_1}{12\ma}\, ,\nonumber\\
&\tp{4}=\frac{R_0}{192 \zh \ma^3\wtq{1}}\,(\zh^3+ 12 \ma \wtq{1})-R_1\left(\frac{1}{24\ma}+\frac{ \wtq{1}}{\zh^3}\right)-\frac{R_1^2\zh \wtq{2}}{48\ma \wtq{1}}\,.\label{TauCoefDistance}
\end{align}
The form of $\tp{4}$ suggests that the coefficients $\tp{n}$ for $n> 4$ are more complicated, however, we remark that (\ref{tpCoefsForm}) implies that they depend linearly on $\wtq{n-2}$ in the following fashion
\begin{align}
&\tp{n}=-\frac{R_1^{n-2} \zh}{2\ma n!}\,\frac{\wtq{n-2}}{\wtq{1}}+\tau^{\text{sub-} \widetilde{R}}_n(R_0,R_1,\wtq{0},\ldots,\wtq{n-3})\,,&&\forall n\geq 4\,,\label{TauExpR}
\end{align}
where $\tau^{\text{sub-} \widetilde{R}}_n$ is independent of $\wtq{k\geq n-2}$.
\subsubsection{\scheme\, identification}
To relate the (redefined) Ricci scalar- and distance- \scheme, we next identify the deformation functions $\Phi^{(\rho)}$ in~\eqref{SeriesDistanceRho} with $\Phi^{(\widetilde{R})}$ in~\eqref{SeriesExpansionRtzh}
\begin{align}
\Phi^{(\rho)}=\sum_{p=0}^\infty \tp{p}\,(\co-\zh)^p=\Phi^{(\widetilde{R})}\ , \label{SchemeIdentificationRrho}
\end{align}
To this end we express the coefficients $\tp{p}$ in terms of $\{\xi_{2n}\}_{n\in\mathbb{N}_0}$ using the result of the previous Section and taking into account the full regular function at the horizon with the conditions~\eqref{XiHigherRel},~\eqref{ThetaHigherRel} and~\eqref{ABcondition}
\begin{align}
&\tp{0}=\xi_0\,,\hspace{1cm}\tp{1}=\xi_2 b_1^2\,,\nonumber\\
&\tp{p}=\sum_{\genfrac{}{}{0pt}{}{0\leq \ell_{p-1}\leq \ldots\leq \ell_1\leq 2n\leq 2p}{n+\ell_1+\ell_2+\ldots+\ell_{p-1}=p}}\frac{(2n)!\,\xi_{2n} b_1^{2n-\ell_1}b_3^{\ell_1-\ell_2}\ldots b_{2p-3}^{\ell_{p-2}-\ell_{p-1}}b_{2p-1}^{\ell_{p-1}}}{(2n-\ell_1)!(\ell_1-\ell_2)!\ldots (\ell_{p-2}-\ell_{p-1})!\ell_{p-1}!}\hspace{1cm}\forall p\geq 2\,.\label{TpDistance}
\end{align}
Using (\ref{Coeffsb}) to express the coefficients $\{b_{2n-1}\}_{n\in\mathbb{N}}$ in terms of the $\{a_{2n-2}\}_{n\in\mathbb{N}}$ and (\ref{Coeffsa}) to express the latter in terms of $\{\xi_{2n-2}\}_{n\in\mathbb{N}}$, we find for the coefficients $\tp{0,\ldots,3}$ of (\ref{TpDistance})
\begin{align}
&\tp{0}=\xi_0\,,\hspace{1cm} \tp{1}=\frac{8 \zh \xi_2}{1+\varpi}\,,\hspace{1cm}\tp{2}=\frac{4\xi_2}{1+2\varpi}+\frac{64\zh^2(1+\varpi(3+2\varpi)+8\zh\xi_2)\xi_4}{(1+\varpi)^3(1+2\varpi)}\,,\nonumber\\
&\tp{3}=\frac{1280\zh^3 \xi_6}{(1+\varpi)^2(2+3\varpi)}+\frac{4096\zh^4(7+8\varpi) \xi_4^2}{(1+\varpi)^3(1+2\varpi)(2+3\varpi)}+\frac{48\zh(5+9\varpi+6\varpi^2)\xi_4}{(1+\varpi)(1+2\varpi)^2(2+3\varpi)}\nonumber\\
&\hspace{1cm}+\frac{(1+\varpi)^2(1-3\varpi+2\varpi^2)}{8\zh^2(1+2\varpi)^2(2+3\varpi)}\,.\label{TauCoefRicci}
\end{align}
The general form (\ref{ConditionExpansionBs}) of the coefficients $b_{2n-1}$ implies
\begin{align}
\tp{p}=\frac{2^{3p} \zh^p }{(1+\varpi)^p}\left(\frac{8\xi_2 \zh}{(1+\varpi)((p-1)+p\varpi)}+1\right)\,\xi_{2p}+\tau^{\text{sub-}\rho}_{p}\left(\xi_0,\ldots,\xi_{2p-2}\right)\,,\label{TauExpd}
\end{align}
where $\tau^{\text{sub-}\rho}_{p}$ is independent of $\xi_{2k}$ for $k\geq p+1$. Identifying (\ref{TauExpR}) with (\ref{TauExpd}) allows to relate the coefficients $\{\xi_{2n}\}_{n\in\mathbb{N}_0}$ to $(R_0,R_1,\{\wtq{n}\}_{n\in\mathbb{N}_0})$. Concretely, identifying $\tau_{0,\ldots,3}$ gives relations between $(\xi_0,\xi_2,\xi_4,\xi_6)$ and $(R_0,R_1,\wtq{0},\wtq{1})$, {{which are explicitly given in Appendix~\ref{App:SolutionOfEquationDistRicci}.}} Once these parameters are fixed, the identification of $\tp{p\geq 4}$ (\emph{i.e.} (\ref{TauExpR}) and (\ref{TauExpd})) is a linear relation relating $\xi_{2p}$ and $\wtq{p-2}$, which therefore allows to completely express the distance \scheme\,in terms of the Ricci-scalar \scheme\,and vice versa. Notice that this identification also allows to write physical quantities unambiguously in either \scheme. For example, the Hawking temperature in the distance scheme (\ref{HawkingTempFirst}) can be given more generally in terms of $\tp{0,1}$ as follows
\begin{align}
T_H=\frac{1-2\ma\, \tp{1}}{8\pi \ma\, \tp{0}}\,.\label{HawkingTempGen}
\end{align}
Using (\ref{cp coeff}) and (\ref{TauCoefDistance}) we can therefore write the Hawking temperature in terms of {{the free parameters}} of the Ricci-scalar \scheme
\begin{align}
T_H=\frac{1-2 \ma R_1\,\sum_{n=1}^\infty\frac{q_n\,R_0^{n-1}}{(n-1)!}}{8\pi \ma\sum_{m=0}^\infty \frac{q_m R_0^m}{m!}}=\frac{1-2\ma R_1 \wtq{1}}{8\pi\ma \wtq{0}}\,.
\end{align}
Before closing this Section, we briefly remark that the condition $\xi_2\leq \frac{M}{8\zh}$ for the coefficient $\xi_2$ in the distance \scheme\,is automatically satisfied in the Ricci scalar \scheme. Indeed, for any value of $\wtq{1}\,R_1\in\mathbb{R}$, $\xi_2\leq\frac{M}{8\zh}$ is automatically satisfied in (\ref{RicciSchemeInverse}).


\section{Kretschmann Scalar \scheme}\label{Sect:KretschmannScalar}
\subsection{Definition of the \scheme}
Following the example of the Ricci scalar, we can define further \scheme s based on other curvature scalars. As an example of a second-order curvature invariant, we shall consider the Kretschmann scalar \cite{Kretschmann}, which in terms of the Riemann tensor ${R^\mu}_{\nu\rho\lambda}$ is defined as $K=R_{\mu\nu\rho\lambda}R^{\mu\nu\rho\lambda}$, and define the deformed metric function
\begin{align}
f(r)=1-\frac{\Phi^{(K)}(K(r))}{r}\ .
\end{align}
The Kretschmann scalar is related to the deformation function through the following differential equation
\begin{align}
K(\co)&=\frac{1}{\co^4}\left(4-8f+4f^2+4\co^2 (f')^2+\co^4 (f'')^2\right)\nonumber\\
&=\frac{2}{\co^2} \left[\frac{\dd^2}{\dd\co^2}\left(\frac{(\Phi^{(K)})^2}{\co^2}\right)+\frac{\co^4}{2}\left(\frac{\dd}{\dd \co}\left(\frac{1}{\co^2}\frac{\dd \Phi^{(K)}}{\dd\co}\right)\right)^2\right]\,.\label{SelfConsistencyKretschmann}
\end{align}
Following the example of (\ref{SeriesExpansionszh}), in order to solve this self-consistency relation, we expand the function $v^{(K)}$ as 
\begin{align}
&\Phi^{(K)}(K)=2M\sum_{n=0}^\infty \frac{\si{n}}{n!}\, K^n\,,&&\text{and} &&K(\co)=\sum_{n=0}^\infty \frac{\kc{n}}{n!}\,(\co-\zh)^n\,,\label{KretschmannDofs}
\end{align}
where the coefficients $\{\si{n}\}_{n\in\mathbb{N}_0}$ are external input parameters that {{define}} the \scheme. We choose them in such a way to satisfy the normalisation condition $\sum_{n=0}^\infty \frac{\si{n}}{n!} \kc{0}^n =\frac{\hr}{2\ma}$. As we shall explain, the coefficients $\{K_n\}_{n\in\mathbb{N}_0}$ are a priori not all independent, but related to the $\{\si{n}\}_{n\in\mathbb{N}_0}$ via (\ref{SelfConsistencyKretschmann}). To study these dependencies, we expand $\Phi^{(K)}$ in the Schwarzschild coordinate\footnote{In view of later identifying this \scheme\,with the ones discussed previously, we use the same coefficients $\{\tp{n}\}_{n\in\mathbb{N}_0}$ as in (\ref{v exp r}) and (\ref{SchemeIdentificationRrho}).}
\begin{equation}\label{sigma series sp}
\Phi^{(K)}(K(\co)) = 2M\sum_{p=0}^\infty \tp{p} (\co- \hr)^p \ ,
\end{equation}
where the coefficients $\tp{n}$ can be expressed in terms of the $\{\kc{n}\}_{n\in\mathbb{N}_0}$ as follows
\begin{align}
\tp{0}&=\sum_{n=0}^\infty \frac{\si{n}}{n!}\,(K_0)^n\,,\nonumber\\
 \tp{p}&= \sum_{n=0}^{\infty} \si{n} \sum_{\genfrac{}{}{0pt}{}{0\leq \ell_p\leq \ell_{p-1}\leq\ldots\leq \ell_1\leq n}{\ell_1+\ell_2+\ldots+\ell_p=p}}\frac{\left(\kc{0}\right)^{n-\ell_1}\,\left(\kc{1}\right)^{\ell_1-\ell_2}\ldots\left(\frac{\kc{p}}{p!}\right)^{\ell_p}}{(n-\ell_1)!\,(\ell_1-\ell_2)!\ldots (\ell_{p-1}-\ell_p)!\ell_p!}\,.\label{coeff s}
\end{align}
Furthermore, from (\ref{SelfConsistencyKretschmann}) we obtain
{\allowdisplaybreaks
\begin{align}
&\frac{\kc{p}}{p!}=8\ma^2\,\sum_{m=0}^p\sum_{n=0}^{m+2}\sum_{\ell=0}^{m+2-n}\frac{(-1)^{p+\ell-m}(p-m+1)(\ell+1)(m+2)(m+1)}{\zh^{p-m+\ell+4}}\,\tp{n}\,\tp{m+2-n-\ell}\nonumber\\
&+4\ma^2\sum_{n=1}^{p-1}\sum_{\ell=0}^{p-n}\sum_{k=0}^n\frac{(-1)^{\ell+k}(\ell+1)(k+1)(p-n)n(p-n+1-\ell)(n+1-k)}{\zh^{k+\ell+4}}\,\tp{p-n-\ell+1}\,\tp{n-k+1}\nonumber\\
&+8\ma^2\sum_{n=1}^{p}\sum_{\ell=0}^{p-n+1}\sum_{k=0}^n\frac{(-1)^{\ell+k}(\ell+1)(k+1)(p-n+1)n(p-n+2-\ell)(n+1-k)}{\zh^{k+\ell+3}}\,\tp{p-n-\ell+2}\,\tp{n-k+1}\nonumber\\
&+4\ma^2\sum_{n=1}^{p+1}\sum_{\ell=0}^{p-n+2}\sum_{k=0}^n\frac{(-1)^{\ell+k}(\ell+1)(k+1)(p-n+2)n(p-n+3-\ell)(n+1-k)}{\zh^{k+\ell+2}}\,\tp{p-n-\ell+3}\,\tp{n-k+1}\,.\label{FullRecursionKretschmann}
\end{align}}
Notice that for $p>0$ 
\begin{align}
&\frac{\kc{p}}{p!}=\frac{16\ma^2(p+1)(p+2)}{\zh^4}\,\left(\tp{0}-\zh\,\tp{1}+\zh^2\,\tp{2}\right)\,\tp{p+2}+\widehat{\tau}^{\text{sub-}K}_{p+2}(\tp{0},\ldots,\tp{p+1})\nonumber\\
&=\frac{16\ma^2 \kc{p+2}}{\zh^4 p!}\,\left(\sum_{m=0}^\infty\si{m} \kc{0}^{m-2}\left(\frac{\kc{0}^2}{m!}-\frac{\zh \kc{0} \kc{1}}{(m-1)!}+\frac{\zh^2 \kc{0} \kc{2}}{2(m-1)!}+\frac{\zh^2 \kc{1}^2}{2(m-2)!}\right)\right)\,\left(\sum_{n=1}^\infty \frac{\si{n} (\kc{0})^{n-1}}{(n-1)!}\right)\nonumber\\
&\hspace{3cm}+\tau_{p+2}^{\text{sub-}K}(\kc{0},\ldots,\kc{p+1},\{\si{n}\})\,,\label{KretschmannRecursion}
\end{align}
where $\widehat{\tau}^{\text{sub-}K}_{p+2}$ and $\tau_{p+2}^{\text{sub-}K}$ denote terms independent of $\tp{n>p+2}$ and $\kc{n>p+2}$ respectively. Furthermore, in the second line we have used an expansion of the coefficients $\tp{p+2}$ similar to (\ref{Leading1}) or (\ref{TauExpR}). We therefore find that (\ref{KretschmannRecursion}) is a linear equation for $\kc{p+2}$, which, similar to the case of the Ricci scalar \scheme, allows to express the $\kc{p+2}$ uniquely in terms of $\kc{p'<p+2}$ for all $p\geq 3$. For $p=0$, (\ref{FullRecursionKretschmann}) is a quadratic equation for $\kc{2}$ with the following solutions 
\begin{align}
\kc{2}=\frac{2\zeta_1(\kc{1}(2\zeta_1-\kc{1}\zh \zeta_2)\ma-1)+u|\zeta_1|\sqrt{16 \kc{1}\zeta_1\ma(1-\kc{1}\zeta_1\ma)-8+\kc{0}\zh^4}}{2\ma\zh \zeta_1^2}\,,\label{K2solt}
\end{align}
where we have introduced a phase factor $u=\pm1$ and defined the following shorthand notation
\begin{align}
&\zeta_1=\sum_{n=1}^\infty \si{n}\frac{\kc{0}^{n-1}}{(n-1)!} \qq{and} \zeta_2=\sum_{n=2}^\infty \si{n}\frac{\kc{0}^{n-2}}{(n-2)!}\,.\label{ZetaDefs}
\end{align}
Furthermore, we have used $\tp{0}=\frac{\zh}{2\ma}$ to simplify the expression. Based on this solution, the Kretschmann scalar \scheme\,is summarized in Figure~\ref{Fig:Overview}. Here we consider as {{free parameters}} $(\kc{0},\kc{1},\{\si{n}\}_{n\in\mathbb{N}_0})$ (which are real parameters), while the phase factor $u=\pm1$ is an additional condition. This is similar to our choice for dealing with a sign ambiguity when resolving the self-consistency condition in the distance \scheme. Finally, we remark that with (\ref{coeff s}) the Hawking temperature (\ref{HawkingTempGen}) in the Kretschmann-scalar \scheme\,takes the following form
\begin{align}
T_H=\frac{1-2 \ma \kc{1}\,\sum_{n=1}^\infty\frac{\si{n}\,\kc{0}^{n-1}}{(n-1)!}}{8\pi \ma\sum_{m=0}^\infty \frac{\si{m} \kc{0}^m}{m!}}=\frac{1-2\ma \kc{1}\,\zeta_1}{4\pi \zh}\,.
\end{align}

\subsection{Relation to other \scheme s}
After having introduced the Kretschmann scalar \scheme, we now relate it to other \scheme s. To this end, following the example of the Ricci scalar \scheme, we shall first propose a slight redefinition, before making contact to the {{distance- and}} Ricci scalar \scheme.

Instead of working with the Kretschmann scalar directly, we instead subtract from it its value at the horizon, \emph{i.e.} we introduce $\widetilde{K}(r):=K(r)-\kc{0}$ and define
\begin{align}
&\Phi^{(\widetilde{K})}(\widetilde{K})=2M\sum_{n=0}^\infty \frac{\sit{n}}{n!}\, \widetilde{K}^n\,,&&\text{and} &&\widetilde{K}(\co)=\sum_{n=1}^\infty \frac{\kc{n}}{n!}\,(\co-\zh)^n\,,
\end{align}
The relation between the coefficients $\{\si{n}\}_{n\in\mathbb{N}_0}$ and $\{\sit{n}\}_{n\in\mathbb{N}_0}$ follows in the same manner as in (\ref{SchemeReductionR}), namely
\begin{align}
&\si{p}=\sum_{n=p}^\infty\frac{\sit{n}\,(-\kc{0})^{n-p}}{(n-p)!}\,,&&\text{and}&&\sit{p}=\sum_{n=p}^\infty\frac{\si{n}\,\kc{0}^{n-p}}{(n-p)!}\,.\label{SchemeReductionK}
\end{align}
To make contact with other \scheme s, we expand $v^{(\widetilde{K})}$ in terms of the Schwarzschild coordinate
\begin{align}
\Phi^{(\widetilde{K})}=2M\sum_{p=0}^\infty \tp{p}\,(\co-\zh)^p\,. \label{ExpansionDeformationFunctionKretschmann}
\end{align}
and interpret the coefficients $\tp{p}$ to be the same as in (\ref{v exp r}), (\ref{SchemeIdentificationRrho}) and (\ref{sigma series sp}), while being expanded as a function of the $\{\sit{n}\}_{n\in\mathbb{N}_0}$ and $\{\kc{n}\}_{n\in\mathbb{N}}$ 
\begin{align}
\tp{0}&=\sit{0}\,,\nonumber\\
\tp{p}&=\sum_{\genfrac{}{}{0pt}{}{0\leq \ell_{p-1}\leq\ldots\leq \ell_1\leq n\leq p}{\ell_1+\ell_2+\ldots+\ell_{p-1}+n=p}} \frac{\sit{n}\,\left(\tfrac{\kc{1}}{1!}\right)^{n-\ell_1}\left(\tfrac{\kc{2}}{2!}\right)^{\ell_1-\ell_2}\ldots \left(\tfrac{\kc{p-1}}{(p-1)!}\right)^{\ell_{p-2}-\ell_{p-1}}\left(\tfrac{\kc{p}}{p!}\right)^{\ell_{p-1}}}{(n-\ell_1)!\,(\ell_1-\ell_2)!\ldots (\ell_{p-2}-\ell_{p-1})!\ell_{p-1}!}\hspace{0.5cm} \forall p\geq 1\,.\label{tpCoefsFormKretschmann}
\end{align}
Using (\ref{FullRecursionKretschmann}) we can explicitly express the $\tp{p}$ in terms of $\{\sit{n}\}_{n\in\mathbb{N}_0}$ and $(\kc{0},\kc{1},u)$, \emph{e.g.} we find for the first few instances
\begin{align}
&\tp{0}=\sit{0}\,,\hspace{1cm}\tp{1}=\kc{1}\,\sit{1}\,,\hspace{1cm}\tp{2}=\frac{4\ma \sit{1}(\kc{1}\zh \sit{1}-\sit{0})+\mathfrak{p}}{4\ma\sit{1}\zh}\,,\nonumber\\
&\tp{3}=\frac{6 \kc{0} \sit{1} \zh ^6-64 \kc{1}^2 \sit{1}^3 \ma^2 \zh ^2+\kc{1} \sit{1} \zh  \left(-16
   \mathfrak{p} \ma+96 \sit{0} \sit{1} \ma^2+\zh ^6\right)+16 \sit{0} \ma (\mathfrak{p}-4 \sit{0}
   \sit{1} \ma)}{24 \mathfrak{p} \ma \zh ^3}\,,\label{TpKretschLow}
\end{align}
with the abbreviation $\mathfrak{p}=u |\sit{1}|\sqrt{\kc{0}\zh^6-16(2\sit{0}^2-2\kc{1}\zh\sit{0}\sit{1}+\kc{1}^2\zh^2\sit{1}^2)\ma^2}$. Higher orders in $\tp{n}$ are more complicated, however, we remark that they depend linearly on $\sit{n-2}$
\begin{align}
&\tp{n}=-\frac{\kc{1}^{n-2} \zh^4}{4\ma \, \mathfrak{p}\, n! }\,\sit{n-2}+\tau^{\text{sub-} \widetilde{K}}_n(\kc{0},\kc{1},u,\sit{0},\ldots,\sit{n-3})\,,&&\forall n\geq 4\,,\label{TauExpKtilde}
\end{align}
where $\tau^{\text{sub-} \widetilde{K}}_n$ is independent of $\{\sit{k\geq n-1}\}$. Identifying the $\tp{p}$ in (\ref{TpKretschLow}) and (\ref{TauExpKtilde}) with (\ref{v exp r}), (\ref{SchemeIdentificationRrho}) and (\ref{sigma series sp}) yields equations that allow to relate the Kretschmann scalar \scheme\,to the distance- and Ricci scalar \scheme. Indeed, identifying $\tp{0,1,2,3}$ fixes $(\kc{0},\kc{1},\sit{0},\sit{1})$ in terms of $(R_0,R_1,\wtq{0},\wtq{1})$ or $(\xi_0,\xi_2,\xi_4,\xi_6)$ (and vice-versa). Notice, that the space of solutions for these equations is fairly restricted: for example, for given $(\tp{0},\tp{1},\tp{2},\tp{3})$ (either given in terms of $(\xi_0,\xi_2,\xi_4,\xi_6)$ or $(R_0,R_1,\wtq{0},{\wtq{1}})$), consistent solutions of $(\kc{0},\kc{1},\sit{0},\sit{1})$ not only require a specific choice of $u$, but they are in fact uniquely fixed. To see this, we realise that the third equation in (\ref{TpKretschLow}) can be rewritten in the form
\begin{align}
\frac{u\,\text{sign}(\sit{1})}{4M \zh}\,\sqrt{\kc{0}\zh^6-16(2\tp{0}^2-2\tp{1}\tp{0}\zh+\tp{1}^2\zh^2)\ma^2}=\tp{2}-\tp{1}+\frac{\tp{0}}{\zh}\,,\label{RecSignDem}
\end{align}
which only has a solution if $u\, \text{sign}(\sit{1})=\text{sign}\left(\tp{2}-\tp{1}+1\right)$, in which case $K_0$ is uniquely fixed. This, however, renders the last equation in (\ref{TpKretschLow}) linear in $\sit{1}$, with the solution
\begin{align}
\sit{1}=\frac{\zh^6 \tp{1}}{32\zh(\tp{1}-2\tp{2}+2\tp{1}\tp{2}+3\zh^2\tp{3}(1-\tp{1}+\tp{2})-6K_0}\,.
\end{align}
Since this fixes (the sign of) $\sit{1}$, (\ref{RecSignDem}) is consistent only for a specific value of $u$. Finally, as in the previous cases, once (\ref{TpKretschLow}) is solved, (\ref{TauExpKtilde}) are linear equations for $\sit{n-2}$ in terms of $\tp{n}$. Due to the length of the ensuing expressions, we refrain from reproducing the explicit form of the solution,\footnote{For the same reason, we also do not provide the (unique) inverse transformations that express $(\xi_0,\xi_2,\xi_4,\xi_6)$ or $(R_0,R_1,\wtq{0},\wtq{1})$ in terms of $(K_0,K_1,\sit{0},\sit{1})$)): they follow from replacing $\tp{n}$ in (\ref{TauCoefRicci}), (\ref{TauExpd}) and (\ref{TauCoefDistance}), (\ref{TauExpR}) respectively with (\ref{TpKretschLow}) and (\ref{TauExpKtilde}). We furthermore remark in passing that, similar to the case of the Ricci scalar \scheme, the condition $\xi_2\leq \frac{M}{8\zh}$ also poses no further conditions on $(K_0,K_1,\{\sit{n}\}_{n\in\mathbb{N}_0})$.}, however, we provide a concrete mapping from the distance to the Kretschmann scalar \scheme\,for a particular small deformation of the Schwarzschild solution in Appendix~\ref{App:KretschmannSigns}. Higher orders $\tp{n\geq 4}$ are linear in $\sit{n-2}$ and thus uniquely allow to fix them in terms of $\wtq{n-2}$ or $\xi_{2n}$ (or vice-versa). 


\subsection{Overview of the relations}\label{Sect:Conclusions}
In the previous three Sections, we have defined \scheme s as coherent frameworks to describe (quantum) deformedm relative to the classical Schwarzschild geometry \cite{Schwarzschild:1916uq}, spherically symmetric and static black holes in 4 dimensions. Such black hole geometries are captured by one\footnote{In Sections~\ref{Sect:RicciScalar\scheme} and~\ref{Sect:KretschmannScalar} we only focused on the case of a single deformation function. The more general case of two such functions, which is compatible with spherical symmetry and static behaviour of the black hole, can be tackled using the same methods described here.} deformation function of a physical quantity $\mathcal{X}$ to ensure compatibility of the quantum geometry with the same symmetries as its classical counterpart. 

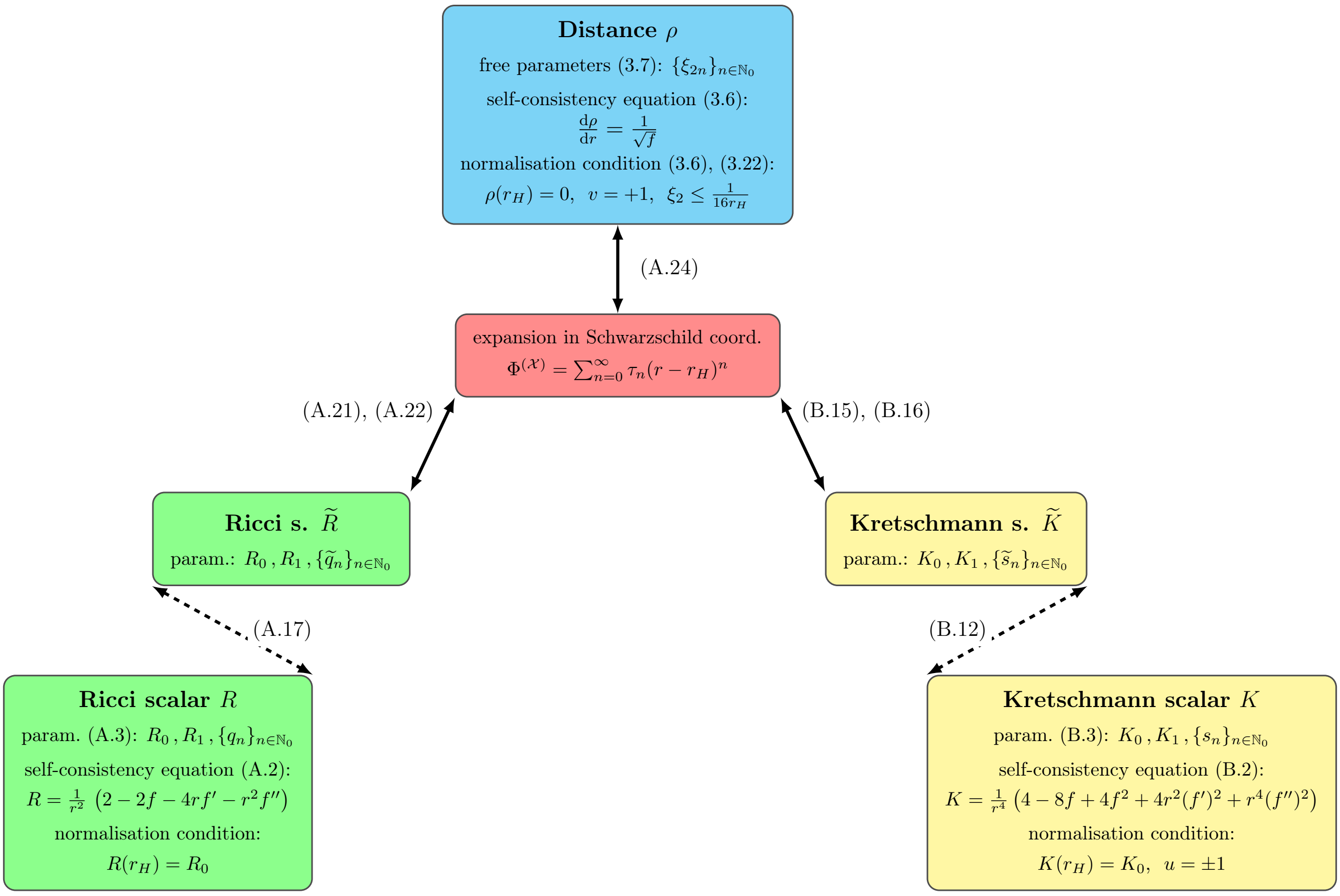
\begin{figure}[t]
\begin{center}
\scalebox{0.70}{\begin{tikzpicture}[
  embox/.style={draw=black!70, thick, rounded corners=6pt, align=center, inner xsep=9pt, inner ysep=8pt},
  lbl/.style={fill=white, inner sep=2.5pt, font=\small},
  link/.style={ultra thick, <->}
]
\node[embox, fill=cyan!45!white] (dist) at (0,0) {%
  {\bf Distance $\rho$}\\[3pt]
  {\footnotesize free parameters (\ref{SeriesDistanceRho}): $\{\xi_{2n}\}_{n\in\mathbb{N}_0}$}\\[3pt]
  {\footnotesize self-consistency equation (\ref{DiffEqDistance}):}\\[1pt]
  {$\frac{\dd\rho}{\dd\co}=\frac{1}{\sqrt{f}}$}\\[3pt]
  {\footnotesize normalisation condition (\ref{DiffEqDistance}), (\ref{Coeffsa}):}\\[1pt]
  {\footnotesize $\rho(\zh)=0$, \ $v=+1$, \ $\xi_2\leq\frac{1}{16\zh}$}};
\node[embox, fill=red!45!white, below=1.6cm of dist] (exp) {%
  {\footnotesize expansion in Schwarzschild coord.}\\[2pt]
  {\footnotesize $\Phi^{(\mathcal{X})}=\sum_{n=0}^\infty \tp{n}(\co-\zh)^n$}};
\node[embox, fill=green!45!white, below left=1.7cm and 0.8cm of exp] (rictil) {%
  {\bf Ricci s. $\widetilde{R}$}\\[3pt]
  {\footnotesize param.: $R_0\,,R_1\,,\{\wtq{n}\}_{n\in\mathbb{N}_0}$}};
\node[embox, fill=yellow!45!white, below right=1.7cm and 0.8cm of exp] (kretil) {%
  {\bf Kretschmann s. $\widetilde{K}$}\\[3pt]
  {\footnotesize param.: $\kc{0}\,,\kc{1}\,,\{\sit{n}\}_{n\in\mathbb{N}_0}$}};
\node[embox, fill=green!45!white, below left=1.6cm and -2.9cm of rictil] (ricci) {%
  {\bf Ricci scalar $R$}\\[3pt]
  {\footnotesize param.\ (\ref{SeriesExpansionszh}): $R_0\,,R_1\,,\{q_n\}_{n\in\mathbb{N}_0}$}\\[3pt]
  {\footnotesize self-consistency equation (\ref{FormRicciEq}):}\\[1pt]
  {\footnotesize $R=\frac{1}{\co^2}\,\left(2-2f -4\co f'-\co^2 f''\right)$}\\[3pt]
  {\footnotesize normalisation condition:}\\[1pt]
  {\footnotesize $R(\zh)=R_0$}};
\node[embox, fill=yellow!45!white, below right=1.6cm and -2.9cm of kretil] (kret) {%
  {\bf Kretschmann scalar $K$}\\[3pt]
  {\footnotesize param.\ (\ref{KretschmannDofs}): $\kc{0}\,,\kc{1}\,,\{\si{n}\}_{n\in\mathbb{N}_0}$}\\[3pt]
  {\footnotesize self-consistency equation (\ref{SelfConsistencyKretschmann}):}\\[1pt]
  {\footnotesize $K=\frac{1}{\co^4}\left(4-8f+4f^2+4\co^2 (f')^2+\co^4 (f'')^2\right)$}\\[3pt]
  {\footnotesize normalisation condition:}\\[1pt]
  {\footnotesize $K(\zh)=\kc{0}$, \ $u=\pm1$}};
\draw[link] (dist) -- (exp) node[lbl, midway, right=8pt] {(\ref{TpDistance})};
\draw[link] (exp.south west) -- (rictil.north east)
  node[lbl, midway, above left=8pt and -4pt] {(\ref{TauCoefDistance}), (\ref{TauExpR})};
\draw[link] (exp.south east) -- (kretil.north west)
  node[lbl, midway, above right=8pt and -4pt] {(\ref{TpKretschLow}), (\ref{TauExpKtilde})};
\draw[link, dashed] (rictil.south west) -- (ricci.north east)
  node[lbl, midway, right=7pt] {(\ref{SchemeReductionR})};
\draw[link, dashed] (kretil.south east) -- (kret.north west)
  node[lbl, midway, left=7pt] {(\ref{SchemeReductionK})};
\end{tikzpicture}}
\end{center}
\caption{Relation between the three EMDs for a deformed Schwarzschild metric discussed in this Chapter.}
\label{Fig:Overview}
\end{figure}

In~\ref{Sect:ReviewDistance} the physical distance, measured from the (horizon of the) black hole, was chosen. However, other physical quantities (\emph{e.g.} curvature scalars of different orders) are equally viable. In each case, in order to remain self-consistent, $\mathcal{X}$ is computed from the deformed metric, thus leading to an involved equation to tackle (see eqs.~\eqref{DiffEqDistance},~\eqref{FormRicciEq} and~\eqref{SelfConsistencyKretschmann} respectively for the distance, Ricci scalar and Kretschmann scalar). To solve this equation, we follow the general framework developed in~\ref{Sect:ReviewDistance} and consider series expansions in the vicinity of the event horizon of the black hole. In this way, we not only provide self-consistent solutions for metric deformations based on the physical distance, Ricci- and Kretschmann scalar, but also propose clear definitions of each of the \scheme s, as shown in the overview in Figure~\ref{Fig:Overview}. Moreover, as also indicated in the schematic drawing, by expressing the deformation functions as expansions in the Schwarzschild coordinate, we find equations which allow to relate the different \scheme s to one another. In the case of the \scheme s based on curvature invariants, we first perform a slight re-definition of the \scheme\,(~\eqref{SchemeReductionR}) and~\eqref{SchemeReductionK} respectively), by subtracting the value at the horizon: in this way, the equations relating the (modified) \scheme\, to the distance \scheme, can be solved in an iterative fashion.

From a conceptual perspective, the relation between the distance \scheme\, and \scheme s based on curvature scalars is non-trivial: indeed, the former is a non-local physical quantity (since it is defined as an integral between two points in space-time), while the latter are local invariants. Nevertheless, at least sufficiently close to the horizon, they are both expected to be monotonic functions of the Schwarzschild coordinate, which thus allows the mapping between the \scheme s.

We furthermore also remark that a basic assumption of our work is a sufficient level of regularity of the geometry close to the horizon, such that series expansions remain well-defined. While here we have defined \scheme s based on full (\emph{i.e.} infinite) series expansions with suitable radius of convergence, in practice the relations among different \scheme s remain (approximately) valid if these series are truncated after certain orders. This property is useful for practical computations, when geometries are approximated by their leading coefficients (see \emph{e.g.} \cite{DAlise:2023hls,Hohenegger:2025dic} for an example).

\section{Series relations}\label{App:SerExpansions}

\subsection{Series expansion of inverse radial coordinate}\label{app:inverse r to rho}
In this Section, starting from the first series in Eq. (\ref{SeriesDistance}), we express the inverse radial coordinate as a series expansion in $\rho$. More precisely, we determine the coefficients $\zi{m}$ in
\begin{align}
P(\rho):=\frac{1}{r(\rho)} =\sum_{m=0}^\infty \zi{m}\,\left(\frac{\rho}{2M}\right)^m\,,\, &&\text{with} &&\begin{array}{l}\zi{m}\in\mathbb{R}\hspace{0.5cm}\forall m\in\mathbb{N}\,,\\ \zi{0}=1/\zh\,.\end{array}\label{SeriesFormaInvZ}
\end{align}
Let us remark that, in contrast to the coefficients in Eqs.~\eqref{SeriesDistanceRho} and \eqref{SeriesDistance}, the series coefficients have dimensions $[\mathfrak{p}_n] = [L]^{-1} \, , \ \forall n \geq 0 $.
Differentiating Eq.~\eqref{SeriesFormaInvZ} both sides with respect to $\rho$ and multiplying by $r$, we find
\begin{align}
r(\rho)\,P(\rho)'&=-P(\rho)\,\dv{r}{\rho}\,,
\end{align}
which with (\ref{SeriesDistance}) becomes the following series identity
\begin{multline}
\left(\zh+2M\sum_{n=2}^\infty a_n\,\left(\frac{\rho}{2M}\right)^n\right)\left(\sum_{m=1}^\infty m\,\frac{\zi{m}}{2M}\,\left(\frac{\rho}{2M}\right)^{m-1}\right)=\\-\left(\sum_{m=0}^\infty\zi{m}\,\left(\frac{\rho}{2M}\right)^m\right)\left(\sum_{n=2}^\infty n\, a_n\,\left(\frac{\rho}{2M}\right)^{n-1}\right)\ ,
\end{multline}
note that we have used the condition in $a_1=0$. Re-arranging both sides of this equation, we find
\begin{align}
\frac{\zh\,\zi{1}}{2M}+(\frac{\zh}{M}\,\zi{2}+2a_2\,\zi{0})\,\frac{\rho}{2M}+\sum_{p=2}^\infty(p+1)\,\left[\frac{\zh}{2M}\,\,\zi{p+1}+\zi{0}\,a_{p+1}+\sum_{m=1}^{p-1}\zi{m}\,a_{p+1-m}\right]\,\left( \frac{\rho}{2M}\right)^p=0\,. 
\end{align}
Order by order we therefore obtain the relations
\begin{align}
&\zi{1}=0\,,&&\zi{2}=-\frac{2M a_2}{\zh^2}\,,&&\zi{p}=-\frac{2Ma_p}{\zh^2}-\frac{2M }{\zh}\sum_{m=1}^{p-2}\,\zi{m}\,a_{p-m}\hspace{0.5cm}\forall p\geq 3\,.\label{1zRecursive}
\end{align}

\section{Series expansions for \texorpdfstring{$\xi_1\neq 0$}{xi1}}\label{App:SerNonZeroXi}
For completeness, we shall generalise the approach of Section~\ref{Sect:DistanceFunctionExpansion} also to accommodate functions $\Phi$ such that the coefficient $\xi_1\neq 0$ in Eq.~(\ref{SeriesDistanceRho}) is non-zero. Indeed, assuming that $r \sim \zh+\mathcal{O}(\rho^p)$ for some $p\in\mathbb{R}_+$, for the left-hand-side of the equation (\ref{DistanceDiffEq}) to have a term of order $\mathcal{O}(\rho)$ requires either $p=1$ or $p=3/2$. Since we have seen in Section~\ref{Sect:DistanceFunctionExpansion} that $p=1$ still requires $\xi_1=0$, we shall here explore the case $p=3/2$, \emph{i.e.} instead of (\ref{SeriesFormaInt}) we shall consider the expansion
\begin{align}
&\Phi(\rho)= \hr + \sum_{n=1}^\infty \xi_n \rho^n \qq{and} r=\zh+\sum_{n=1}^{\infty}\ha{n}\,\rho^{\frac{n+2}{2}}\,, \qq{with} \ha{n}\in \mathbb{R}\hspace{0.2cm}\forall n\in\mathbb{N}\,,\label{SeriesOdd}
\end{align}
which as before we assume to have an interval of convergence $\rho\in[0,\rho_A)$ (with some $\rho_A>0$). We then obtain
\begin{align}
r\left(1-\left(\dv{r}{\rho}\right)^2\right)&=\zh-\frac{9}{4}\,\rho\,\zh\,\ha{1}^2+\rho^{3/2}\left(\ha{1}-6\,\zh\,\ha{1}\,\ha{2}\right)+\rho^2\,\left[\ha{2}-\zh\left(4\,\ha{2}^2+\frac{15}{2}\,\ha{1}\,\ha{3}\right)\right]\nonumber\\
&\hspace{0.5cm}+\sum_{p=5}^\infty\rho^{p/2}\bigg[\ha{p-2}-\zh\sum_{m=1}^{p-1}\frac{(m+2)(p-m+2)}{4}\,\ha{m}\,\ha{p-m}\nonumber\\
&\hspace{2.6cm}-\sum_{k=1}^{p-4}\ha{k}\sum_{m=1}^{p-k-3}\frac{(m+2)(p-k-m)}{4}\,\ha{m}\,\ha{p-k-m-2}\bigg]\,.\label{SerExpOddPower}
\end{align}
Comparing the coefficients of $\rho^0$, $\rho^{1}$, $\rho^{3/2}$ and $\rho^2$ to the series expansion of $\Phi$ in Eq.~\eqref{SeriesOdd}, yields the relations
\begin{align}
&\zh=\xi_0\,,&&\xi_1=-\frac{9}{4}\,\zh\,\ha{1}^2\,,&&0=\ha{1}-6\,\zh\,\ha{1}\,\ha{2}\,,&&\xi_2=\ha{2}-\zh\left(4\,\ha{2}^2+\frac{15}{2}\,\ha{1}\,\ha{3}\right)\,,
\end{align}
which has solution
\begin{align}
&\ha{1}=\pm \frac{2}{3}\sqrt{-\frac{\xi_1}{\zh}}\,,&&\ha{2}=\frac{1}{6\zh}\,,&&\ha{3}=\frac{1-18\,\zh\,\xi_2}{135\,\ha{1}\,\zh^2}\,.\label{LowSolutionsHa}
\end{align}
The coefficient $\ha{1}$ is real only for $\xi_1<0$. In the case $\xi_1>0$, there exists no real solution of (\ref{DistanceDiffEq}), which is of the form (\ref{SeriesOdd}). In the following, we shall assume $\xi_1<0$ and furthermore pick the positive sign for $\ha{1}$ in (\ref{LowSolutionsHa}): indeed, for the negative sign, the function $z(\rho)$ would not be monotonically growing for $\rho>0$.

Comparing the remaining terms in (\ref{SerExpOddPower}) order by order, we can express $\ha{p-1}$ in terms of $\ha{k}$ with $k<p-1$
\begin{align}
\ha{p-1}&=\frac{2}{3\,\ha{1}\,\zh\,(p+1)}\bigg[\ha{p-2}-\zh\sum_{m=2}^{p-2}\frac{(m+2)(p-m+2)}{4}\,\ha{m}\,\ha{p-m}\nonumber\\
&\hspace{0.5cm}-\sum_{k=1}^{p-4}\ha{k}\sum_{m=1}^{p-k-3}\frac{(m+2)(p-k-m)}{4}\,\ha{m}\,\ha{p-k-m-2}\bigg]-\left\{\begin{array}{lcl} 0 & \text{if} & p\in\mathbb{N}_{\text{odd}} \\ \frac{2\xi_{p/2}}{3\ha{1}\,\zh\,(p+1)} & \text{if} & p\in\mathbb{N}_{\text{even}} \end{array}\right.\label{SolutionsHa}
\end{align}

\section{Transition from Ricci to Distance \scheme}\label{App:SolutionOfEquationDistRicci}
Identifying the coefficients $\{\tp{n}\}_{n=0,1,2,3}$ in (\ref{TauCoefDistance}) with those in (\ref{TauCoefRicci}) leads to the following solutions of $(R_0,R_1,\wtq{0},\wtq{1})$ in terms of $(\xi_0,\xi_2,\xi_4,\xi_6)$ 
{\allowdisplaybreaks
\begin{align}
R_0&=\frac{3+5\varpi-3\varpi^2-5\varpi^3+384 \zh^3\xi_4}{2\zh^2(1+\varpi)(1+2\varpi)}\,,\nonumber\\
R_1&=\frac{7680\zh^2}{(1+\varpi)^2(2+3\varpi)}\,\xi_6-\frac{9+61\varpi+73\varpi^2-53\varpi^3-90\varpi^4}{4\zh^3(1+2\varpi)^2(2+3\varpi)}\nonumber\\
&\hspace{0.5cm}+\frac{96(19+41\varpi+30\varpi^2)}{(1+\varpi)(1+2\varpi)^2(2+3\varpi)}\,\xi_4+\frac{24576\zh^3(7+8\varpi)}{(1+\varpi)^3(1+2\varpi)^2(2+3\varpi)}\,\xi_4^2\,,\nonumber\\
\wtq{0}&=\frac{\xi_0}{2\ma}\,,\nonumber\\
\wtq{1}&=\bigg[\frac{30720 \zh^2\ma \xi_6}{(1-\varpi)(1+\varpi)^2(2+3\varpi)\xi_2}-\frac{(9+70\varpi+143\varpi^2+90\varpi^3)\ma}{\zh^3(1+2\varpi)^2(2+3\varpi)}\nonumber\\
&\hspace{0.5cm}+\frac{384(19+41\varpi+30\varpi^2)\ma \xi_4}{(1-\varpi)(1+\varpi)(1+2\varpi)^2(2+3\varpi)}+\frac{98304\zh^3(7+8\varpi)\ma \xi_4^2}{(1-\varpi)(1+\varpi)^3(1+2\varpi)^2(2+3\varpi)}\bigg]^{-1}
\end{align}}
These relations can also be inverted to express $(\xi_0,\xi_2,\xi_4,\xi_6)$ in terms of $(R_0,R_1,\wtq{0},\wtq{1})$ as follows
{\allowdisplaybreaks
\begin{align}
\xi_0&=2\ma\,\wtq{0}\,,\nonumber\\
\xi_2&=\frac{\wtq{1} R_1 \ma}{2\ma}\,(1-2\wtq{1}\,R_1\,\ma)\,,\nonumber\\
\xi_4&=\frac{(1+\sigma)^3\left(R_0\zh^2(1+2\sigma)-8\wtq{1} R_1\ma(1+\sigma-\wtq{1}R_1 \ma)\right)}{384\zh^3\left(1+\sigma-4\wtq{1}R_1\ma+8\wtq{1}R_1^2\ma^2\right)}\,,\nonumber\\
\xi_6&=\frac{(1+\sigma)^4(2+3\sigma)\ma}{46080\zh^3 u}\bigg[\frac{36\wtq{1} R_1}{\zh^2}-\frac{3R_0}{\ma}+\frac{3R_1\zh}{\ma}-\frac{(1+\sigma)^2}{\zh^2\ma(1+2\sigma)u^2}\nonumber\\
&\hspace{3cm}\times \left(R_0\zh^2(1+2\sigma)-8\wtq{1}R_1\ma(1+\sigma+\wtq{1}R_1\ma)\right)\nonumber\\
&\hspace{3cm}\times\left(R_0\zh^2(1+2\sigma)+(1+\sigma-4\wtq{1}R_1\ma)(3-8\wtq{1}R_1\ma)\right)\nonumber\\
&\hspace{1cm}-\frac{2\wtq{1}R_1(1+\sigma)(1-2\wtq{1}R_1\ma)}{u^2 \zh^2(1+2\sigma)^2(2+3\sigma)}\bigg(2 R_0 ^2 \zh^4 \left(32 \wtq{1} R_1  (4 \sigma +7) \ma -32 \sigma
   -31-64 \wtq{1}^2 R_1 ^2 (4 \sigma +7) \ma ^2\right)\nonumber\\
   &\hspace{3cm}+R_0  \zh^2 (\sigma  (16 \wtq{1} R_1  \ma  (2 \wtq{1} R_1  \ma  (128 \wtq{1} R_1  \ma -107)+75)-117)\nonumber\\
   &\hspace{3cm}+4
   \wtq{1} R_1  \ma  (417-2 \wtq{1} R_1  \ma  (64 \wtq{1} R_1  \ma  (17 \wtq{1} R_1  \ma -31)+1013))-117)\nonumber\\
   &\hspace{3cm}-2 (3-8
   \wtq{1} R_1  \ma )^2 (4 \wtq{1} R_1  \ma  (17 \wtq{1} R_1  \ma -7)-1) (-4 \wtq{1} R_1  \ma +\sigma +1)\bigg)\bigg]\label{RicciSchemeInverse}
\end{align}}
with $\sigma=|1-4\wtq{1}R_1 \ma|$ and $u=1+\sigma-4\wtq{1}R_1\ma(1-\wtq{1}R_1\ma)$.
\section{Signs in the Kretschmann Scalar \scheme}\label{App:KretschmannSigns}
The sign choice appearing in the definition of the Kretschmann scalar \scheme\,in Section~\ref{Sect:KretschmannScalar} a priori have an impact on the type of solutions that can be described within this \scheme. Indeed, in Section~\ref{Sect:ReviewDistance}, following \cite{DelPiano:2023fiw}, we have fixed a similar choice in Eq.~(\ref{Coeffsa}) by demanding that the corresponding geometry allows a smooth limit recovering the Schwarzschild space-time. A similar condition also fixes certain signs in the Kretschmann scalar \scheme~(and its connection to the distance \scheme). To illustrate this idea, we start with the distance \scheme~for the particular choice $\xi_{2n}=0$ $\forall n>1$, such that the only non-trivial remaining parameter is $\xi_2$, or equivalently $\varpi$. The classical limit, which indeed reproduces the Schwarzschild space-time, is $\varpi\to 1$. For this particular choice of parameters we find
\begin{align}
&\tp{0}=\frac{\zh}{2\ma}\,,&&\tp{1}=\frac{1-\varpi^2}{4\ma(1+\varpi)}\,,&&\tp{2}=\frac{1-\varpi^2}{8\ma\zh(1+2\varpi)}\,,&&\tp{3}=\frac{(1-\varpi)(1+\varpi)^2(1-2\varpi)}{16\ma\zh^2(2+3\varpi)(1+2\varpi)^2}\,.\nonumber
\end{align}
Thus, identification with the Kretschmann scalar \scheme\,leads to
\begin{align}
K_0&=\frac{29+124\varpi+170\varpi^2+84\varpi^3+25\varpi^4}{4\zh^4(1+2\varpi)^2}=\frac{12}{\zh^4}+\mathcal{O}(\varpi-1)\,,\nonumber\\
K_1&=-\frac{323+2593\varpi+8128\varpi^2+12658\varpi^3+10235\varpi^4+4157\varpi^5+786\varpi^6}{4\zh^5(1+2\varpi)^3(2+3\varpi)}\nonumber\\
&=-\frac{72}{\zh^5}+\mathcal{O}(\varpi-1)\, ,\label{ExK0K1}
\end{align}
and for $\zeta_{1,2}$ in (\ref{ZetaDefs}) we obtain
\begin{align}
\zeta_1&=-\frac{\zh^5(1+2\varpi)^3(2+\varpi-3\varpi^2)}{\ma(323+2593\varpi+8128\varpi^2+12658\varpi^3+10235\varpi^4+4157\varpi^5+786\varpi^6)}\nonumber\\
&=\frac{\zh^5}{288\ma}\,(\varpi-1)+\mathcal{O}((\varpi-1)^2)\,,\nonumber\\
\zeta_2&=\frac{(2 \varpi +1)^5 \zh^9}{\ma (3+4 \varpi) \left(786 \varpi ^6+4157
   \varpi ^5+10235 \varpi ^4+12658 \varpi ^3+8128 \varpi ^2+2593 \varpi +323\right)^3 }\nonumber\\
   &\hspace{0.5cm}\times\big(68694+901511 \varpi+5091799 \varpi ^2+15949627 \varpi ^3+29316755 \varpi ^4+28823217
   \varpi ^5\nonumber\\
   &\hspace{0.8cm}+4572149 \varpi ^6-24913951 \varpi ^7-32758909 \varpi ^8-19894452 \varpi ^9-6289848 \varpi ^{10}-866592 \varpi ^{11}\big) \nonumber\\
   &=-\frac{25\zh^9}{124416\ma}\,(\varpi-1)+\mathcal{O}((\varpi-1)^2)\,.\label{Exzeta}
\end{align}
Inserting (\ref{ExK0K1}) and (\ref{Exzeta}) into (\ref{K2solt}) we find
\begin{align}
K_2&=\begin{cases}
\frac{252}{\hr^6}+\frac{5116}{35\hr^6}\,(\varpi-1)+\mathcal{O}(\varpi-1) & \qq{if}  u=+1 \qq{and }\varpi\geq 1\,,\\[6pt] 
-\frac{576}{\hr^6(\varpi-1)}-\frac{4028}{15\hr^6}+\mathcal{O}(\varpi-1) & \qq{if} u=-1 \qq{and }\varpi<1\, . \\ 
\end{cases}
\label{LimitAmbiguity}
\end{align}
We remark that Eq.~(\ref{LimitAmbiguity}) is the corrected version of Eq.~(74) of the published version of~\cite{DelPiano:2024gvw}.
Thus, depending on how the limit $\varpi\to 1$ is taken, \emph{i.e.} depending on the sign of $\xi_2$, only one choice of the sign $u$ in (\ref{K2solt}) yields a viable solution, which moreover leads to a finite limit for the Kretschmann scalar as a function of $\varpi$. Notice also, that in this case (\ref{ExK0K1}) and (\ref{LimitAmbiguity}) are compatible with the expansion of the Kretschmann scalar for the (classical) Schwarzschild space-time
\begin{align}
K_{\text{class}}=\frac{48\ma^2}{\co^6}=\frac{12}{\zh^4}-\frac{72}{\zh^5}\,(\co-\zh)+\frac{252}{\zh^6}\,(\co-\zh)^2+\mathcal{O}((r-\zh)^3)\,,
\end{align}
where we have used $\zh=2\ma$ for the classical Schwarzschild black hole.




\section{Conditions in the extremal case with \texorpdfstring{$Q^2 \neq M \, r_H$}{Q\texttwosuperior{} != M r\textunderscore H}}\label{Appx:A}
Here, we collect all conditions on the deformation functions coefficients for an extremal black hole with $\widetilde{\Gamma} \neq \widetilde{\Omega}$ and $Q^2 \neq M r_H$ discussed in Subsection \ref{Sect:GeneralCharge}. 
In particular, requiring $\widetilde{a}_1 \in \mathbb{R}$ and \eqref{PositiveSignature}, we have:

\begin{itemize}
    \item[$\displaystyle \widetilde{\xi}_0$] $ \displaystyle = \widetilde{\theta}_0 = \frac{r_H^2}{2 M r_H - Q^2} \ , \quad \widetilde{\xi}_1 = \widetilde{\theta}_1 = \widetilde{a}_1 \frac{2 \widetilde{\xi}_0^2 \left(M  r_H-Q^2\right)}{ r_H^3}  \ , \quad \widetilde{\xi}_2 > \widetilde{\theta} _2 \ ,$
    \item  $\displaystyle \frac{Q^2}{2}<  M \, r_H < Q^2 \ , \ \widetilde{a}_2<\frac{r_H \left(4 \widetilde{\theta}_2 r_H^2 \left(Q^2- M \, r_H\right)^2+\widetilde{\xi}_1^2 Q^4 \left( Q^2 - 2 M \, r_H \right)\right)}{4 \widetilde{\xi}_0^2 \left(M \, r_H - Q^2\right)^3} \ ,$
    \item $\displaystyle M \, r_H<\frac{Q^2}{2} \ , \ \widetilde{a}_2>\frac{r_H \left(4 \widetilde{\xi}_2 r_H^2 \left(Q^2- M r_H\right)^2+\widetilde{\xi} _1^2 Q^4 \left( Q^2 - 2 M r_H \right)\right)}{4 \widetilde{\xi}_0^2 \left( M r_H-Q^2\right)^3}  \ , \nonumber$
    \item $\displaystyle M \, r_H>Q^2 \ , \ \widetilde{a}_2>\frac{r_H \left(4 \widetilde{\theta} _2 r_H^2 \left(Q^2-M r_H\right)^2+\widetilde{\xi} _1^2 Q^4 \left( Q^2 - 2 M r_H \right)\right)}{4 \widetilde{\xi}_0^2 \left( M r_H-Q^2\right)^3} \ . \nonumber$
\end{itemize}

\noindent
If $\left.\dv[2]{f}{r}\right|_{r=\zh} = \left.\dv[2]{h}{r}\right|_{r=\zh}$, then $\widetilde{a}_1=1$, and the relation $\widetilde{\xi}_2>\widetilde{\theta} _2$ becomes $\widetilde{\xi}_2 = \widetilde{\theta} _2$, such that:

\begin{itemize}
    \item[$\displaystyle \widetilde{\xi}_0$] $ \displaystyle = \widetilde{\theta}_0 = \frac{z_H^2}{2 M r_H - Q^2} \ , \quad \widetilde{\xi}_1 = \widetilde{\theta}_1 = \frac{2 \widetilde{\xi}_0^2 \left( M r_H-Q^2\right)}{ r_H^3}  \ , \quad \widetilde{\xi}_2 = \widetilde{\theta} _2 \ , $
    \item $\displaystyle \frac{Q^2}{2 }< M \, r_H \ , \quad \widetilde{\xi} _2<-\frac{Q^4}{\left(Q^2-2 M  r_H \right)^3}  , \nonumber$
    \item $ \displaystyle M \,  r_H<\frac{Q^2}{2} \ ,\quad \widetilde{\xi} _2>-\frac{Q^4}{\left(Q^2-2 M r_H \right)^3}  \  . \nonumber $
\end{itemize}

\section{Padé Approximant of a Series Expansion}\label{App:DefPadeApproximant}
Let $U:\,[0,x_0]\rightarrow \mathbb{R}$ be a function with series expansion
\begin{align}
U(x)=\sum_{k=0}^\infty  \mathfrak{u}_k\,x^k\,,
\end{align}
that has radius of convergence $x_0>0$. We then introduce the \emph{Padé approximant} \cite{bender78:AMM} as the quotient of two polynomials 
\begin{align} \label{padedef}
&U_{N,M}:=\frac{P_N(x)}{Q_M(x)}\,,&&\text{with} &&\begin{array}{l}P_N=\sum_{n=0}^N p_n\,x^n\,,\\ Q_M=1+\sum_{m=1}^Mq_m\,x^m\,,\end{array}
\end{align}
such that
\begin{align}
U_{N,M}(x)=\sum_{k=0}^{N+M}\mathfrak{u}_k\,x^k+\mathcal{O}(x^{N+M+1})\,.
\end{align}
If $U_{N,M}$ exists, the coefficients $\{p_n\}$ and $\{q_m\}$ are unique and given by the linear equations
\begin{align}
&\mathcal{M}_1\cdot
\begin{pmatrix}
q_M\\
q_{M-1}\\
\vdots\\
q_1
\end{pmatrix}
=-\left(\begin{array}{c}\mathfrak{u}_{N+1} \\ \mathfrak{u}_{N+2}\\ \vdots \\ \mathfrak{u}_{N+M}\end{array}\right)\,,&&\text{with} &&\mathcal{M}_1=\left(\begin{array}{cccc} \mathfrak{u}_{N-M+1} & \mathfrak{u}_{N-M+2} & \cdots   & \mathfrak{u}_{N} \\ \mathfrak{u}_{N-M+2} & \mathfrak{u}_{N-M+3} & \cdots & \mathfrak{u}_{N+1} \\ \vdots & \vdots  & & \vdots \\ \mathfrak{u}_{N} & \mathfrak{u}_{N+1} &\cdots  & \mathfrak{u}_{N+M-1} \end{array}\right)\,,\nonumber\\
&\mathcal{M}_2\cdot\left(\begin{array}{c}1 \\ q_{1}\\ \vdots \\ q_M\end{array}\right)=\left(\begin{array}{c} p_0 \\ p_1 \\  \vdots \\ p_N\end{array}\right)\,,&&\text{with} &&\mathcal{M}_2=\left(\begin{array}{ccccccc} 0 & 0 & \cdots   & & & & 0 \\ \mathfrak{u}_1 & 0 & 0 & \cdots   & & & 0 \\ \mathfrak{u}_2 & \mathfrak{u}_1 & 0 & 0 &\cdots   & & 0 \\ \vdots & \vdots & \vdots & \vdots & \ddots & &\vdots \\ \mathfrak{u}_N & \mathfrak{u}_{N-1} & \mathfrak{u}_{N-2} & \cdots &  & \mathfrak{u}_{N-M-1} &\mathfrak{u}_{N-M} \end{array}\right)\,,
\end{align}
If it exists, an explicit form of $U_{N,M}$ can be given in the form
\begin{align}
U_{N,M}(x)=\frac{\left|\begin{array}{cccc}\mathfrak{u}_{N-M+1} & \mathfrak{u}_{N-M+2} & \cdots & \mathfrak{u}_{N+1} \\ \vdots & \vdots & \ddots & \vdots \\ \mathfrak{u}_N & \mathfrak{u}_{N+1} & \cdots & \mathfrak{u}_{N+M} \\ \sum_{j=M}^N \mathfrak{u}_{j-M}x^j & \sum_{j=M-1}^N \mathfrak{u}_{j-M+1}x^j & \cdots & \sum_{j=0}^N \mathfrak{u}_j x^j\end{array}\right|}{\left|\begin{array}{cccc}\mathfrak{u}_{N-M+1} & \mathfrak{u}_{N-M+2} & \cdots & \mathfrak{u}_{N+1} \\ \vdots & \vdots & \ddots & \vdots \\ \mathfrak{u}_{N} & \mathfrak{u}_{N+1} & \cdots & \mathfrak{u}_{N+M} \\ x^M & x^{M-1} & \cdots & 1\end{array}\right|}\,.
\end{align}

%% file: references.bib
@article{Binetti:2022xdi,
    author = "Binetti, Emanuele and Del Piano, Manuel and Hohenegger, Stefan and Pezzella, Franco and Sannino, Francesco",
    title = "{Effective theory of quantum black holes}",
    eprint = "2203.13515",
    archivePrefix = "arXiv",
    primaryClass = "gr-qc",
    reportNumber = "CERN-TH-2022-048",
    doi = "10.1103/PhysRevD.106.046006",
    journal = "Phys. Rev. D",
    volume = "106",
    number = "4",
    pages = "046006",
    year = "2022"
}

@article{DelPiano:2023fiw,
    author = "Del Piano, Manuel and Hohenegger, Stefan and Sannino, Francesco",
    title = "{Quantum black hole physics from the event horizon}",
    eprint = "2307.13489",
    archivePrefix = "arXiv",
    primaryClass = "gr-qc",
    doi = "10.1103/PhysRevD.109.024045",
    journal = "Phys. Rev. D",
    volume = "109",
    number = "2",
    pages = "024045",
    year = "2024"
}

@article{DelPiano:2024nrl,
    author = "Del Piano, Manuel and Hohenegger, Stefan and Sannino, Francesco",
    title = "{Black hole shadow and other observables away from the horizon: Extending the effective metric descriptions}",
    eprint = "2412.13673",
    archivePrefix = "arXiv",
    primaryClass = "gr-qc",
    doi = "10.1103/PhysRevD.111.064070",
    journal = "Phys. Rev. D",
    volume = "111",
    number = "6",
    pages = "064070",
    year = "2025"
}

@article{Alonso-Serrano:2025spn,
    author = "Alonso-Serrano, Ana and de Cesare, Marco and Del Piano, Manuel",
    title = "{Black hole solutions in quantum phenomenological gravitational dynamics}",
    eprint = "2504.01539",
    archivePrefix = "arXiv",
    primaryClass = "gr-qc",
    doi = "10.1103/4vd6-dml9",
    journal = "Phys. Rev. D",
    volume = "112",
    number = "8",
    pages = "084031",
    year = "2025"
}

@article{DamiaPaciarini:2025xjc,
    author = "Damia Paciarini, Mattia and Del Piano, Manuel and Hohenegger, Stefan and Sannino, Francesco",
    title = "{Effective metric description of charged black holes}",
    eprint = "2504.20810",
    archivePrefix = "arXiv",
    primaryClass = "gr-qc",
    doi = "10.1140/epjc/s10052-025-14551-w",
    journal = "Eur. Phys. J. C",
    volume = "85",
    number = "8",
    pages = "848",
    year = "2025"
}

@article{DeFalco:2025xxx,
    author = "Del Piano, Manuel and De Simone, Ciro and Damia Paciarini, Mattia and De Falco, Vittorio and Myszkowski, Miko{\l}aj and Sannino, Francesco and Vellucci, Vania",
    title = "{Toward a unified view of agnostic parametrizations for deformed black holes}",
    eprint = "2510.14707",
    archivePrefix = "arXiv",
    primaryClass = "gr-qc",
    month = "10",
    year = "2025",
    journal = "Submitted to Phys. Rev. D"
}

@article{DamiaPaciarini:2024ssf,
    author = "Damia Paciarini, Mattia and Hohenegger, Stefan and Myszkowski, Mikolaj and Sannino, Francesco",
    title = "{Effective metric description of 2+1 dimensional quantum black holes}",
    eprint = "2412.15960",
    archivePrefix = "arXiv",
    primaryClass = "gr-qc",
    doi = "10.1140/epjc/s10052-025-14286-8",
    journal = "Eur. Phys. J. C",
    volume = "85",
    number = "6",
    pages = "611",
    year = "2025"
}

@article{DamiaPaciarini:2025qhq,
    author = "Damia Paciarini, Mattia and Myszkowski, Miko{\l}aj and Sannino, Francesco and Vellucci, Vania",
    title = "{Regular Spacetimes in the Effective Metric Description}",
    eprint = "2506.12620",
    archivePrefix = "arXiv",
    primaryClass = "gr-qc",
    month = "6",
    year = "2025"
}

@article{DAlise:2023hls,
    author = "D'Alise, Alessandra and Fabiano, Giuseppe and Frattulillo, Domenico and Hohenegger, Stefan and Iacobacci, Davide and Pezzella, Franco and Sannino, Francesco",
    title = "{Positivity conditions for generalized Schwarzschild space-times}",
    eprint = "2305.12965",
    archivePrefix = "arXiv",
    primaryClass = "gr-qc",
    reportNumber = "LYCEN 2023-01",
    doi = "10.1103/PhysRevD.108.084042",
    journal = "Phys. Rev. D",
    volume = "108",
    number = "8",
    pages = "084042",
    year = "2023"
}

@article{Kerr:1963ud,
    author = "Kerr, Roy P.",
    title = "{Gravitational field of a spinning mass as an example of algebraically special metrics}",
    doi = "10.1103/PhysRevLett.11.237",
    journal = "Phys. Rev. Lett.",
    volume = "11",
    pages = "237--238",
    year = "1963"
}

@article{Synge1966,
    author = {Synge, J. L.},
    title = "{The Escape of Photons from Gravitationally Intense Stars}",
    journal = {Monthly Notices of the Royal Astronomical Society},
    volume = {131},
    number = {3},
    pages = {463-466},
    year = {1966},
    month = {02},
    issn = {0035-8711},
    doi = {10.1093/mnras/131.3.463},
    url = {https://doi.org/10.1093/mnras/131.3.463},
    eprint = {https://academic.oup.com/mnras/article-pdf/131/3/463/8076763/mnras131-0463.pdf},
}

@article{Gan:2021pwu,
    author = "Gan, Qingyu and Wang, Peng and Wu, Houwen and Yang, Haitang",
    title = "{Photon spheres and spherical accretion image of a hairy black hole}",
    eprint = "2104.08703",
    archivePrefix = "arXiv",
    primaryClass = "gr-qc",
    doi = "10.1103/PhysRevD.104.024003",
    journal = "Phys. Rev. D",
    volume = "104",
    number = "2",
    pages = "024003",
    year = "2021"
}

@article{DelPiano:2024gvw,
    author = "Del Piano, Manuel and Hohenegger, Stefan and Sannino, Francesco",
    title = "{Effective metric descriptions of quantum black holes}",
    eprint = "2403.12679",
    archivePrefix = "arXiv",
    primaryClass = "gr-qc",
    doi = "10.1140/epjc/s10052-024-13609-5",
    journal = "Eur. Phys. J. C",
    volume = "84",
    number = "12",
    pages = "1273",
    year = "2024"
}

@article{Dondi:2019ivp,
    author = "Dondi, Nicola Andrea and Dunne, Gerald V. and Reichert, Manuel and Sannino, Francesco",
    title = "{Analytic Coupling Structure of Large $N_f$ (Super) QED and QCD}",
    eprint = "1903.02568",
    archivePrefix = "arXiv",
    primaryClass = "hep-th",
    doi = "10.1103/PhysRevD.100.015013",
    journal = "Phys. Rev. D",
    volume = "100",
    number = "1",
    pages = "015013",
    year = "2019"
}

@article{Li:2023djs,
    author = "Li, Yong-Zhuang and Kuang, Xiao-Mei",
    title = "{Precession of bounded orbits and shadow in quantum black hole spacetime}",
    doi = "10.1103/PhysRevD.107.064052",
    journal = "Phys. Rev. D",
    volume = "107",
    number = "6",
    pages = "064052",
    year = "2023"
}

@article{ruse1931,
author = {Ruse, H. S.},
title = {Taylor's Theorem in the Tensor Calculus},
journal = {Proceedings of the London Mathematical Society},
volume = {s2-32},
number = {1},
pages = {87-92},
doi = {https://doi.org/10.1112/plms/s2-32.1.87},
url = {https://londmathsoc.onlinelibrary.wiley.com/doi/abs/10.1112/plms/s2-32.1.87},
eprint = {https://londmathsoc.onlinelibrary.wiley.com/doi/pdf/10.1112/plms/s2-32.1.87},
year = {1931}
}

@article{synge1931,
author = {Synge, J. L.},
title = {A Characteristic Function in Riemannian Space and its Application to the Solution of Geodesic Triangles},
journal = {Proceedings of the London Mathematical Society},
volume = {s2-32},
number = {1},
pages = {241-258},
doi = {https://doi.org/10.1112/plms/s2-32.1.241},
url = {https://londmathsoc.onlinelibrary.wiley.com/doi/abs/10.1112/plms/s2-32.1.241},
eprint = {https://londmathsoc.onlinelibrary.wiley.com/doi/pdf/10.1112/plms/s2-32.1.241},
year = {1931}
}

@article{dewitt1960radiation,
  title={Radiation damping in a gravitational field},
  author={DeWitt, Bryce S and Brehme, Robert W},
  journal={Annals of Physics},
  volume={9},
  number={2},
  pages={220--259},
  year={1960},
  publisher={Elsevier}
}

@book{Carroll:2004st,
    author = "Carroll, Sean M.",
    title = "{Spacetime and Geometry}",
    isbn = "978-0-8053-8732-2",
    publisher = "Cambridge University Press",
    month = "7",
    year = "2019"
}

@article{Whittaker,
    author = "Edmund Taylor Whittaker",
    title = "{On the reversion of series}",
    journal = "Gaz. Mat. Lisboa.",
    volume = "12(50)",
    pages = "1",
    year = "1950"
}

@article{Hayward_2006,
	doi = {10.1103/physrevlett.96.031103},
	url = {https://doi.org/10.1103%2Fphysrevlett.96.031103},
	year = 2006,
	month = {1},
	publisher = {American Physical Society ({APS})},
	volume = {96},
	number = {3},
	author = {Sean A. Hayward},
	title = {Formation and Evaporation of Nonsingular Black Holes},
	journal = {Physical Review Letters}
}

@article{Dymnikova:1992ux,
    author = "Dymnikova, I.",
    title = "{Vacuum nonsingular black hole}",
    doi = "10.1007/BF00760226",
    journal = "Gen. Rel. Grav.",
    volume = "24",
    pages = "235--242",
    year = "1992"
}

@article{Perlick:2021aok,
    author = "Perlick, Volker and Tsupko, Oleg Yu.",
    title = "{Calculating black hole shadows: Review of analytical studies}",
    eprint = "2105.07101",
    archivePrefix = "arXiv",
    primaryClass = "gr-qc",
    doi = "10.1016/j.physrep.2021.10.004",
    journal = "Phys. Rept.",
    volume = "947",
    pages = "1--39",
    year = "2022"
}

@book{misner1973gravitation,
  title={Gravitation},
  author={Misner, Charles W and Thorne, Kip S and Wheeler, John Archibald},
  year={1973},
  publisher={Macmillan}
}

@article{Bozza:2002zj,
    author = "Bozza, V.",
    title = "{Gravitational lensing in the strong field limit}",
    eprint = "gr-qc/0208075",
    archivePrefix = "arXiv",
    doi = "10.1103/PhysRevD.66.103001",
    journal = "Phys. Rev. D",
    volume = "66",
    pages = "103001",
    year = "2002"
}

@article{Perlick:2015vta,
    author = "Perlick, Volker and Tsupko, Oleg Yu. and Bisnovatyi-Kogan, Gennady S.",
    title = "{Influence of a plasma on the shadow of a spherically symmetric black hole}",
    eprint = "1507.04217",
    archivePrefix = "arXiv",
    primaryClass = "gr-qc",
    doi = "10.1103/PhysRevD.92.104031",
    journal = "Phys. Rev. D",
    volume = "92",
    number = "10",
    pages = "104031",
    year = "2015"
}

@book{bender78:AMM,
  author = {Bender, C. M. and Orszag, S. A.},
  publisher = {McGraw-Hill},
  title = {{Advanced Mathematical Methods for Scientists and Engineers}},
  year = 1978
}

@article{Rezzolla:2014mua,
    author = "Rezzolla, Luciano and Zhidenko, Alexander",
    title = "{New parametrization for spherically symmetric black holes in metric theories of gravity}",
    eprint = "1407.3086",
    archivePrefix = "arXiv",
    primaryClass = "gr-qc",
    doi = "10.1103/PhysRevD.90.084009",
    journal = "Phys. Rev. D",
    volume = "90",
    number = "8",
    pages = "084009",
    year = "2014"
}

@article{Calmet:2021lny,
    author = "Calmet, Xavier and Kuipers, Folkert",
    title = "{Quantum gravitational corrections to the entropy of a Schwarzschild black hole}",
    eprint = "2108.06824",
    archivePrefix = "arXiv",
    primaryClass = "hep-th",
    doi = "10.1103/PhysRevD.104.066012",
    journal = "Phys. Rev. D",
    volume = "104",
    number = "6",
    pages = "066012",
    year = "2021"
}

@article{Battista:2023iyu,
    author = "Battista, Emmanuele",
    title = "{Quantum Schwarzschild geometry in effective field theory models of gravity}",
    eprint = "2312.00450",
    archivePrefix = "arXiv",
    primaryClass = "gr-qc",
    doi = "10.1103/PhysRevD.109.026004",
    journal = "Phys. Rev. D",
    volume = "109",
    number = "2",
    pages = "026004",
    year = "2024"
}

@article{Cardoso:2014rha,
    author = "Cardoso, Vitor and Pani, Paolo and Rico, Joao",
    title = "{On generic parametrizations of spinning black-hole geometries}",
    eprint = "1401.0528",
    archivePrefix = "arXiv",
    primaryClass = "gr-qc",
    doi = "10.1103/PhysRevD.89.064007",
    journal = "Phys. Rev. D",
    volume = "89",
    pages = "064007",
    year = "2014"
}

@article{Konoplya:2016jvv,
    author = "Konoplya, Roman and Rezzolla, Luciano and Zhidenko, Alexander",
    title = "{General parametrization of axisymmetric black holes in metric theories of gravity}",
    eprint = "1602.02378",
    archivePrefix = "arXiv",
    primaryClass = "gr-qc",
    doi = "10.1103/PhysRevD.93.064015",
    journal = "Phys. Rev. D",
    volume = "93",
    number = "6",
    pages = "064015",
    year = "2016"
}

@article{Simpson:2019mud,
    author = "Simpson, Alex and Visser, Matt",
    title = "{Regular black holes with asymptotically Minkowski cores}",
    eprint = "1911.01020",
    archivePrefix = "arXiv",
    primaryClass = "gr-qc",
    doi = "10.3390/universe6010008",
    journal = "Universe",
    volume = "6",
    number = "1",
    pages = "8",
    year = "2019"
}

@article{EventHorizonTelescope:2021dqv,
    author = "Kocherlakota, Prashant and others",
    collaboration = "Event Horizon Telescope",
    title = "{Constraints on black-hole charges with the 2017 EHT observations of M87*}",
    eprint = "2105.09343",
    archivePrefix = "arXiv",
    primaryClass = "gr-qc",
    reportNumber = "FERMILAB-PUB-21-847-PPD",
    doi = "10.1103/PhysRevD.103.104047",
    journal = "Phys. Rev. D",
    volume = "103",
    number = "10",
    pages = "104047",
    year = "2021"
}

@article{Vagnozzi:2022moj,
    author = "Vagnozzi, Sunny and others",
    title = "{Horizon-scale tests of gravity theories and fundamental physics from the Event Horizon Telescope image of Sagittarius A}",
    eprint = "2205.07787",
    archivePrefix = "arXiv",
    primaryClass = "gr-qc",
    reportNumber = "UCI-HEP-TR-2022-07",
    doi = "10.1088/1361-6382/acd97b",
    journal = "Class. Quant. Grav.",
    volume = "40",
    number = "16",
    pages = "165007",
    year = "2023"
}

@article{Cadoni:2022vsn,
    author = "Cadoni, M. and De Laurentis, M. and De Martino, I. and Della Monica, R. and Oi, M. and Sanna, A. P.",
    title = "{Are nonsingular black holes with super-Planckian hair ruled out by S2 star data?}",
    eprint = "2211.11585",
    archivePrefix = "arXiv",
    primaryClass = "gr-qc",
    doi = "10.1103/PhysRevD.107.044038",
    journal = "Phys. Rev. D",
    volume = "107",
    number = "4",
    pages = "044038",
    year = "2023"
}

@article{Garcia95,
  title = {Class of Stationary Axisymmetric Solutions of the Einstein-Maxwell-Dilaton-Axion Field Equations},
  author = {Garc\'{\i}a, Alberto and Galtsov, Dmitri and Kechkin, Oleg},
  journal = {Phys. Rev. Lett.},
  volume = {74},
  issue = {8},
  pages = {1276--1279},
  numpages = {0},
  year = {1995},
  month = {2},
  publisher = {American Physical Society},
  doi = {10.1103/PhysRevLett.74.1276},
  url = {https://link.aps.org/doi/10.1103/PhysRevLett.74.1276}
}

@article{Simpson:2018tsi,
    author = "Simpson, Alex and Visser, Matt",
    title = "{Black-bounce to traversable wormhole}",
    eprint = "1812.07114",
    archivePrefix = "arXiv",
    primaryClass = "gr-qc",
    doi = "10.1088/1475-7516/2019/02/042",
    journal = "JCAP",
    volume = "02",
    pages = "042",
    year = "2019"
}

@article{Cunha:2020azh,
    author = "Cunha, Pedro V. P. and Herdeiro, Carlos A. R.",
    title = "{Stationary black holes and light rings}",
    eprint = "2003.06445",
    archivePrefix = "arXiv",
    primaryClass = "gr-qc",
    doi = "10.1103/PhysRevLett.124.181101",
    journal = "Phys. Rev. Lett.",
    volume = "124",
    number = "18",
    pages = "181101",
    year = "2020"
}

@book{Wald:1984rg,
    author = "Wald, Robert M.",
    title = "{General Relativity}",
    doi = "10.7208/chicago/9780226870373.001.0001",
    publisher = "Chicago Univ. Pr.",
    address = "Chicago, USA",
    year = "1984"
}

@inproceedings{bardeen1968,
       author = {{Bardeen}, James},
        title = "{Non-singular general relativistic gravitational collapse}",
    booktitle = {Proceedings of the 5th International Conference on Gravitation and the Theory of Relativity},
         year = 1968,
        month = {9},
        pages = {87},
       adsurl = {https://ui.adsabs.harvard.edu/abs/1968qtr..conf...87B}
}

@article{Culetu:2014lca,
    author = "Culetu, Hristu",
    title = "{On a regular charged black hole with a nonlinear electric source}",
    eprint = "1408.3334",
    archivePrefix = "arXiv",
    primaryClass = "gr-qc",
    doi = "10.1007/s10773-015-2521-6",
    journal = "Int. J. Theor. Phys.",
    volume = "54",
    number = "8",
    pages = "2855--2863",
    year = "2015"
}

@article{LIGOScientific:2020tif,
    author = "Abbott, R. and others",
    collaboration = "LIGO Scientific, Virgo",
    title = "{Tests of general relativity with binary black holes from the second LIGO-Virgo gravitational-wave transient catalog}",
    eprint = "2010.14529",
    archivePrefix = "arXiv",
    primaryClass = "gr-qc",
    reportNumber = "LIGO-P2000091",
    doi = "10.1103/PhysRevD.103.122002",
    journal = "Phys. Rev. D",
    volume = "103",
    number = "12",
    pages = "122002",
    year = "2021"
}

@article{LIGOScientific:2020stg,
    author = "Abbott, R. and others",
    collaboration = "LIGO Scientific, Virgo",
    title = "{GW190412: Observation of a Binary-Black-Hole Coalescence with Asymmetric Masses}",
    eprint = "2004.08342",
    archivePrefix = "arXiv",
    primaryClass = "astro-ph.HE",
    reportNumber = "LIGO-P190412",
    doi = "10.1103/PhysRevD.102.043015",
    journal = "Phys. Rev. D",
    volume = "102",
    number = "4",
    pages = "043015",
    year = "2020"
}

@article{LIGOScientific:2016aoc,
    author = "Abbott, B. P. and others",
    collaboration = "LIGO Scientific, Virgo",
    title = "{Observation of Gravitational Waves from a Binary Black Hole Merger}",
    eprint = "1602.03837",
    archivePrefix = "arXiv",
    primaryClass = "gr-qc",
    reportNumber = "LIGO-P150914",
    doi = "10.1103/PhysRevLett.116.061102",
    journal = "Phys. Rev. Lett.",
    volume = "116",
    number = "6",
    pages = "061102",
    year = "2016"
}

@article{EventHorizonTelescope:2019dse,
    author = "Akiyama, Kazunori and others",
    collaboration = "Event Horizon Telescope",
    title = "{First M87 Event Horizon Telescope Results. I. The Shadow of the Supermassive Black Hole}",
    eprint = "1906.11238",
    archivePrefix = "arXiv",
    primaryClass = "astro-ph.GA",
    doi = "10.3847/2041-8213/ab0ec7",
    journal = "Astrophys. J. Lett.",
    volume = "875",
    pages = "L1",
    year = "2019"
}

@article{EventHorizonTelescope:2022wkp,
    author = "Akiyama, Kazunori and others",
    collaboration = "Event Horizon Telescope",
    title = "{First Sagittarius A* Event Horizon Telescope Results. I. The Shadow of the Supermassive Black Hole in the Center of the Milky Way}",
    eprint = "2311.08680",
    archivePrefix = "arXiv",
    primaryClass = "astro-ph.HE",
    doi = "10.3847/2041-8213/ac6674",
    journal = "Astrophys. J. Lett.",
    volume = "930",
    number = "2",
    pages = "L12",
    year = "2022"
}

@article{EventHorizonTelescope:2022xqj,
    author = "Akiyama, Kazunori and others",
    collaboration = "Event Horizon Telescope",
    title = "{First Sagittarius A* Event Horizon Telescope Results. VI. Testing the Black Hole Metric}",
    eprint = "2311.09484",
    archivePrefix = "arXiv",
    primaryClass = "astro-ph.HE",
    reportNumber = "FERMILAB-PUB-22-422-PPD",
    doi = "10.3847/2041-8213/ac6756",
    journal = "Astrophys. J. Lett.",
    volume = "930",
    number = "2",
    pages = "L17",
    year = "2022"
}

@article{Schwarzschild:1916uq,
    author = "Schwarzschild, Karl",
    title = "{On the gravitational field of a mass point according to Einstein's theory}",
    eprint = "physics/9905030",
    archivePrefix = "arXiv",
    journal = "Sitzungsber. Preuss. Akad. Wiss. Berlin (Math. Phys. )",
    volume = "1916",
    pages = "189--196",
    year = "1916"
}

@book{Chandrasekhar:1985kt,
    author = "Chandrasekhar, Subrahmanyan",
    title = "{The mathematical theory of black holes}",
    isbn = "978-0-19-850370-5",
    year = "1985"
}

@article{Wei:2013kza,
    author = "Wei, Shao-Wen and Liu, Yu-Xiao",
    title = "{Observing the shadow of Einstein-Maxwell-Dilaton-Axion black hole}",
    eprint = "1311.4251",
    archivePrefix = "arXiv",
    primaryClass = "gr-qc",
    doi = "10.1088/1475-7516/2013/11/063",
    journal = "JCAP",
    volume = "11",
    pages = "063",
    year = "2013"
}

@article{Palti:2019pca,
    author = "Palti, Eran",
    title = "{The Swampland: Introduction and Review}",
    eprint = "1903.06239",
    archivePrefix = "arXiv",
    primaryClass = "hep-th",
    reportNumber = "MPP-2019-53",
    doi = "10.1002/prop.201900037",
    journal = "Fortsch. Phys.",
    volume = "67",
    number = "6",
    pages = "1900037",
    year = "2019"
}

@article{Niedermaier:2006wt,
    author = "Niedermaier, Max and Reuter, Martin",
    title = "{The Asymptotic Safety Scenario in Quantum Gravity}",
    doi = "10.12942/lrr-2006-5",
    journal = "Living Rev. Rel.",
    volume = "9",
    pages = "5--173",
    year = "2006"
}

@article{Belfiglio:2024qsa,
    author = "Belfiglio, Alessio and Luongo, Orlando and Mancini, Stefano and Tomasi, Sebastiano",
    title = "{Entanglement entropy in quantum black holes}",
    eprint = "2404.00715",
    archivePrefix = "arXiv",
    primaryClass = "gr-qc",
    doi = "10.1088/1361-6382/ad9e66",
    journal = "Class. Quant. Grav.",
    volume = "42",
    number = "3",
    pages = "035006",
    year = "2025"
}

@article{Claudel:2000yi,
    author = "Claudel, Clarissa-Marie and Virbhadra, K. S. and Ellis, G. F. R.",
    title = "{The Geometry of photon surfaces}",
    eprint = "gr-qc/0005050",
    archivePrefix = "arXiv",
    doi = "10.1063/1.1308507",
    journal = "J. Math. Phys.",
    volume = "42",
    pages = "818--838",
    year = "2001"
}

@article{PhysRevD.65.103004,
  title = {Gravitational lensing by naked singularities},
  author = {Virbhadra, K. S. and Ellis, G. F. R.},
  journal = {Phys. Rev. D},
  volume = {65},
  issue = {10},
  pages = {103004},
  numpages = {10},
  year = {2002},
  month = {5},
  publisher = {American Physical Society},
  doi = {10.1103/PhysRevD.65.103004},
  url = {https://link.aps.org/doi/10.1103/PhysRevD.65.103004}
}

@article{Almheiri:2020cfm,
    author = "Almheiri, Ahmed and Hartman, Thomas and Maldacena, Juan and Shaghoulian, Edgar and Tajdini, Amirhossein",
    title = "{The entropy of Hawking radiation}",
    eprint = "2006.06872",
    archivePrefix = "arXiv",
    primaryClass = "hep-th",
    doi = "10.1103/RevModPhys.93.035002",
    journal = "Rev. Mod. Phys.",
    volume = "93",
    number = "3",
    pages = "035002",
    year = "2021"
}

@article{Bonanno:2000ep,
    author = "Bonanno, Alfio and Reuter, Martin",
    title = "{Renormalization group improved black hole space-times}",
    eprint = "hep-th/0002196",
    archivePrefix = "arXiv",
    reportNumber = "INFN-CT-03-00, MZ-TH-00-04",
    doi = "10.1103/PhysRevD.62.043008",
    journal = "Phys. Rev. D",
    volume = "62",
    pages = "043008",
    year = "2000"
}

@book{birrell_davies_1982,
    place={Cambridge},
    series={Cambridge Monographs on Mathematical Physics},
    title={Quantum Fields in Curved Space},
    DOI={10.1017/CBO9780511622632},
    publisher={Cambridge University Press},
    author={Birrell, N. D. and Davies, P. C. W.},
    year={1982},
    collection={Cambridge Monographs on Mathematical Physics}
}

@article{Fursaev:1994te,
    author = "Fursaev, Dmitri V.",
    title = "{Temperature and entropy of a quantum black hole and conformal anomaly}",
    eprint = "hep-th/9412161",
    archivePrefix = "arXiv",
    reportNumber = "JINR-E2-94-484",
    doi = "10.1103/PhysRevD.51.R5352",
    journal = "Phys. Rev. D",
    volume = "51",
    pages = "5352--5355",
    year = "1995"
}

@article{Xiao:2021zly,
    author = "Xiao, Yong and Tian, Yu",
    title = "{Logarithmic correction to black hole entropy from the nonlocality of quantum gravity}",
    eprint = "2104.14902",
    archivePrefix = "arXiv",
    primaryClass = "gr-qc",
    doi = "10.1103/PhysRevD.105.044013",
    journal = "Phys. Rev. D",
    volume = "105",
    number = "4",
    pages = "044013",
    year = "2022"
}

@article{Solodukhin:2011gn,
    author = "Solodukhin, Sergey N.",
    title = "{Entanglement entropy of black holes}",
    eprint = "1104.3712",
    archivePrefix = "arXiv",
    primaryClass = "hep-th",
    doi = "10.12942/lrr-2011-8",
    journal = "Living Rev. Rel.",
    volume = "14",
    pages = "8",
    year = "2011"
}

@article{Solodukhin:1994yz,
    author = "Solodukhin, Sergei N.",
    title = "{The Conical singularity and quantum corrections to entropy of black hole}",
    eprint = "hep-th/9407001",
    archivePrefix = "arXiv",
    reportNumber = "JINR-E2-94-246",
    doi = "10.1103/PhysRevD.51.609",
    journal = "Phys. Rev. D",
    volume = "51",
    pages = "609--617",
    year = "1995"
}

@article{Sen:2012dw,
    author = "Sen, Ashoke",
    title = "{Logarithmic Corrections to Schwarzschild and Other Non-extremal Black Hole Entropy in Different Dimensions}",
    eprint = "1205.0971",
    archivePrefix = "arXiv",
    primaryClass = "hep-th",
    doi = "10.1007/JHEP04(2013)156",
    journal = "JHEP",
    volume = "04",
    pages = "156",
    year = "2013"
}

@article{El-Menoufi:2015cqw,
    author = "El-Menoufi, Basem Kamal",
    title = "{Quantum gravity of Kerr-Schild spacetimes and the logarithmic correction to Schwarzschild black hole entropy}",
    eprint = "1511.08816",
    archivePrefix = "arXiv",
    primaryClass = "hep-th",
    reportNumber = "NSF-KITP-15-152",
    doi = "10.1007/JHEP05(2016)035",
    journal = "JHEP",
    volume = "05",
    pages = "035",
    year = "2016"
}

@article{Cai:2009ua,
    author = "Cai, Rong-Gen and Cao, Li-Ming and Ohta, Nobuyoshi",
    title = "{Black Holes in Gravity with Conformal Anomaly and Logarithmic Term in Black Hole Entropy}",
    eprint = "0911.4379",
    archivePrefix = "arXiv",
    primaryClass = "hep-th",
    reportNumber = "KU-TP-039",
    doi = "10.1007/JHEP04(2010)082",
    journal = "JHEP",
    volume = "04",
    pages = "082",
    year = "2010"
}

@article{Carlip:2000nv,
    author = "Carlip, Steven",
    title = "{Logarithmic corrections to black hole entropy from the Cardy formula}",
    eprint = "gr-qc/0005017",
    archivePrefix = "arXiv",
    reportNumber = "UCD-00-11, UCD-2000-11",
    doi = "10.1088/0264-9381/17/20/302",
    journal = "Class. Quant. Grav.",
    volume = "17",
    pages = "4175--4186",
    year = "2000"
}

@article{Banerjee:2008cf,
    author = "Banerjee, Rabin and Majhi, Bibhas Ranjan",
    title = "{Quantum Tunneling Beyond Semiclassical Approximation}",
    eprint = "0805.2220",
    archivePrefix = "arXiv",
    primaryClass = "hep-th",
    doi = "10.1088/1126-6708/2008/06/095",
    journal = "JHEP",
    volume = "06",
    pages = "095",
    year = "2008"
}

@article{Banerjee:2008fz,
    author = "Banerjee, Rabin and Majhi, Bibhas Ranjan",
    title = "{Quantum Tunneling, Trace Anomaly and Effective Metric}",
    eprint = "0808.3688",
    archivePrefix = "arXiv",
    primaryClass = "hep-th",
    doi = "10.1016/j.physletb.2009.03.019",
    journal = "Phys. Lett. B",
    volume = "674",
    pages = "218--222",
    year = "2009"
}

@article{Kaul:2000kf,
    author = "Kaul, Romesh K. and Majumdar, Parthasarathi",
    title = "{Logarithmic correction to the Bekenstein-Hawking entropy}",
    eprint = "gr-qc/0002040",
    archivePrefix = "arXiv",
    doi = "10.1103/PhysRevLett.84.5255",
    journal = "Phys. Rev. Lett.",
    volume = "84",
    pages = "5255--5257",
    year = "2000"
}

@article{Kretschmann,
author = {Kretschmann, Erich},
title = {\"Uber die prinzipielle Bestimmbarkeit der
berechtigten Bezugssysteme beliebiger Relativit\"atstheorien (I), (II)},
journal = {Annalen der Physik},
volume = 48,
year = 1915,
pages =  {907-942, 943-982}
}

@article{Held:2021vwd,
    author = "Held, Aaron",
    title = "{Invariant Renormalization-Group improvement}",
    eprint = "2105.11458",
    archivePrefix = "arXiv",
    primaryClass = "gr-qc",
    reportNumber = "Imperial/TP/2021/AH/04",
    month = "5",
    year = "2021"
}

@ARTICLE{Bogoliubov1958,
       author = {{Bogoljubov}, N.~N.},
        title = "{On a new method in the theory of superconductivity}",
      journal = {Il Nuovo Cimento},
         year = 1958,
        month = {3},
       volume = {7},
       number = {6},
        pages = {794-805},
          doi = {10.1007/BF02745585},
       adsurl = {https://ui.adsabs.harvard.edu/abs/1958NCim....7..794B}
}

@article{Frolov:2016pav,
    author = "Frolov, Valeri P.",
    title = "{Notes on nonsingular models of black holes}",
    eprint = "1609.01758",
    archivePrefix = "arXiv",
    primaryClass = "gr-qc",
    doi = "10.1103/PhysRevD.94.104056",
    journal = "Phys. Rev. D",
    volume = "94",
    number = "10",
    pages = "104056",
    year = "2016"
}

@article{Dymnikova:2004qg,
    author = "Dymnikova, Irina and Galaktionov, Evgeny",
    title = "{Stability of a vacuum nonsingular black hole}",
    eprint = "gr-qc/0409049",
    archivePrefix = "arXiv",
    doi = "10.1088/0264-9381/22/12/003",
    journal = "Class. Quant. Grav.",
    volume = "22",
    pages = "2331--2358",
    year = "2005"
}

@article{Donoghue:1993eb,
    author = "Donoghue, John F.",
    title = "{Leading quantum correction to the Newtonian potential}",
    eprint = "gr-qc/9310024",
    archivePrefix = "arXiv",
    reportNumber = "UMHEP-396",
    doi = "10.1103/PhysRevLett.72.2996",
    journal = "Phys. Rev. Lett.",
    volume = "72",
    pages = "2996--2999",
    year = "1994"
}

@article{Donoghue:1994dn,
    author = "Donoghue, John F.",
    title = "{General relativity as an effective field theory: The leading quantum corrections}",
    eprint = "gr-qc/9405057",
    archivePrefix = "arXiv",
    reportNumber = "UMHEP-408",
    doi = "10.1103/PhysRevD.50.3874",
    journal = "Phys. Rev. D",
    volume = "50",
    pages = "3874--3888",
    year = "1994"
}

@article{Hamber:1995cq,
    author = "Hamber, H. W. and Liu, S.",
    title = "{On the quantum corrections to the Newtonian potential}",
    eprint = "hep-th/9505182",
    archivePrefix = "arXiv",
    reportNumber = "UCI-95-17",
    doi = "10.1016/0370-2693(95)00790-R",
    journal = "Phys. Lett. B",
    volume = "357",
    pages = "51--56",
    year = "1995"
}

@article{Bjerrum-Bohr:2002gqz,
    author = "Bjerrum-Bohr, N. E. J and Donoghue, John F. and Holstein, Barry R.",
    title = "{Quantum gravitational corrections to the nonrelativistic scattering potential of two masses}",
    eprint = "hep-th/0211072",
    archivePrefix = "arXiv",
    doi = "10.1103/PhysRevD.71.069903",
    journal = "Phys. Rev. D",
    volume = "67",
    pages = "084033",
    year = "2003",
    note = "[Erratum: Phys.Rev.D 71, 069903 (2005)]"
}

@article{Donoghue:2022chi,
    author = "Donoghue, John F.",
    title = "{Nonlocal partner to the cosmological constant}",
    eprint = "2201.12217",
    archivePrefix = "arXiv",
    primaryClass = "hep-th",
    reportNumber = "ACFI-T22-01",
    doi = "10.1103/PhysRevD.105.105025",
    journal = "Phys. Rev. D",
    volume = "105",
    number = "10",
    pages = "105025",
    year = "2022"
}

@article{Foong_2008,
doi = {10.1088/0143-0807/29/5/012},
url = {https://dx.doi.org/10.1088/0143-0807/29/5/012},
year = {2008},
month = {8},
publisher = {},
volume = {29},
number = {5},
pages = {987},
author = {S K Foong},
title = {From Moon-fall to motions under inverse square laws},
journal = {European Journal of Physics}
}

@article{Meissel,
    author = "Meissel, E.",
    title = "{{\"U}ber Reihen, denen man bei der numerischen L{\"o}sung des Dreik{\"o}rperproblems begegnet, wenn die Anfangsgeschwindigkeiten Null sind}",
    journal = "Jahresbericht {\"u}ber die Realschule in Kiel w{\"a}hrend des Schuljahres 1881/82",
    pages = "1-11",
    year = "1882"
}

@inbook{Peetre1997ErnstMA,
  title={Ernst Meissel and the Pythagorean Problem - the Dreik{\"o}rper-Problem in the Nachlass Meissel},
  author={Jaak Peetre},
  year={1997}
}

@article{Sarkar:2007uz,
    author = "Sarkar, Sudipta and Shankaranarayanan, S. and Sriramkumar, L.",
    title = "{Sub-leading contributions to the black hole entropy in the brick wall approach}",
    eprint = "0710.2013",
    archivePrefix = "arXiv",
    primaryClass = "gr-qc",
    reportNumber = "AEI-2007-139",
    doi = "10.1103/PhysRevD.78.024003",
    journal = "Phys. Rev. D",
    volume = "78",
    pages = "024003",
    year = "2008"
}

@article{Kontou:2020bta,
    author = "Kontou, Eleni-Alexandra and Sanders, Ko",
    title = "{Energy conditions in general relativity and quantum field theory}",
    eprint = "2003.01815",
    archivePrefix = "arXiv",
    primaryClass = "gr-qc",
    doi = "10.1088/1361-6382/ab8fcf",
    journal = "Class. Quant. Grav.",
    volume = "37",
    number = "19",
    pages = "193001",
    year = "2020"
}

@book{Hawking:1973uf,
    author = "Hawking, S. W. and Ellis, G. F. R.",
    title = "{The Large Scale Structure of Space-Time}",
    doi = "10.1017/CBO9780511524646",
    isbn = "978-0-521-20016-5",
    publisher = "Cambridge University Press",
    series = "Cambridge Monographs on Mathematical Physics",
    month = "2",
    year = "2011"
}

@article{Curiel:2014zba,
    author = "Curiel, Erik",
    title = "{A Primer on Energy Conditions}",
    eprint = "1405.0403",
    archivePrefix = "arXiv",
    primaryClass = "physics.hist-ph",
    doi = "10.1007/978-1-4939-3210-8_3",
    journal = "Einstein Stud.",
    volume = "13",
    pages = "43--104",
    year = "2017"
}

@article{Nicolini:2019irw,
    author = "Nicolini, Piero and Spallucci, Euro and Wondrak, Michael F.",
    title = "{Quantum Corrected Black Holes from String T-Duality}",
    eprint = "1902.11242",
    archivePrefix = "arXiv",
    primaryClass = "gr-qc",
    doi = "10.1016/j.physletb.2019.134888",
    journal = "Phys. Lett. B",
    volume = "797",
    pages = "134888",
    year = "2019"
}

@article{Bjerrum-Bohr:2002fji,
    author = "Bjerrum-Bohr, Niels Emil Jannik and Donoghue, John F. and Holstein, Barry R.",
    title = "{Quantum corrections to the Schwarzschild and Kerr metrics}",
    eprint = "hep-th/0211071",
    archivePrefix = "arXiv",
    doi = "10.1103/PhysRevD.68.084005",
    journal = "Phys. Rev. D",
    volume = "68",
    pages = "084005",
    year = "2003",
    note = "[Erratum: Phys.Rev.D 71, 069904 (2005)]"
}

@article{Platania:2019kyx,
    author = "Platania, Alessia",
    title = "{Dynamical renormalization of black-hole spacetimes}",
    eprint = "1903.10411",
    archivePrefix = "arXiv",
    primaryClass = "gr-qc",
    doi = "10.1140/epjc/s10052-019-6990-2",
    journal = "Eur. Phys. J. C",
    volume = "79",
    number = "6",
    pages = "470",
    year = "2019"
}

@article{Penrose:1964wq,
    author = "Penrose, Roger",
    title = "{Gravitational collapse and space-time singularities}",
    doi = "10.1103/PhysRevLett.14.57",
    journal = "Phys. Rev. Lett.",
    volume = "14",
    pages = "57--59",
    year = "1965"
}

@article{Penrose:1973um,
    author = "Penrose, R.",
    title = "{Naked singularities}",
    doi = "10.1111/j.1749-6632.1973.tb41447.x",
    journal = "Annals N. Y. Acad. Sci.",
    volume = "224",
    pages = "125--134",
    year = "1973"
}

@article{Penrose:1969pc,
    author = "Penrose, R.",
    title = "{Gravitational collapse: The role of general relativity}",
    doi = "10.1023/A:1016578408204",
    journal = "Riv. Nuovo Cim.",
    volume = "1",
    pages = "252--276",
    year = "1969"
}

@article{Knorr:2022kqp,
    author = "Knorr, Benjamin and Platania, Alessia",
    title = "{Sifting quantum black holes through the principle of least action}",
    eprint = "2202.01216",
    archivePrefix = "arXiv",
    primaryClass = "hep-th",
    doi = "10.1103/PhysRevD.106.L021901",
    journal = "Phys. Rev. D",
    volume = "106",
    number = "2",
    pages = "L021901",
    year = "2022"
}

@article{Hawking:1976ra,
    author = "Hawking, S. W.",
    title = "{Breakdown of Predictability in Gravitational Collapse}",
    doi = "10.1103/PhysRevD.14.2460",
    journal = "Phys. Rev. D",
    volume = "14",
    pages = "2460--2473",
    year = "1976"
}

@article{Ryu:2006bv,
    author = "Ryu, Shinsei and Takayanagi, Tadashi",
    title = "{Holographic derivation of entanglement entropy from AdS/CFT}",
    eprint = "hep-th/0603001",
    archivePrefix = "arXiv",
    reportNumber = "NSF-KITP-06-11",
    doi = "10.1103/PhysRevLett.96.181602",
    journal = "Phys. Rev. Lett.",
    volume = "96",
    pages = "181602",
    year = "2006"
}

@article{Bekenstein:1973ur,
    author = "Bekenstein, Jacob D.",
    title = "{Black holes and entropy}",
    doi = "10.1103/PhysRevD.7.2333",
    journal = "Phys. Rev. D",
    volume = "7",
    pages = "2333--2346",
    year = "1973"
}

@article{Maldacena:1997re,
    author = "Maldacena, Juan Martin",
    title = "{The Large N limit of superconformal field theories and supergravity}",
    eprint = "hep-th/9711200",
    archivePrefix = "arXiv",
    reportNumber = "HUTP-97-A097, HUTP-98-A097",
    doi = "10.4310/ATMP.1998.v2.n2.a1",
    journal = "Adv. Theor. Math. Phys.",
    volume = "2",
    pages = "231--252",
    year = "1998"
}

@article{Platania:2023srt,
    author = "Platania, Alessia",
    title = "{Black Holes in Asymptotically Safe Gravity}",
    eprint = "2302.04272",
    archivePrefix = "arXiv",
    primaryClass = "gr-qc",
    reportNumber = "NORDITA 2022-085",
    month = "2",
    year = "2023"
}

@article{Calmet:2017qqa,
    author = "Calmet, Xavier and El-Menoufi, Basem Kamal",
    title = "{Quantum Corrections to Schwarzschild Black Hole}",
    eprint = "1704.00261",
    archivePrefix = "arXiv",
    primaryClass = "hep-th",
    doi = "10.1140/epjc/s10052-017-4802-0",
    journal = "Eur. Phys. J. C",
    volume = "77",
    number = "4",
    pages = "243",
    year = "2017"
}

@article{Eichhorn:2021iwq,
    author = "Eichhorn, Astrid and Held, Aaron",
    title = "{From a locality-principle for new physics to image features of regular spinning black holes with disks}",
    eprint = "2103.13163",
    archivePrefix = "arXiv",
    primaryClass = "gr-qc",
    reportNumber = "Imperial/TP/2021/AH/02",
    doi = "10.1088/1475-7516/2021/05/073",
    journal = "JCAP",
    volume = "05",
    pages = "073",
    year = "2021"
}

@article{Nordstrom,
author = "Nordstrom, G.",
title = "On the Energy of the Gravitational Field in Einstein's Theory",
year =  "1918",
journal =  "Verhandl. Koninkl. Ned. Akad. Wetenschap., Afdel. Natuurk. (Amsterdam)",
volume = "26",
pages = "1201–1208"
}

@article{Donoghue:2001qc,
    author = {Donoghue, John F. and Holstein, Barry R. and Garbrecht, Bj\"orn and Konstandin, Thomas},
    title = {{Quantum corrections to the Reissner-Nordstr\"om and Kerr-Newman metrics}},
    eprint = "hep-th/0112237",
    archivePrefix = "arXiv",
    doi = "10.1016/S0370-2693(02)01246-7",
    journal = "Phys. Lett. B",
    volume = "529",
    pages = "132--142",
    year = "2002",
    note = "[Erratum: Phys.Lett.B 612, 311--312 (2005)]"
}

@article{Kirilin:2006en,
    author = "Kirilin, G. G.",
    title = "{Quantum corrections to the Schwarzschild metric and reparametrization transformations}",
    eprint = "gr-qc/0601020",
    archivePrefix = "arXiv",
    doi = "10.1103/PhysRevD.75.108501",
    journal = "Phys. Rev. D",
    volume = "75",
    pages = "108501",
    year = "2007"
}

@article{Calmet:2019eof,
    author = "Calmet, Xavier and Casadio, Roberto and Kuipers, Folkert",
    title = "{Quantum Gravitational Corrections to a Star Metric and the Black Hole Limit}",
    eprint = "1909.13277",
    archivePrefix = "arXiv",
    primaryClass = "hep-th",
    doi = "10.1103/PhysRevD.100.086010",
    journal = "Phys. Rev. D",
    volume = "100",
    number = "8",
    pages = "086010",
    year = "2019"
}

@book{Carroll,
  asin = {0805387323},
  author = {Carroll, Sean},
  description = {Spacetime and Geometry: An Introduction to General Relativity (9780805387322): Sean Carroll: Books},
  dewey = {530.11},
  ean = {9780805387322},
  isbn = {0805387323},
  publisher = {Benjamin Cummings},
  title = {Spacetime and Geometry: An Introduction to General Relativity},
  url = {http://www.amazon.com/Spacetime-Geometry-Introduction-General-Relativity/dp/0805387323},
  year = 2003
}

@article{Campos_Delgado_2022,
	doi = {10.1140/epjc/s10052-022-10232-0},
  
	url = {https://doi.org/10.1140%2Fepjc%2Fs10052-022-10232-0},
  
	year = 2022,
	month = {3},
  
	publisher = {Springer Science and Business Media {LLC}
},
  
	volume = {82},
  
	number = {3},
  
	author = {Ruben Campos Delgado},
  
	title = {Quantum gravitational corrections to the entropy of a Reissner{\textendash}Nordström black hole},
  
	journal = {The European Physical Journal C}
}

@article{Eichhorn:2022bgu,
    author = "Eichhorn, Astrid and Held, Aaron",
    title = "{Black holes in asymptotically safe gravity and beyond}",
    eprint = "2212.09495",
    archivePrefix = "arXiv",
    primaryClass = "gr-qc",
    month = "12",
    year = "2022"
}

@article{Ishibashi:2021kmf_ImprovedRN,
    author = "Ishibashi, Akihiro and Ohta, Nobuyoshi and Yamaguchi, Daiki",
    title = "{Quantum improved charged black holes}",
    eprint = "2106.05015",
    archivePrefix = "arXiv",
    primaryClass = "hep-th",
    doi = "10.1103/PhysRevD.104.066016",
    journal = "Phys. Rev. D",
    volume = "104",
    number = "6",
    pages = "066016",
    year = "2021"
}

@article{Mazharimousavi:2012ca_FofR,
    author = "Mazharimousavi, S. Habib and Kerachian, M. and Halilsoy, M.",
    title = "{Existence of Reissner-Nordstrom type black holes in $f(R)$ gravity}",
    eprint = "1210.4696",
    archivePrefix = "arXiv",
    primaryClass = "gr-qc",
    doi = "10.1142/S0218271813500570",
    journal = "Int. J. Mod. Phys. D",
    volume = "22",
    pages = "1350057",
    year = "2013"
}

@article{Vafa:2005ui,
    author = "Vafa, Cumrun",
    title = "{The String landscape and the swampland}",
    eprint = "hep-th/0509212",
    archivePrefix = "arXiv",
    reportNumber = "HUTP-05-A043",
    month = "9",
    year = "2005"
}

@article{Arkani-Hamed:2006emk,
    author = "Arkani-Hamed, Nima and Motl, Lubos and Nicolis, Alberto and Vafa, Cumrun",
    title = "{The String landscape, black holes and gravity as the weakest force}",
    eprint = "hep-th/0601001",
    archivePrefix = "arXiv",
    reportNumber = "HUTP-05-A0057",
    doi = "10.1088/1126-6708/2007/06/060",
    journal = "JHEP",
    volume = "06",
    pages = "060",
    year = "2007"
}

@article{Harlow:2022ich,
    author = "Harlow, Daniel and Heidenreich, Ben and Reece, Matthew and Rudelius, Tom",
    title = "{Weak gravity conjecture}",
    eprint = "2201.08380",
    archivePrefix = "arXiv",
    primaryClass = "hep-th",
    reportNumber = "ACFI-T22-01",
    doi = "10.1103/RevModPhys.95.035003",
    journal = "Rev. Mod. Phys.",
    volume = "95",
    number = "3",
    pages = "035003",
    year = "2023"
}

@article{Carroll:2009maa,
    author = "Carroll, Sean M. and Johnson, Matthew C. and Randall, Lisa",
    title = "{Extremal limits and black hole entropy}",
    eprint = "0901.0931",
    archivePrefix = "arXiv",
    primaryClass = "hep-th",
    reportNumber = "CALT-68-2717",
    doi = "10.1088/1126-6708/2009/11/109",
    journal = "JHEP",
    volume = "11",
    pages = "109",
    year = "2009"
}

@article{vanBeest:2021lhn,
    author = "van Beest, Marieke and Calder\'on-Infante, Jos\'e and Mirfendereski, Delaram and Valenzuela, Irene",
    title = "{Lectures on the Swampland Program in String Compactifications}",
    eprint = "2102.01111",
    archivePrefix = "arXiv",
    primaryClass = "hep-th",
    doi = "10.1016/j.physrep.2022.09.002",
    journal = "Phys. Rept.",
    volume = "989",
    pages = "1--50",
    year = "2022"
}

@article{Reissner,
author = {Reissner, H.},
title = {Über die Eigengravitation des elektrischen Feldes nach der Einsteinschen Theorie},
journal = {Annalen der Physik},
volume = {355},
number = {9},
pages = {106-120},
doi = {https://doi.org/10.1002/andp.19163550905},
url = {https://onlinelibrary.wiley.com/doi/abs/10.1002/andp.19163550905},
eprint = {https://onlinelibrary.wiley.com/doi/pdf/10.1002/andp.19163550905},
year = {1916}
}

@article{Weyl,
author = {Weyl, Hermann},
title = {Zur Gravitationstheorie},
journal = {Annalen der Physik},
volume = {359},
number = {18},
pages = {117-145},
doi = {https://doi.org/10.1002/andp.19173591804},
url = {https://onlinelibrary.wiley.com/doi/abs/10.1002/andp.19173591804},
eprint = {https://onlinelibrary.wiley.com/doi/pdf/10.1002/andp.19173591804},
year = {1917}
}

@article{Baker,
author = {Jeffery, George Barker  and Filon, Louis Napoleon George },
title = {The field of an electron on Einstein’s theory of gravitation},
journal = {Proceedings of the Royal Society of London. Series A, Containing Papers of a Mathematical and Physical Character},
volume = {99},
number = {697},
pages = {123-134},
year = {1921},
doi = {10.1098/rspa.1921.0028},
URL = {https://royalsocietypublishing.org/doi/abs/10.1098/rspa.1921.0028}
}

@article{Balbinot:2007kr,
    author = "Balbinot, R. and Fabbri, A. and Farese, S. and Parentani, R.",
    title = "{Hawking radiation from extremal and non-extremal black holes}",
    eprint = "0710.0388",
    archivePrefix = "arXiv",
    primaryClass = "hep-th",
    doi = "10.1103/PhysRevD.76.124010",
    journal = "Phys. Rev. D",
    volume = "76",
    pages = "124010",
    year = "2007"
}

@article{Maselli:2023khq,
    author = "Maselli, Andrea and Yi, Sophia and Pierini, Lorenzo and Vellucci, Vania and Reali, Luca and Gualtieri, Leonardo and Berti, Emanuele",
    title = "{Black hole spectroscopy beyond Kerr: Agnostic and theory-based tests with next-generation interferometers}",
    eprint = "2311.14803",
    archivePrefix = "arXiv",
    primaryClass = "gr-qc",
    doi = "10.1103/PhysRevD.109.064060",
    journal = "Phys. Rev. D",
    volume = "109",
    number = "6",
    pages = "064060",
    year = "2024"
}

@article{Antonelli:2025zwc,
    author = "Antonelli, Tommaso and Calmet, Xavier",
    title = "{The weak gravity conjecture in the Vilkovisky-DeWitt effective action of quantum gravity}",
    eprint = "2503.09227",
    archivePrefix = "arXiv",
    primaryClass = "hep-th",
    doi = "10.1016/j.physletb.2025.139403",
    journal = "Phys. Lett. B",
    volume = "864",
    pages = "139403",
    year = "2025"
}

@article{Heusler:1996ft,
    author = "Heusler, Markus",
    editor = "Straumann, N. and Jetzer, P. and Lavrelashvili, George V.",
    title = "{No hair theorems and black holes with hair}",
    eprint = "gr-qc/9610019",
    archivePrefix = "arXiv",
    reportNumber = "ZU-TH-28-96",
    journal = "Helv. Phys. Acta",
    volume = "69",
    number = "4",
    pages = "501--528",
    year = "1996"
}

@article{Johannsen:2010gy,
    author = "Johannsen, Tim and Psaltis, Dimitrios",
    title = "{Testing the No-Hair Theorem with Observations of Black Holes in the Electromagnetic Spectrum}",
    eprint = "1008.3902",
    archivePrefix = "arXiv",
    primaryClass = "astro-ph.HE",
    doi = "10.1016/j.asr.2010.10.019",
    journal = "Adv. Space Res.",
    volume = "47",
    pages = "528--532",
    year = "2011"
}

@article{Johannsen:2010xs,
    author = "Johannsen, Tim and Psaltis, Dimitrios",
    title = "{Testing the No-Hair Theorem with Observations in the Electromagnetic Spectrum: I. Properties of a Quasi-Kerr Spacetime}",
    eprint = "1003.3415",
    archivePrefix = "arXiv",
    primaryClass = "astro-ph.HE",
    doi = "10.1088/0004-637X/716/1/187",
    journal = "Astrophys. J.",
    volume = "716",
    pages = "187--197",
    year = "2010"
}

@article{Glampedakis:2005cf,
    author = "Glampedakis, Kostas and Babak, Stanislav",
    title = "{Mapping spacetimes with LISA: Inspiral of a test-body in a `quasi-Kerr' field}",
    eprint = "gr-qc/0510057",
    archivePrefix = "arXiv",
    doi = "10.1088/0264-9381/23/12/013",
    journal = "Class. Quant. Grav.",
    volume = "23",
    pages = "4167--4188",
    year = "2006"
}

@article{Konoplya2017,
    author = "Konoplya, R. A. and Stuchl\'\i{}k, Z.",
    title = "Are eikonal quasinormal modes linked to the unstable circular null geodesics?",
    eprint = "1705.05928",
    archivePrefix = "arXiv",
    primaryClass = "gr-qc",
    doi = "10.1016/j.physletb.2017.06.015",
    journal = "Phys. Lett. B",
    volume = "771",
    pages = "597--602",
    year = "2017"
}

@book{Maggiore:2007ulw,
    author = "Maggiore, Michele",
    title = "{Gravitational Waves. Vol. 1: Theory and Experiments}",
    doi = "10.1093/acprof:oso/9780198570745.001.0001",
    isbn = "978-0-19-171766-6",
    publisher = "Oxford University Press",
    year = "2007"
}

@article{Konoplya:2011qq,
    author = "Konoplya, R. A. and Zhidenko, A.",
    title = "{Quasinormal modes of black holes: From astrophysics to string theory}",
    eprint = "1102.4014",
    archivePrefix = "arXiv",
    primaryClass = "gr-qc",
    doi = "10.1103/RevModPhys.83.793",
    journal = "Rev. Mod. Phys.",
    volume = "83",
    pages = "793--836",
    year = "2011"
}

@article{Dolan:2010wr,
    author = "Dolan, Sam R.",
    title = "{The Quasinormal Mode Spectrum of a Kerr Black Hole in the Eikonal Limit}",
    eprint = "1007.5097",
    archivePrefix = "arXiv",
    primaryClass = "gr-qc",
    doi = "10.1103/PhysRevD.82.104003",
    journal = "Phys. Rev. D",
    volume = "82",
    pages = "104003",
    year = "2010"
}

@article{Cardoso:2008bp,
    author = "Cardoso, Vitor and Miranda, Alex S. and Berti, Emanuele and Witek, Helvi and Zanchin, Vilson T.",
    title = "{Geodesic stability, Lyapunov exponents and quasinormal modes}",
    eprint = "0812.1806",
    archivePrefix = "arXiv",
    primaryClass = "hep-th",
    doi = "10.1103/PhysRevD.79.064016",
    journal = "Phys. Rev. D",
    volume = "79",
    number = "6",
    pages = "064016",
    year = "2009"
}

@article{Johannsen:2011dh,
    author = "Johannsen, Tim and Psaltis, Dimitrios",
    title = "{A Metric for Rapidly Spinning Black Holes Suitable for Strong-Field Tests of the No-Hair Theorem}",
    eprint = "1105.3191",
    archivePrefix = "arXiv",
    primaryClass = "gr-qc",
    doi = "10.1103/PhysRevD.83.124015",
    journal = "Phys. Rev. D",
    volume = "83",
    pages = "124015",
    year = "2011"
}

@article{Will2014kxa,
    author = "Will, Clifford M.",
    title = "{The Confrontation between General Relativity and Experiment}",
    eprint = "1403.7377",
    archivePrefix = "arXiv",
    primaryClass = "gr-qc",
    doi = "10.12942/lrr-2014-4",
    journal = "Living Rev. Rel.",
    volume = "17",
    pages = "4",
    year = "2014"
}

@article{Berti:2015itd,
    author = "Berti, Emanuele and others",
    title = "{Testing General Relativity with Present and Future Astrophysical Observations}",
    eprint = "1501.07274",
    archivePrefix = "arXiv",
    primaryClass = "gr-qc",
    doi = "10.1088/0264-9381/32/24/243001",
    journal = "Class. Quant. Grav.",
    volume = "32",
    pages = "243001",
    year = "2015"
}

@article{Berti:2022xfj,
    author = "Berti, Emanuele and Cardoso, Vitor and Cheung, Mark Ho-Yeuk and Di Filippo, Francesco and Duque, Francisco and Martens, Paul and Mukohyama, Shinji",
    title = "{Stability of the fundamental quasinormal mode in time-domain observations against small perturbations}",
    eprint = "2205.08547",
    archivePrefix = "arXiv",
    primaryClass = "gr-qc",
    doi = "10.1103/PhysRevD.106.084011",
    journal = "Phys. Rev. D",
    volume = "106",
    number = "8",
    pages = "084011",
    year = "2022"
}

@article{Stefanov:2010xz,
    author = "Stefanov, Ivan Zh. and Yazadjiev, Stoytcho S. and Gyulchev, Galin G.",
    title = "{Connection between Black-Hole Quasinormal Modes and Lensing in the Strong Deflection Limit}",
    eprint = "1003.1609",
    archivePrefix = "arXiv",
    primaryClass = "gr-qc",
    doi = "10.1103/PhysRevLett.104.251103",
    journal = "Phys. Rev. Lett.",
    volume = "104",
    pages = "251103",
    year = "2010"
}

@article{Williams:2004qba,
    author = "Williams, James G. and Turyshev, Slava G. and Boggs, Dale H.",
    title = "{Progress in lunar laser ranging tests of relativistic gravity}",
    eprint = "gr-qc/0411113",
    archivePrefix = "arXiv",
    doi = "10.1103/PhysRevLett.93.261101",
    journal = "Phys. Rev. Lett.",
    volume = "93",
    pages = "261101",
    year = "2004"
}

@article{Bardeen:1972fi,
    author = "Bardeen, James M. and Press, William H. and Teukolsky, Saul A",
    title = "{Rotating black holes: Locally nonrotating frames, energy extraction, and scalar synchrotron radiation}",
    reportNumber = "OAP-288",
    doi = "10.1086/151796",
    journal = "Astrophys. J.",
    volume = "178",
    pages = "347",
    year = "1972"
}

@article{Dondi2020,
  title = {Towards the QED beta function and renormalons at $1/{N}_{f}^{2}$ and $1/{N}_{f}^{3}$},
  author = {Dondi, Nicola Andrea and Dunne, Gerald V. and Reichert, Manuel and Sannino, Francesco},
  journal = {Phys. Rev. D},
  volume = {102},
  issue = {3},
  pages = {035005},
  numpages = {15},
  year = {2020},
  month = {8},
  publisher = {American Physical Society},
  doi = {10.1103/PhysRevD.102.035005},
  url = {https://link.aps.org/doi/10.1103/PhysRevD.102.035005}
}

@article{Gruber:2013wua,
    author = "Gruber, Christine and Luongo, Orlando",
    title = "{Cosmographic analysis of the equation of state of the universe through Pad{\'e} approximations}",
    eprint = "1309.3215",
    archivePrefix = "arXiv",
    primaryClass = "gr-qc",
    doi = "10.1103/PhysRevD.89.103506",
    journal = "Phys. Rev. D",
    volume = "89",
    number = "10",
    pages = "103506",
    year = "2014"
}

@article{Sultan:2025mvc,
    author = "Sultan, Abdul Malik and Batool, Aliya and Abbas, G. and Jawad, Abdul and Shaymatov, Sanjar",
    title = "{Pad{\'e} approximated traversable wormholes in f(R,T) gravity}",
    doi = "10.1016/j.nuclphysb.2025.116838",
    journal = "Nucl. Phys. B",
    volume = "1013",
    pages = "116838",
    year = "2025"
}

@article{Mroue:2008fu,
    author = "Mroue, Abdul H. and Kidder, Lawrence E. and Teukolsky, Saul A.",
    title = "{Ineffectiveness of Pade resummation techniques in post-Newtonian approximations}",
    eprint = "0805.2390",
    archivePrefix = "arXiv",
    primaryClass = "gr-qc",
    doi = "10.1103/PhysRevD.78.044004",
    journal = "Phys. Rev. D",
    volume = "78",
    pages = "044004",
    year = "2008"
}

@article{Berti:2004md,
    author = "Berti, Emanuele",
    title = "{Black hole quasinormal modes: Hints of quantum gravity?}",
    eprint = "gr-qc/0411025",
    archivePrefix = "arXiv",
    journal = "Conf. Proc. C",
    volume = "0405132",
    pages = "145--186",
    year = "2004"
}

@article{Press:1972zz,
    author = "Press, William H. and Teukolsky, Saul A.",
    title = "{Floating Orbits, Superradiant Scattering and the Black-hole Bomb}",
    doi = "10.1038/238211a0",
    journal = "Nature",
    volume = "238",
    pages = "211--212",
    year = "1972"
}

@article{Damour:1976jd,
    author = "Damour, T. and Ruffini, R.",
    title = "{Black Hole Evaporation in the Klein-Sauter-Heisenberg-Euler Formalism}",
    doi = "10.1103/PhysRevD.14.332",
    journal = "Phys. Rev. D",
    volume = "14",
    pages = "332--334",
    year = "1976"
}

@article{Brito:2015oca,
    author = "Brito, Richard and Cardoso, Vitor and Pani, Paolo",
    title = "{Superradiance}: {New Frontiers in Black Hole
Physics}",
    eprint = "1501.06570",
    archivePrefix = "arXiv",
    primaryClass = "gr-qc",
    doi = "10.1007/978-3-319-19000-6",
    journal = "Lect. Notes Phys.",
    volume = "906",
    pages = "pp.1--237",
    year = "2015"
}

@article{Alonso-Serrano:2020dcz,
    author = "Alonso-Serrano, Ana and Li{\v{s}}ka, Marek",
    title = "{Quantum phenomenological gravitational dynamics: A general view from thermodynamics of spacetime}",
    eprint = "2009.03826",
    archivePrefix = "arXiv",
    primaryClass = "gr-qc",
    doi = "10.1007/JHEP12(2020)196",
    journal = "JHEP",
    volume = "12",
    pages = "196",
    year = "2020"
}

@article{Wald:1993nt,
    author = "Wald, Robert M.",
    title = "{Black hole entropy is the Noether charge}",
    eprint = "gr-qc/9307038",
    archivePrefix = "arXiv",
    reportNumber = "EFI-93-42",
    doi = "10.1103/PhysRevD.48.R3427",
    journal = "Phys. Rev. D",
    volume = "48",
    number = "8",
    pages = "R3427--R3431",
    year = "1993"
}

@article{Jacobson:1995ab,
    author = "Jacobson, Ted",
    title = "{Thermodynamics of space-time: The Einstein equation of state}",
    eprint = "gr-qc/9504004",
    archivePrefix = "arXiv",
    reportNumber = "UMDGR-95-114",
    doi = "10.1103/PhysRevLett.75.1260",
    journal = "Phys. Rev. Lett.",
    volume = "75",
    pages = "1260--1263",
    year = "1995"
}

@article{Jacobson:1993vj,
    author = "Jacobson, Ted and Kang, Gungwon and Myers, Robert C.",
    title = "{On black hole entropy}",
    eprint = "gr-qc/9312023",
    archivePrefix = "arXiv",
    reportNumber = "MCGILL-93-22, NSF-ITP-93-152, UMDGR-94-75",
    doi = "10.1103/PhysRevD.49.6587",
    journal = "Phys. Rev. D",
    volume = "49",
    pages = "6587--6598",
    year = "1994"
}

@article{Eling:2006aw,
    author = "Eling, Christopher and Guedens, Raf and Jacobson, Ted",
    title = "{Non-equilibrium thermodynamics of spacetime}",
    eprint = "gr-qc/0602001",
    archivePrefix = "arXiv",
    doi = "10.1103/PhysRevLett.96.121301",
    journal = "Phys. Rev. Lett.",
    volume = "96",
    pages = "121301",
    year = "2006"
}

@article{Jacobson:2015hqa,
    author = "Jacobson, Ted",
    title = "{Entanglement Equilibrium and the Einstein Equation}",
    eprint = "1505.04753",
    archivePrefix = "arXiv",
    primaryClass = "gr-qc",
    doi = "10.1103/PhysRevLett.116.201101",
    journal = "Phys. Rev. Lett.",
    volume = "116",
    number = "20",
    pages = "201101",
    year = "2016"
}

@article{Matzner:1979zz,
    author = "Matzner, Richard A. and Zamorano, Nelson and Sandberg, Vernon D.",
    title = "{Instability of the Cauchy horizon of Reissner-Nordstrom black holes}",
    doi = "10.1103/PhysRevD.19.2821",
    journal = "Phys. Rev. D",
    volume = "19",
    pages = "2821--2826",
    year = "1979"
}

@article{Poisson:1989zz,
    author = "Poisson, Eric and Israel, W.",
    title = "{Inner-horizon instability and mass inflation in black holes}",
    doi = "10.1103/PhysRevLett.63.1663",
    journal = "Phys. Rev. Lett.",
    volume = "63",
    pages = "1663--1666",
    year = "1989"
}

@article{Chandrasekhar1982,
 ISSN = {00804630},
 URL = {http://www.jstor.org/stable/2397225},
 author = {S. Chandrasekhar and J. B. Hartle},
 journal = {Proceedings of the Royal Society of London. Series A, Mathematical and Physical Sciences},
 number = {1787},
 pages = {301--315},
 publisher = {Royal Society},
 title = {On Crossing the Cauchy Horizon of a Reissner-Nordström Black-Hole},
 urldate = {2025-10-28},
 volume = {384},
 year = {1982}
}

@article{Hollands:2019whz,
    author = "Hollands, Stefan and Wald, Robert M. and Zahn, Jochen",
    title = {{Quantum instability of the Cauchy horizon in Reissner{\textendash}Nordstr{\"o}m{\textendash}deSitter spacetime}},
    eprint = "1912.06047",
    archivePrefix = "arXiv",
    primaryClass = "gr-qc",
    doi = "10.1088/1361-6382/ab8052",
    journal = "Class. Quant. Grav.",
    volume = "37",
    number = "11",
    pages = "115009",
    year = "2020"
}

@inproceedings{Page:1993up,
    author = "Page, Don N.",
    title = "{Black hole information}",
    booktitle = "{5th Canadian Conference on General Relativity and Relativistic Astrophysics (5CCGRRA)}",
    eprint = "hep-th/9305040",
    archivePrefix = "arXiv",
    reportNumber = "ALBERTA-THY-23-93",
    month = "5",
    year = "1993"
}

@article{Bianchi:2018mml,
    author = "Bianchi, Eugenio and Christodoulou, Marios and D'Ambrosio, Fabio and Haggard, Hal M. and Rovelli, Carlo",
    title = "{White Holes as Remnants: A Surprising Scenario for the End of a Black Hole}",
    eprint = "1802.04264",
    archivePrefix = "arXiv",
    primaryClass = "gr-qc",
    doi = "10.1088/1361-6382/aae550",
    journal = "Class. Quant. Grav.",
    volume = "35",
    number = "22",
    pages = "225003",
    year = "2018"
}

@article{Chen:2014jwq,
    author = "Chen, Pisin and Ong, Yen Chin and Yeom, Dong-han",
    title = "{Black Hole Remnants and the Information Loss Paradox}",
    eprint = "1412.8366",
    archivePrefix = "arXiv",
    primaryClass = "gr-qc",
    reportNumber = "NORDITA-2014-149",
    doi = "10.1016/j.physrep.2015.10.007",
    journal = "Phys. Rept.",
    volume = "603",
    pages = "1--45",
    year = "2015"
}

@article{Percacci:2007sz,
  author = {Percacci, Roberto},
  title  = {Asymptotic Safety},
  journal= {In *Oriti, D. (ed.): Approaches to Quantum Gravity* 111–128},
  year   = {2009},
  eprint = {0709.3851},
  archivePrefix = {arXiv},
  primaryClass = {hep-th},
}

@article{Eichhorn:2018yfc,
  author = {Eichhorn, Astrid},
  title  = {An asymptotically safe guide to quantum gravity and matter},
  journal= {Front.\ Astron.\ Space Sci.},
  year   = {2018},
  volume = {5},
  pages  = {47},
  doi    = {10.3389/fspas.2018.00047},
  eprint = {1810.07615},
  archivePrefix = {arXiv},
  primaryClass = {hep-th},
}

@article{Weinberg:1979,
  author = {Weinberg, Steven},
  title  = {Ultraviolet divergences in quantum theories of gravitation},
  journal= {In *General Relativity: An Einstein Centenary Survey*, eds. S. W. Hawking and W. Israel},
  year   = {1979},
}

@article{Reichert:20200Q,
  author = "Reichert, Manuel",
  title = "{Lecture notes: Functional Renormalisation Group and Asymptotically Safe Quantum Gravity}",
  doi = "10.22323/1.384.0005",
  journal = "PoS",
  year = 2020,
  volume = "Modave2019",
  pages = "005"
}

@article{Reuter:1996cp,
    author = "Reuter, M.",
    title = "{Nonperturbative evolution equation for quantum gravity}",
    eprint = "hep-th/9605030",
    archivePrefix = "arXiv",
    reportNumber = "DESY-96-080",
    doi = "10.1103/PhysRevD.57.971",
    journal = "Phys. Rev. D",
    volume = "57",
    pages = "971--985",
    year = "1998"
}

@article{Gubser:1998bc,
    author         = "Gubser, Steven S. and Klebanov, Igor R. and Polyakov, Alexander M.",
    title          = "{Gauge theory correlators from noncritical string theory}",
    journal        = "Phys. Lett. B",
    volume         = "428",
    pages          = "105--114",
    year           = "1998",
    eprint         = "hep-th/9802109",
    archivePrefix  = "arXiv",
    primaryClass   = "hep-th",
    doi            = "10.1016/S0370-2693(98)00377-3"
}

@article{Witten:1998qj,
    author         = "Witten, Edward",
    title          = "{Anti-de Sitter space and holography}",
    journal        = "Adv. Theor. Math. Phys.",
    volume         = "2",
    pages          = "253--291",
    year           = "1998",
    eprint         = "hep-th/9802150",
    archivePrefix  = "arXiv",
    primaryClass   = "hep-th"
}

@article{Ryu:2006ef,
    author         = "Ryu, Shinsei and Takayanagi, Tadashi",
    title          = "{Aspects of Holographic Entanglement Entropy}",
    journal        = "JHEP",
    volume         = "08",
    pages          = "045",
    year           = "2006",
    eprint         = "hep-th/0605073",
    archivePrefix  = "arXiv",
    primaryClass   = "hep-th",
    doi            = "10.1088/1126-6708/2006/08/045"
}

@article{Hubeny:2007xt,
    author         = "Hubeny, Veronika E. and Rangamani, Mukund and Takayanagi, Tadashi",
    title          = "{A Covariant Holographic Entanglement Entropy Proposal}",
    journal        = "JHEP",
    volume         = "07",
    pages          = "062",
    year           = "2007",
    eprint         = "0705.0016",
    archivePrefix  = "arXiv",
    primaryClass   = "hep-th",
    doi            = "10.1088/1126-6708/2007/07/062"
}

@article{Nishioka:2009un,
    author         = "Nishioka, Tatsuma and Ryu, Shinsei and Takayanagi, Tadashi",
    title          = "{Holographic Entanglement Entropy: An Overview}",
    journal        = "J. Phys. A",
    volume         = "42",
    pages          = "504008",
    year           = "2009",
    eprint         = "0905.0932",
    archivePrefix  = "arXiv",
    primaryClass   = "hep-th",
    doi            = "10.1088/1751-8113/42/50/504008"
}

@book{Rangamani:2016dms,
    author         = "Rangamani, Mukund and Takayanagi, Tadashi",
    title          = "{Holographic Entanglement Entropy}",
    publisher      = "Springer",
    series         = "Lecture Notes in Physics",
    volume         = "931",
    year           = "2017",
    eprint         = "1609.01287",
    archivePrefix  = "arXiv",
    primaryClass   = "hep-th",
    doi            = "10.1007/978-3-319-52573-0"
}

@article{Ooguri:2006in,
    author = "Ooguri, Hirosi and Vafa, Cumrun",
    title = "{On the Geometry of the String Landscape and the Swampland}",
    eprint = "hep-th/0605264",
    archivePrefix = "arXiv",
    reportNumber = "CALT-68-2600, HUTP-06-A017",
    doi = "10.1016/j.nuclphysb.2006.10.033",
    journal = "Nucl. Phys. B",
    volume = "766",
    pages = "21--33",
    year = "2007"
}

@article{Agmon:2022thq,
    author = "Agmon, Nathan Benjamin and Bedroya, Alek and Kang, Monica Jinwoo and Vafa, Cumrun",
    title = "{Lectures on the string landscape and the Swampland}",
    eprint = "2212.06187",
    archivePrefix = "arXiv",
    primaryClass = "hep-th",
    month = "12",
    year = "2022"
}

@article{BenAchour:2025uzp,
    author = "Ben Achour, Jibril and Gourgoulhon, Eric and Roussille, Hugo",
    title = "{Black hole photon ring beyond General Relativity: an integrable parametrization}",
    eprint = "2506.09882",
    archivePrefix = "arXiv",
    primaryClass = "gr-qc",
    doi = "10.1088/1475-7516/2025/10/012",
    journal = "JCAP",
    volume = "10",
    pages = "012",
    year = "2025"
}

@article{Kumar:2024utz,
    author = "Kumar, Shailesh and Singh, Rishabh Kumar and Chowdhuri, Abhishek and Bhattacharyya, Arpan",
    title = "{Exploring waveforms with non-GR deviations for extreme mass-ratio inspirals}",
    eprint = "2405.18508",
    archivePrefix = "arXiv",
    primaryClass = "gr-qc",
    doi = "10.1088/1475-7516/2024/10/047",
    journal = "JCAP",
    volume = "10",
    pages = "047",
    year = "2024"
}

@article{AbhishekChowdhuri:2023gvu,
    author = "Chowdhuri, Abhishek and Bhattacharyya, Arpan and Kumar, Shailesh",
    title = "{Prospects of detecting deviations to Kerr geometry with radiation reaction effects in EMRIs}",
    eprint = "2311.05983",
    archivePrefix = "arXiv",
    primaryClass = "gr-qc",
    doi = "10.1088/1475-7516/2024/04/001",
    journal = "JCAP",
    volume = "04",
    pages = "001",
    year = "2024"
}

@article{Antipin:2021rsh,
    author = "Antipin, Oleg and Bersini, Jahmall and Sannino, Francesco and Wang, Zhi-Wei and Zhang, Chen",
    title = "{More on the weak gravity conjecture via convexity of charged operators}",
    eprint = "2109.04946",
    archivePrefix = "arXiv",
    primaryClass = "hep-th",
    reportNumber = "RBI-ThPhys-2021-35",
    doi = "10.1007/JHEP12(2021)204",
    journal = "JHEP",
    volume = "12",
    pages = "204",
    year = "2021"
}

@article{Regge:1957td,
    author = "Regge, T. and Wheeler, J. A.",
    title = "{Stability of a Schwarzschild singularity}",
    journal = "Phys. Rev.",
    volume = "108",
    pages = "1063--1069",
    year = "1957",
    doi = "10.1103/PhysRev.108.1063"
}

@article{Zerilli:1970se,
    author = "Zerilli, F. J.",
    title = "{Effective potential for even parity Regge-Wheeler gravitational perturbation equations}",
    journal = "Phys. Rev. Lett.",
    volume = "24",
    pages = "737--738",
    year = "1970",
    doi = "10.1103/PhysRevLett.24.737"
}

@article{Kokkotas:1999bd,
    author = "Kokkotas, Kostas D. and Schmidt, Bernd G.",
    title = "{Quasi-normal modes of stars and black holes}",
    journal = "Living Rev. Rel.",
    volume = "2",
    pages = "2",
    year = "1999",
    eprint = "gr-qc/9909058",
    archivePrefix = "arXiv",
    doi = "10.12942/lrr-1999-2"
}

@article{Berti:2009kk,
    author = "Berti, Emanuele and Cardoso, Vitor and Starinets, Andrei O.",
    title = "{Quasinormal modes of black holes and black branes}",
    journal = "Class. Quant. Grav.",
    volume = "26",
    pages = "163001",
    year = "2009",
    eprint = "0905.2975",
    archivePrefix = "arXiv",
    primaryClass = "gr-qc",
    doi = "10.1088/0264-9381/26/16/163001"
}

@article{Iyer:1986np,
    author = "Iyer, S. and Will, Clifford M.",
    title = "{Black-hole normal modes: A WKB approach. I. Foundations and application of a higher-order WKB analysis of potential-barrier scattering}",
    journal = "Phys. Rev. D",
    volume = "35",
    pages = "3621--3631",
    year = "1987",
    doi = "10.1103/PhysRevD.35.3621"
}

@article{Konoplya:2003ii,
    author = "Konoplya, Roman A.",
    title = "{Quasinormal behavior of the D-dimensional Schwarzschild black hole and higher order WKB approach}",
    journal = "Phys. Rev. D",
    volume = "68",
    pages = "024018",
    year = "2003",
    eprint = "gr-qc/0303052",
    archivePrefix = "arXiv",
    doi = "10.1103/PhysRevD.68.024018"
}

@article{Leaver:1985ax,
    author = "Leaver, E. W.",
    title = "{An analytic representation for the quasi-normal modes of Kerr black holes}",
    journal = "Proc. Roy. Soc. Lond. A",
    volume = "402",
    pages = "285--298",
    year = "1985",
    doi = "10.1098/rspa.1985.0119"
}

@article{Abbott:2016nmj,
  author = "Abbott, B. P. and others",
  collaboration = "LIGO Scientific and Virgo",
  title = "{Tests of general relativity with GW150914}",
  journal = "Phys. Rev. Lett.",
  volume = "116",
  number = "22",
  pages = "221101",
  year = "2016",
  eprint = "1602.03841",
  archivePrefix = "arXiv",
  primaryClass = "gr-qc",
  doi = "10.1103/PhysRevLett.116.221101"
}

@article{Vishveshwara:1970zz,
  author = "Vishveshwara, C. V.",
  title = "{Scattering of gravitational radiation by a Schwarzschild black-hole}",
  journal = "Nature",
  volume = "227",
  pages = "936--938",
  year = "1970",
  doi = "10.1038/227936a0"
}

@article{Press:1971wr,
  author = "Press, W. H.",
  title = "{Long Wave Trains of Gravitational Waves from a Vibrating Black Hole}",
  journal = "Astrophys. J. Lett.",
  volume = "170",
  pages = "L105--L108",
  year = "1971",
  doi = "10.1086/180849"
}

@article{Chandrasekhar:1975zza,
  author = "Chandrasekhar, S. and Detweiler, S.",
  title = "{The quasi-normal modes of the Schwarzschild black hole}",
  journal = "Proc. Roy. Soc. Lond. A",
  volume = "344",
  pages = "441--452",
  year = "1975",
  doi = "10.1098/rspa.1975.0112"
}

@article{Nollert:1999ji,
  author = "Nollert, H. P.",
  title = "{Quasinormal modes: The characteristic `sound' of black holes and neutron stars}",
  journal = "Class. Quant. Grav.",
  volume = "16",
  pages = "R159--R216",
  year = "1999",
  doi = "10.1088/0264-9381/16/12/201"
}

@article{Cardoso:2016rao,
  author = "Cardoso, Vitor and Pani, Paolo",
  title = "{Tests for the existence of horizons through gravitational wave echoes}",
  journal = "Nat. Astron.",
  volume = "1",
  pages = "586--591",
  year = "2017",
  eprint = "1709.01525",
  archivePrefix = "arXiv",
  primaryClass = "gr-qc",
  doi = "10.1038/s41550-017-0225-y"
}

@article{Barack:2018yly,
  author = "Barack, Leor and others",
  title = "{Black holes, gravitational waves and fundamental physics: a roadmap}",
  journal = "Class. Quant. Grav.",
  volume = "36",
  number = "14",
  pages = "143001",
  year = "2019",
  eprint = "1806.05195",
  archivePrefix = "arXiv",
  primaryClass = "gr-qc",
  doi = "10.1088/1361-6382/ab0587"
}

@article{Schutz:1985km,
  author = "Schutz, Bernard F. and Will, Clifford M.",
  title = "{Black Hole Normal Modes: A Semianalytic Approach}",
  journal = "Astrophys. J. Lett.",
  volume = "291",
  pages = "L33--L36",
  year = "1985",
  doi = "10.1086/184453"
}

@article{Matyjasek:2017psv,
  author = "Matyjasek, Jerzy and Opala, Michal",
  title = "{Quasinormal modes of black holes: The improved semianalytic approach}",
  journal = "Phys. Rev. D",
  volume = "96",
  number = "2",
  pages = "024011",
  year = "2017",
  eprint = "1704.00361",
  archivePrefix = "arXiv",
  primaryClass = "gr-qc",
  doi = "10.1103/PhysRevD.96.024011"
}

@article{Berti:2025hly,
    author = "Berti, Emanuele and others",
    title = "{Black hole spectroscopy: from theory to experiment}",
    eprint = "2505.23895",
    archivePrefix = "arXiv",
    primaryClass = "gr-qc",
    month = "5",
    year = "2025"
}

@article{Blome:1981azp,
    author = "Blome, Hans-Joachim and Mashhoon, Bahram",
    title = "{Quasi-normal oscillations of a schwarzschild black hole}",
    doi = "10.1016/0375-9601(84)90769-2",
    journal = "Phys. Lett. A",
    volume = "100",
    number = "5",
    pages = "231--234",
    year = "1981"
}

@article{Gundlach:1993tp,
    author = "Gundlach, Carsten and Price, Richard H. and Pullin, Jorge",
    title = "{Late time behavior of stellar collapse and explosions: 1. Linearized perturbations}",
    eprint = "gr-qc/9307009",
    archivePrefix = "arXiv",
    reportNumber = "NSF-ITP-93-84",
    doi = "10.1103/PhysRevD.49.883",
    journal = "Phys. Rev. D",
    volume = "49",
    pages = "883--889",
    year = "1994"
}

@article{Israel:1967wq,
    author         = "Israel, Werner",
    title          = "{Event horizons in static vacuum space-times}",
    journal        = "Phys. Rev.",
    volume         = "164",
    pages          = "1776--1779",
    year           = "1967",
    doi            = "10.1103/PhysRev.164.1776"
}

@article{Carter:1971zc,
    author         = "Carter, Brandon",
    title          = "{Axisymmetric Black Hole Has Only Two Degrees of Freedom}",
    journal        = "Phys. Rev. Lett.",
    volume         = "26",
    pages          = "331--333",
    year           = "1971",
    doi            = "10.1103/PhysRevLett.26.331"
}

@article{Robinson:1975bv,
    author         = "Robinson, D. C.",
    title          = "{Uniqueness of the Kerr black hole}",
    journal        = "Phys. Rev. Lett.",
    volume         = "34",
    pages          = "905--906",
    year           = "1975",
    doi            = "10.1103/PhysRevLett.34.905"
}

@article{Wheeler:1971ba,
    author         = "Wheeler, John Archibald",
    title          = "{No Hair Conjecture and the Physics of Black Holes}",
    journal        = "Proc. Ecole d'Eté de Physique Théorique, Les Houches",
    volume         = "1",
    pages          = "121–208",
    year           = "1971",
    note           = "Reprinted in: Black Holes (Les Houches Lectures), C. DeWitt and B. DeWitt (eds.), Gordon and Breach, New York (1973)"
}

@article{Bekenstein:1972ky,
    author         = "Bekenstein, Jacob D.",
    title          = "{Transcendence of the law of baryon-number conservation in black hole physics}",
    journal        = "Phys. Rev. Lett.",
    volume         = "28",
    pages          = "452--455",
    year           = "1972",
    doi            = "10.1103/PhysRevLett.28.452"
}

@article{Bekenstein:1972ny,
    author         = "Bekenstein, Jacob D.",
    title          = "{Nonexistence of baryon number for static black holes}",
    journal        = "Phys. Rev. D",
    volume         = "5",
    pages          = "1239--1246",
    year           = "1972",
    doi            = "10.1103/PhysRevD.5.1239"
}

@article{Chrusciel:2012jk,
    author         = "Chruściel, Piotr T. and Costa, Joao Lopes and Heusler, Markus",
    title          = "{Stationary Black Holes: Uniqueness and Beyond}",
    journal        = "Living Rev. Rel.",
    volume         = "15",
    pages          = "7",
    year           = "2012",
    doi            = "10.12942/lrr-2012-7",
    eprint         = "1205.6112",
    archivePrefix  = "arXiv",
    primaryClass   = "gr-qc"
}

@article{Herdeiro2015,
author = {Herdeiro, Carlos A. R. and Radu, Eugen},
title = {Asymptotically flat black holes with scalar hair: A review},
journal = {International Journal of Modern Physics D},
volume = {24},
number = {09},
pages = {1542014},
year = {2015},
doi = {10.1142/S0218271815420146},
URL = {https://doi.org/10.1142/S0218271815420146},
eprint = { https://doi.org/10.1142/S0218271815420146}
}

@article{Carballo-Rubio:2025ntd,
    author = "Carballo-Rubio, Ra{\'u}l",
    title = "{Master field equations for spherically symmetric gravitational fields beyond general relativity}",
    eprint = "2507.15920",
    archivePrefix = "arXiv",
    primaryClass = "gr-qc",
    month = "7",
    year = "2025"
}

@article{Carballo-Rubio:2025fnc,
    author = "Carballo-Rubio, Ra{\'u}l and others",
    title = "{Towards a non-singular paradigm of black hole physics}",
    eprint = "2501.05505",
    archivePrefix = "arXiv",
    primaryClass = "gr-qc",
    doi = "10.1088/1475-7516/2025/05/003",
    journal = "JCAP",
    volume = "05",
    pages = "003",
    year = "2025"
}

@article{Emparan:2020znc,
    author = "Emparan, Roberto and Frassino, Antonia Micol and Way, Benson",
    title = "{Quantum BTZ black hole}",
    eprint = "2007.15999",
    archivePrefix = "arXiv",
    primaryClass = "hep-th",
    doi = "10.1007/JHEP11(2020)137",
    journal = "JHEP",
    volume = "11",
    pages = "137",
    year = "2020"
}

@article{Emparan:2021hyr,
    author = "Emparan, Roberto and Frassino, Antonia Micol and Sasieta, Martin and Toma{\v{s}}evi{\'c}, Marija",
    title = "{Holographic complexity of quantum black holes}",
    eprint = "2112.04860",
    archivePrefix = "arXiv",
    primaryClass = "hep-th",
    doi = "10.1007/JHEP02(2022)204",
    journal = "JHEP",
    volume = "02",
    pages = "204",
    year = "2022"
}

@article{Frassino:2024bjg,
    author = "Frassino, Antonia M. and Hennigar, Robie A. and Pedraza, Juan F. and Svesko, Andrew",
    title = "{Quantum Inequalities for Quantum Black Holes}",
    eprint = "2406.17860",
    archivePrefix = "arXiv",
    primaryClass = "hep-th",
    reportNumber = "IFT-UAM/CSIC-24-94",
    doi = "10.1103/PhysRevLett.133.181501",
    journal = "Phys. Rev. Lett.",
    volume = "133",
    number = "18",
    pages = "181501",
    year = "2024"
}

@article{Banados:1992wn,
    author = "Banados, Maximo and Teitelboim, Claudio and Zanelli, Jorge",
    title = "{The Black hole in three-dimensional space-time}",
    eprint = "hep-th/9204099",
    archivePrefix = "arXiv",
    reportNumber = "PRINT-92-0151 (CHILE), IASSNS-HEP-92-29",
    doi = "10.1103/PhysRevLett.69.1849",
    journal = "Phys. Rev. Lett.",
    volume = "69",
    pages = "1849--1851",
    year = "1992"
}

@article{Banados:1992gq,
    author = "Banados, Maximo and Henneaux, Marc and Teitelboim, Claudio and Zanelli, Jorge",
    title = "{Geometry of the (2+1) black hole}",
    eprint = "gr-qc/9302012",
    archivePrefix = "arXiv",
    reportNumber = "IASSNS-HEP-92-81",
    doi = "10.1103/PhysRevD.48.1506",
    journal = "Phys. Rev. D",
    volume = "48",
    pages = "1506--1525",
    year = "1993",
    note = "[Erratum: Phys.Rev.D 88, 069902 (2013)]"
}

@article{Carlip:1995qv,
    author = "Carlip, Steven",
    title = "{The (2+1)-Dimensional black hole}",
    eprint = "gr-qc/9506079",
    archivePrefix = "arXiv",
    reportNumber = "UCD-95-15",
    doi = "10.1088/0264-9381/12/12/005",
    journal = "Class. Quant. Grav.",
    volume = "12",
    pages = "2853--2880",
    year = "1995"
}

@article{Hohenegger:2025dic,
    author = "Hohenegger, Stefan",
    title = "{Probing effective black hole deformations}",
    eprint = "2508.17781",
    archivePrefix = "arXiv",
    primaryClass = "gr-qc",
    doi = "10.1140/epjc/s10052-025-15126-5",
    journal = "Eur. Phys. J. C",
    volume = "85",
    number = "12",
    pages = "1413",
    year = "2025"
}

@article{Hawking:1975vcx,
    author = "Hawking, S. W.",
    title = "{Particle creation by black holes}",
    doi = "10.1007/BF02345020",
    journal = "Commun. Math. Phys.",
    volume = "43",
    pages = "199--220",
    year = "1975",
    note = "[Erratum: Commun.Math.Phys. 46, 206 (1976)]"
}

@article{Page:1993wv,
    author = "Page, Don N.",
    title = "{Information in black hole radiation}",
    eprint = "hep-th/9306083",
    archivePrefix = "arXiv",
    doi = "10.1103/PhysRevLett.71.3743",
    journal = "Phys. Rev. Lett.",
    volume = "71",
    pages = "3743--3746",
    year = "1993"
}

@article{Poisson:1990eh,
    author = "Poisson, Eric and Israel, Werner",
    title = "{Internal structure of black holes}",
    doi = "10.1103/PhysRevD.41.1796",
    journal = "Phys. Rev. D",
    volume = "41",
    pages = "1796--1809",
    year = "1990"
}

@article{Carballo-Rubio:2018pmi,
    author = "Carballo-Rubio, Ra\'ul and Di Filippo, Francesco and Liberati, Stefano and Visser, Matt",
    title = "{On the viability of regular black holes}",
    eprint = "1805.02675",
    archivePrefix = "arXiv",
    primaryClass = "gr-qc",
    doi = "10.1007/JHEP07(2018)023",
    journal = "JHEP",
    volume = "07",
    pages = "023",
    year = "2018"
}

@article{Carballo-Rubio:2021bpr,
    author = "Carballo-Rubio, Ra\'ul and Di Filippo, Francesco and Liberati, Stefano and Pacilio, Costantino and Visser, Matt",
    title = "{Inner horizon instability and the unstable cores of regular black holes}",
    eprint = "2101.05006",
    archivePrefix = "arXiv",
    primaryClass = "gr-qc",
    doi = "10.1007/JHEP05(2021)132",
    journal = "JHEP",
    volume = "05",
    pages = "132",
    year = "2021"
}

@article{Penington:2019npb,
    author = "Penington, Geoffrey",
    title = "{Entanglement Wedge Reconstruction and the Information Paradox}",
    eprint = "1905.08255",
    archivePrefix = "arXiv",
    primaryClass = "hep-th",
    doi = "10.1007/JHEP09(2020)002",
    journal = "JHEP",
    volume = "09",
    pages = "002",
    year = "2020"
}

@article{Almheiri:2019psf,
    author = "Almheiri, Ahmed and Engelhardt, Netta and Marolf, Donald and Maxfield, Henry",
    title = "{The entropy of bulk quantum fields and the entanglement wedge of an evaporating black hole}",
    eprint = "1905.08762",
    archivePrefix = "arXiv",
    primaryClass = "hep-th",
    doi = "10.1007/JHEP12(2019)063",
    journal = "JHEP",
    volume = "12",
    pages = "063",
    year = "2019"
}

@article{Dvali:2011aa,
    author = "Dvali, Gia and Gomez, Cesar",
    title = "{Black Hole's Quantum N-Portrait}",
    eprint = "1112.3359",
    archivePrefix = "arXiv",
    primaryClass = "hep-th",
    doi = "10.1002/prop.201300001",
    journal = "Fortsch. Phys.",
    volume = "61",
    pages = "742--767",
    year = "2013"
}

@article{Dvali:2018xpy,
    author = "Dvali, Gia",
    title = "{A Microscopic Model of Holography: Survival by the Burden of Memory}",
    eprint = "1810.02336",
    archivePrefix = "arXiv",
    primaryClass = "hep-th",
    year = "2018"
}

@article{Dvali:2020wft,
    author = "Dvali, Gia and Eisemann, Lukas and Michel, Marco and Zell, Sebastian",
    title = "{Black hole metamorphosis and stabilization by memory burden}",
    eprint = "2006.00011",
    archivePrefix = "arXiv",
    primaryClass = "hep-th",
    doi = "10.1103/PhysRevD.102.103523",
    journal = "Phys. Rev. D",
    volume = "102",
    number = "10",
    pages = "103523",
    year = "2020"
}

@article{Alexandre:2024nuo,
    author = "Alexandre, Ana and Dvali, Gia and Koutsangelas, Emmanouil",
    title = "{New mass window for primordial black holes as dark matter from the memory burden effect}",
    eprint = "2402.14069",
    archivePrefix = "arXiv",
    primaryClass = "hep-ph",
    doi = "10.1103/PhysRevD.110.036004",
    journal = "Phys. Rev. D",
    volume = "110",
    number = "3",
    pages = "036004",
    year = "2024"
}

@article{Thoss:2024hsr,
    author = "Thoss, Valentin and Burkert, Andreas and Kohri, Kazunori",
    title = "{Breakdown of Hawking evaporation opens new mass window for primordial black holes as dark matter candidate}",
    eprint = "2402.17823",
    archivePrefix = "arXiv",
    primaryClass = "astro-ph.CO",
    doi = "10.1093/mnras/stae1098",
    journal = "Mon. Not. Roy. Astron. Soc.",
    volume = "532",
    number = "1",
    pages = "451--459",
    year = "2024"
}

@article{Gibbons:1987ps,
    author = "Gibbons, G. W. and Maeda, Kei-ichi",
    title = "{Black Holes and Membranes in Higher Dimensional Theories with Dilaton Fields}",
    doi = "10.1016/0550-3213(88)90006-5",
    journal = "Nucl. Phys. B",
    volume = "298",
    pages = "741--775",
    year = "1988"
}

@article{Garfinkle:1991bv,
    author = "Garfinkle, David and Horowitz, Gary T. and Strominger, Andrew",
    title = "{Charged black holes in string theory}",
    doi = "10.1103/PhysRevD.43.3140",
    journal = "Phys. Rev. D",
    volume = "43",
    pages = "3140",
    year = "1991",
    note = "[Erratum: Phys.Rev.D 45, 3888 (1992)]"
}

@article{Ashtekar:2018lag,
    author = "Ashtekar, Abhay and Olmedo, Javier and Singh, Parampreet",
    title = "{Quantum Transfiguration of Kruskal Black Holes}",
    eprint = "1806.00648",
    archivePrefix = "arXiv",
    primaryClass = "gr-qc",
    doi = "10.1103/PhysRevLett.121.241301",
    journal = "Phys. Rev. Lett.",
    volume = "121",
    number = "24",
    pages = "241301",
    year = "2018"
}

@article{Sotiriou:2008rp,
    author = "Sotiriou, Thomas P. and Faraoni, Valerio",
    title = "{f(R) Theories Of Gravity}",
    eprint = "0805.1726",
    archivePrefix = "arXiv",
    primaryClass = "gr-qc",
    doi = "10.1103/RevModPhys.82.451",
    journal = "Rev. Mod. Phys.",
    volume = "82",
    pages = "451--497",
    year = "2010"
}

@article{Clifton:2011jh,
    author = "Clifton, Timothy and Ferreira, Pedro G. and Padilla, Antonio and Skordis, Constantinos",
    title = "{Modified Gravity and Cosmology}",
    eprint = "1106.2476",
    archivePrefix = "arXiv",
    primaryClass = "astro-ph.CO",
    doi = "10.1016/j.physrep.2012.01.001",
    journal = "Phys. Rept.",
    volume = "513",
    pages = "1--189",
    year = "2012"
}

@article{Horndeski:1974wa,
    author = "Horndeski, Gregory Walter",
    title = "{Second-order scalar-tensor field equations in a four-dimensional space}",
    doi = "10.1007/BF01807638",
    journal = "Int. J. Theor. Phys.",
    volume = "10",
    pages = "363--384",
    year = "1974"
}

@article{Zwiebach:1985uq,
    author = "Zwiebach, Barton",
    title = "{Curvature Squared Terms and String Theories}",
    doi = "10.1016/0370-2693(85)91616-8",
    journal = "Phys. Lett. B",
    volume = "156",
    pages = "315--317",
    year = "1985"
}

@article{Callan:1988hs,
    author = "Callan, Curtis G., Jr. and Myers, Robert C. and Perry, Malcolm J.",
    title = "{Black Holes in String Theory}",
    doi = "10.1016/0550-3213(89)90172-7",
    journal = "Nucl. Phys. B",
    volume = "311",
    pages = "673--698",
    year = "1989"
}

@article{Gambini:2013ooa,
    author = "Gambini, Rodolfo and Pullin, Jorge",
    title = "{Loop quantization of the Schwarzschild black hole}",
    eprint = "1302.5265",
    archivePrefix = "arXiv",
    primaryClass = "gr-qc",
    doi = "10.1103/PhysRevLett.110.211301",
    journal = "Phys. Rev. Lett.",
    volume = "110",
    number = "21",
    pages = "211301",
    year = "2013"
}

@article{Starobinsky:1980te,
    author = "Starobinsky, Alexei A.",
    title = "{A New Type of Isotropic Cosmological Models Without Singularity}",
    doi = "10.1016/0370-2693(80)90670-X",
    journal = "Phys. Lett. B",
    volume = "91",
    pages = "99--102",
    year = "1980"
}

@article{Hui:2012qt,
    author = "Hui, Lam and Nicolis, Alberto",
    title = "{No-Hair Theorem for the Galileon}",
    eprint = "1202.1296",
    archivePrefix = "arXiv",
    primaryClass = "hep-th",
    doi = "10.1103/PhysRevLett.110.241104",
    journal = "Phys. Rev. Lett.",
    volume = "110",
    number = "24",
    pages = "241104",
    year = "2013"
}

@article{Nicolini:2005vd,
    author = "Nicolini, Piero and Smailagic, Anais and Spallucci, Euro",
    title = "{Noncommutative geometry inspired Schwarzschild black hole}",
    eprint = "gr-qc/0510112",
    archivePrefix = "arXiv",
    doi = "10.1016/j.physletb.2005.11.004",
    journal = "Phys. Lett. B",
    volume = "632",
    pages = "547--551",
    year = "2006"
}

@article{AyonBeato:1998ub,
    author = "Ayon-Beato, Eloy and Garcia, Alberto",
    title = "{Regular black hole in general relativity coupled to nonlinear electrodynamics}",
    eprint = "gr-qc/9911046",
    archivePrefix = "arXiv",
    doi = "10.1103/PhysRevLett.80.5056",
    journal = "Phys. Rev. Lett.",
    volume = "80",
    pages = "5056--5059",
    year = "1998"
}

@article{AyonBeato:2000zs,
    author = "Ayon-Beato, Eloy and Garcia, Alberto",
    title = "{The Bardeen model as a nonlinear magnetic monopole}",
    eprint = "gr-qc/0009077",
    archivePrefix = "arXiv",
    doi = "10.1016/S0370-2693(00)01125-4",
    journal = "Phys. Lett. B",
    volume = "493",
    pages = "149--152",
    year = "2000"
}

@article{Randall:1999vf,
    author = "Randall, Lisa and Sundrum, Raman",
    title = "{An Alternative to compactification}",
    eprint = "hep-th/9906064",
    archivePrefix = "arXiv",
    doi = "10.1103/PhysRevLett.83.4690",
    journal = "Phys. Rev. Lett.",
    volume = "83",
    pages = "4690--4693",
    year = "1999"
}

@article{Herrera-Aguilar:2015kea,
    author = "Herrera-Aguilar, Alfredo and Nucamendi, Ulises",
    title = "{Kerr black hole parameters in terms of the redshift/blueshift of photons emitted by geodesic particles}",
    eprint = "1506.05182",
    archivePrefix = "arXiv",
    primaryClass = "gr-qc",
    doi = "10.1103/PhysRevD.92.045024",
    journal = "Phys. Rev. D",
    volume = "92",
    number = "4",
    pages = "045024",
    year = "2015"
}

@article{Herrera-Aguilar:2012jhe,
    author = {Herrera-Aguilar, A. and Nucamendi, U. and Santos, E. and Corradini, O. and Álvarez, C.},
    title = {On the galactic rotation curves problem within an axisymmetric approach},
    journal = {Monthly Notices of the Royal Astronomical Society},
    volume = {432},
    number = {1},
    pages = {301-306},
    year = {2013},
    month = {04},
    issn = {0035-8711},
    doi = {10.1093/mnras/stt461},
    url = {https://doi.org/10.1093/mnras/stt461},
    eprint = {https://academic.oup.com/mnras/article-pdf/432/1/301/3899773/stt461.pdf},
}

@article{Becerril:2016qxf,
    author = "Becerril, Ricardo and Valdez-Alvarado, Susana and Nucamendi, Ulises",
    title = "{Obtaining mass parameters of compact objects from redshifts and blueshifts emitted by geodesic particles around them}",
    eprint = "1610.01718",
    archivePrefix = "arXiv",
    primaryClass = "gr-qc",
    doi = "10.1103/PhysRevD.94.124024",
    journal = "Phys. Rev. D",
    volume = "94",
    number = "12",
    pages = "124024",
    year = "2016"
}

@article{ShankarKuniyal:2017sxc,
    author = "Shankar Kuniyal, Ravi and Uniyal, Rashmi and Biswas, Anindya and Nandan, Hemwati and Purohit, K. D.",
    title = "{Null geodesics and red\textendash{}blue shifts of photons emitted from geodesic particles around a noncommutative black hole space\textendash{}time}",
    eprint = "1703.07921",
    archivePrefix = "arXiv",
    primaryClass = "gr-qc",
    doi = "10.1142/S0217751X18500987",
    journal = "Int. J. Mod. Phys. A",
    volume = "33",
    number = "16",
    pages = "1850098",
    year = "2018"
}

@article{Uniyal:2017yll,
    author = "Uniyal, Rashmi and Nandan, Hemwati and Purohit, K. D.",
    title = "{Null geodesics and observables around the Kerr\textendash{}Sen black hole}",
    eprint = "1703.07510",
    archivePrefix = "arXiv",
    primaryClass = "gr-qc",
    doi = "10.1088/1361-6382/aa9ad9",
    journal = "Class. Quant. Grav.",
    volume = "35",
    number = "2",
    pages = "025003",
    year = "2018"
}

@article{Sheoran:2017dwb,
    author = "Sheoran, Pankaj and Herrera-Aguilar, Alfredo and Nucamendi, Ulises",
    title = "{Mass and spin of a Kerr black hole in modified gravity and a test of the Kerr black hole hypothesis}",
    eprint = "1712.03344",
    archivePrefix = "arXiv",
    primaryClass = "gr-qc",
    doi = "10.1103/PhysRevD.97.124049",
    journal = "Phys. Rev. D",
    volume = "97",
    number = "12",
    pages = "124049",
    year = "2018"
}

@article{Kraniotis:2019ked,
    author = "Kraniotis, G. V.",
    title = "{Gravitational redshift/blueshift of light emitted by geodesic test particles, frame-dragging and pericentre-shift effects, in the Kerr\textendash{}Newman\textendash{}de Sitter and Kerr\textendash{}Newman black hole geometries}",
    eprint = "1912.10320",
    archivePrefix = "arXiv",
    primaryClass = "gr-qc",
    doi = "10.1140/epjc/s10052-021-08911-5",
    journal = "Eur. Phys. J. C",
    volume = "81",
    number = "2",
    pages = "147",
    year = "2021"
}

@article{Becerril:2020fek,
    author = "Becerril, Ricardo and Valdez-Alvarado, Susana and Nucamendi, Ulises and Sheoran, Pankaj and D\'avila, J. M.",
    title = "{Mass parameter and the bounds on redshifts and blueshifts of photons emitted from geodesic particle orbiting in the vicinity of regular black holes}",
    eprint = "2007.15300",
    archivePrefix = "arXiv",
    primaryClass = "gr-qc",
    doi = "10.1103/PhysRevD.103.084054",
    journal = "Phys. Rev. D",
    volume = "103",
    number = "8",
    pages = "084054",
    year = "2021"
}

@article{Banerjee:2022him,
    author = "Banerjee, Pritam and Herrera-Aguilar, Alfredo and Momennia, Mehrab and Nucamendi, Ulises",
    title = "{Mass and spin of Kerr black holes in terms of observational quantities: The dragging effect on the redshift}",
    eprint = "2203.09092",
    archivePrefix = "arXiv",
    primaryClass = "gr-qc",
    doi = "10.1103/PhysRevD.105.124037",
    journal = "Phys. Rev. D",
    volume = "105",
    number = "12",
    pages = "124037",
    year = "2022"
}

@article{Herrera-Aguilar:2018uxd,
    author = "Herrera-Aguilar, Alfredo and Lizardo-Castro, Ra\'ul and Nucamendi, Ulises",
    editor = "Zen Vasconcellos, C. A. and Hess, P. O. and Piccinelli, G. and Vargas Maga\~na, M. and Ure\~na-Lopez, L. and Gonz\'alez Felipe, R. and Boller, Th. and Gullberg, S.",
    title = "{General relativistic formulas for mass and spin of a Kerr black hole in terms of redshifts and orbital parameters}",
    eprint = "1805.10512",
    archivePrefix = "arXiv",
    primaryClass = "gr-qc",
    doi = "10.1002/asna.202113904",
    journal = "Astron. Nachr.",
    volume = "342",
    number = "1-2",
    pages = "198--204",
    year = "2021"
}

@article{Shakura:1972te,
    author = "Shakura, N. I. and Sunyaev, R. A.",
    title = "{Black holes in binary systems. Observational appearance}",
    journal = "Astron. Astrophys.",
    volume = "24",
    pages = "337--355",
    year = "1973"
}

@inproceedings{Novikov:1973kta,
    author = "Novikov, I. D. and Thorne, K. S.",
    title = "{Astrophysics and black holes}",
    booktitle = "{Les Houches Summer School of Theoretical Physics}: {Black Holes}",
    pages = "343--550",
    year = "1973"
}

@article{Bambi:2015kza,
    author = "Bambi, Cosimo",
    title = "{Testing black hole candidates with electromagnetic radiation}",
    eprint = "1509.03884",
    archivePrefix = "arXiv",
    primaryClass = "gr-qc",
    doi = "10.1103/RevModPhys.89.025001",
    journal = "Rev. Mod. Phys.",
    volume = "89",
    number = "2",
    pages = "025001",
    year = "2017"
}

@article{Tripathi:2021rqs,
    author = "Tripathi, Ashutosh and Abdikamalov, Askar B. and Ayzenberg, Dimitry and Bambi, Cosimo and Grinberg, Victoria and Liu, Honghui and Zhou, Menglei",
    title = "{Testing the Kerr black hole hypothesis with the continuum-fitting and the iron line methods: the case of GRS~1915+105}",
    eprint = "2106.10982",
    archivePrefix = "arXiv",
    primaryClass = "astro-ph.HE",
    doi = "10.1088/1475-7516/2022/01/019",
    journal = "JCAP",
    volume = "01",
    number = "01",
    pages = "019",
    year = "2022"
}

@article{Stella:1997tc,
    author = "Stella, Luigi and Vietri, Mario",
    title = "{Lense-Thirring precession and QPOS in low mass x-ray binaries}",
    eprint = "astro-ph/9709085",
    archivePrefix = "arXiv",
    doi = "10.1086/311075",
    journal = "Astrophys. J. Lett.",
    volume = "492",
    pages = "L59",
    year = "1998"
}

@article{Basile:2025zjc,
    author = "Basile, Ivano and Knorr, Benjamin and Platania, Alessia and Schiffer, Marc",
    title = "{Asymptotic safety, quantum gravity, and the swampland: A conceptual assessment}",
    eprint = "2502.12290",
    archivePrefix = "arXiv",
    primaryClass = "hep-th",
    reportNumber = "MPP-2025-37",
    doi = "10.21468/SciPostPhys.20.2.027",
    journal = "SciPost Phys.",
    volume = "20",
    number = "2",
    pages = "027",
    year = "2026"
}

@article{Basile:2021xxx,
    author = "Basile, Ivano and Platania, Alessia",
    title = "{Asymptotic Safety: Swampland or Wonderland?}",
    eprint = "2107.06897",
    archivePrefix = "arXiv",
    primaryClass = "hep-th",
    journal = "Universe",
    volume = "7",
    pages = "389",
    year = "2021"
}

@article{Eichhorn:2024xxx,
    author = "Eichhorn, Astrid and Hebecker, Arthur and Pawlowski, Jan M. and Walcher, Johannes",
    title = "{The absolute swampland}",
    eprint = "2405.20386",
    archivePrefix = "arXiv",
    primaryClass = "hep-th",
    journal = "EPL",
    volume = "149",
    pages = "39001",
    year = "2025"
}

@article{Knorr:2024xxx,
    author = "Knorr, Benjamin and Platania, Alessia",
    title = "{Unearthing the intersections: positivity bounds, weak gravity conjecture, and asymptotic safety landscapes from photon-graviton flows}",
    eprint = "2405.08860",
    archivePrefix = "arXiv",
    primaryClass = "hep-th",
    journal = "JHEP",
    volume = "03",
    pages = "003",
    year = "2025"
}

@article{deAlwis:2019aud,
    author = "de Alwis, S. and Eichhorn, Astrid and Held, Aaron and Pawlowski, Jan M. and Schiffer, Marc and Versteegen, Fleur",
    title = "{Asymptotic safety, string theory and the weak gravity conjecture}",
    eprint = "1907.07894",
    archivePrefix = "arXiv",
    primaryClass = "hep-th",
    journal = "Phys. Lett. B",
    volume = "798",
    pages = "134991",
    year = "2019"
}

@article{Banks:2010zn,
    author = "Banks, Tom and Seiberg, Nathan",
    title = "{Symmetries and Strings in Field Theory and Gravity}",
    eprint = "1011.5120",
    archivePrefix = "arXiv",
    primaryClass = "hep-th",
    journal = "Phys. Rev. D",
    volume = "83",
    pages = "084019",
    year = "2011"
}

@article{Harlow:2018tng,
    author = "Harlow, Daniel and Ooguri, Hirosi",
    title = "{Symmetries in quantum field theory and quantum gravity}",
    eprint = "1810.05338",
    archivePrefix = "arXiv",
    primaryClass = "hep-th",
    journal = "Commun. Math. Phys.",
    volume = "383",
    pages = "1669",
    year = "2021"
}

@article{Kats:2006xp,
    author = "Kats, Yevgeny and Motl, Lubos and Padi, Megha",
    title = "{Higher-order corrections to mass-charge relation of extremal black holes}",
    eprint = "hep-th/0606100",
    archivePrefix = "arXiv",
    journal = "JHEP",
    volume = "12",
    pages = "068",
    year = "2007"
}

@article{Cheung:2018cwt,
    author = "Cheung, Clifford and Liu, Junyu and Remmen, Grant N.",
    title = "{Proof of the Weak Gravity Conjecture from Black Hole Entropy}",
    eprint = "1801.08546",
    archivePrefix = "arXiv",
    primaryClass = "hep-th",
    journal = "JHEP",
    volume = "10",
    pages = "004",
    year = "2018"
}

@article{SHlinear,
    author = "Hohenegger, Stefan",
    title = "{Linear Response to Black Hole Deformations}",
    eprint = "2610.18521",
    archivePrefix = "arXiv",
    primaryClass = "gr-qc",
    month = "10",
    year = "2026"
}

@article{Israel:1967za,
    author = "Israel, Werner",
    title = "{Event horizons in static electrovac space-times}",
    doi = "10.1007/BF01645859",
    journal = "Commun. Math. Phys.",
    volume = "8",
    pages = "245--260",
    year = "1968"
}

@article{Taub:1950ez,
    author = "Taub, A. H.",
    title = "{Empty space-times admitting a three parameter group of motions}",
    doi = "10.2307/1969567",
    journal = "Annals Math.",
    volume = "53",
    pages = "472--490",
    year = "1951"
}

@article{Newman:1963yy,
    author = "Newman, E. and Tamburino, L. and Unti, T.",
    title = "{Empty space generalization of the Schwarzschild metric}",
    doi = "10.1063/1.1704018",
    journal = "J. Math. Phys.",
    volume = "4",
    pages = "915",
    year = "1963"
}

@article{Kundt:1961xxx,
    author = "Kundt, W.",
    title = "{The plane-fronted gravitational waves}",
    doi = "10.1007/BF01328918",
    journal = "Z. Phys.",
    volume = "163",
    pages = "77--86",
    year = "1961"
}

@article{Podolsky:2004xxx,
    author = "Podolsk{\'y}, J. and Belan, M.",
    title = "{Geodesic motion in the Kundt spacetimes and the character of envelope singularity}",
    doi = "10.1088/0264-9381/21/12/003",
    journal = "Class. Quant. Grav.",
    volume = "21",
    pages = "2811--2830",
    year = "2004"
}

@article{Geroch:1968ut,
    author = "Geroch, Robert P.",
    title = "{What is a singularity in general relativity?}",
    doi = "10.1016/0003-4916(68)90144-9",
    journal = "Annals Phys.",
    volume = "48",
    pages = "526--540",
    year = "1968"
}

@article{Booth:2005qc,
    author = "Booth, Ivan",
    title = "{Black hole boundaries}",
    eprint = "gr-qc/0508107",
    archivePrefix = "arXiv",
    doi = "10.1139/p05-063",
    journal = "Can. J. Phys.",
    volume = "83",
    pages = "1073--1099",
    year = "2005"
}

@article{Newman:1965my,
    author = "Newman, E. T. and Couch, E. and Chinnapared, K. and Exton, A. and Prakash, A. and Torrence, R.",
    title = "{Metric of a Rotating, Charged Mass}",
    doi = "10.1063/1.1704351",
    journal = "J. Math. Phys.",
    volume = "6",
    number = "6",
    pages = "918--919",
    year = "1965"
}

@article{Hayward:1993wb,
    author = "Hayward, Sean A.",
    title = "{General laws of black hole dynamics}",
    eprint = "gr-qc/9303006",
    archivePrefix = "arXiv",
    doi = "10.1103/PhysRevD.49.6467",
    journal = "Phys. Rev. D",
    volume = "49",
    pages = "6467--6474",
    year = "1994"
}

@article{Ashtekar:2004cn,
    author = "Ashtekar, Abhay and Krishnan, Badri",
    title = "{Isolated and dynamical horizons and their applications}",
    eprint = "gr-qc/0407042",
    archivePrefix = "arXiv",
    doi = "10.12942/lrr-2004-10",
    journal = "Living Rev. Rel.",
    volume = "7",
    pages = "10",
    year = "2004"
}

@article{Held:2019xde,
    author = "Held, Aaron and Gold, Roman and Eichhorn, Astrid",
    title = "{Asymptotic safety casts its shadow}",
    eprint = "1904.07133",
    archivePrefix = "arXiv",
    primaryClass = "gr-qc",
    doi = "10.1088/1475-7516/2019/06/029",
    journal = "JCAP",
    volume = "06",
    pages = "029",
    year = "2019"
}

@article{Gaiotto:2014kfa,
    author = "Gaiotto, Davide and Kapustin, Anton and Seiberg, Nathan and Willett, Brian",
    title = "{Generalized Global Symmetries}",
    eprint = "1412.5148",
    archivePrefix = "arXiv",
    primaryClass = "hep-th",
    doi = "10.1007/JHEP02(2015)172",
    journal = "JHEP",
    volume = "02",
    pages = "172",
    year = "2015"
}

@article{Cordova:2022ruw,
    author = "Cordova, Clay and Dumitrescu, Thomas T. and Intriligator, Kenneth and Shao, Shu-Heng",
    title = "{Snowmass White Paper: Generalized Symmetries in Quantum Field Theory and Beyond}",
    eprint = "2205.09545",
    archivePrefix = "arXiv",
    primaryClass = "hep-th",
    year = "2022"
}

@article{Benedetti:2021lxj,
    author = "Benedetti, Valentin and Casini, Horacio and Magan, Javier M.",
    title = "{Generalized symmetries of the graviton}",
    eprint = "2111.12089",
    archivePrefix = "arXiv",
    primaryClass = "hep-th",
    doi = "10.1007/JHEP05(2022)045",
    journal = "JHEP",
    volume = "05",
    pages = "045",
    year = "2022"
}

@article{Hinterbichler:2022agn,
    author = "Hinterbichler, Kurt and Hofman, Diego M. and Joyce, Austin and Mathys, Gr\'egoire",
    title = "{Gravity as a gapless phase and biform symmetries}",
    eprint = "2205.12272",
    archivePrefix = "arXiv",
    primaryClass = "hep-th",
    doi = "10.1007/JHEP02(2023)151",
    journal = "JHEP",
    volume = "02",
    pages = "151",
    year = "2023"
}

@article{Bondi:1962px,
    author = "Bondi, H. and van der Burg, M. G. J. and Metzner, A. W. K.",
    title = "{Gravitational waves in general relativity. 7. Waves from axisymmetric isolated systems}",
    doi = "10.1098/rspa.1962.0161",
    journal = "Proc. Roy. Soc. Lond. A",
    volume = "269",
    pages = "21--52",
    year = "1962"
}

@article{Sachs:1962wk,
    author = "Sachs, R. K.",
    title = "{Gravitational waves in general relativity. 8. Waves in asymptotically flat space-times}",
    doi = "10.1098/rspa.1962.0206",
    journal = "Proc. Roy. Soc. Lond. A",
    volume = "270",
    pages = "103--126",
    year = "1962"
}

@article{Brown:1986nw,
    author = "Brown, J. D. and Henneaux, M.",
    title = "{Central Charges in the Canonical Realization of Asymptotic Symmetries: An Example from Three-Dimensional Gravity}",
    doi = "10.1007/BF01211590",
    journal = "Commun. Math. Phys.",
    volume = "104",
    pages = "207--226",
    year = "1986"
}

@book{Strominger:2017zoo,
    author = "Strominger, Andrew",
    title = "{Lectures on the Infrared Structure of Gravity and Gauge Theory}",
    eprint = "1703.05448",
    archivePrefix = "arXiv",
    primaryClass = "hep-th",
    publisher = "Princeton University Press",
    year = "2018"
}

@article{Hawking:2016msc,
    author = "Hawking, Stephen W. and Perry, Malcolm J. and Strominger, Andrew",
    title = "{Soft Hair on Black Holes}",
    eprint = "1601.00921",
    archivePrefix = "arXiv",
    primaryClass = "hep-th",
    doi = "10.1103/PhysRevLett.116.231301",
    journal = "Phys. Rev. Lett.",
    volume = "116",
    number = "23",
    pages = "231301",
    year = "2016"
}

@article{Donnay:2015abr,
    author = "Donnay, Laura and Giribet, Gaston and Gonzalez, Hernan A. and Pino, Miguel",
    title = "{Supertranslations and Superrotations at the Black Hole Horizon}",
    eprint = "1511.08687",
    archivePrefix = "arXiv",
    primaryClass = "hep-th",
    doi = "10.1103/PhysRevLett.116.091101",
    journal = "Phys. Rev. Lett.",
    volume = "116",
    number = "9",
    pages = "091101",
    year = "2016"
}

@article{LISA:2017pwj,
    author = "Amaro-Seoane, Pau and others",
    collaboration = "LISA",
    title = "{Laser Interferometer Space Antenna}",
    eprint = "1702.00786",
    archivePrefix = "arXiv",
    primaryClass = "astro-ph.IM",
    year = "2017"
}

@article{LISA:2024hlh,
    author = "Colpi, Monica and others",
    collaboration = "LISA",
    title = "{LISA Definition Study Report}",
    eprint = "2402.07571",
    archivePrefix = "arXiv",
    primaryClass = "astro-ph.CO",
    year = "2024"
}

@article{LISA:2022kgy,
    author = "Arun, K. G. and others",
    collaboration = "LISA",
    title = "{New horizons for fundamental physics with LISA}",
    eprint = "2205.01597",
    archivePrefix = "arXiv",
    primaryClass = "gr-qc",
    doi = "10.1007/s41114-022-00036-9",
    journal = "Living Rev. Rel.",
    volume = "25",
    number = "1",
    pages = "4",
    year = "2022"
}

@article{Punturo:2010zz,
    author = "Punturo, M. and others",
    title = "{The Einstein Telescope: A third-generation gravitational wave observatory}",
    doi = "10.1088/0264-9381/27/19/194002",
    journal = "Class. Quant. Grav.",
    volume = "27",
    pages = "194002",
    year = "2010"
}

@article{Maggiore:2019uih,
    author = "Maggiore, Michele and others",
    title = "{Science Case for the Einstein Telescope}",
    eprint = "1912.02622",
    archivePrefix = "arXiv",
    primaryClass = "astro-ph.CO",
    doi = "10.1088/1475-7516/2020/03/050",
    journal = "JCAP",
    volume = "03",
    pages = "050",
    year = "2020"
}

@article{ET:2025xjr,
    author = "Abac, Adrian and others",
    collaboration = "ET",
    title = "{The Science of the Einstein Telescope}",
    eprint = "2503.12263",
    archivePrefix = "arXiv",
    primaryClass = "gr-qc",
    doi = "10.1088/1475-7516/2026/03/081",
    journal = "JCAP",
    volume = "03",
    pages = "081",
    year = "2026"
}

@article{Johnson:2023ynn,
    author = "Johnson, Michael D. and others",
    title = "{Key Science Goals for the Next-Generation Event Horizon Telescope}",
    eprint = "2304.11188",
    archivePrefix = "arXiv",
    primaryClass = "astro-ph.HE",
    doi = "10.3390/galaxies11030061",
    journal = "Galaxies",
    volume = "11",
    number = "3",
    pages = "61",
    year = "2023"
}

@article{Hu:2007pj,
    author = "Hu, Wayne and Sawicki, Ignacy",
    title = "{A Parameterized Post-Friedmann Framework for Modified Gravity}",
    eprint = "0708.1190",
    archivePrefix = "arXiv",
    primaryClass = "astro-ph",
    doi = "10.1103/PhysRevD.76.104043",
    journal = "Phys. Rev. D",
    volume = "76",
    pages = "104043",
    year = "2007"
}

@article{Gubitosi:2012hu,
    author = "Gubitosi, Giulia and Piazza, Federico and Vernizzi, Filippo",
    title = "{The Effective Field Theory of Dark Energy}",
    eprint = "1210.0201",
    archivePrefix = "arXiv",
    primaryClass = "hep-th",
    doi = "10.1088/1475-7516/2013/02/032",
    journal = "JCAP",
    volume = "02",
    pages = "032",
    year = "2013"
}

@article{Bellini:2014fua,
    author = "Bellini, Emilio and Sawicki, Ignacy",
    title = "{Maximal freedom at minimum cost: linear large-scale structure in general modifications of gravity}",
    eprint = "1404.3713",
    archivePrefix = "arXiv",
    primaryClass = "astro-ph.CO",
    doi = "10.1088/1475-7516/2014/07/050",
    journal = "JCAP",
    volume = "07",
    pages = "050",
    year = "2014"
}

@article{Ishak:2018his,
    author = "Ishak, Mustapha",
    title = "{Testing General Relativity in Cosmology}",
    eprint = "1806.10122",
    archivePrefix = "arXiv",
    primaryClass = "astro-ph.CO",
    doi = "10.1007/s41114-018-0017-4",
    journal = "Living Rev. Rel.",
    volume = "22",
    number = "1",
    pages = "1",
    year = "2019"
}

@article{Maharana:2025lbm,
    author = "Maharana, Swayamsiddha and Simovic, Fil and Soranidis, Ioannis and Terno, Daniel R.",
    title = "{Pad{\'e} metrics for black hole perturbations and light rings}",
    eprint = "2503.06752",
    archivePrefix = "arXiv",
    primaryClass = "gr-qc",
    doi = "10.1103/PhysRevD.111.104063",
    journal = "Phys. Rev. D",
    volume = "111",
    number = "10",
    pages = "104063",
    year = "2025"
}
